\documentclass[useAMS,usenatbib]{mn2e}
\usepackage{txfonts}
\usepackage{graphicx}
\usepackage{pdflscape}
\usepackage{threeparttable}
\def\mum{\,$\mu$m}
\def\deg{^{\circ}}
\def\sun{$_{\odot}$}
\def\Hyp{\textit{Hyper}}

\def\Her{\textit{Herschel}}

\def\Msun{\,M$_{\odot}$}

\def\12co{$^{12}$CO}
\def\co13{$^{13}$CO}
\def\NH2{N$_{\mathrm{H}_{2}}$}
\def\n2h{N$_{2}$H$^{+}$}
\def\d2n{N$_{2}$D$^{+}$}
\def\nh3{NH$_{3}$}
\def\hco{HCO$^{+}$}
\def\h20{H$_{2}$O}

\def\rms{\textit{r.m.s.}}

\def\0avir{$\alpha_{0}$}

\def\tex{T$_{ex}$}
\def\vinf{v$_{in}$}

\title{Massive 70\mum\ quiet clumps I: evidence of embedded low/intermediate-mass star formation activity} 

\author[A. Traficante, G. A. Fuller et al.]{A. Traficante$^{1,2}$\thanks{e-mail:alessio.traficante@manchester.ac.uk}, 
G.A. Fuller$^{1}$, N. Billot$^{3}$, A. Duarte-Cabral$^{4}$, M. Merello$^{2}$,
\newauthor
S. Molinari$^{2}$, N. Peretto$^{4}$ and E. Schisano$^{2}$\\
$^{1}$Jodrell Bank Centre for Astrophysics, School of Physics and Astronomy, University of Manchester, Oxford Road, 
Manchester M13 9PL, UK \\
$^{2}$IAPS - INAF, via Fosso del Cavaliere, 100, I-00133 Roma, Italy \\
$^{3}$ Observatoire astronomique de l'universit\'{e} de Geneve, Chemin des Maillettes, 51, CH-1290 Versoix, Suisse \\
$^{4}$School of Physics and Astronomy, Cardiff University, Queens Buildings, The Parade, Cardiff CF24 3AA, UK}

\begin{document}
\maketitle

\label{firstpage}

\begin{abstract}
Massive clumps, prior to the formation of any visible protostars, are the best candidates to search for the elusive massive starless cores. In this work we investigate the dust and gas properties of massive clumps selected to be 70\mum\ quiet, therefore good starless candidates. Our sample of 18 clumps has masses $300\lesssim\mathrm{M}\lesssim3000$ M\sun, radius $0.54\leq\mathrm{R}\leq1.00$ pc, surface densities $\Sigma\geq0.05$ g cm$^{-2}$ and luminosity/mass ratio L/M$\leq0.3$. We show that half of these 70\mum\ quiet clumps embed faint 24\mum\ sources. Comparison with GLIMPSE counterparts shows that 5 clumps embed young stars of intermediate stellar mass up to $\simeq5.5$ M\sun. We study the clump dynamics with observations of \n2h\ (1$-$0), HNC (1$-$0) and \hco\ (1$-$0) made with the IRAM 30m telescope. Seven clumps have blue-shifted spectra compatible with infall signatures, for which we estimate a mass accretion rate $0.04\lesssim\mathrm{\dot{M}}\lesssim2.0\times10^{-3}$ M\sun\ yr$^{-1}$, comparable with values found in high-mass protostellar regions, and free-fall time of the order of $t_{ff}\simeq3\times10^{5}$ yr. The only appreciable difference we find between objects with and without embedded 24\mum\ sources is that the infall rate appears to increase from 24\mum\ dark to 24\mum\ bright objects. We conclude that all 70\mum\ quiet objects have similar properties on clump scales, independently of the presence of an embedded protostar. Based on our data we speculate that the majority, if not all of these clumps may already embed faint, low-mass protostellar cores. If these clumps are to form massive stars, this must occur after the formation of these lower mass stars.


\end{abstract}


\section{Introduction}
Massive stars play a crucial role in the formation and gas enrichment of the hosting Galaxy, and yet the formation mechanism of these extreme objects is unclear \citep{Beuther07,Zinnecker07,Tan14}. The massive star formation begins in molecular clouds with sufficient density to form massive objects \citep{Tan14}. Some of these regions have been detected in absorption against the strong 8 and 24 \mum\ background, the so-called IRDCs \citep{Perault96,Carey98}. Massive stars form in the densest part of the natal molecular cloud, within condensations 
that are called clumps  \citep{Blitz93,Zinnecker07,Tan14}, objects with size $\simeq0.5-2$ pc \citep{Urquhart14,Traficante15b}. The most massive of these regions, with surface density in excess of $\Sigma=0.05$ 
g cm$^{-2}$ \citep{Urquhart14}, and with no signatures of on-going star formation activity are ideal massive starless clump candidates. 

These extremely young clumps may be the precursors of massive starless cores, an initial condition required in core accretion models of star formation \citep{Tan14}. It is however not well determined if massive clumps embed massive starless cores, or if they fragment into a number of low-mass cores. In a sample of 9 high-mass infared-quiet cores in Cygnus X, \citet{Duarte-Cabral13} found that 8 out of 9 of these cores are driving outflows, therefore must be protostellar. The remaining one has only a tentative outflow detection, and could potentially be in a prestellar phase.

Starless clump candidates are hard to find, in particular for their short lifetime, of the order t$\simeq10^{4}$ yr \citep{Motte07}. The identification of a statistically significant number of these candidates in the Galaxy requires unbiased surveys of the 
Galactic Plane at wavelengths which allows us to trace the cold dust envelopes of these star forming regions, which emit 
principally in the far-infrared (FIR)/sub-mm.

The ATLASGAL survey \citep{Schuller09} observed a wide portion of the I and IV quadrant of the Galactic plane at 870\mum\ and produced a survey of starless clumps in the region $10\deg\leq l\leq20\deg$, $\vert b\vert\leq1\deg$ \citep{Tackenberg12}. This survey identified 210 starless clumps, but only 14 which may form stars more massive than 20 M\sun. The search for young massive cluster (YMC) precursors in the range $20\deg\geq l\geq280\deg$, combining ATLASGAL data with methanol emission, found only 7 potential YMC candidates \citep{Urquhart13}. In the characterization of the properties of cluster progenitors combining the MALT90 \citep{Jackson13} and the ATLASGAL surveys, \citet{Contreras17} identified 24 over 1244 sources as potential starless candidates and only 1 clump with properties consistent with a YMC precursor. A recent search for young massive cluster progenitors in the Galactic center using ATLASGAL and the \h20\ southern Galactic plane survey (HOPS) found that all YMC candidates are already forming stars \citep{Longmore17}. These results are in agreement with the finding of \citet{Ginsburg12} using the Bolocam Galactic Plane survey \citep[BGPS,][]{Aguirre10}. These authors searched for massive clumps in the first quadrant and found that none was effectively starless. More recently \citet{Svoboda16} identified over 2000 starless clump candidates in the entire BGPS survey and a lack of candidates with masses in excess of $10^4$ M\sun.

A major contribution to the field comes from the \textit{Herschel} survey of the Galactic Plane, Hi-GAL \citep{Molinari10_PASP}. Hi-GAL observed the entire Plane in the wavelength range $70\leq\lambda\leq500$\mum\ allowing a direct estimation of temperature, mass and luminosity of the clumps with known distances. In particular, since the presence of a 70\mum\ source is interpreted as a signpost of protostellar activity \citep{Dunham08}, the Hi-GAL survey allows a characterization of hundreds of 70\mum\ quiet clumps, i.e. starless clump candidates \citep{Veneziani12, Elia13, Elia17}. A first search of starless clumps with Hi-GAL was carried out by \citet{Veneziani12}. These authors analyzed Hi-GAL science verification data taken in two $2\deg\times2\deg$ wide regions centered on $l=30\deg$ and $l=59\deg$ and found hundreds of starless clump candidates. In a $\simeq10\deg$ wide region of the outer Galaxy \citet{Elia13} identified 688 starless clumps, the majority of them gravitationally bound sources. 

Combining the Hi-GAL data with the comprehensive catalogue of IRDCs of \citet{Peretto09}, \citet{Traficante15b} have performed a survey of starless and protostellar clumps associated with IRDCs with known distances ($\simeq3500$) in the Galactic range $15\deg\leq l\leq55\deg$. These authors found 667 starless clump candidates with masses up to $10^{4}$ M\sun, $\simeq240$ of which with surface density $\Sigma\geq0.05$ g cm$^{-2}$, so potentially forming massive stars. 

In this paper we present a detailed study of a sample of 70\mum\ quiet clumps mostly extracted from the \citet{Traficante15b} catalogue for which we made follow-up observations in the dense molecular tracers \n2h\ (1$-$0), HNC (1$-$0) and \hco\ (1$-$0) with IRAM 30m telescope\footnote{IRAM is supported by INSU/CNRS (France), MPG (Germany) and IGN (Spain).}. A second paper in this series \citep[][submitted; hereafter, Paper II]{Traficante17_PII} is dedicated to the study of the properties of the non-thermal motions of these clumps. 

This paper is divided as follow: in Section \ref{sec:observations} we present the observations of the dust continuum and 
the line emission; in Section \ref{sec:photometry} we describe the photometry steps that we follow to 
obtain the fluxes of these clumps combining the Hi-GAL, ATLASGAL and BGPS datasets. In Section \ref{sec:dust_properties} we 
analyze the spectral energy distribution (SED) of the clumps and derive the main properties of their dust emission. In the same Section we also analyze the Mid-infrared (MIR) counterparts to identify clumps with faint 24\mum\ emission and GLIMPSE counterparts. In Section \ref{sec:line_analysis} we analyze the spectra of the dense gas tracers and we 
derive gas column densities and abundances. In Section 
\ref{sec:evolution} we identify clumps with evidence of infalling motions and
explore the relations between dust and dense gas tracer properties, comparing the properties of the 24\mum\ dark and 24\mum\ bright sources; In Section \ref{sec:conclusion} we summarize our results.

\section{Observations}\label{sec:observations}
A sample of 17 starless clump candidates has been selected from the \citet{Traficante15b} as objects with $\Sigma\geq0.05$ g cm$^{-2}$, mass $\mathrm{M}\geq300$\Msun, bolometric luminosity over envelope mass ratio $\mathrm{L/M}\leq0.3$ and very low dust temperature (T$<$15 K), indicative of very young stage of evolution (see Section \ref{sec:dust_properties}), no (or faint) emission at 70\mum\ after visual inspection of each source and no counterparts in the MSX and WISE catalogues in correspondence of the \Her\ dust column density peak. In addition, we checked for different masers emission associated with these clumps, as they are an indication of on-going star formation activity. We searched in the methanol multibeam survey \citep[MMB,][]{Green09} and found no Class II CH$_{3}$OH masers in the sources of our sample \citep{Breen15}; from the MMB survey we also searched for hydroxyl (OH) masers at 6035 MHz \citep{Avison16}, a transition often associated with high-mass star forming regions, and also found no associations. We searched for CH$_{3}$OH and OH masers using also the Arecibo surveys of \citet{Olmi14}, which is more sensitive than the MMB survey and it is targeted to identify weak masers associated with Hi-GAL high-mass objects. We found no CH$_{3}$OH masers at distances less then 100\arcsec\ from the source centroids. We found one source, 34.131+0.075, with a weak OH maser \citep[peak emission of 20 mJy, $\simeq3\sigma$ above the \rms\ of the observations made with the Arecibo telescope,][]{Olmi14}. Finally, we checked several surveys of water masers in the first quadrant \citep[][and references therein]{Merello17} and found that only one source, 23.271-0.263, has a \h20\ maser association (at $\simeq3$\arcsec\ from the source centroid), identified in the survey of \citet{Svoboda16}. Note that the source 18.787-0.286 is classified as starless in \citet{Traficante15b} catalogue and has no maser associations, although it shows a 70\mum\ counterpart. The 70\mum\ source however is faint, with a peak emission of $\simeq60$ mJy/pixel compared to a background of $\simeq130$ mJy/pixel. The clump follows all the other selection criteria and has very low dust temperature (T=10.6 K, see Section \ref{sec:dust_properties}), so we include it in the analysis. The clump embedded in the cloud SDC19.281-0.387, which follows the same criteria but it is not in the \citet{Traficante15b} catalogue, was also included in our selection. The final sample of 18 clumps presented here contains some of the most massive 70\mum\ quiet clumps observed in the Galaxy.

\subsection{Dust continuum datasets}
The dust continuum properties of the clumps have been evaluated from the Hi-GAL fluxes at 160, 250 350 and 500\mum. We combined these data with fluxes at 870\mum\ from the ATLASGAL survey \citep{Schuller09} and at 1.1 mm from the BGPS survey \citep{Aguirre10}. 

The Hi-GAL survey \citep{Molinari10_PASP} observed the whole Galactic plane ($\vert b\vert\leq1\deg$, and following the Galactic warp) at wavelengths of 70, 160, 250, 350 and 500\mum\ using both PACS \citep{Griffin10} and SPIRE \citep{Poglitsch10} instruments in parallel mode. The nominal Hi-GAL spatial resolution is $\simeq[5,10.2,18,24,34.5]$\arcsec\ at [70, 160, 250, 350, 500]\mum\ respectively. However, due to the fast scan speed mode adopted in the parallel mode, the 70 and 160\mum\ beams are degraded down to $\simeq10.2$\arcsec\ and $13.5$\arcsec\ respectively. The sensitivity is $\simeq[27,70,16,6,6]$ MJy/sr at [70, 160, 250, 350, 500]\mum\ respectively \citep[][]{Traficante11}. The data reduction follows the standard Hi-GAL data reduction pipeline \citep{Traficante11}, and the final maps have been corrected following the weighted-GLS procedure described in \citet{Piazzo11}. The maps have been calibrated in comparison with IRAS and Planck data as described in \citet{Bernard10}.

The ATLASGAL survey \citep{Schuller09} covers a total of $\simeq420$ square degrees of the Galactic Plane in the longitude range $-80\deg\leq l\leq60\deg$ and has been carried out with the LABOCA instrument installed in the APEX 12m telescope. The survey has a spatial resolution of $\simeq19.2\arcsec$ and a sensitivity of $\simeq70$ mJy/beam in the $\vert l\vert\leq60\deg$\ longitude region \citep{Csengeri14}.

The BGPS survey has covered $\simeq170$ square degrees of the inner Galaxy in the range $-10.5\deg\leq l\leq90.5\deg$, $\vert b\vert\leq1\deg$ and has mapped the emission at 1.1 mm with a spatial resolution of 33\arcsec\ and a sensitivity of 30-100 mJy/beam \citep{Aguirre10,Ginsburg13}. 

We made dedicated photometry measurements for each clump directly on the maps instead of using the existent catalogues in order to minimize the uncertainties arising from the combination of different surveys. The method is described in in Section \ref{sec:photometry}.

\subsubsection{Mid-infrared sources}
We searched for sources associated with each clump in the mid-infrared (MIR) using the Spitzer surveys of the Galactic Plane at 24\mum\ \citep[MIPSGAL,][]{Carey09}, and in the range $3.6-8$\mum\ \citep[GLIMPSE,][]{Benjamin03}. The MIPSGAL sensitivity is $\simeq$2.5 mJy/beam, while the GLIMPSE sensitivity is [0.5,0.5,2.0,5.0] at [3.6,4.5,5.8,8.0]\mum\ respectively \citep[][and references therein]{Carey09}. Details of the data reduction are in \citet{Benjamin03} and \citep{Carey09} for GLIMPSE and MIPSGAL respectively.

\subsection{Line data}  
Molecular line data were acquired at the IRAM 30m telescope in June 2014 under the project 034-14. The observations have been carried out with the On the Fly observing mode to map a $2\arcmin\times2\arcmin$ wide region which covers the entire extension of each clump. Off-positions has been chosen within 30$\arcmin$ from the source centroids and checked with single pointings to verify that they were emission-free. The EMIR receiver at 3 mm was tuned at the \n2h\ (1--0) central frequency (93.17346 GHz). This tuning includes the simultaneous observations of the HNC (1$-$0) and \hco\ 1($-$0) emission lines. The VESPA backend was tuned at the maximum spectral resolution, 20 kHz ($\simeq0.06$ km/s), to resolve the \n2h\ hyperfine components and covers only the \n2h\ (1$-$0) emission line. The Fast Fourier Transform Spectrometer (FTS) was tuned to cover a wider range of frequencies and was used to trace the HNC and \hco emission with a spectral resolution of 50 kHz  ($\simeq0.17$ km/s). The system temperature varied in the range $92\leq\mathrm{T_{sys}}\leq162$ K. The data have been reduced with the standard GILDAS CLASS\footnote{\texttt{http://www.iram.fr/IRAMFR/GILDAS}} software. The average sensitivity per channel of the reduced spectra has been evaluated after smoothing the data to $\simeq0.2$ km/s and measuring the r.m.s. in 20 emission-free channels for each source. The 1-sigma r.m.s. per $\simeq0.2$ km/s channel varies in the range $0.13\leq\mathrm{\sigma}\leq0.32$  K. The beam FWHM at this frequency is $\simeq27\arcsec$.

\section{Far infrared dust photometry}\label{sec:photometry}
The far infrared (FIR) dust photometry at wavelengths $160\leq\lambda\leq1100$\mum\ has been done using \Hyp, an enhanced aperture photometry algorithm specifically designed for crowded regions, blended sources and multi-wavelength analysis \citep{Traficante15a}. The photometry process is the same adopted in \citet{Traficante15b}. For each source, a 2d-Gaussian fit at 250\mum\ defines the clumps. The fit can vary to encompass a region of at least 1 FWHM at 250\mum\ ($18\arcsec$) and can be up to twice the 250\mum\ FWHM in each direction, to avoid the contribution from underlying filamentary structures. The FWHMs of the Gaussian fit define the aperture radius. This definition of the aperture region includes at least one 500\mum\ beam. The aperture region is used to estimate the flux at 160, 250, 350 and 500\mum\ for Hi-GAL and it is also used to estimate the flux of the ATLASGAL and BGPS counterparts directly from the maps. With this choice we estimate the flux consistently at all wavelengths.

\subsection{Hi-GAL clumps}
The Hi-GAL fluxes of the clumps in the \citet{Traficante15b} catalogue have been re-evaluated with \Hyp\ parameters tuned specifically for each clump, in order to maximize the photometry accuracy of these highly confused regions. For 6 sources, the adapted photometry coincides with the photometry of the \citet{Traficante15b} catalogue. For 9 clumps we perform a different source deblending with respect to the source catalogue. In 5 cases we deblended more sources to account for faint sources not identified in \citet{Traficante15b}, and for the other 4 cases we did not include any companion subtraction since the regions are highly confused and the background estimation dominates the emission surrounding the clump. Due to the complexity of the local background emission associated with each source, in most cases the fit reaches the maximum allowed size (FWHM=36\arcsec) along one direction. However, the \Hyp\ fit converges for all sources but one, 23.271-0.263. We manually forced the source aperture for 28.792+0.141 and 23.271-0.263. In 28.792+0.141 the \Hyp\ fit converges but the region is heavily confused and we forced the aperture to be circular. In 23.271-0.263 the automatic fit did not converge and we manually force the aperture region in order to encompass at least a $\simeq30\arcsec$ region in one direction. The fluxes we estimate differ for $\simeq25\%$ on average with the fluxes in \citet{Traficante15b}. 

The Hi-GAL fluxes have been corrected for both aperture and colour corrections as described in \citet{Traficante15b}. For the colour correction we consider a clump temperature of T=11 K, the average temperature of the clumps (see Section \ref{sec:dust_properties}).

The coordinates and photometry for all the clumps are in Table\,\ref{tab:photometry}.

\begin{center}
\begin{table*}
\centering
\begin{tabular}{c|c|c|c|c|c|c|c|c|c|c|c|c|c}
\hline
\hline
Clump & Source & RA & Dec & F$_{160\mu m}$ & F$_{250\mu m}$ & F$_{350\mu m}$ & F$_{500\mu m}$ & F$_{870\mu m}$ & F$_{1100\mu m}$ & FWHM$_{min}$ & FWHM$_{max}$	&	PA	& Deblend \\
 & & ($\deg$) & ($\deg$)  & (Jy)  & (Jy)   & (Jy)   & (Jy)   & (Jy)   & (Jy)   & (\arcsec) & (\arcsec) & ($\deg$) & \\
\hline
15.631-0.377 & 1	&	18:20:29:1	&	-15:31:26	&	1.21	& 3.98	&	4.66	&	2.68	&	0.61	&	0.29	&	28.37	&	36.00	&  126.59	&	 0	\\
18.787-0.286 & 1	&	18:26:15.3	&	-12:41:33	&	9.50	&	32.76	&	28.67	&	15.41	&	3.40	&	1.17	&	29.58	&	36.00	&	 223.89	&	 0	   \\
19.281-0.387 & 1	&	18:27:33.9	&	-12:18:17	&   11.38	&	25.53	&	21.38	&	8.38	&	2.23	&	0.88	&	36.00	&	36.00	& 90.0 &  1	\\
22.53-0.192$^{1}$ & 1	&	18:32:59.7	&	-09:20:03	& 14.61 &	33.21	&	22.02	&	8.90	&	2.43	&	...	&	22.81	&	36.00	&  263.65	&	0	 \\
22.756-0.284 & 1	&	18:33:49.1	&	-09:13:04	&  10.66	&	21.25	&	15.23	&	7.10	&	1.36	&	0.67	&	18.04	&	36.00	&  237.51	&	0	 \\
23.271-0.263 & 7	&	18:34:38.0	&	-08:40:45	&   14.77	&	34.53	&	21.09	&	7.10	&	1.93	&	1.16	&	27.36	&	30.00	&	232.83	&  0	\\
24.013+0.488 & 1	&	18:33:18.5	&	-07:42:23	&  10.81	&	33.36	&	32.25	&	13.31	&	1.97	&	1.04	&	28.86	&	36.00	&	262.17	&  1	 \\
24.528-0.136 & 1	&	18:36:31.0	&	-07:32:24	&  11.87	&	 20.93	&	21.12	&	13.80	&	3.24	&	1.38	&	27.88	&	36.00	&	193.65	&  0	\\
25.609+0.228$^{1}$ & 3	&	18:37:10.6	&	-06:23:32	&  10.13	&	26.27	&	29.51	&	17.11	&	2.34	&	...	&	36.00	&	36.00	&	90.0	&	1	   \\
25.982-0.056$^{1}$ & 1	&	18:38:54.5	&	-06:12:31	&	 11.84	&	24.95	&	16.92	&	7.05	&	1.68	&	...	&	30.51	&	36.00	&	251.49	&	 1	   \\
28.178-0.091 & 2	&	18:43:02.7	&	-04:14:52	&	 34.49	&	71.11	&	47.44	&	29.17	&	2.79	&	1.10	&	29.58	&	36.00	&	115.91	&  0	   \\
28.537-0.277 & 1	&	18:44:22.0	&	-04:01:40	&  3.01	&	19.65	&	15.67	&	5.78	&	1.81	&	0.46	&	21.48	&	36.00	&	143.86	&	1   \\
28.792+0.141 & 2	&	18:43:08.8	&	-03:36:16	&  2.44	&	11.79	&	6.49	&	3.10	&	0.69	&	0.30	&	27.05	&	27.05	&	90.0	&	1	   \\
30.357-0.837 & 2	&	18:45:40.6	&	-02:39:45	&  9.70	&	15.13	&	11.89	&	6.48	&	1.00	&	0.31	&	28.46	&	36.00	&	217.28	&  0   \\
30.454-0.135 & 1	&	18:47:24.0	&	-02:16:01	&  13.15	&	18.72	&	15.76	&	8.17	&	2.85	&	0.90	&	36.00	&	36.00	&	90.0	& 0  \\
31.946+0.076 & 2	&	18:49:22.2	&	-00:50:32	&  7.37	&	13.90	&	13.81	&	8.24	&	1.70	&	0.91	&	26.41	&	36.00	&	262.13	&  0 \\
32.006-0.51 & 1	&	18:51:34.1	&	-01:03:24	&  1.88	&	8.44	&	9.09	&	3.83	&	0.54	&	0.31	&	31.72	&	36.00	&	212.56	&	1   \\
34.131+0.075 & 2	&	18:53:21.5	&	+01:06:14	&  13.59	&	30.01	&	19.09	&	6.70	&	1.51	&	0.49	&	27.91	&	36.00	&	158.51	& 0  \\
\hline
\end{tabular}
\begin{tablenotes}
\scriptsize
\item $^{1}$ \textbf{These sources are not covered by the BGPS observations.}
\end{tablenotes}
\caption{Photometry results of the 18 clumps studied in this work. Col.1: Clump name; Col.2: source id number as in \citet{Traficante15b} catalogue; Cols. 3$-$4: Coordinates of the clump centroids obtained from the Gaussian fit done at 250\mum; Cols. 5$-$10: Clump fluxes at 160, 250, 350, 500, 870 and 1100\mum\ respectively; Cols. 11$-$13: minimum, maximum FWHMs and PA of the 2d-Gaussian fit. When the fit gives FWHM$_{min}$=FWHM$_{max}$ the source is circular and the PA if fixed to $90\deg$; Col. 14: deblend parameter. 1 means that one (or more) source companion has been deblended before measuring the clump flux.}
\label{tab:photometry}
\end{table*}
\end{center}

\subsection{ATLASGAL clump counterparts}
\begin{figure}
\centering
\includegraphics[width=8cm]{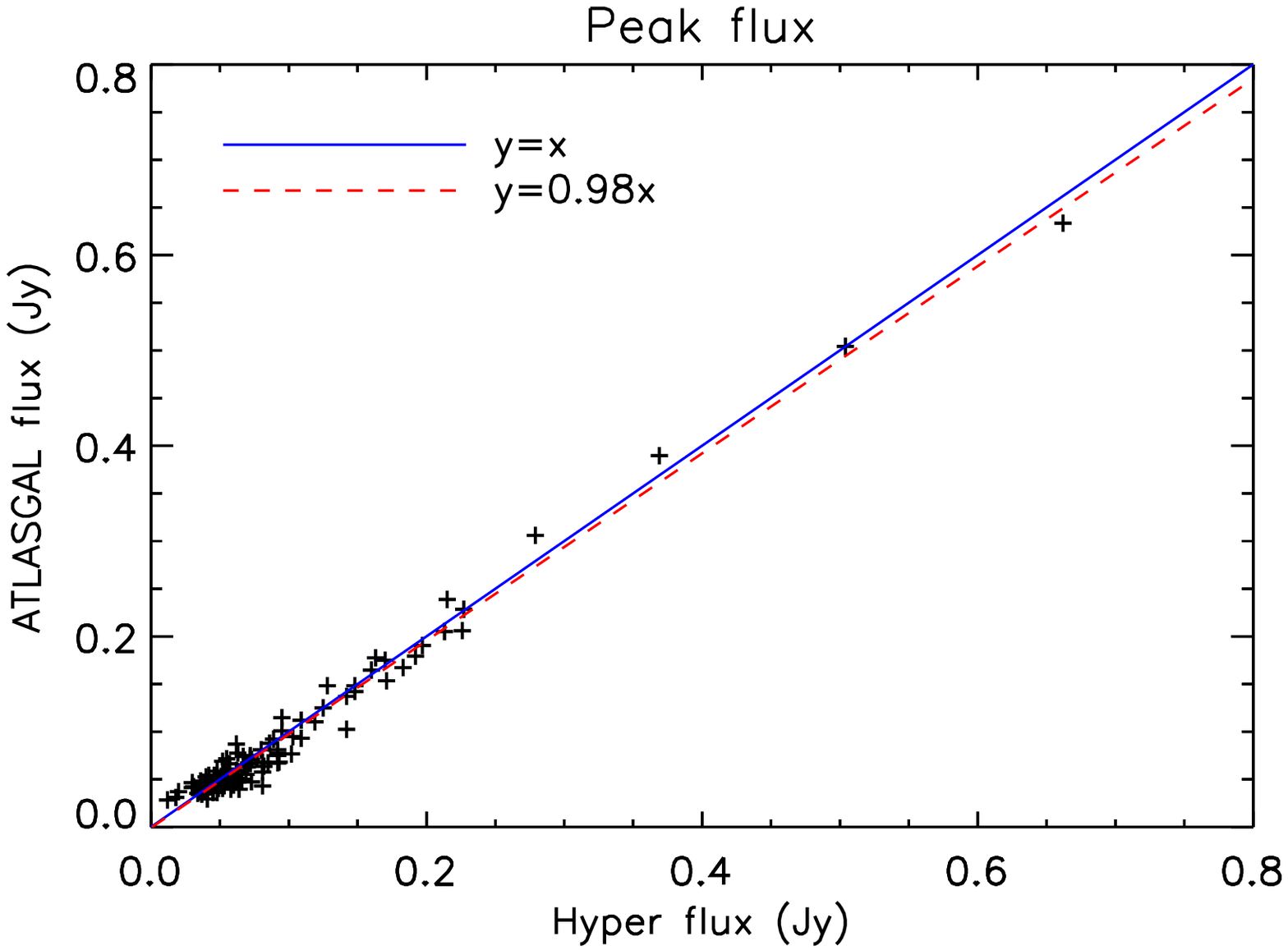} \qquad
\includegraphics[width=8cm]{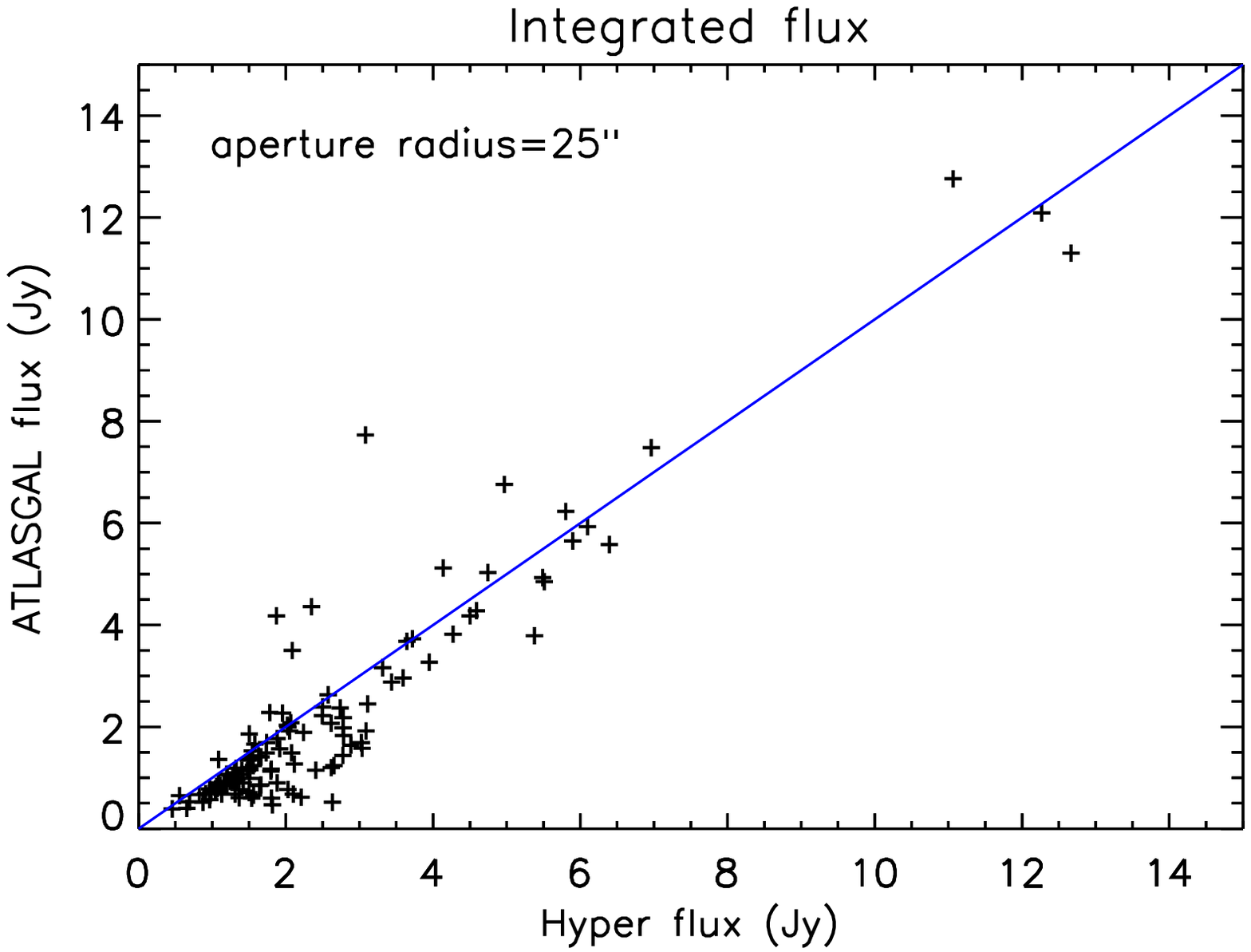} \qquad
\includegraphics[width=8cm]{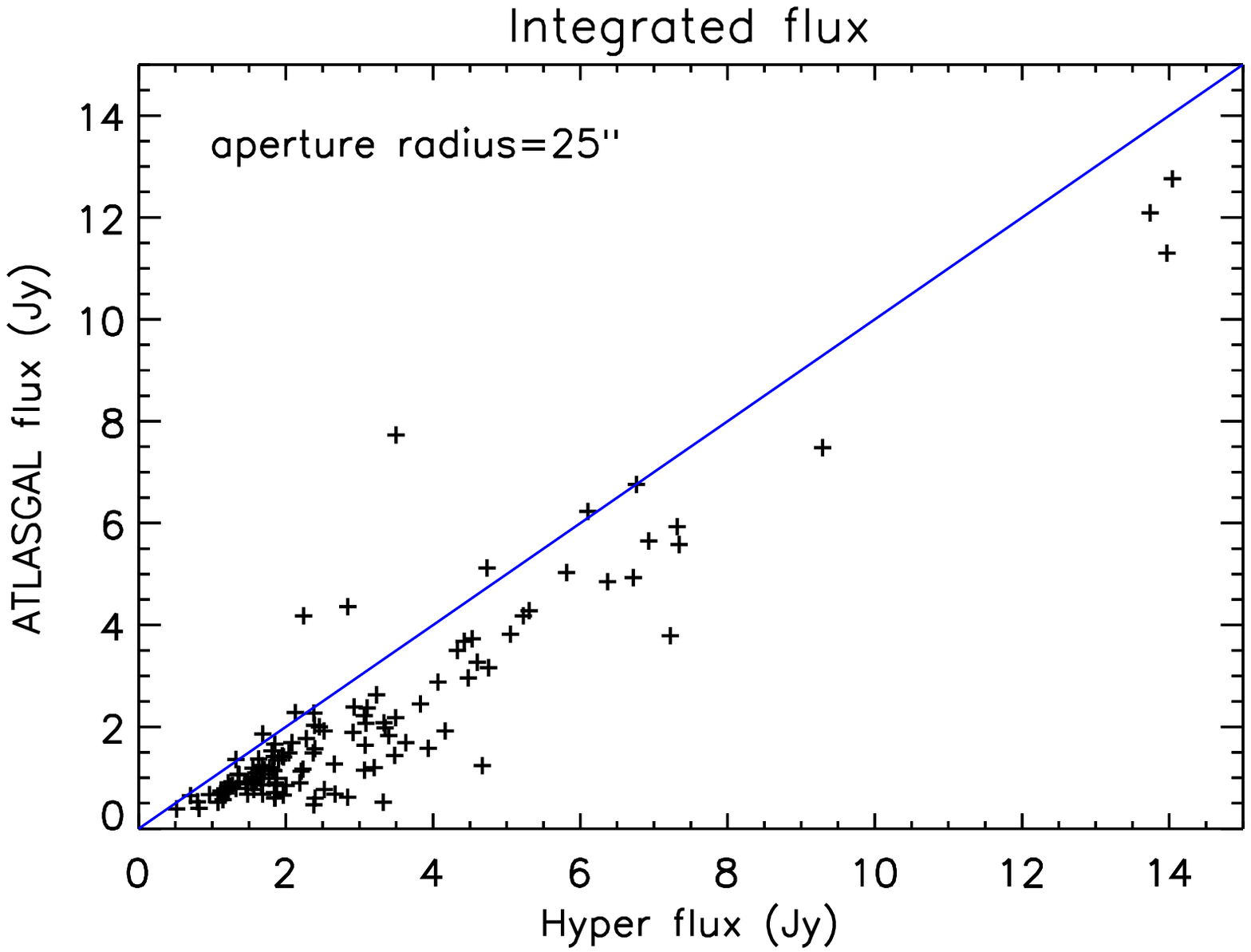}
\caption{Flux comparison between the ATLASGAL catalogue of \citet{Csengeri14} and the \Hyp\ photometry for 122 sources identified in the Galactic region  $21\deg\leq l\leq24\deg$. \textit{Upper panel}: peak flux comparison. The agreement is excellent, with a mean difference of $\simeq2\%$. The red line is the fit, and the blue line is the y=x line. \textit{Central panel}: integrated flux comparison using an aperture of 25\arcsec for the \Hyp\ photometry. The blue line is the y=x line. \textit{Lower panel}: the same of the central panel but using an aperture radius of 30\arcsec.}
\label{fig:hyper_atlasgal_peak}
\end{figure}

We evaluated the 870\mum\ fluxes from the ATLASGAL calibrated map for each source. For consistency, we compare our photometry with the fluxes presented in the ATLASGAL compact sources catalogue \citep{Csengeri14}. This catalogue contains $\simeq10000$ centrally condensed, compact objects \citep{Csengeri14}. Thirteen clumps of our sample have been identified in the ATLASGAL catalogue. The mean difference in the flux estimation of these sources between the \Hyp\ integrated fluxes and the fluxes in the ATLASGAL catalogue is $\simeq40\%$, most likely for the different approaches used to evaluate the flux. The extraction method adopted by \citet{Csengeri14}, a multi-scale wavelet filtering of the large scale structures, preserves the compact dust condensations but filters out the emission arising from scales larger than $\simeq50$\arcsec. The aperture size chosen for our clumps is up to $72\arcsec$, so it is likely that  part of the flux in the ATLASGAL catalogue has been filtered out. To test for this effect and also to check the reliability of \Hyp\ on the ATLASGAL maps we compare the \Hyp\ photometry with the photometry of the ATLASGAL catalogue in a random region of the Galactic Plane. We chose a 3 degree wide region, $21\deg\leq l\leq24\deg$, and we measure with \Hyp\ the integrated flux in a circular region of radius R=25\arcsec\ and R=30\arcsec, similar to the median of the source size of the ATLASGAL sources \citep[27\arcsec,][]{Csengeri14} and of our 18 clumps (30.8\arcsec) respectively. We identified 122 sources in common in the region $21\deg\leq l\leq24\deg$. The peak fluxes, less sensitive to the chosen algorithm, are in excellent agreement between \Hyp\ and the ATLASGAL catalogue (Figure \ref{fig:hyper_atlasgal_peak}, upper panel). The average flux difference is only $\simeq2\%$. The integrated fluxes are similar using an aperture radius of 25\arcsec\ and slightly different using an aperture radius of 30\arcsec\ (Figure \ref{fig:hyper_atlasgal_peak}, central and lower panel, respectively), with an average difference of $\simeq17\%$ and $\simeq32\%$ respectively. These tests show that the \Hyp\ photometry on the ATLASGAL maps is reliable and that part of the large-scale flux may be filtered out in the ATLASGAL catalogue. The source fluxes at 870\mum\ are in Table \ref{tab:photometry}.

\subsection{BGPS clump counterparts}
The 1100\mum\ flux has been evaluated from the BGPS maps, which cover 15 out of 18 sources. Three sources are not covered by the BGPS survey (22.53-0.192, 25.609+0.228 and 25.982-0.056). We corrected the \Hyp\ photometry for a factor 1.46, the suggested aperture correction for an aperture radius of 20 \arcsec\ \citep{Aguirre10}, since the aperture radii vary between 20\arcsec\ and 35\arcsec (see Table \ref{tab:photometry}). Ten of the fifteen sources have been identified in the most recent version of the catalogue, the BGPSv2.1, which contains 8594 compact sources \citep{Ginsburg13}. The \Hyp\ fluxes of these 10 sources in common are in good agreement with the fluxes in the BGPS catalogue, with a mean difference of $\simeq13\%$. The good agreement between the \Hyp\ photometry and the BGPS catalogue has been also showed in \citet{Traficante15a}, with mean flux differences of $\simeq9\%$. The \Hyp\ fluxes at 1.1 mm are in Table \ref{tab:photometry}.

\subsection{Photometry sources of uncertainties}
One of the major sources of uncertainties in the flux estimation arises from the background emission associated with each clump. The background identification and removal is particularly critical for Hi-GAL data, where the cold dust emission associated with the background structures has the peak of its emission \citep[e.g.][]{Peretto10}. The uncertainties in the background estimation are less important in the ATLASGAL and BGPS data, since most of the extended emission is filtered out in these ground-based experiments. Furthermore, these clumps are not isolated but found in proximity of other contaminating sources. \Hyp\ does a 2d-Gaussian modeling of the source companions which are then subtracted prior to the flux evaluation. As indicated in Table \ref{tab:photometry}, we performed the source deblending in seven clumps. Finally, another source of uncertainty arises from the comparison of surveys with different sensitivities and spatial resolutions, as showed in Table \ref{tab:sensitivities}, which could potentially affect the background estimation and the flux estimation of the clumps. Also, in ground-based experiments part of the extended emission may be filtered-out, resulting in a potential underestimation of the source fluxes. We have assumed an error on the Hi-GAL fluxes of 20\% \citep[as in][]{Traficante15b} and of 40\% on the ATLASGAL and BGPS fluxes to account for the uncertainties arising from the combination of different surveys, namely due to the different beam responses and filtering associated with space-based and ground-based surveys.

\begin{center}
\begin{table}
\centering
\begin{tabular}{c|c|c|c}
\hline
\hline
Survey & Wavelength & Sensitivity & FWHM\\
       &  (\mum) & (MJy/sr) & (\arcsec)\\
\hline
MIPSGAL & 24 & 2.7$^{1}$ & 5.9\\
Hi-GAL & 70 & 27$^{2}$ & 10.0\\
Hi-GAL & 160 & 70$^{2}$ & 13.5\\
Hi-GAL & 250  & 16$^{2}$ & 18.0\\
Hi-GAL & 350  & 6$^{2}$ &24.0\\
Hi-GAL & 500  & 6$^{2}$ & 34.5\\
ATLASGAL & 870 & 7$^{3}$ & 19.2\\
BGPS & 1100 & 2.5$^{4}$ & 33.0\\
\hline
\end{tabular}
\begin{tablenotes}
\scriptsize
\item $^{1}$ adapted from \citet{Carey09}
\item $^{2}$ adapted from \citet{Traficante11}
\item $^{3}$ adapted from \citet{Csengeri14}
\item $^{4}$ adapted from \citet{Ginsburg13}
\end{tablenotes}

\caption{Sensitivity and FWHMs comparison between the surveys used in this work: MIPSGAL, Hi-GAL, ATLASGAL and the BGPS.}
\label{tab:sensitivities}
\end{table}
\end{center}

\section{Dust properties}\label{sec:dust_properties}
We adopted a single-temperature greybody model to fit the source SEDs. The model assumes that the temperature gradient across the clump is small due to the absence of significance internal sources and strong interstellar radiation fields outside the clump. However, the error associated with the mass estimation assuming a single-temperature greybody model instead of solving a complete radiative transfer model is negligible for starless clumps \citep{Wilcock11}. 

The source flux $S_{\nu}$ at frequency $\nu$ is: 

\begin{equation}
 S_{\nu}=\frac{\mathrm{M}\kappa_{0}}{d^2}\bigg(\frac{\nu}{\nu_{0}}\bigg)^{\beta}B_{\nu}(T)\Omega
\end{equation}

where M is the source mass, $d$ the heliocentric distance, $\kappa_{0}$ the dust mass absorption coefficient at reference frequency $\nu_{0}=230$ GHz, fixed to 0.5 g cm$^{-2}$ and assuming a gas/dust mass ratio of 100 \citep{Preibisch93}. $B_{\nu}(T)$ is the blackbody value at temperature $T$ and frequency $\nu$, and $\Omega$ is the solid angle of the source. The free parameters of the fit are mass and temperature. We fixed the spectral index $\beta$ for all the sources to $\beta=2.0$, appropriate for dense, cold clumps \citep[e.g.][]{Sadavoy13}. 

\begin{figure*}
\centering
\includegraphics[width=5cm]{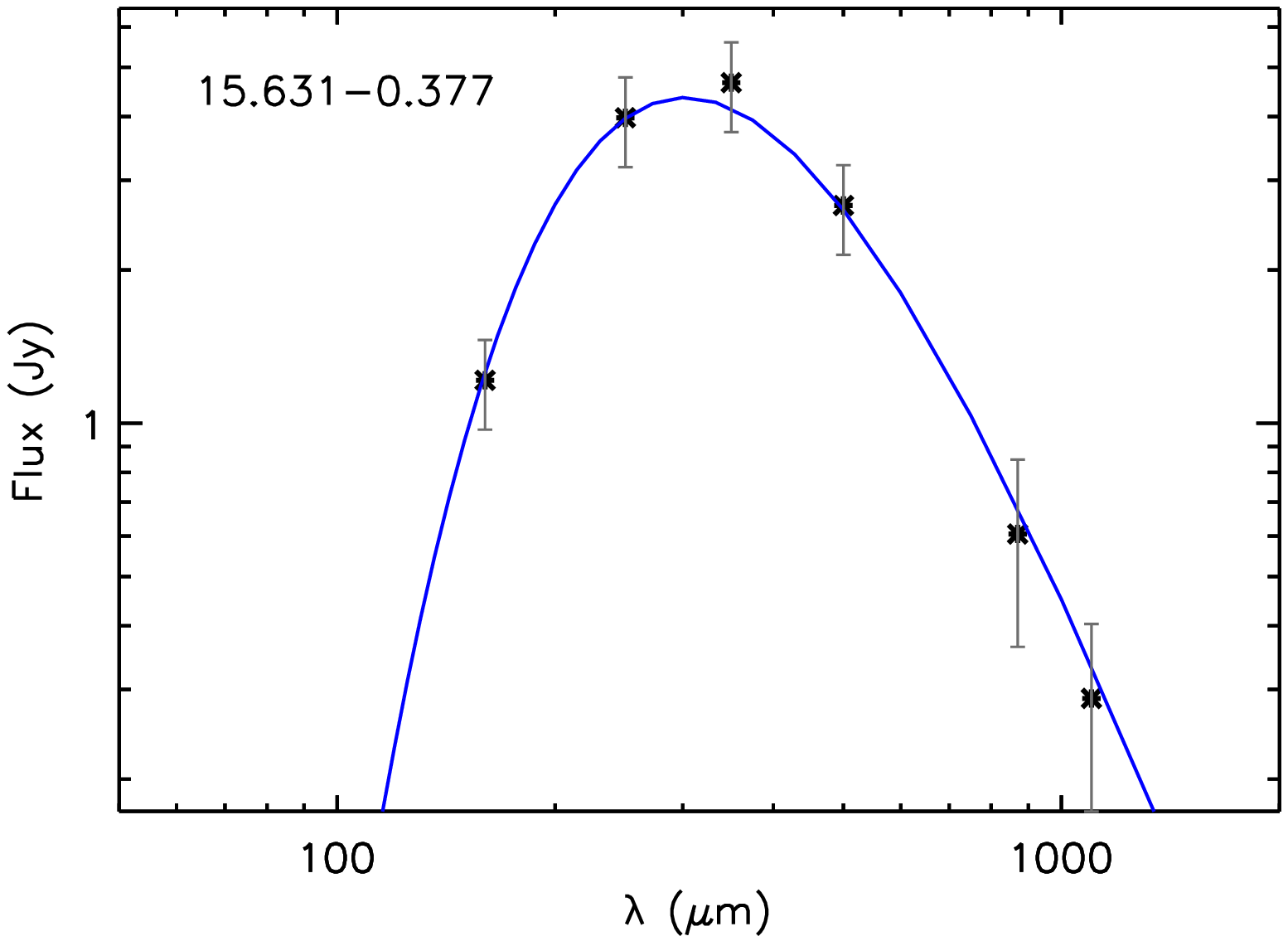}
\includegraphics[width=5cm]{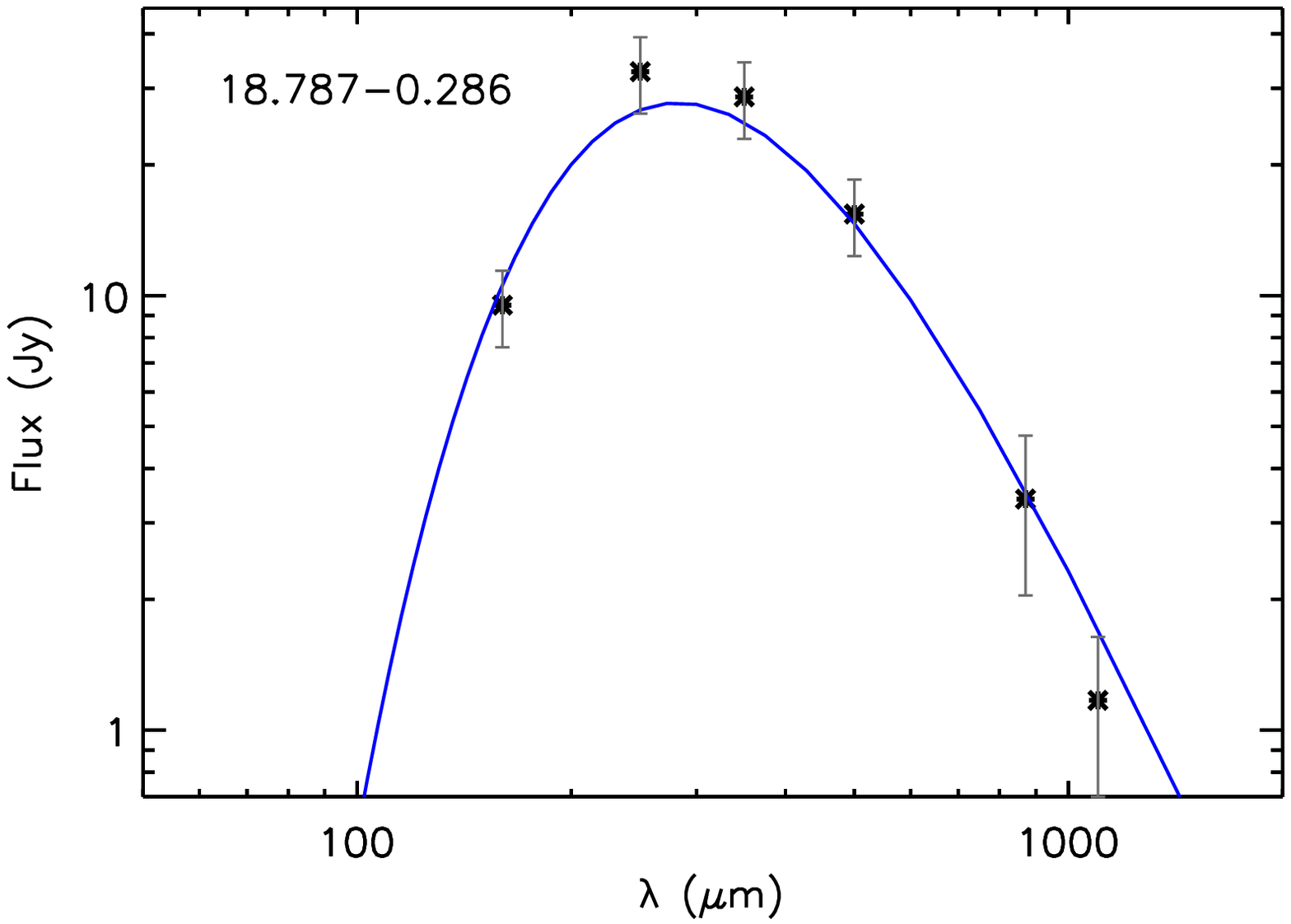} 
\includegraphics[width=5cm]{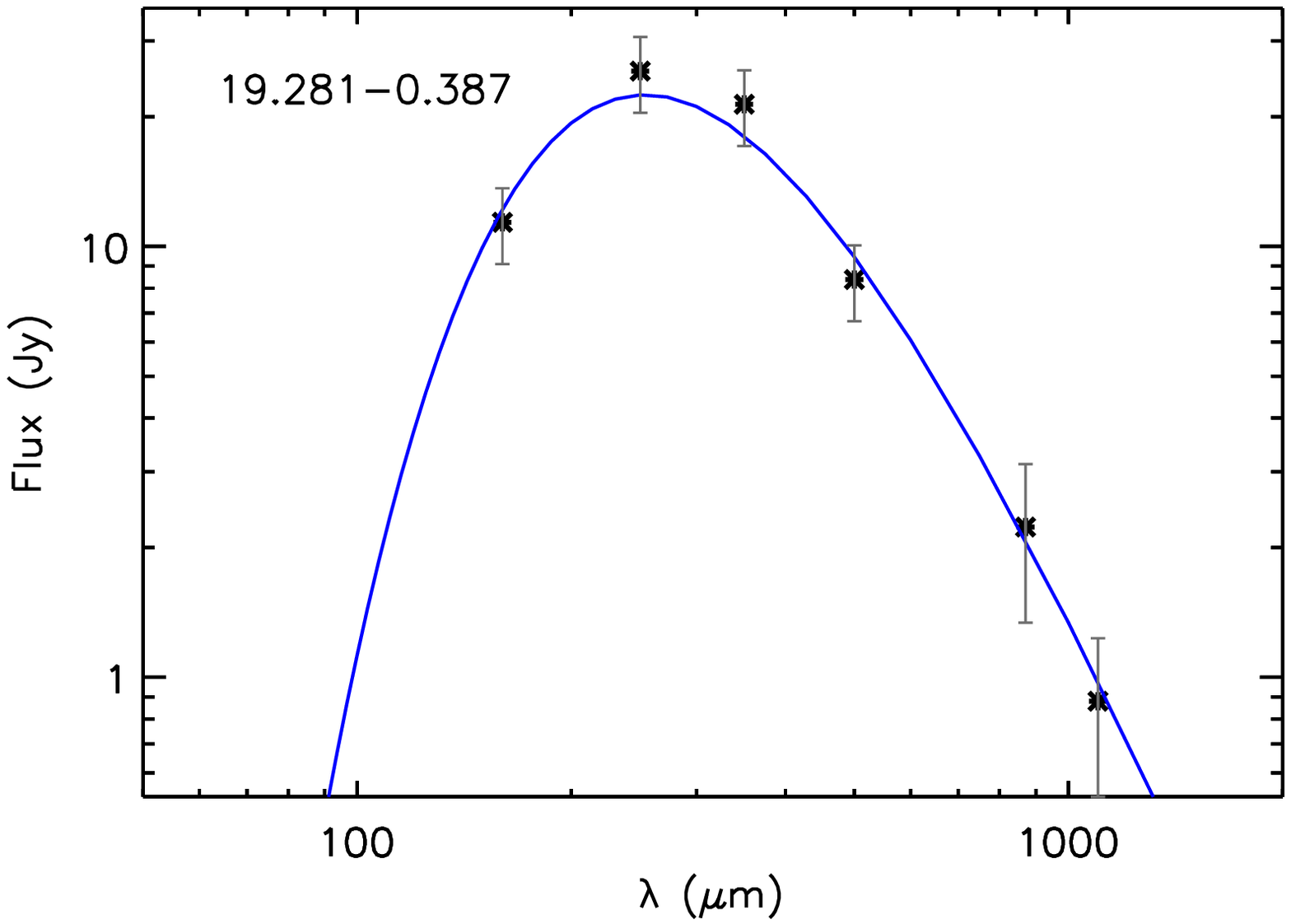}
\includegraphics[width=5cm]{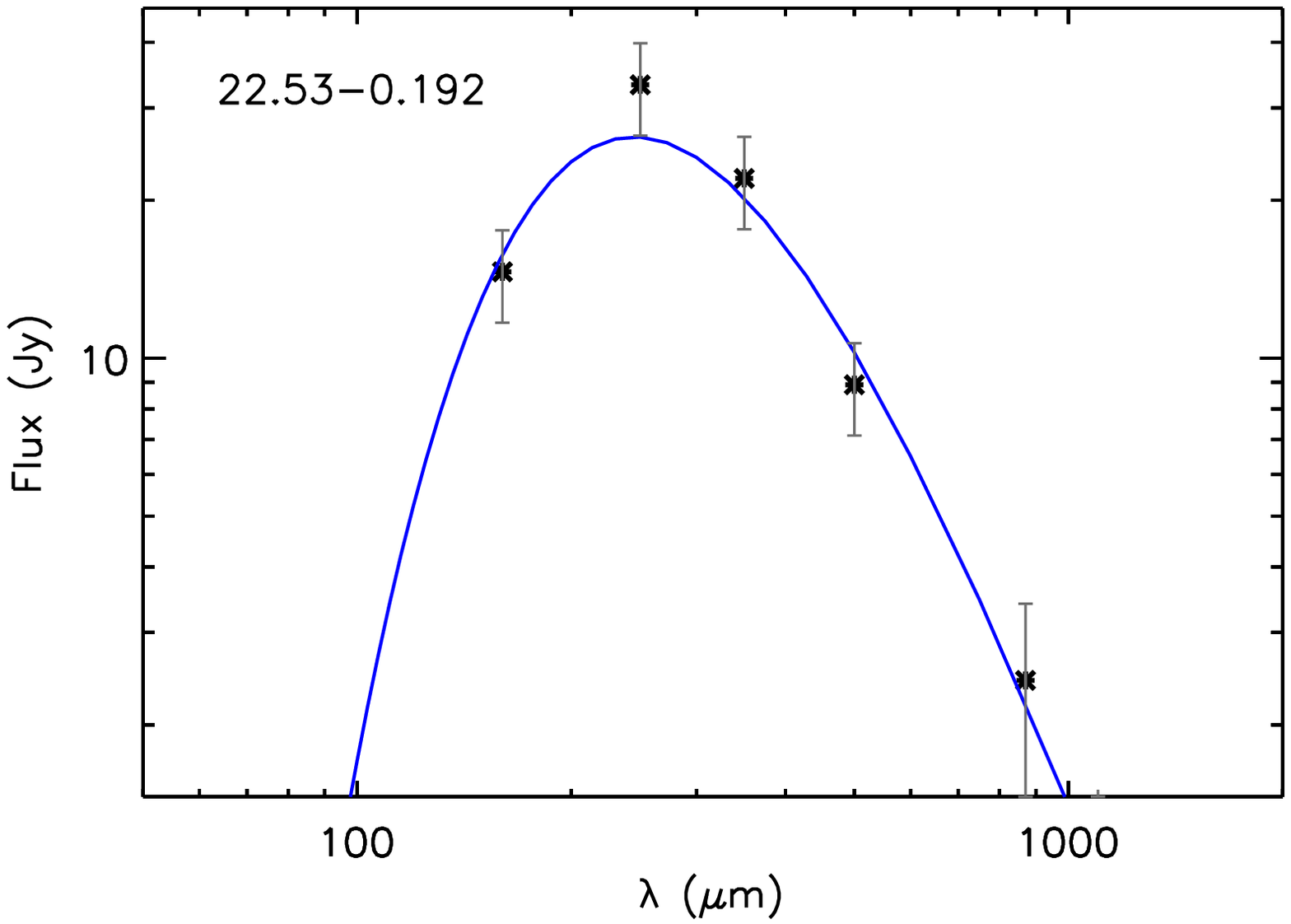} 
 \includegraphics[width=5cm]{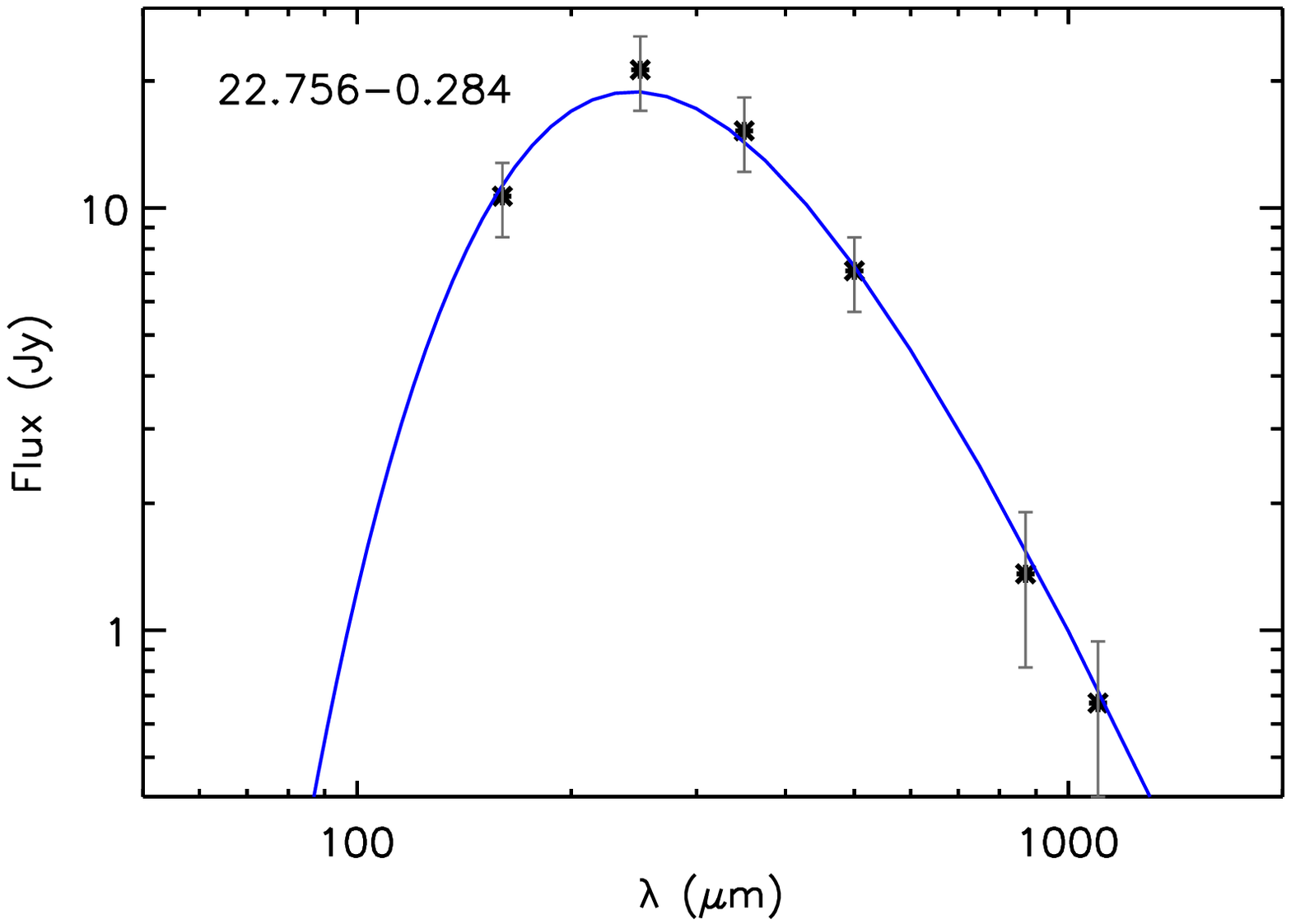} 
\includegraphics[width=5cm]{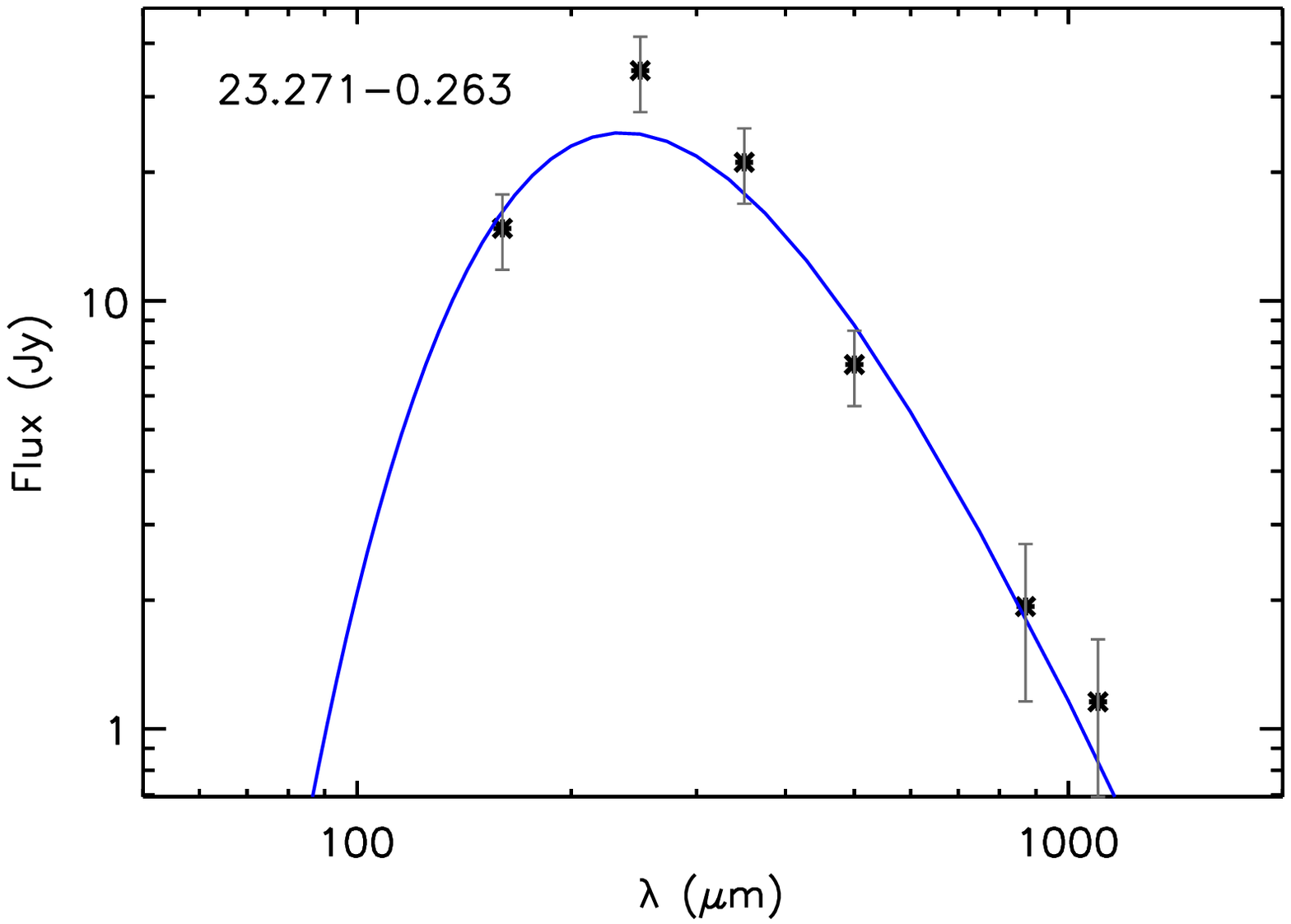}
\includegraphics[width=5cm]{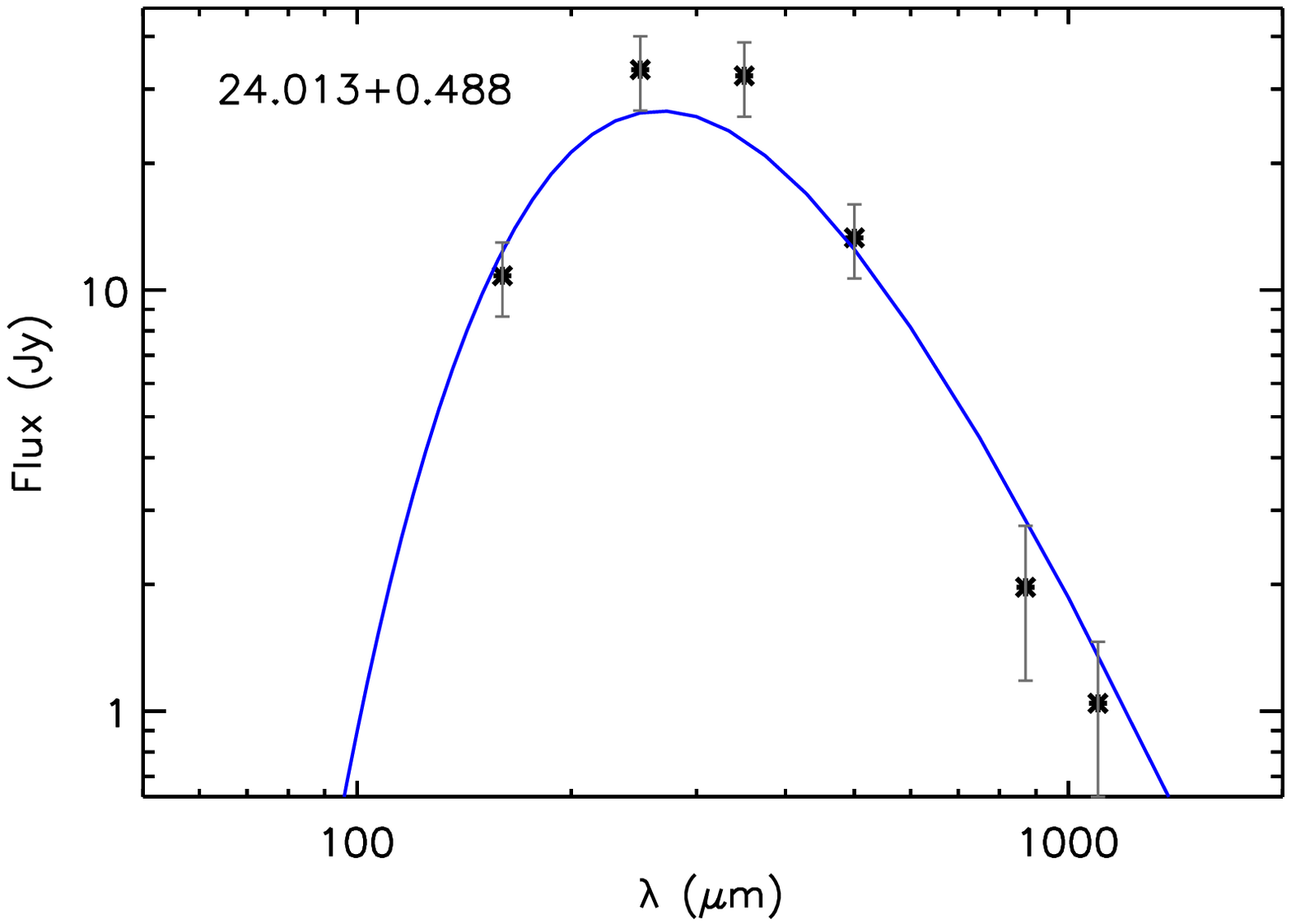}
\includegraphics[width=5cm]{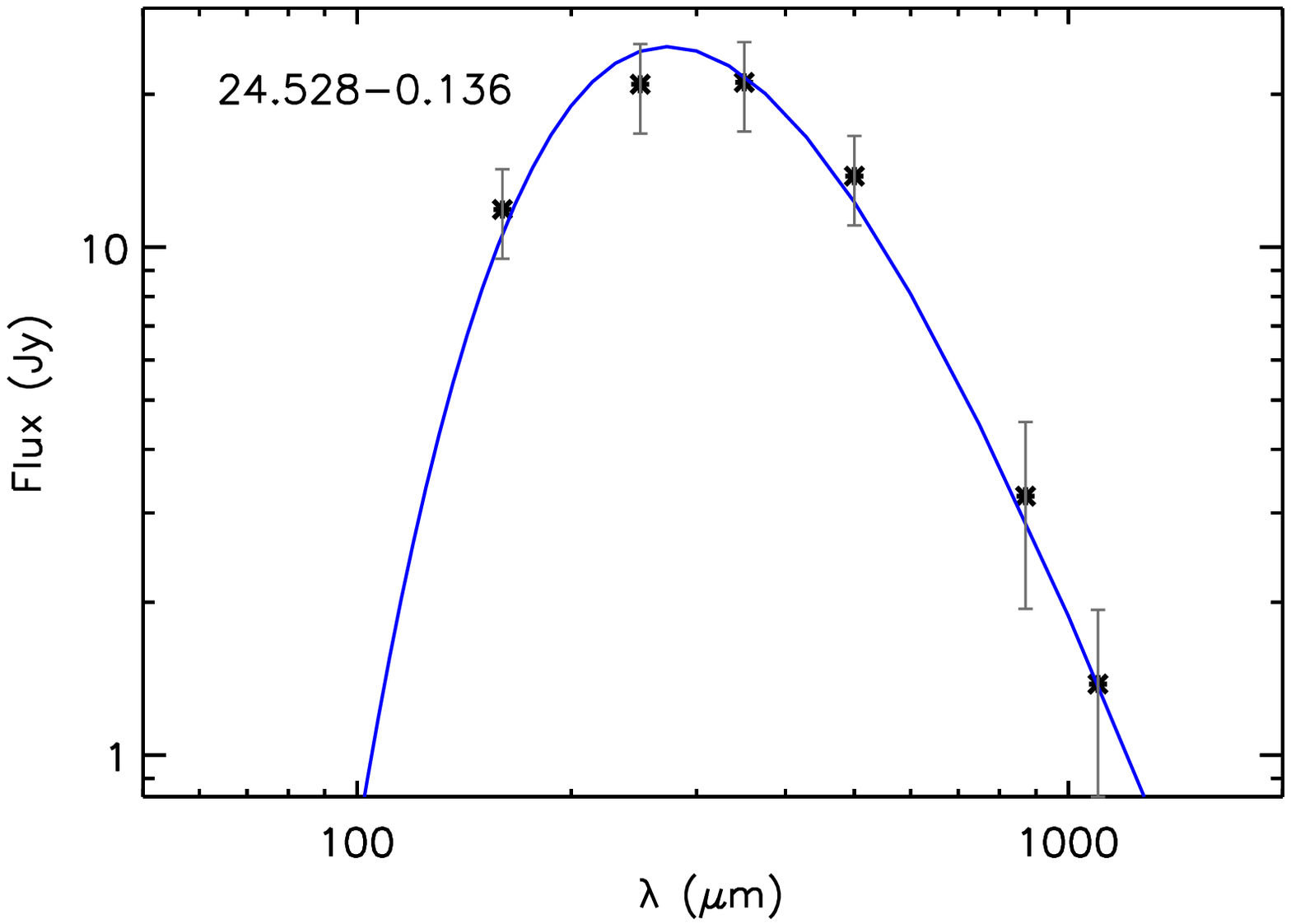} 
 \includegraphics[width=5cm]{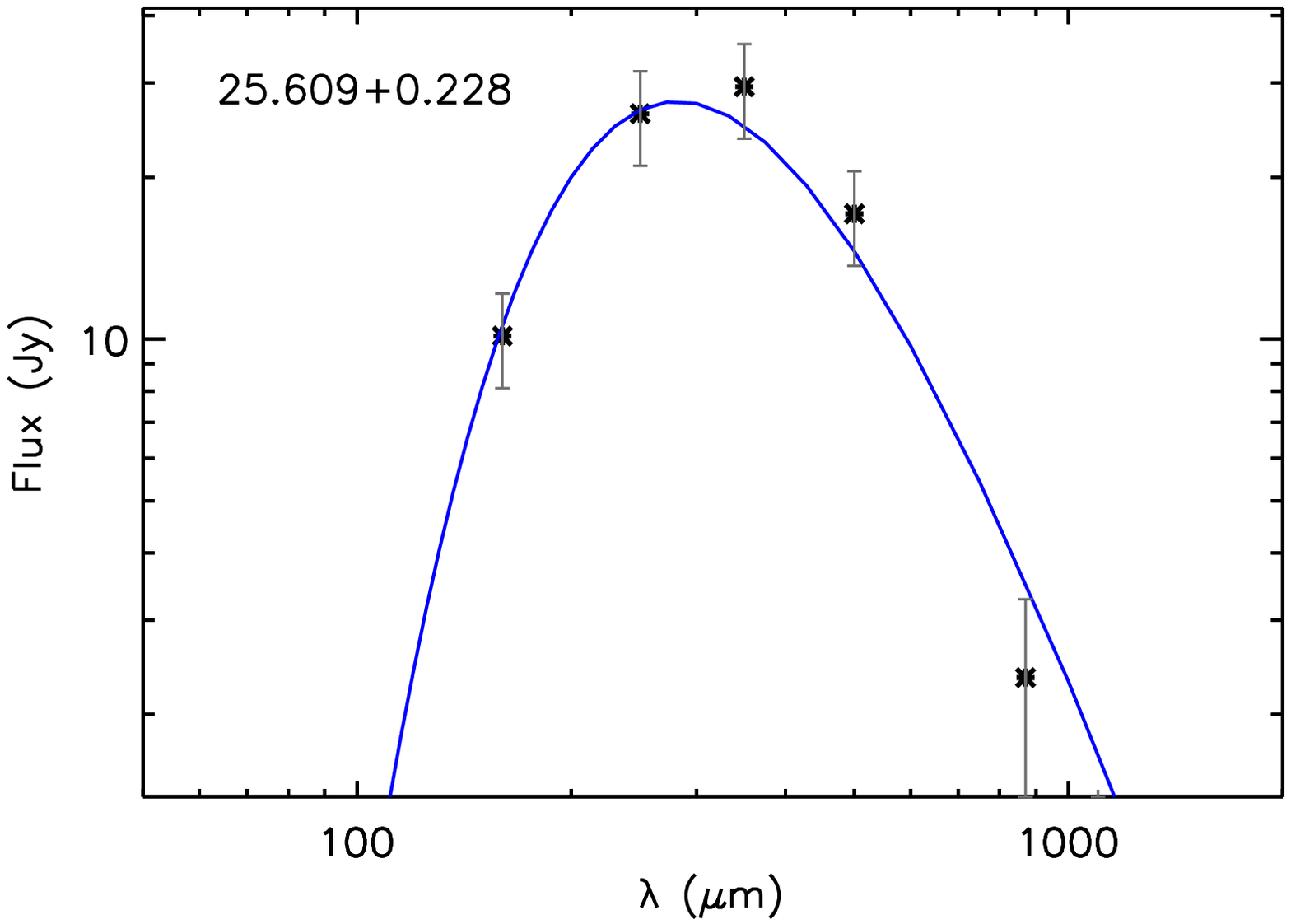}
 \includegraphics[width=5cm]{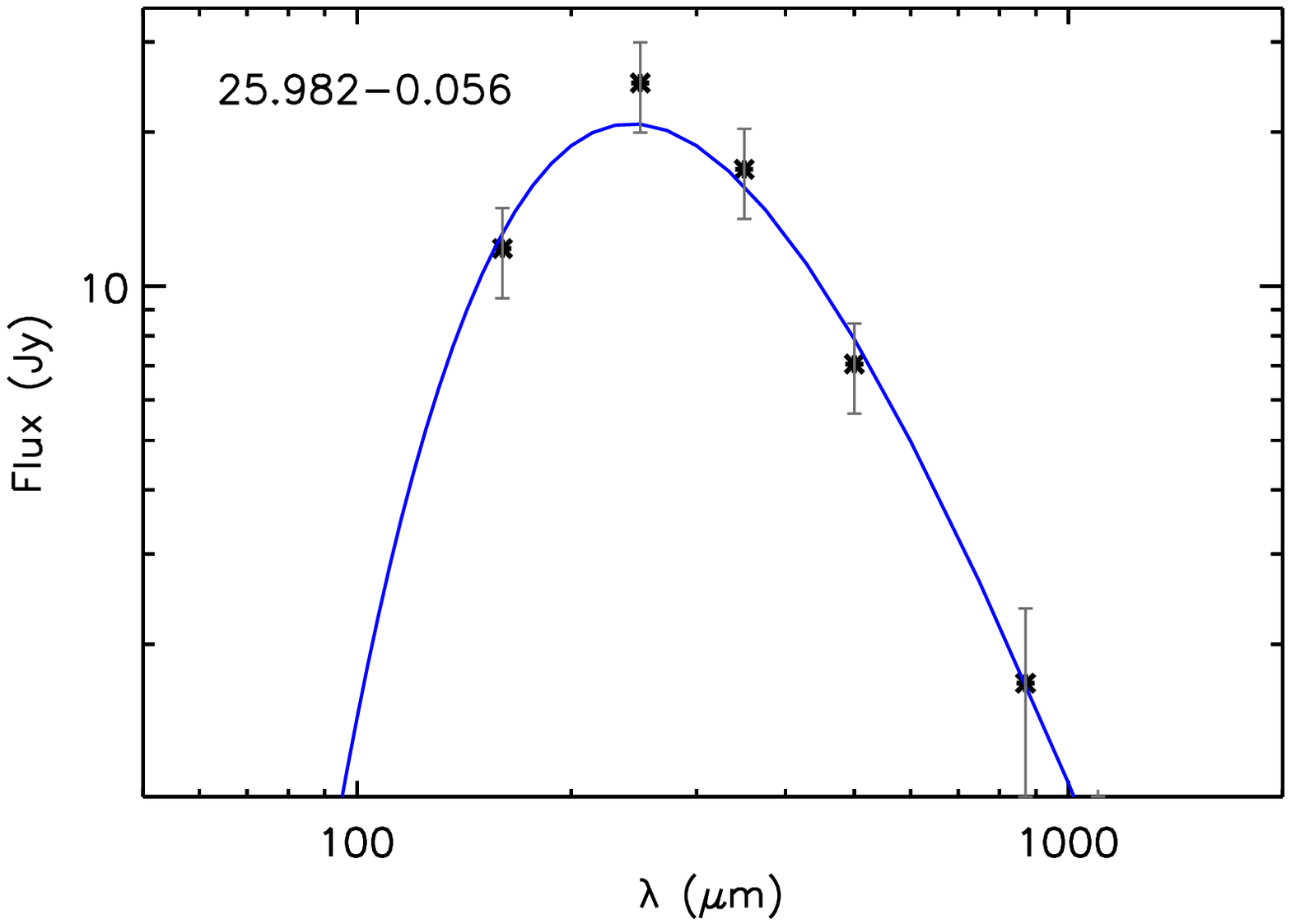}
 \includegraphics[width=5cm]{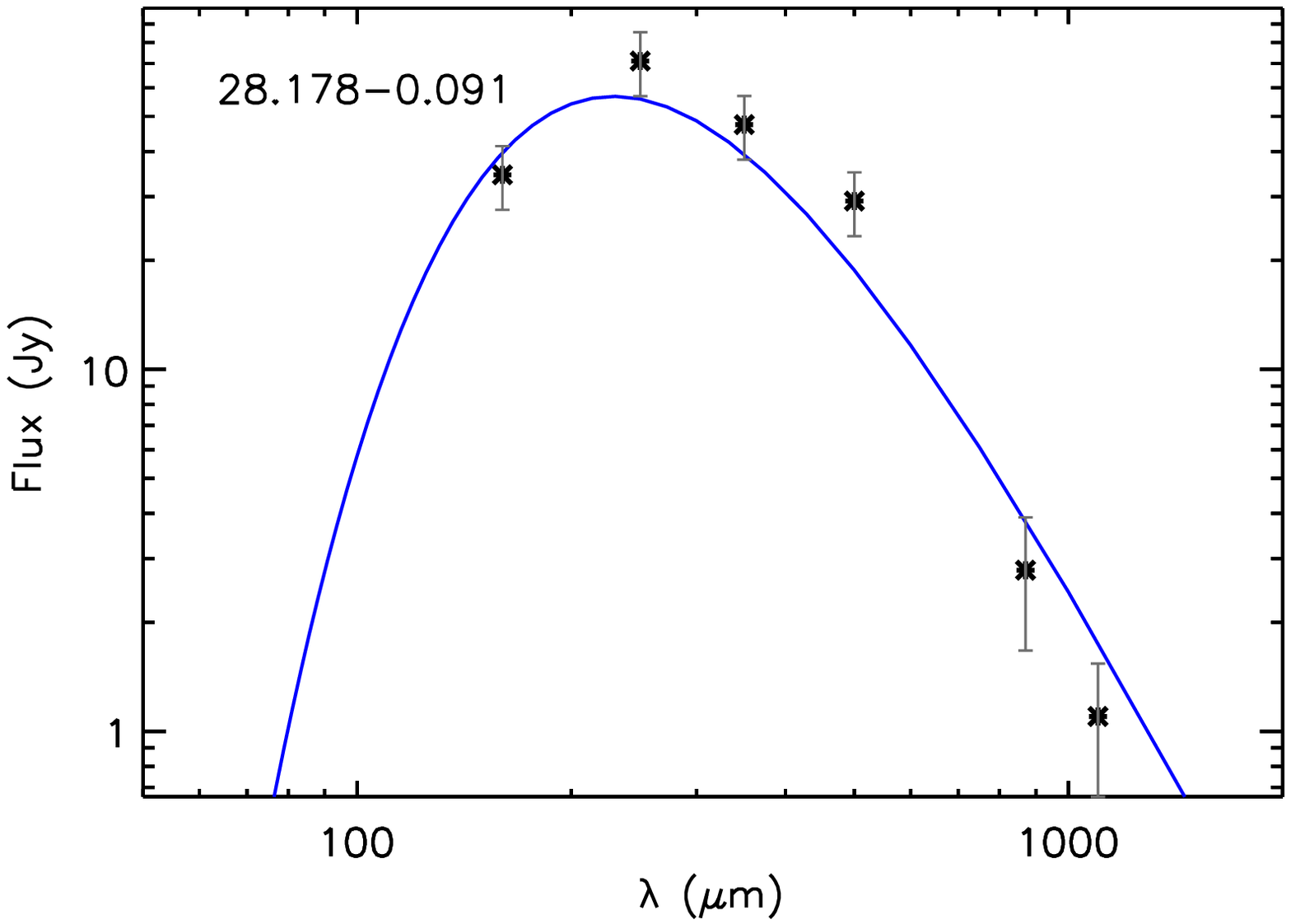} 
\includegraphics[width=5cm]{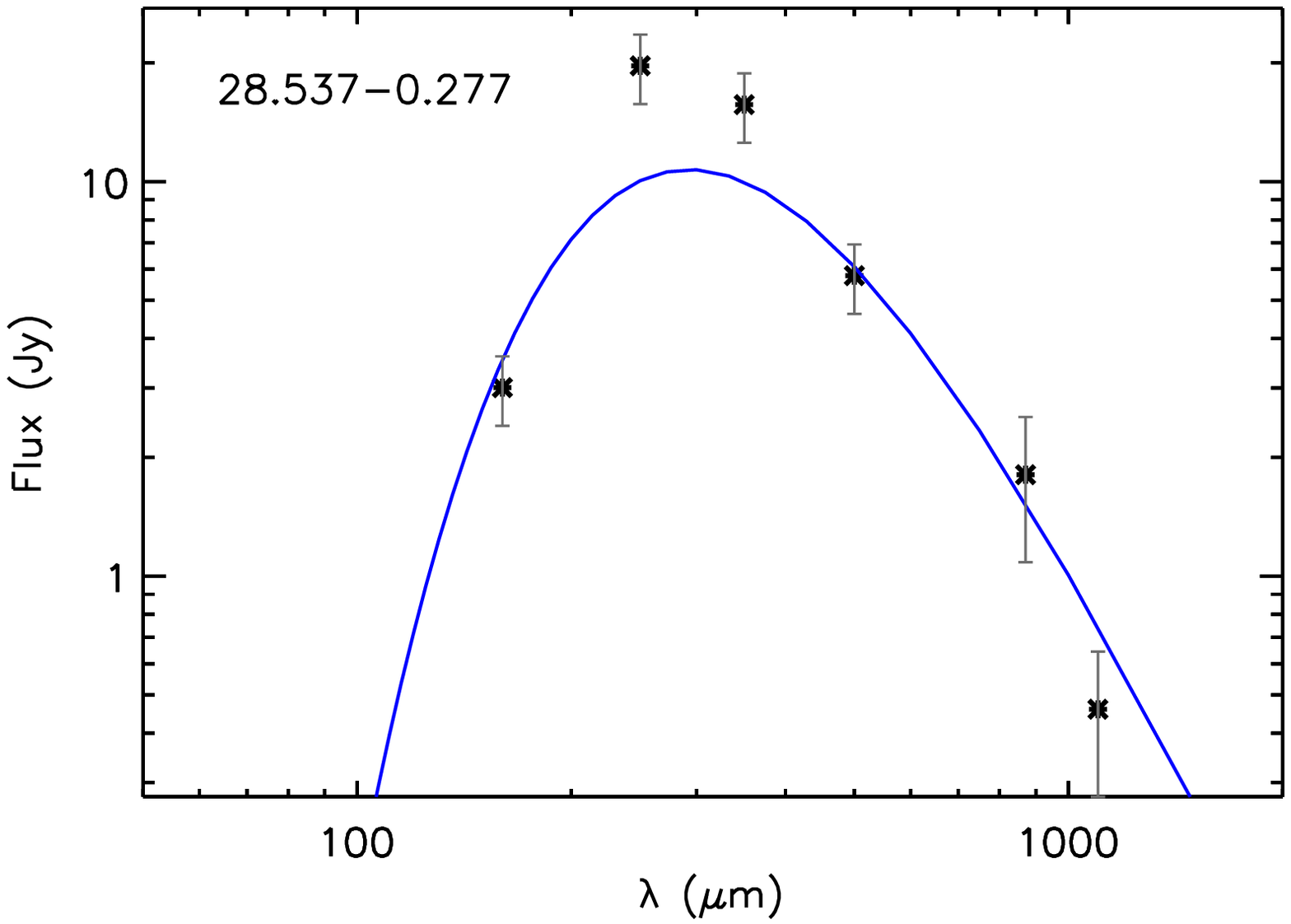} \includegraphics[width=5cm]{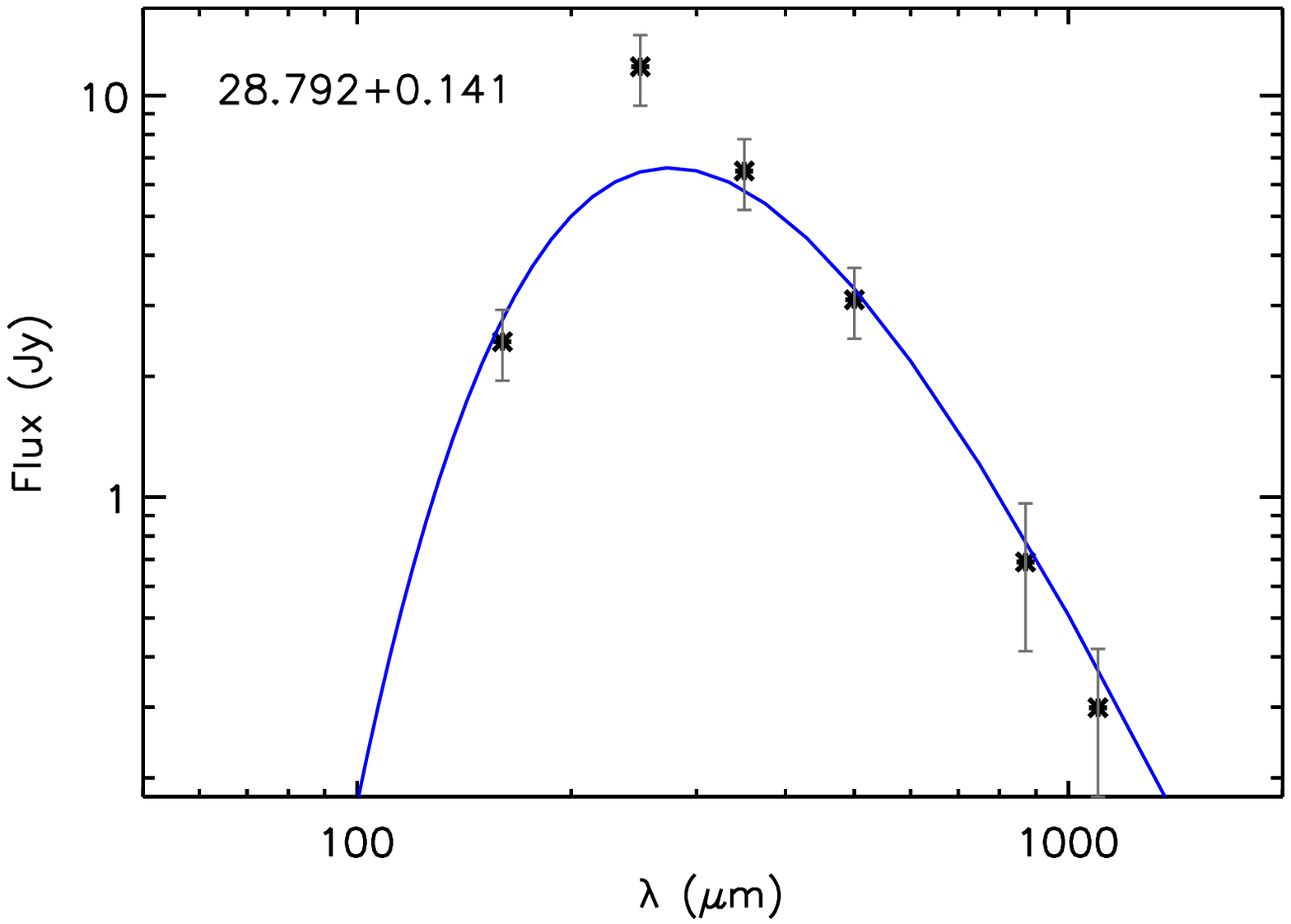} \includegraphics[width=5cm]{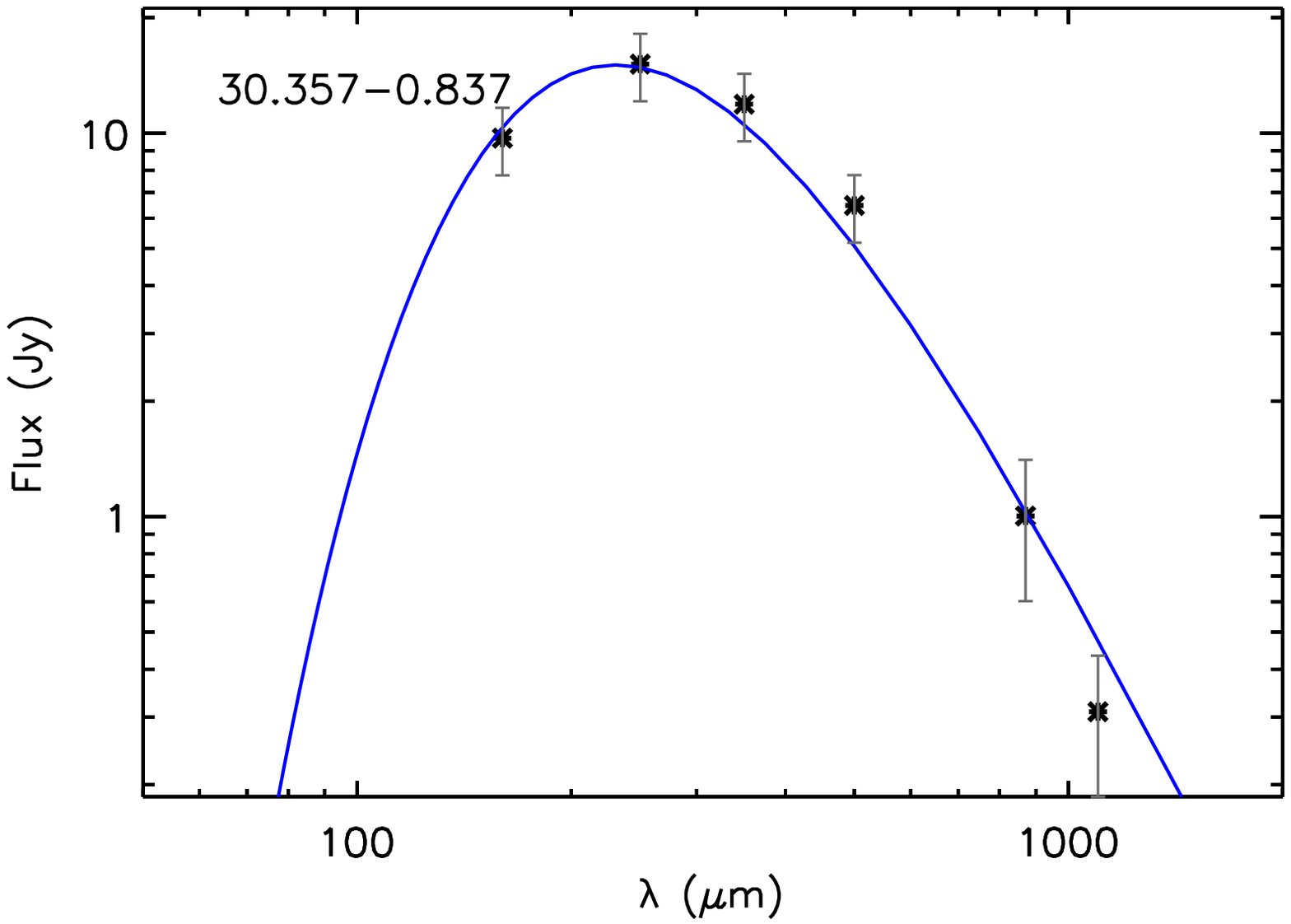} 
\includegraphics[width=5cm]{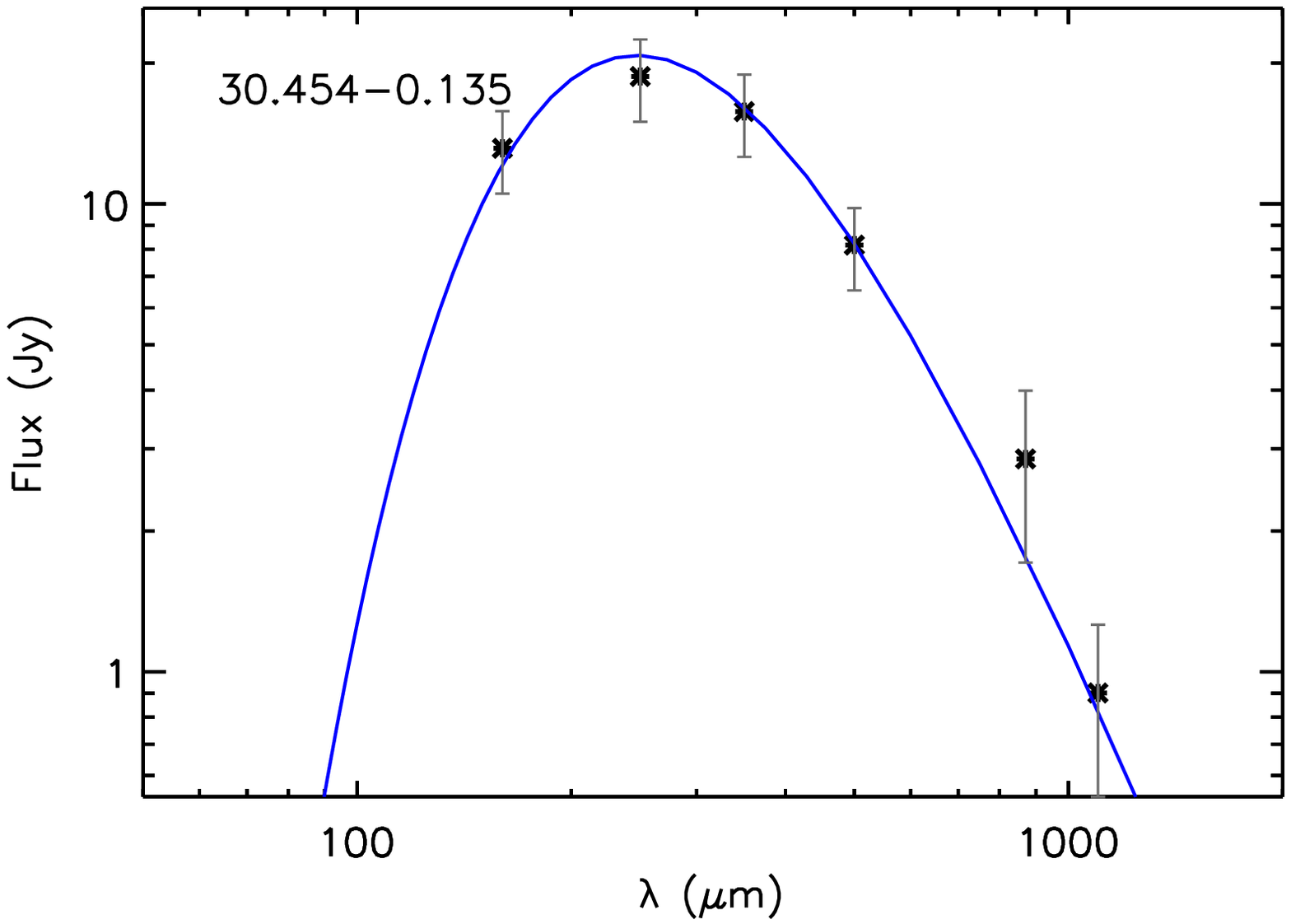} 
\includegraphics[width=5cm]{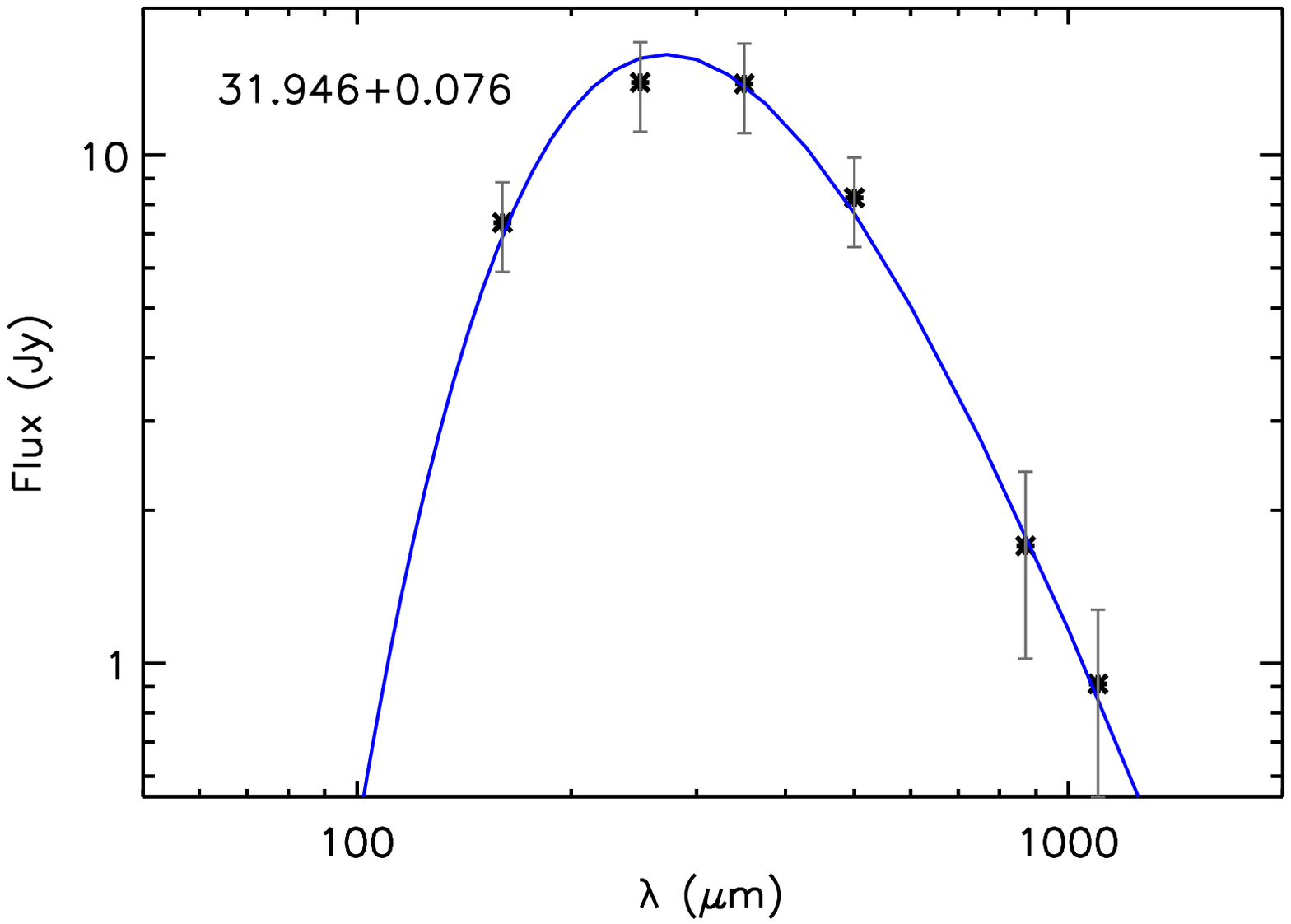} 
\includegraphics[width=5cm]{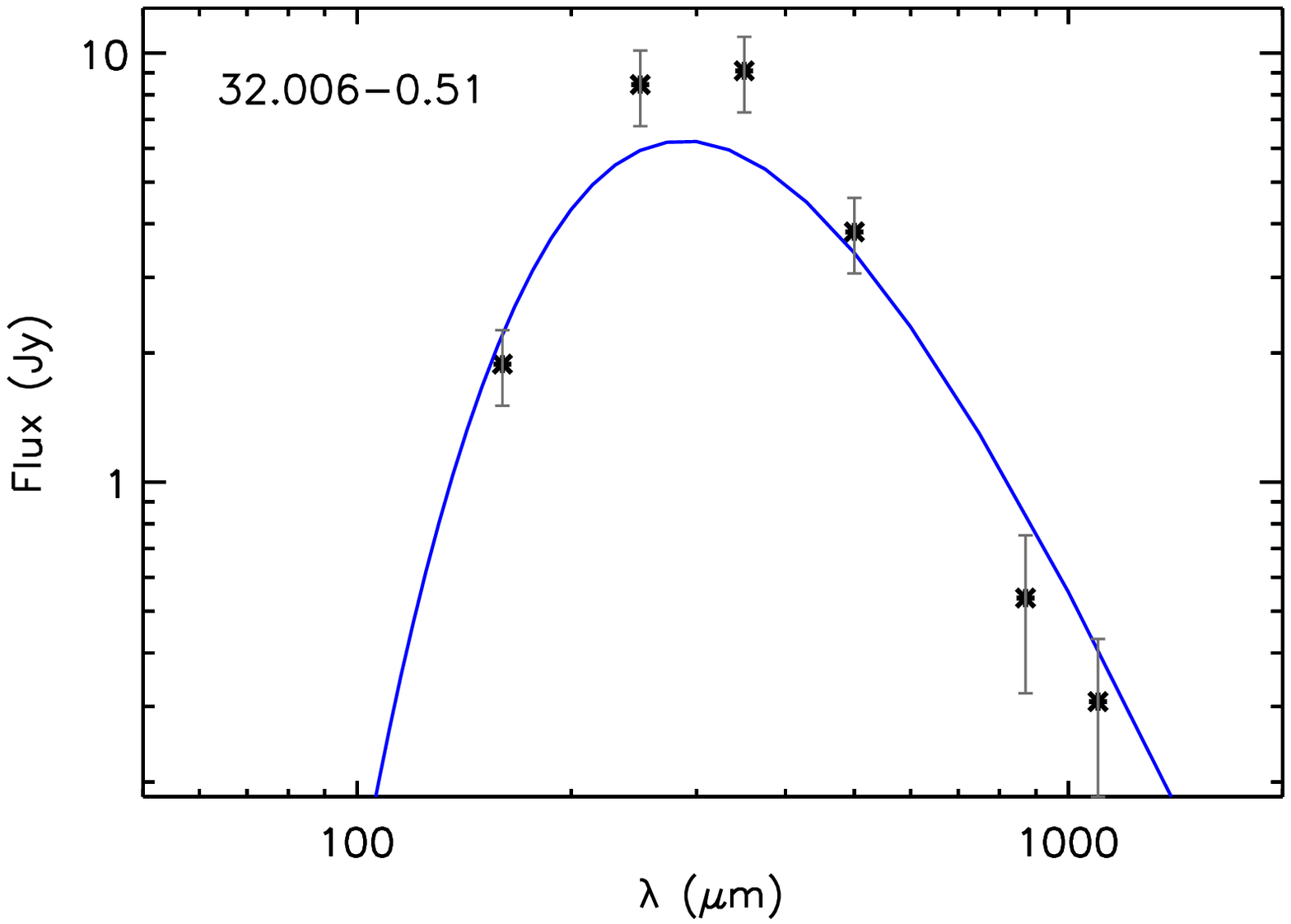} \includegraphics[width=5cm]{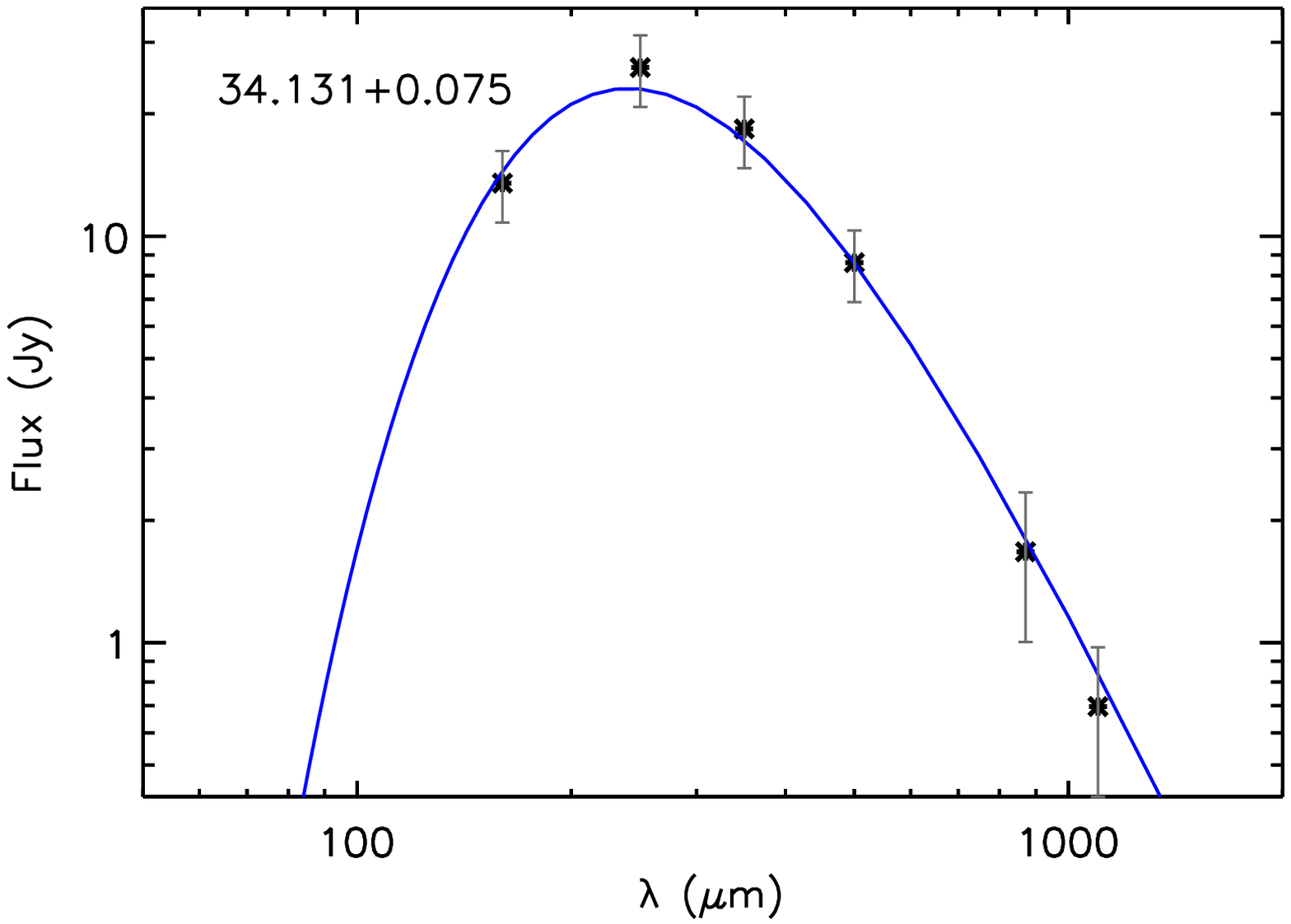} 

\caption{SED fitting for the 18 clumps studied in this work.}
\label{fig:SEDs_all_sources}
\end{figure*}

We derived the physical parameters of each clump using Hi-GAL plus ATLASGAL and (when available) BGPS fluxes. The fits to the SEDs have been performed with the \texttt{mpfit} routine \citep{Markwardt09}. The SEDs are shown in Figure \ref{fig:SEDs_all_sources} and the physical parameters are summarized in Table \ref{tab:SED_parameters}.

The mean temperature of these clumps is $<\mathrm{T}>=11.2\pm1.0$\,K. This temperature is significantly lower than the average temperature of starless clump candidates \citep[$<\mathrm{T}>\simeq15.5$\,K,][]{Traficante15b}. Massive clumps prior to the ignition of any protostars are mostly warmed-up by the external radiation field and show a temperature gradient, from about 18-28 K at the edges towards about 8-11 K at the center, where the region is shielded by external radiation fields \citep{Peretto10,Wilcock11}. The adopted greybody model does not take into account this variation, however the very low temperatures we measure are compatible with a cold central core and relatively low contribution of any external radiation field. 

\begin{figure}
\centering
\includegraphics[width=8cm]{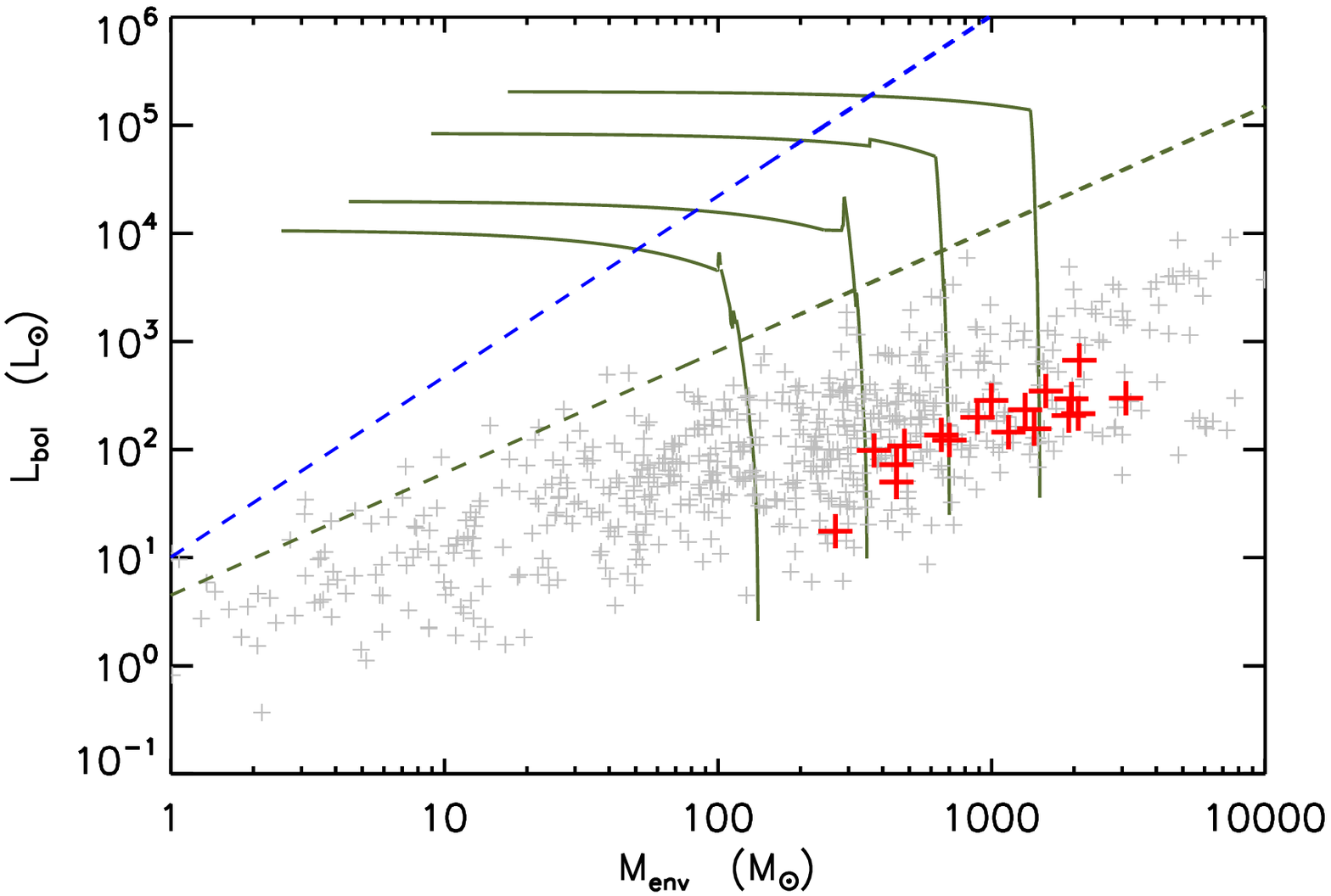} 
\caption{Bolometric luminosity vs. envelope mass. The red points are the clumps presented in this work. The grey points are all the starless clump candidates identified in \citet{Traficante15b}. The four tracks correspond to the evolutionary tracks for massive cores with final masses of 8, 13, 18 and 28 M\sun, from left to right, respectively. The green-dotted line is the fit to Class 0 objects as in \citet{Molinari08}, L$\propto$M$^{1.13}$. The blue-dotted line is the empirical border between individual Class 0 and Class I protostellar objects discussed in \citet[][and references therein]{Duarte-Cabral13}, L$\propto$M$^{1.67}$.}
\label{fig:molplot}
\end{figure}

The mass of the clumps covers the range $269\leq$M$\leq3098$ M\sun, with a bolometric luminosity (evaluated in the range $24\,\leq\lambda\leq1100$\mum) $18\leq\mathrm{L}\leq669$ L\sun. The mean mass is $\simeq1.2\times10^{3}$ M\sun\ for a mean bolometric luminosity of $\simeq$200 L\sun. The average L/M ratio, a good indicator of the evolution of massive regions \citep{Molinari16_l_m, Cesaroni15}, is only $<\mathrm{L/M}>\simeq0.17$, $\simeq5$ times lower than the mean L/M of starless clump candidates \citep[1.1,][]{Traficante15b}. $\mathrm{L/M}<<1$ is a strong indication that the regions are still quiescent \citep{Molinari16_l_m}. In Figure \ref{fig:molplot} we show the clump distribution in a L$-$M diagram against the sample of starless clump candidates in \citet{Traficante15b}. The green tracks are the \citet{Molinari08} evolutionary tracks for single high-mass cores. This model describes the evolution of massive cores from the beginning of the star formation process prior to the formation of a zero-age main sequence star in the L$-$M diagram, following the \citet{McKee03} accretion model. The high-mass cores follow a two-phases model. During the initial accretion phase the luminosity increases sustained by the collapse and the mass slightly decreases due to outflows (the analogous of a Class 0 object in the low-mass regime). The sources follow an almost vertical path in the diagram, up to a turnover point (the analogous of a class I object) after which the sources follow an almost horizontal path. This second phase begins with the formation of an HII region, the luminosity remains roughly constant while the mass is expelled through radiation and molecular outflows. The dotted line in Figure \ref{fig:molplot} corresponds to the best fit of the analogous of Class 0 objects for massive cores \citep{Molinari08}. This model assumes that the accretion rate onto the central star increases with time. For comparison, we also show the empirical border between Class 0 and Class I sources \citep[discussed in, e.g.][]{Andre00,Duarte-Cabral13}, which instead considers a decreasing accretion rate (blue-dotted line). In both cases, the clumps distribution lie well below the Class 0 regime, in a region characterized by extremely young objects.

\begin{center}
\begin{table*}
\centering
\begin{tabular}{c|c|c|c|c|c|c|c}
\hline
\hline
Clump & Mass & Luminosity & Radius & $\Sigma$ & Temperature & $\chi^{2}$ & Distance\\
       &  (M\sun) & (L\sun) & (pc) & (g cm$^{-2}$) & (K) &  & (kpc) \\
\hline
     15.631-0.377  &          269(79)  &           18  &          0.54(0.05)  &     0.06(0.01)  &      9.6(0.4) &        0.60  &         3.47(0.35)  \\
     18.787-0.286  &         1915(550)  &          206  &          0.69(0.07)  &    0.27(0.06) &      10.3(0.4) &        2.96  &         4.36(0.44)  \\
     19.281-0.387  &          701(206) &          123  &          0.67(0.07)  &   0.10(0.02) &       11.4(0.5) &        1.61  &         3.82(0.38)  \\
     22.53-0.192  &       1579(488)   &  349          &  0.80(0.08)   &    0.16(0.04) &  11.8(0.6)      &  2.01        &         5.77(0.58)  \\
     22.756-0.284  &          655(194) &          136  &          0.55(0.06)  &     0.14(0.03)  &    11.8(0.5) &        0.67  &         4.43(0.44)  \\
     23.271-0.263  &         997(297)  &          285  &          0.72(0.07)  &    0.13(0.03) &      12.3(0.6) &        4.76  &         5.21(0.52)  \\
     24.013+0.488  &         1957(559) &          294  &          0.81(0.08)  &       0.20(0.04)  &  10.9(0.4)  &        5.83  &         5.18(0.05)  \\
     24.528-0.136  &         2074(650) &          215  &          0.80(0.08)  &     0.22(0.05)   &    10.6(0.5)  &        1.34  &         5.19(0.52)  \\
     25.609+0.228  &       3098(953)   &        299    &         0.97(0.10)  &   0.22(0.05)  &      10.3(0.4)  &        2.81  &         5.57(0.56)  \\
     25.982-0.056  &        889(275)   &          200  &         0.80(0.08)  &    0.09(0.02)  &     11.9(0.6)  &        1.33  &         5.00(0.50)  \\
     28.178-0.091  &         2092(610) &          669  &          0.85(0.09)  &         0.19(0.04) &  12.6(0.6) &        8.72  &         5.35(0.54)  \\
     28.537-0.277  &         1153(327) &          145  &          0.67(0.07)  &         0.17(0.03) & 9.9(0.4)  &        12.63  &         4.96(0.50)  \\
     28.792+0.141  &          449(128) &           72  &          0.61(0.06)  &         0.08(0.02)  & 10.6(0.4)  &        6.46  &         4.62(0.46)  \\
     30.357-0.837  &          372(111) &           98  &          0.67(0.07)  &          0.06(0.01)  & 12.6(0.6)  &        3.46  &         4.30(0.43)  \\
     30.454-0.135  &         1326(419) &          233  &          1.0(0.10)  &     0.08(0.02)  &     11.7(0.6) &        1.40  &         5.84(0.58)  \\
     31.946+0.076  &         1431(440) &          156  &          0.82(0.08)  &    0.14(0.03)   &  10.7(0.5) &        0.59  &         5.51(0.55)  \\
     32.006-0.51  &          449(127) &           50  &          0.70(0.07)  &         0.06(0.01)  &  10.0(0.3)  &        9.39  &         4.24(0.42)  \\
     34.131+0.075  &          480(142) &          108  &          0.55(0.06)  &    0.11(0.02)  &     12.0(0.6) &        0.86  &         3.56(0.36)  \\
         
\hline
\end{tabular}

\caption{Clumps properties derived from the SEDs. Col.1: Clump name; Col.2: Clumps mass with errors estimated from the SED fit and assuming a flux uncertainties of 20\% for Hi-GAL fluxes \citep{Traficante15b} and of 40\% on ATLASGAL and BGPS fluxes, plus an uncertainties on distance estimation of 10\%; Col.3: Bolometric luminosity estimated integrating the flux in the range 160-1100\mum; Col. 4: Source radius derived from the geometric mean of the FWHMs in Table \ref{tab:photometry}; Col. 5: Temperatures and associated uncertainties; Col. 6: $\chi^{2}$ of the SED fits done using Hi-GAL, ATLASGAL and, where available, BGPS datasets; Col.7: Source distance taken from \citet{Traficante15b}.}
\label{tab:SED_parameters}
\end{table*}
\end{center}

\subsection{Massive stars from massive clumps}\label{sec:massive_stars_from_dust}
\begin{figure}
\centering
\includegraphics[width=8cm]{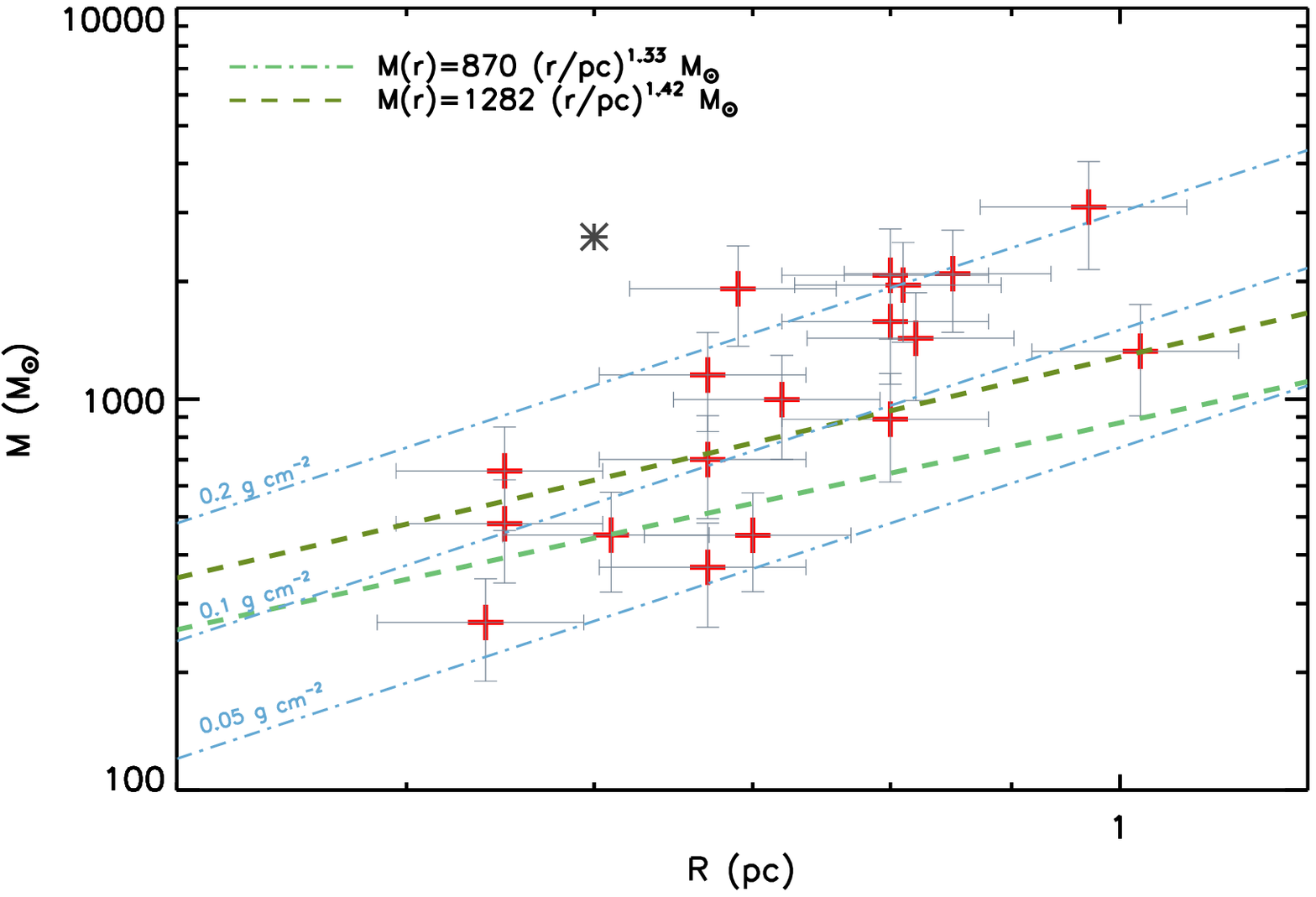} 
\caption{Mass vs. radius distribution of the 70\mum\ quiet clumps. The light green dotted line delimits the empirical KP threshold for high-mass star formation in IRDCs. All but three sources lie above the threshold. The dark green dotted line is the \citet{Baldeschi17} threshold. Seven clumps are below this more stringent threshold. The grey asterisk marks the position of the clump embedding the massive protostar in SDC335 \citep{Peretto13}.}
\label{fig:mass_radius}
\end{figure}

In order to investigate if our clumps are likely going to form massive stars we first explore their mass-radius relationship. An empirical high-mass star formation threshold in this diagram has been proposed by \citet[][KP]{Kauffmann10}, which identified as potentially high-mass star forming regions all the clusters with $\mathrm{M(r)}>870$ M\sun\ ($\mathrm{r/pc})^{1.33}$. Recently, \citet{Baldeschi17} analyzed the bias in the estimation of the physical parameters of massive clumps and found a relationship $\mathrm{M(r)}>1282$ M\sun\ ($\mathrm{r/pc})^{1.42}$, which is more stringent then the KP threshold. As shown in Figure \ref{fig:mass_radius}, following the KP criterion all but three clumps are above the threshold, 15.631-0.377, 30.357-0.837 and 32.006-0.51. Conversely, following the \citet{Baldeschi17} criterion 7 clumps may not form high-mass stars: 15.631-0.377, 30.357-0.837 and 32.006-0.51 plus 19.281-0.387, 25.982-0.056, 28.792+0.141 and 34.131+0.075.
 
Another criterion to identify massive star forming regions is assuming that there is a mass surface density threshold $\Sigma_{t}$ below which clumps may not form massive stars. The value of $\Sigma_{t}$ is still debated. \citet{Tan14} assumes $0.1\leq\Sigma_{t}\leq1$ g cm$^{-2}$ as the range of values for the threshold, while e.g. \citet{Urquhart14} identified $\Sigma_{t}=0.05$ g cm$^{-2}$ based on the analysis of massive clumps in ATLASGAL. In our sample there are no sources with $\Sigma\leq0.05$ g cm$^{-2}$, and 7 sources with $\Sigma\leq0.1$ g cm$^{-2}$ (Table  \ref{tab:SED_parameters}). In Figure \ref{fig:mass_radius} we also show lines of constant surface density. The clumps 15.631-0.377, 30.357-0.837 and 32.006-0.51 are the sources with the lowest mass and surface density. Four sources, 18.787-0.286, 24.013+0.488, 24.528-0.136 and 25.609+0.228 have $\Sigma\geq0.2$ g cm$^{-2}$ and $\mathrm{M}\geq2\times10^{3}$ M\sun. These clumps are among the most massive 70\mum\ quiet clumps observed to date and they will potentially form stars with mass comparable with the most massive protostars observed in the Galaxy \citep{Peretto13,Avison15}.

Combining these two criterion the majority of these clumps will likely produce massive stars.

\subsection{MIR counterparts}\label{sec:24mu_counterparts}

Although 70\mum\ quiet clumps are good candidates to be starless \citep{Dunham08,Giannini12,Veneziani12}, it may happen that some of these clumps already embed 24\mum\ sources \citep{Elia17}, identified thanks to the better MIPSGAL sensitivity compared with Hi-GAL (Table \ref{tab:sensitivities}). In the \citet{Gutermuth15} 24\mum\ source catalogue, 5 clumps have indeed a 24\mum\ counterpart. Two of them (18.787-0.286 and 30.454-0.135) however are likely to be foreground sources, as they are the only to have a 2MASS counterpart and they are not located at the Hi-GAL column density peak.

In order to look for faint 24\mum\ sources not identified in the \citet{Gutermuth15} catalogue, we visually inspected the MIPSGAL counterparts of each clump. This inspection reveals that in 50$\%$ of the sample at least one 24\mum\ counterpart is present within a 250\mum\ beam centered in the clump centroid. Four sources (18.787-0.286, 24.013+0.488, 30.357-0.837 and 34.131+0.075) have more than one 24\mum\ counterpart within the clump region. The complete list of sources identified by eye, with their positions, is in Table \ref{tab:24mu_parameters}.

To investigate the properties of these counterparts we performed a dedicated photometry on the MIPSGAL maps with the Aperture Photometry Tool package\footnote{\texttt{http://www.aperturephotometry.org/aptool/}}. We estimate the photometry in a circular region with a radius of 3.5\arcsec, which includes $\simeq1$ MIPS 24\mum\ beam (5.9\arcsec) in the aperture, and applied the corresponding aperture correction of 2.78, as suggested in the MIPS instrument handbook\footnote{\texttt{http://irsa.ipac.caltech.edu/data/SPITZER/docs/\ mips/mipsinstrumenthandbook/50/}}. The background has been estimated as the median value of the emission measured in a circular annulus surrounding each source. These 24\mum\ counterparts are very faint, with fluxes $F_{24}$ in the range $4.45\leq F_{24}\leq29.31$ mJy. The source photometry is in Table \ref{tab:24mu_parameters}.

To further characterize and classify these sources we look for counterparts in the GLIMPSE survey \citep{Benjamin03}. Eight sources have GLIMPSE counterparts within a radius of 3\arcsec\ from the 24\mum\ centroid (the MIPSGAL beam), associated with 5 different clumps, showed in Table \ref{tab:glimpse_robitaille}. One source, 19.287-0.386, has 2 GLIMPSE sources associated with the 24\mum\ counterpart. All but 1 source have GLIMPSE counterparts at all the four IRAC bands: 3.6, 4.5, 5.8 and 8.0\mum. 18.787-0.286\_2 does not have a counterpart at 8.0\mum. We use the GLIMPSE fluxes to classify these clumps according to the prescriptions of \citet{Lada87} and \citet{Gutermuth09}. The \citet{Lada87} classification scheme is based on the slope  $\alpha$ of spectral index in the IRAC bands: $0\leq\alpha\leq3$ for Class I, $-2\leq\alpha\leq0$ for Class II and $-3\leq\alpha\leq-2$ for Class III objects. The scheme proposed by \citet{Gutermuth09} is a IRAC colour-colour classification (Phase 1) plus a refining using a JHK$_{S}$[3.6][4.5] YSO classification (Phase 2). The two classification schemes agree well and we found 6 Class I sources and 3 Class II sources. 18.787-0.286\_2  has a Class I and Class II source associated with the same 24\mum\ counterpart. The classification is in Table \ref{tab:glimpse_robitaille}. At least five clumps embed Class I or Class II sources but are 70\mum\ quiet in the Hi-GAL maps.

\begin{center}
\begin{table}
\centering
\begin{tabular}{c|c|c|c}
\hline
\hline
Clump	&	24\mum\ RA & 24\mum\ Dec  &  24\mum\ flux  \\

        & ($\deg$) &  ($\deg$)     &   (mJy)    \\
\hline            
18.787-0.286\_1$^{1}$   &    18:26:15.2   & -12:41:30  &  28.2    \\ 
18.787-0.286\_2$^{1}$   &     18:26:15.4   & -12:41:38  &   4.45  \\ 
19.281-0.387   &    18:27:33.9   & -12:18:23  &    18.68 \\ 
22.53-0.192   &    18:32:59.7   &  -09:19:59 &     29.31 \\ 
24.013+0488\_1  & 18:33:18.6   &  -07:42:24 &        9.46 \\ 
24.013+0488\_2   &   18:33:18.1   &  -07:42:33 &      18.62 \\ 
24.013+0488\_3   &    18:33:17.6   &  -07:42:44 &     13.11 \\ 
28.178-0091   &    18:43:02.2   &  -04:14:32 &     24.08 \\ 
30.357-0.837\_1   &    18:49:40.9   &  -02:39:47  &     15.61 \\ 
30.357-0.837\_2   &    18:49:40.4   &  -02:39:50  &     5.49 \\ 
30.454-0.135$^{2}$   &   18:47:23.9   &  -02:15:55  &      9.27 \\ 
31.946+0.076   &    18:49:22.3   &   -00:50:35 &     19.49 \\ 
34.131+0.075\_1   &  18:53:21.1   &   01:06:13  &       8.95 \\ 
34.131+0.075\_2   &    18:53:21.1   &   01:06:23  &     6.13 \\ 
			
\hline
\end{tabular}
\begin{tablenotes}
\scriptsize
\item $^{1}$ There is also a foreground 24\mum\ source with a 2MASS counterpart and a 24\mum\  flux of  560 mJy \citep{Gutermuth15}. 
\item $^{2}$ There is also a foreground 24\mum\ source with a 2MASS counterpart and a 24\mum\  flux of  82 mJy \citep{Gutermuth15}. 
\end{tablenotes}
\caption{Emission of the 24\mum\ source for the clumps with a faint 24\mum\ counterpart identified. Col.1: Clump name; Col. 2-3: Coordinates of the 24\mum\ source counterpart; Col. 6: Integrated flux of the 24\mum\ source. The flux has been estimated in a radius of 3.5\arcsec\ and corrected for an aperture correction of 2.78.}
\label{tab:24mu_parameters}
\end{table}
\end{center}

\begin{figure*}
\centering
\includegraphics[width=6cm]{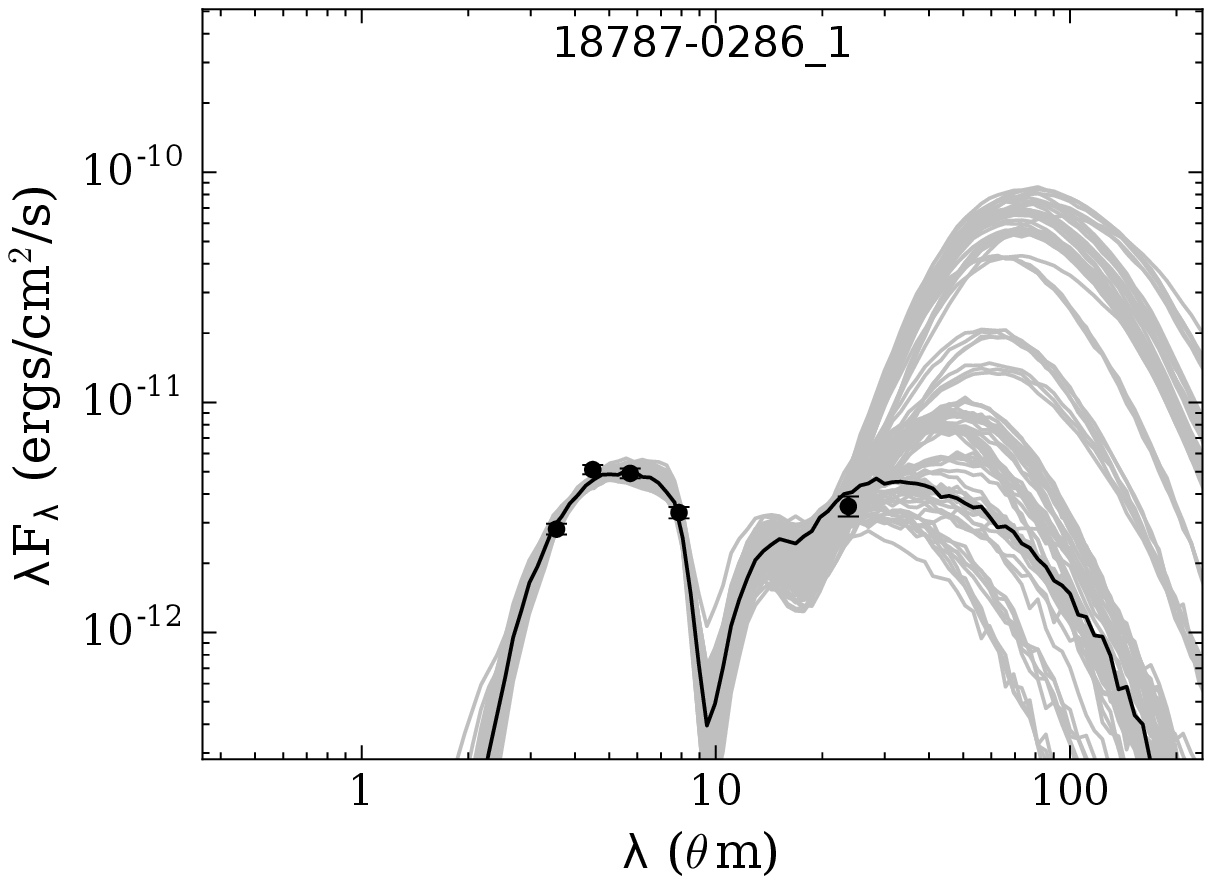}  
\includegraphics[width=6cm]{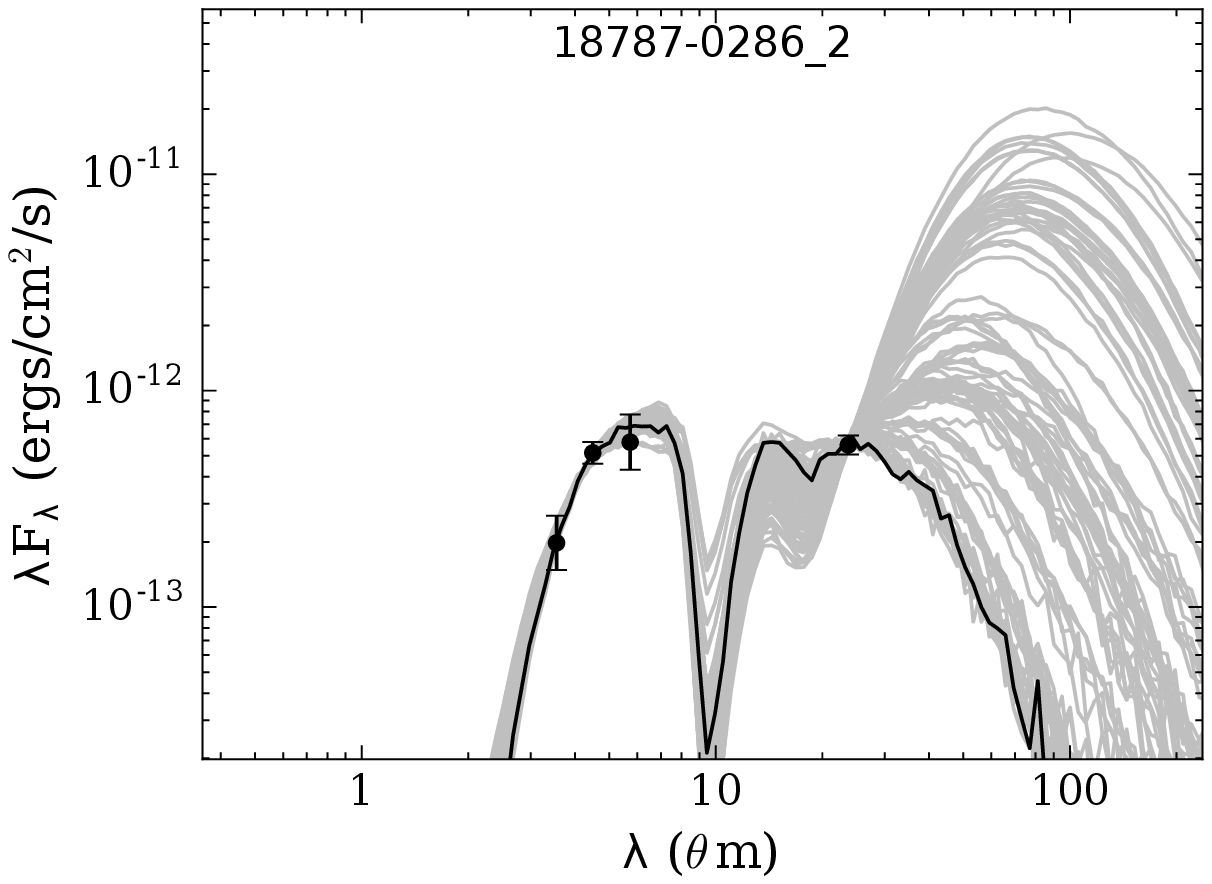}
\includegraphics[width=6cm]{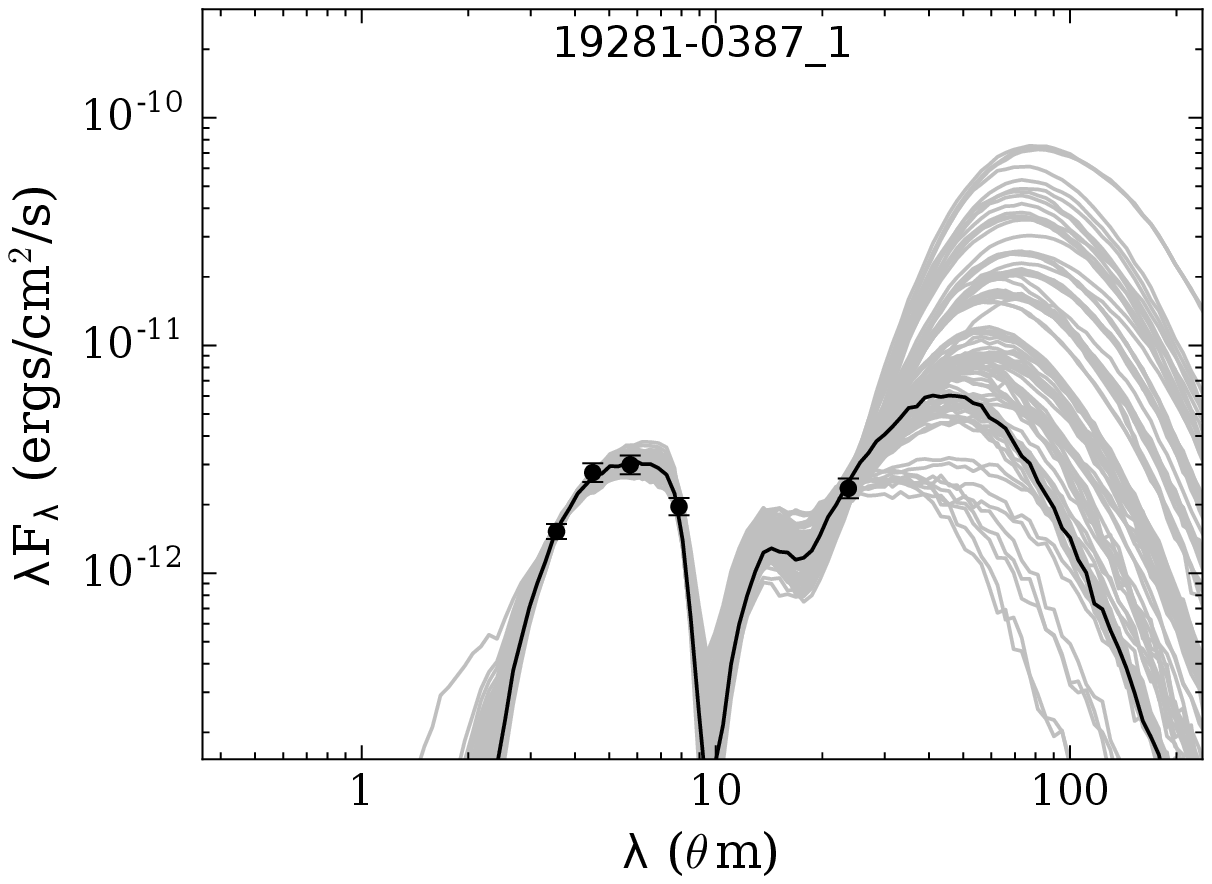}  
\includegraphics[width=6cm]{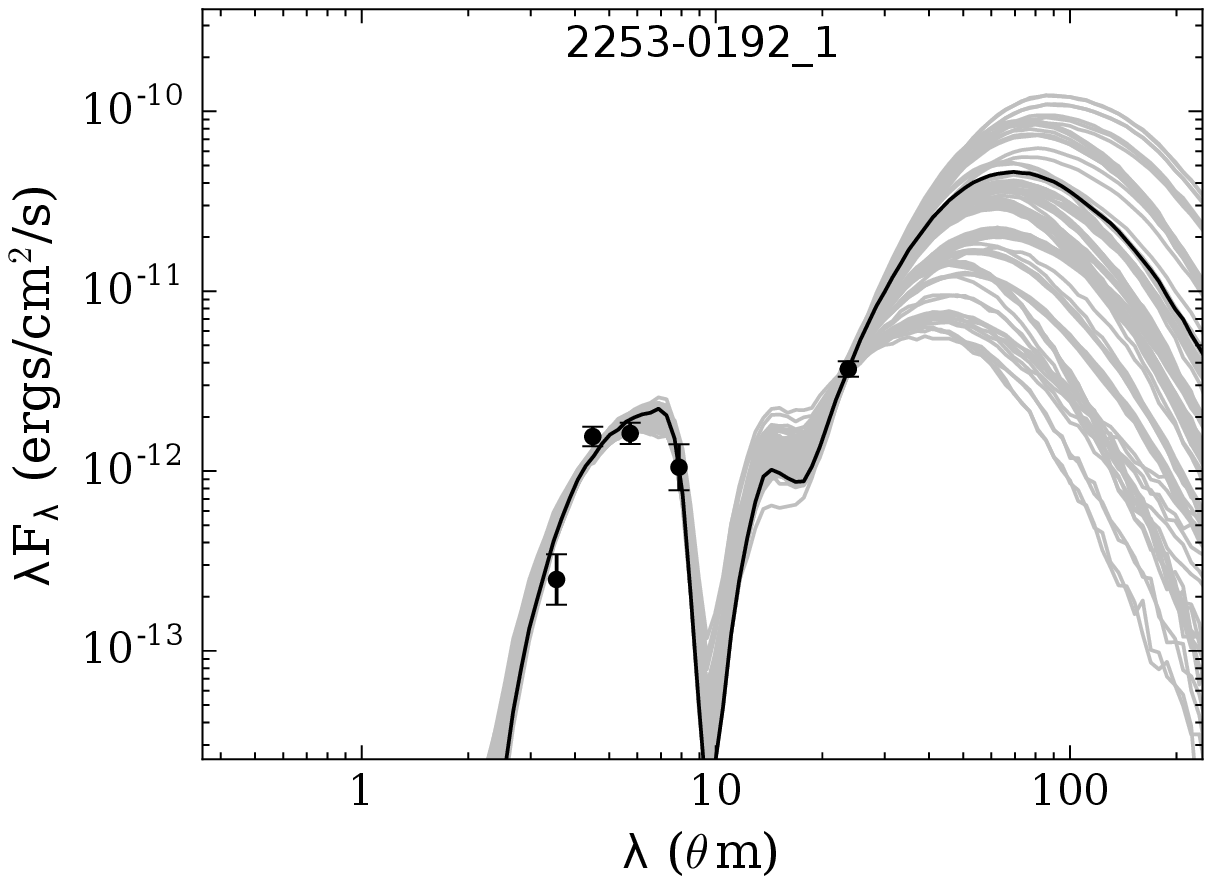}  
\includegraphics[width=6cm]{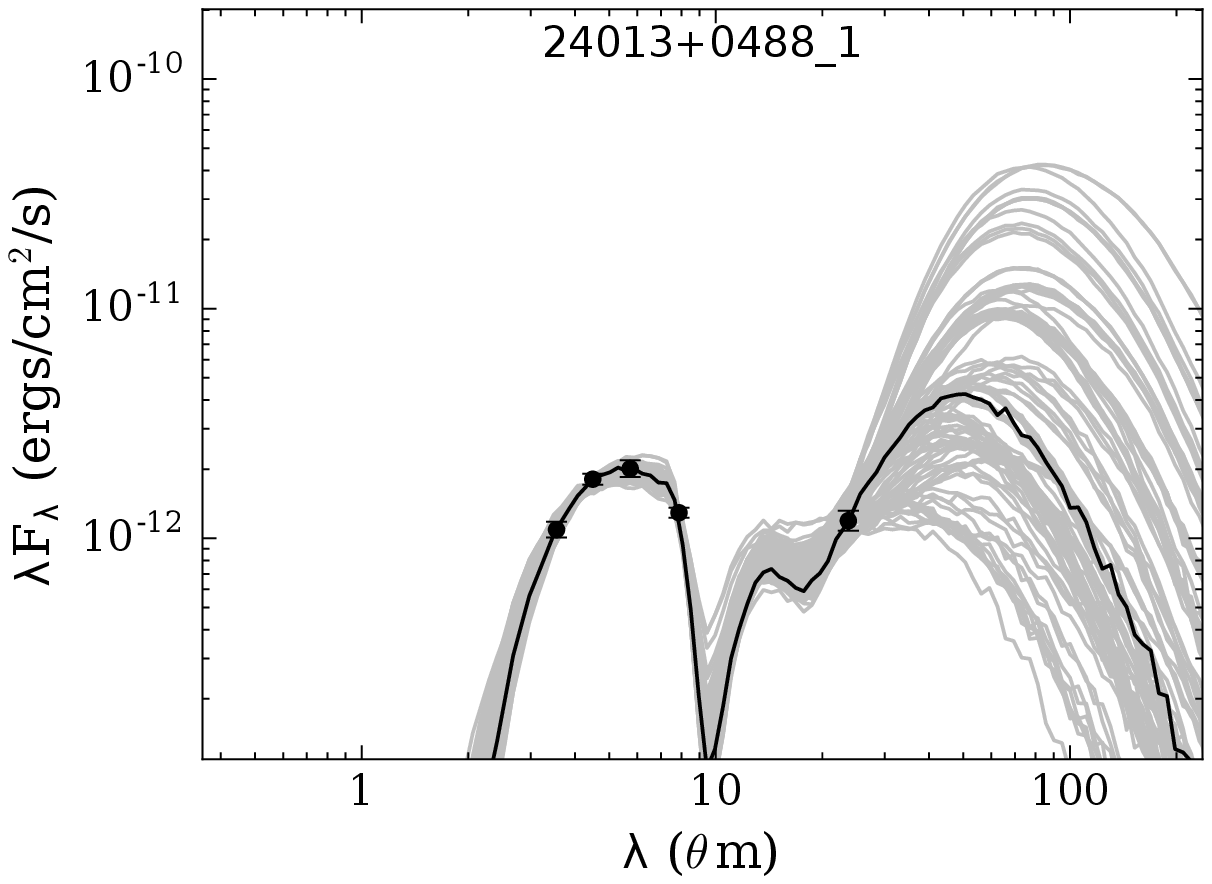}  
\includegraphics[width=6cm]{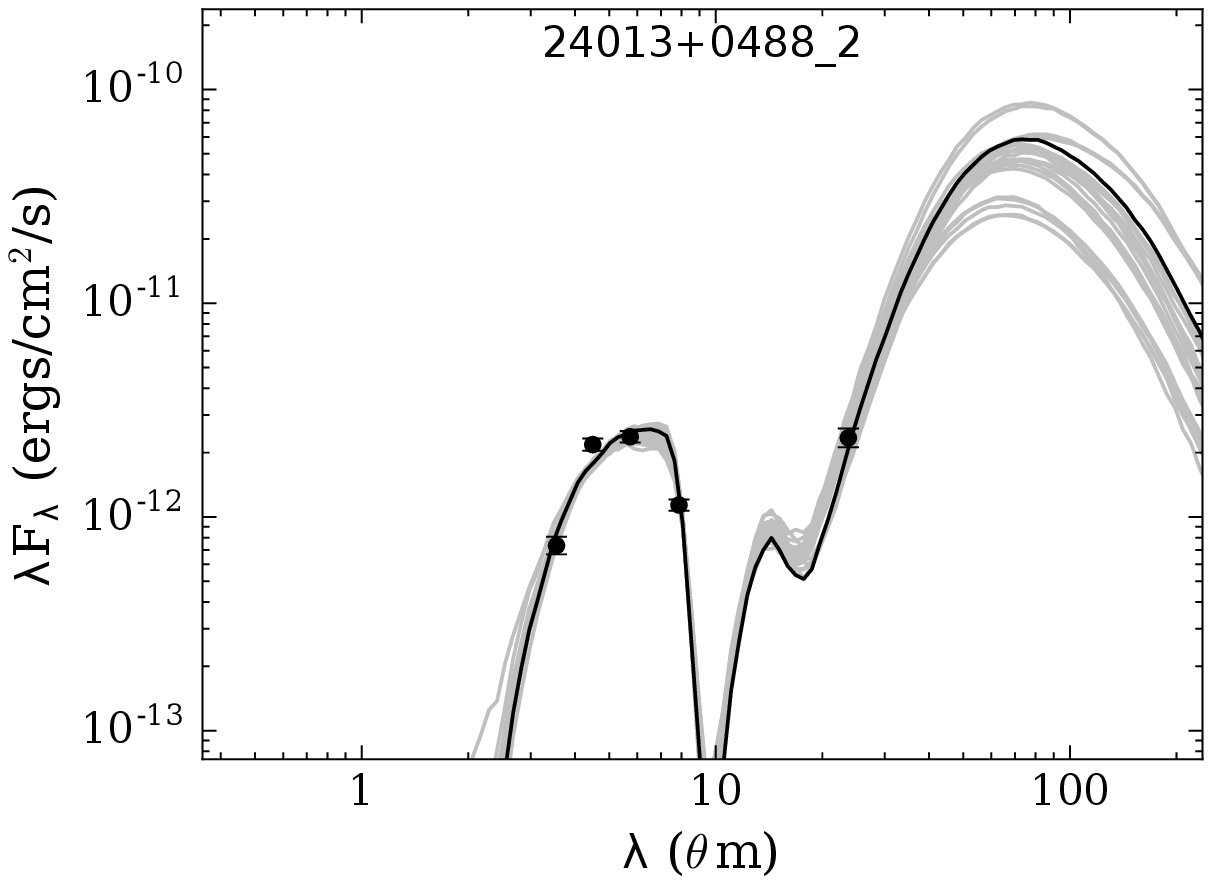}  
\includegraphics[width=6cm]{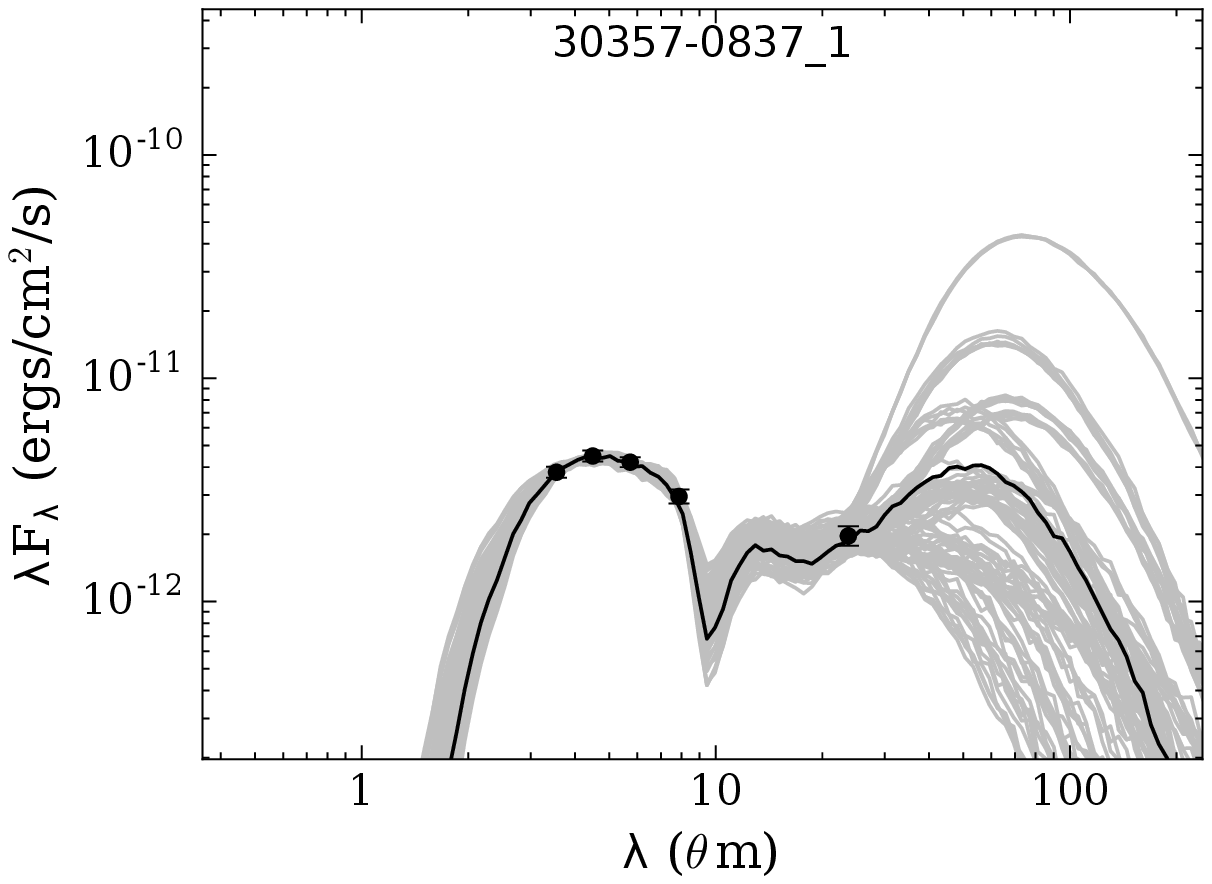}  
\includegraphics[width=6cm]{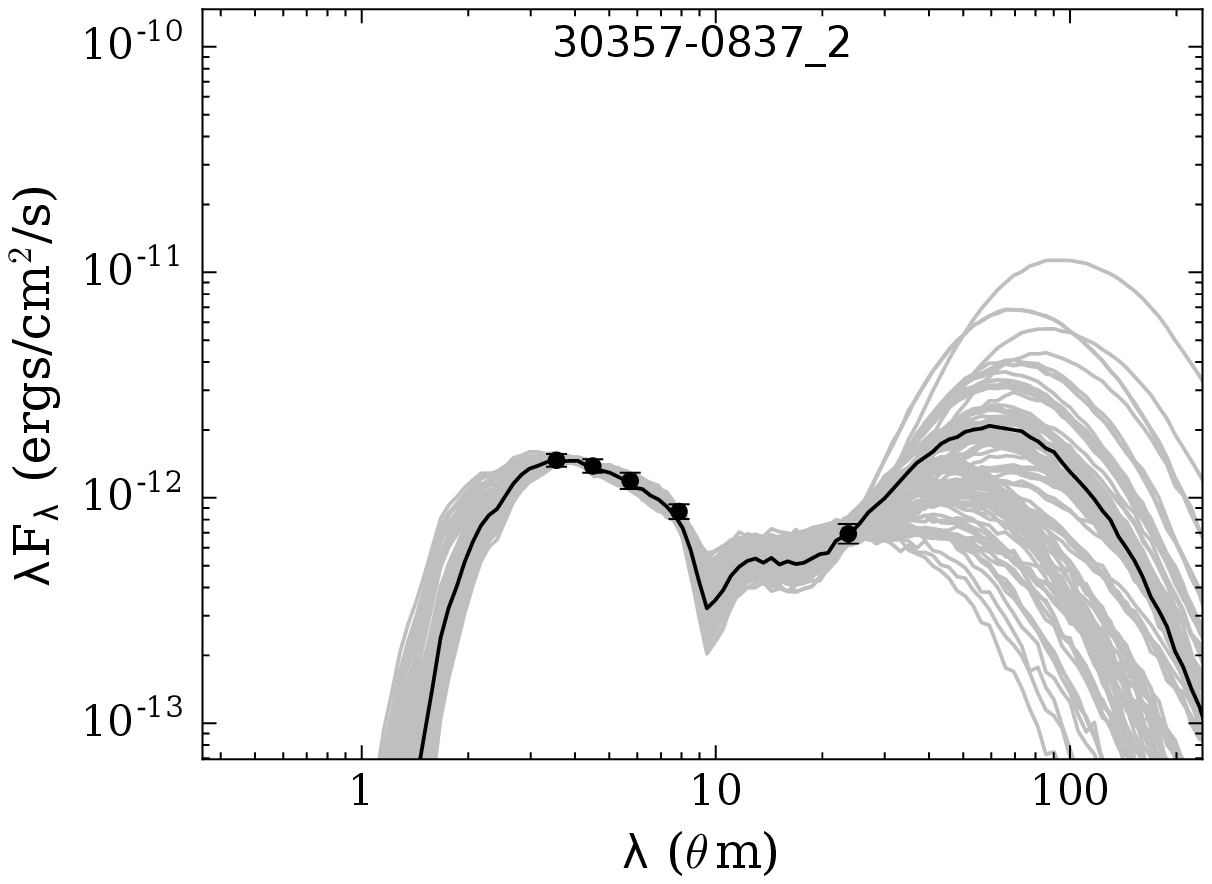}  
\caption{Grid of models for each clump with an embedded MIR source as obtained from the \citet{Robitaille06} SED fitter tool. The black curve in each panel is the best-fit model.}
\label{fig:Robitaille_fit}
\end{figure*}

The MIR fluxes can be used to estimate the properties of the central stars with the \citet{Robitaille06} SED fitting tool. This tool computes radiative transfer models of young stellar objects in a range of masses and evolutionary stages to model a central star, an accretion disk and an envelope \citep{Robitaille06,Robitaille07}. The tool provides several hundreds of models, each one describing a set of physical parameters with a $\chi^{2}$ value describing the goodness of the fit. The grid of models for each source are shown in Figure \ref{fig:Robitaille_fit}. In order to obtain a representative value for the physical parameters, we average the fit results for all models with $\chi^{2}\leq5$. The representative values are obtained with a weighted mean, with the weight being the inverse of the $\chi^{2}$ value, similar to the procedure adopted in \citet{Grave09}. If the number of models with $\chi^{2}\leq5$ is less then 50, or the $\chi^{2}$ values were always above 5, we average the results of the best 50 models. The weighted values of the extinction, column density and mass of the central star for each source are in Table \ref{tab:glimpse_robitaille}. All these sources are highly extincted, with A$_{V}$ in the range $25.7\leq\mathrm{A}_{V}\leq93.4$ mag, corresponding to surface densities $5.8\leq\Sigma\leq21.0$ g cm$^{-2}$ applying the conversion factor described in \citet{Bohlin78}. The best fit models are compatible with stars of intermediate mass, in the range $2.7\leq\mathrm{M}_{*}\leq5.5$ M\sun.

\begin{figure*}
\centering
\includegraphics[width=6cm]{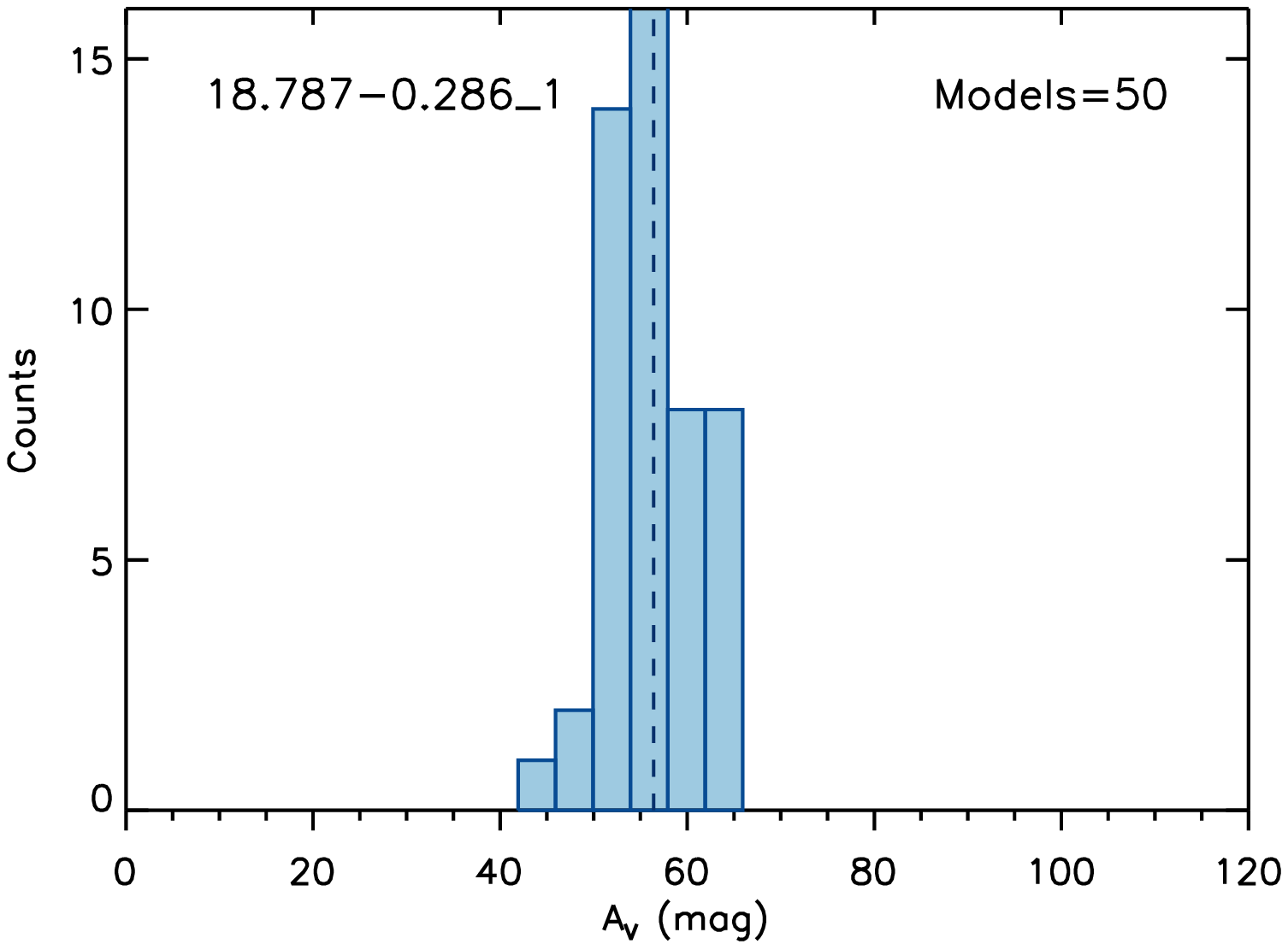}  
\includegraphics[width=6cm]{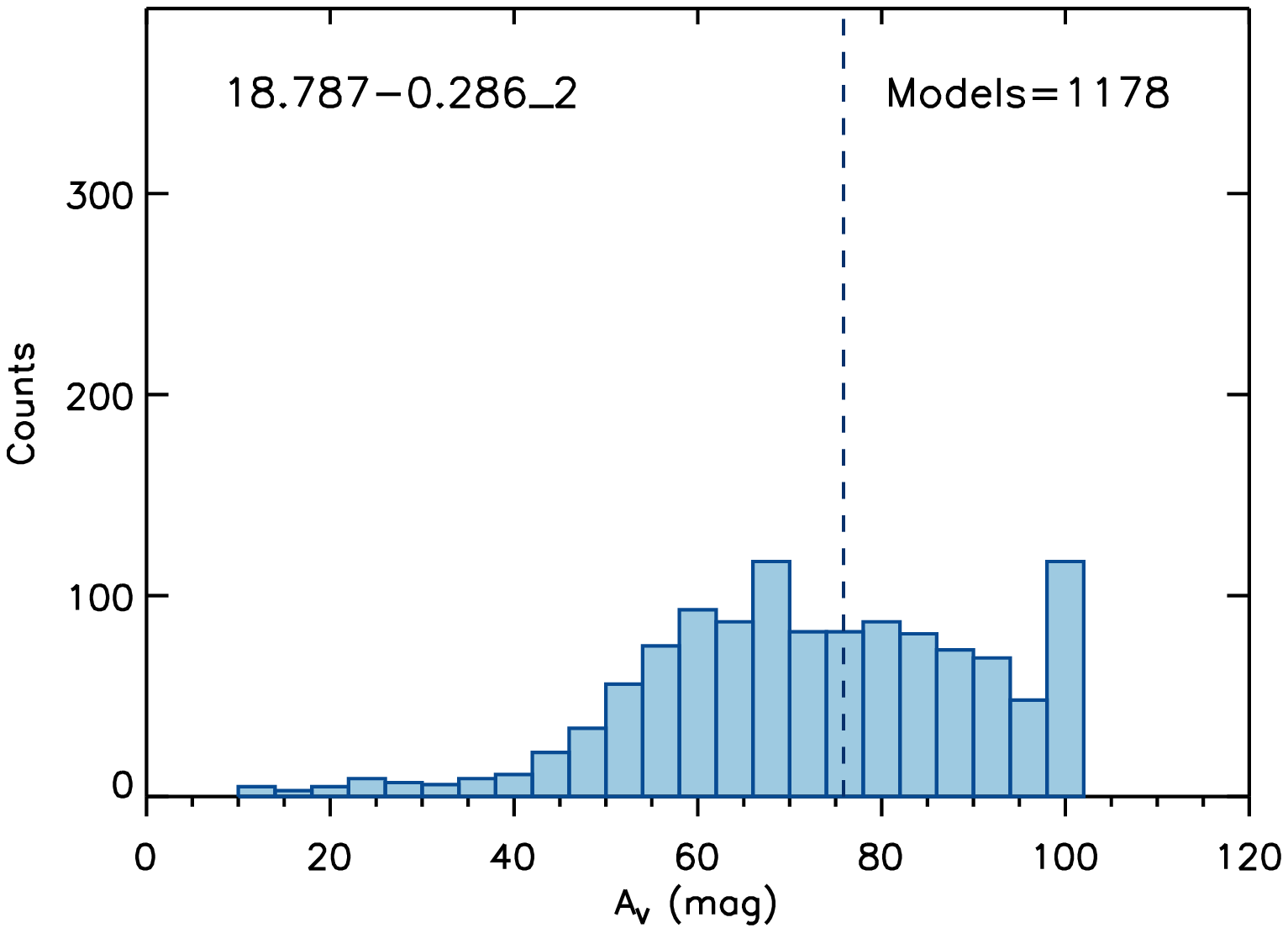}
\includegraphics[width=6cm]{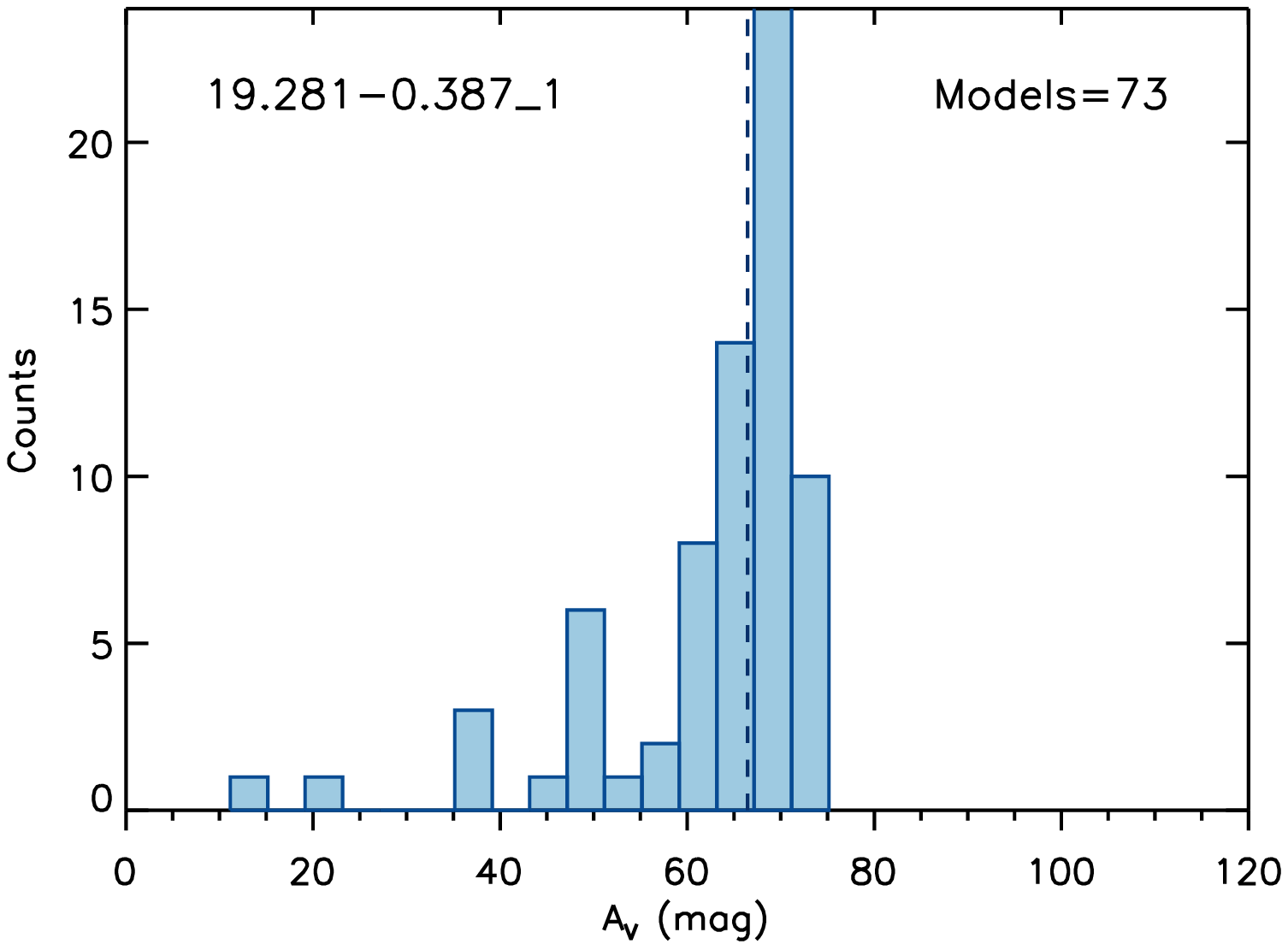}  
\includegraphics[width=6cm]{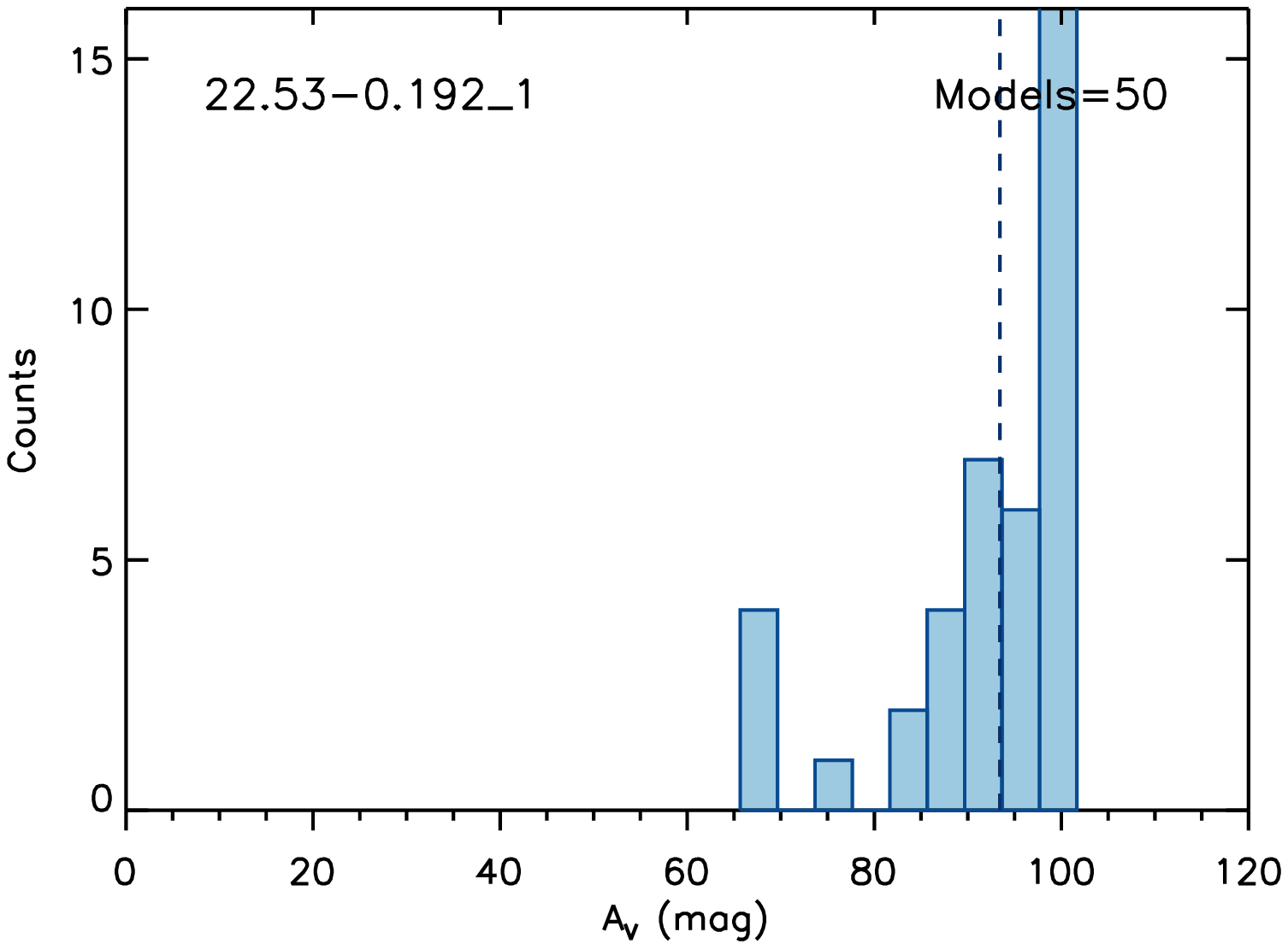}  
\includegraphics[width=6cm]{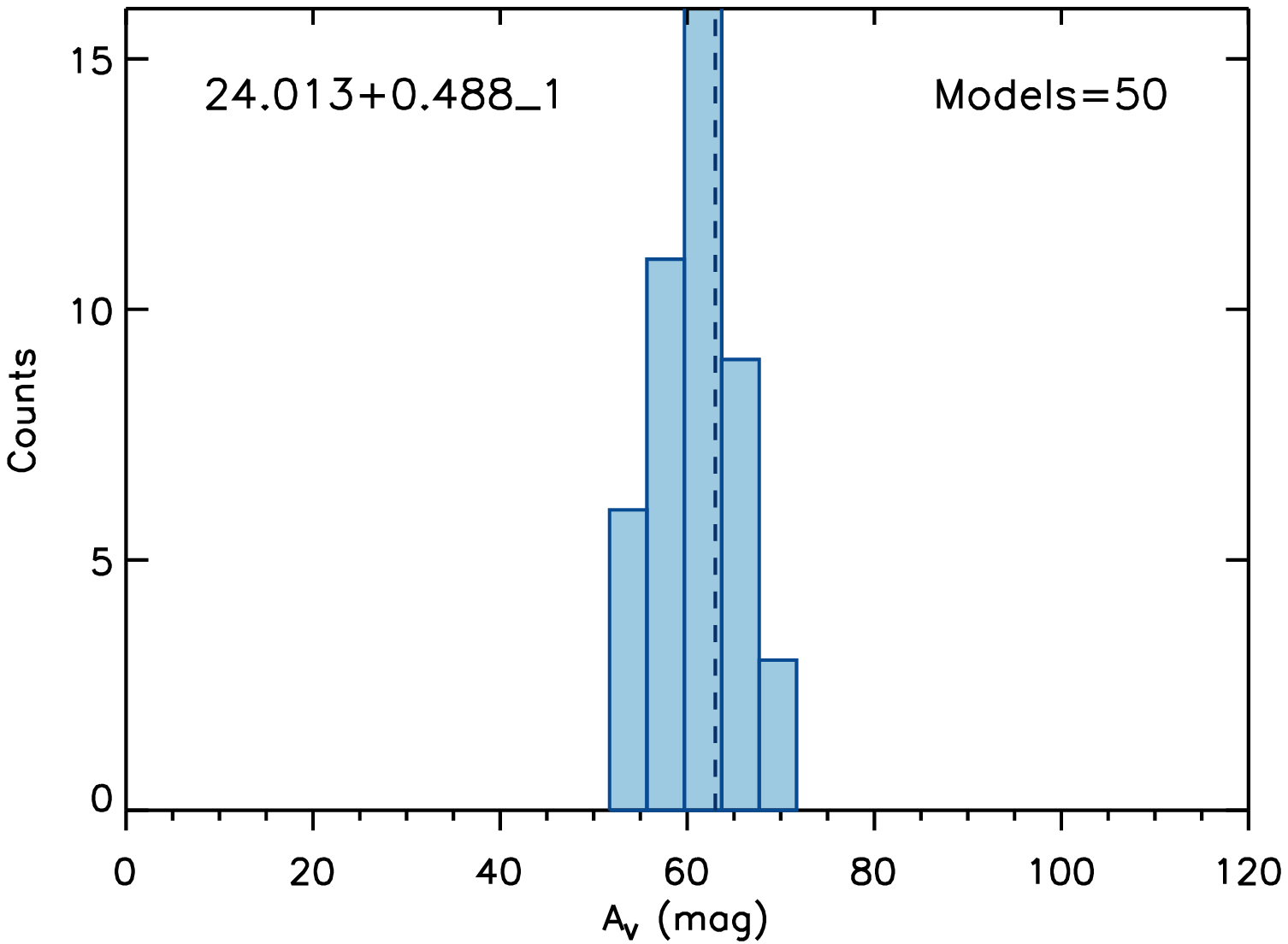}  
\includegraphics[width=6cm]{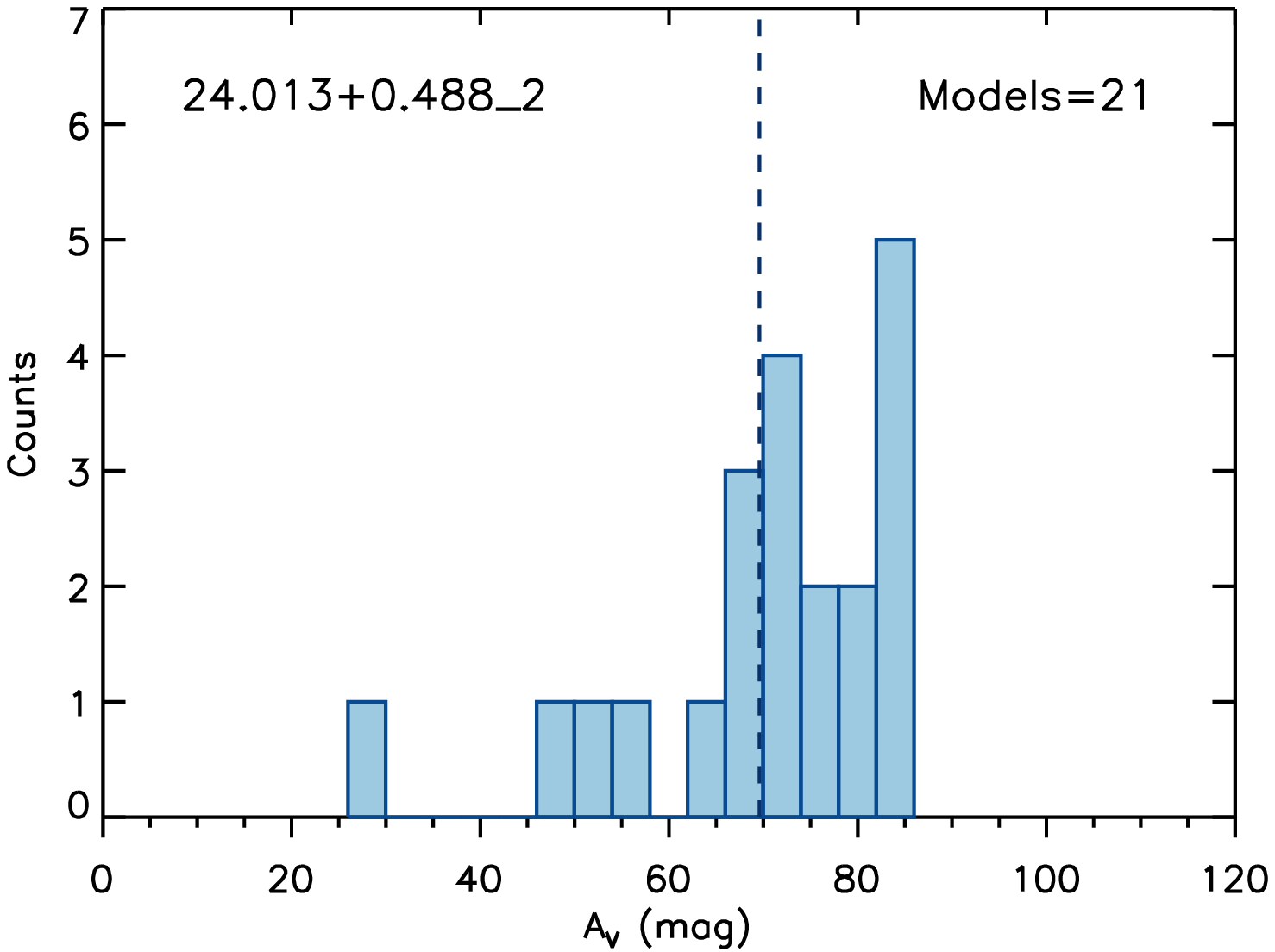}  
\includegraphics[width=6cm]{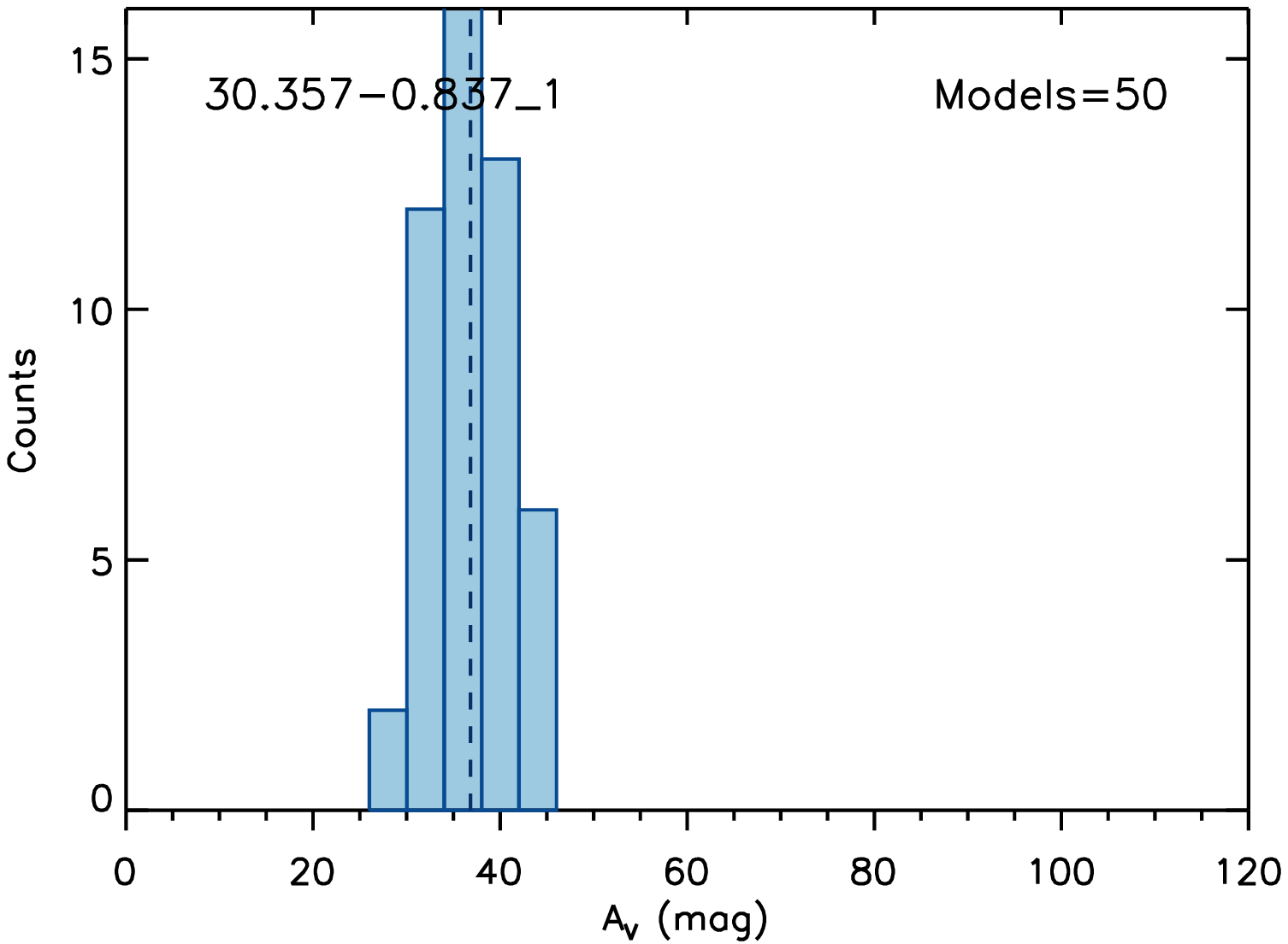}  
\includegraphics[width=6cm]{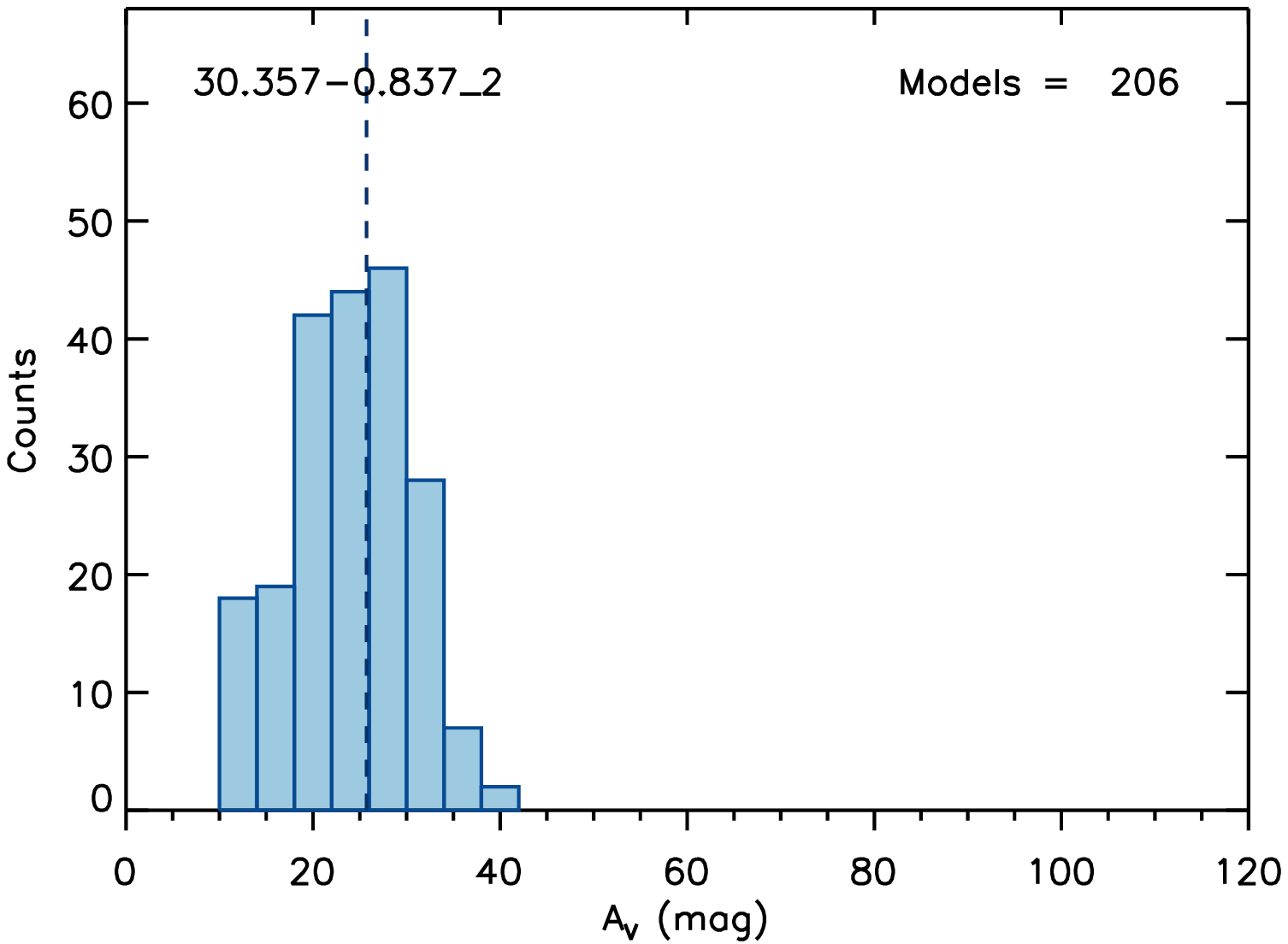}  
\caption{Distribution of A$_{V}$ for the best-fit models of each source embedded in clumps with MIR counterparts. The blue vertical line is the weighted average mean with weight equal to the $\chi^{2}$ value of each fit.}
\label{fig:Robitaille_fit_Av}
\end{figure*}

\begin{figure*}
\centering
\includegraphics[width=6cm]{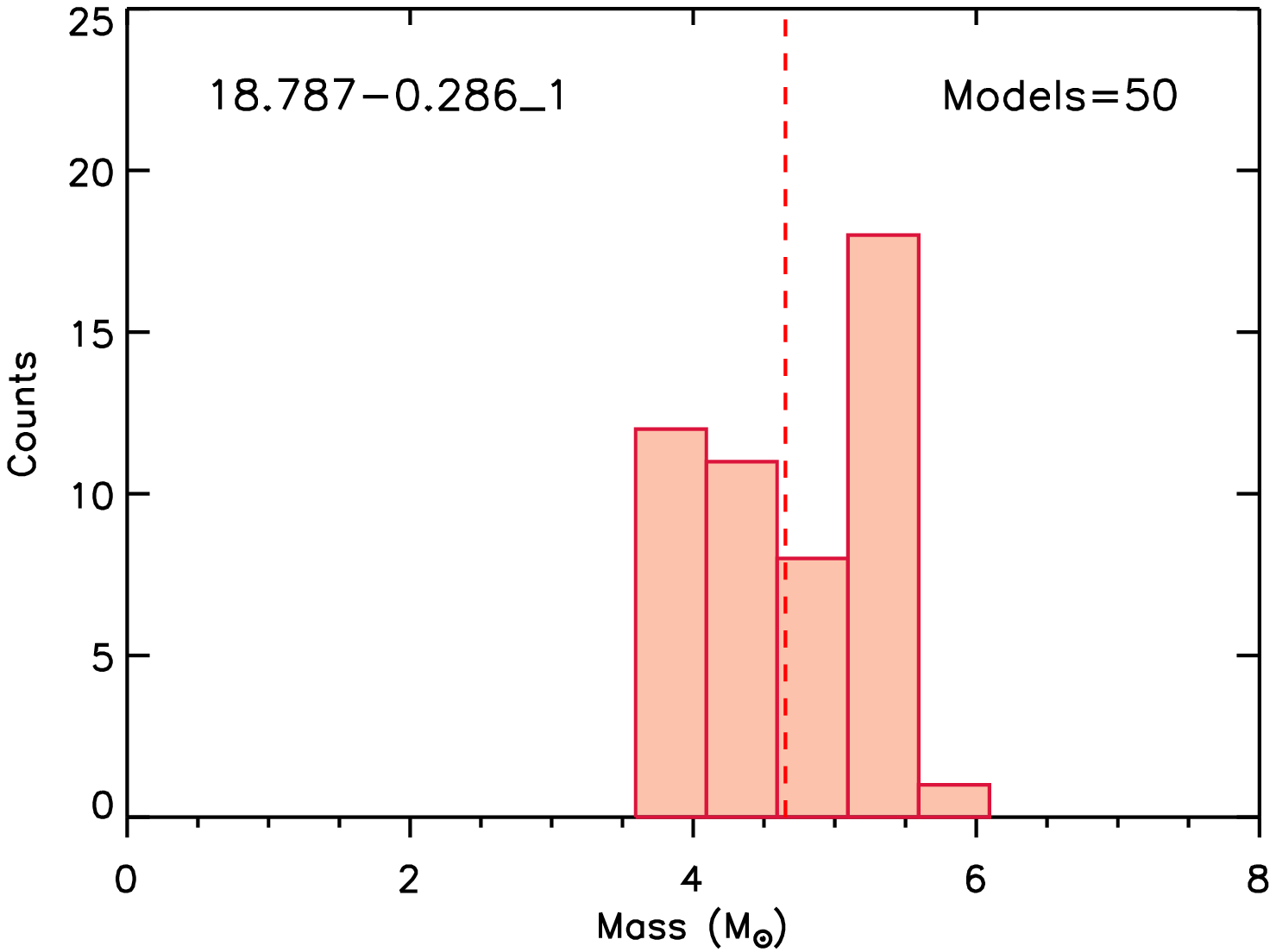}  
\includegraphics[width=6cm]{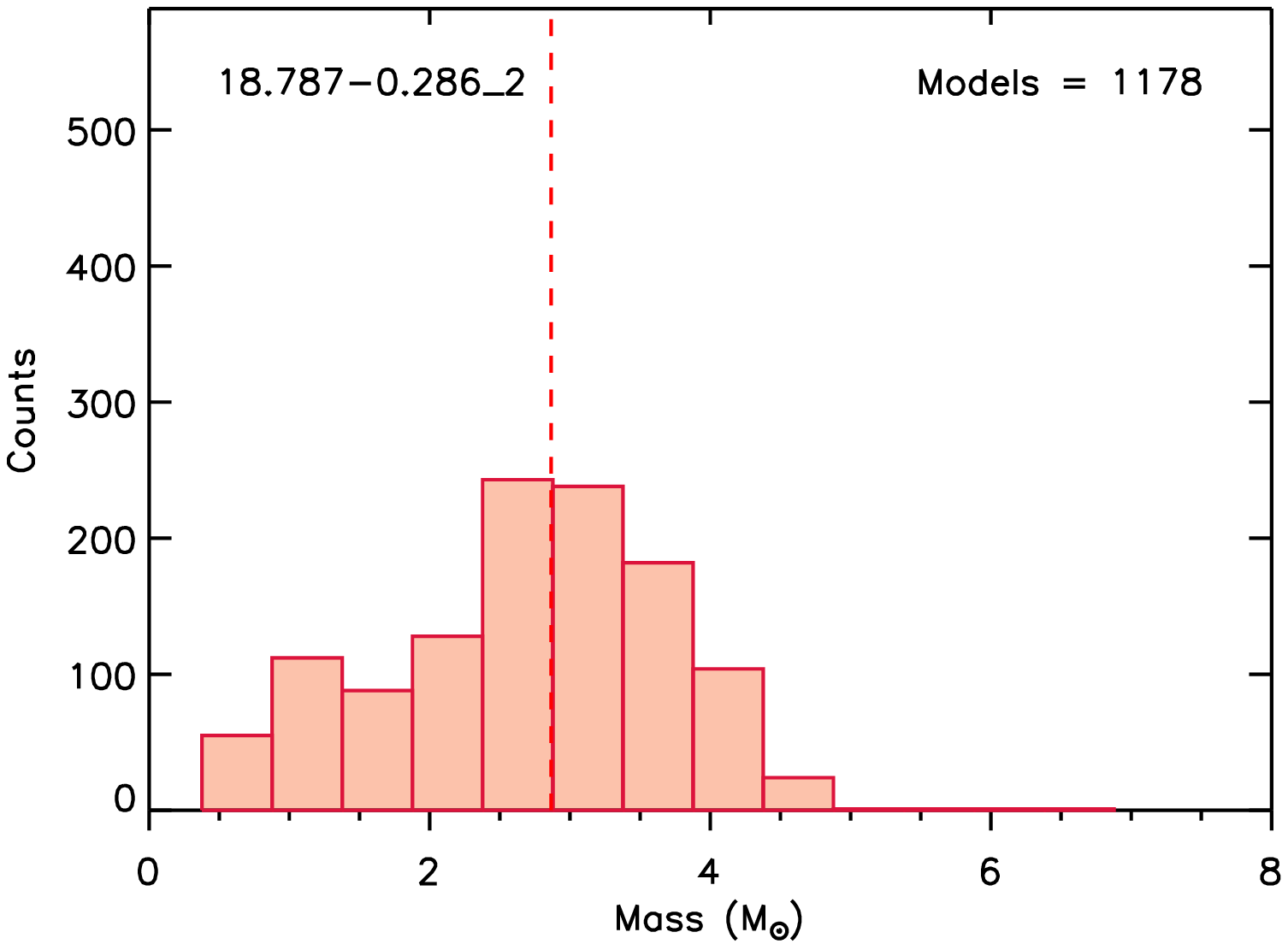}
\includegraphics[width=6cm]{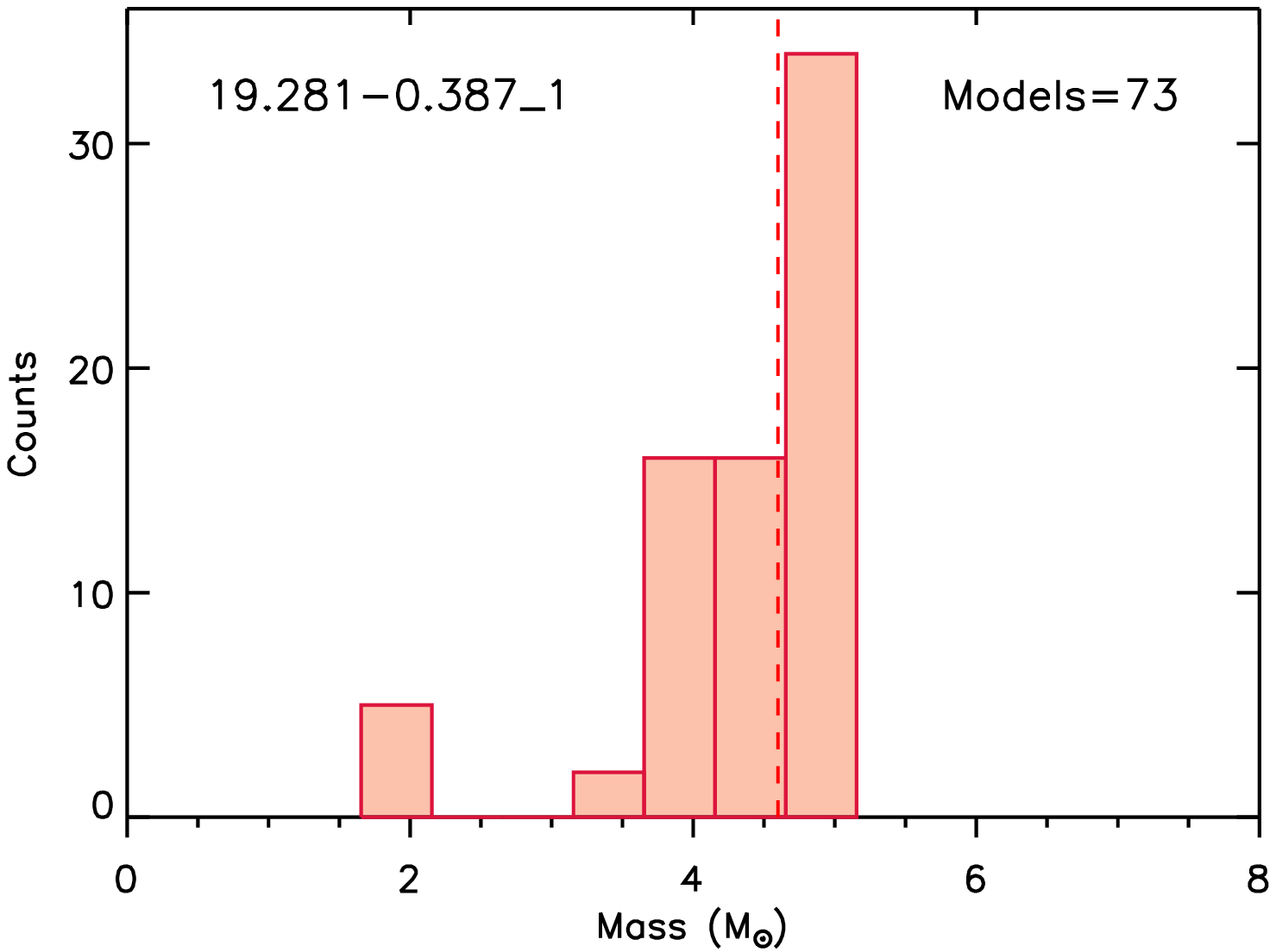}  
\includegraphics[width=6cm]{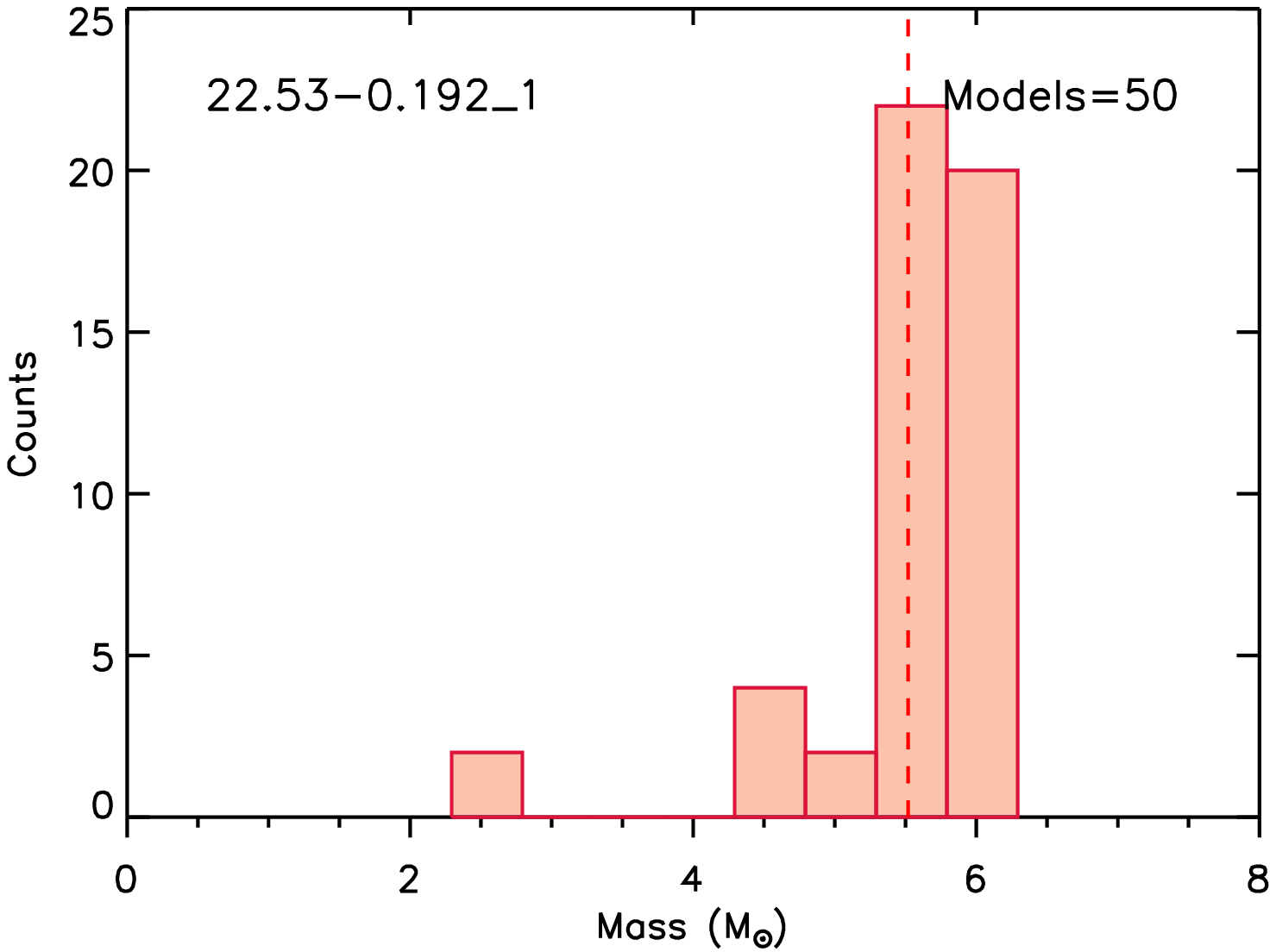}  
\includegraphics[width=6cm]{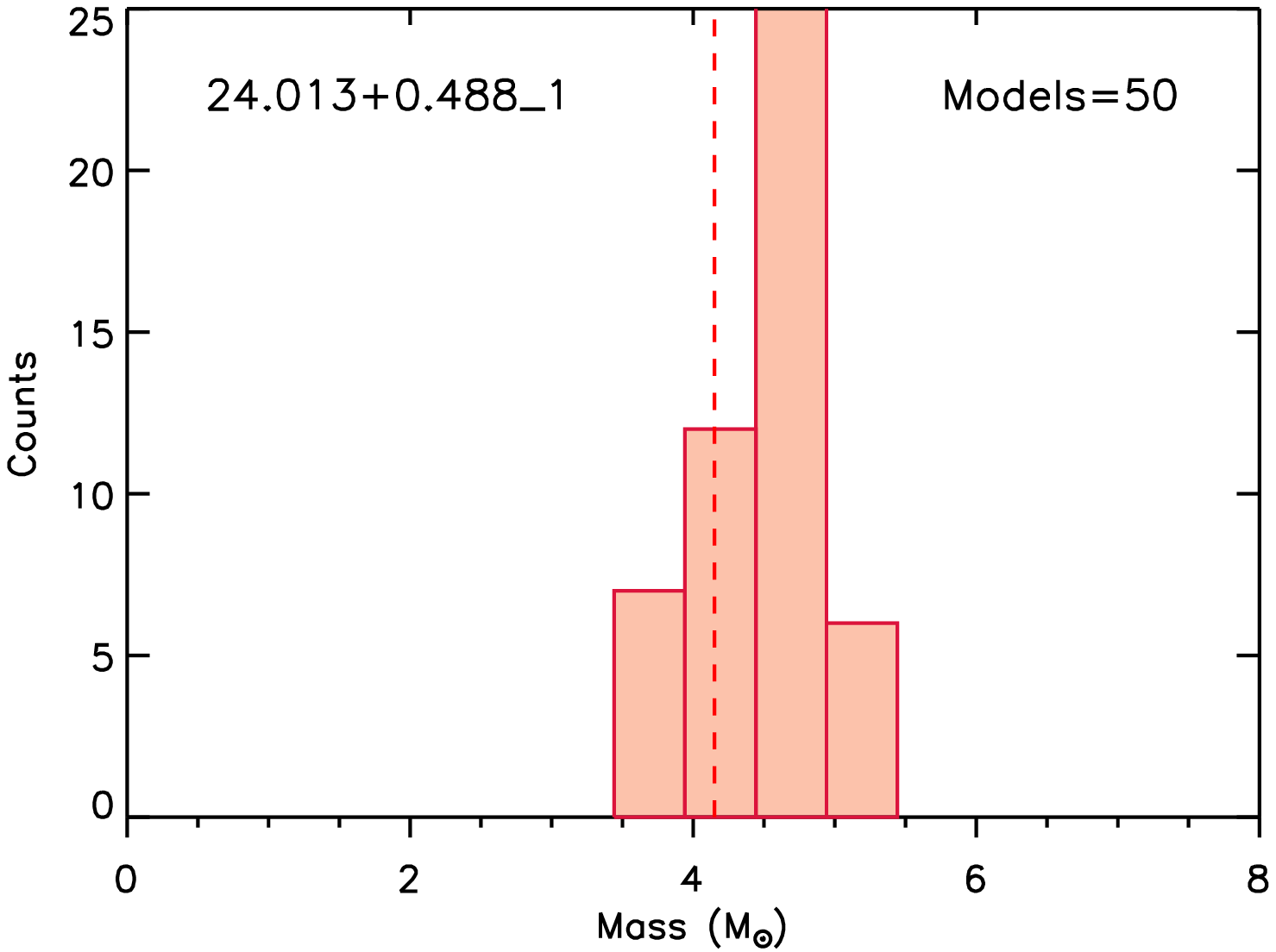}  
\includegraphics[width=6cm]{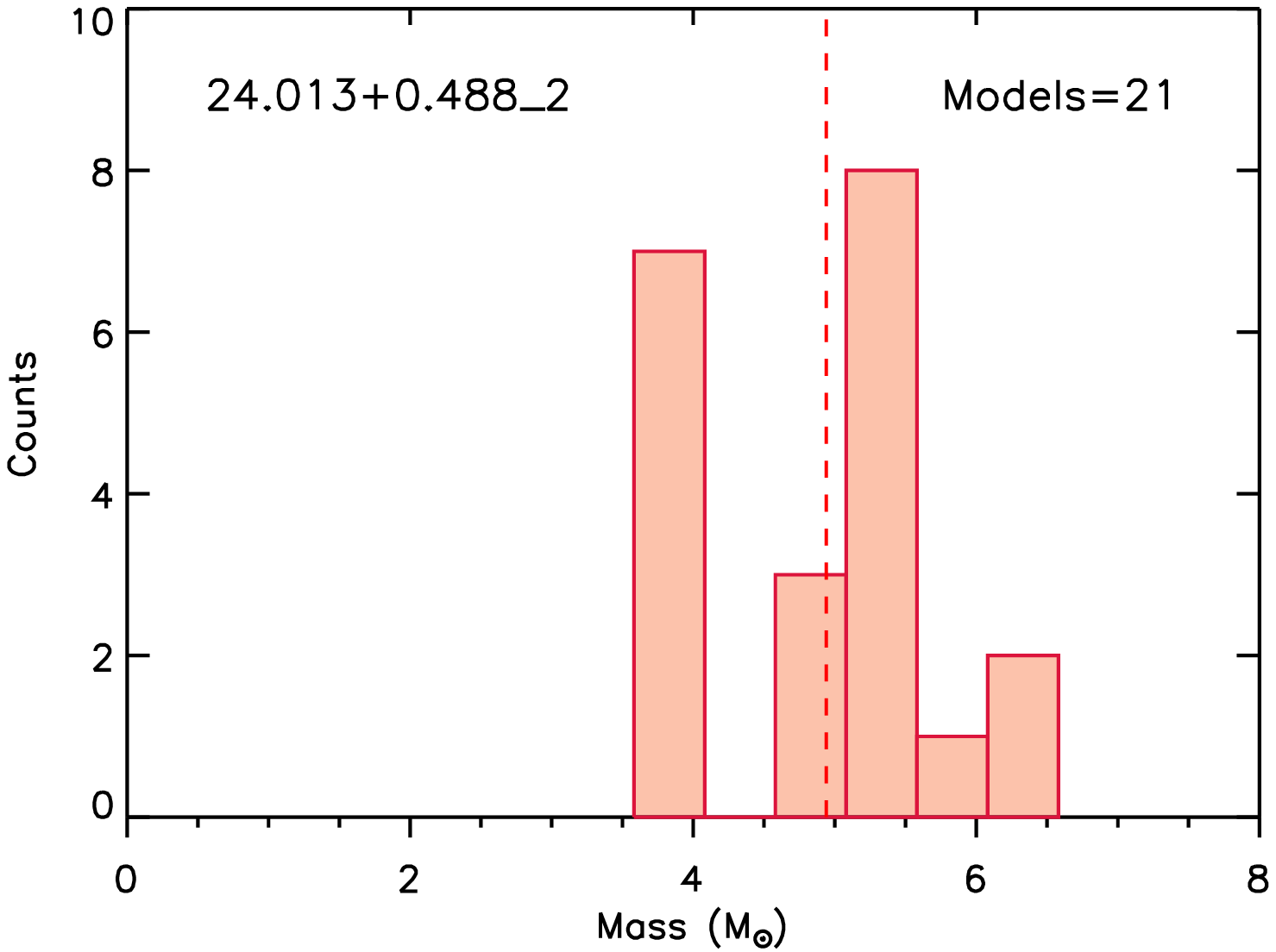}  
\includegraphics[width=6cm]{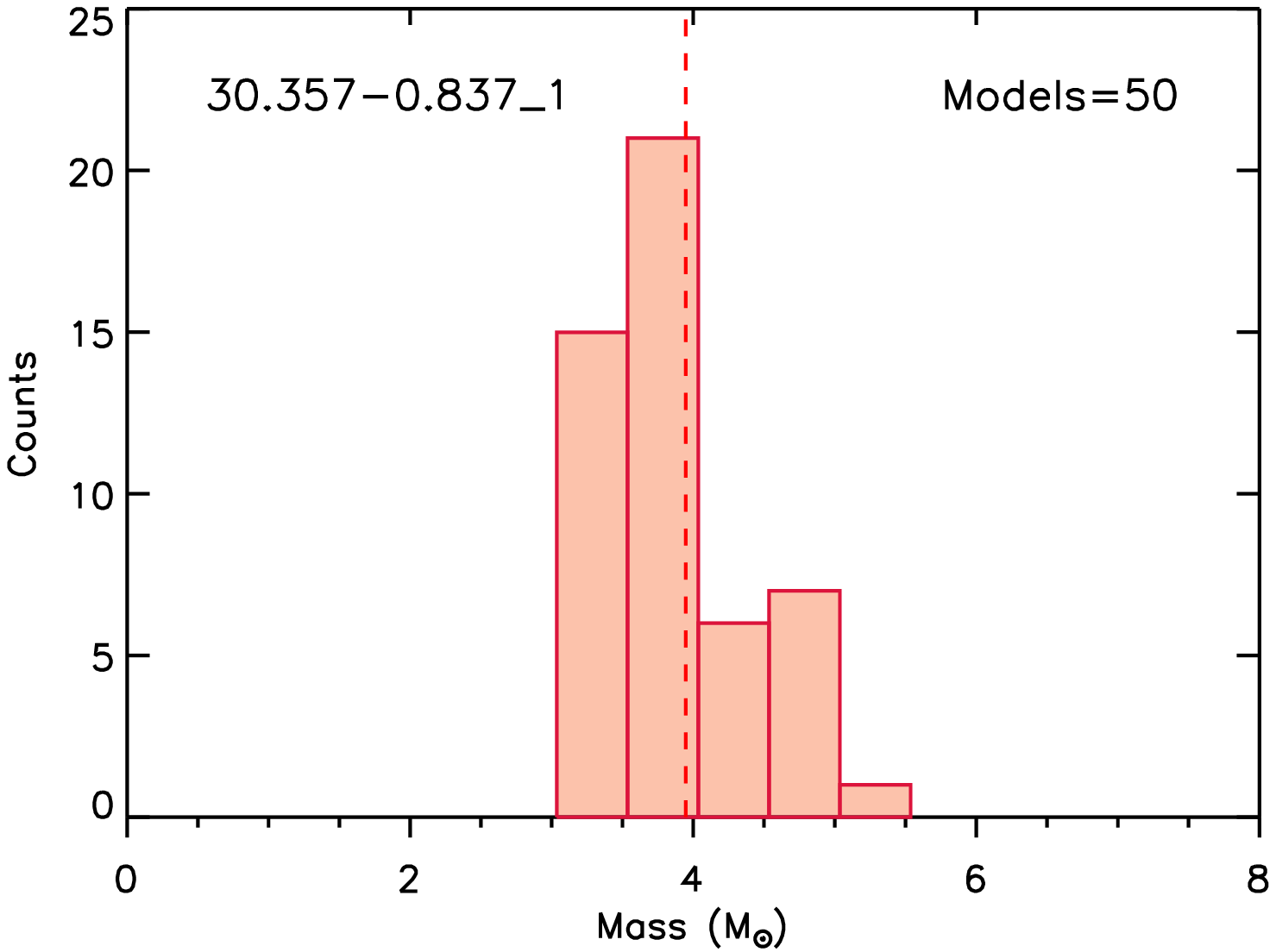}  
\includegraphics[width=6cm]{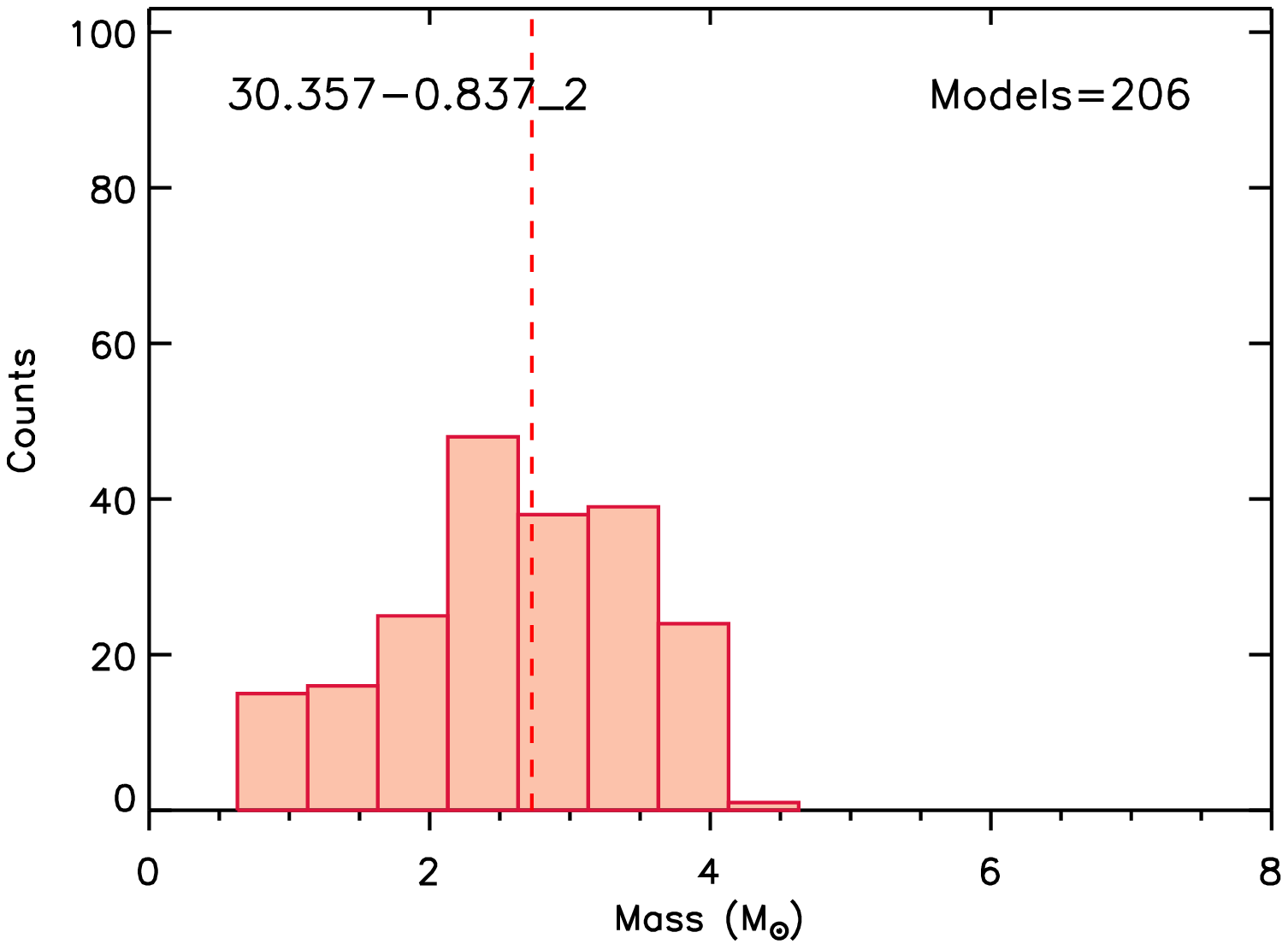}  
\caption{Same of Figure \ref{fig:Robitaille_fit_Av}, but for the mass of central stars.}
\label{fig:Robitaille_fit_mass}
\end{figure*}

The absence of a 70\mum\ counterpart in a FIR/submm bright clump is not a good indicator that the clumps are starless. Some of these clumps embed already formed intermediate mass stars, most of which may still be in the process of accreting. 

In the next Sections we analyze the data obtained from the dense gas tracers and we explore if these data may help to distinguish between clumps with and without 24\mum\ counterparts.

\begin{center}
\begin{table*}
\centering
\begin{tabular}{c|c|c|c|c|c|c}
\hline
\hline
Clump &	Class & A$_{V}$	& $\Sigma$	&	M$_{*}$	&	$\chi^{2}$	& Models \\
			&	 &   (mag)		&   (g cm$^{-2}$)	& (M\sun)	&	 & \\
\hline
18.787-0.286\_1		&	I &	56.4$\pm$4.6	&	12.7$\pm$1.0	&	4.65$\pm$0.62	& 7.69$-$16.42 & 50 \\
18.787-0.286\_2		&	I & 	75.9$\pm$16.9	&	17.0$\pm$3.8	&	2.87$\pm$0.98	&	0.38$-$5.00 & 1178 \\
19.281-0.387$^{1}$		&		I-II & 66.5$\pm$8.3	&	14.9$\pm$1.9	&	4.60$\pm$0.55	&	0.36$-$5.00 & 73 \\
22.53-0.192		&		I & 93.4$\pm$9.3	&	21.0$\pm$2.1	&	5.52$\pm$0.73	&	8.68$-$11.00 & 50 \\
24.013+0.488\_1		&	I & 63.0$\pm$2.2	&	14.1$\pm$0.5	&	4.15$\pm$0.35	&	0.02$-$8.10& 50 \\
24.013+0.488\_2		&	I & 69.6$\pm$14.0 &	15.6$\pm$3.1	&	4.94$\pm$0.96	&	12.90$-$27.70 & 21 \\
30.357-0.837\_1		&	II & 36.8$\pm$4.1	&	8.3$\pm$0.9	&	3.95$\pm$0.45	&	0.69$-$5.00 & 50 \\
30.357-0.837\_2		&	II & 25.7$\pm$5.9	&	5.8$\pm$1.3	&	2.73$\pm$0.82	&	0.12$-$5.00 & 206 \\
\hline
\end{tabular}
\begin{tablenotes}
\scriptsize
\item $^{1}$ 19.281-0.387 has a Class I and Class II source associated with the same 24\mum\ counterpart. 
\end{tablenotes}
\caption{Physical parameters and classification of the central star in each clump with NIR/MIR counterparts. Col 1: Clump name; Col. 2: Source classification obtained as described in the text; Col. 3: Average magnitude; Col. 4: Average column density, obtained from A$_{V}$ using the conversion described in \citet{Bohlin78}; Col. 5: Average mass of the central star; Col.6: $\chi^{2}$ range of the models used to estimate the weighted parameters as described in the text; Col. 6: Number of models used to estimate the physical parameters. The average values and uncertainties associated with the estimation of A$_{V}$, column density and mass of the central star have been evaluated as the weighted mean and variance of the distributions with weights equal to the inverse of the $\chi^{2}$ value of each fit.}
\label{tab:glimpse_robitaille}
\end{table*}
\end{center}

\section{Line analysis}\label{sec:line_analysis}
We detected \n2h (1$-$0), HNC (1$-$0) ad \hco\ (1$-$0) emission in all our clumps. We excluded from the line analysis the clump 24.528-0.136 as it shows absorption features due to the contamination of the off-position and 30.454-0.135 because the spectra show at least two components along the line of sight  with similar intensities but we cannot separate the two sources in the dust continuum data. The spectra of the 16 clumps are presented in Appendix \ref{sec:app_spectra}.

\subsection{N$_{2}$H$^{+}$}
The \n2h (1$-$0) spectra were fitted using the CLASS \texttt{hfs} task, which takes into account all the hyperfine components. Total optical depth $\tau_{tot}$, central \n2h\ velocity v$_{LSR}$ and velocity dispersion $\sigma$ are estimated directly from the fit, following the prescription of the CLASS manual. These are shown in Table \ref{tab:n2h_fit}. The total optical depth varies in the range $0.3\leq\tau_{tot}\leq12.2$. In four clumps the uncertainties associated with the estimation of the optical depth, $\sigma_{\tau}$, are such that $\tau_{tot}\leq3\sigma_{\tau}$. In two of these clumps, 28.178-0.091 and 28.792+0.141, the fit gives $\tau_{tot}=0.10$ which is the lowest allowed value permitted in the \texttt{hfs} routine. For these four clumps we evaluated the \textit{r.m.s.} of the spectrum in CLASS before and after the subtraction of the line fit. These values differ for less than 10\%, suggesting that the fits are well constrained, so we accepted the parameters estimated from the fit. The optical depth of the main component can be recovered from $\tau_{tot}$ as $\tau_{main}=r_{i}*\tau_{tot}$, with $r_{i}$ the relative intensity of the main hyperfine component (0.259). The majority of the clumps for which the fit converges are moderately optically thin with $\tau_{main}\lesssim1$. The average value is $<\tau_{main}>=0.61$, in agreement with the average optical depth of quiescent clumps associated with IRDCs \citep{Sanhueza12}. Two sources have $\tau_{main}\geq1$. There is an indication that the \n2h\ $(1-0)$ optical depth decreases as the clump evolves from a quiescent to an active phase \citep[from $\overline{\tau}_{quiescent}=0.8$ to $\overline{\tau}_{active}=0.5$,][]{Sanhueza12}, to increase again when the protostellar core forms an active HII region  identified by a bright 8\mum\ emission \citep[$\overline{\tau}_{red}=1.8$,][]{Sanhueza12}. Lower optical depths in protostellar cores with respect to prestellar cores are also observed in low-mass star forming regions with comparable column densities \citep{Crapsi05,Emprechtinger09}.

The excitation temperature \tex\ is derived assuming LTE:

\begin{equation}
\mathrm{T}_{ex}=\frac{\mathrm{T}_{0}}{\mathrm{ln(A^{-1}+1)}} \qquad \mathrm{with}
\end{equation}

\begin{equation}
A=\frac{\mathrm{T}_{b}}{\mathrm{T}_{0}(1-e^{-\tau})}+\frac{1}{e^{\mathrm{T}_{0}/\mathrm{T}_{bg}}-1}
\end{equation}

where $\mathrm{T}_{0}=h\nu/k$, $\mathrm{T}_{bg}=2.7$ K and $\mathrm{T}_{b}=\mathrm{T}^{*}_{A}\ F_{eff}/B_{eff}$. $F_{eff}$ and $B_{eff}$ are the telescope forward and beam efficiency respectively. We assume $F_{eff}=0.98$ and $B_{eff}=0.78$ \citep{Rygl13}.

\tex\ varies in the range $3.0\lesssim\mathrm{T}_{ex}\lesssim22.7$ K, as shown in Table \ref{tab:n2h_fit}. Assuming a kinetic temperature  T$_{k}=10$ K, similar to the average dust temperature of these sources, all but three clumps (28.178-0.091, 28.792+0.141, 31.946+0.076) are not thermalised, with \tex$\lesssim$T$_{k}$. However, we assumed a filling factor of 1 for all the observations which may overestimate the region of \n2h\ emission and underestimate \tex. 

The \n2h\ column density has been derived as in \citet{Caselli02} and \citet{Pineda13}, with a dipole magnetic moment $\mu_{D}=3.4$ Debye and a rigid rotor rotational constant of $B=46.58687$ GHz \citep[][and references therein]{Sanhueza12}. The \n2h\ column densities are in Table \ref{tab:gas_abundance}. The average \n2h\ column density is 9.2$\times10^{12}$ cm$^{-2}$, with a maximum of $\simeq1.7\times10^{13}$ cm$^{-2}$ in 24.013+0.488. The \n2h\ column density estimates are similar to what is observed in other star forming regions \citep{Sanhueza12,Rygl13}.

The \n2h\ abundance with respect to the H$_{2}$, $X($\n2h), the latter derived from the dust surface density and assuming a mean molecular weight of 2.33, is in the range $0.5\times10^{-10}\leq X($\n2h) $\leq9.4\times10^{-10}$ (Table \ref{tab:gas_abundance}), with an average value of $<X$(\n2h$)>=2.9\times10^{-10}$. These values are likely to be underestimated as we are assuming a filling factor of 1, however they are in agreement with the findings in massive dense cores/clumps \citep{Pirogov07,Rygl13} and in massive clumps associated with IRDCs \citep{Ragan06,Sanhueza12}. The source with the highest \n2h\ abundance, $X($\n2h)$\simeq10^{-9}$, is 32.006-0.51. This is one of the sources with the lowest mass surface density, is below the KP threshold for the formation of massive stars (see Section \ref{sec:massive_stars_from_dust}) and does not embed any 24\mum\ source. However 15.631-0.377, one of the other source below the KP threshold and with no 24\mum\ counterparts, has a \n2h\ abundance close to the average. There is no clear indication that $X($\n2h) is different between sources with and without a 24\mum\ counterpart.

\begin{center}
\begin{table}
\centering
\begin{tabular}{c|c|c|c|c}
\hline
\hline
Clump & v$_{LSR}$	&	$\sigma$	&	T$_{ex}$  &	$\tau_{tot}$ \\
		  &	(km s$^{-1}$)  		&		(km s$^{-1}$)  &  (K)    &     \\
\hline
		  
15.631-0.377  &  40.2(0.1) &  0.30(0.01)  &  3.41(0.70)  &    3.75(0.73) \\
18.787-0.286   & 65.7(0.1)  &  1.07(0.01)  &  5.20(0.39)  &   1.86(0.13)\\
19.281-0.387  &  53.6(0.1)  &  0.47(0.03)  &  3.15(0.80)  &   5.33(1.17)\\
22.53-0.192  &  76.3(0.1)  &  1.25(0.01)  &  5.01(0.13)   &  1.22(0.03)\\
22.756-0.284  &  105.1(0.1)  &  0.95(0.01)  &  4.25(0.52)  &  1.62(0.19)\\
23.271-0.263  &  82.3(0.1)  &  0.94(0.01)  &  8.32(2.03)  &  0.62(0.15)\\
24.013+0488 &   94.5(0.1) &   0.91(0.02)  &  3.76(0.36) &  3.67(0.31)\\
25.609+0.288  &  113.6(0.1)  &  1.05(0.01)  &  6.64(1.43)  &   0.61(0.13)\\
25.982-0.056  &  89.8(0.1)  &  0.69(0.01)  &  4.19(0.80)   &  1.58(0.29)\\
28.178-0.091  &  98.2(0.1)  &  1.07(0.02)  &  22.71(20.44)  &   0.10(0.09)\\
28.537-0.877  &  88.3(0.1) &   0.78(0.01)  &  4.37(0.35)    &   2.28(0.17)\\
28.792+0.141  &  107.2(0.1)  &   0.99(0.03)  &  13.91(12.52)  &    0.10(0.09)\\
30.357-0.837  &  78.8(0.1)  &  0.57(0.13) &   4.06(4.25)  &   0.90(0.81)\\
31.946+0.076 &   96.4(0.1)  &  1.19(0.01)  & 11.52(0.36)   &   0.32(0.01)\\
32.006-0.51 &  71.6(0.1)  &  0.32(0.02)  &  3.02(0.70)    &   12.19(2.19)\\
34.131+0.075  &  56.8(0.1)  &  0.74(0.06)  &  3.85(3.36)  &    1.34(0.90)\\
\hline
\end{tabular}
\caption{\n2h\ parameters. Col. 1: Clump name; Col. 2: \n2h\ central velocity; Col. 3: Velocity dispersion measured as $1/(8\mathrm{ln}2)^{1/2}\times$FWHM of the hyperfine fit; Col. 4: Excitation temperature; Col. 5: Total optical depth.} 
\label{tab:n2h_fit}
\end{table}
\end{center}

\subsection{HNC and HCO$^{+}$}\label{sec:abundances}
The HNC and \hco\ optical depth and gas column density cannot be directly estimated from the data as we have only observed the HNC and \hco\ ($1-0$) line, which we expect to be optically thick within these cold, dense regions \citep[e.g][]{Sanhueza12}. 

We have estimated the \hco\ and HNC column densities with RADEX \citep{Van_der_Tak07} in a similar fashion to \citet{Peretto13}. We have well constrained measurements of the dust column density, gas temperature (assumed equal to the dust temperature) and velocity dispersion (from \n2h\ emission) for our clumps, and the only unknown variable is the gas column density. We run RADEX iteratively assuming different values of the \hco\ and HNC column densities until the evaluated radiation temperature matched the measured peak temperature. We consider this temperature as the temperature of the main peak in the \hco\ and HNC spectra. RADEX allows also the estimation of the optical depth of the lines. As showed in Table \ref{tab:gas_abundance}, both the \hco\ and HNC lines are, within the uncertainties, optically thick in all clumps. Therefore, the measured temperature is a lower limit for the temperature of the main peak in both \hco\ and HNC spectra, which gives a lower limit to the estimated gas column density. The \hco\ and HNC column densities with the uncertainties and the abundances relative to the dust and to the \n2h\ are in Table \ref{tab:gas_abundance}. Although with the strong caveat that this procedure gives only an estimate of the gas column density using a single optically thick line observation, the values we found are comparable to those found in massive clumps \citep[e.g.][]{Miettinen14,Zhang16}. We do not find any significant differences in the HNC or \hco\ abundances between clumps that host a 24\mum\ counterpart and clumps without any MIR counterparts. Due to the uncertainties in these measurements however, we cannot give definitive conclusions on the observed trends.

\begin{center}
\begin{table*}
\centering
\begin{tabular}{c|c|c|c|c|c|c|c|c|c}
\hline
\hline
Clump &	N(H$_{2}$)	&	N(\n2h)		& N(\hco)	& N(HNC)	& $X($\n2h)  & $X($\hco)  & $X($HNC)  & $\tau$(\hco) & $\tau$(HNC)\\
		  &	(10$^{22}$ cm$^{-2}$)		&			(10$^{12}$ cm$^{-2}$)  &  	(10$^{12}$ cm$^{-2}$)  &  (10$^{12}$ cm$^{-2}$)    &   	(10$^{-10}$)    &  (10$^{-10}$)    & 	(10$^{-10}$ )  &  &   \\
\hline
15.631-0.377   &  1.58(0.26)   &   4.97(0.55)   &   2.00$^{+2.40}_{-0.90}$   &         12.70$^{+13.70}_{-5.40}$ & 3.14(0.62)   &   1.26$^{+1.53}_{-0.60}$     &  8.02$^{+8.76}_{-3.66}$ &  4.77$^{+10.23}_{-2.63}$  & 17.29$^{+34.89}_{-10.17}$  \\
18.787-0.286   &    6.91(1.55)  &   15.94(0.94)  &    6.20$^{+7.40}_{-2.70}$  &    29.10$^{+32.70}_{-12.70}$ &  2.31(0.53)  &    0.90$^{+1.09}_{-0.44}$  &  4.21$^{+4.82}_{-2.07}$ & 3.03$^{+6.57}_{-1.70}$  & 8.79$^{+17.90}_{-5.14}$  \\
19.281-0.387  &   2.68(0.52)   &  10.00(1.21)   &   2.30$^{+2.70}_{-1.00}$   &   9.60$^{+10.40}_{-4.10}$ &  3.73(0.85)  &    0.86$^{+1.02}_{-0.41}$   &  3.58$^{+3.94}_{-1.68}$  & 3.15$^{+6.86}_{-1.75}$  & 8.33$^{+16.90}_{-4.82}$ \\
22.53-0.192   &   4.24(1.03)   &  11.55(0.23)  &   22.70$^{+28.20}_{-10.00}$  &    84.50$^{+97.30}_{-37.20}$ &  2.72(0.67)  &    5.35$^{+6.78}_{-2.70}$   &  19.93$^{+23.46}_{-10.03}$  &  8.14$^{+17.25}_{-4.73}$  & 19.56$^{+40.15}_{-11.51}$ \\
22.756-0.1284   &    3.72(0.78)  &    9.18(0.74)  &    0.90$^{+0.70}_{-0.40}$  &    9.40$^{+10.60}_{-4.10}$ &  2.47(0.55)  &    0.24$^{+0.19}_{-0.12}$    &  2.53$^{+2.90}_{-1.22}$  &  0.62$^{+1.26}_{-0.37}$  & 4.14$^{+8.72}_{-2.34}$ \\
23.271-0.263   &   3.31(0.78)    &  9.88(2.62)   &   10.00$^{+11.80}_{-4.50}$   &   37.30$^{+40.00}_{-16.40}$ &  2.99(1.06)  &    3.02$^{+3.64}_{-1.53}$   & 11.28$^{+12.38}_{-5.62}$ &  5.74$^{+12.04}_{-3.26}$ & 13.70$^{+27.41}_{-8.02}$ \\
24.013+0.488   &     5.13(1.03)   &  16.81(0.93)  &    6.20$^{+7.40}_{-2.70}$  &    26.40$^{+27.20}_{-11.90}$ &  3.28(0.69)   &   1.21$^{+1.46}_{-0.58}$  &   5.15$^{+5.41}_{-2.54}$ &  3.86$^{+8.27}_{-3.16}$  & 10.24$^{+20.21}_{-5.89}$ \\
25.609+0.228   &      5.66(1.29)  &    7.50(1.53)  &    5.70$^{+7.00}_{-2.50}$  &    17.30$^{+20.00}_{-7.60}$ &  1.33(0.41)  &    1.01$^{+1.26}_{-0.50}$  &   3.06$^{+3.60}_{-1.51}$  &  3.33$^{+7.31}_{-1.83}$  & 6.65$^{+14.01}_{-3.81}$ \\
25.982-0.056   &     2.39(0.52)   &   6.37(0.80)  &   19.10$^{+23.60}_{-8.20}$  &    46.40$^{+52.70}_{-19.10}$ &  2.67(0.67)  &    8.00$^{+10.04}_{-3.85}$  &   19.44$^{+22.48}_{-9.04}$ & 13.08$^{+27.68}_{-7.78}$ & 22.88$^{+47.06}_{-13.89}$ \\
28.178-0.091   &    4.98(1.03)   &   2.24(2.24)   &   6.30$^{+7.30}_{-2.80}$   &   24.50$^{+27.30}_{-10.90}$ &  0.45(0.45)   &   1.27$^{+1.49}_{-0.62}$    &  4.92$^{+5.58}_{-2.42}$ & 3.36$^{+7.21}_{-1.88}$ & 8.46$^{+17.29}_{-4.86}$ \\
28.537-0.277   &     4.41(0.78)  &   11.04(0.60)   &   2.50$^{+3.00}_{-1.00}$  &    20.90$^{+21.80}_{-9.10}$ &  2.50(0.46)  &    0.57$^{+0.69}_{-0.25}$   &  4.74$^{+5.01}_{-2.22}$  & 2.12$^{+4.66}_{-1.18}$  & 9.77$^{+19.39}_{-5.71}$ \\
28.792+0.141   &     2.08(0.52)   &   1.53(1.53)  &   16.40$^{+18.10}_{-7.00}$  &    56.40$^{+61.80}_{-23.70}$ &  0.74(0.74)  &    7.90$^{+8.94}_{-3.90}$  &   27.17$^{+30.53}_{-13.27}$  & 9.07$^{+18.45}_{-5.36}$  & 20.06$^{+40.49}_{-12.07}$\\
30.357-0.837   &     1.42(0.26)   &  2.87(1.85)   &   1.90$^{+2.30}_{-0.80}$  &   9.00$^{+9.20}_{-3.60}$ &  2.02(1.35)  &    1.34$^{+1.63}_{-0.61}$  &   6.33$^{+6.57}_{-2.58}$ & 3.00$^{+6.54}_{-1.70}$ & 8.69$^{+17.33}_{-5.24}$  \\
31.946+0.076   &     3.66(0.78)   &   11.28(0.45)   &  17.30$^{+20.00}_{-8.00}$  &    66.40$^{+79.10}_{-29.10}$ &  3.08(0.67)  &    4.73$^{+5.56}_{-2.41}$  &   18.15$^{+21.96}_{-8.84}$  & 7.62$^{+15.89}_{-4.25}$ & 18.40$^{+38.58}_{-10.78}$ \\
32.006-0.51   &     1.57(0.26)  &   14.78(1.49)  &    3.20$^{+3.40}_{-1.40}$  &    10.90$^{+11.80}_{-4.30}$ &  9.39(1.81)  &    2.03$^{+2.19}_{-0.95}$  &   6.93$^{+7.58}_{-2.96}$   & 6.76$^{+13.82}_{-3.92}$ & 14.58$^{+29.68}_{-9.00}$ \\
34.131+0.075   &     2.73(0.52)   &   11.81(6.01)   &   1.70$^{+2.10}_{-0.60}$   &   10.00$^{+11.80}_{-4.20}$ &  4.33(2.35)   &   0.62$^{+0.78}_{-0.25}$    &  3.67$^{+4.38}_{-1.69}$ & 1.60$^{+3.60}_{-0.96}$ & 5.71$^{+12.15}_{-3.35}$ \\
\hline	  
\hline

\end{tabular}
\caption{\n2h\ and \hco\ column density and abundances of our clumps. Col. 1: Clump name; Col. 2: H$_{2}$ column density derived from the parameters in Table \ref{tab:SED_parameters}; Col. 3: \n2h\ column density; Col. 4-5: \hco\ and HCN column densities estimated iterating RADEX as described in Section \ref{sec:abundances}; Col. 6-8: \n2h, \hco\ and HNC abundances with respect to H$_{2}$;  Col. 9-10: Optical depth of \hco\ and HNC spectra as obtained from the RADEX run.}
\label{tab:gas_abundance}
\end{table*}
\end{center}

\subsection{Skewness in HNC and \hco\ spectra}
Optically thick line profiles can be used to identify signatures of dynamical activity in star forming regions. The difference in velocity between the brightest peaks of an optically thick line and an optically thin line can be used to compute the the skewness parameter $\delta$v \citep{Mardones97}, defined as

\begin{equation}
\delta\mathrm{v}=\frac{\mathrm{v}_{\mathrm{thick}}-\mathrm{v}_{\mathrm{thin}}}{\Delta\mathrm{v}_{\mathrm{thin}}}
\end{equation}
where $\mathrm{v_{thick}}$ is the LSR velocity of the brightest \hco\ or HNC peak, $\mathrm{v_{thin}}$ and $\Delta\mathrm{v_{thin}}$ are respectively the LSR velocity and FWHM of \n2h\, assumed as an optically thin line. We estimate $\mathrm{v_{thick}}$ fitting 2 Gaussians to each \hco\  and HNC spectrum with the $\texttt{mpfitfun}$ IDL routine \citep{Markwardt09}. In clumps without a well defined double peak in the \hco\ or HNC spectra, $\mathrm{v_{thick}}$ was estimated with a single Gaussian. The skewness parameters are in Table \ref{tab:skewness}. \citet{Mardones97} define $\vert\delta\mathrm{v}\vert>0.25$ as a significant detection of skewness.

A positive skewness parameter indicates a red asymmetry in the spectrum, which could be interpreted as signature of outflows activity \citep[e.g.][]{Peretto06}.  A negative $\delta\mathrm{v}$ conversely is indicative of blue-asymmetric spectrum, a signature of infall motions in both \hco\ \citep[][]{Fuller05} and HNC \citep{Kirk13} spectra. Simulations of infalling high-mass star forming regions showed however that red-asymmetric \hco\ $(1-0)$ spectra may be observed also in absence of outflows activity \citep{Smith13}. Outflows are more reliably traced by looking for high-velocity wing emission away from the systemic velocity of the cloud. Since this is outside the scope of this paper, for the purpose of this work, we simply interpret a significant value of $\delta\mathrm{v}$ in the \hco\ spectra as an indication of significant dynamical activity, regardless of their origin. This is the case for all but 2 clumps. Five HNC spectra have $\vert\delta\mathrm{v}\vert\leq0.25$, an indication that \hco\ emission traces more dynamically active gas as further explored in Paper II. Note that the clump 25.982-0.056 has symmetric \hco\ and HNC line profiles with peaks shifted from the \n2h\ central velocity. This clumps has a skewness parameter higher than in some asymmetric, double-peaked spectra (e.g. 28.178-0.091, see spectra in Appendix \ref{sec:app_spectra}).

In the following, we restrict the analysis to clumps with asymmetric line profiles and significant blue-shifted peaks (i.e. with $\delta\mathrm{v}\leq-0.25$), as these profiles are the more unambiguously associated with infall motions. We identify 7 clumps with significant infall signatures in either \hco\ and HNC spectra, or both. The \hco\ spectrum of 22.756-0.284 has the blue- and red-shifted peaks of the same intensity. The HNC spectra on the contrary has a blue-shifted peak, compatible with infall signatures. Three of these clumps (22.53-0.192, 24.013+0.488 and 30.357-0.837) embed a Class I or II source (Tables \ref{tab:glimpse_robitaille} and \ref{tab:gas_abundance}). It is very likely that these clumps are already collapsing and these central stars are still accreting from the surrounding clumps.

\begin{center}
\begin{table}
\centering
\begin{tabular}{c|c|c}
\hline
\hline
Clump	&	$\delta$v (\hco) & $\delta$v (HNC)  \\
			&	&  \\	
\hline
15.631-0.837 & -0.06 	& 0.06	\\
18.787-0.286 & 0.71 & 0.58  \\
19.281-0.387 & 0.70 & 0.75   \\
22.53-0.192 &	-0.28 & -0.18   \\
22.756-0.284 &	1.25	& -0.50   \\
23.271-0.263 &	0.28	& 0.31  \\
24.013+0.488 & -0.70	& -0.58   \\
25.609+0.228 & -0.64	&	-0.64  \\
25.982-0.056 & 0.48 & 0.36  \\
28.178-0.091 &	-0.82 & 	0.19   \\
28.537-0.277 &	0.89	&	0.63	 \\
28.792+0.141 & -0.01 & 0.08  \\
30.357-0.837 &	-0.43	& -0.32	  \\
31.946+0.076 &  -0.95	& -0.23   \\ 
32.006-0.51 & 1.42	& 1.53   \\
34.131+0.075 & 1.36	& -0.11   \\
\hline
\end{tabular}
\caption{Skewness parameter $\delta$v evaluated in each clump according to the definition of \citet{Mardones97}. Col. 1: clump name; Col. 2-3: skewness parameter evaluated for the \hco\ and HNC spectra respectively.}
\label{tab:skewness}
\end{table}
\end{center}

\subsection{Infalling properties}\label{sec:infalling_properties}
The spectra of the seven clumps with blue-shifted peaks can be used to determine their infall velocities \vinf and mass accretion rates $\dot{\mathrm{M}}$. 

\begin{figure*}
\centering
\includegraphics[width=6cm]{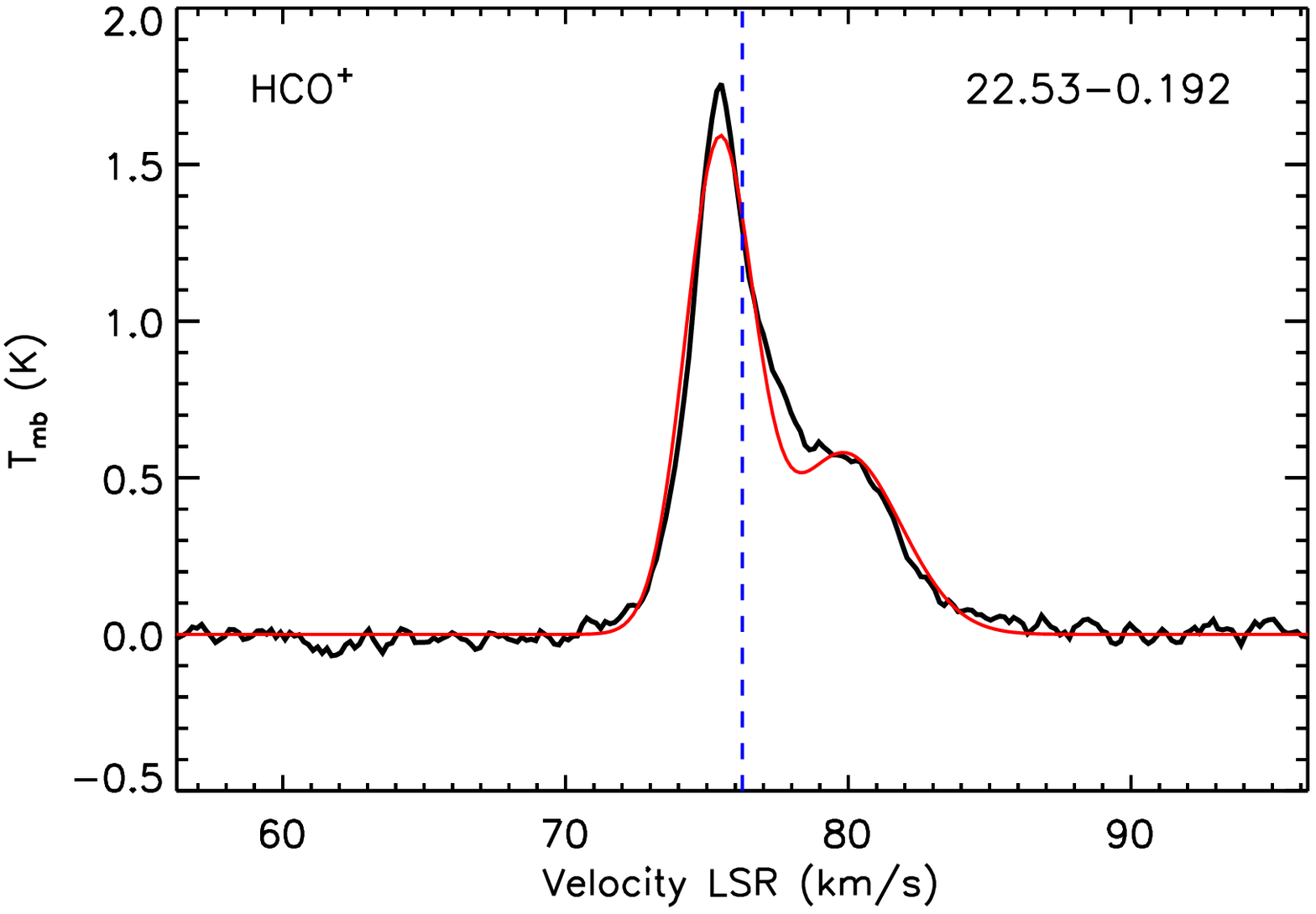}  
\includegraphics[width=6cm]{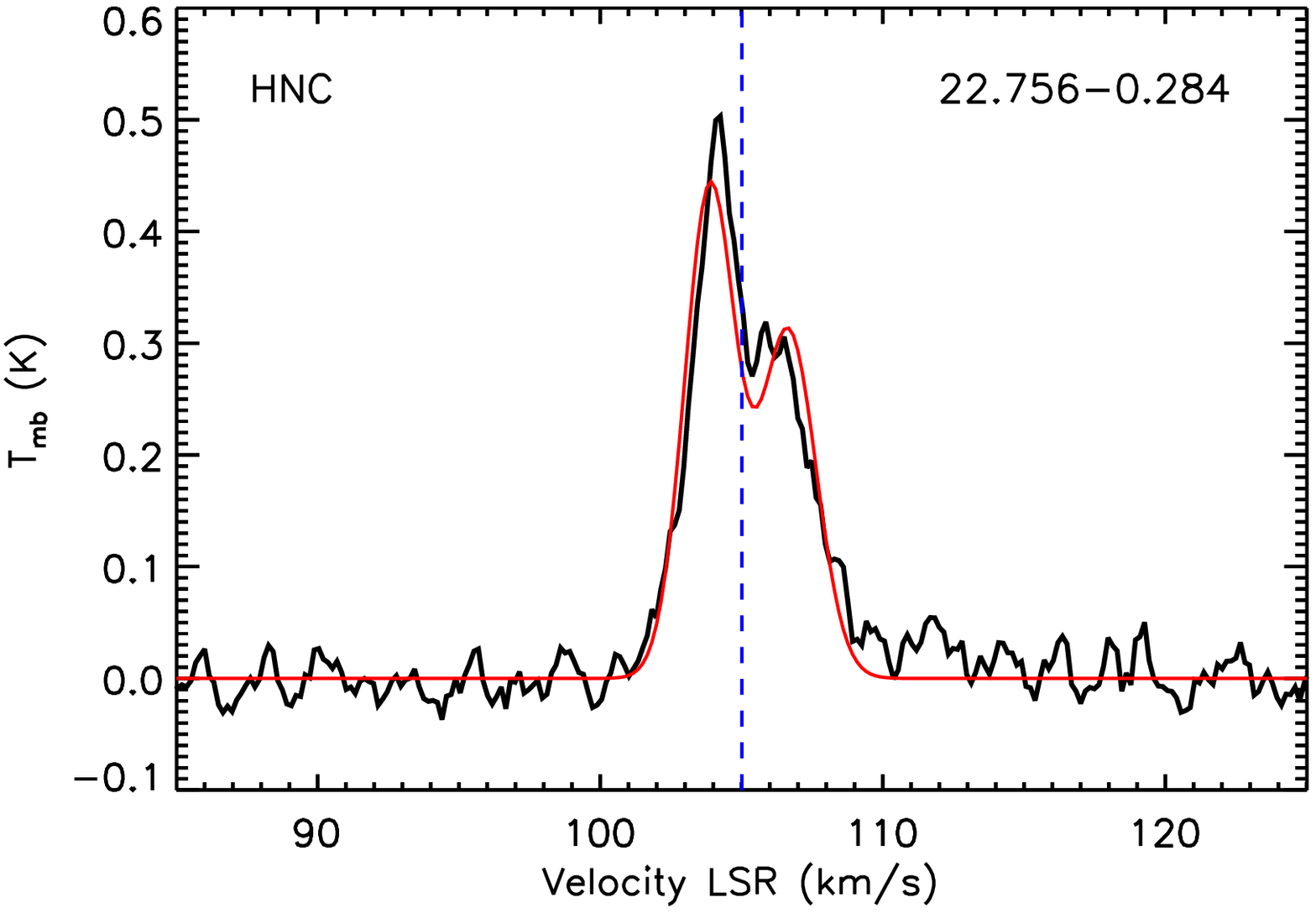}  
\includegraphics[width=6cm]{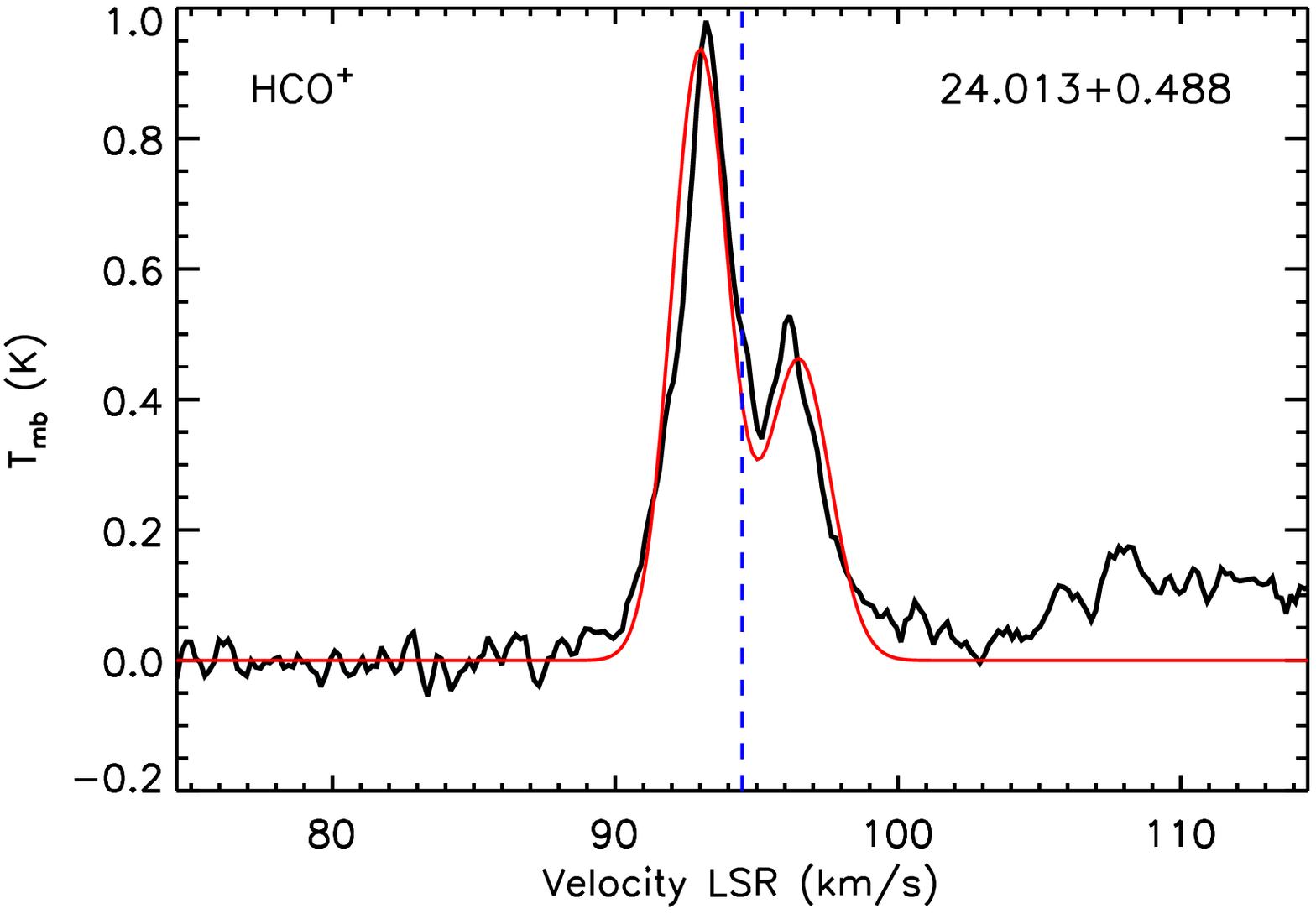}   \includegraphics[width=6cm]{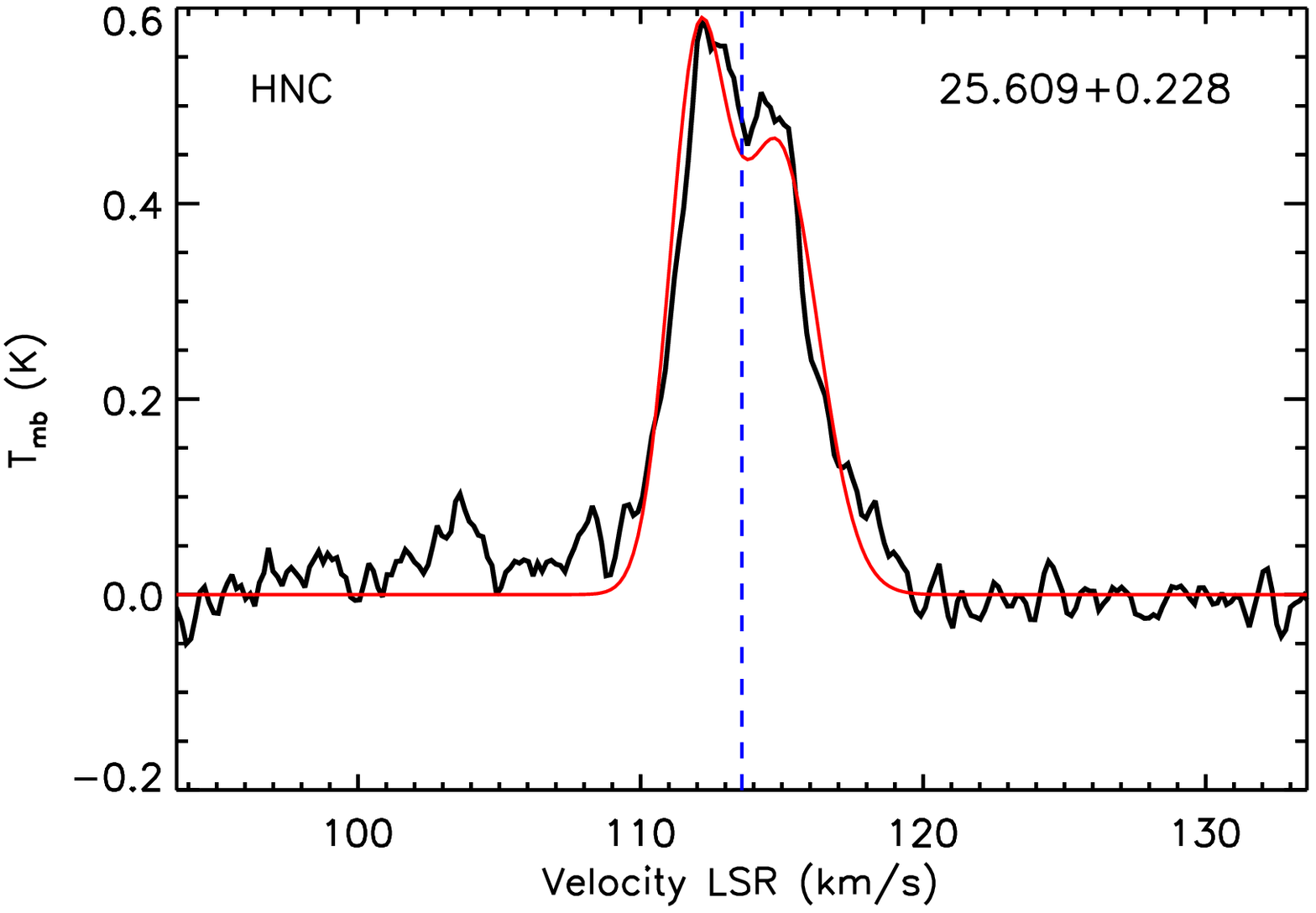}  
\includegraphics[width=6cm]{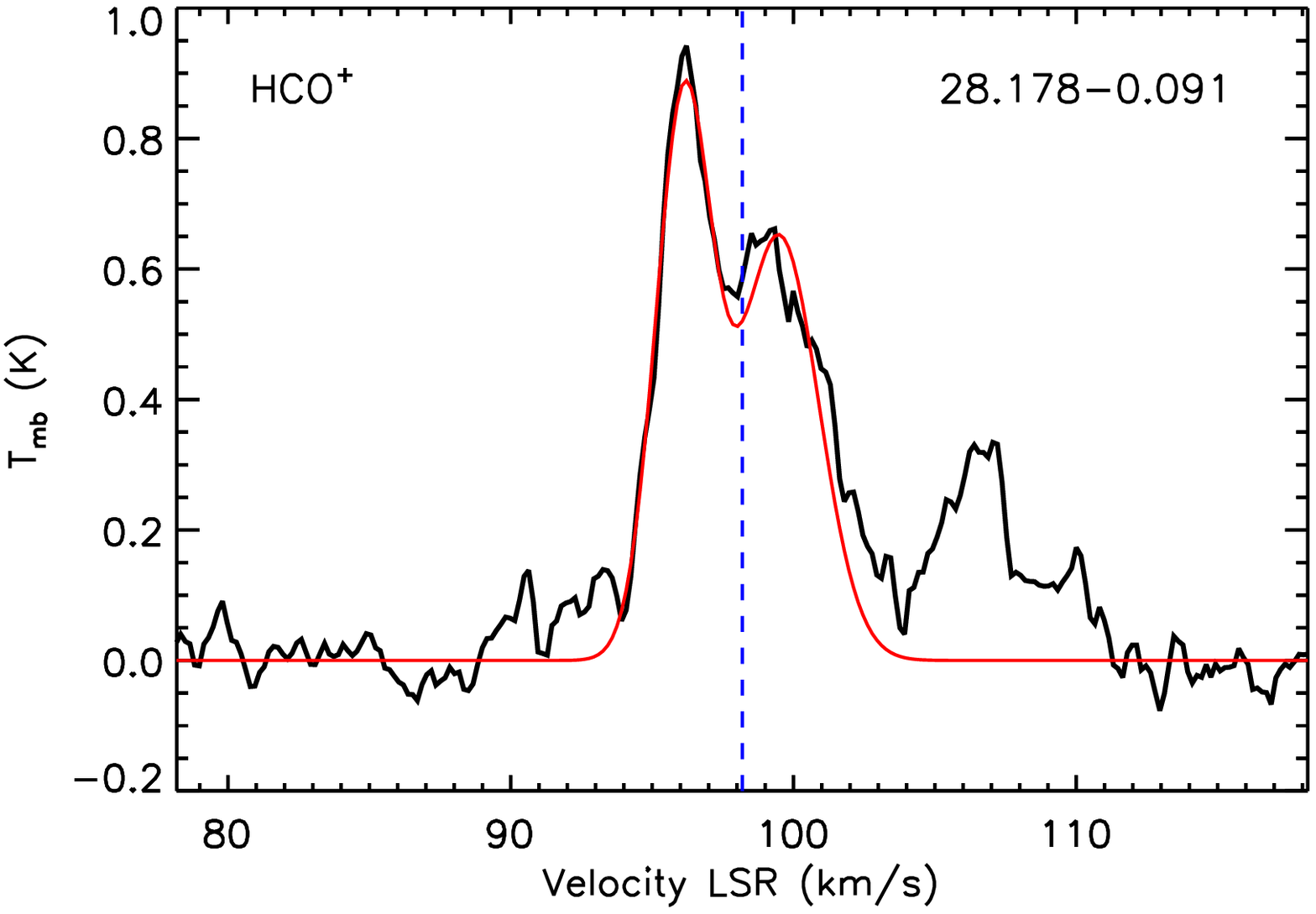}  
\includegraphics[width=6cm]{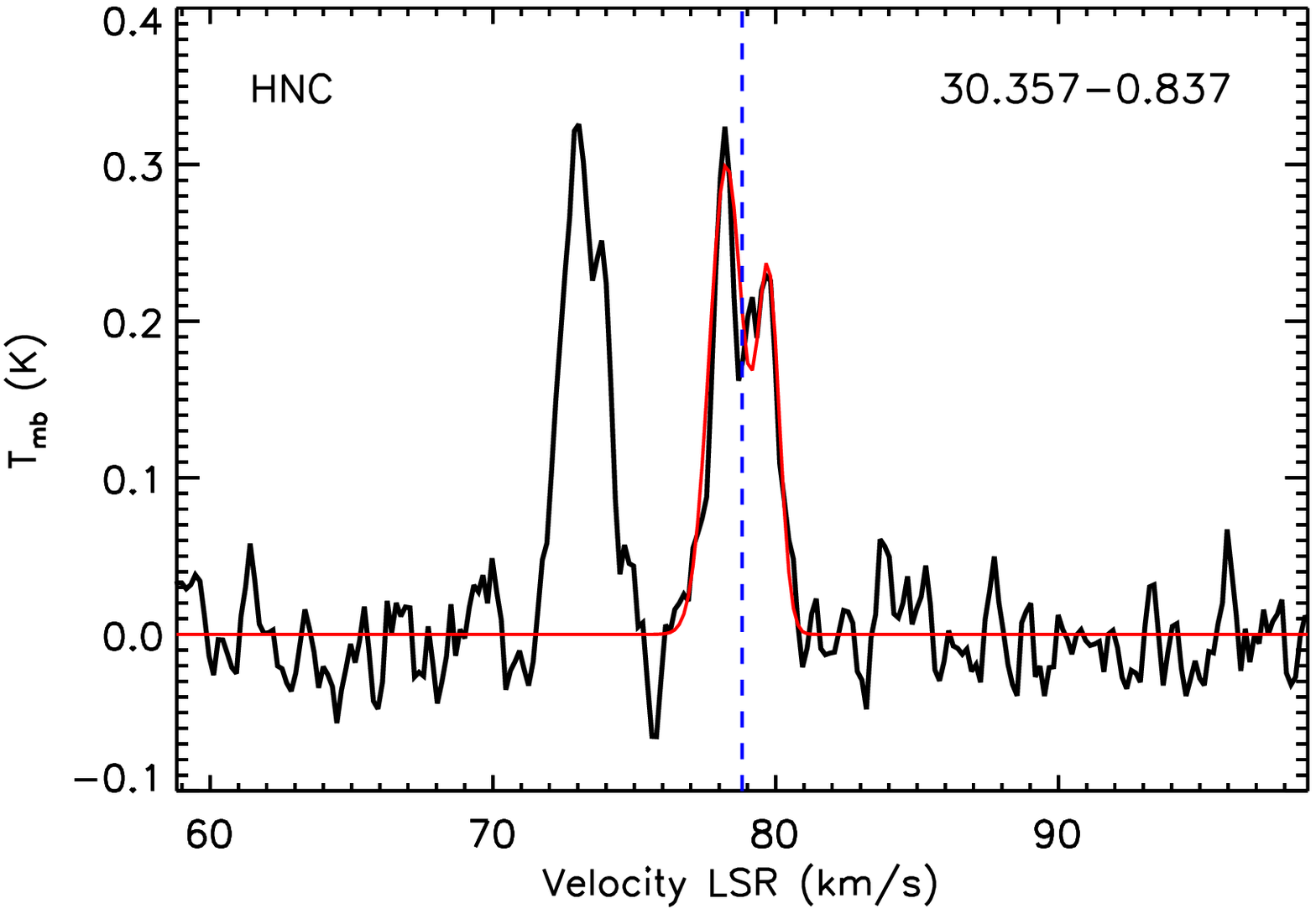}  
\includegraphics[width=6cm]{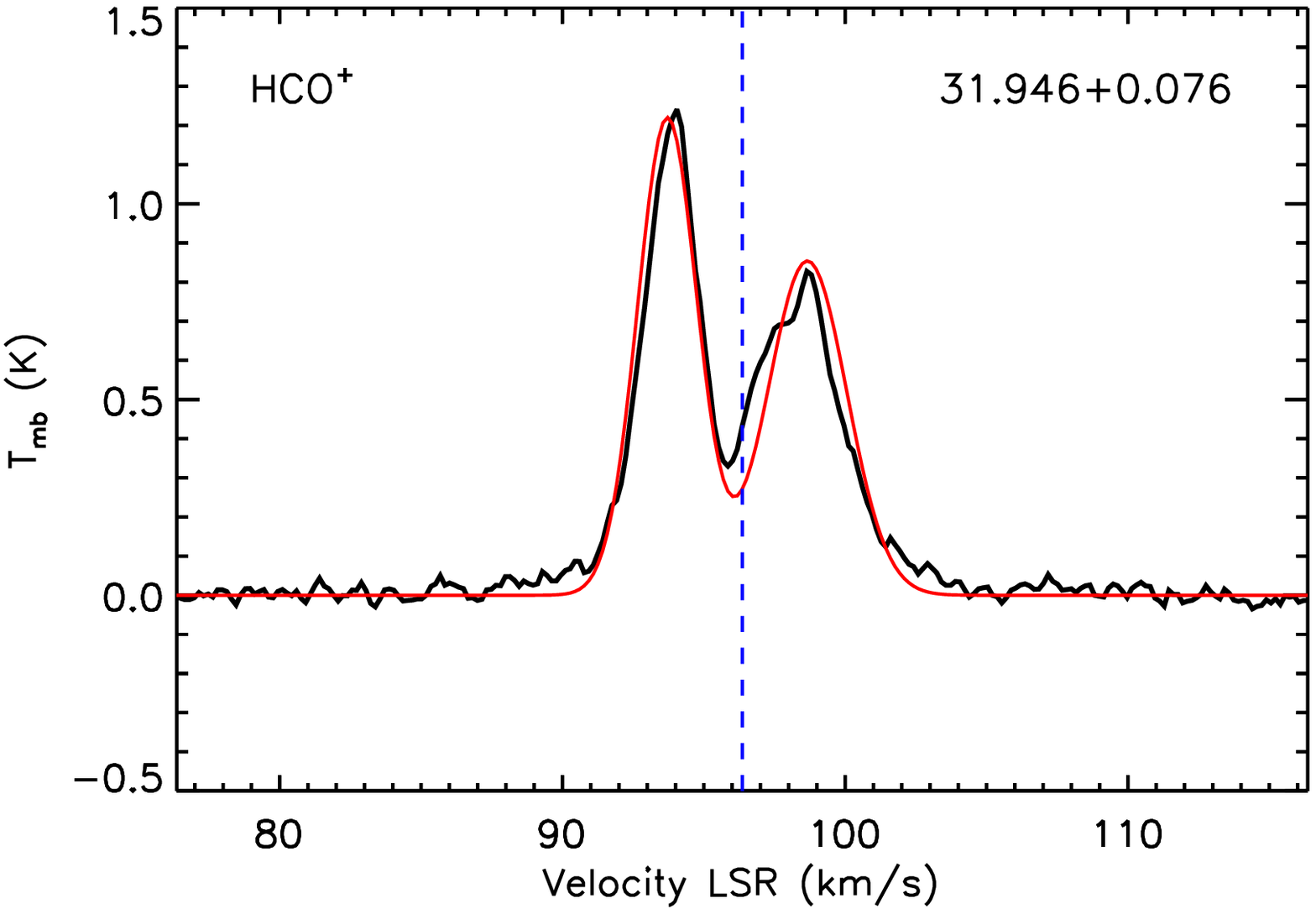}
\caption{Bue-shifted spectra used to estimate the infall parameters. The red line is the result of the IDL \texttt{mpfitfun} routine. The blue-dotted vertical lines are in correspondence of the systemic velocity of the clump determined by the \n2h\ fit.}
\label{fig:spectrum_infall_fit}
\end{figure*}

We calculated the infall velocities following the ``two layers'' model of \citet{Myers96}. According to this model, \vinf\ is:

\begin{equation}
\mathrm{v}_{in}=\frac{\sigma^{2}}{\mathrm{v}_{red}-\mathrm{v}_{blue}}\ \mathrm{ln}\bigg(\frac{1+e^{\mathrm{(T_{blue}-T_{dip})/T_{dip}}}}{1+e^{\mathrm{(T_{red}-T_{dip})/T_{dip}}}}\bigg).
\end{equation}
$\mathrm{v}_{red}$ and $\mathrm{v}_{blue}$ are the velocities of the red and blue peaks respectively, and $T_{red}$ and $T_{blue}$ their main beam temperatures. $T_{dip}$  is the main beam temperature of the valley between the two peaks. The parameters for each source are in Table \ref{tab:infall_delta_v}. These parameters have been obtained fitting 2 Gaussians to either the \hco\ or HNC spectrum, as indicated in Table \ref{tab:infall_delta_v}, with the $\texttt{mpfitfun}$ IDL routine \citep{Markwardt09}. The spectra with the Gaussian fits are in Figure \ref{fig:spectrum_infall_fit}.

\begin{center}
\begin{table*}
\centering
\begin{tabular}{c|c|c|c|c|c|c}
\hline
\hline
Clump	&	v$_{red}$ & v$_{blue}$ & T$_{red}$ & T$_{blue}$ & T$_{dip}$ & Line \\
			&	(km s$^{-1}$)	&(km s$^{-1}$) & K & K & K  & \\	
\hline
22.53-0.192  &  79.86(0.03)	& 75.43(0.02)		& 0.58($<$0.01)	 &	1.54(0.03)	&	0.52(0.01) & \hco \\
22.756-0.284  & 106.66(0.04)	&	103.89(0.02)	&	0.31(0.01)	 &	0.44(0.01)	&	0.24(0.01) & HNC \\
24.013+0.488  &  96.50(0.01)	&	93.00(0.01)	&	0.46(0.01)  &	0.94(0.02)	& 0.31($<$0.01) & \hco \\
25.609+0.228	&	114.87(0.05)	&	112.00(0.01)	&	0.46(0.01)	 &	0.53(0.01) &	0.45(0.01) & HNC \\
28.178-0.091	&	 99.50(0.01)	&	96.12(0.02)	&	0.65(0.01)	 &	0.85(0.02)	&	0.51(0.01) & \hco \\
30.357-0.837 &  79.75(0.03)	&	78.26(0.03)	&	0.22(0.01)	 &	0.30(0.01)	&	0.17(0.01) & HNC \\
31.946+0.076  &  98.67(0.02)	&	93.71(0.02)	&	0.85(0.02)	 &	1.22(0.02)	&	0.25($<$0.01) & \hco \\
\hline

\end{tabular}
\caption{Parameters used to estimate infall velocities adopting the \citet{Myers96} model. The parameters have been derived fitting two Gaussians at each \hco\ spectrum showing infall signatures with the \texttt{mpfitfun} routine \citep{Markwardt09}. The uncertainties comes from the fit for all but T$_{dip}$ for which we assume the same uncertainties of the corresponding T$_{red}$ and T$_{blue}$. Col. 1: Clump name; Cols. 2$-$3: velocities of the red- and blue-shifted peak respectively; Cols. 4$-$5: temperatures of the red- and blue-shifted peak respectively; Col. 6: temperature of the dip between the red- and blue-shifed peaks. Col. 7: line spectrum used to fit the Gaussians.}
\label{tab:infall_delta_v}
\end{table*}
\end{center}

The infall velocities are listed in Table \ref{tab:infall_params}. They are in the range $0.02\leq\mathrm{v}_{in}\leq0.37$ km s$^{-1}$, with an average $\mathrm{\overline{v}}_{in}$=0.16 km s$^{-1}$. Infall velocities of massive star forming regions are in the range $0.1\leq\mathrm{v_{in,mass}}\leq1$ km s$^{-1}$ \citep{Fuller05}, and similar velocities have been observed in massive collapsing clouds \citep[][]{Kirk13,Peretto13}.

The infall velocities allow us to evaluate the mass accretion rate 
$\dot{\mathrm{M}}=4\pi\mathrm{R}^{2}n_{\mathrm{H}_{2}}\mu m_{\mathrm{H}}\mathrm{v}_{in}$ \citep{Myers96}, where 
$m_{\mathrm{H}}$ is the hydrogen mass and $n_{\mathrm{H}_{2}}$ the volume density obtained from the dust mass and assuming spherical clumps for simplicity. The accretion rates we obtain are 
$0.07\leq\dot{\mathrm{M}}\leq2.04\times10^{-3}$ M\sun /yr (see Table \ref{tab:infall_params}). These values are 
comparable with the predicted accretion rates onto massive protostellar cores \citep{McKee03} and with values observed 
in high-mass star forming regions \citep{Fuller05,Rygl13,Peretto13} and individual protostellar sources \citep{Duarte-Cabral13}.

The free-fall time t$_{ff}=(3\pi/(32\mathrm{G}n_{\mathrm{H}_{2}}))^{1/2}$, with G gravitational constant and $n_{\mathrm{H}_{2}}$ gas column density, is $2.6-4.7\times10^{5}$ yr (see Table 
\ref{tab:infall_params}), significantly higher than the massive starless clump candidates lifetime 
\citep[$\lesssim10^4$ yr,][]{Motte07,Tackenberg12,Svoboda16} but consistent with the accretion timescales of \citet{Duarte-Cabral13}. Within 1 free-fall time, assuming constant accretion rate equal to the value measured today, the clumps accrete a mass 
$31\lesssim\mathrm{M}_{accr.}\lesssim637$ M\sun. A clump such as 22.53-0.192 already embeds a core with a central star of $\simeq5$ M\sun\ and has the potential to accrete mass comparable with or even higher than the most massive core in SDC335 \citep{Peretto13,Avison15} within one free-fall time.

Two clumps have infall signatures but no visible 24\mum\ counterparts, 22.756-0.284 and 25.609+0.228. We can estimate an upper limit to the mass of an embedded object assuming a constant accretion rate over time into a single protostar equal to the actual clump accretion rate and a clump lifetime prior to the formation of a detectable protostar \citep[$t\sim10^{4}$ yr, the expected lifetime of infrared-quiet high-mass protostars,][]{Motte07}. This is an upper limit as the clump accretion rate measured today may have increased since the start of the collapse, and the clump may fragment in several protostellar cores. We obtain $\textrm{M}<2.5$\Msun\ and $\textrm{M}<1.7$\Msun\ for 22.756-0.284 and 25.609+0.228 respectively. Also, the mass upper limits are below the range of masses estimated for stars with MIPSGAL and GLIMPSE counterparts. We cannot exclude that low-mass stars may have already formed but are deeply embedded in these clumps, and likely not yet visible at 24\mum.

\begin{center}
\begin{table*}
\centering
\begin{tabular}{c|c|c|c|c|c}
\hline
\hline
Clump	& Infall vel. &  Accr. rate  & t$_{ff}$ &   M$_{ff}$ & 24\mum\ \\
			&		(km s$^{-1}$)		&	
(10$^{-3}$ M\sun\ yr$^{-1}$) & ($10^{5}$ yr) & (M\sun) &  \\	
\hline
22.53-0.192  & 0.34($<$0.01) & 2.04(0.66) & 3.0(0.6)	& 610(237)	&  1  \\
22.756-0.284 & 0.07($<$0.01) &0.25(0.08) & 2.6(0.6)	&	65(24)	& 0  \\
24.013+0.488 & 0.21($<$0.01) & 1.52(0.46) & 2.7(0.6)	&	415(152)  & 1  \\
25.609+0.228	&  0.02($<$0.01)	& 0.17(0.06) & 2.8(0.6)	&	50(19)	&  0  \\
28.178-0.091	&  0.05($<$0.01)	& 0.37(0.11) & 2.8(0.6)	&	105(39)	 & 1   \\
30.357-0.837	& 0.04($<$0.01)	& 0.07(0.02) & 4.7(1.0)	&	31(12)	&  1  \\
31.946+0.076 & 0.37($<$0.01)	& 1.96(0.63) & 3.3(0.7)	&	637(247)	&	1  \\

\hline
\end{tabular}
\caption{Infall parameters of the seven clumps with blue-shifted spectra. Col.1: Clump name; Col. 2: Infall velocity; Col. 3: Mass accretion rate derived from the infall velocity; Col. 4: Estimated free-fall time; Col. 5: Mass accreted within 1 free-fall time; Col. 6: Presence (or absence) of a 24\mum\ counterpart.}
\label{tab:infall_params}
\end{table*}
\end{center}

\section{Evolutionary indicators in 70\mum\ quiet clumps}\label{sec:evolution}
In this Section we compare various evolutionary indicators (L/M ratio, dust temperature, surface density, linewidth, \n2h\ abundance and mass accretion rate) to look for differences between clumps with or without a 24\mum\ source. We first divide our 16 clumps with well defined gas emission spectra in two groups: clumps without 24\mum\ counterparts ($N24$, 8 clumps) and clumps with a 24\mum\ counterpart ($Y24$, 8 clumps). We further divide the first group in two sub-samples: clumps with low values of the skewness parameter (2 clumps, $N24\_L$), which may be considered as the less evolved, and clumps with a significant value of the skewness parameter ($N24\_S$). The properties for the three groups are summarized in Table \ref{tab:evolution_params}.

\begin{center}
\begin{table*}
\centering
\begin{tabular}{c|c|c|c|c|c|c}
\hline
\hline
Clump	&	L/M & T &  $\Sigma$  & $\sigma$ & $X($\n2h) & $\mathrm{\dot{M}}$ \\
	group		&	(L\sun/M\sun) & (K) & (g cm$^{-2}$)  & (km s$^{-1}$)	& (10$^{-10}$) & (10$^{-3}$ M\sun\ yr$^{-1}$)  \\	
\hline
$N24\_L$  & 0.11$\pm$0.07 &  10.1$\pm$0.7 &  0.07$\pm$0.01 &  0.65$\pm$0.49 & 1.92$\pm$1.73 & $-$ \\
$N24\_S$  &  0.18$\pm$0.08 &  11.0$\pm$1.1 & 0.14$\pm$0.06 & 0.79$\pm$0.27 &  3.56$\pm$2.91 &  0.21$\pm$0.06  \\
$Y24$  &  0.20$\pm$0.07 &  11.6$\pm$0.9 & 0.15$\pm$0.07 & 0.91$\pm$0.29 & 2.66$\pm$1.04 & 
1.19$\pm$0.92 \\
\hline
\end{tabular}
\caption{Average values of various parameters and the associated dispersion for the three classes of 70\mum\ quiet clumps: objects with no 24\mum\ counterparts and low skewness parameter ($N24\_L$); clumps with no 24\mum\ counterparts but significant value of the skewness parameter ($N24\_S$); clumps with 24\mum\ counterparts ($Y24$). Col. 1: clump phase; Col.2: L/M ratio; Col. 3: dust temperature; Col. 4: surface density; Col. 5: velocity dispersion; Col. 6: \n2h\ abundance relative to the H$_{2}$; Col. 7: mass accretion rate.}
\label{tab:evolution_params}
\end{table*}
\end{center}

\begin{figure}
\centering
\includegraphics[width=8cm]{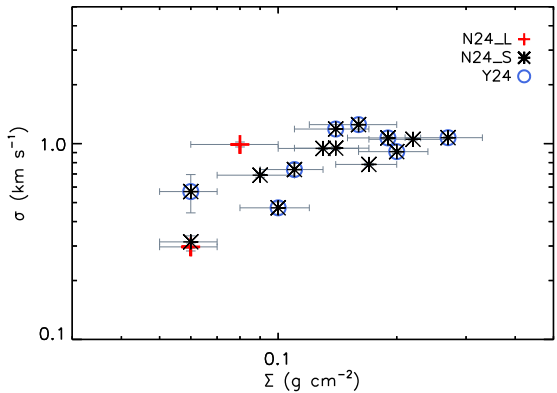} 
\caption{$\Sigma$ vs. $\sigma$ distribution of the clumps. Red crosses indicate sources with no 24\mum\ counterpart. Black crosses mark sources with a faint 24\mum\ counterpart. Green and blue circles mark clumps with infall signatures and red-shifted \hco\ spectra respectively.}
\label{fig:Sigma_sigma}
\end{figure}

\begin{itemize}
\item[$\bullet$] \textit{L/M ratio}: The first indicator, the L/M ratio, is a well-identified indicator of clumps evolution \citep{Molinari08,Molinari16_l_m,Cesaroni15}. The average values of the three groups are L/M$_{N24\_L}=0.11\pm0.07$, L/M$_{N24\_S}=0.18\pm0.08$ and L/M$_{Y24}=0.20\pm0.07$. The L/M ratio is very low in each group and, within the dispersion of the measurements, they all exhibit a very similar value.

\item[$\bullet$] \textit{Dust temperature}: The dust temperature is also thought to increase as the clump evolves, and the inner cores warm-up the dust envelope. The average temperatures of the three groups are T$_{N24\_L}=10.1\pm0.7$, T$_{N24\_S}=11.0\pm1.1$ and T$_{Y24}=11.6\pm0.9$. As for the L/M indicator, although on average the ${N24\_L}$ clumps are slightly colder than the ${N24\_S}$ and $Y24$ clumps, within the dispersion there is no clear indication of a trend among these three groups. The observed 24\mum\ sources may be too young to significantly alter the properties of the surrounding dust on clump scales.

\item[$\bullet$] \textit{Surface density}: A trend of increasing surface density from more quiescent to more evolved clumps has been observed in previous surveys of star forming clumps \citep[e.g.][]{Urquhart14,Svoboda16}, although it is not well established \citep[e.g.][]{Rathborne10}. Combining a large sample of massive clumps in the Galaxy taken from the Hi-GAL survey, \citet{Merello17} showed that there is no evidence of increasing $\Sigma$ with the clumps evolution. Here, we find $\Sigma_{N24\_L}=0.07\pm0.01$, $\Sigma_{N24\_S}=0.14\pm0.06$ and $\Sigma_{Y24}=0.15\pm0.07$ g cm$^{-2}$. The ${N24\_L}$ clumps have the lowest values of surface densities on average, but there is no differences between ${N24\_L}$ and $Y24$ clumps.

\item[$\bullet$] \textit{Velocity dispersion}: An indicator of the evolution of massive clumps and cores that can be derived from the gas properties is the expected increase of the linewidth as the region evolves \citep{Smith13}. Average values for the three groups are $\sigma_{N24\_L}=0.65\pm0.49$, $\sigma_{N24\_S}=0.79\pm0.27$  and $\sigma_{Y24}=0.91\pm0.29$. There is a slight increase going from the first to the third group, however the average values of the velocity dispersion are consistent within the dispersion. 

\item[$\bullet$] \textit{$\Sigma$ vs. $\sigma$}: In Figure \ref{fig:Sigma_sigma} we show the relation between gas velocity dispersion and mass surface density of our clumps to explore if there is an evolutionary trend. The Pearson correlation coefficient is significant, 0.71. However the clumps with a 24\mum\ source do not occupy a specific locus of points in this plot, supporting the hypothesis that this correlation may be due to the dynamical properties of the star forming regions, and not with the clumps evolution, as suggested in some star formation models \citep{Ballesteros-Paredes11}.

\item[$\bullet$] \textit{\n2h\ abundance}: There is evidence that the \n2h\ abundance increases as the clump evolves \citep{Sanhueza12}. The \n2h\ abundance for the three groups are respectively $X($\n2h$)_{N24\_L}=1.92\pm1.73\times10^{-10}$, $X($\n2h$)_{N24\_S}=3.56\pm2.91\times10^{-10}$ and $X($\n2h$)_{Y24}=2.66\pm1.04\times10^{-10}$. Again, there is a weak indication that the abundance of the ${N24\_L}$ clumps is lower than in the other two groups, but consistent within the dispersion.

\item[$\bullet$] \textit{Mass accretion rate}: In some star formation models the accretion rate is expected to increase with time \citep[e.g.][]{McKee03}. We have this information available for only 7 clumps, 2 ${N24\_S}$ and 5 $Y24$ clumps. The mean values are $\mathrm{\dot{M}}_{N24\_S}=0.21\pm0.06\times10^{-3}$ and $\mathrm{\dot{M}}_{Y24}=1.19\pm0.92\times10^{-3}$ M\sun\ yr$^{-1}$ respectively. Despite using only 7 clumps,this is suggestive that the mass accretion rate is higher in clumps with a detectable 24\mum\ source than in regions with still no observable inner cores (but dynamically active at the clump scale). If the accretion rate is increasing with time, this increase must be very rapid based on these (few) points, as all the other indicators do not yet show evidences of evolution among these groups of clumps. 

\end{itemize}

The best candidates to embed massive pre-stellar cores are the two ${N24\_L}$ clumps, 15.631-0-377 and 28.792-0.141. These clumps have on average slightly different values of the evolutionary indicators compared to the other two groups. However, the values are compatible among the three groups within the dispersion of the measurements.

These results suggest that 70\mum\ quiet clumps are all at a very similar (and very early) stage of evolution. The early rise of a visible 24\mum\ source does not alter the properties of star forming regions at the clump scales. 

\section{Summary}\label{sec:conclusion}
We investigated the gas and dust properties of a sample of 18 massive clumps selected to be in a very early stage of massive star formation and 70\mum\ quiet. 

The dust properties have been constrained combining data from the Hi-GAL, ATLASGAL and BGPS surveys. The clumps have mass of $1.2\times10^{3}$ M\sun\ on average with 2 clumps that exceed $2\times10^{3}$ M\sun, and mass surface densities $\Sigma\geq0.05$ g cm$^{-2}$. Based on the analysis of the mass surface density and the KP criterion to identify high-mass stars precursor in IRDCs, the majority of these clumps have the potential to form high-mass stars. The dust temperatures are $\mathrm{T}<13$ K, lower than the average dust temperatures of starless clump candidates \citep[$\mathrm{T}\simeq15$ K][]{Traficante15b}. The luminosity is on average $\simeq2\times10^{2}$ L\sun\ with $	\mathrm{L/M}\simeq0.17$, significantly lower than the L/M ratio below which clumps are thought to be quiescent \citep[L/M=1,][]{Molinari16_l_m}. These values in clumps selected to be 70\mum\ quiet suggest that these massive clumps are at earliest stages of star formation.

The inspection of the 24\mum\ maps shows that half of these clumps have at least one faint 24\mum\ counterpart. Eight 24\mum\ sources associated with 5 different clumps have at least one GLIMPSE counterpart. We used the SED fitter tool of \citet{Robitaille06} to get an estimate of the properties of these MIR sources and found that they all have central stars deeply embedded in the clumps with $25\lesssim\mathrm{A}_{V}\lesssim93$ mag. These are sources with masses $2.7\lesssim\mathrm{M}_{*}\lesssim5.5$ M\sun\ and the equivalent of low-mass Class I and Class II sources.

The gas dynamics has been studied analyzing the emission of the dense gas tracers \n2h ($1-0$), HNC ($1-0$) and \hco\ ($1-0$) in the 16 clumps for which we have well defined spectral line emission. The \n2h\ emission is moderately optically thin ($<\tau>_{main}=0.6$), in line with previous observations of regions at the early stages of star formation. Blue asymmetries in HNC and \hco\ spectra have been used to identify infall signatures. Seven clumps have blue-shifted spectra with skewness parameter $\delta_{\mathrm{v}}\leq-0.25$. Two clumps with no visible 24\mum\ sources have signs of infall, suggesting that they are in a dynamical state at the clump scale prior to the formation of an intermediate/high mass core. 

The infall velocities are $\simeq0.16$ km s$^{-1}$ on average, similar to what is observed in other high-mass star forming regions with hints of protostellar activity. Similarly the mass accretion rate, $0.04\leq\dot{\mathrm{M}}\leq2.0\times10^{-3}$ M\sun /yr, is comparable with other massive star forming regions. With these accretion rates, a clump such as 22.53-0.192 has the potential to form massive stars comparable with the most massive protostellar cores observed in the Galaxy to date within one free-fall time, t$_{ff}\simeq2.5-4.5\times10^{5}$ yr. Assuming a lifetime of $10^{4}$ yr, clumps with infall signatures and no 24\mum\ sources may embed faint, low-mass protostars not detected in the MIPSGAL survey.

Finally we combine the dust properties with the gas dynamics to discuss the evolution of these clumps and to search for differences between clumps with and without 24\mum\ counterparts. We divided the clumps in three groups: clumps with no 24\mum\ counterpart and low values of the \hco\ skewness parameter, ($N24\_L$, 2 clumps), clumps with no 24\mum\ counterparts but significant value of the skewness parameter ($N24\_S$, 6 clumps), and finally objects with at least one 24\mum\ counterpart ($Y24$, 8 clumps). We found no significant differences, within the dispersion, between these three groups from indicators as L/M ratio, dust temperature, surface density, \n2h\ velocity dispersion and gas abundance. The only evidence is that the accretion rate increases from 24\mum\ dark to 24\mum\ bright clumps. This increase of the accretion rate may be the first sign of evolution in massive clumps, as all the other indicators do not show any significant difference between clumps with and without 24\mum\ counterparts.

We conclude that massive starless clumps are extremely rare. The lack of 70\mum\ (and possibly 24\mum) emission is a necessary, but not sufficient condition to identify massive starless clumps. Massive condensations may quickly form deeply embedded protostars, and the majority, if not all of these massive clumps may already harbor low-mass fragments. High resolution observations are needed to reveal the embedded content of these high density regions.

\bibliographystyle{mn2e}
\bibliography{bibliography.bib}

\clearpage

\appendix
\section{24\mum\ and Hi-GAL images}\label{sec:app_maps}
 24\mum\ and Higal 70-500\mum\ images of the 18 clumps. The blue cross in each map is the position of the 250\mum\ source centre.

 \begin{figure*}
 \centering
 \includegraphics[width=8cm]{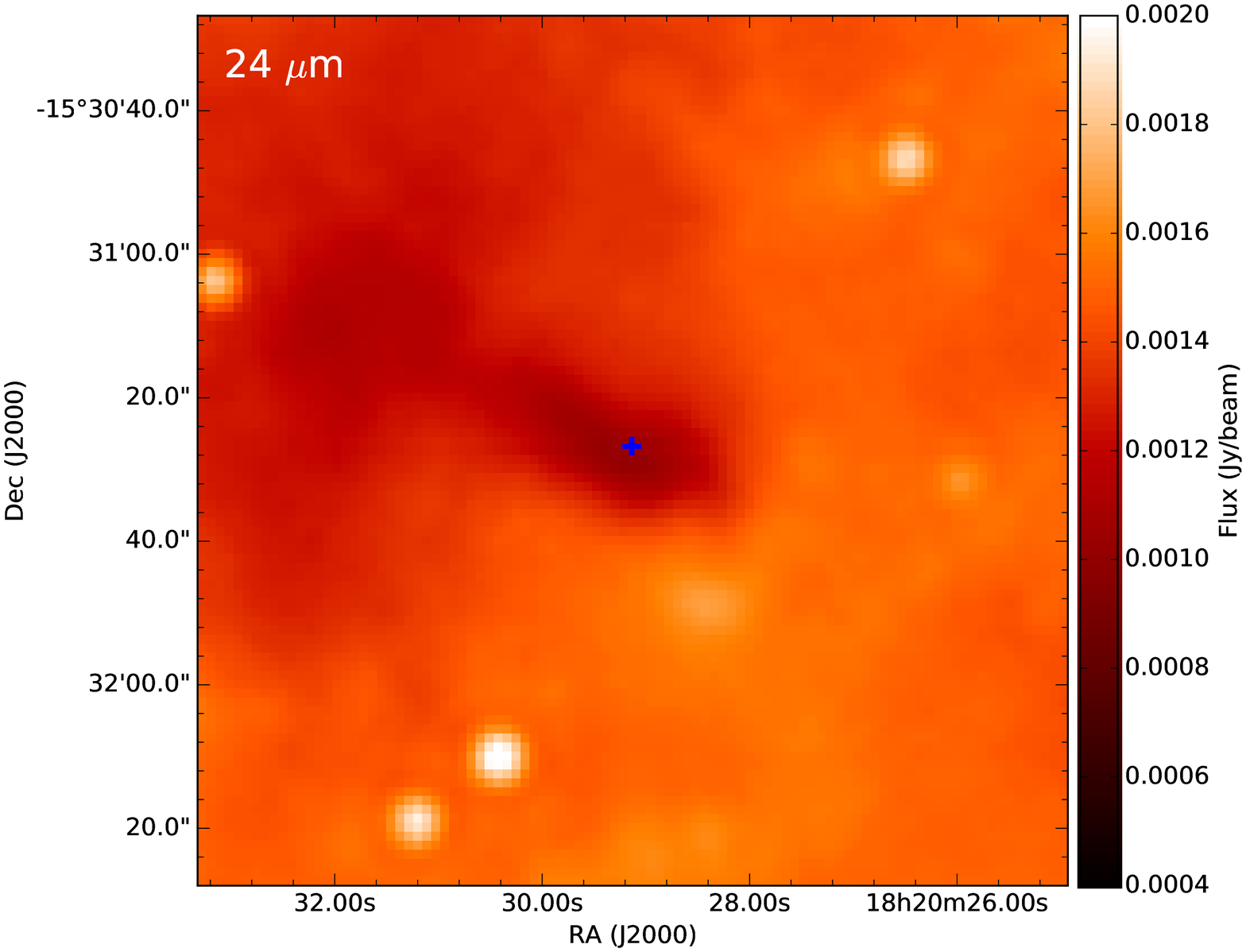}  \includegraphics[width=8cm]{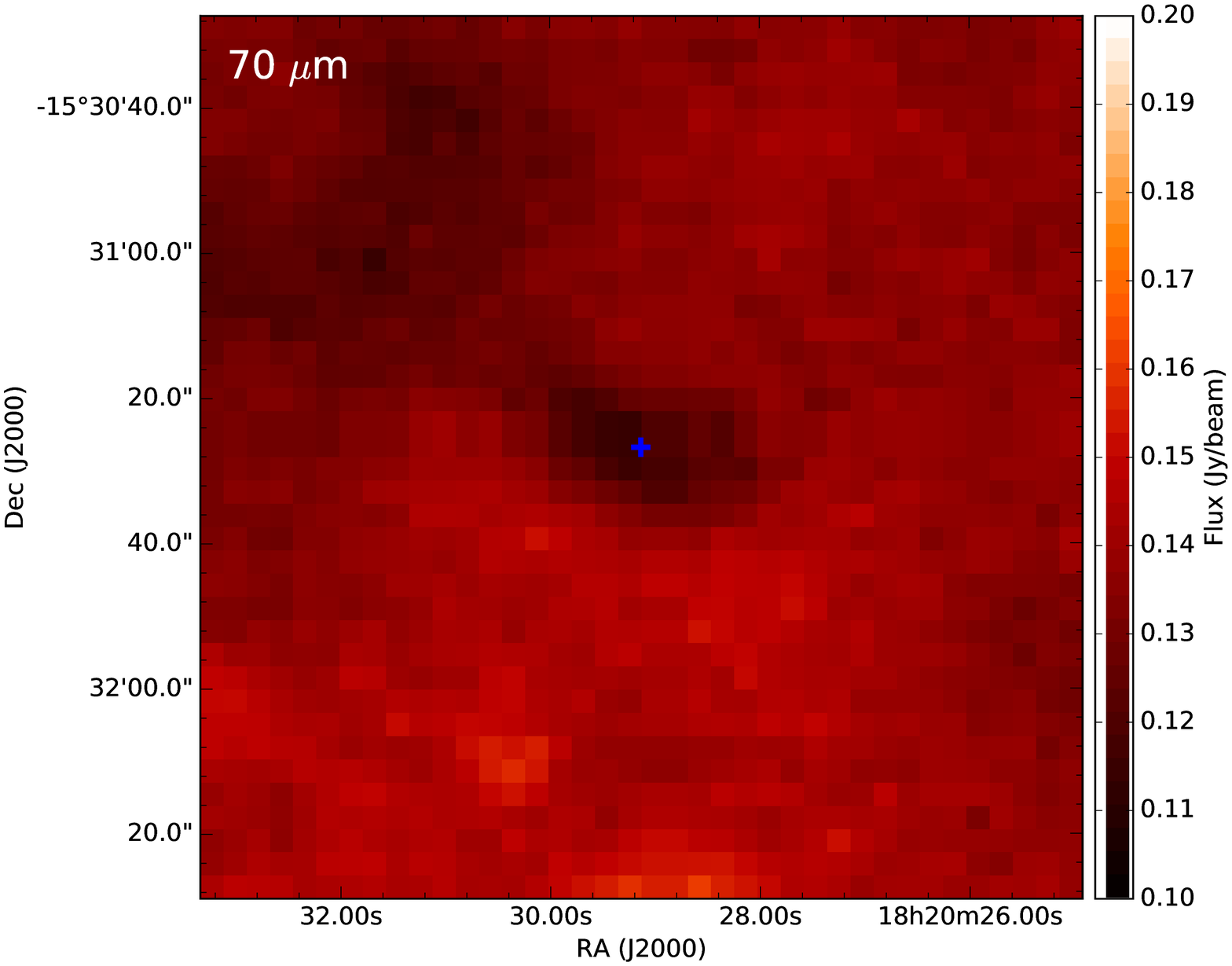} 
 \includegraphics[width=8cm]{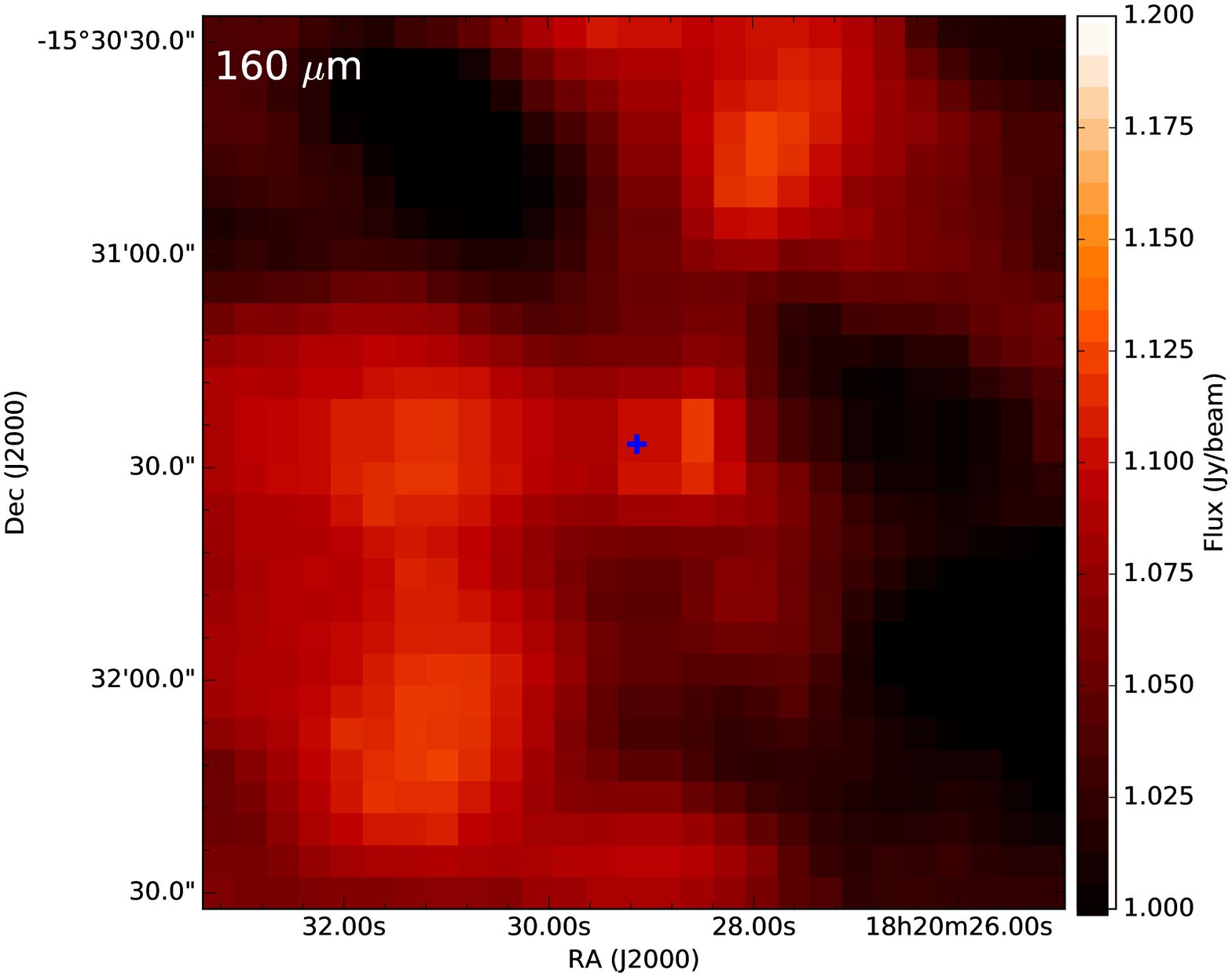}  \includegraphics[width=8cm]{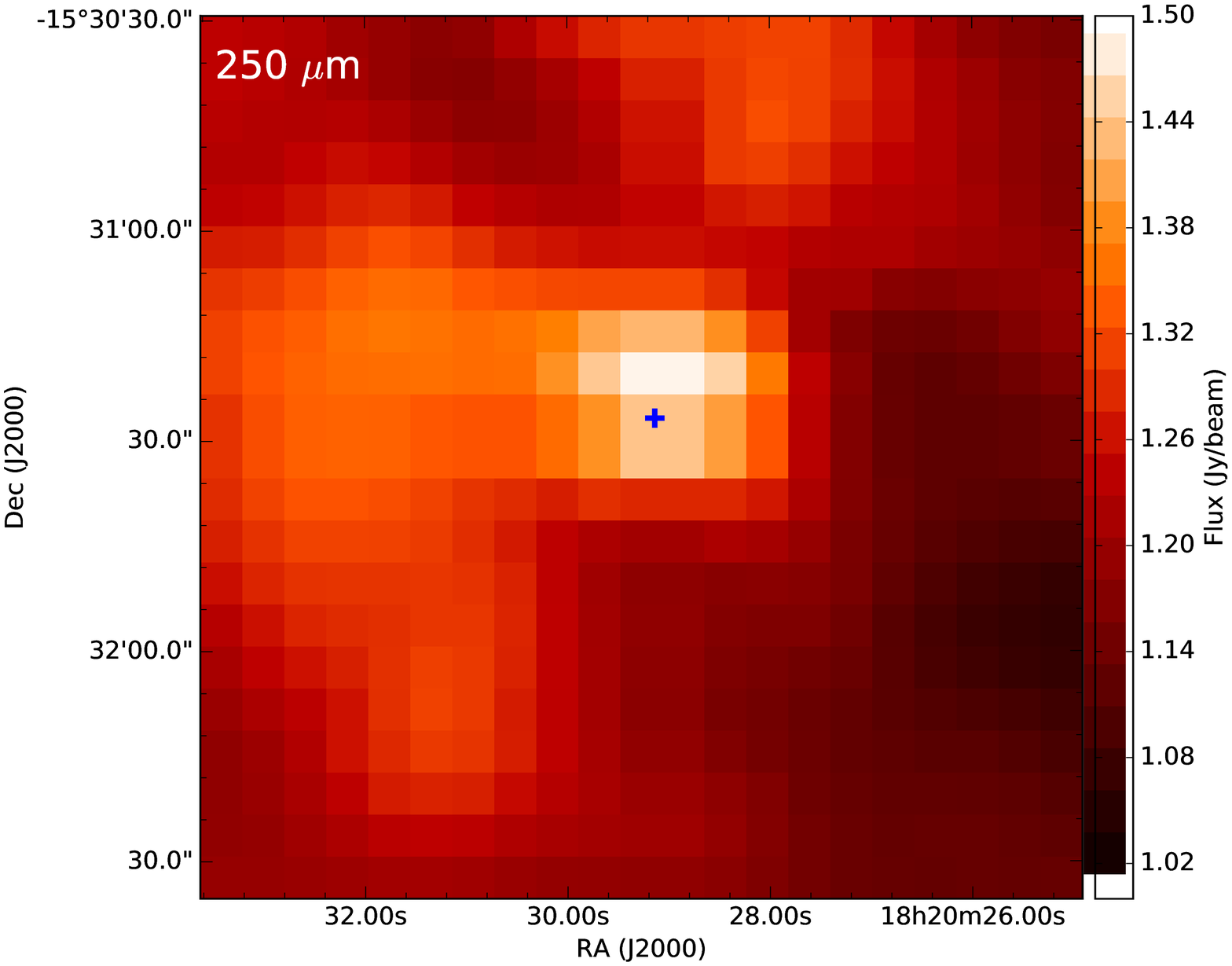} 
 \includegraphics[width=8cm]{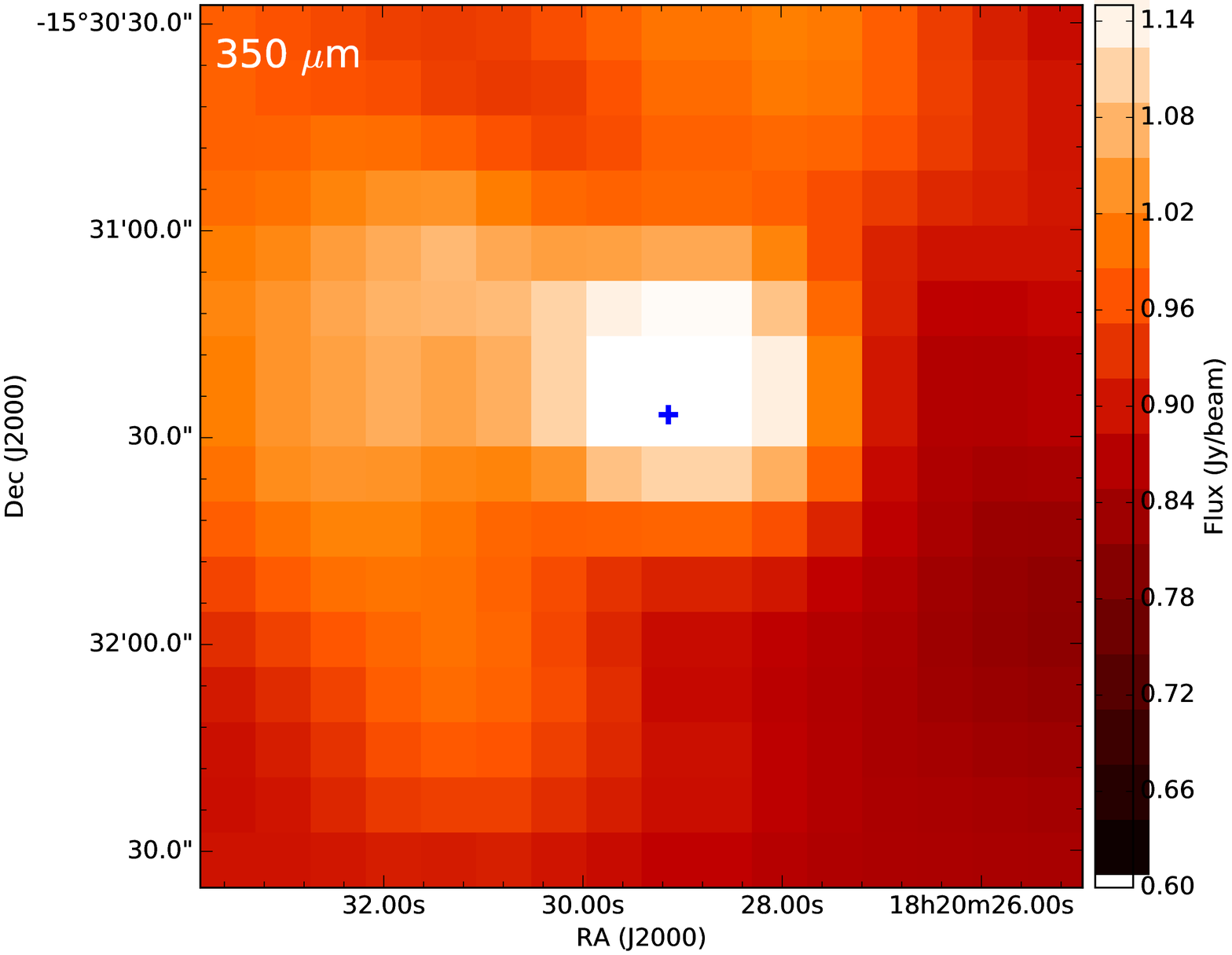}  \includegraphics[width=8cm]{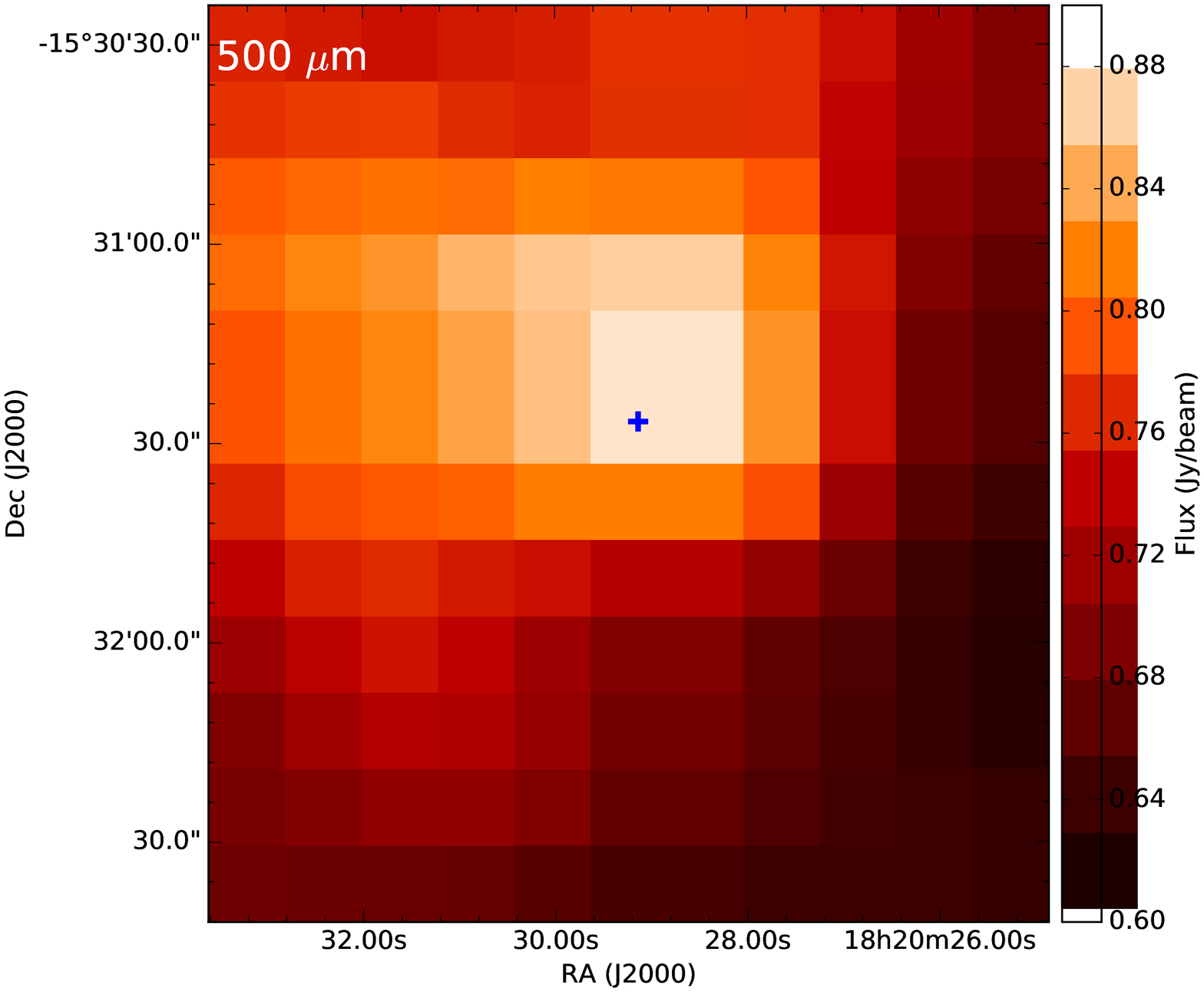} 
 \caption{15.631-0.377}
 \label{fig:sources_all_wave_longitude_distribution}
 \end{figure*}

 \begin{figure*}
 \centering
 \includegraphics[width=8cm]{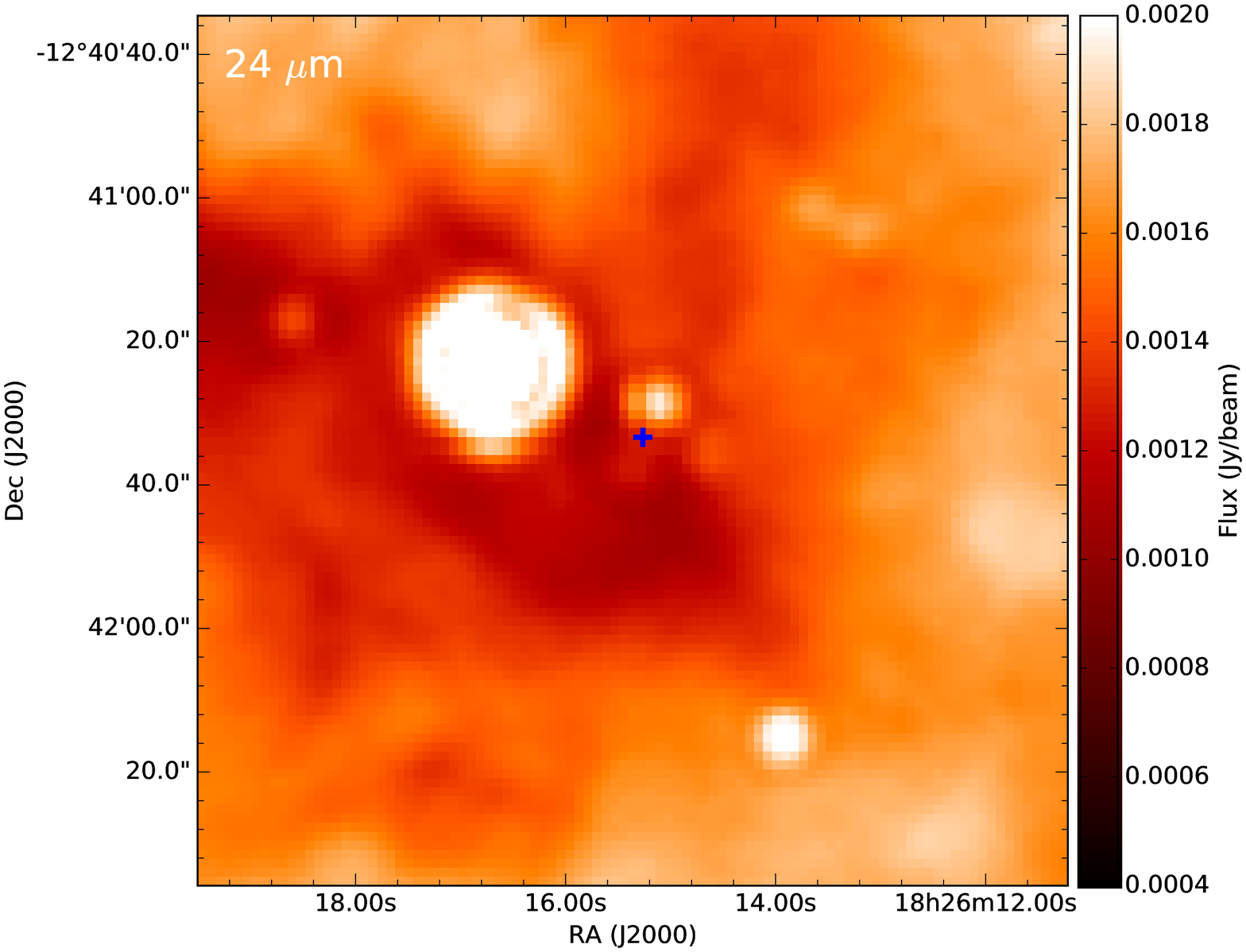}  \includegraphics[width=8cm]{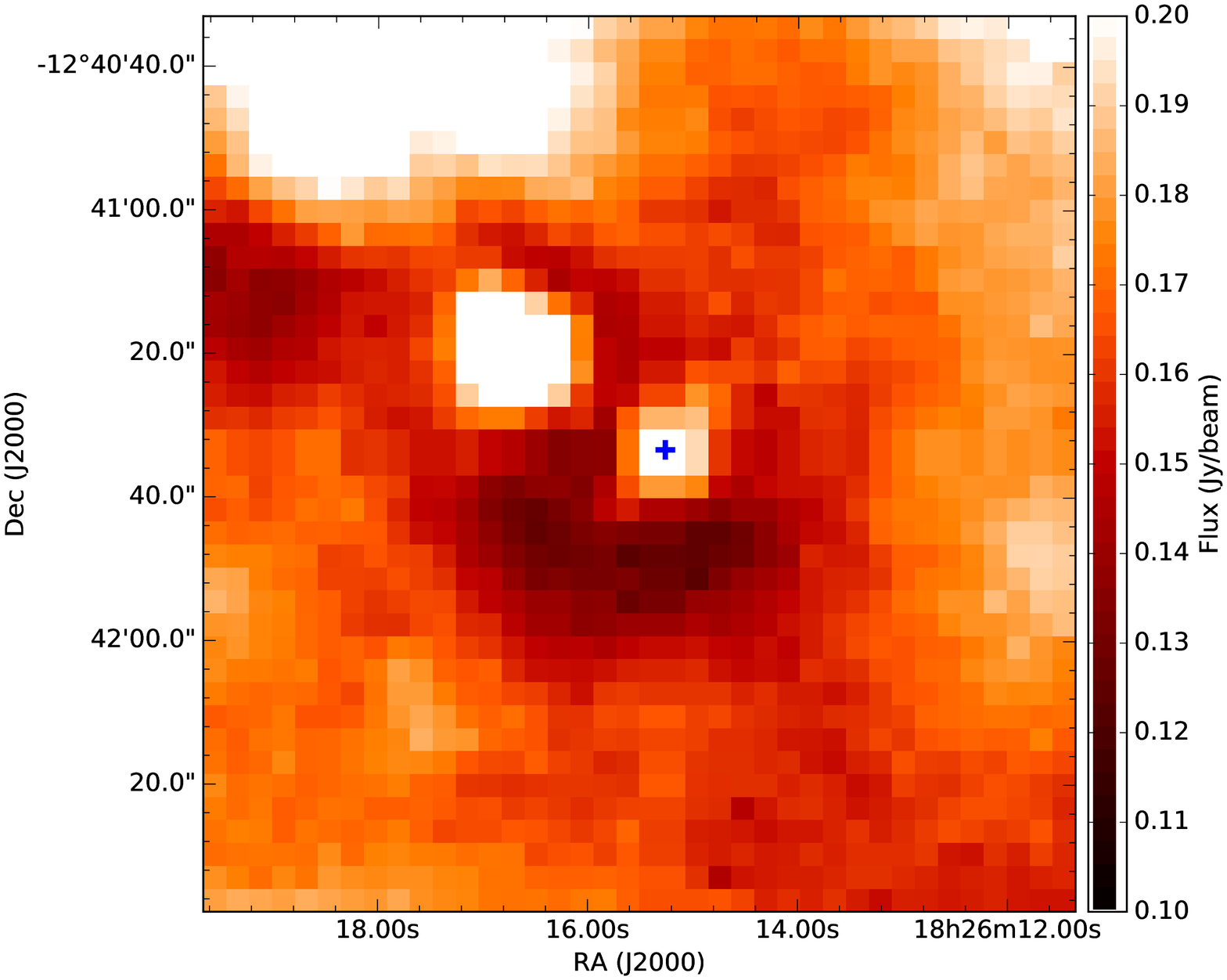} 
 \includegraphics[width=8cm]{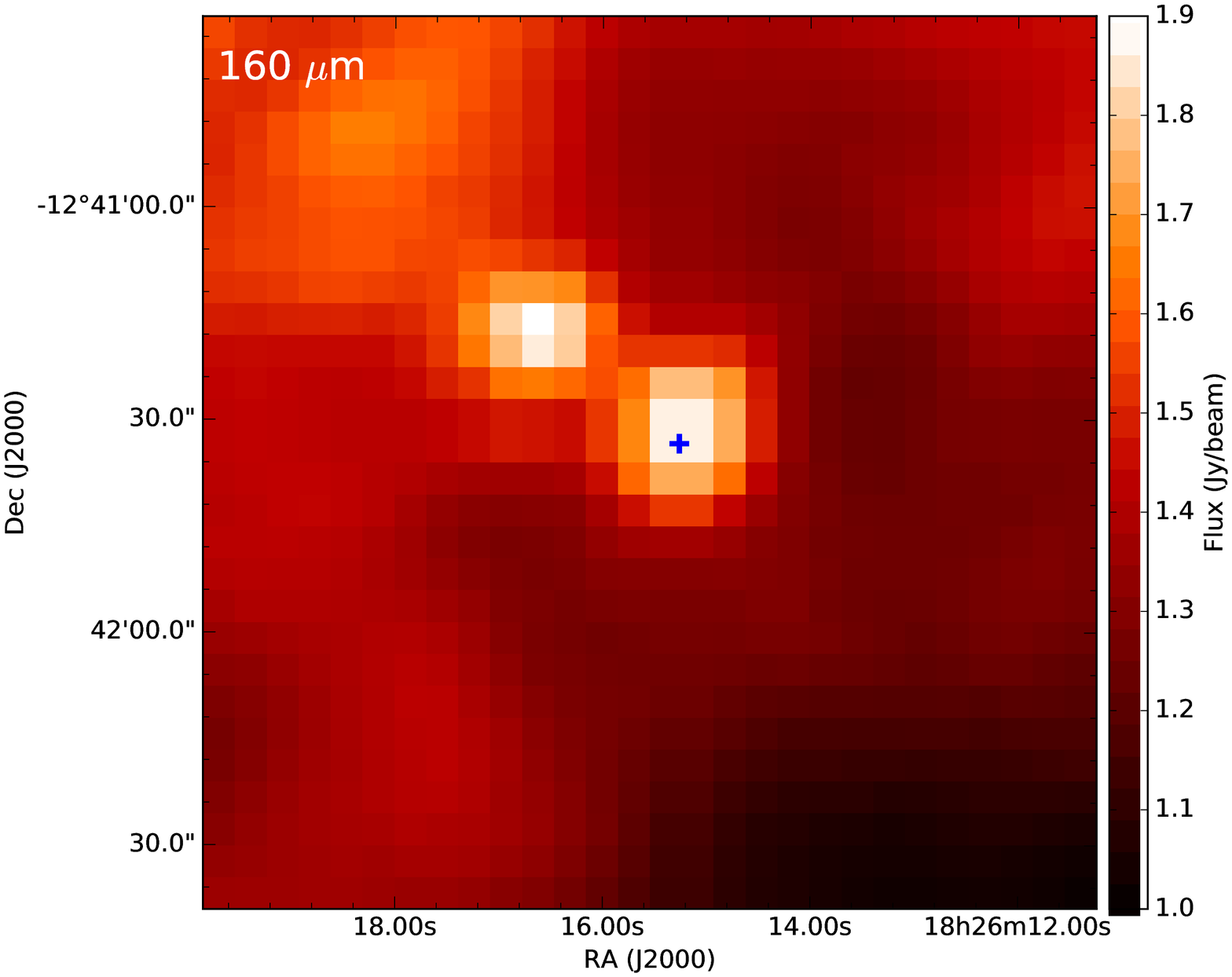}  \includegraphics[width=8cm]{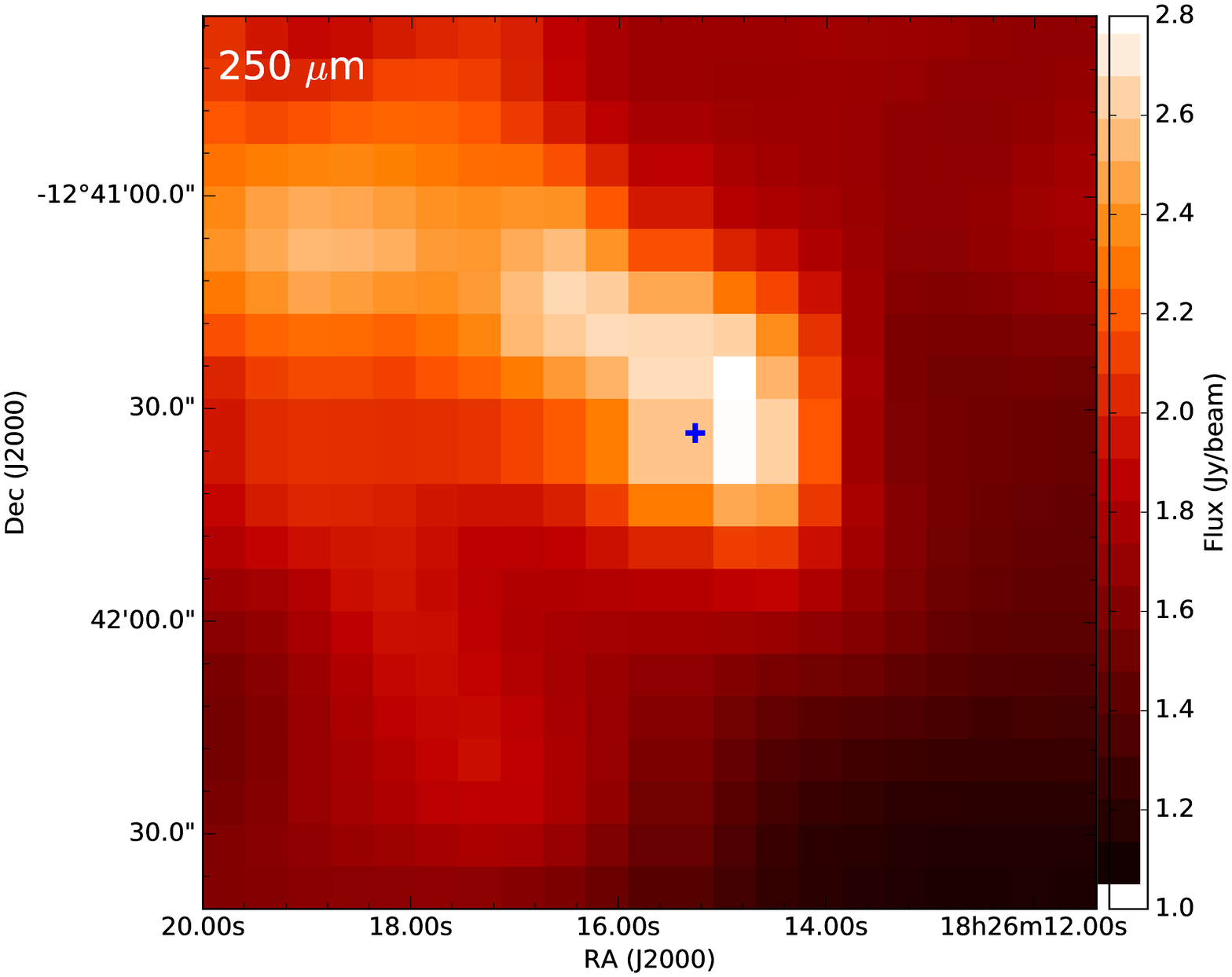} 
 \includegraphics[width=8cm]{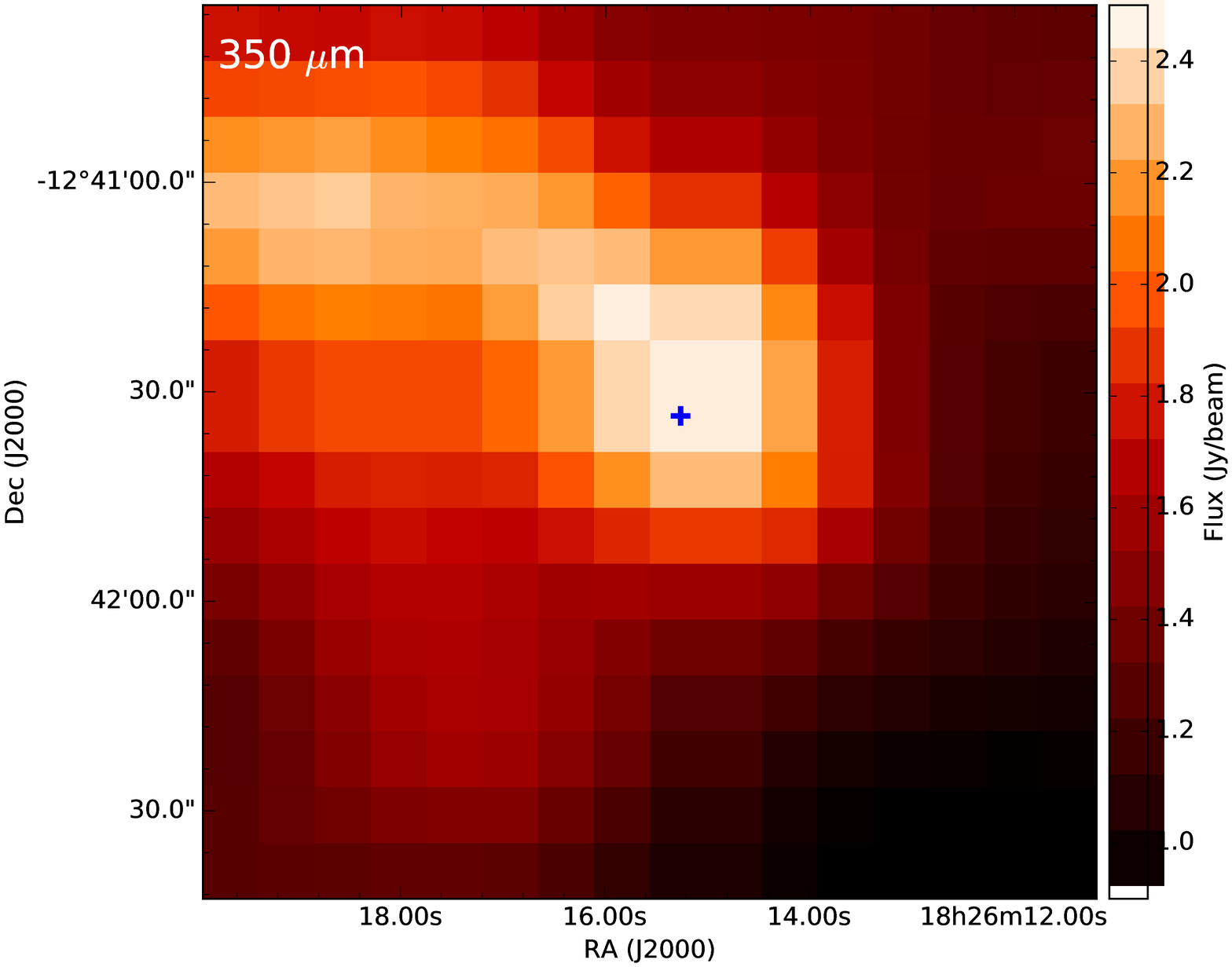}  \includegraphics[width=8cm]{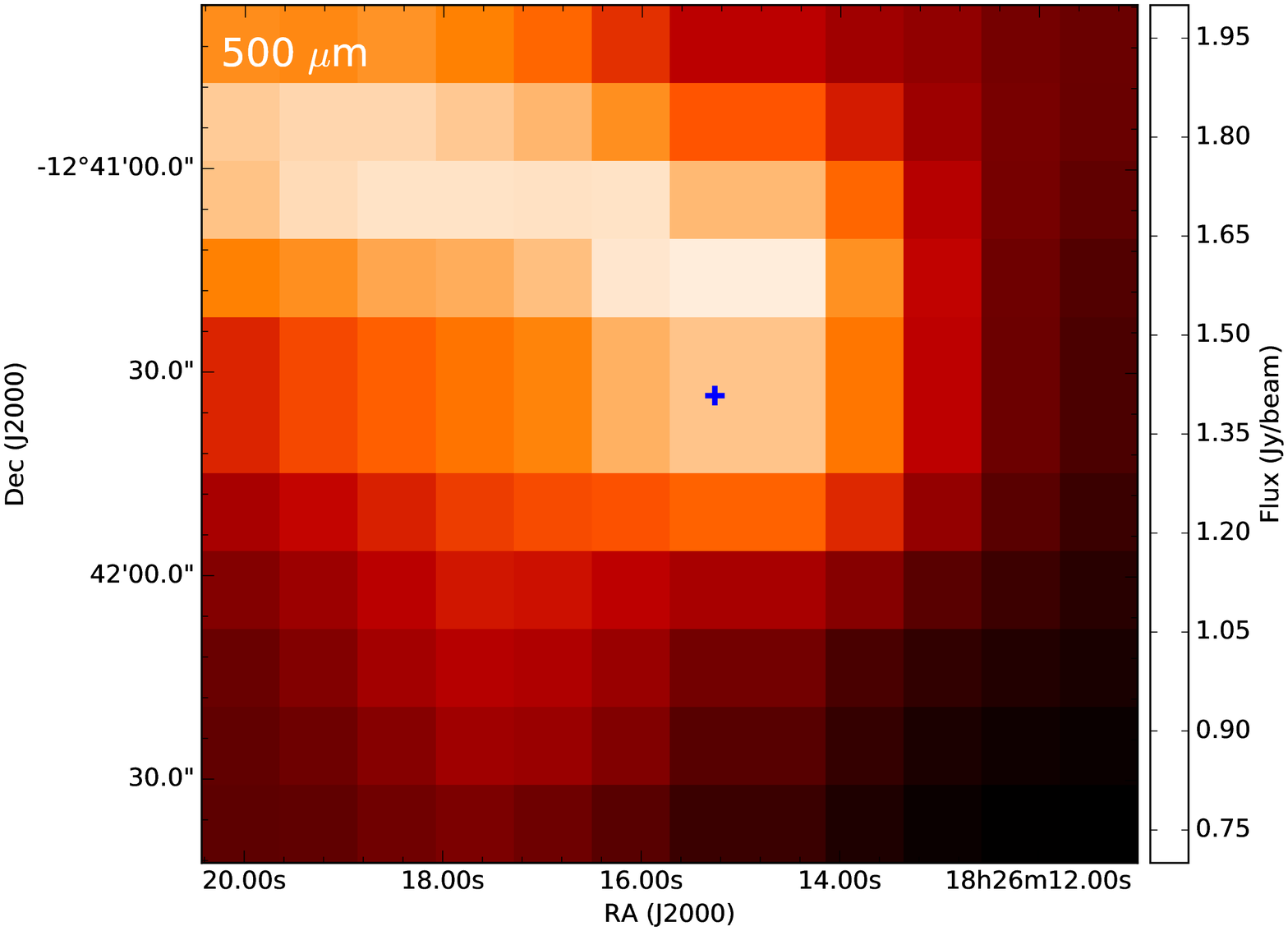} 
 \caption{18.787-0.286}
 \end{figure*}

 \begin{figure*}
 \centering
 \includegraphics[width=8cm]{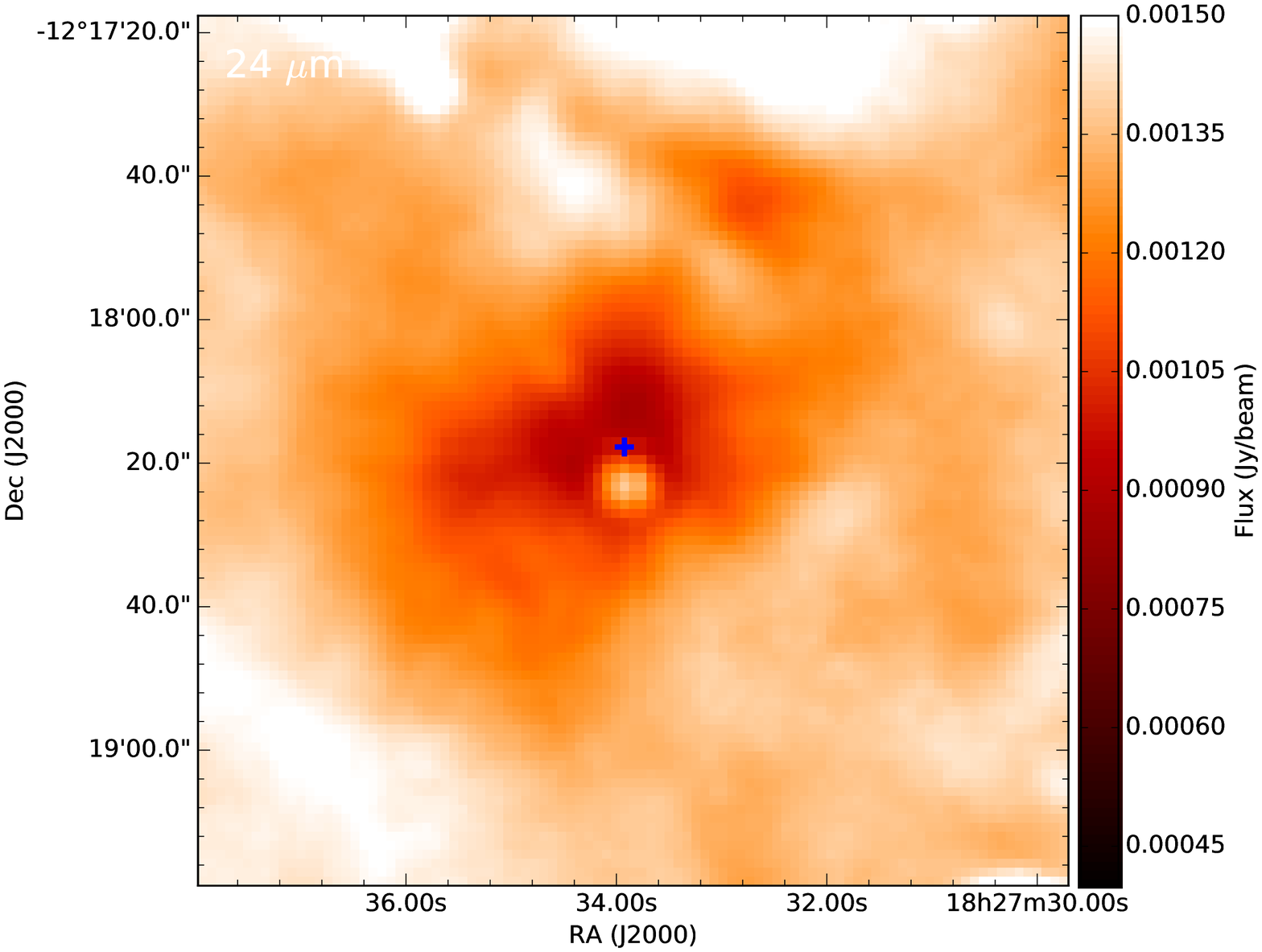}  \includegraphics[width=8cm]{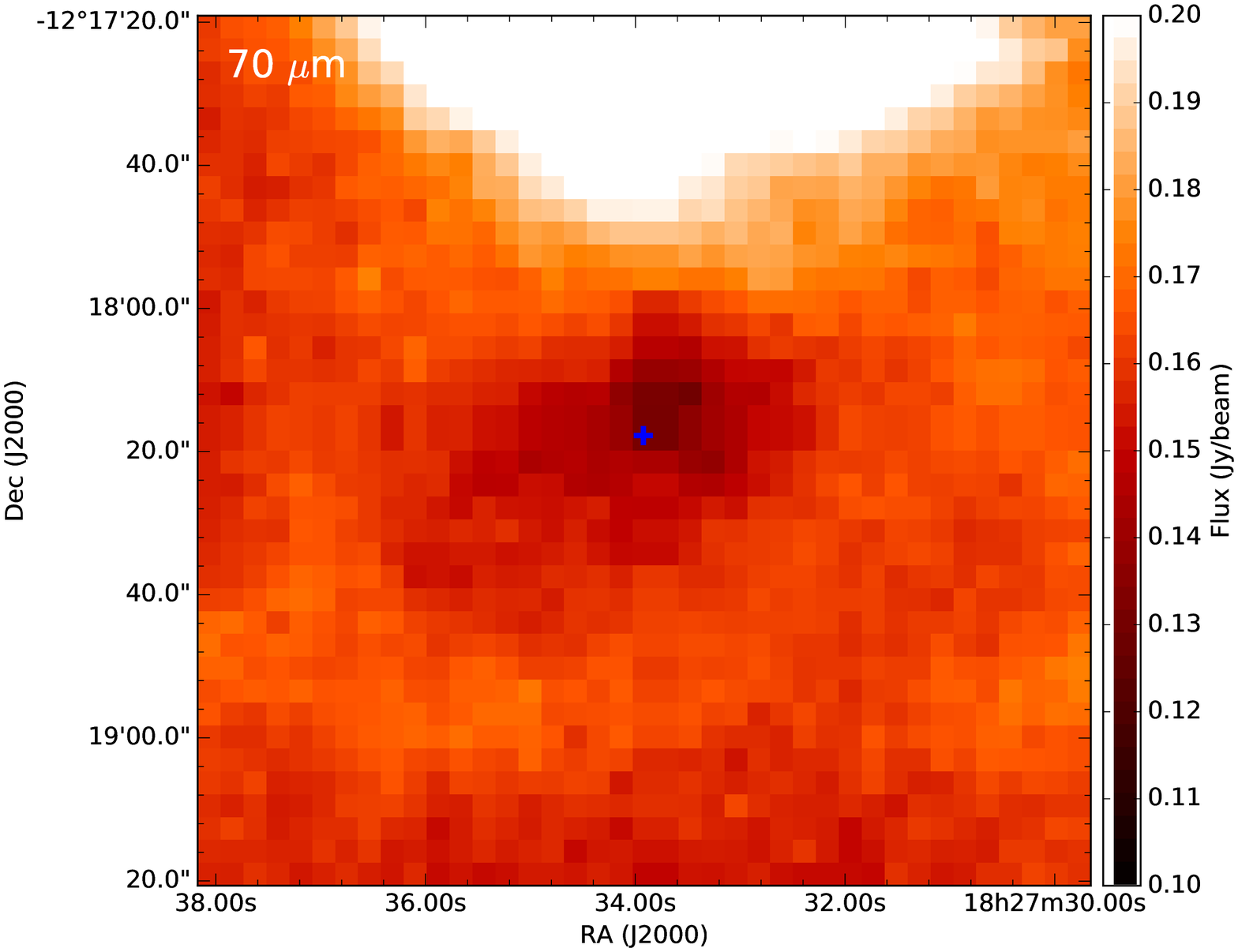} 
 \includegraphics[width=8cm]{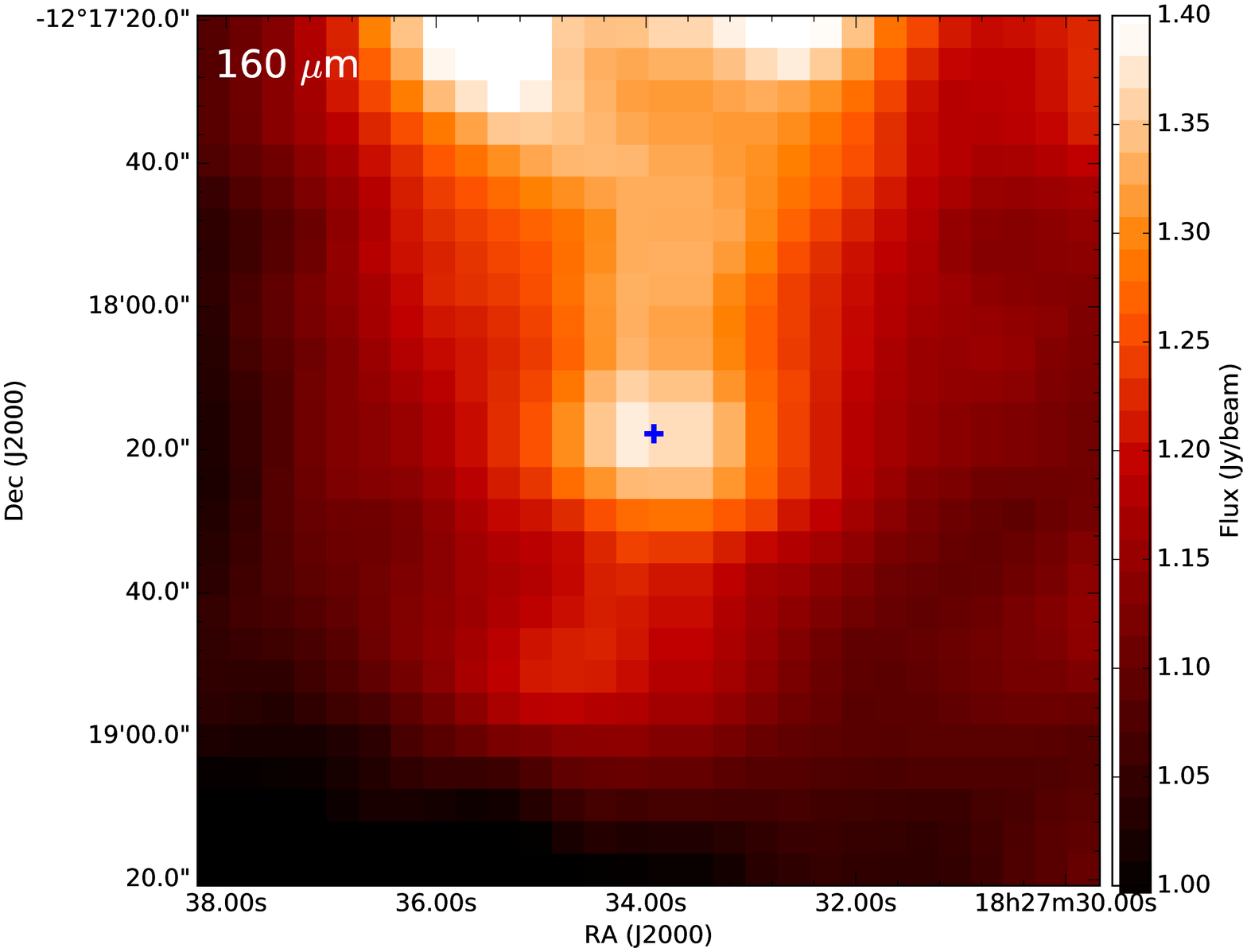}  \includegraphics[width=8cm]{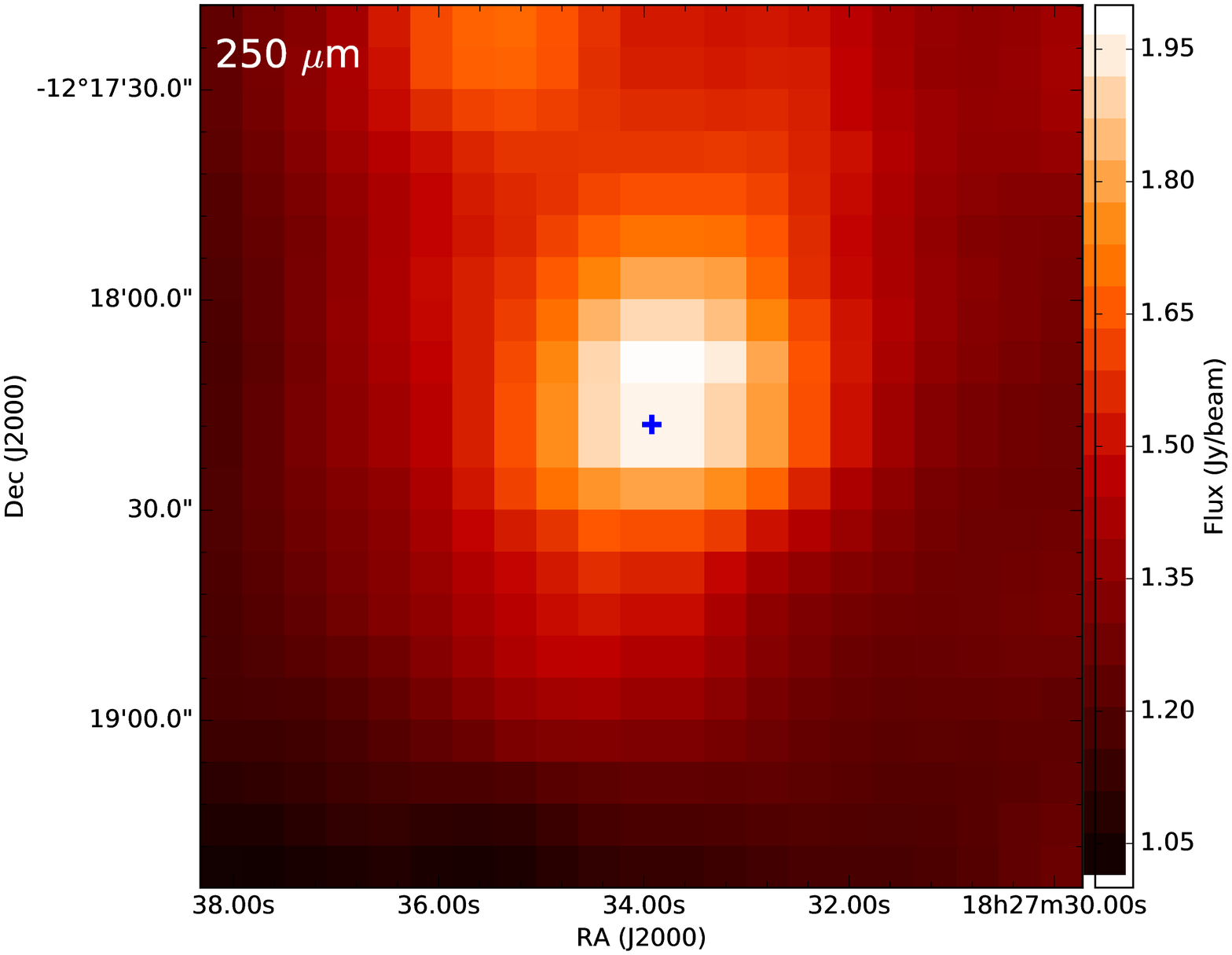} 
 \includegraphics[width=8cm]{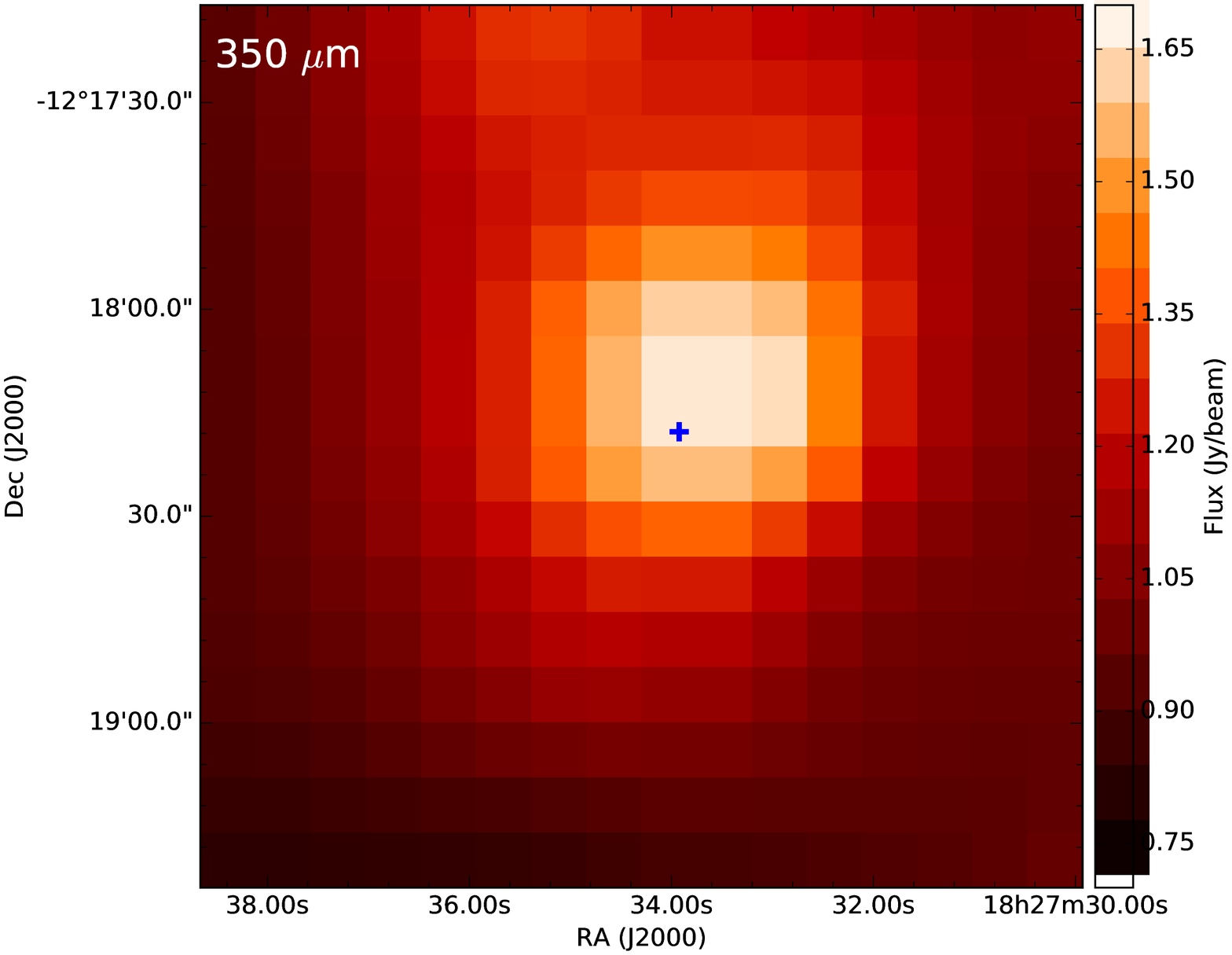}  \includegraphics[width=8cm]{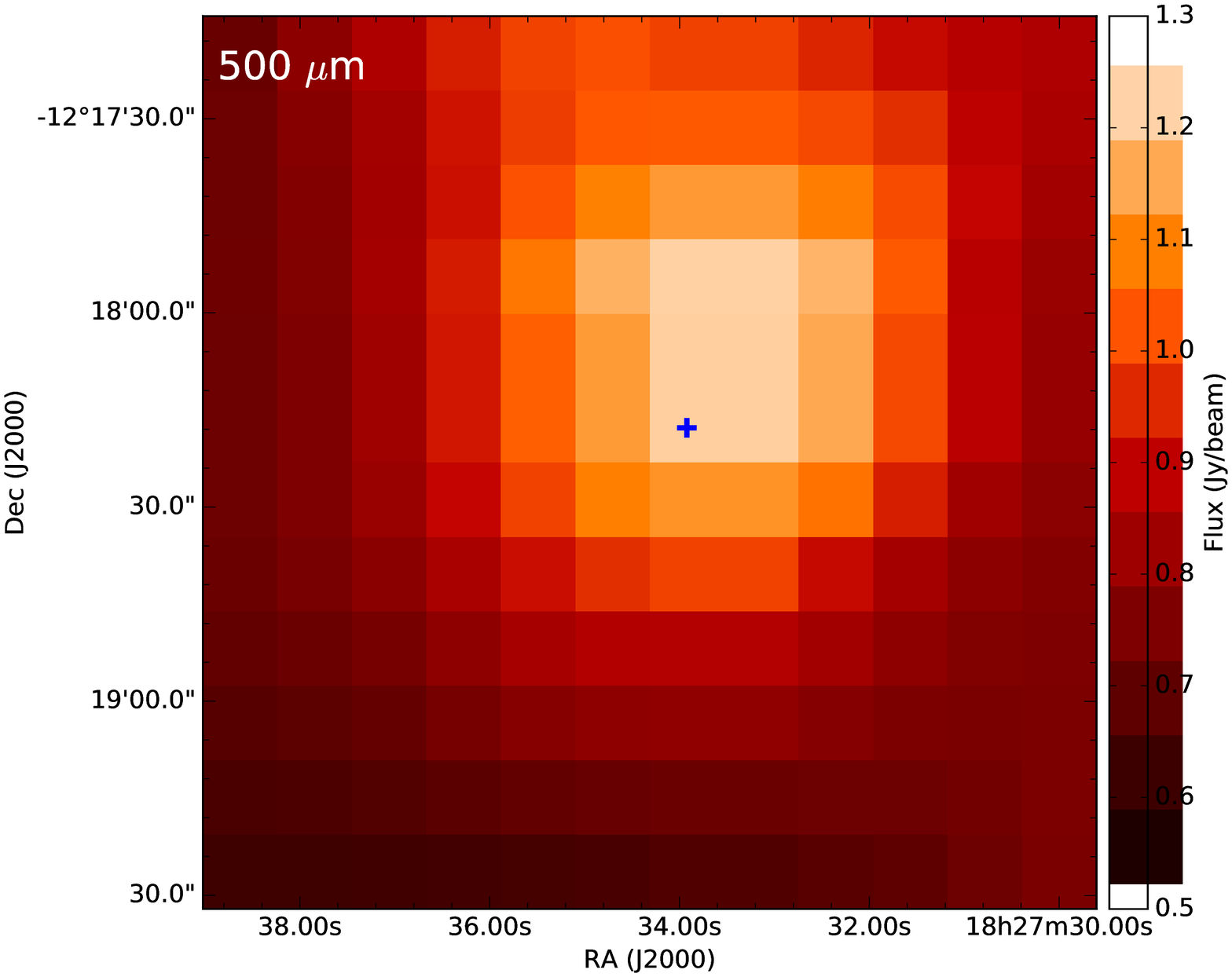} 
 \caption{19.281-0.387}
 \end{figure*}

\begin{figure*}
 \centering
 \includegraphics[width=8cm]{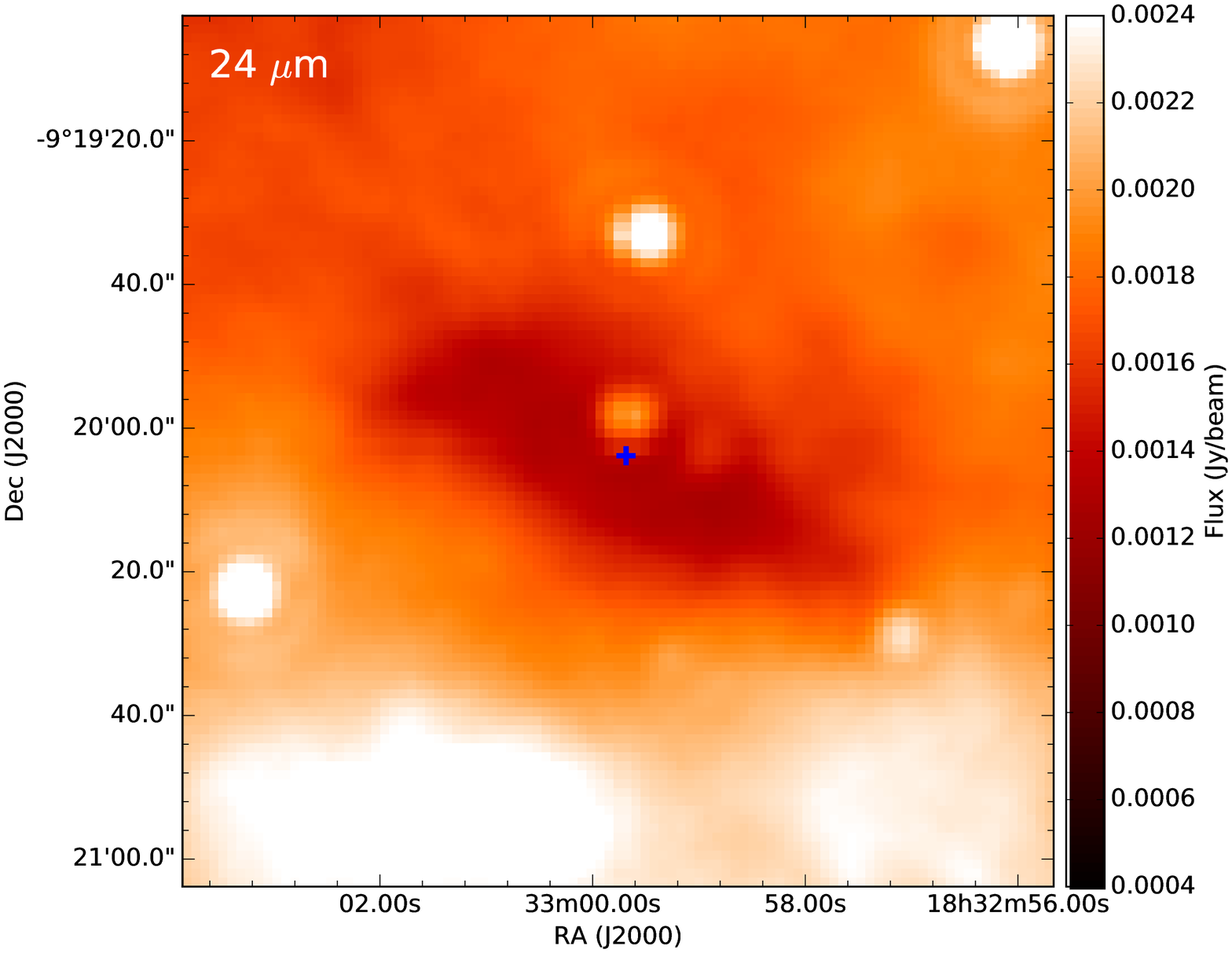}  \includegraphics[width=8cm]{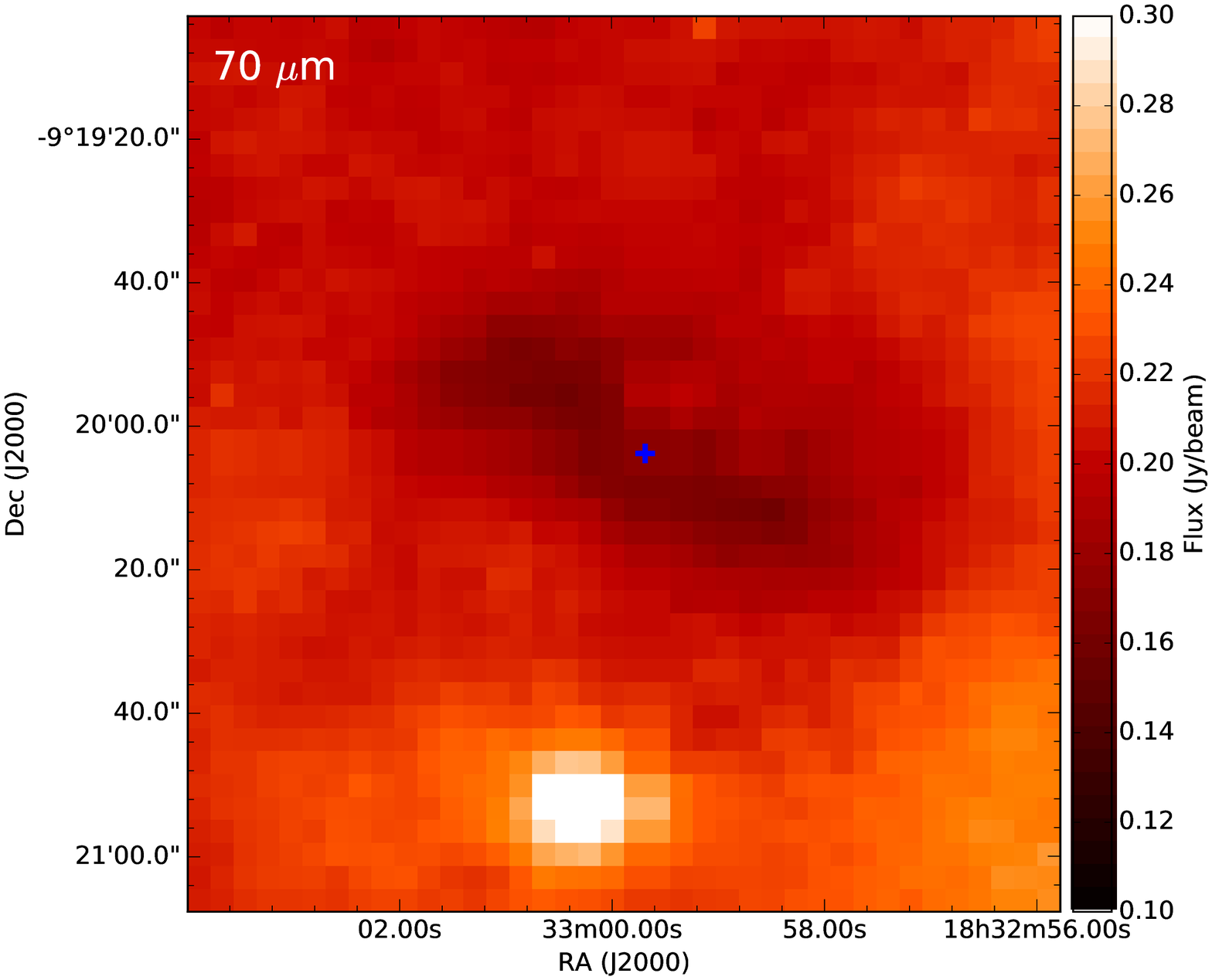} 
 \includegraphics[width=8cm]{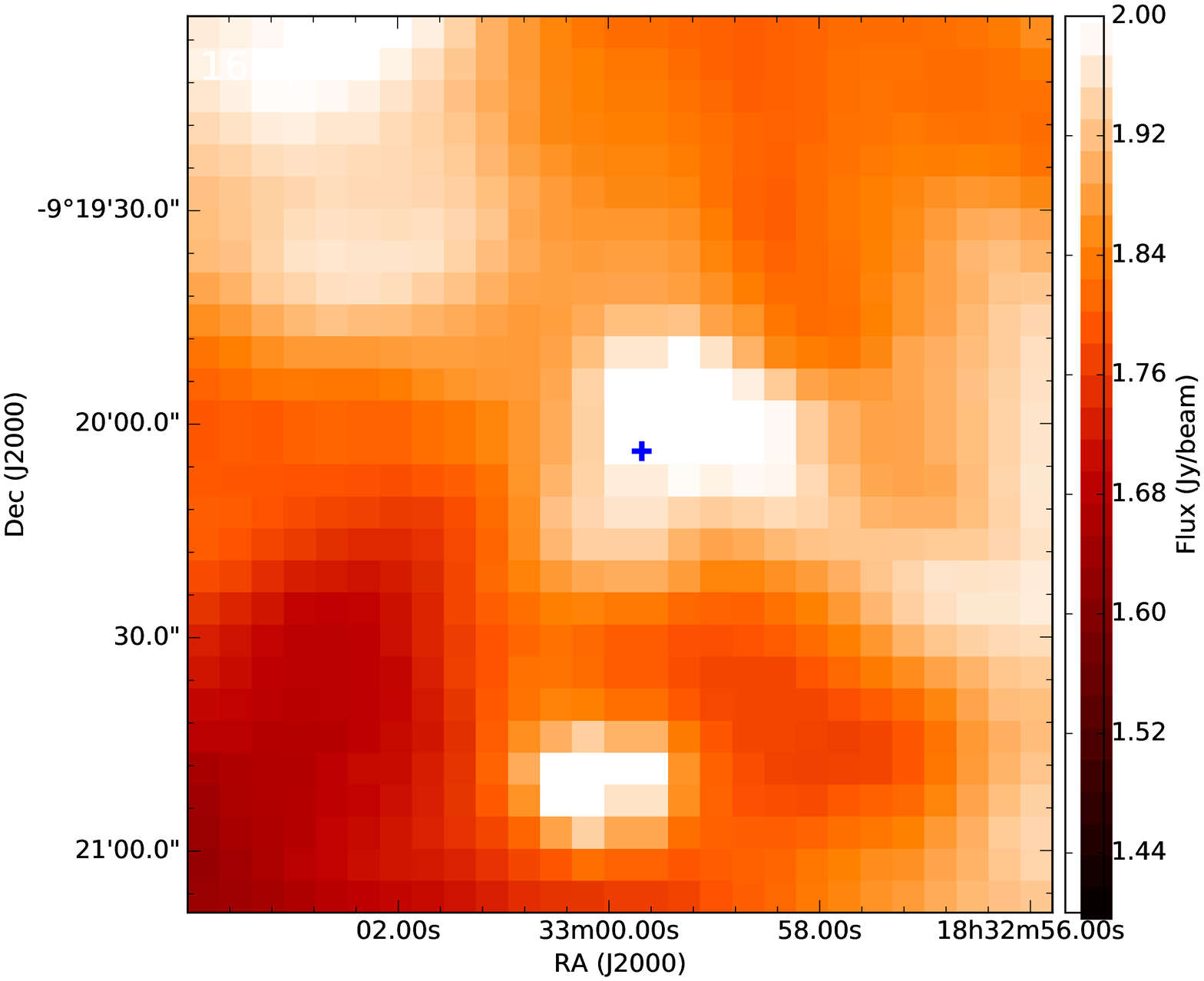}  \includegraphics[width=8cm]{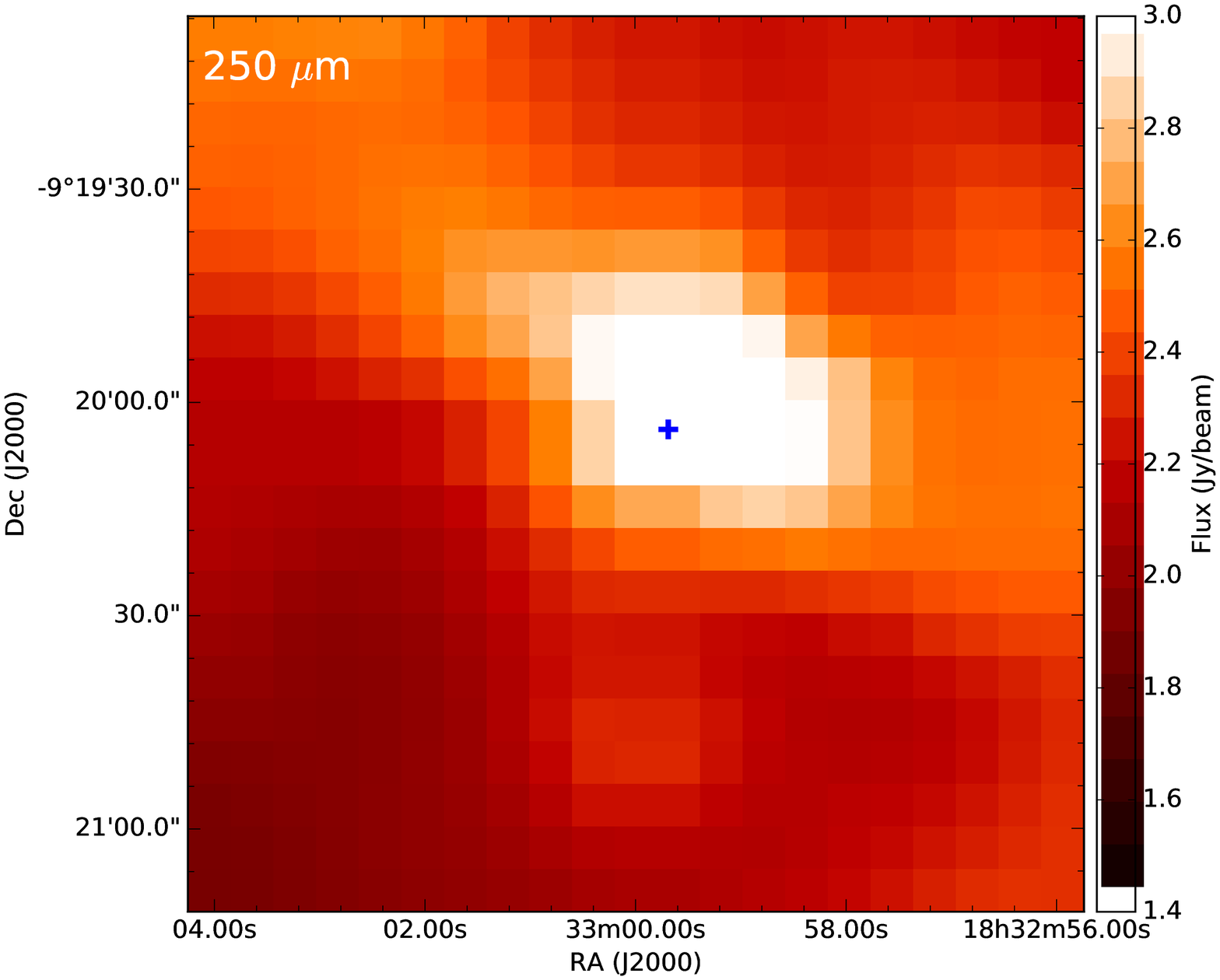} 
 \includegraphics[width=8cm]{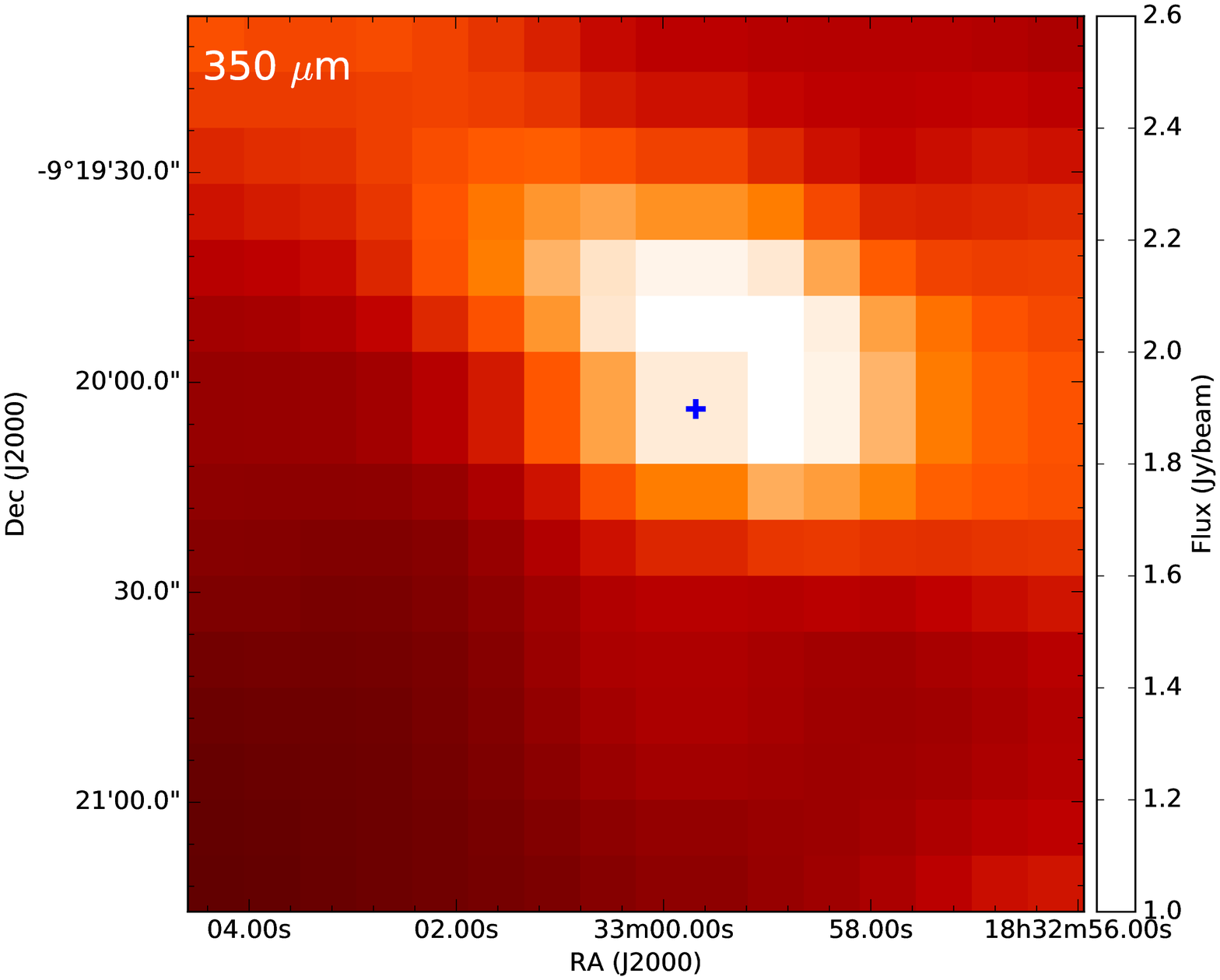}  \includegraphics[width=8cm]{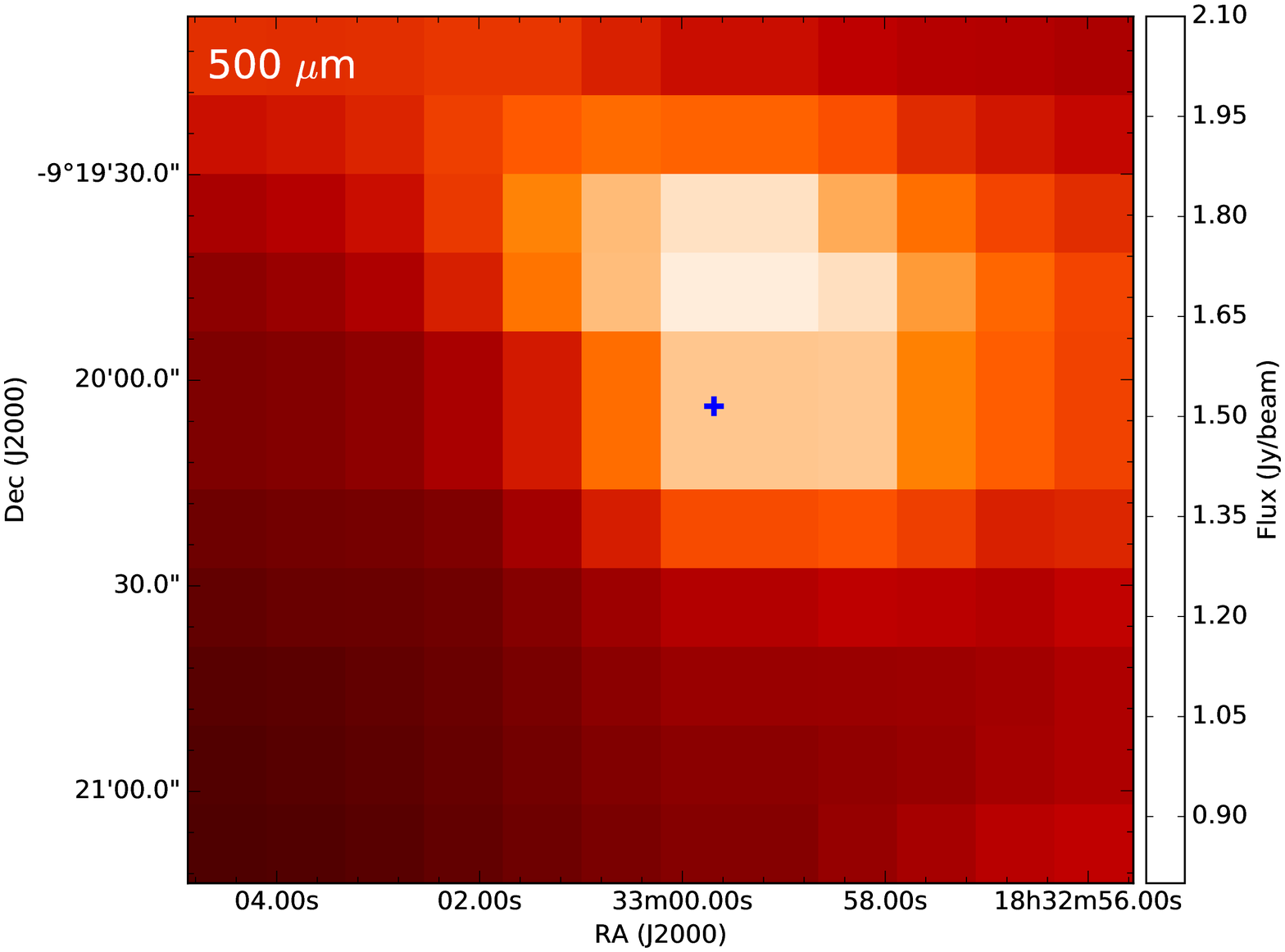} 
 \caption{22.53-0.192}
 \end{figure*}

\begin{figure*}
 \centering
 \includegraphics[width=8cm]{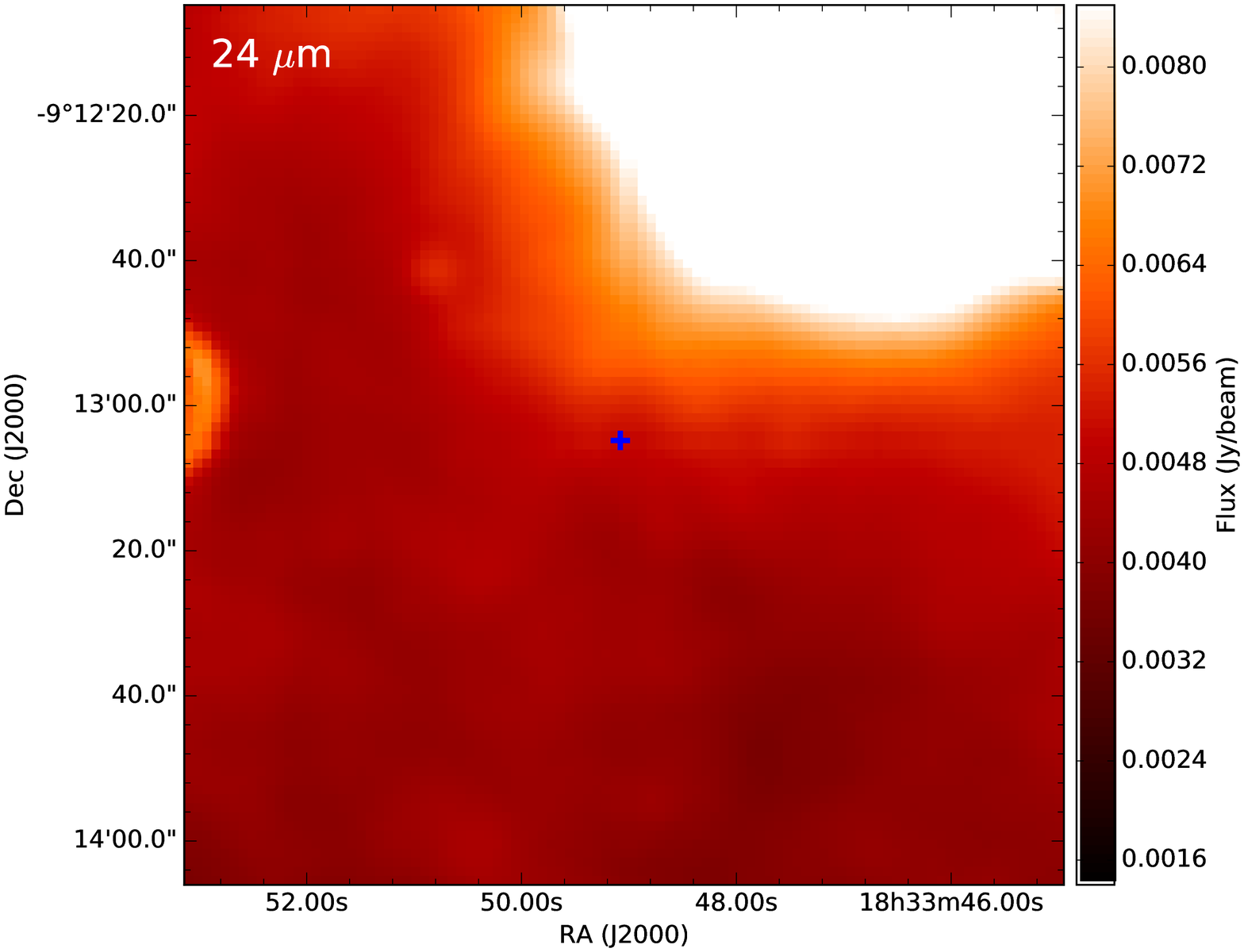}  \includegraphics[width=8cm]{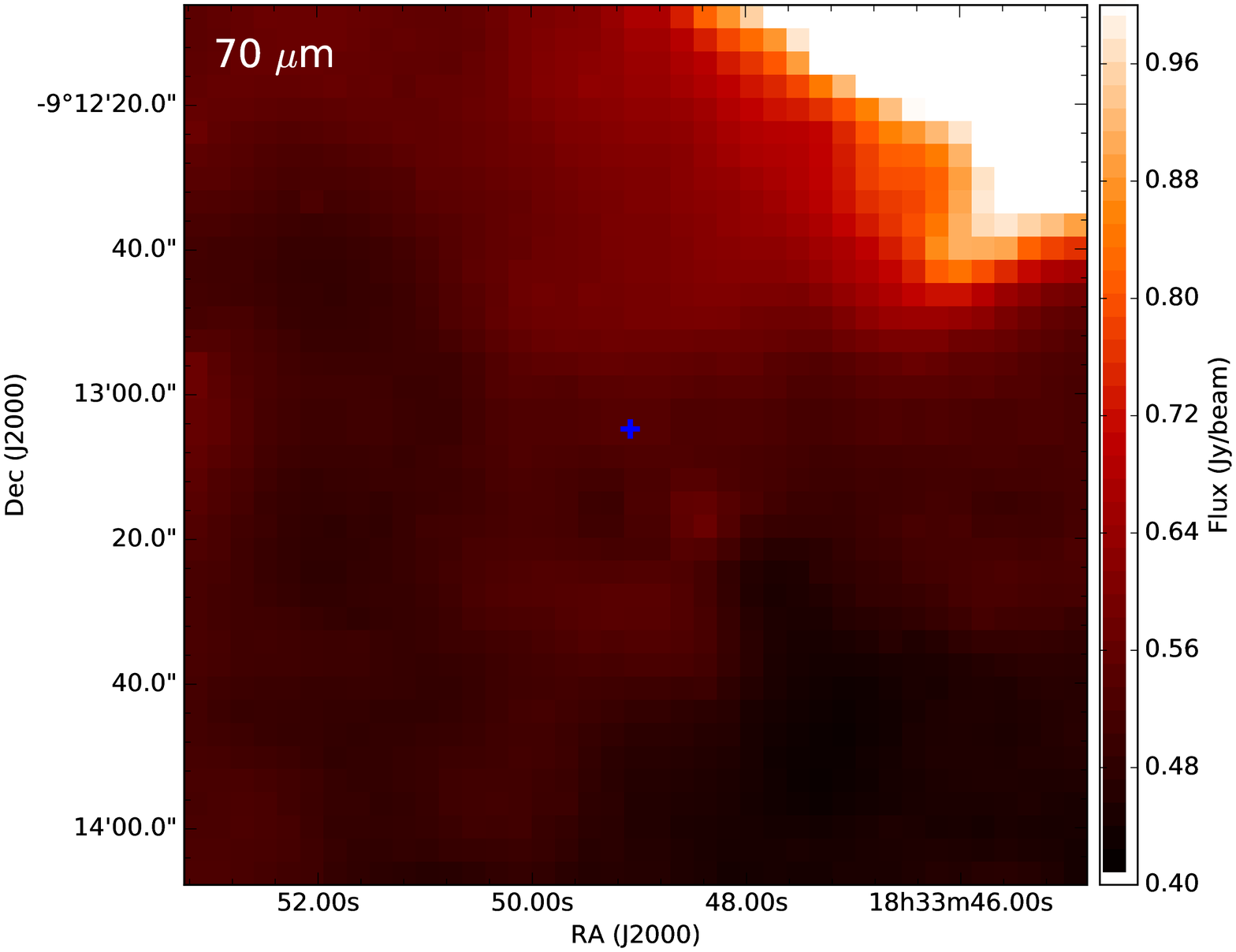} 
 \includegraphics[width=8cm]{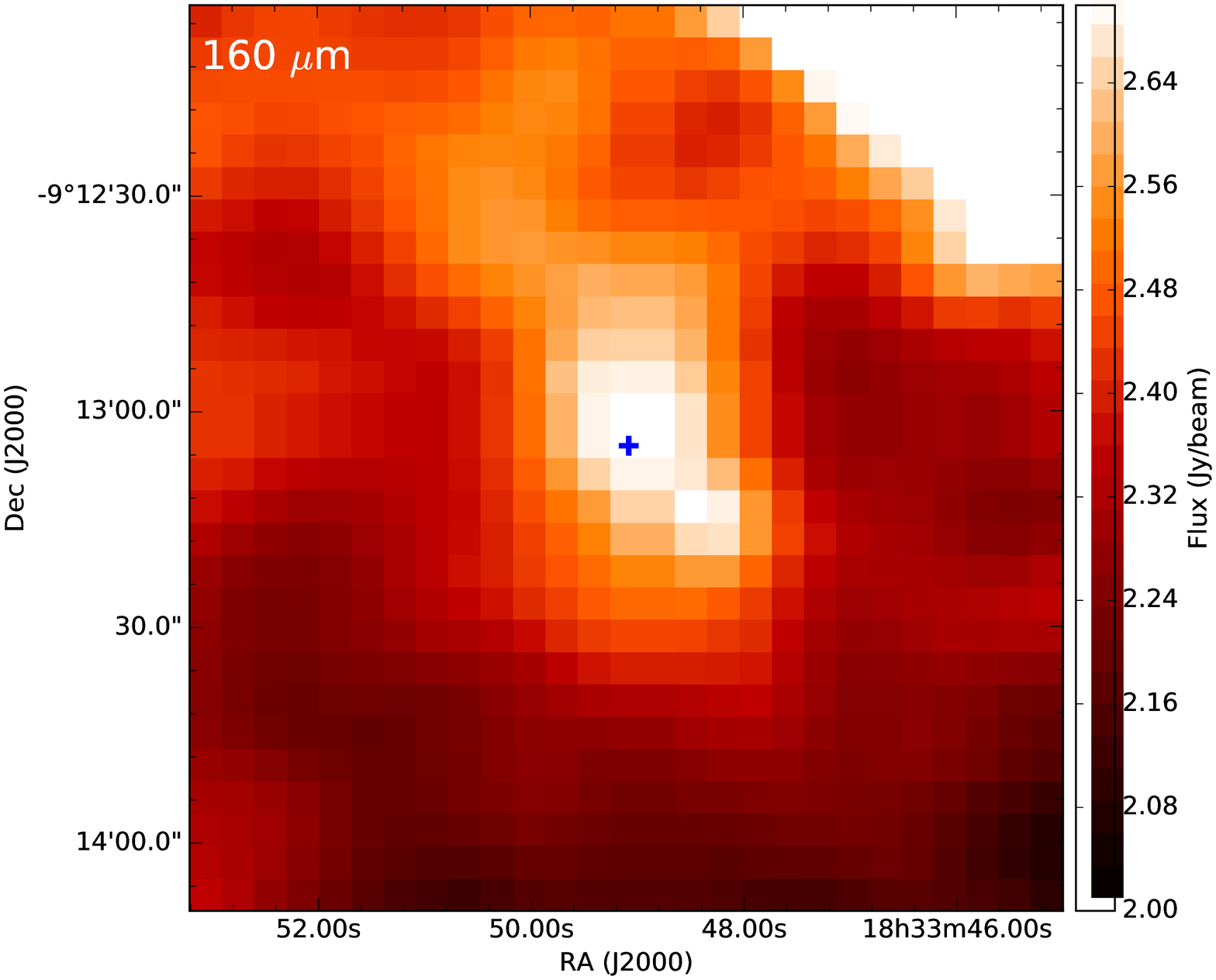}  \includegraphics[width=8cm]{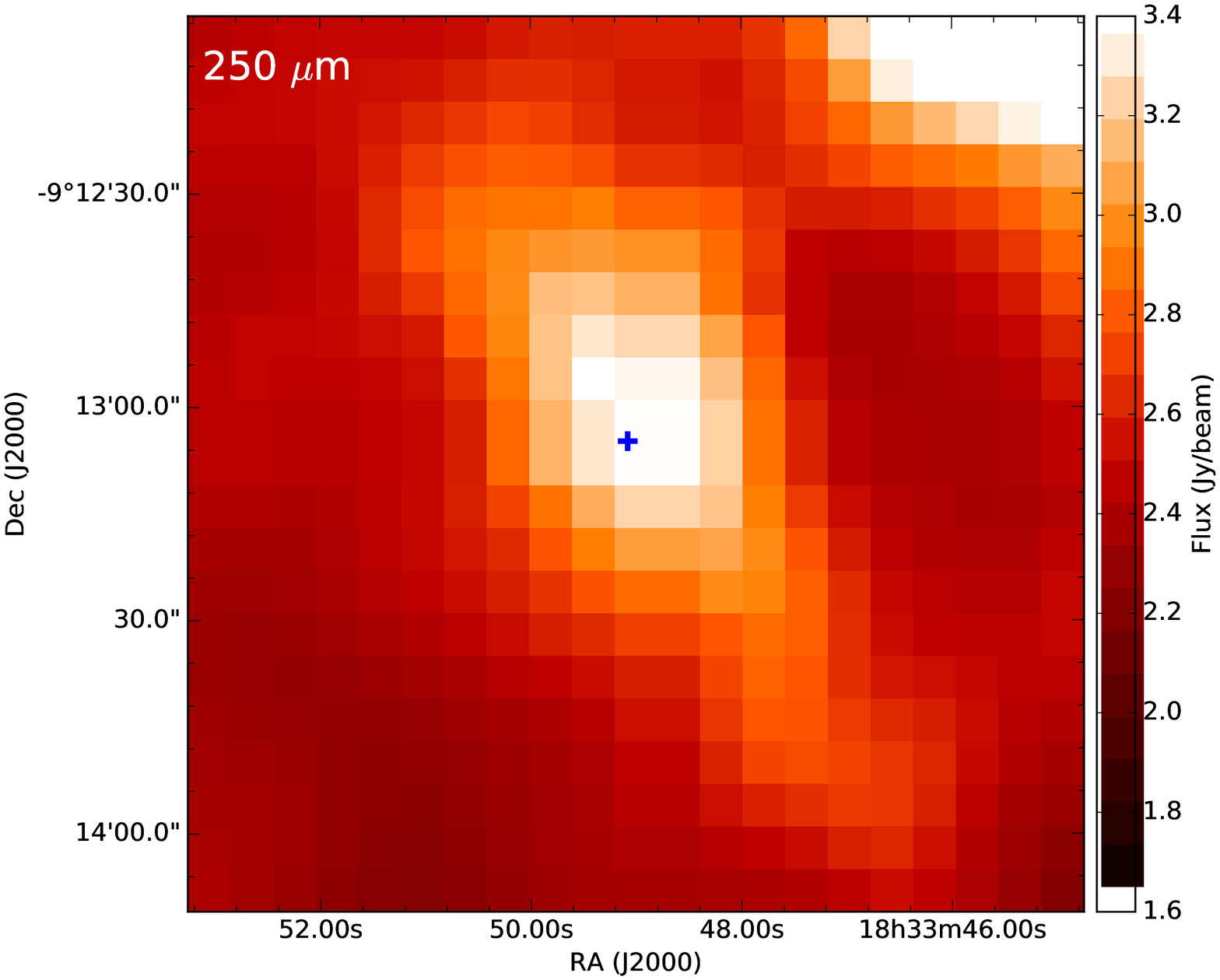} 
 \includegraphics[width=8cm]{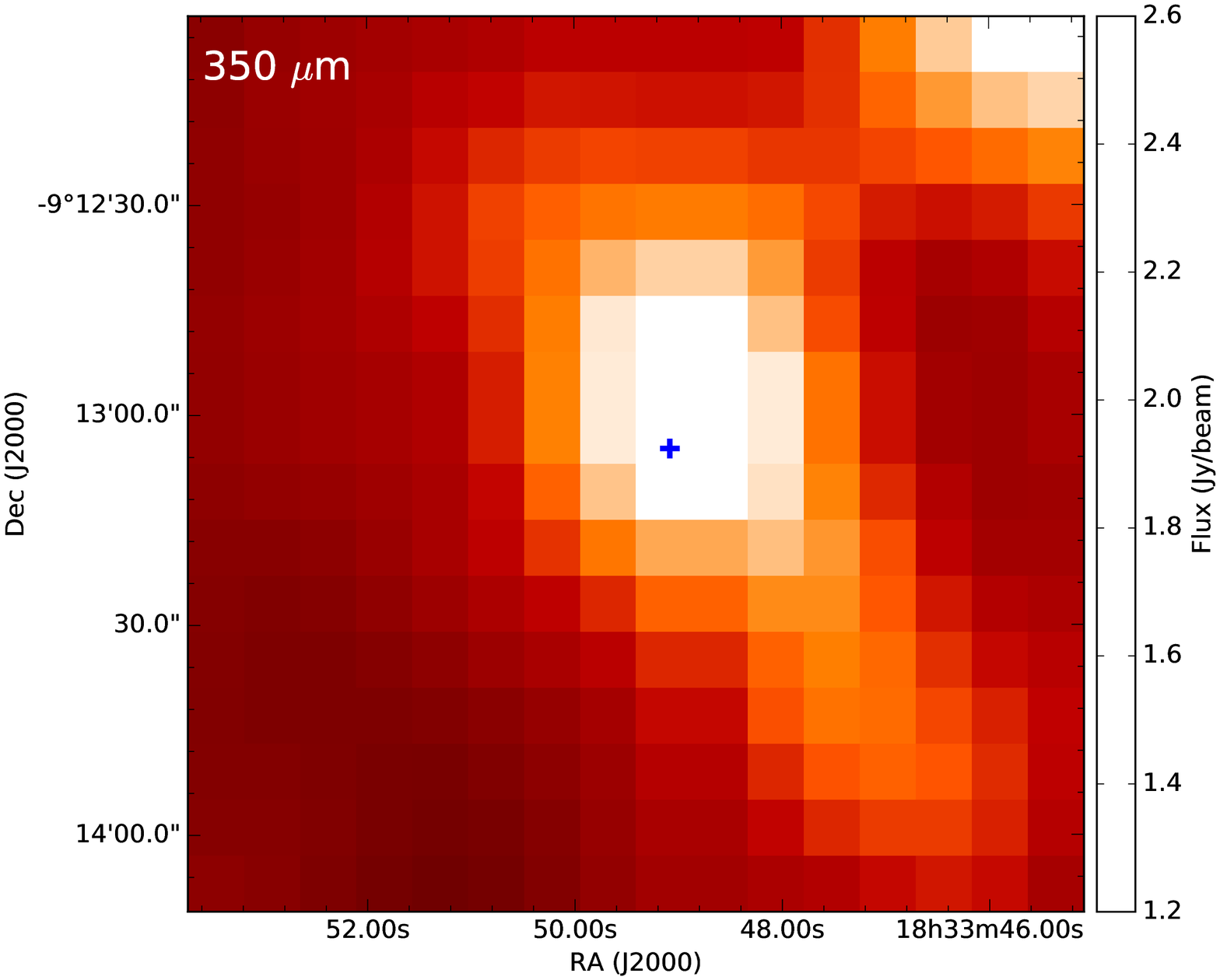}  \includegraphics[width=8cm]{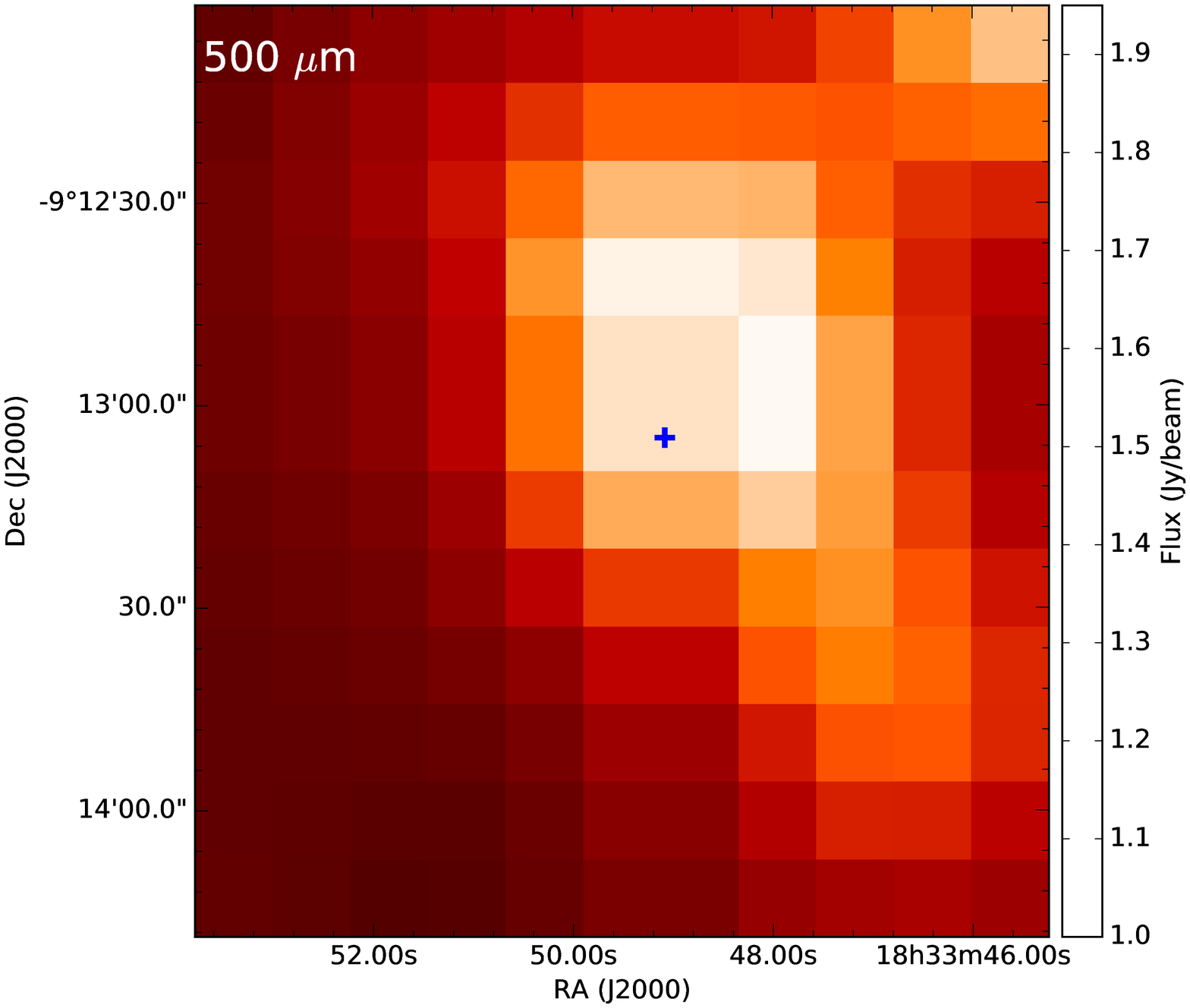} 
 \caption{22.756-0.284}
 \end{figure*}

\begin{figure*}
 \centering
 \includegraphics[width=8cm]{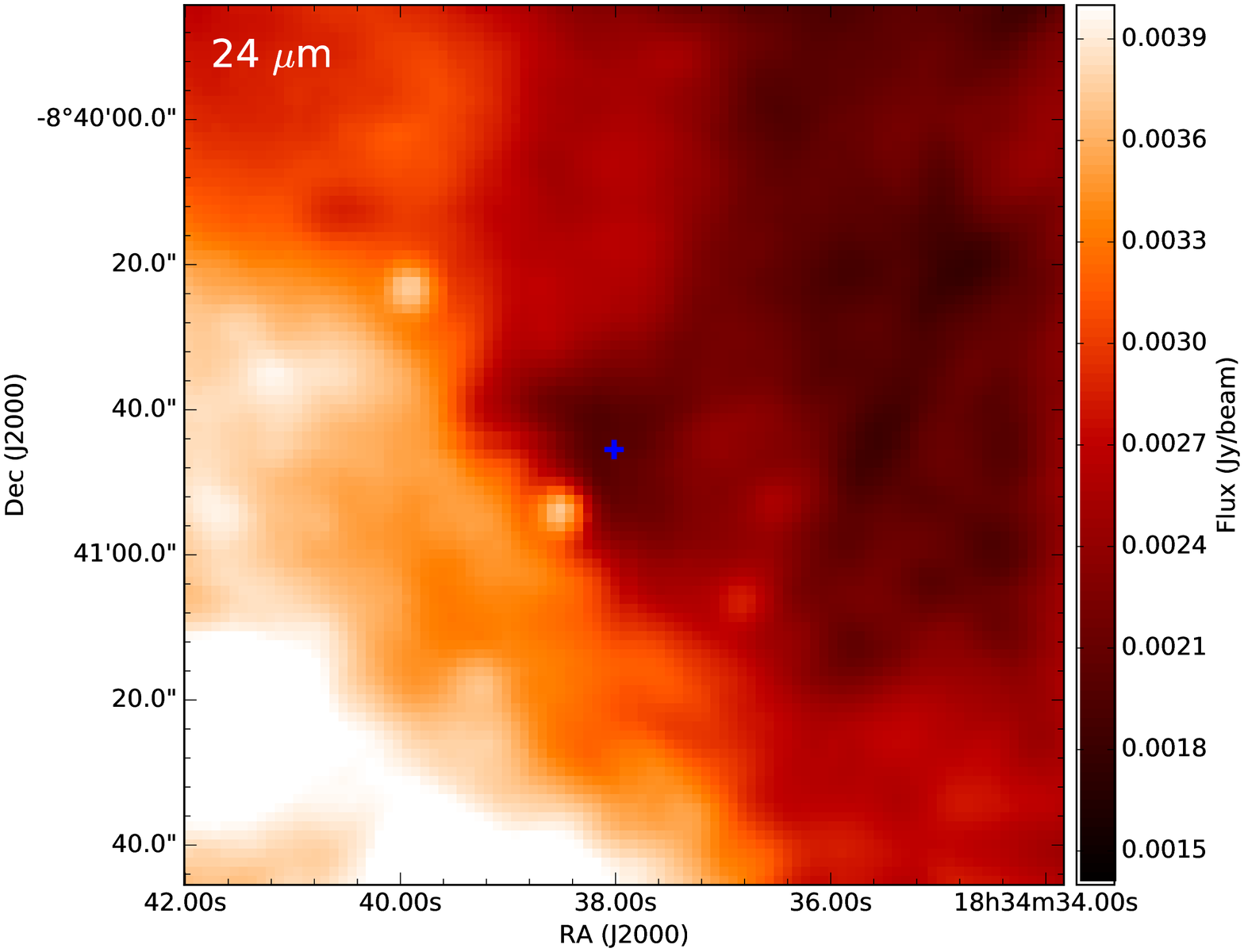}  \includegraphics[width=8cm]{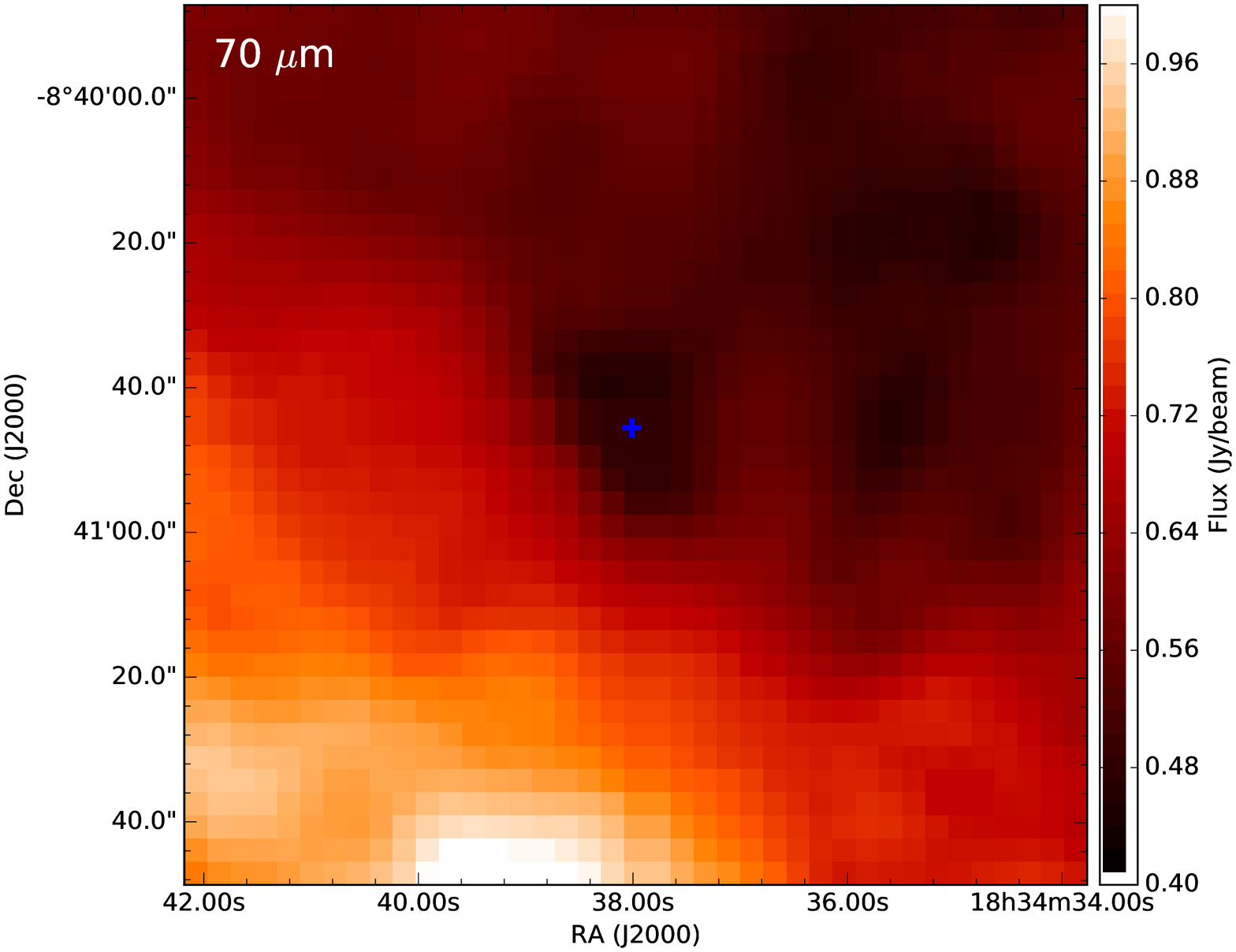} 
 \includegraphics[width=8cm]{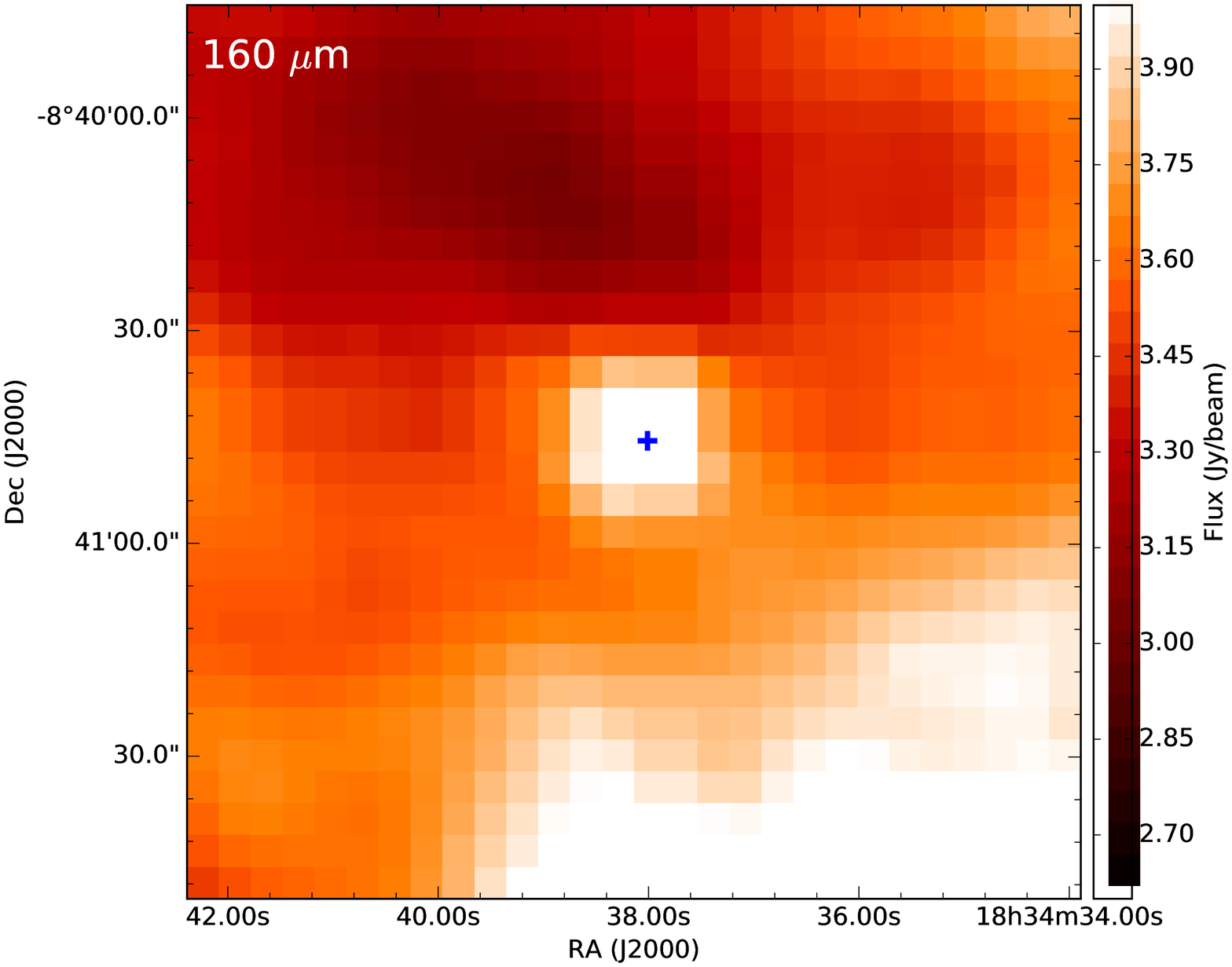}  \includegraphics[width=8cm]{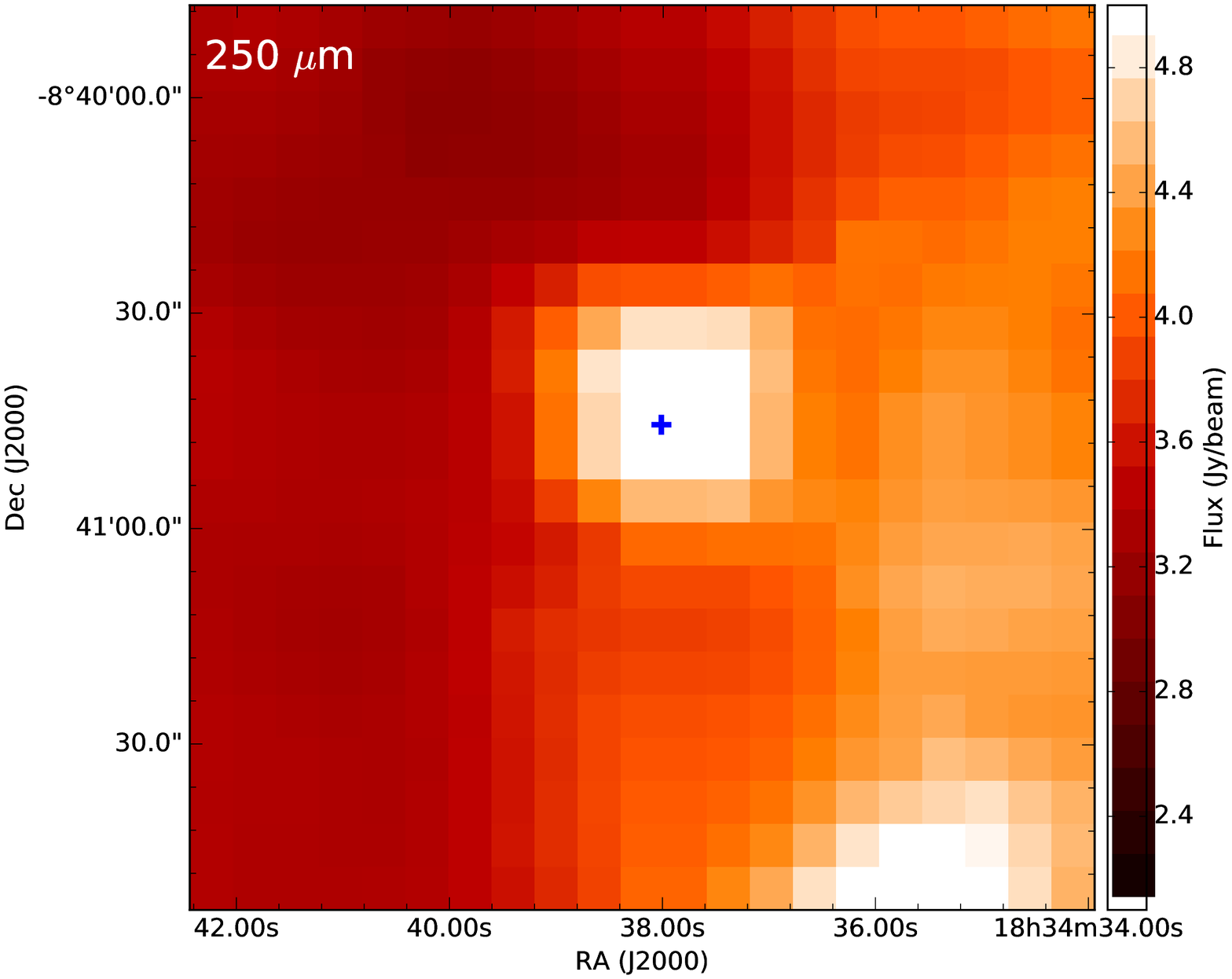} 
 \includegraphics[width=8cm]{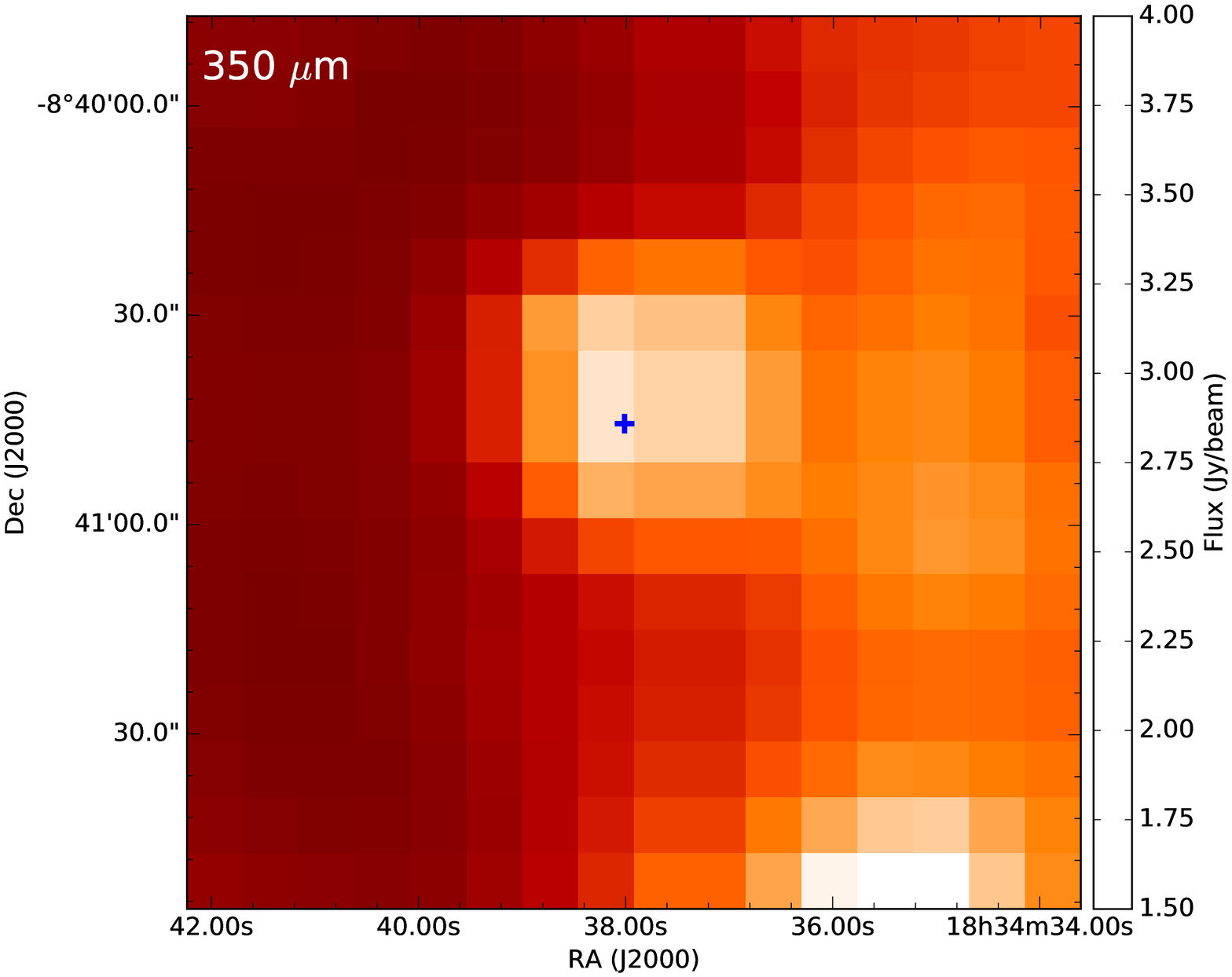}  \includegraphics[width=8cm]{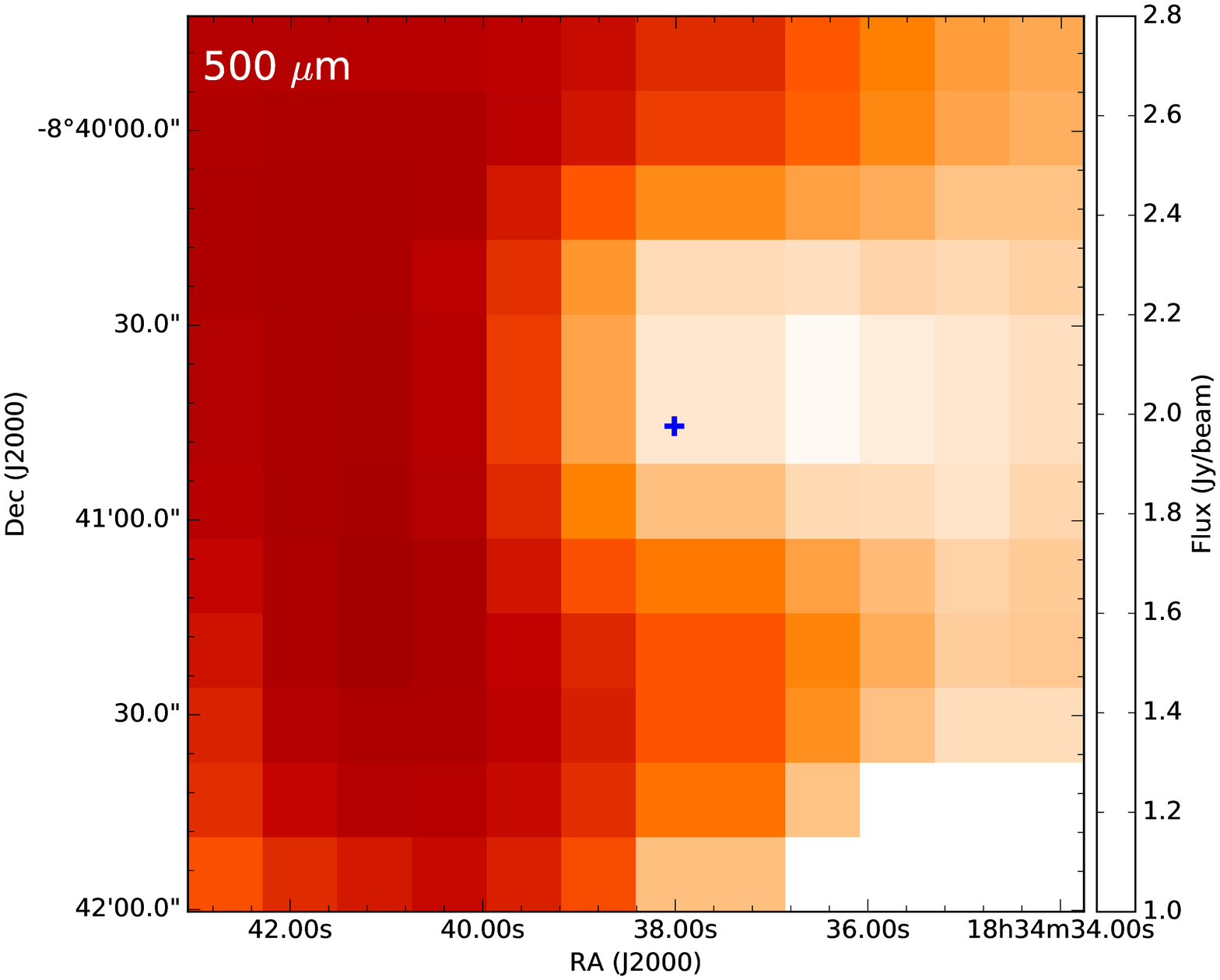} 
 \caption{23.271-0.263}
 \end{figure*}

 \begin{figure*}
 \centering
 \includegraphics[width=8cm]{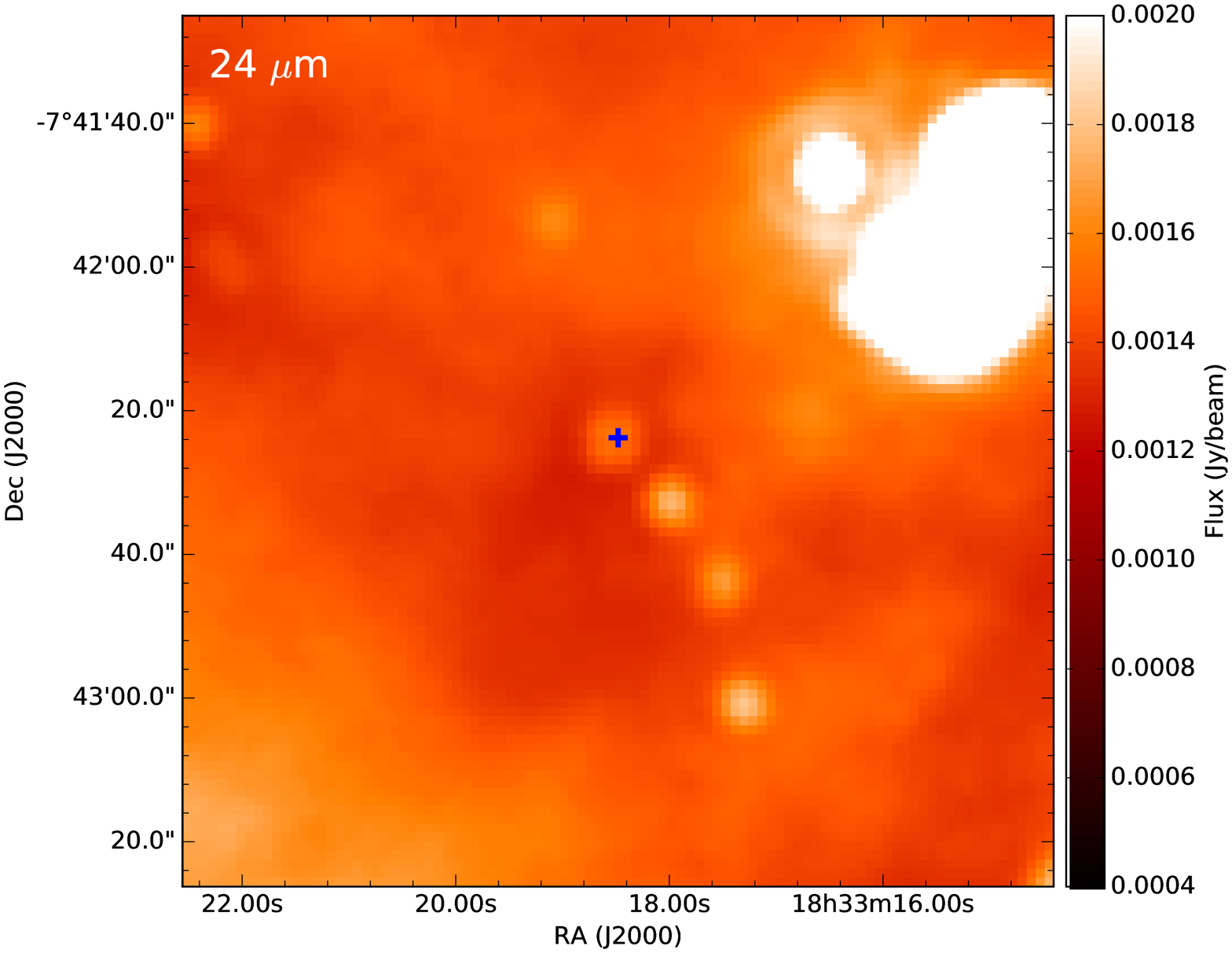}  \includegraphics[width=8cm]{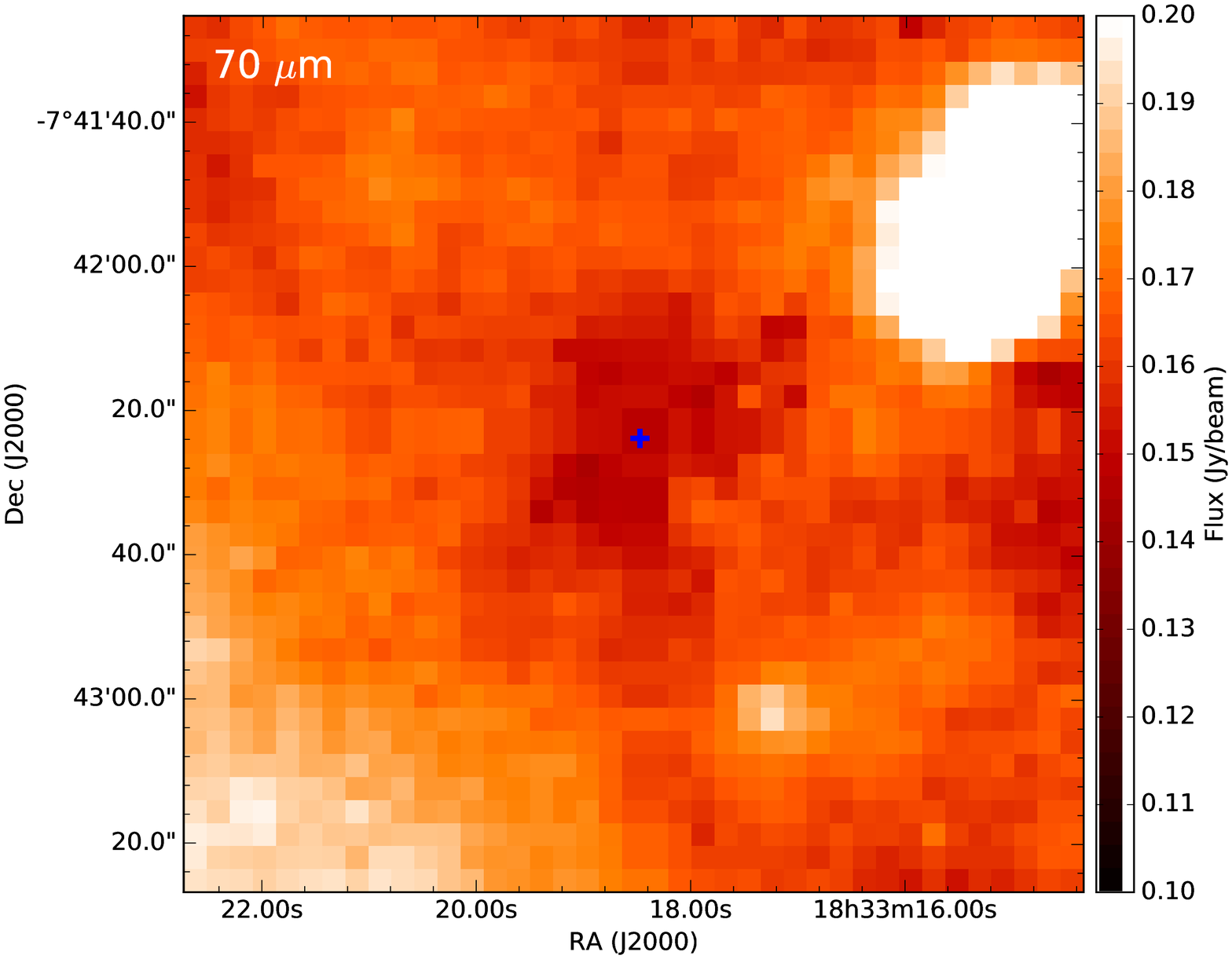} 
 \includegraphics[width=8cm]{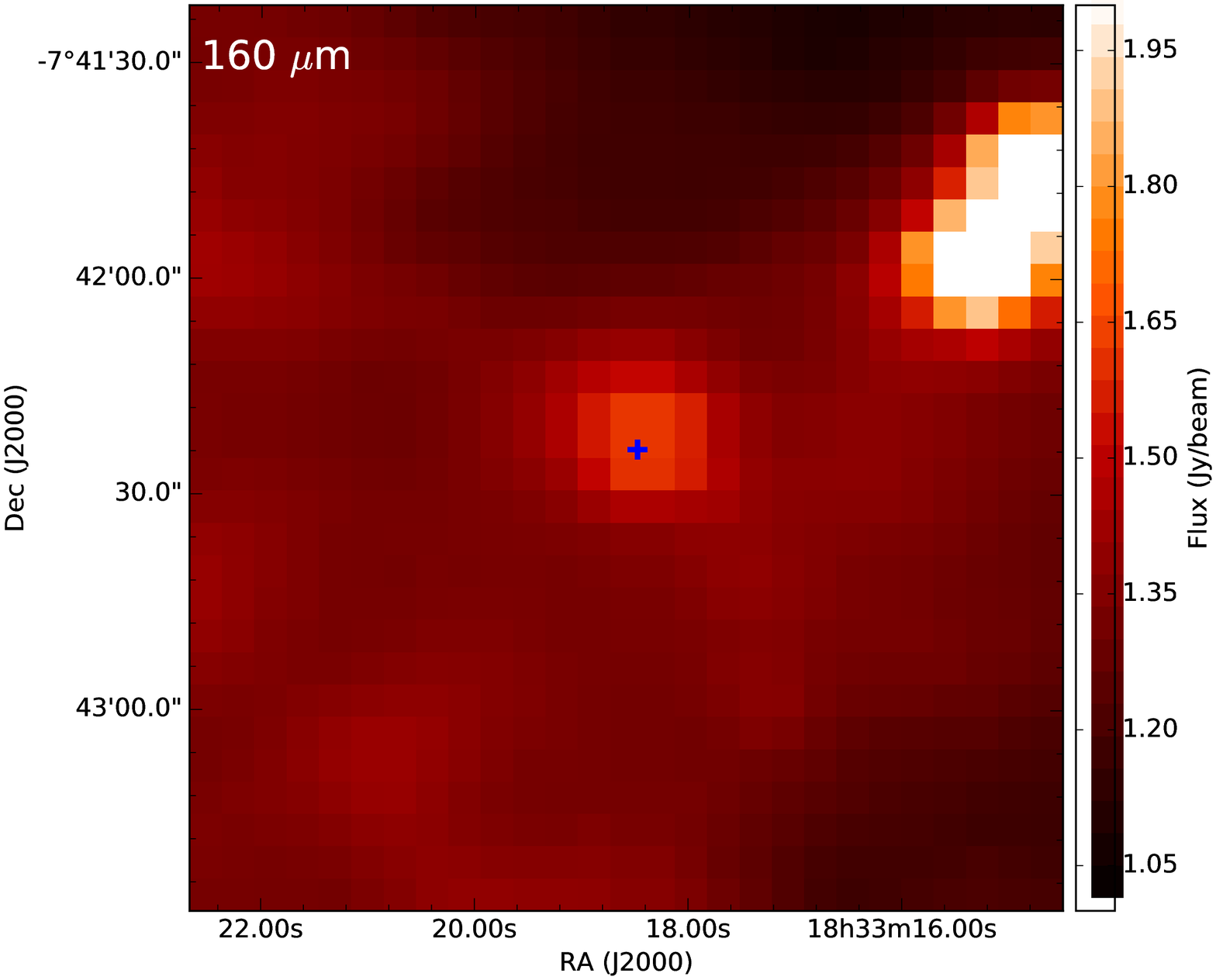}  \includegraphics[width=8cm]{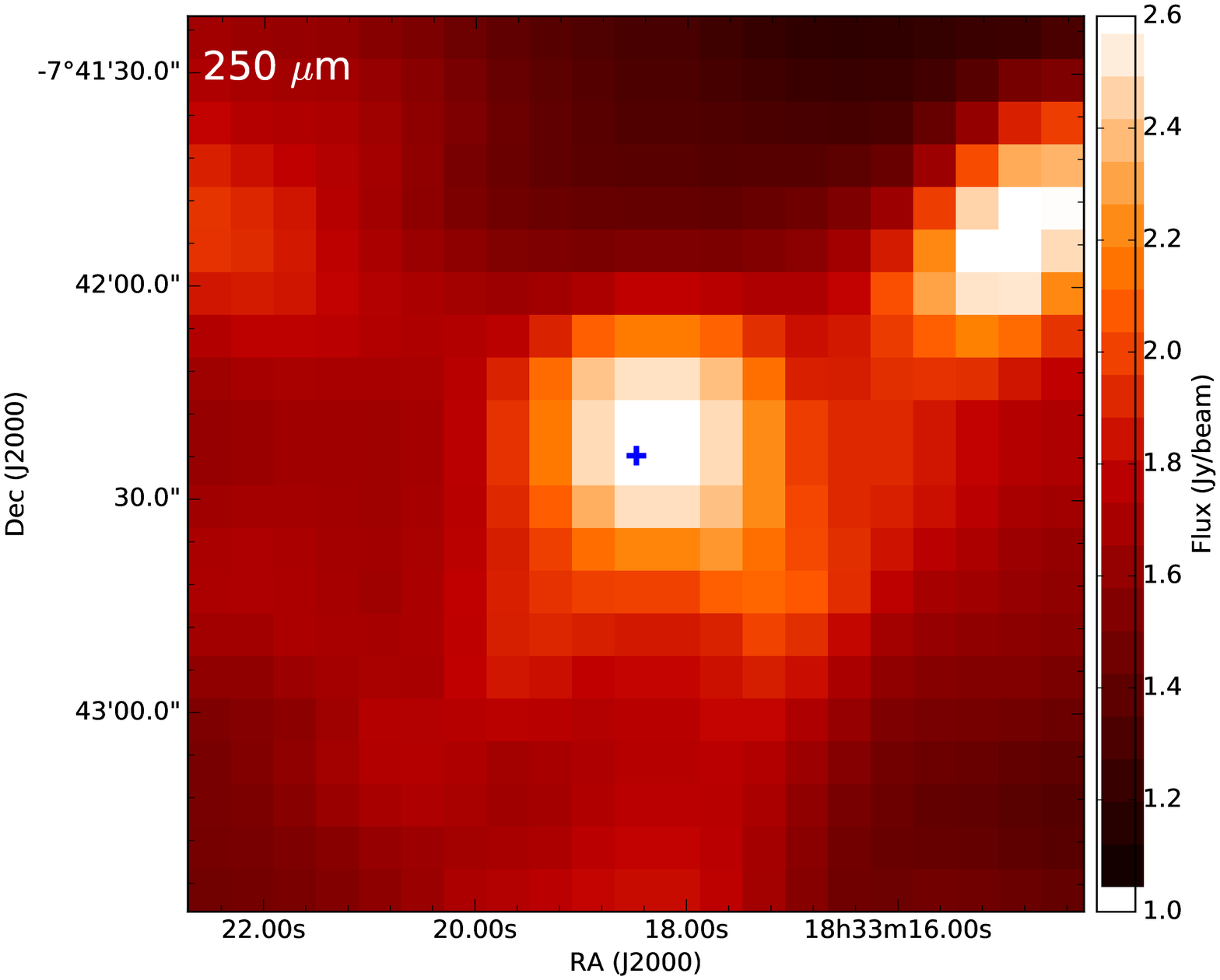} 
 \includegraphics[width=8cm]{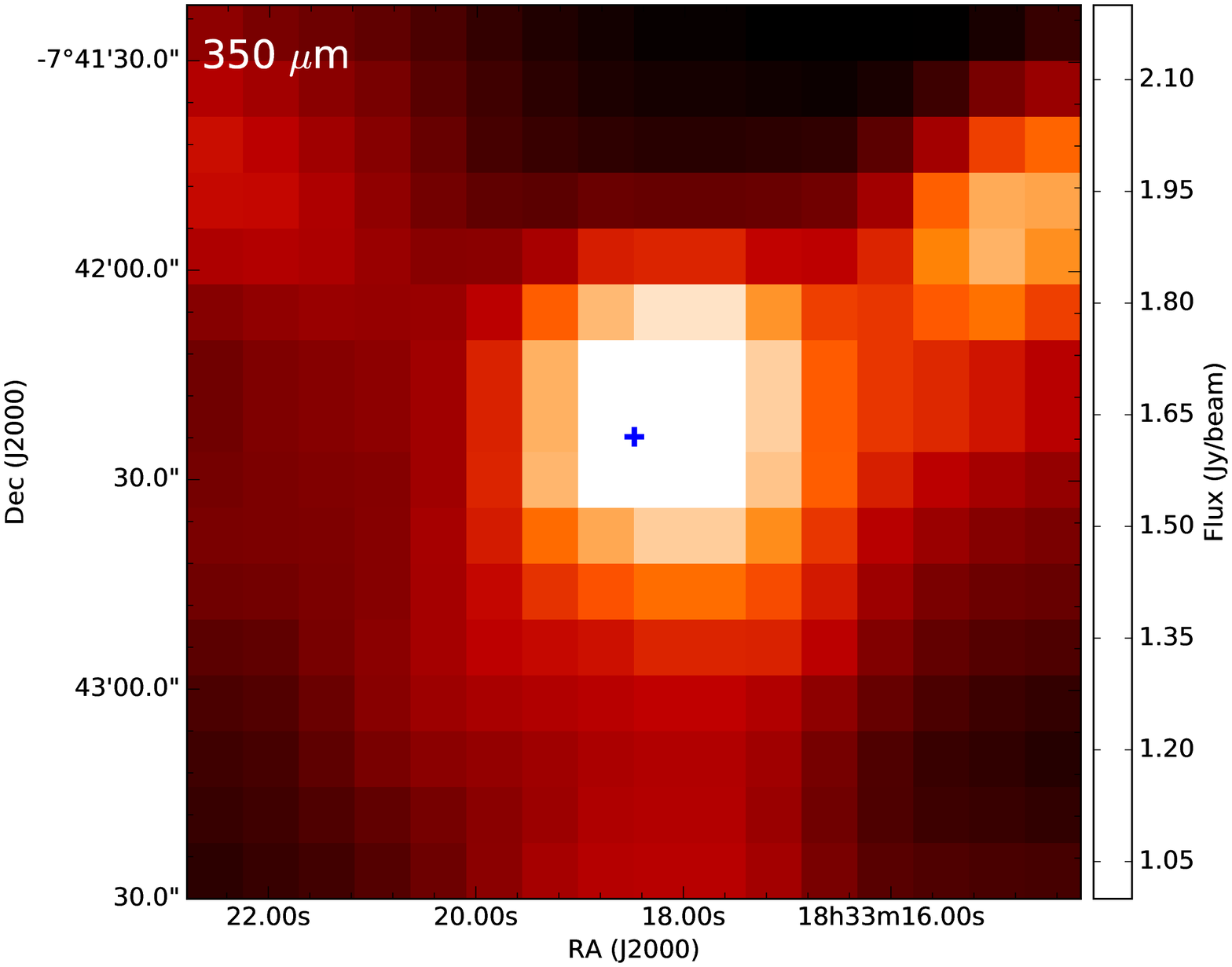}  \includegraphics[width=8cm]{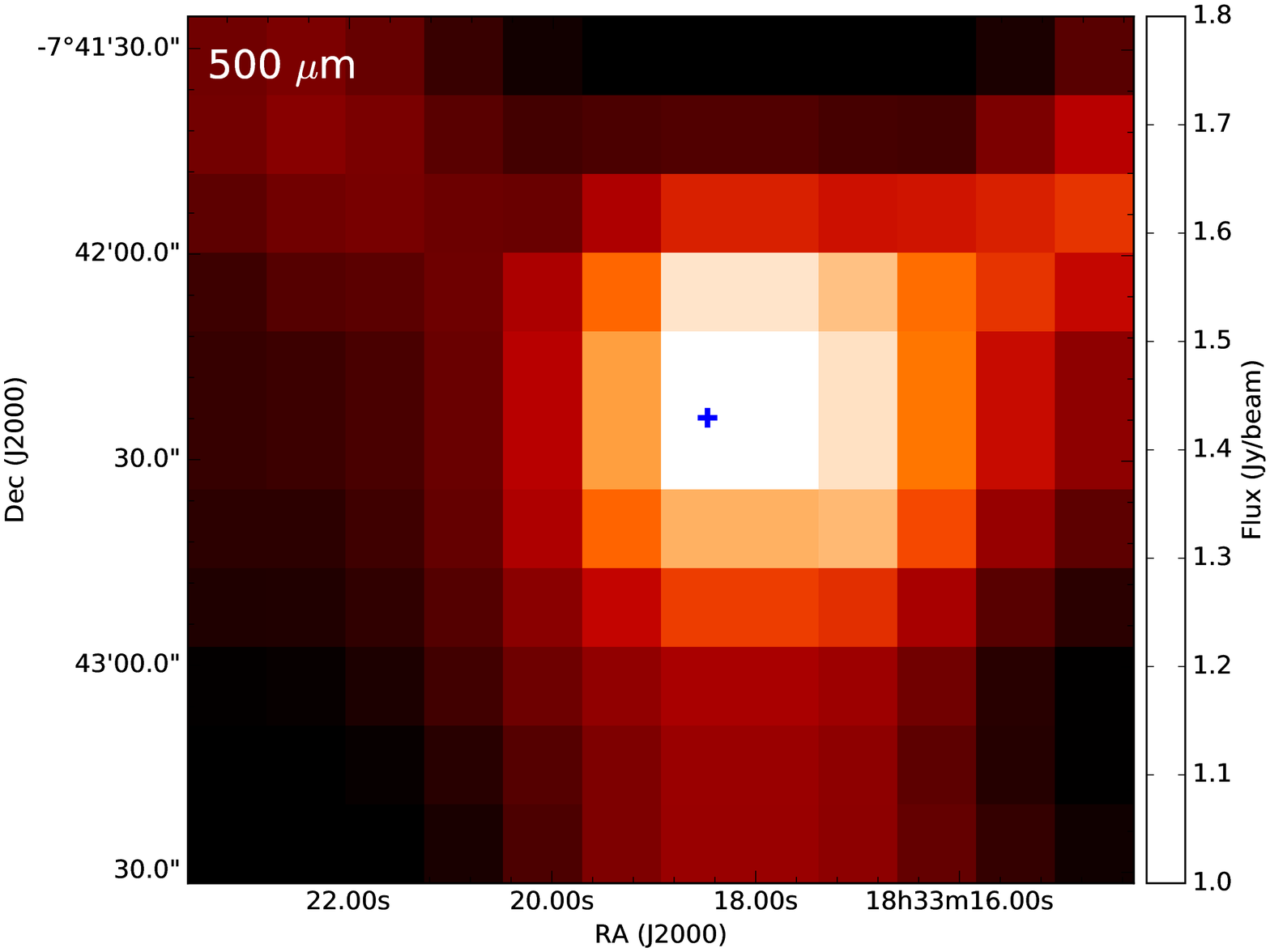} 
 \caption{24.013+0.488}
 \end{figure*}

\begin{figure*}
 \centering
 \includegraphics[width=8cm]{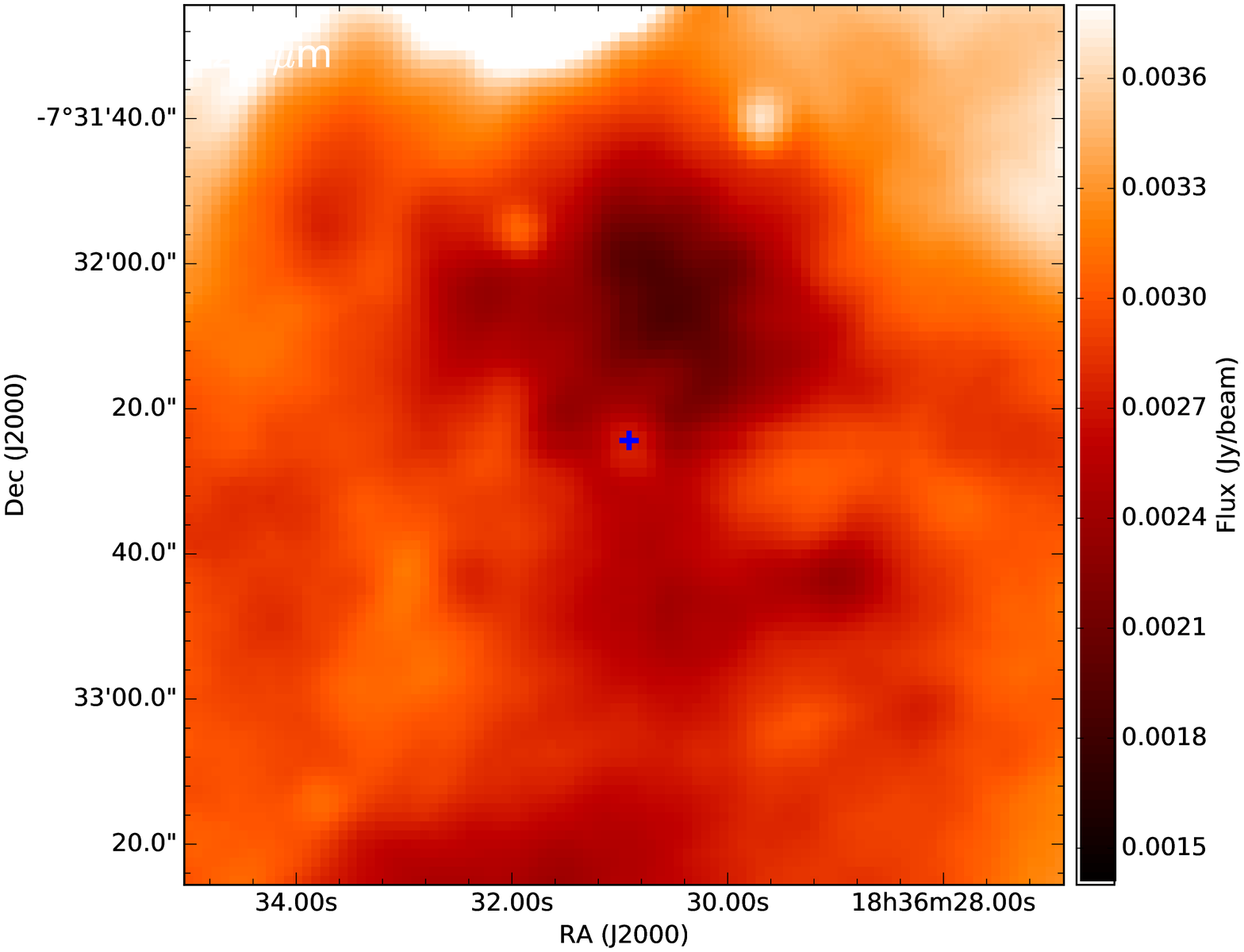}  \includegraphics[width=8cm]{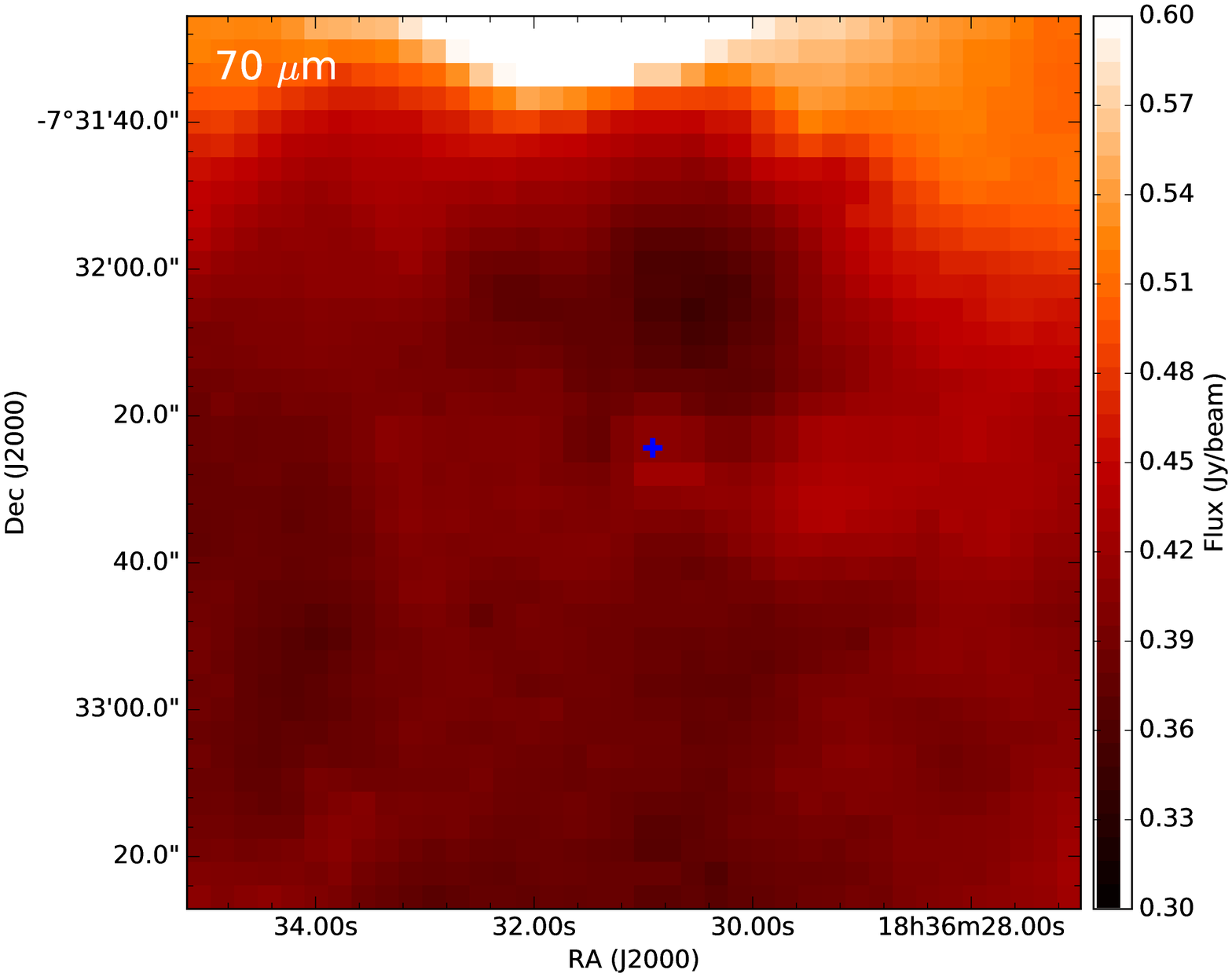} 
 \includegraphics[width=8cm]{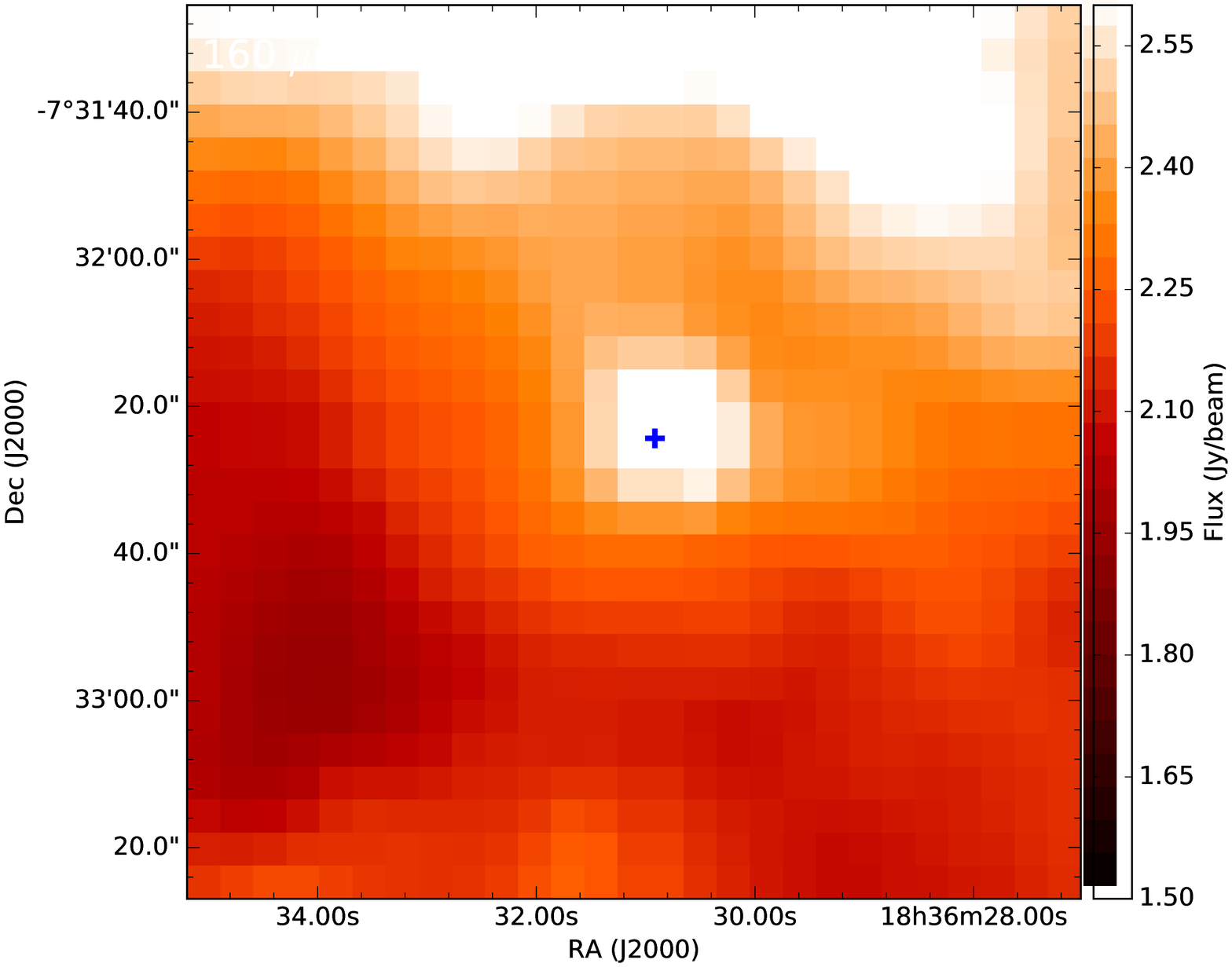}  \includegraphics[width=8cm]{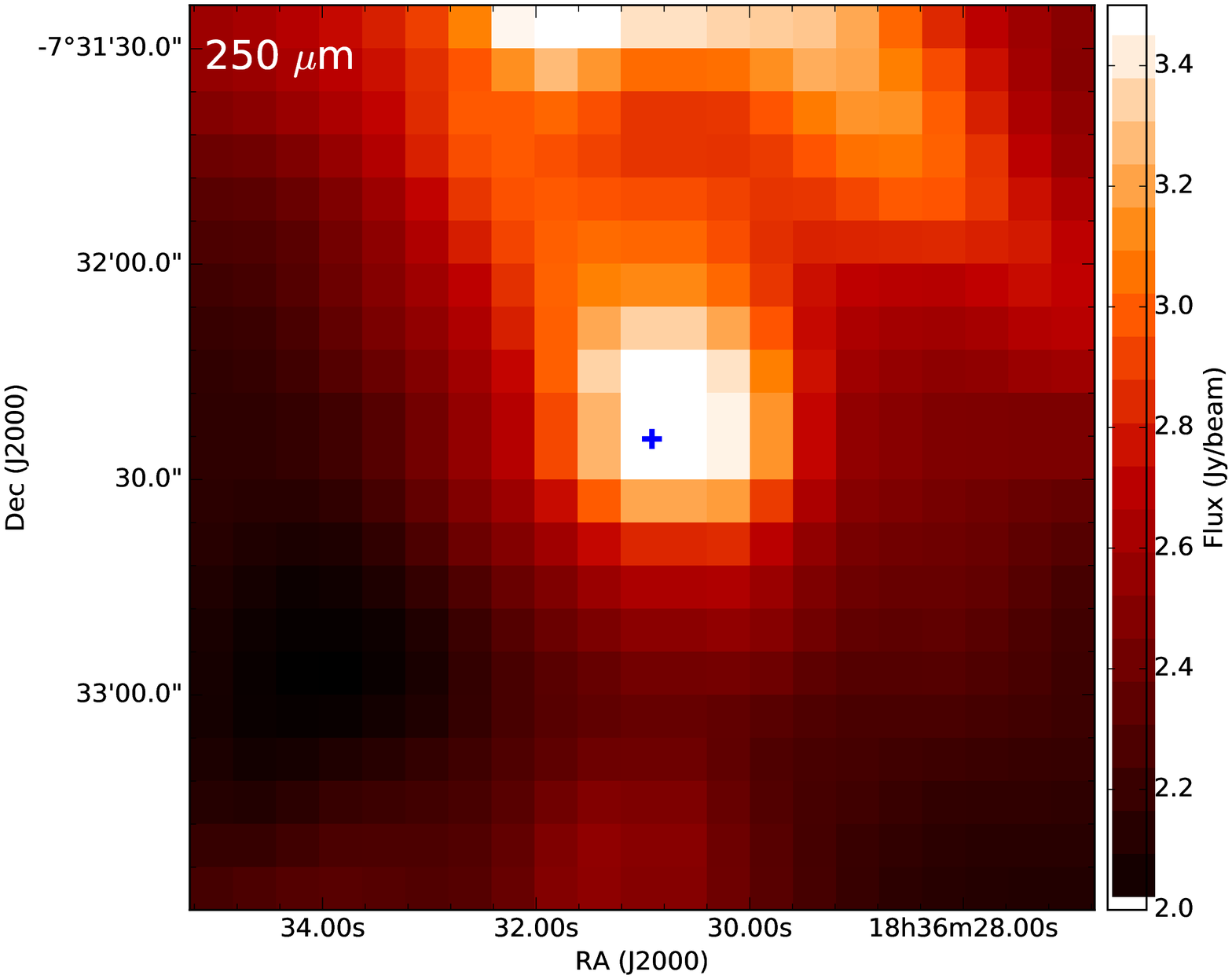} 
 \includegraphics[width=8cm]{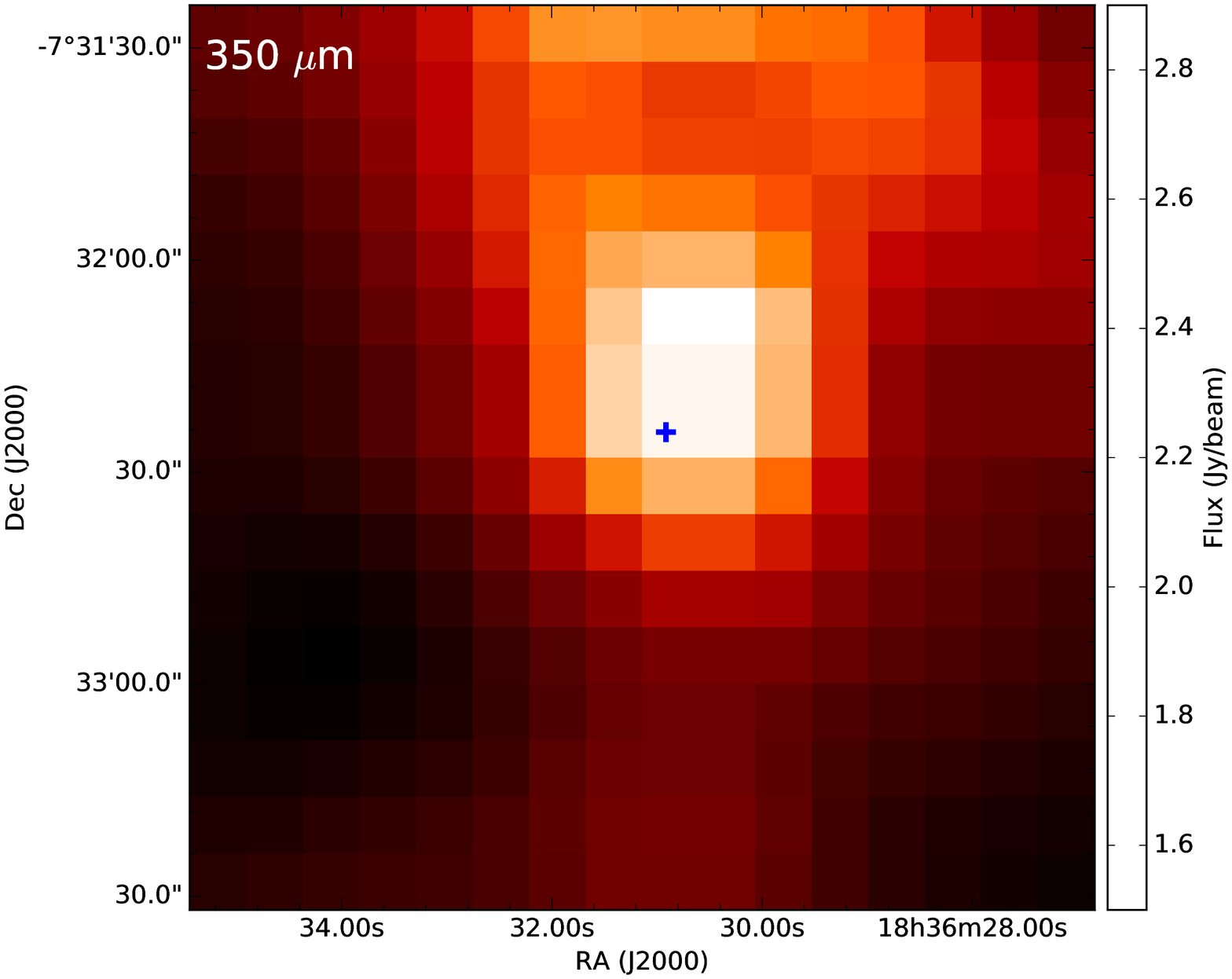}  \includegraphics[width=8cm]{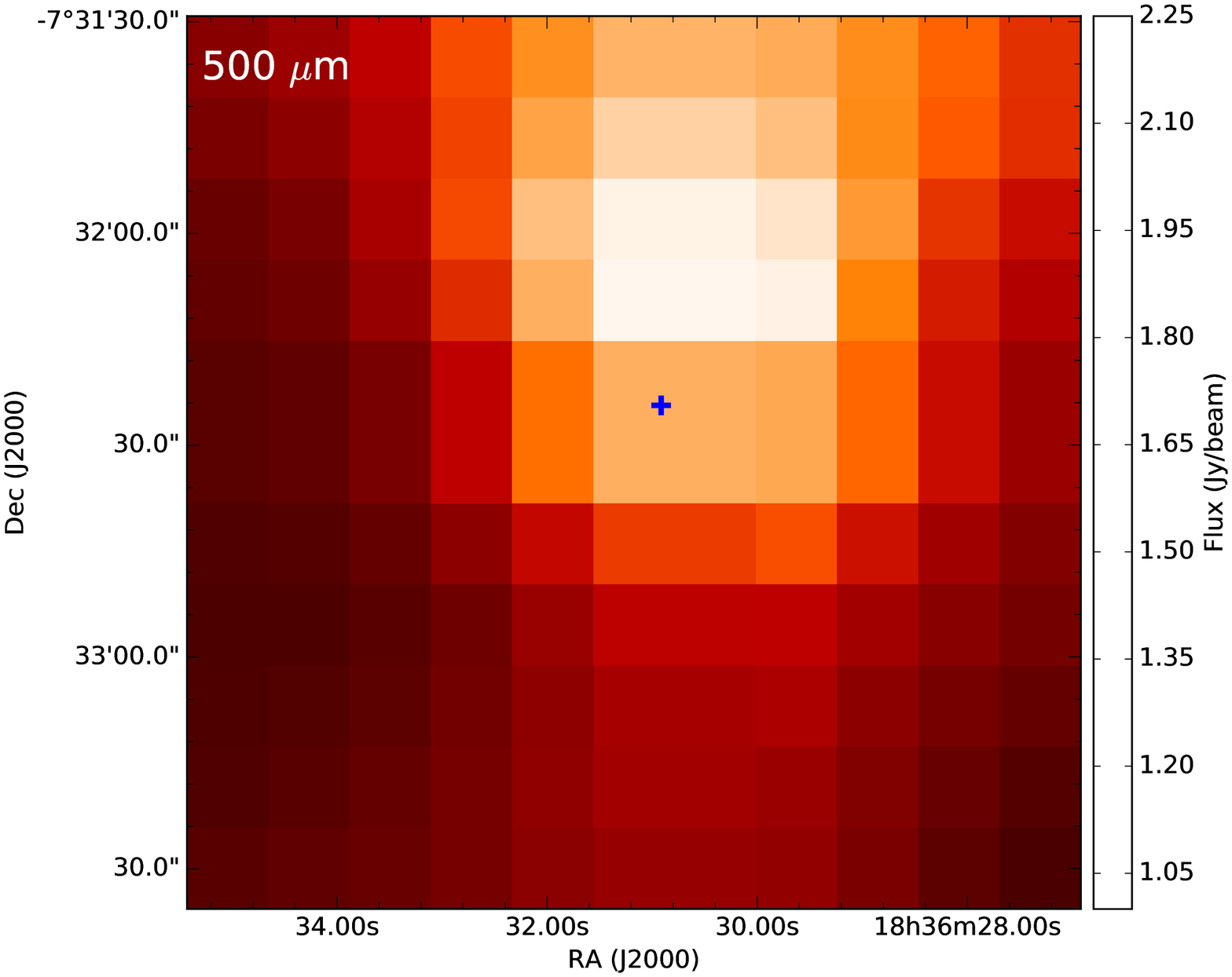} 
 \caption{24.528-0.136}
 \end{figure*}

\begin{figure*}
 \centering
 \includegraphics[width=8cm]{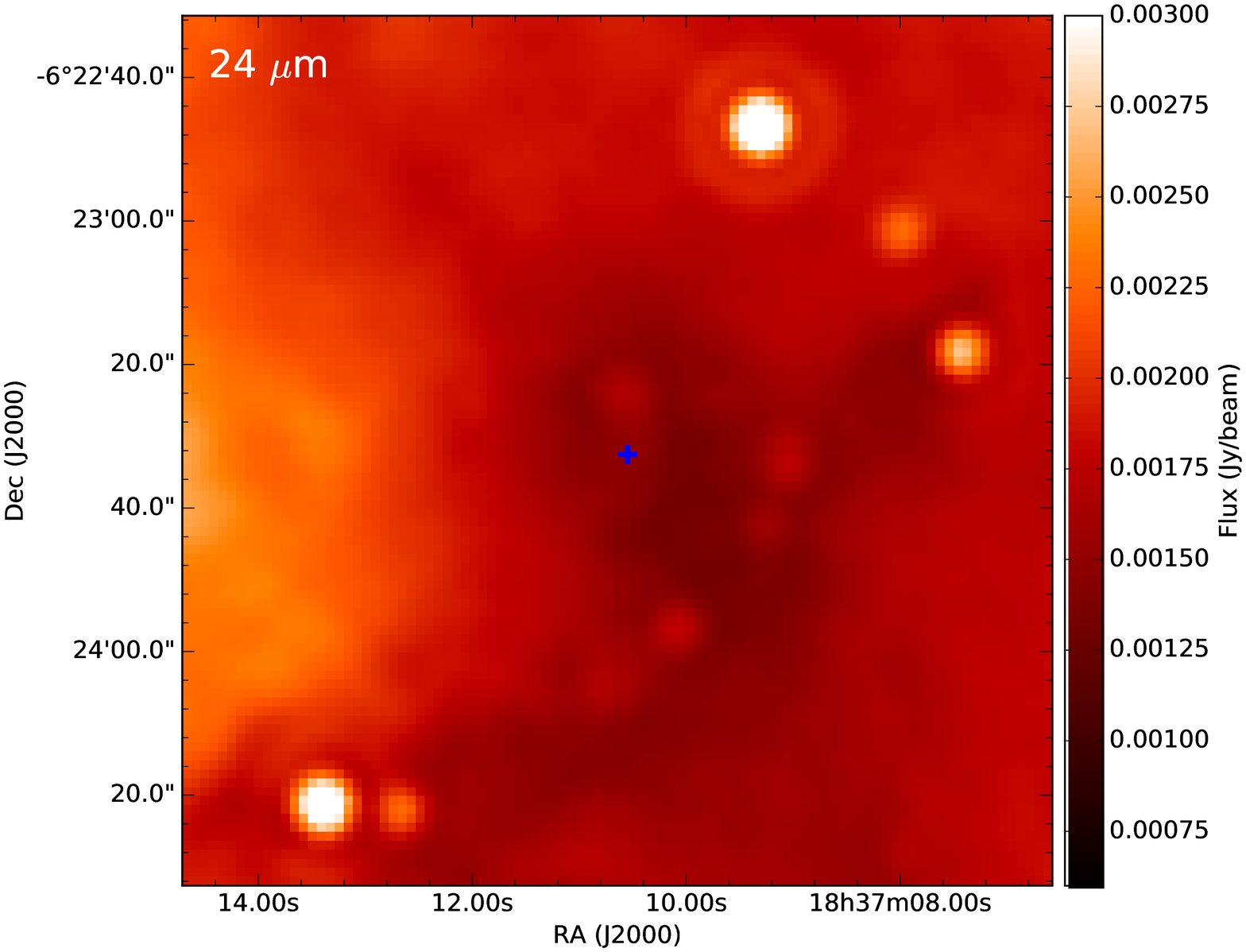}  \includegraphics[width=8cm]{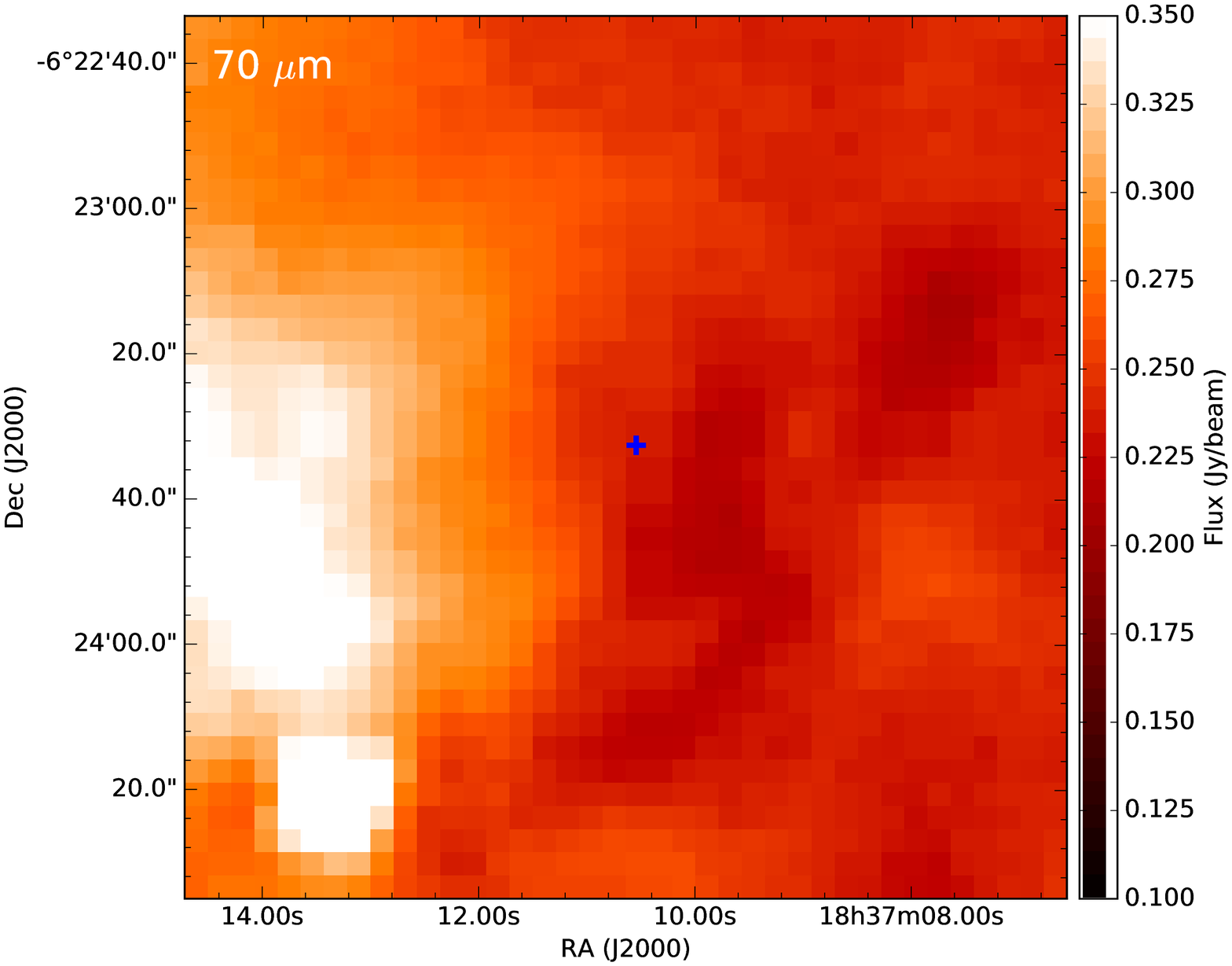} 
 \includegraphics[width=8cm]{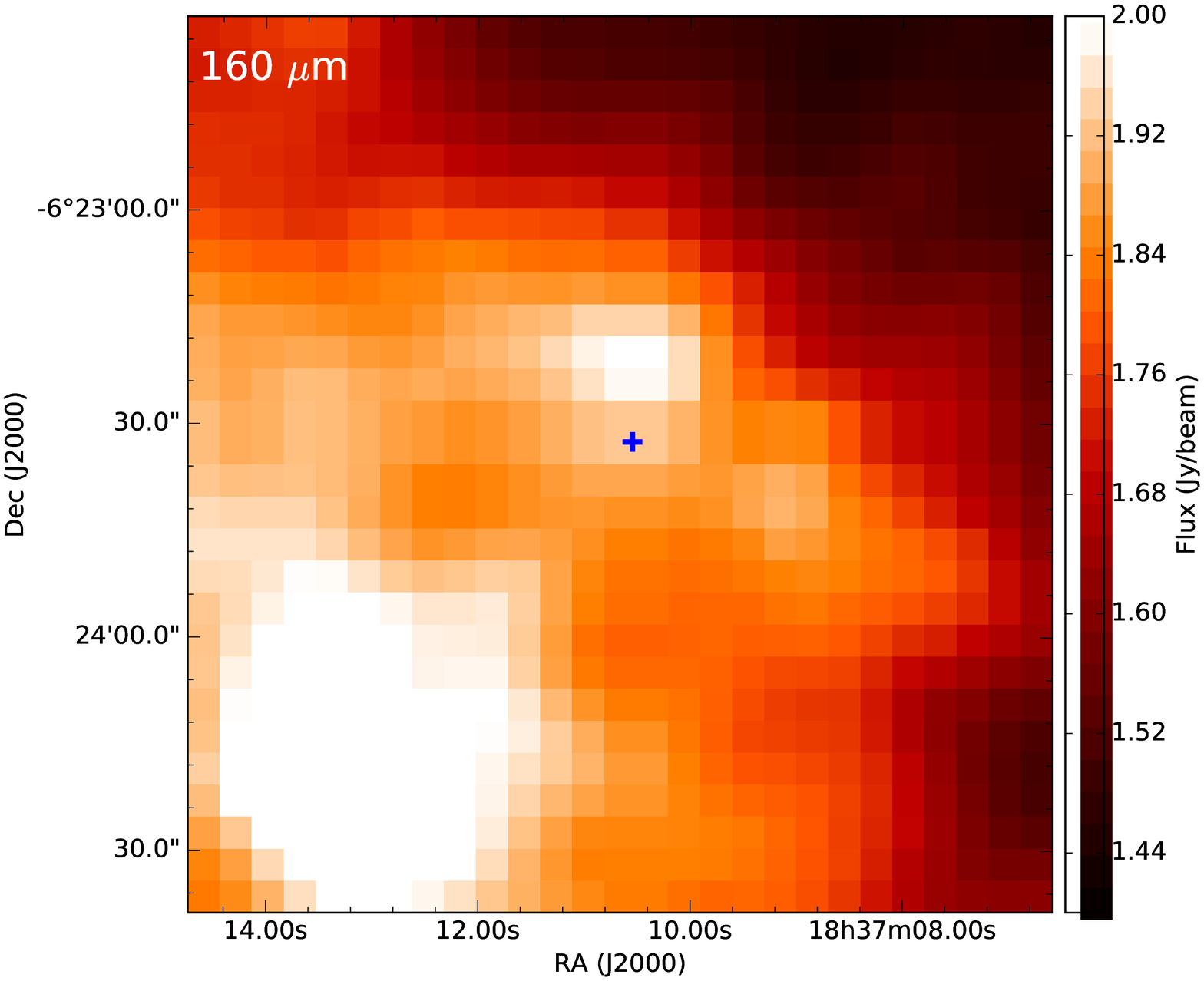}  \includegraphics[width=8cm]{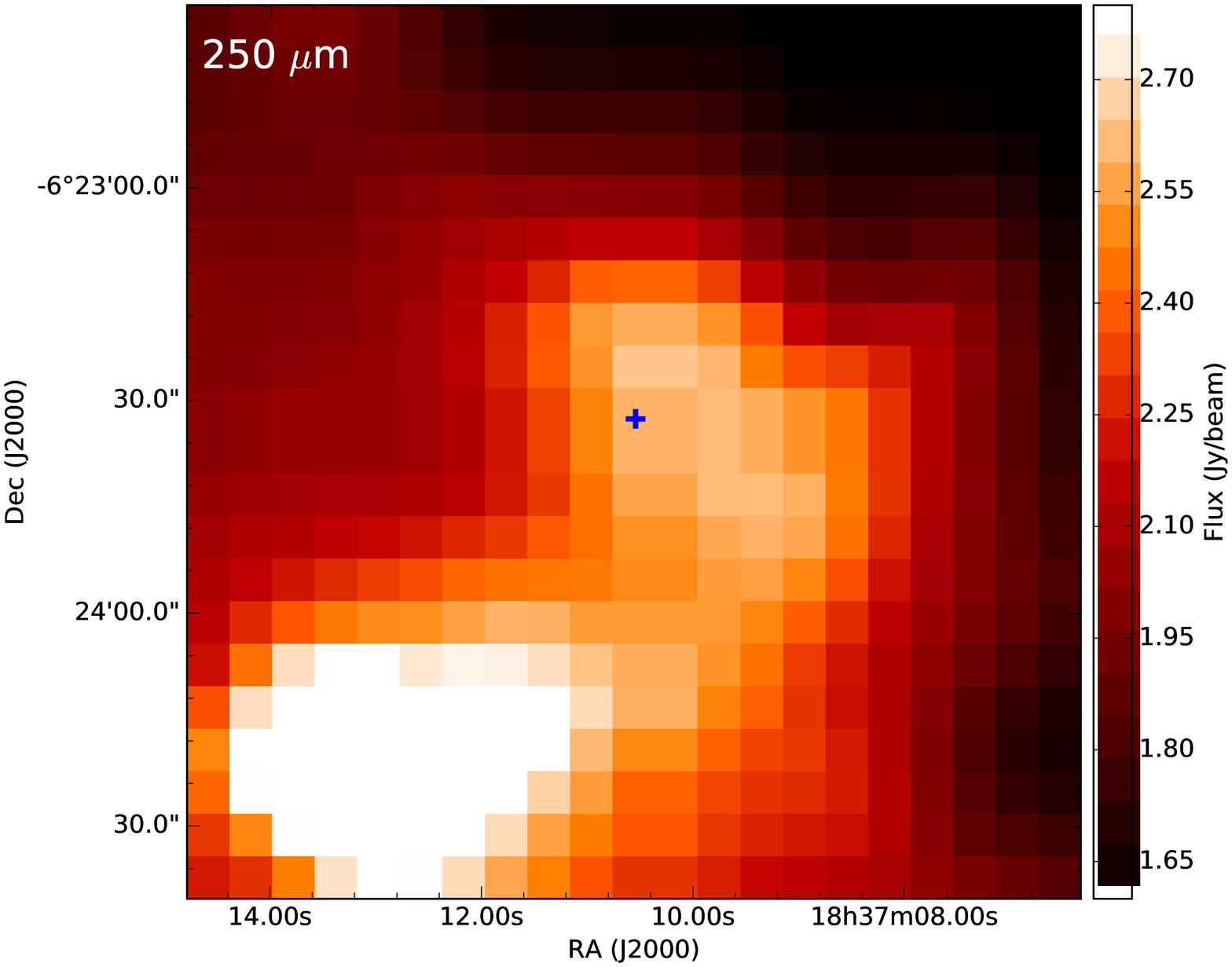} 
 \includegraphics[width=8cm]{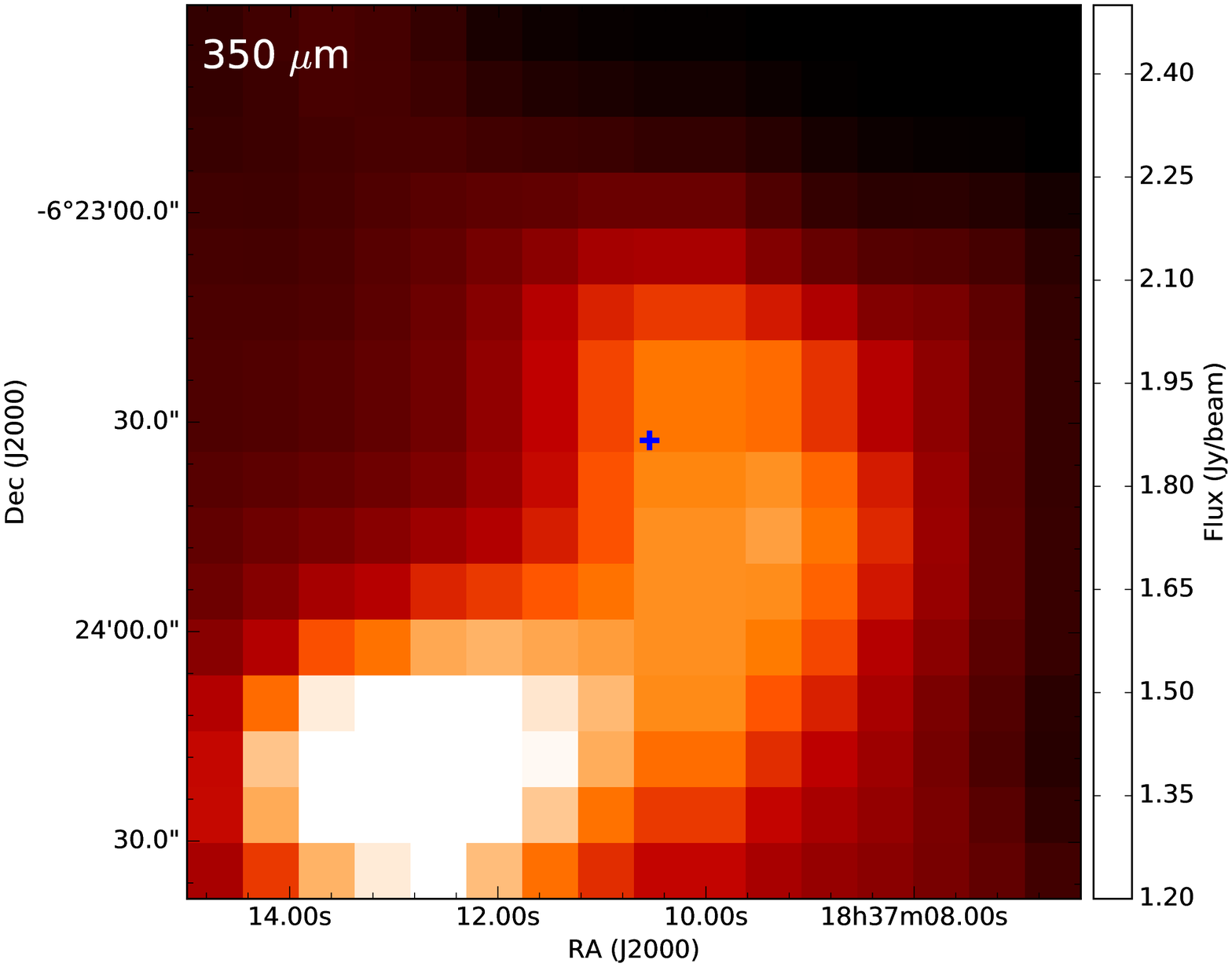}  \includegraphics[width=8cm]{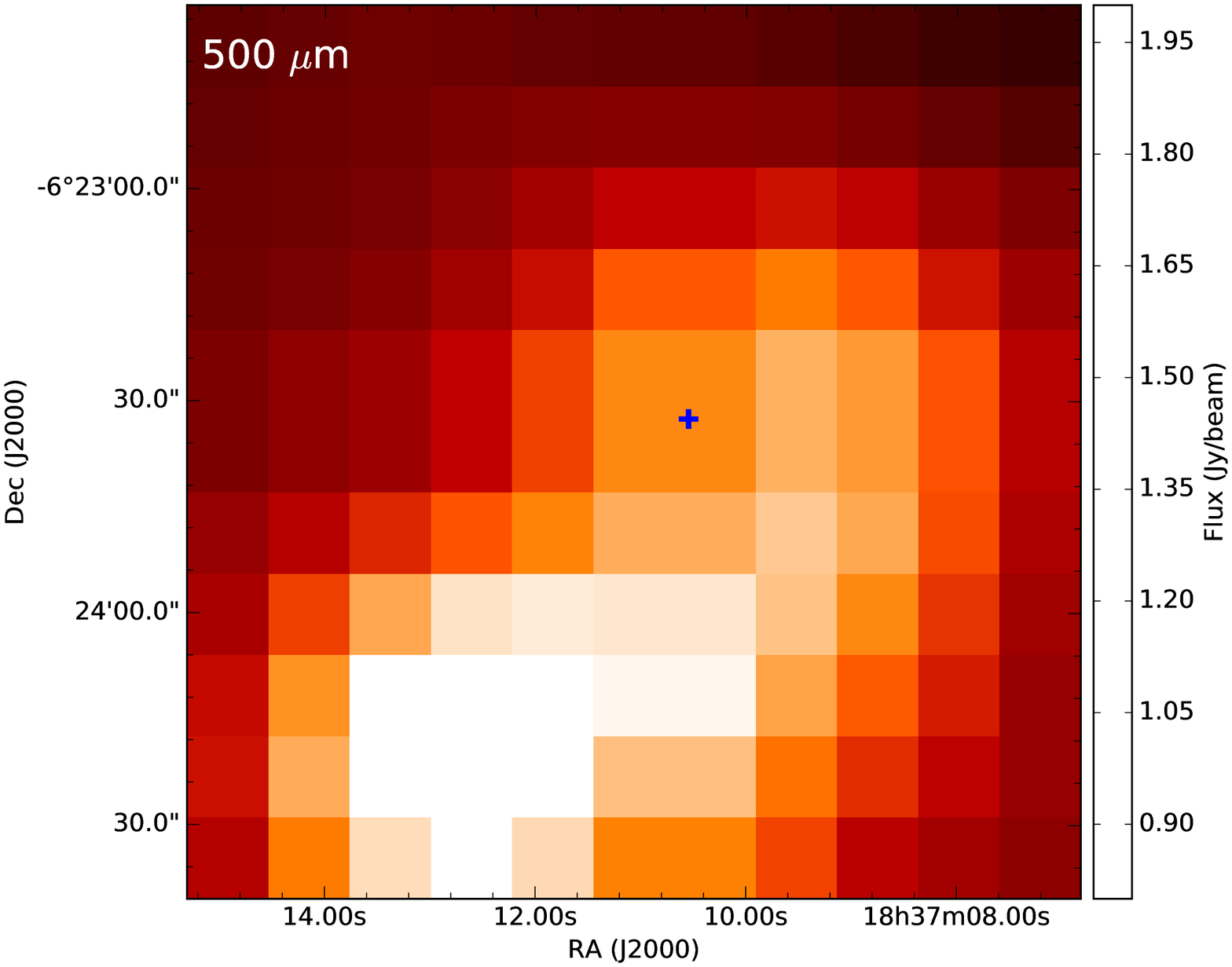} 
 \caption{25.609+0.228}
 \end{figure*}

\begin{figure*}
 \centering
 \includegraphics[width=8cm]{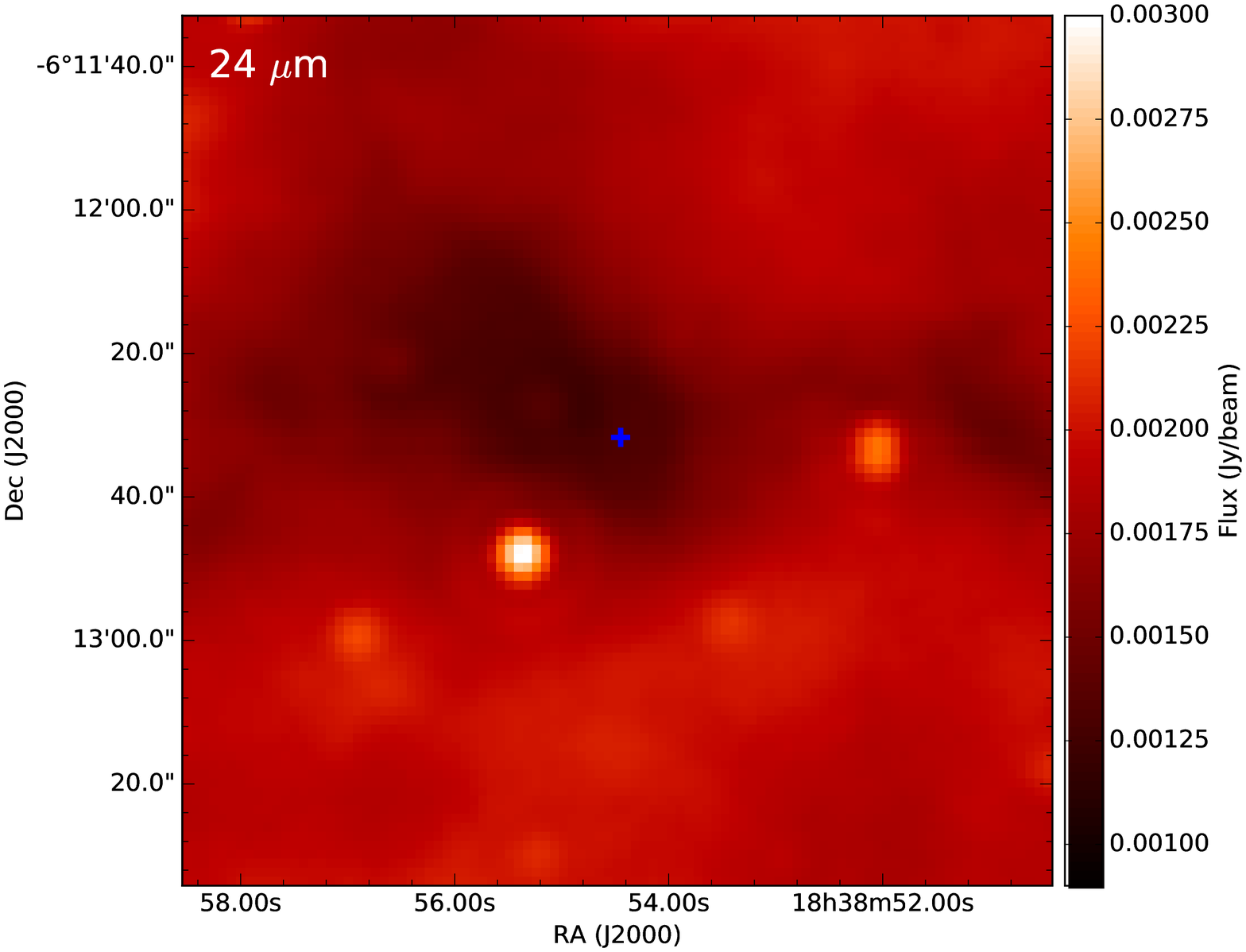}  \includegraphics[width=8cm]{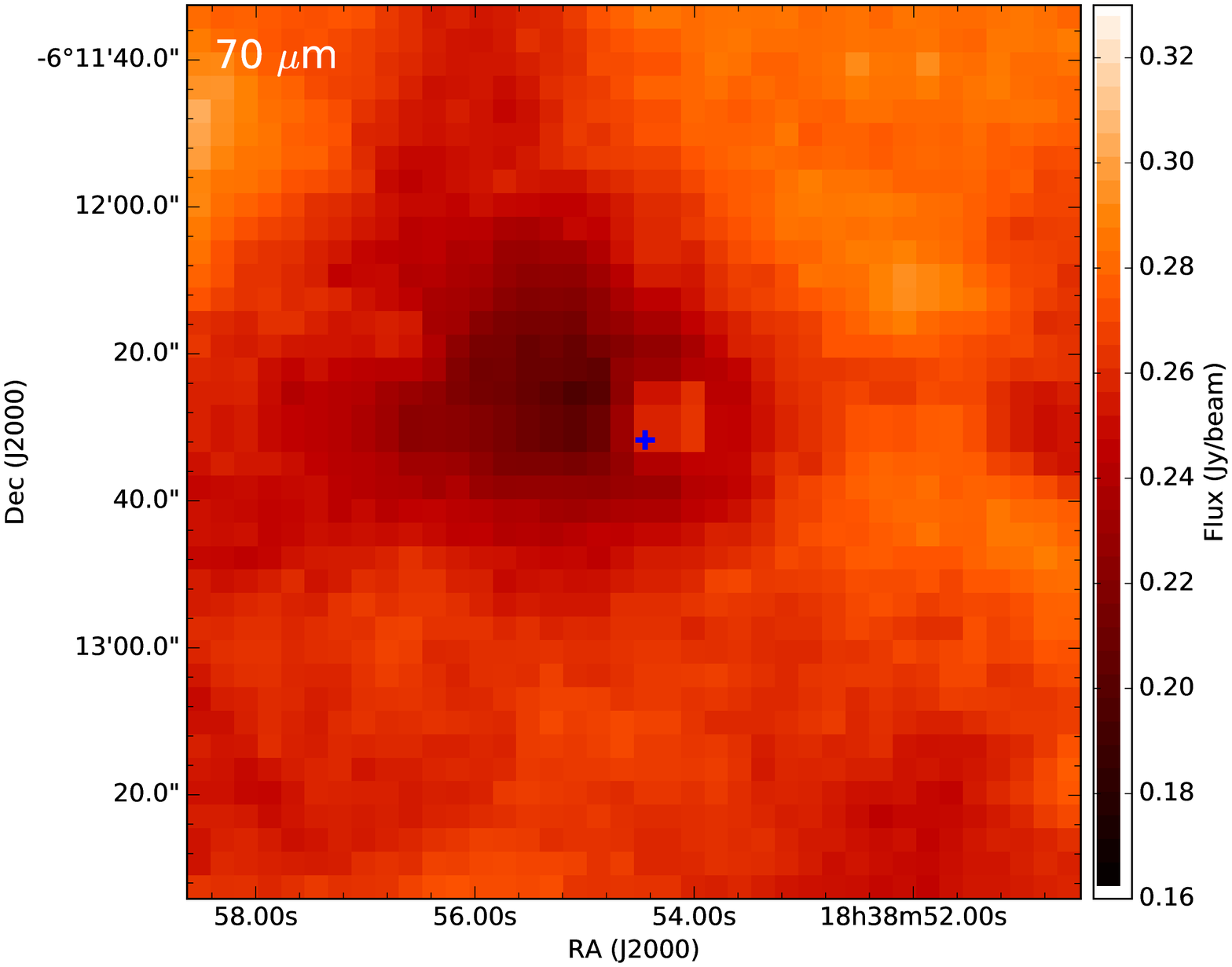} 
 \includegraphics[width=8cm]{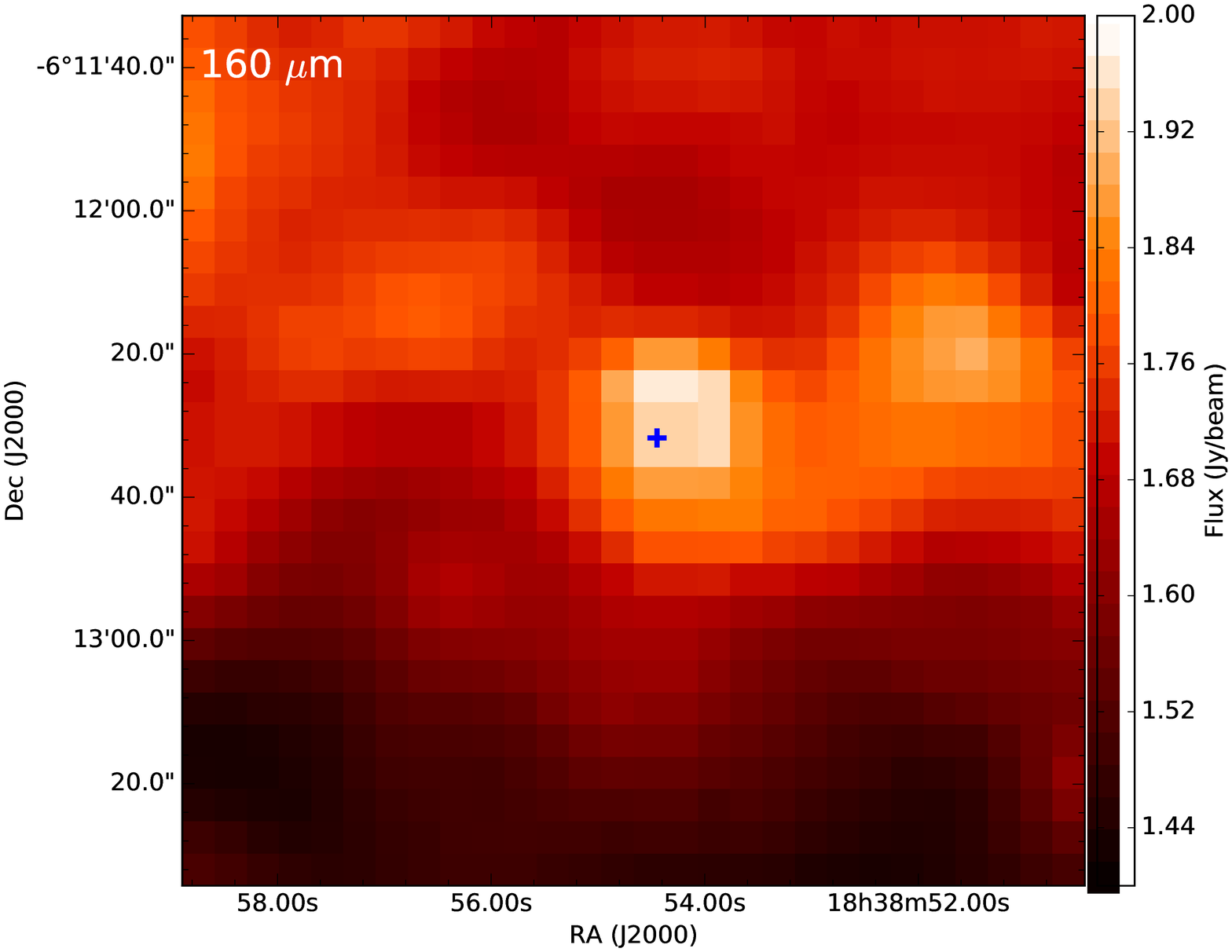}  \includegraphics[width=8cm]{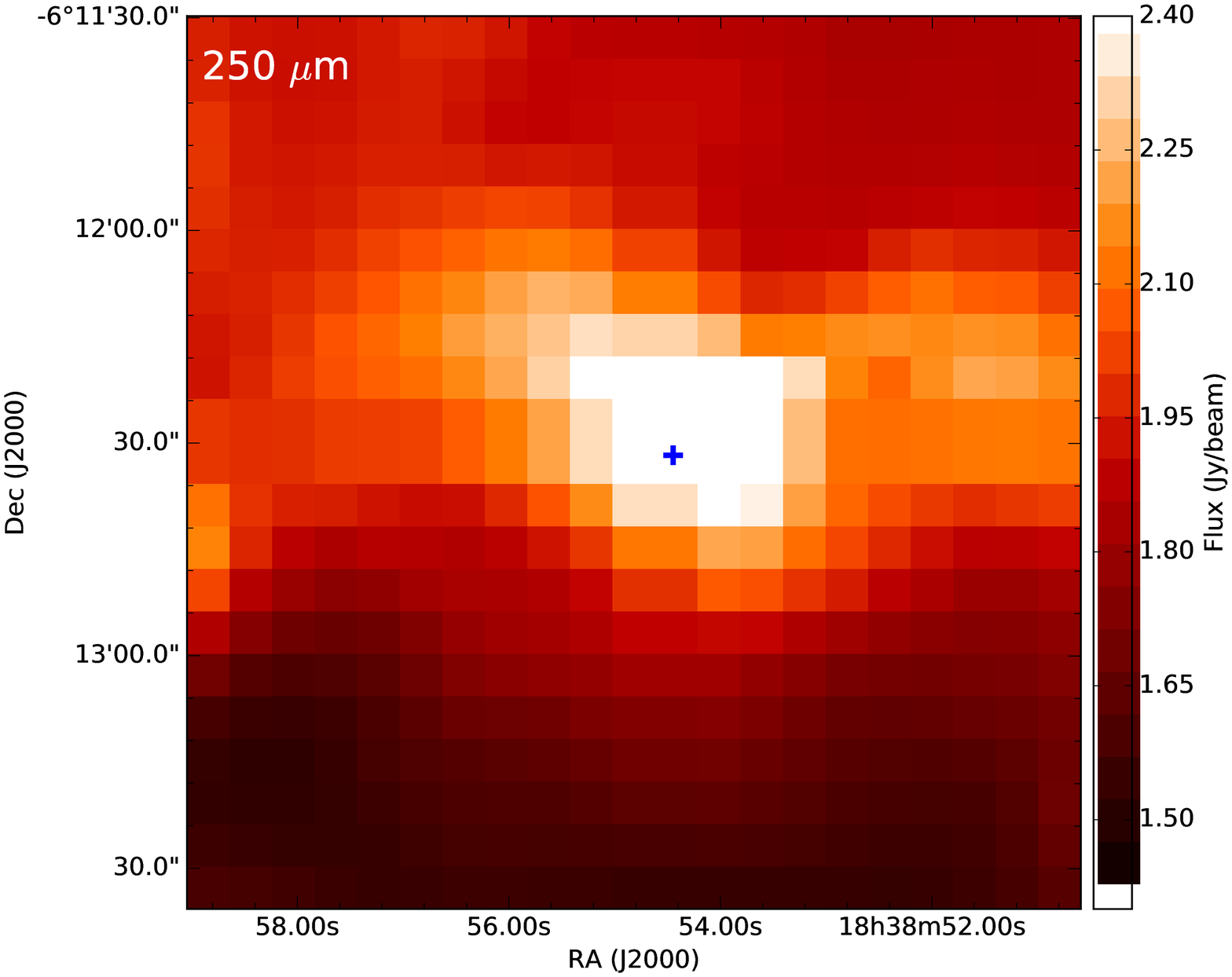} 
 \includegraphics[width=8cm]{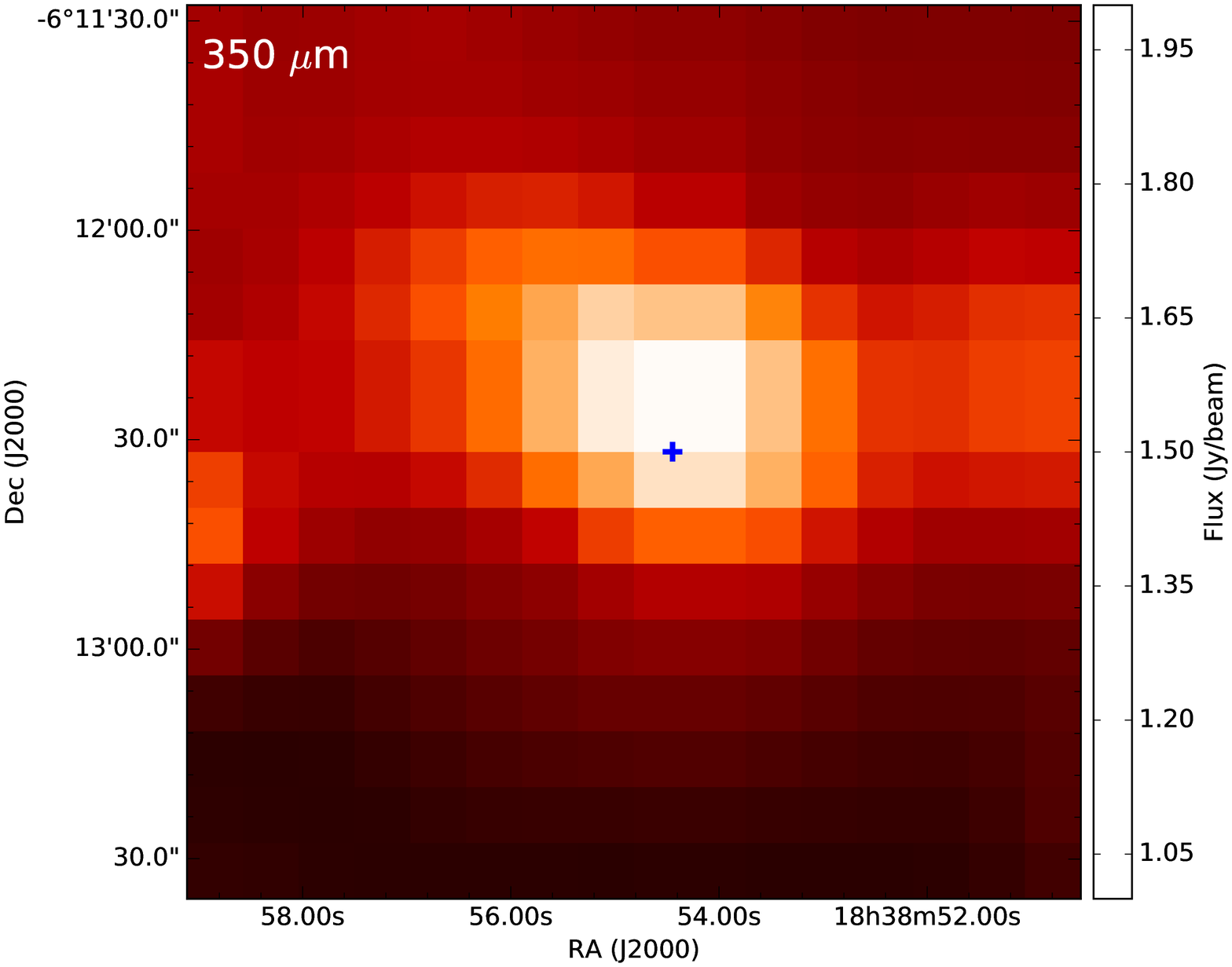}  \includegraphics[width=8cm]{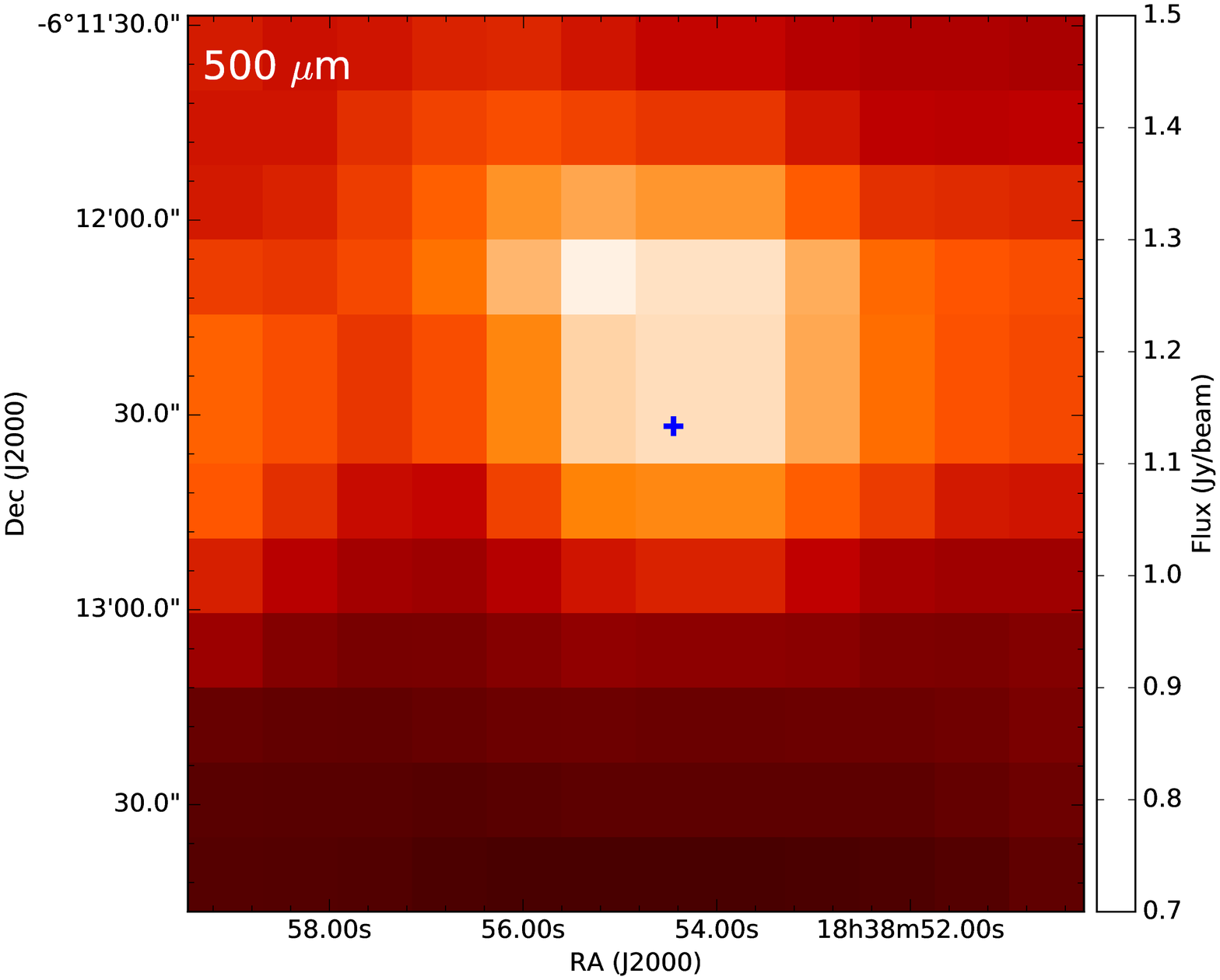} 
 \caption{25.982-0.056}
 \end{figure*}

\begin{figure*}
 \centering
 \includegraphics[width=8cm]{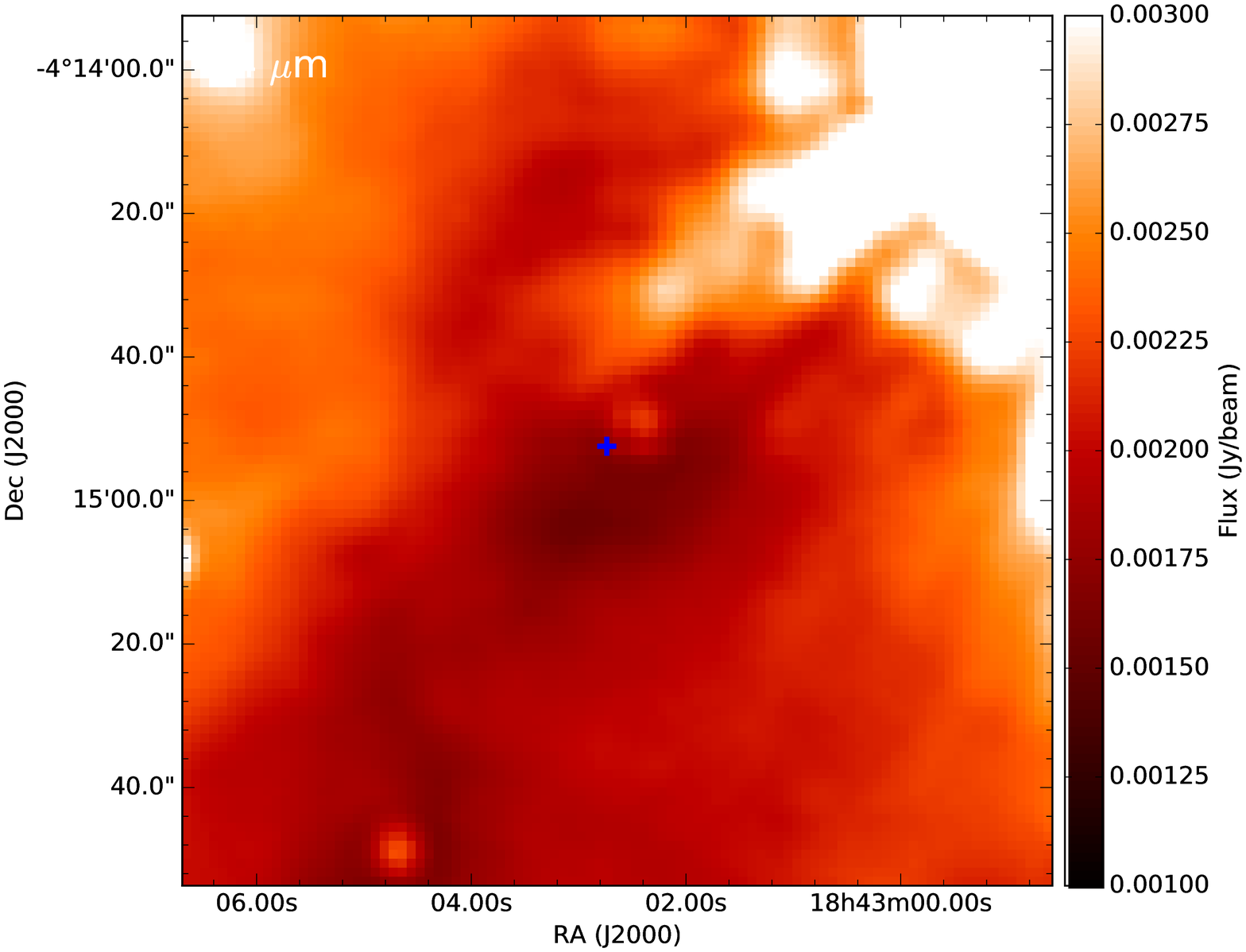}  \includegraphics[width=8cm]{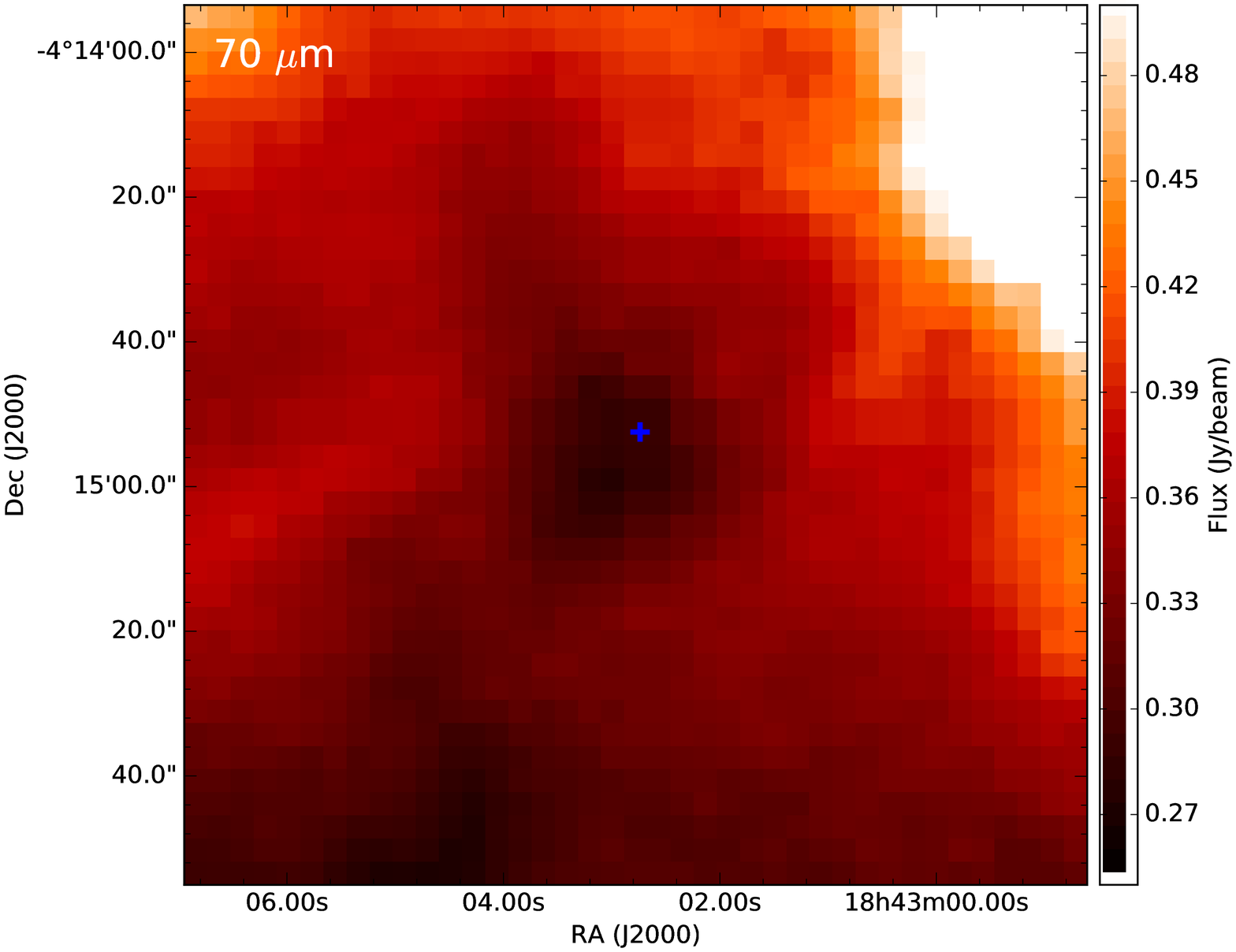} 
 \includegraphics[width=8cm]{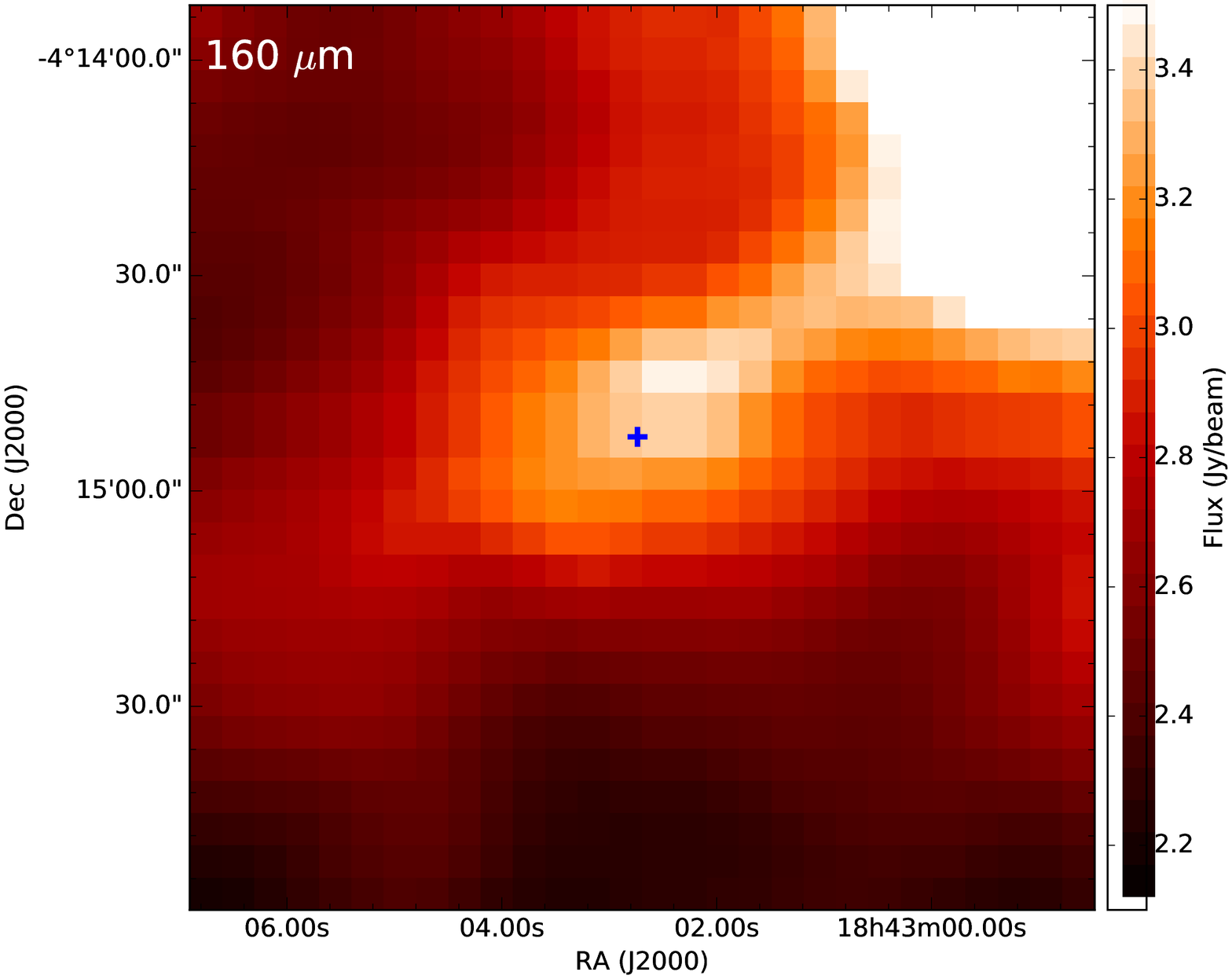}  \includegraphics[width=8cm]{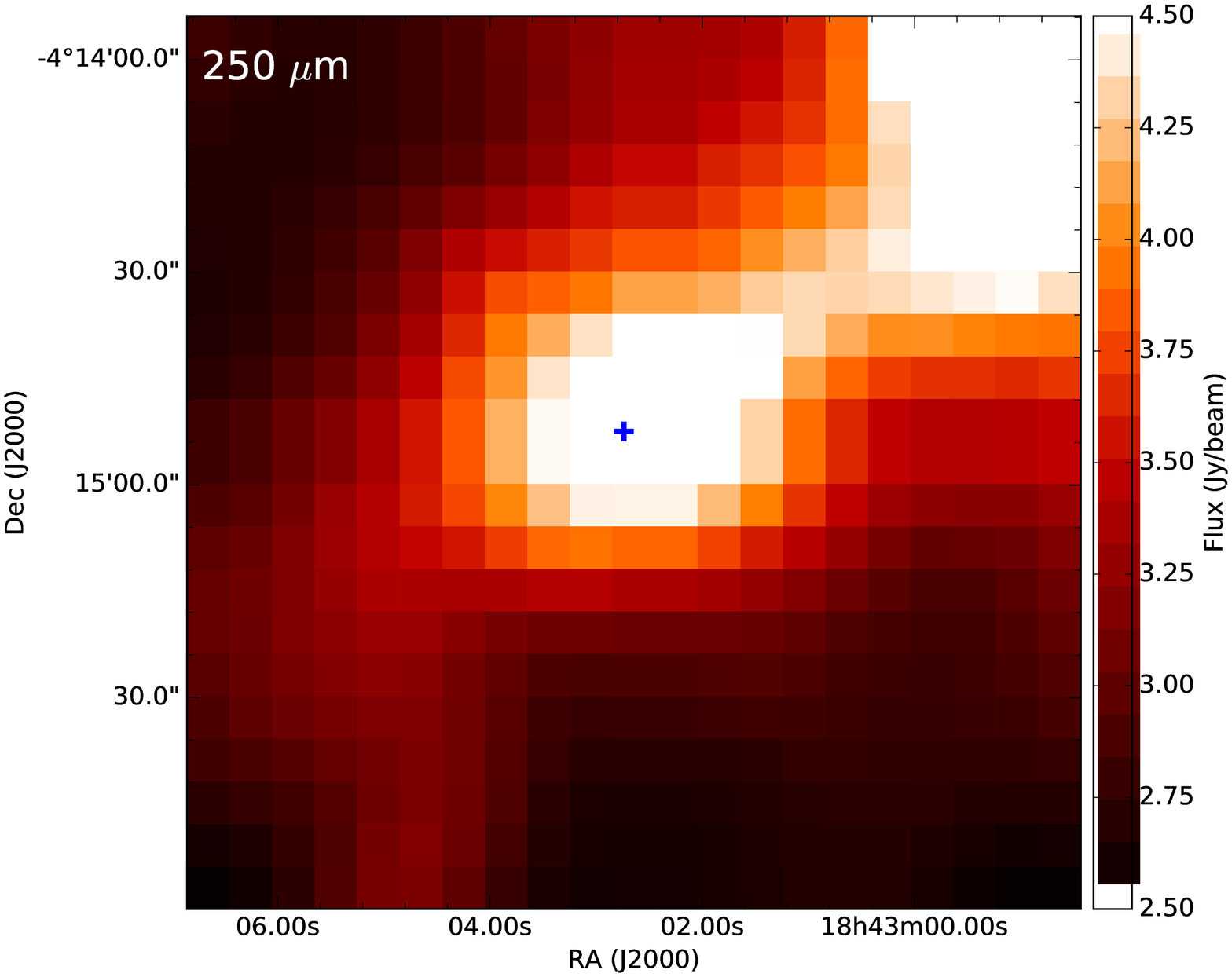} 
 \includegraphics[width=8cm]{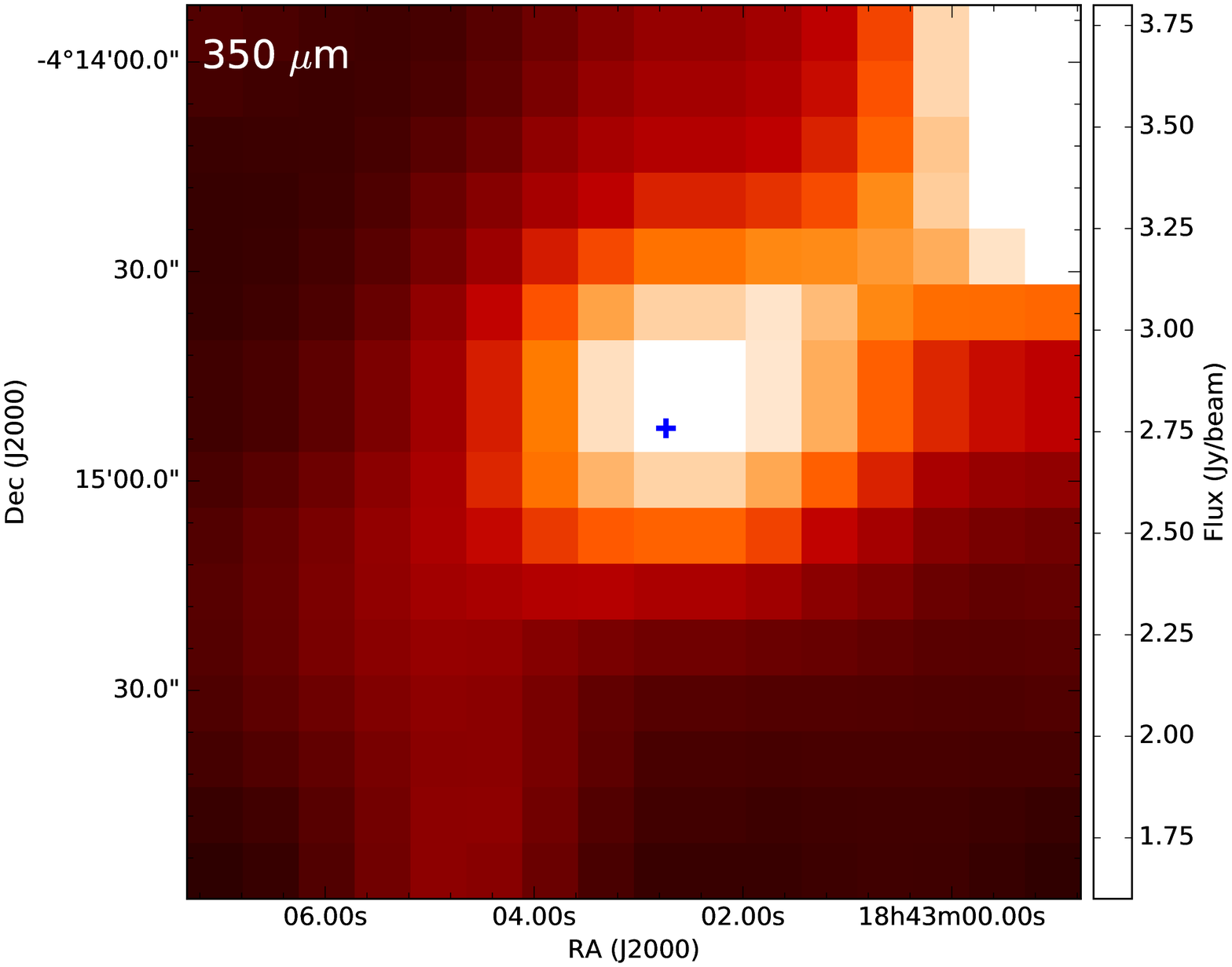}  \includegraphics[width=8cm]{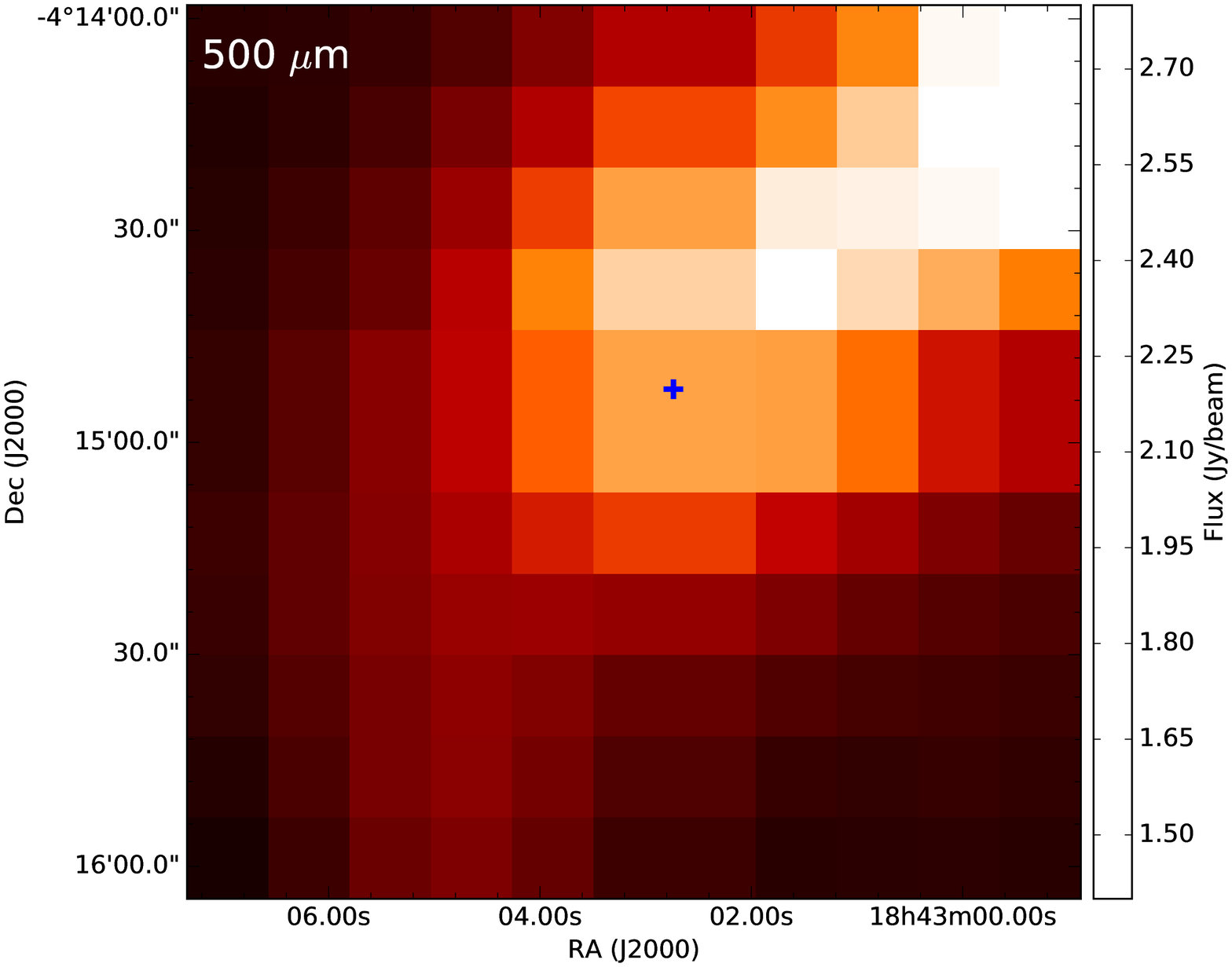} 
 \caption{28.178-0.091}
 \end{figure*}

\begin{figure*}
 \centering
 \includegraphics[width=8cm]{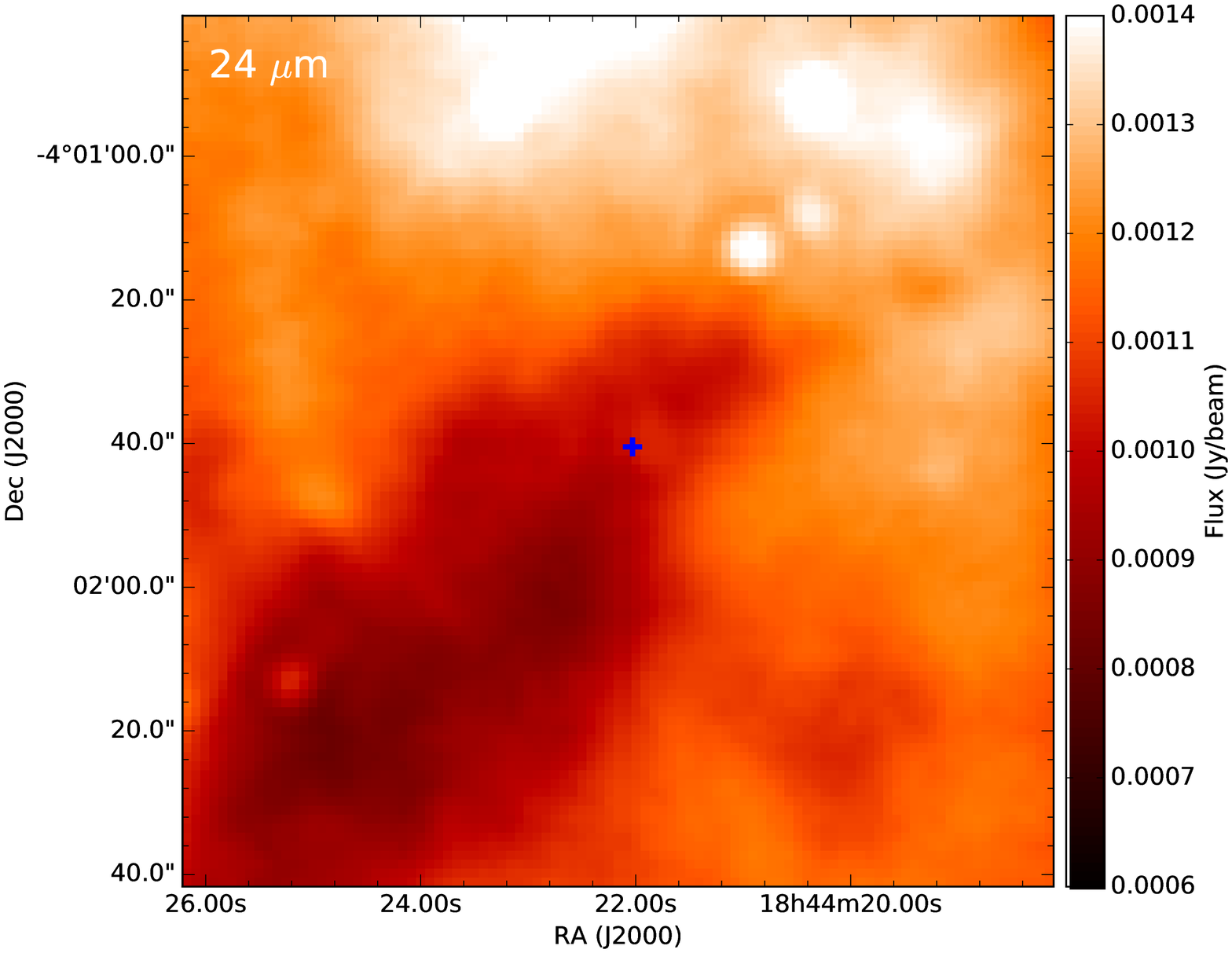}  \includegraphics[width=8cm]{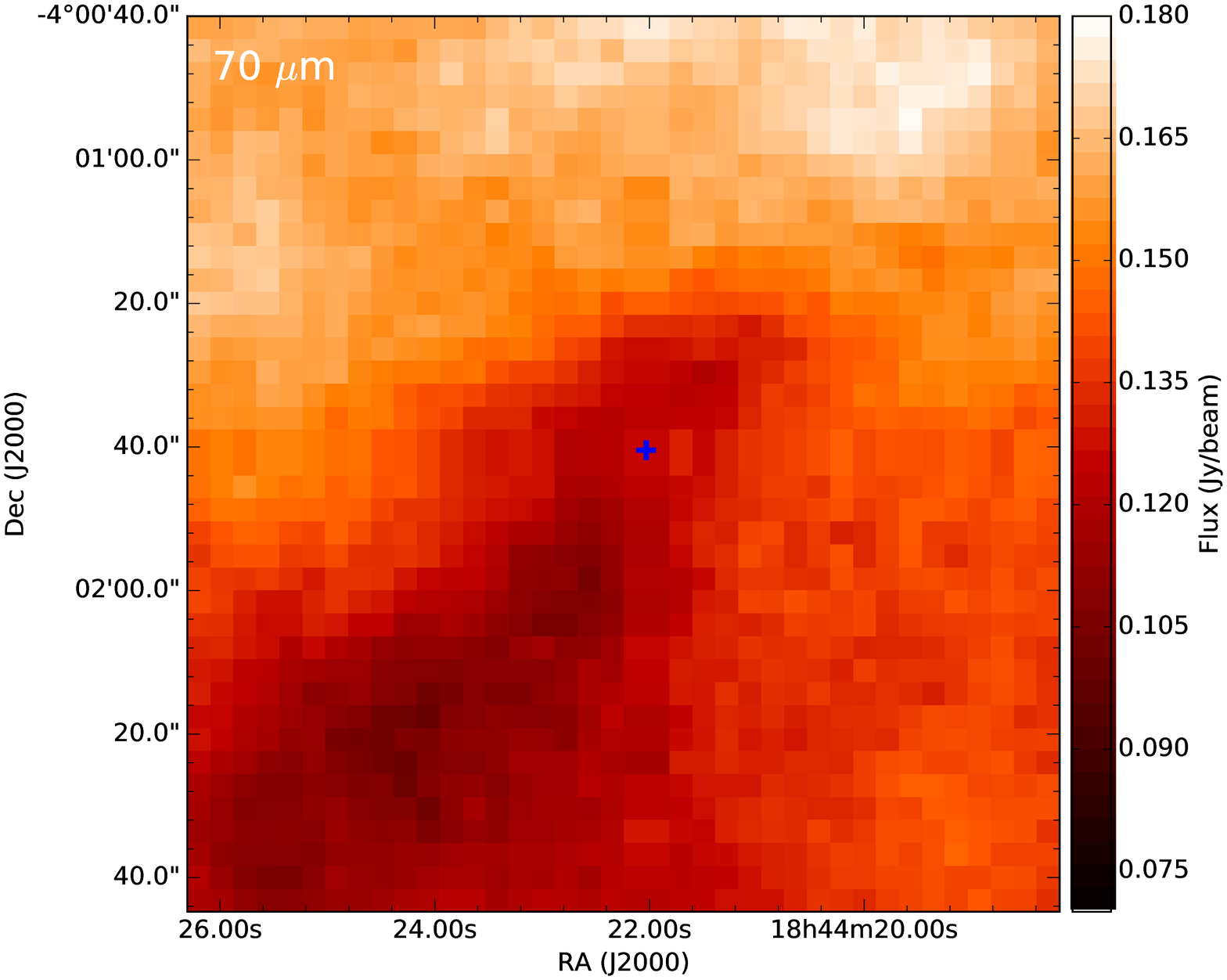} 
 \includegraphics[width=8cm]{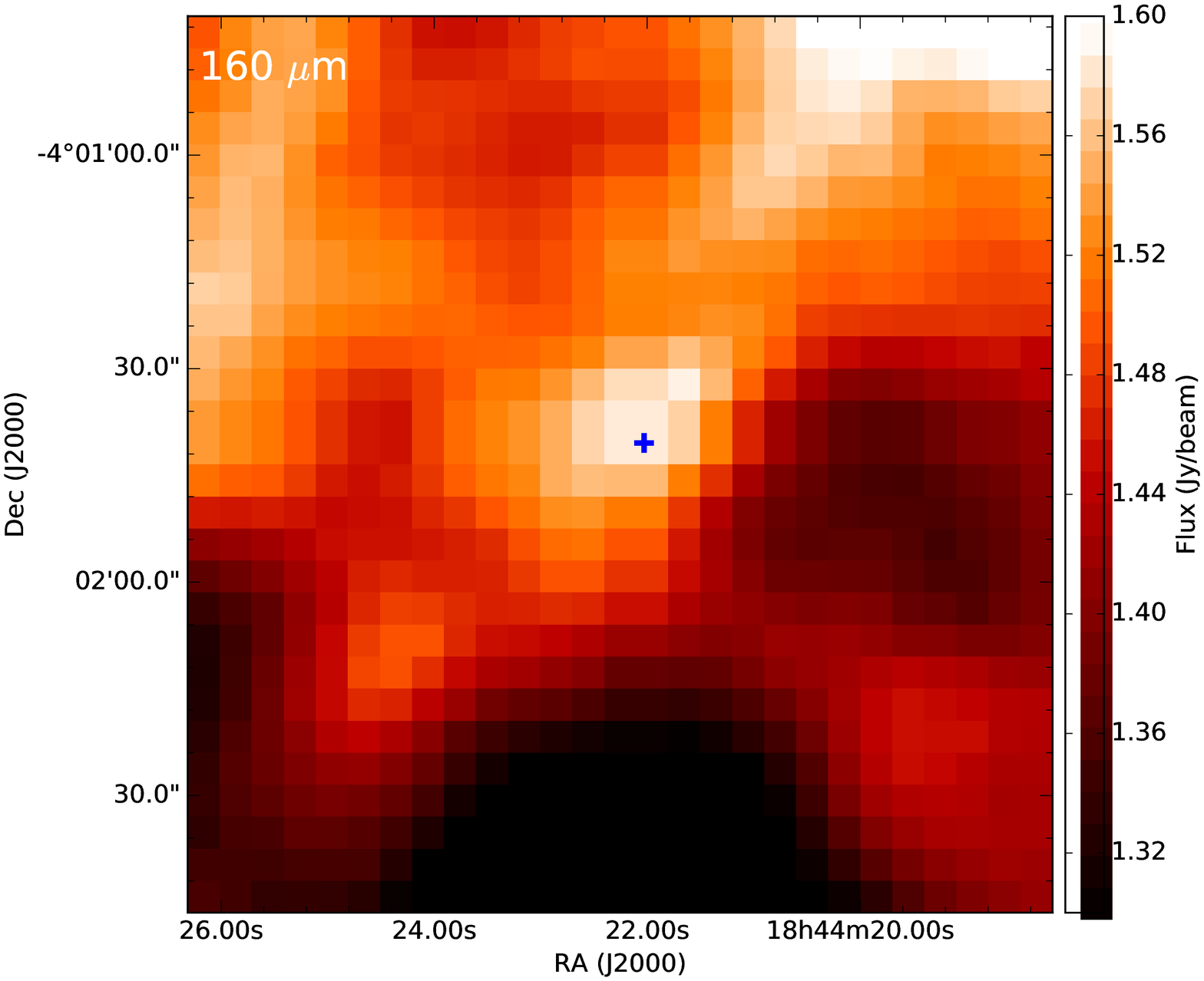}  \includegraphics[width=8cm]{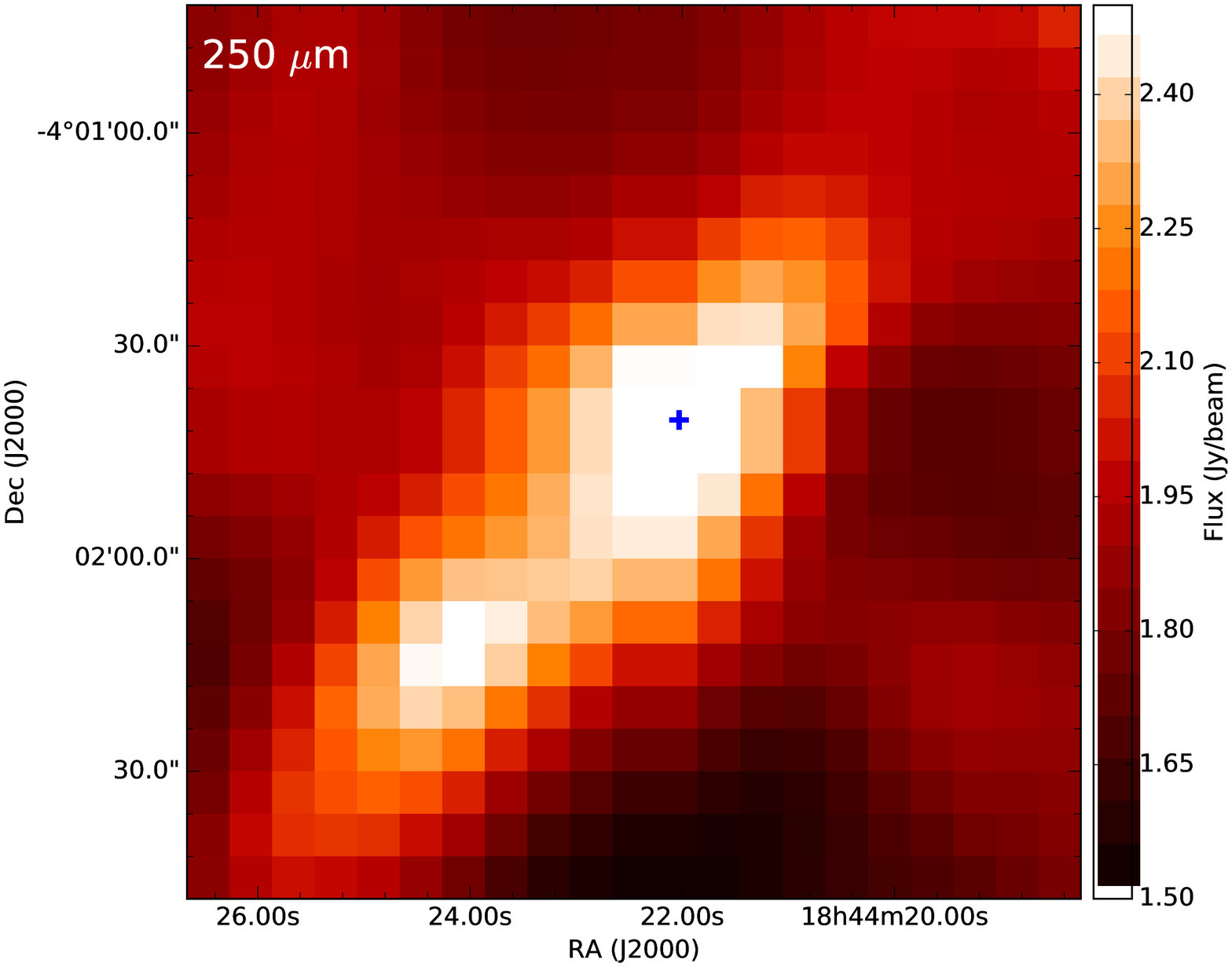} 
 \includegraphics[width=8cm]{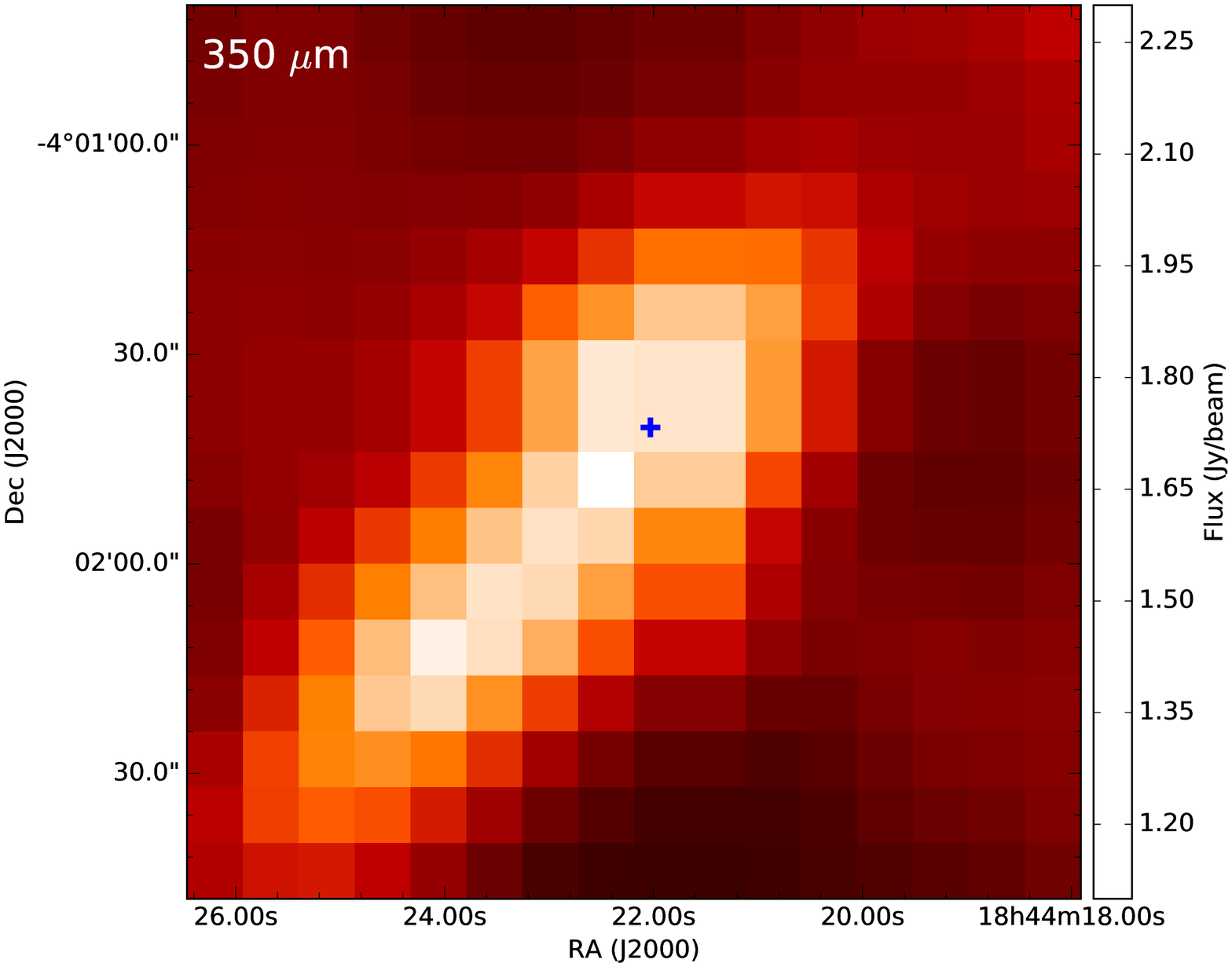}  \includegraphics[width=8cm]{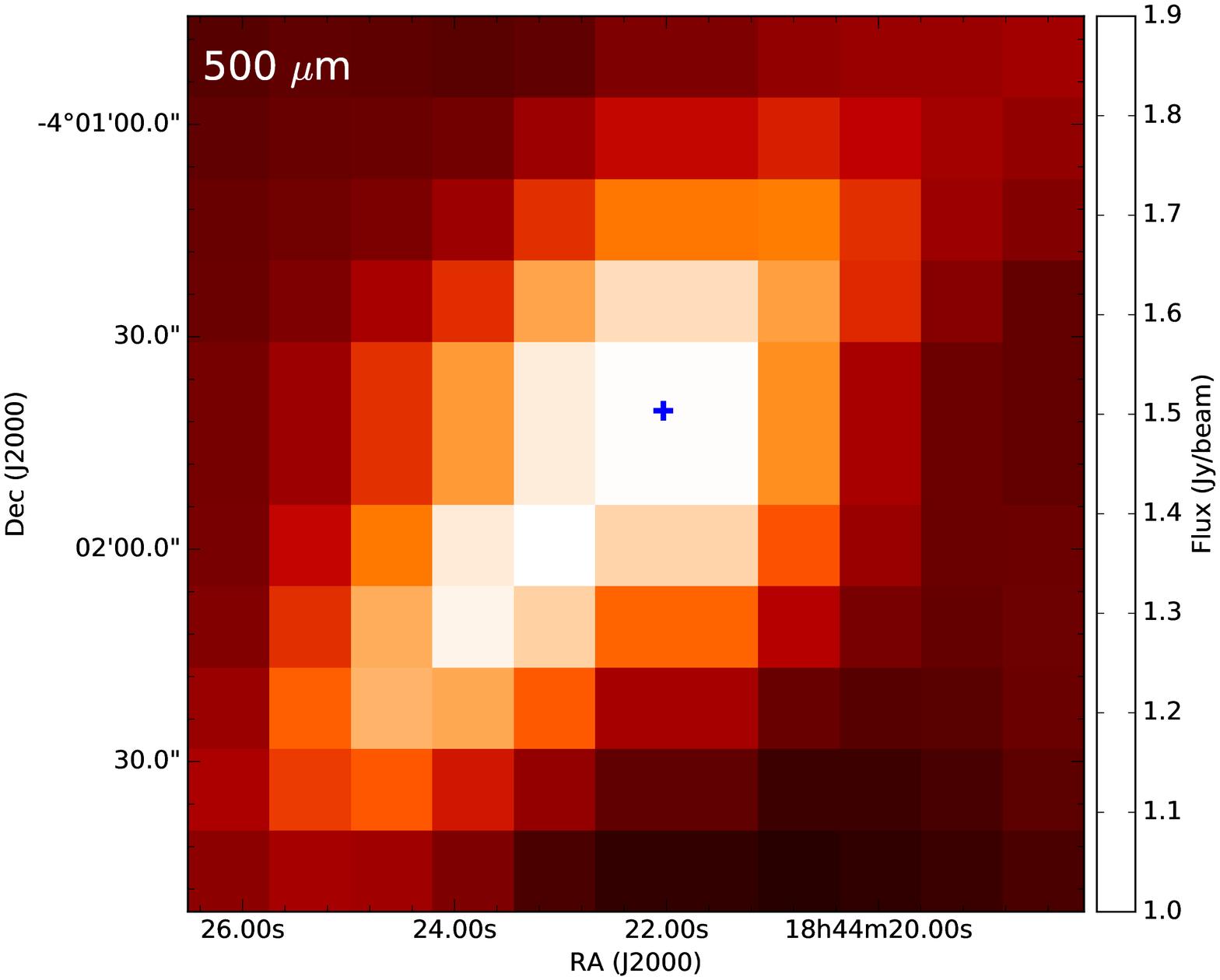} 
 \caption{28.537-0.277}
 \end{figure*}

\begin{figure*}
 \centering
 \includegraphics[width=8cm]{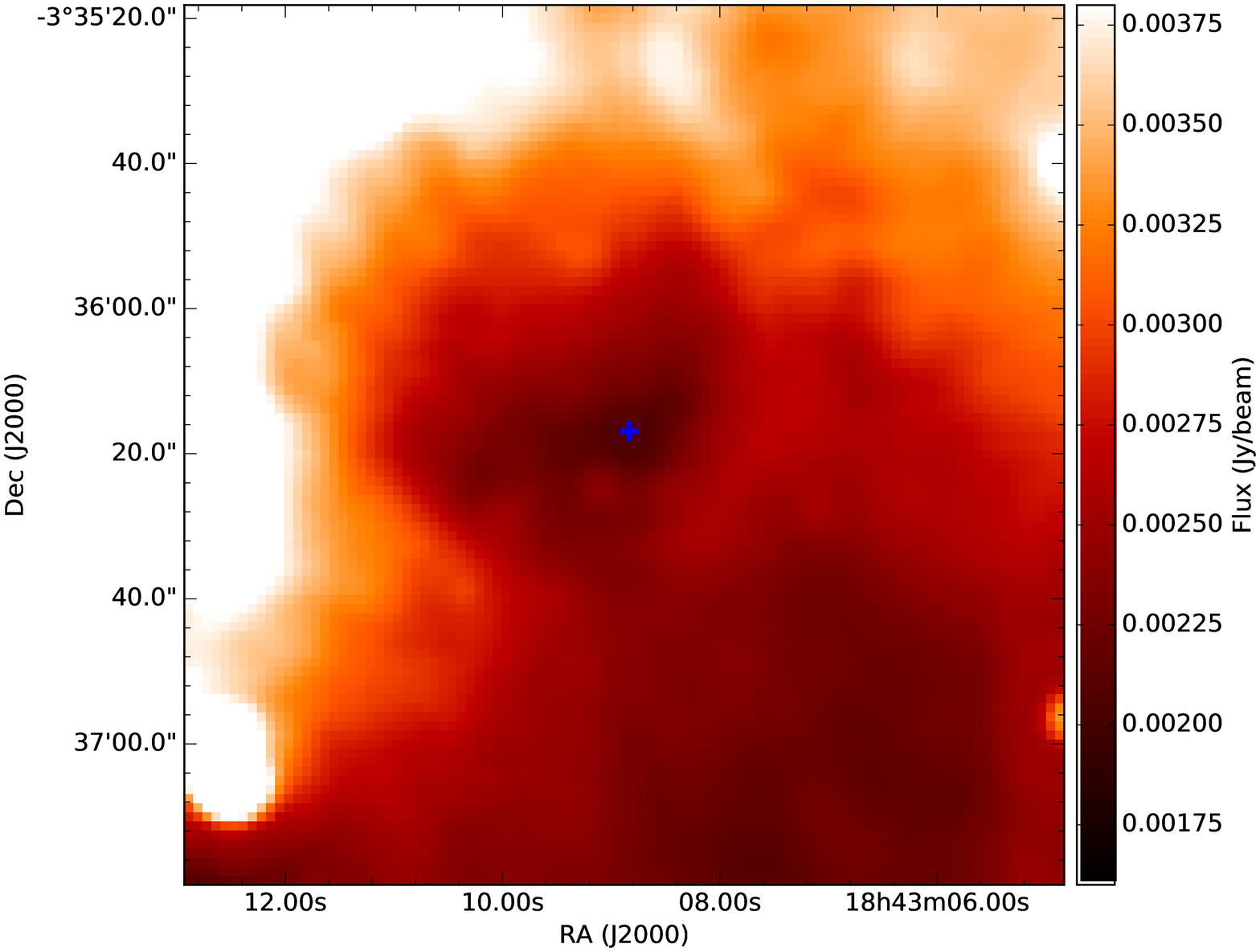}  \includegraphics[width=8cm]{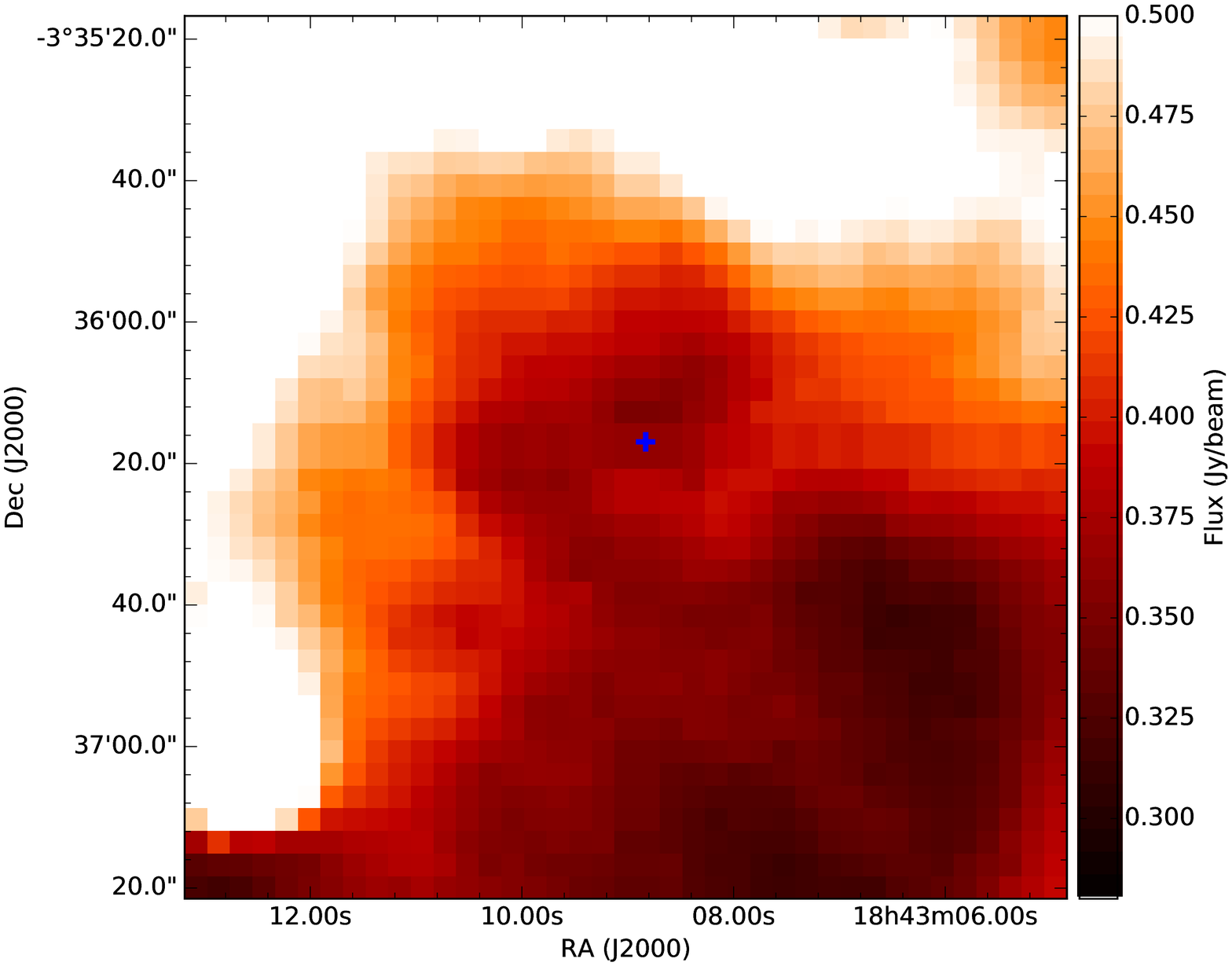} 
 \includegraphics[width=8cm]{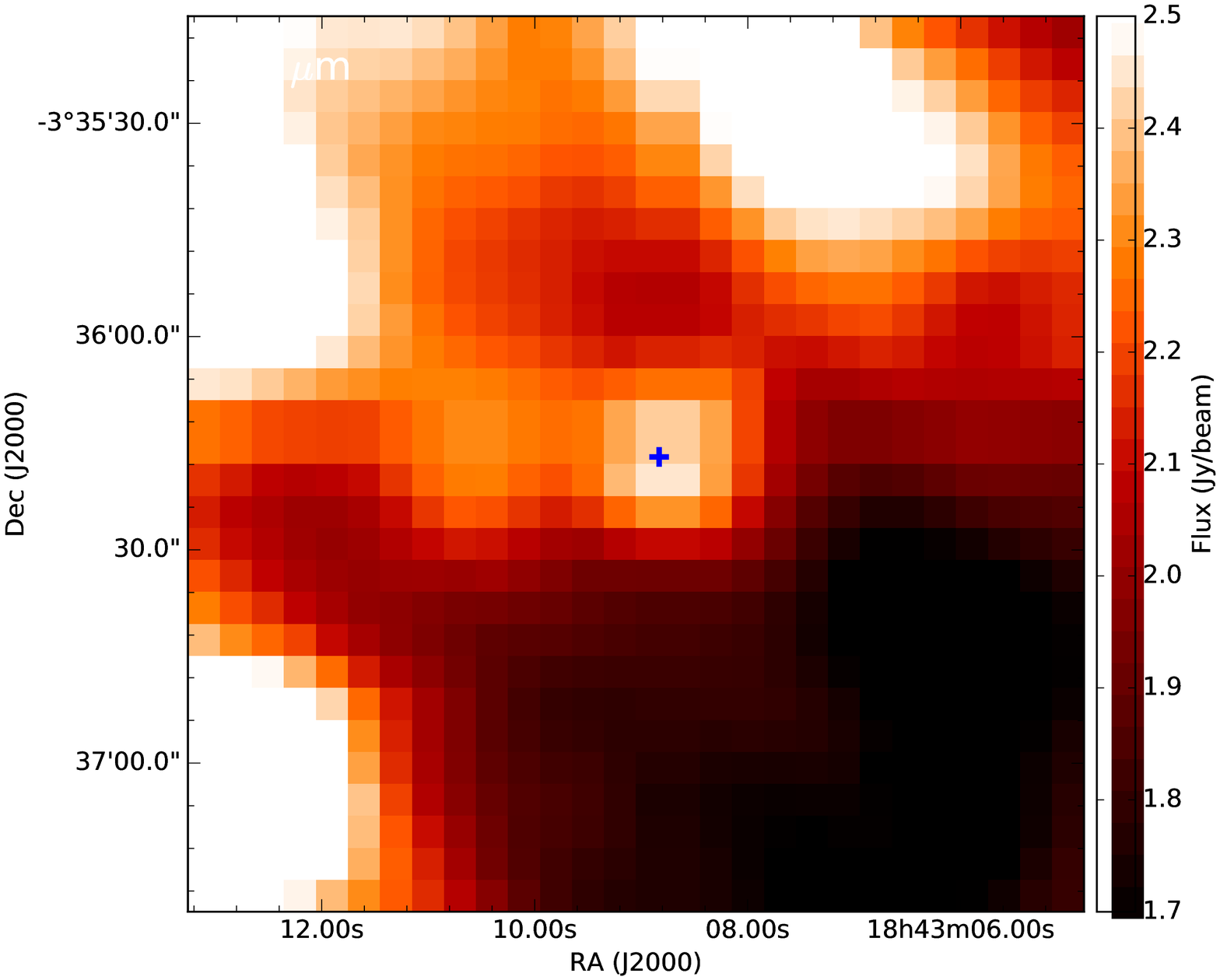}  \includegraphics[width=8cm]{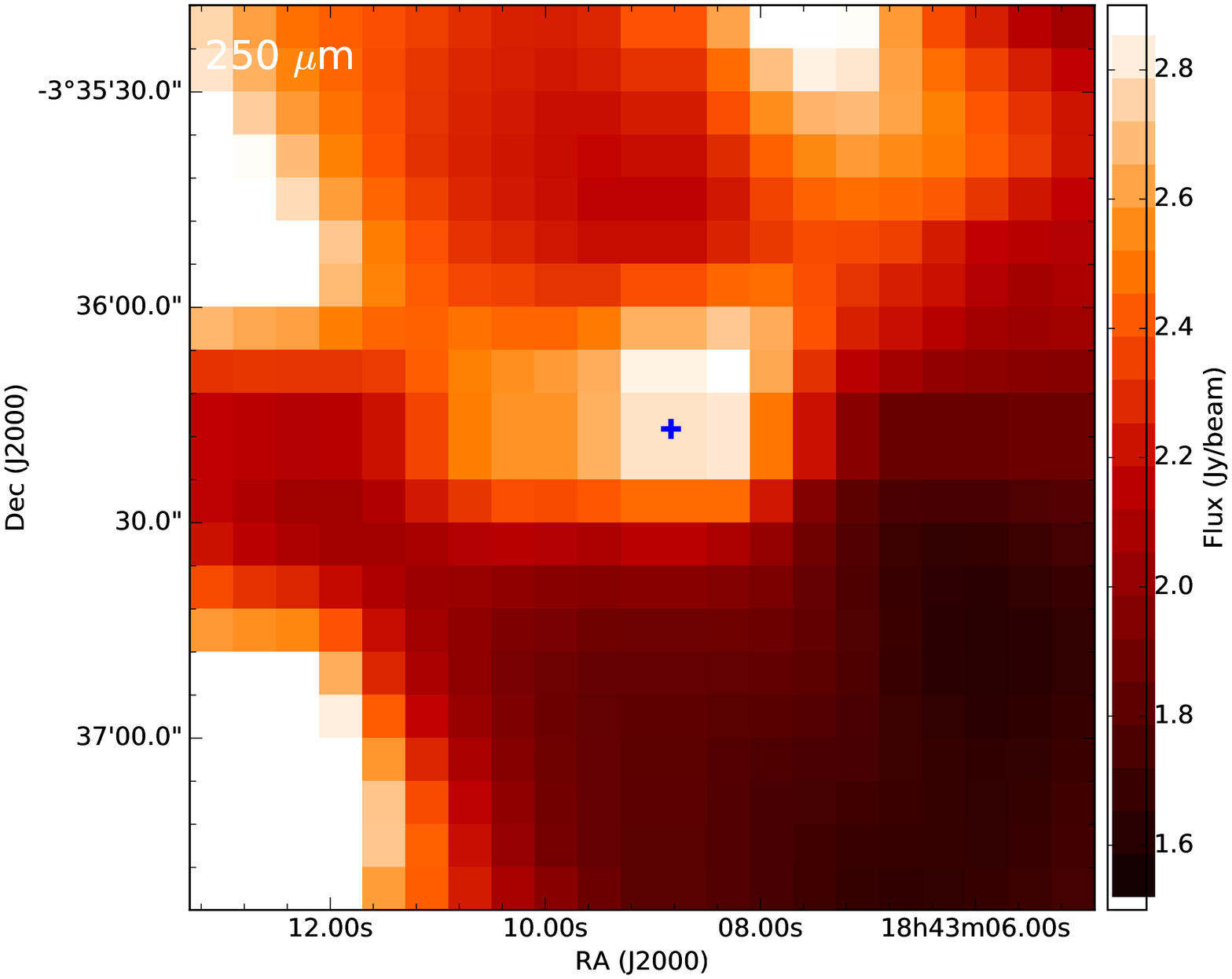} 
 \includegraphics[width=8cm]{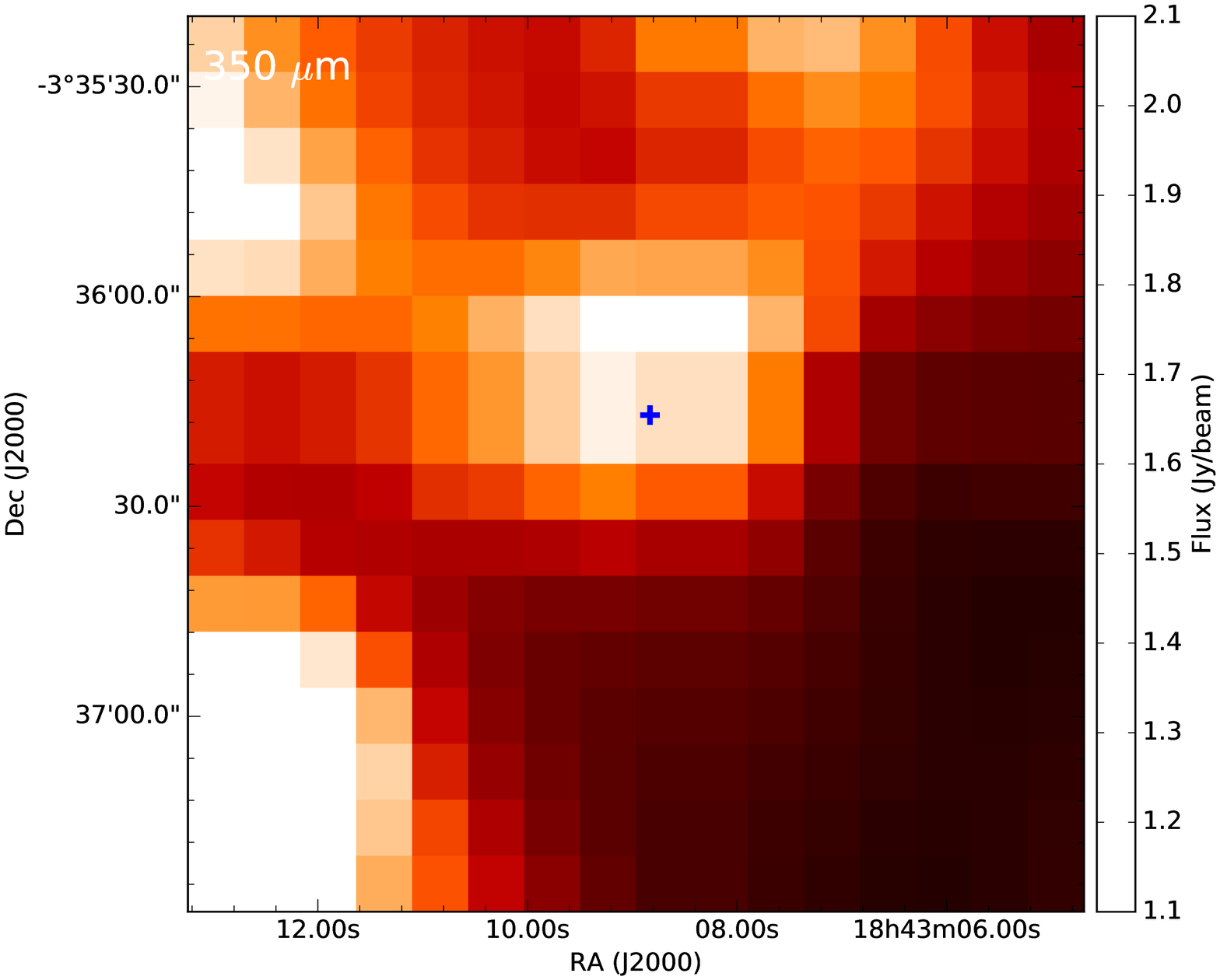}  \includegraphics[width=8cm]{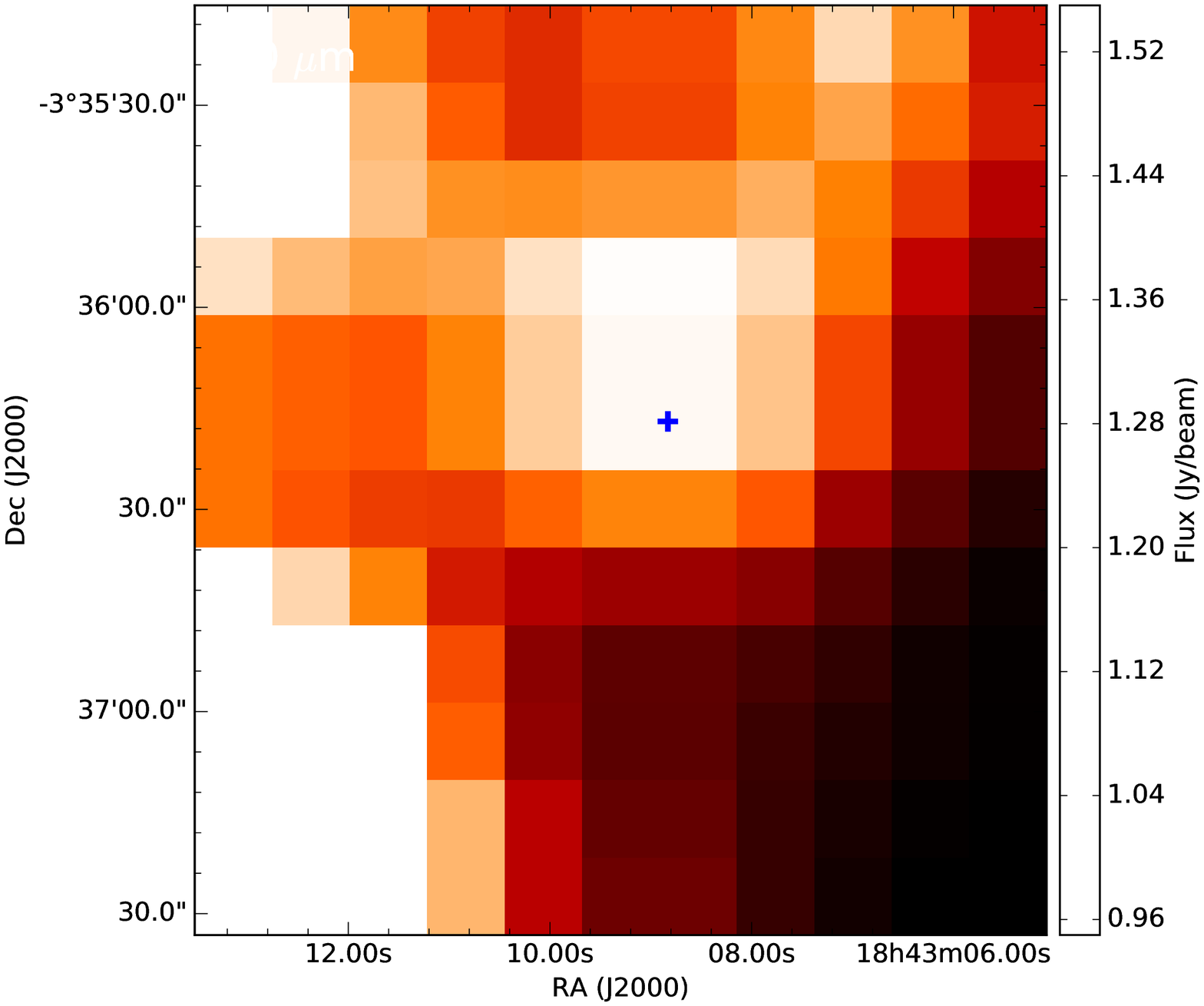} 
 \caption{28.792+0.141}
 \end{figure*}

\begin{figure*}
 \centering
 \includegraphics[width=8cm]{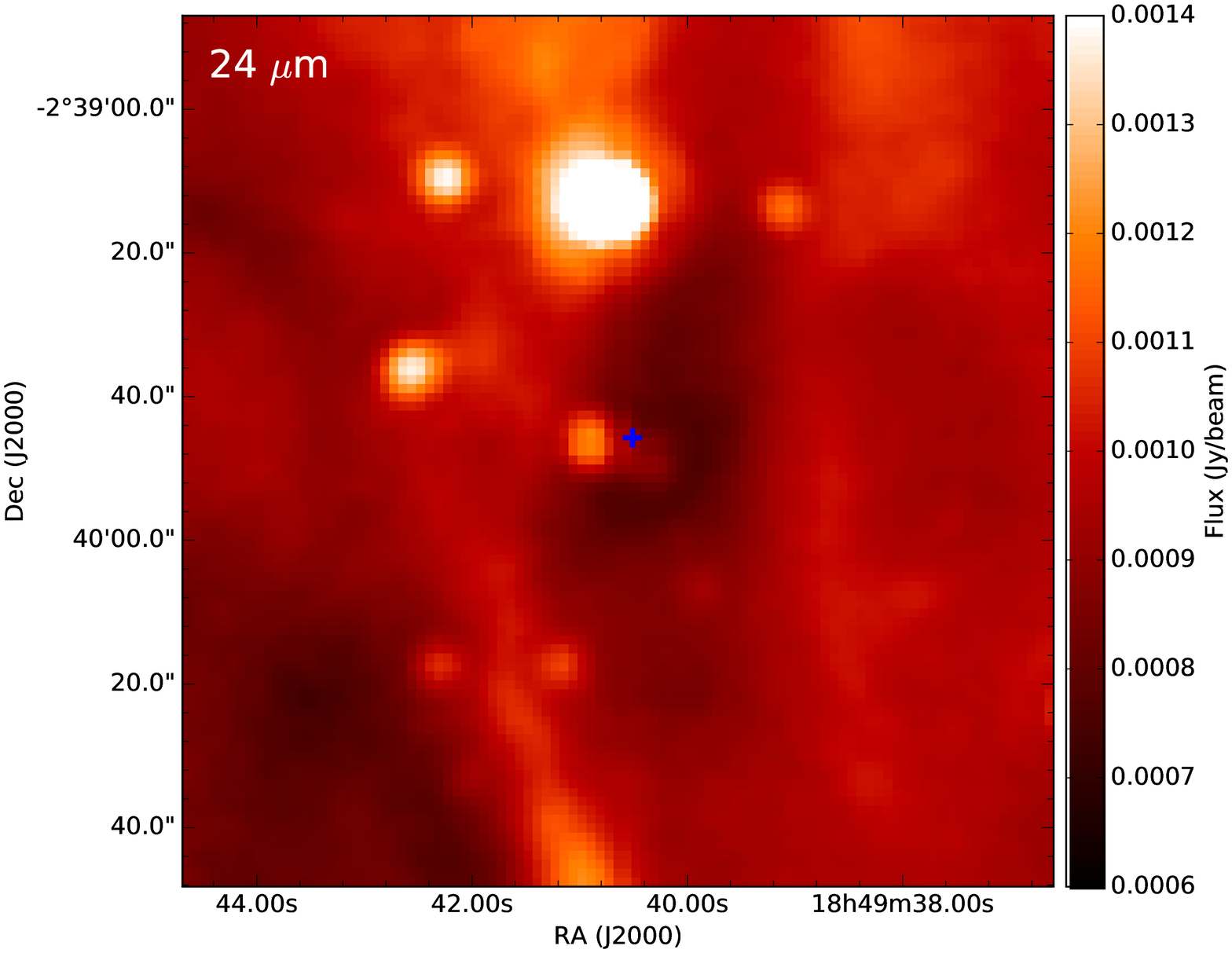}  \includegraphics[width=8cm]{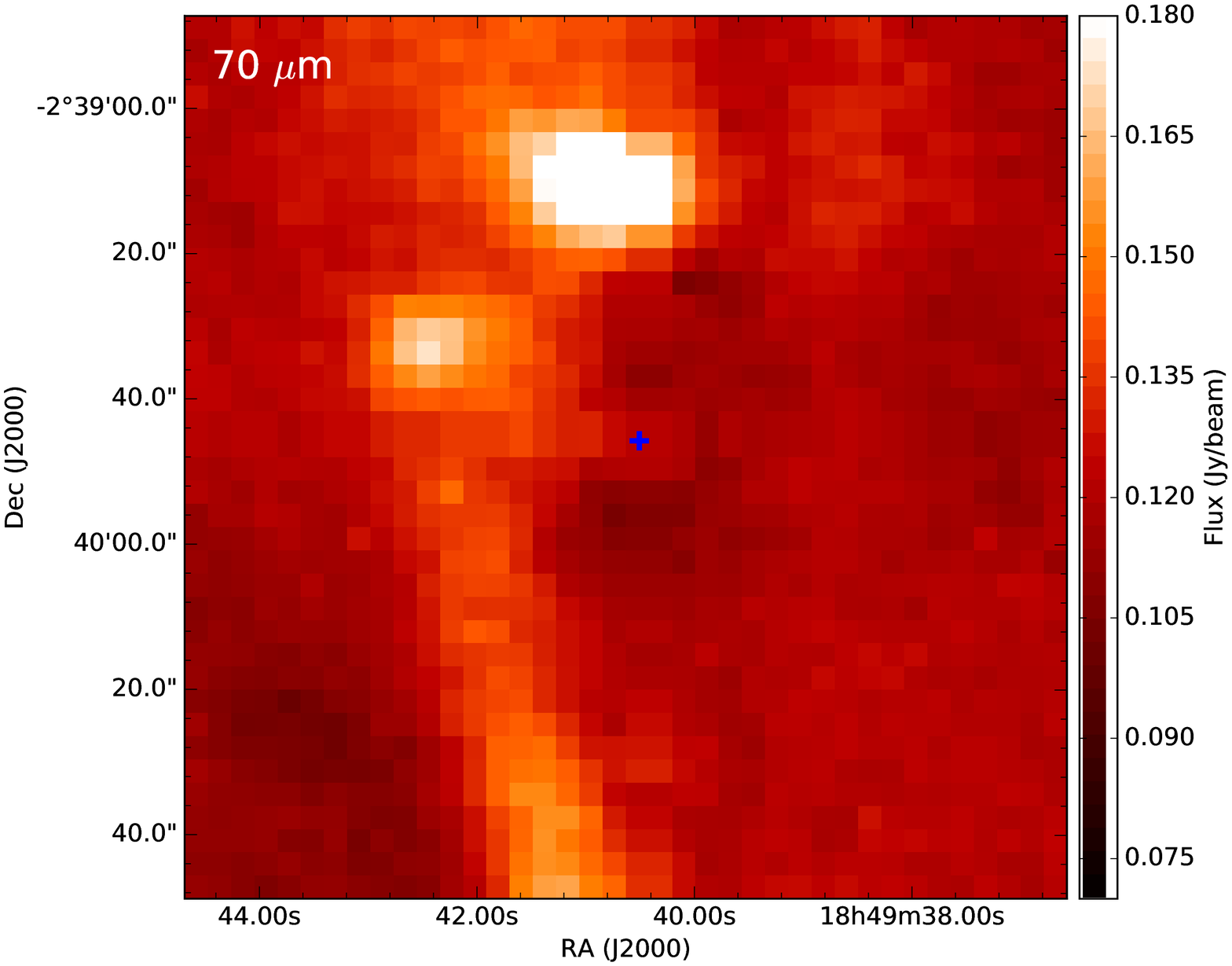} 
 \includegraphics[width=8cm]{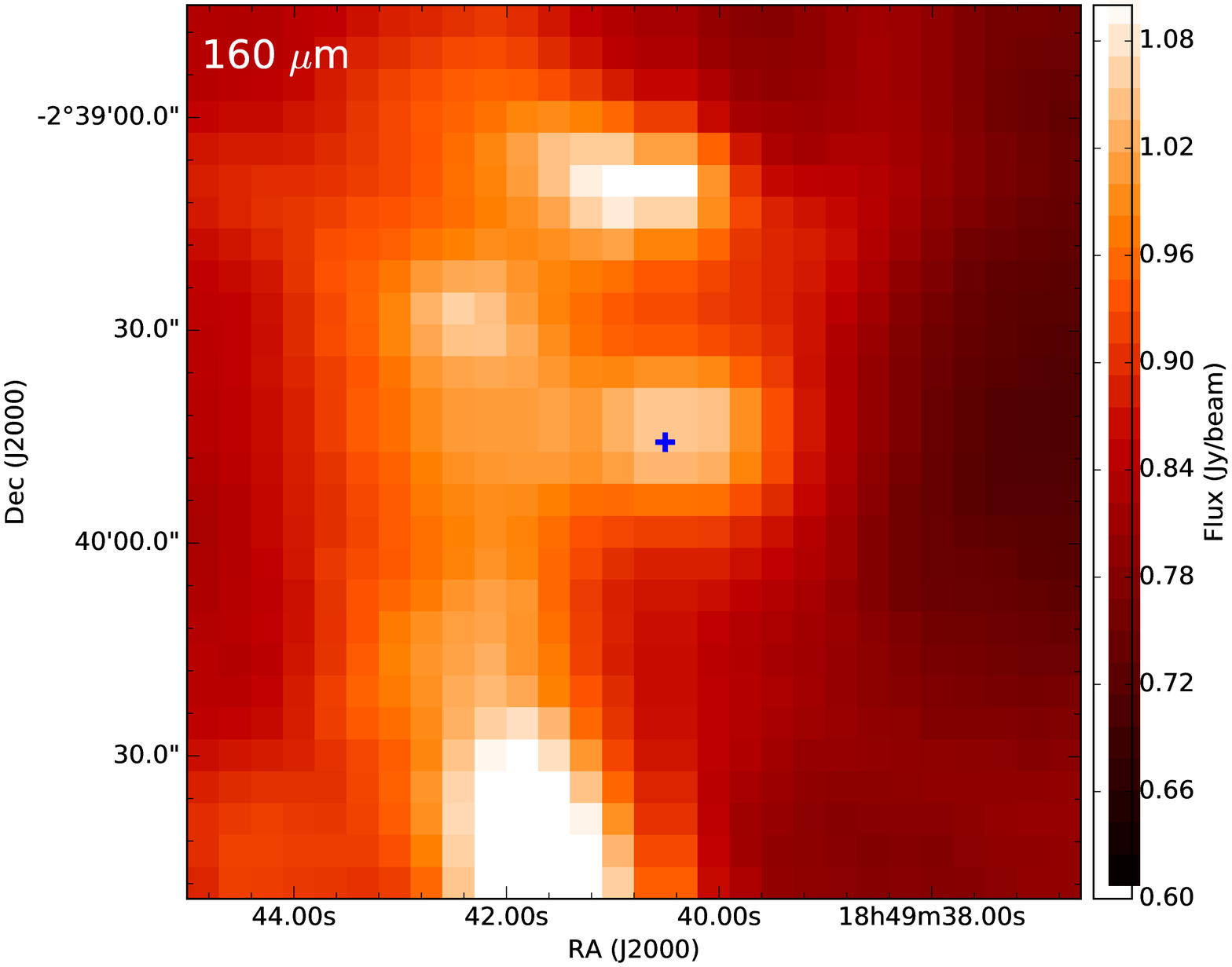}  \includegraphics[width=8cm]{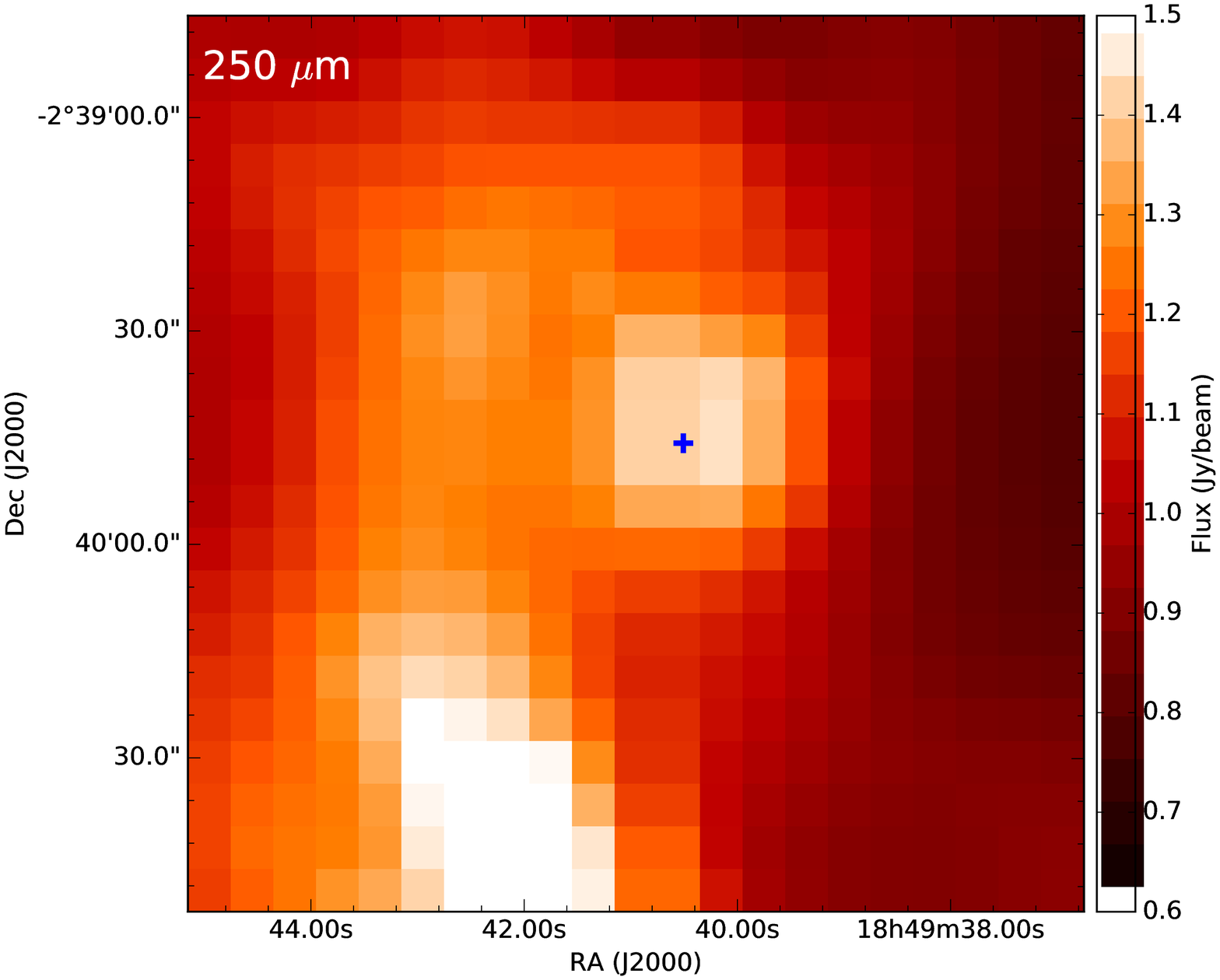} 
 \includegraphics[width=8cm]{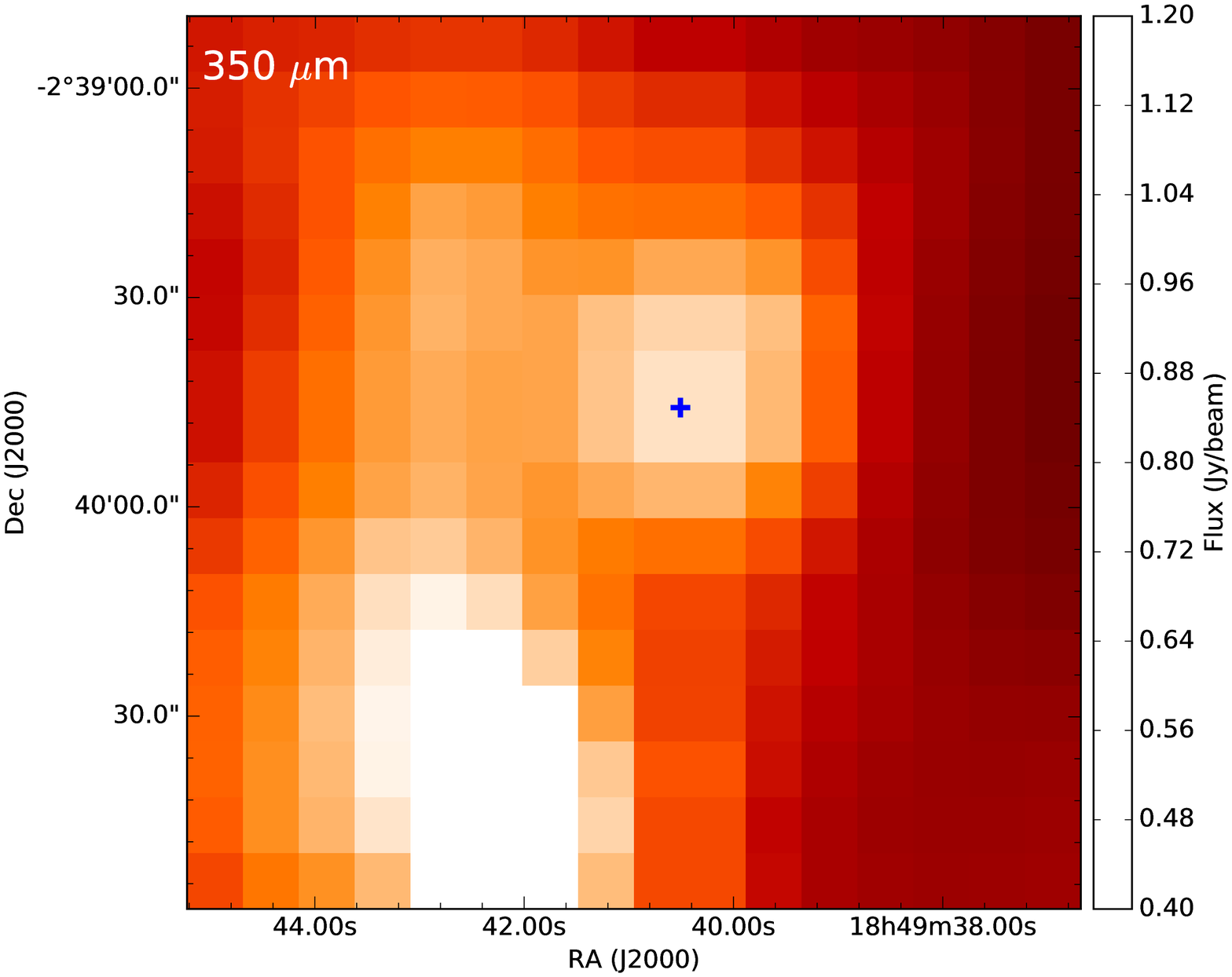}  \includegraphics[width=8cm]{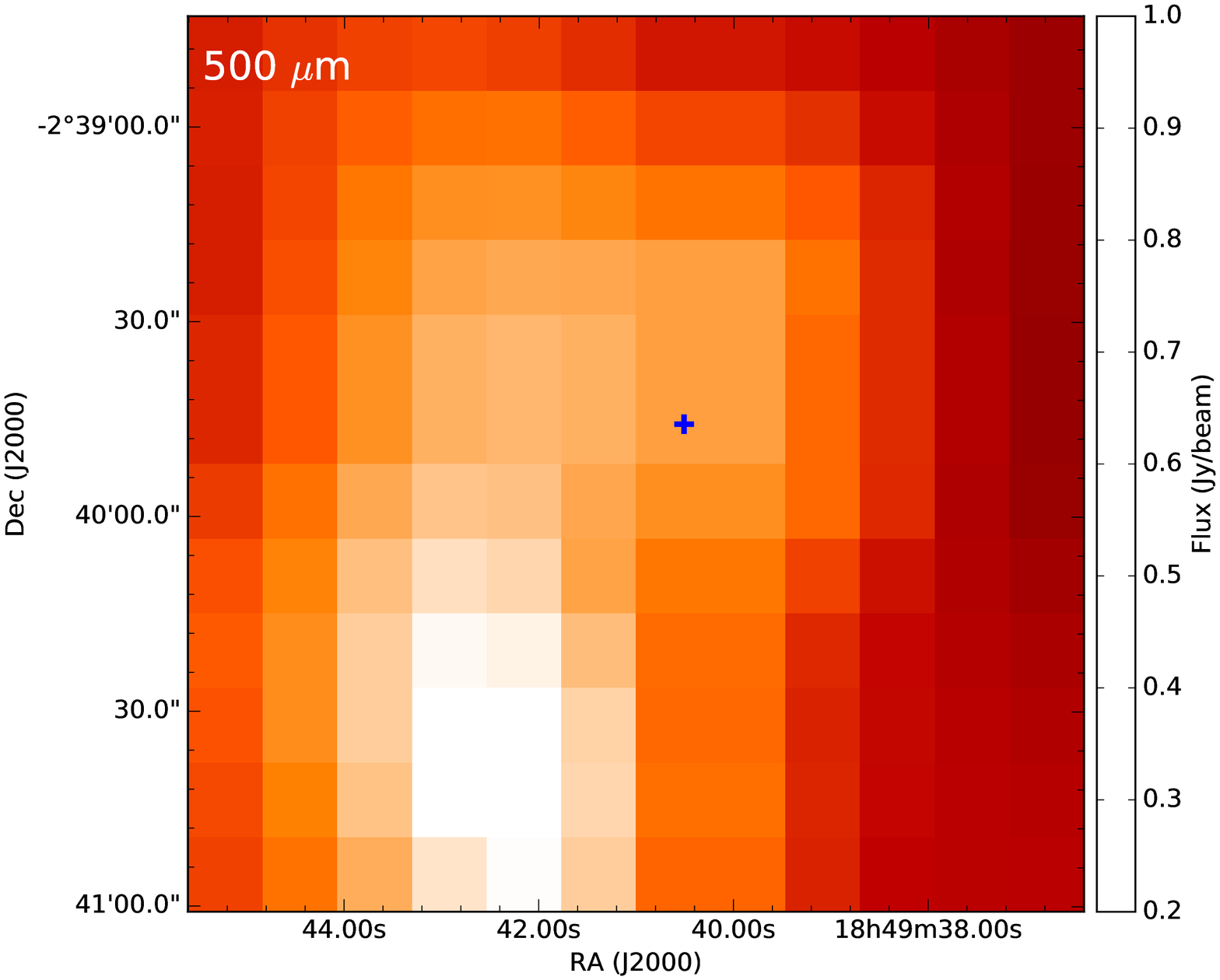} 
 \caption{30.357-0.837}
 \end{figure*}

\begin{figure*}
 \centering
 \includegraphics[width=8cm]{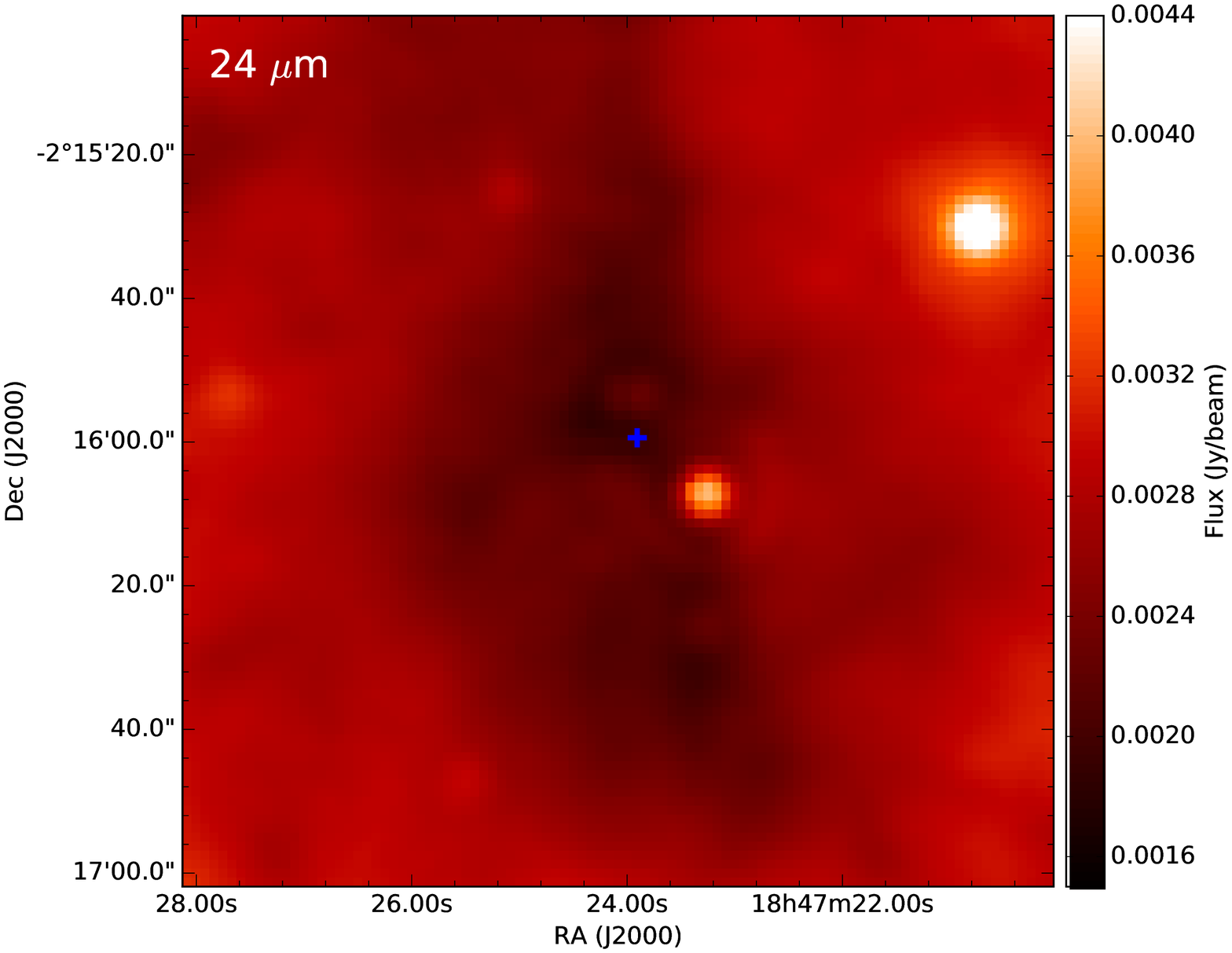}  \includegraphics[width=8cm]{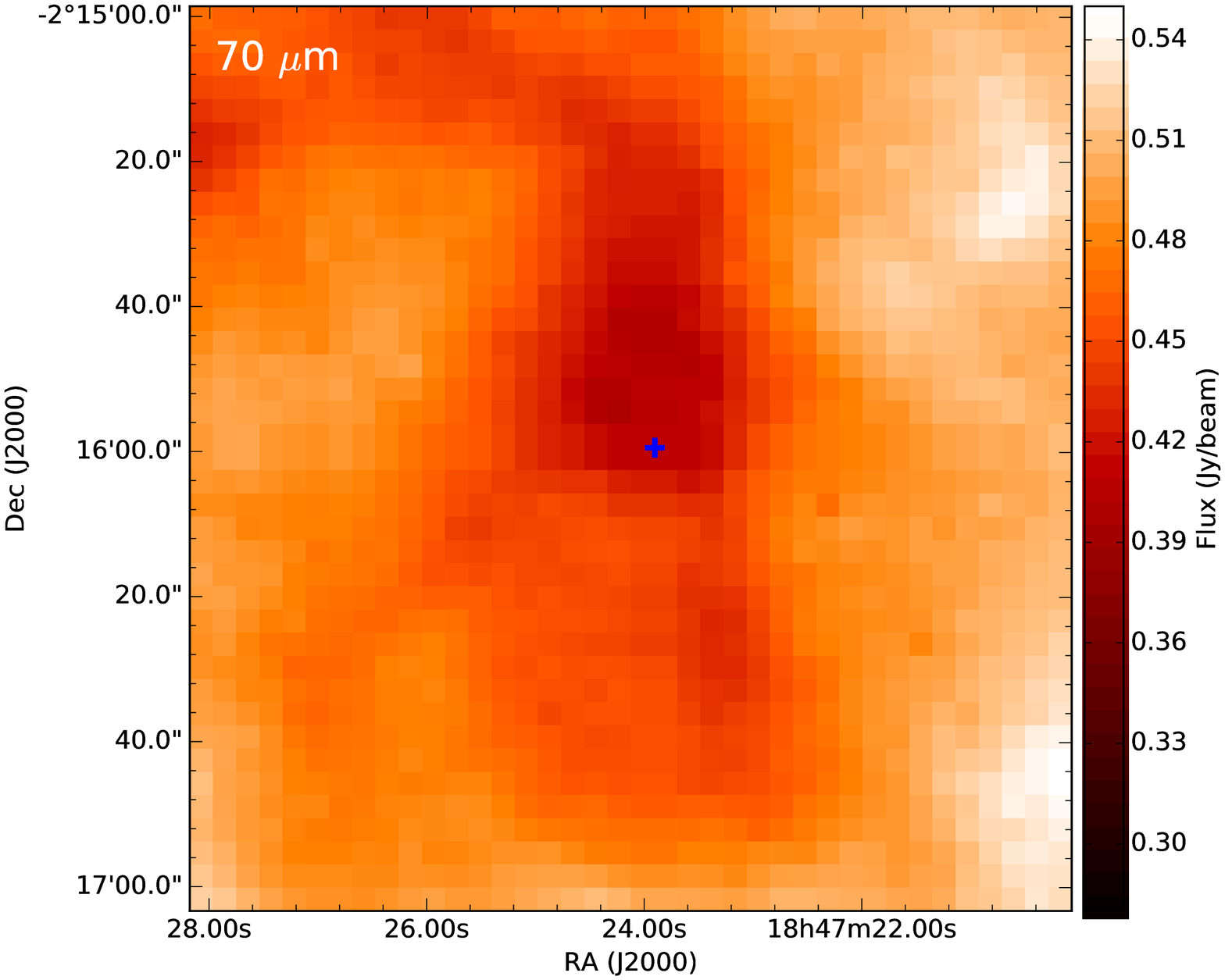} 
 \includegraphics[width=8cm]{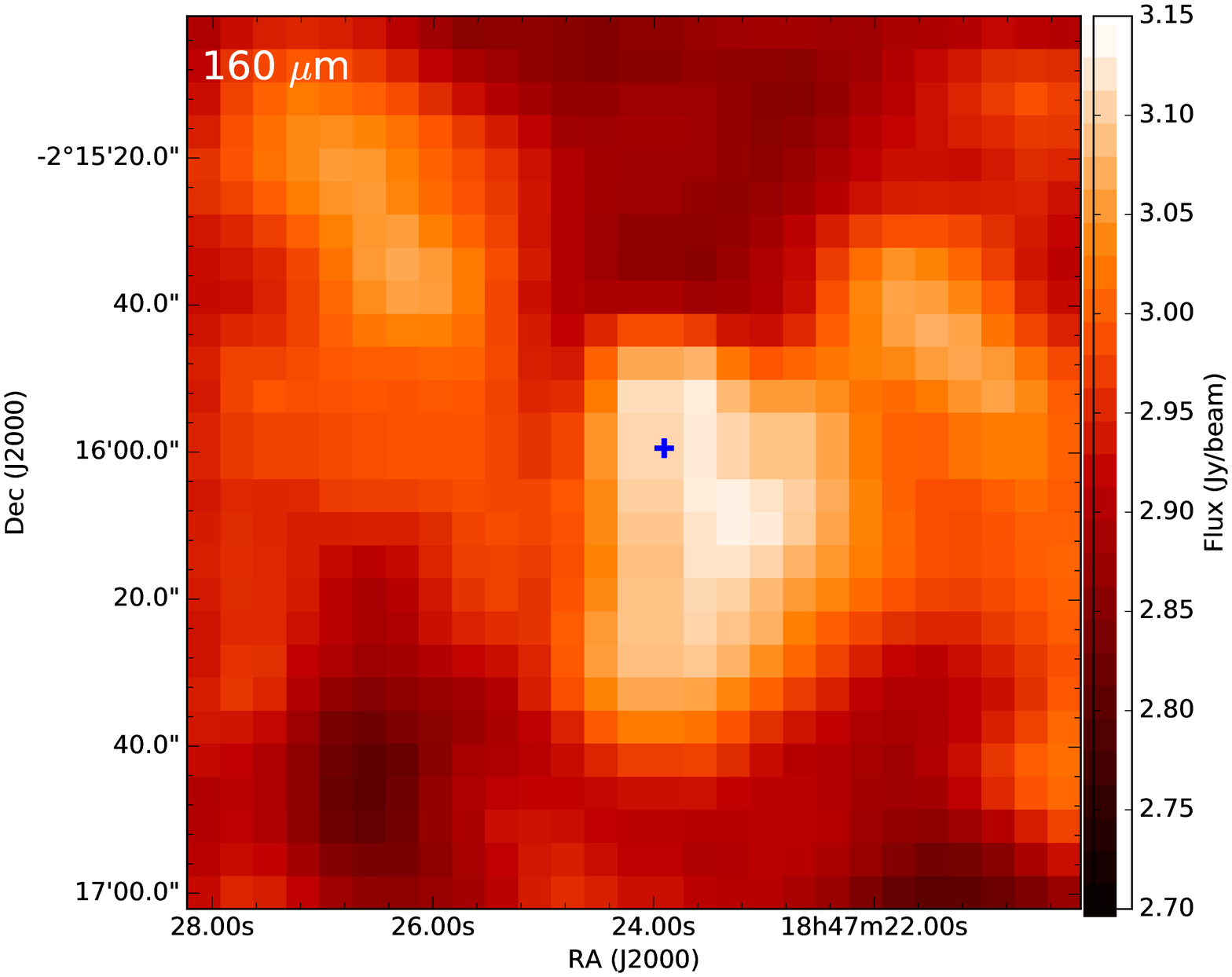}  \includegraphics[width=8cm]{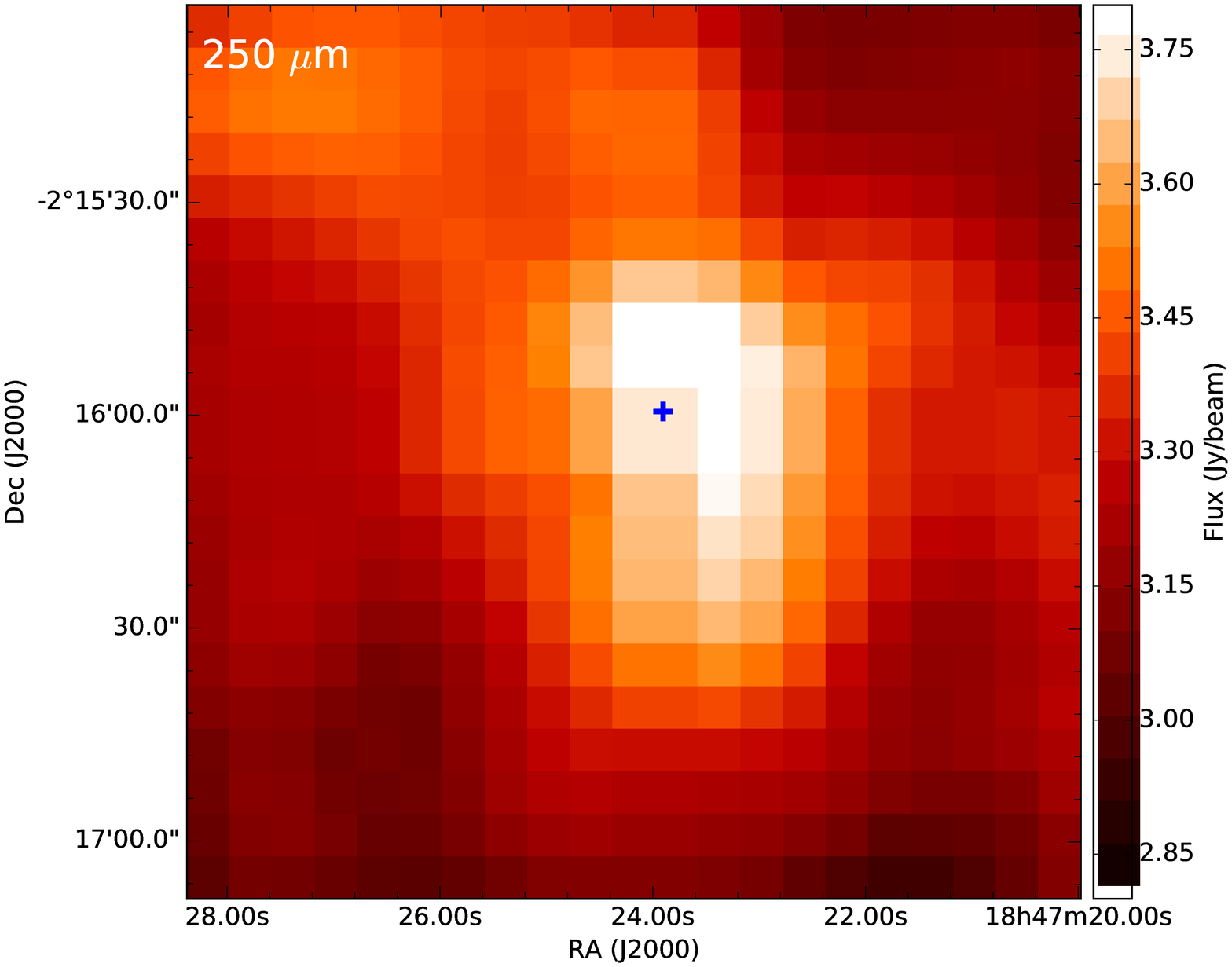} 
 \includegraphics[width=8cm]{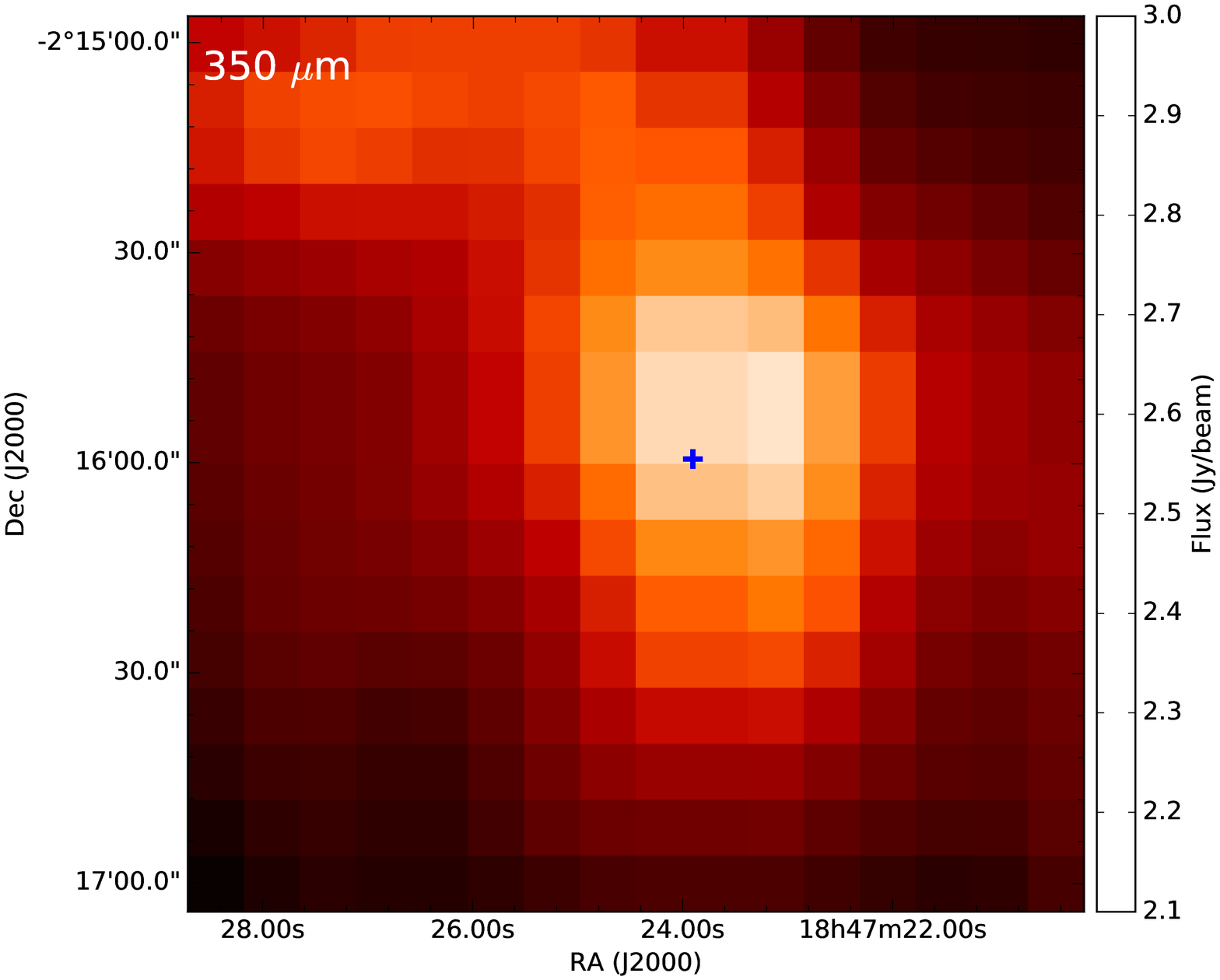}  \includegraphics[width=8cm]{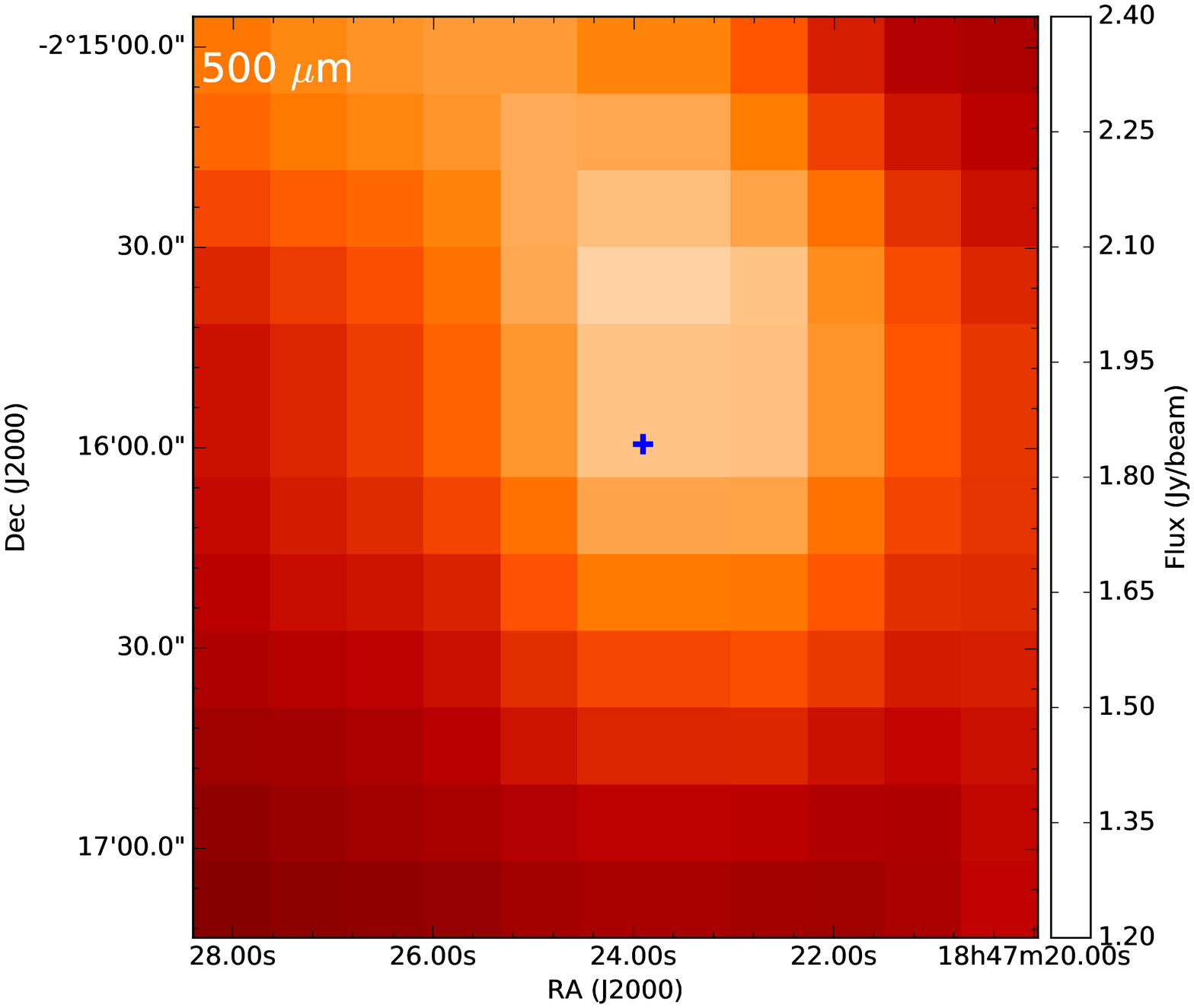} 
 \caption{30.454-0.135}
 \end{figure*}

\begin{figure*}
 \centering
 \includegraphics[width=8cm]{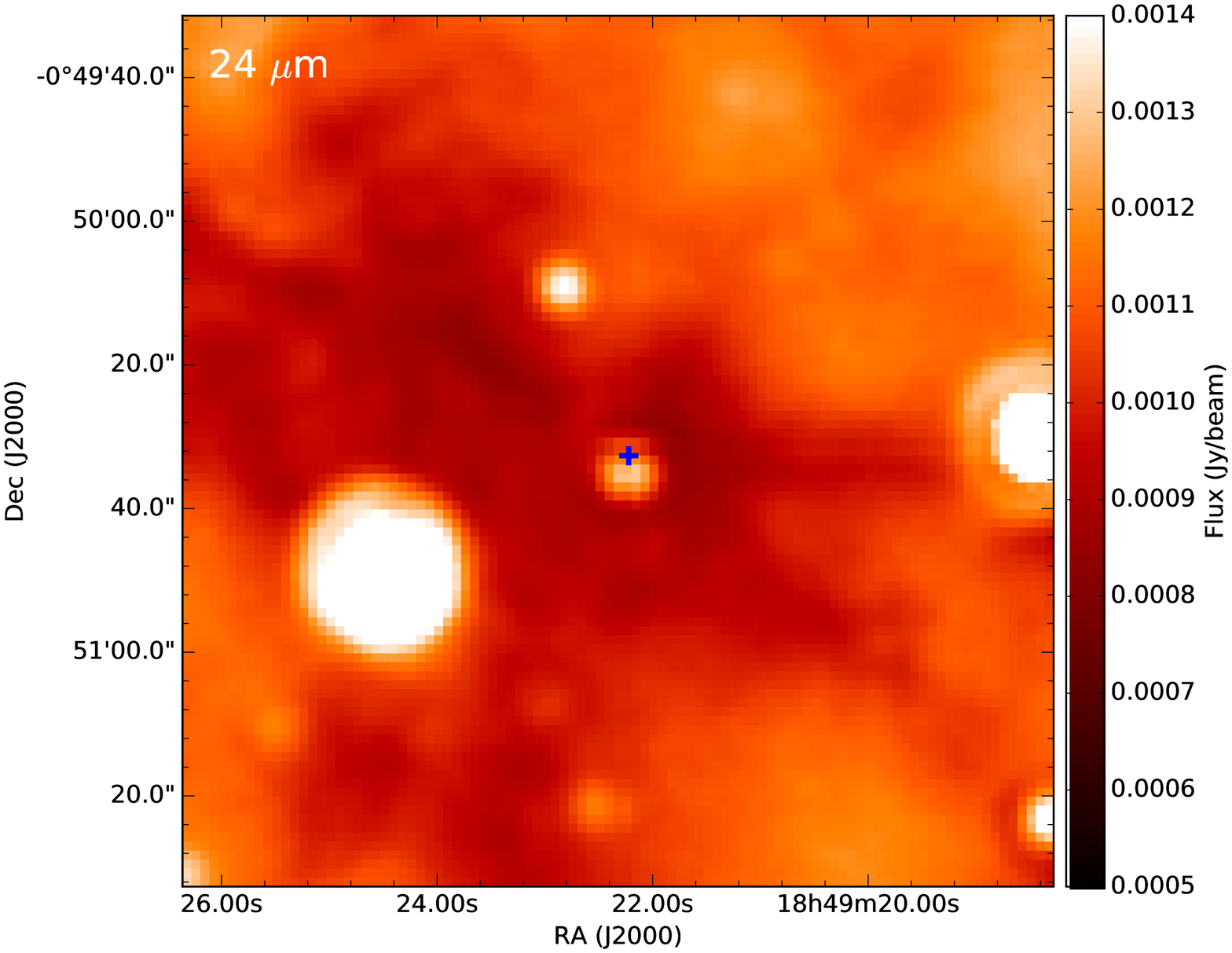}  \includegraphics[width=8cm]{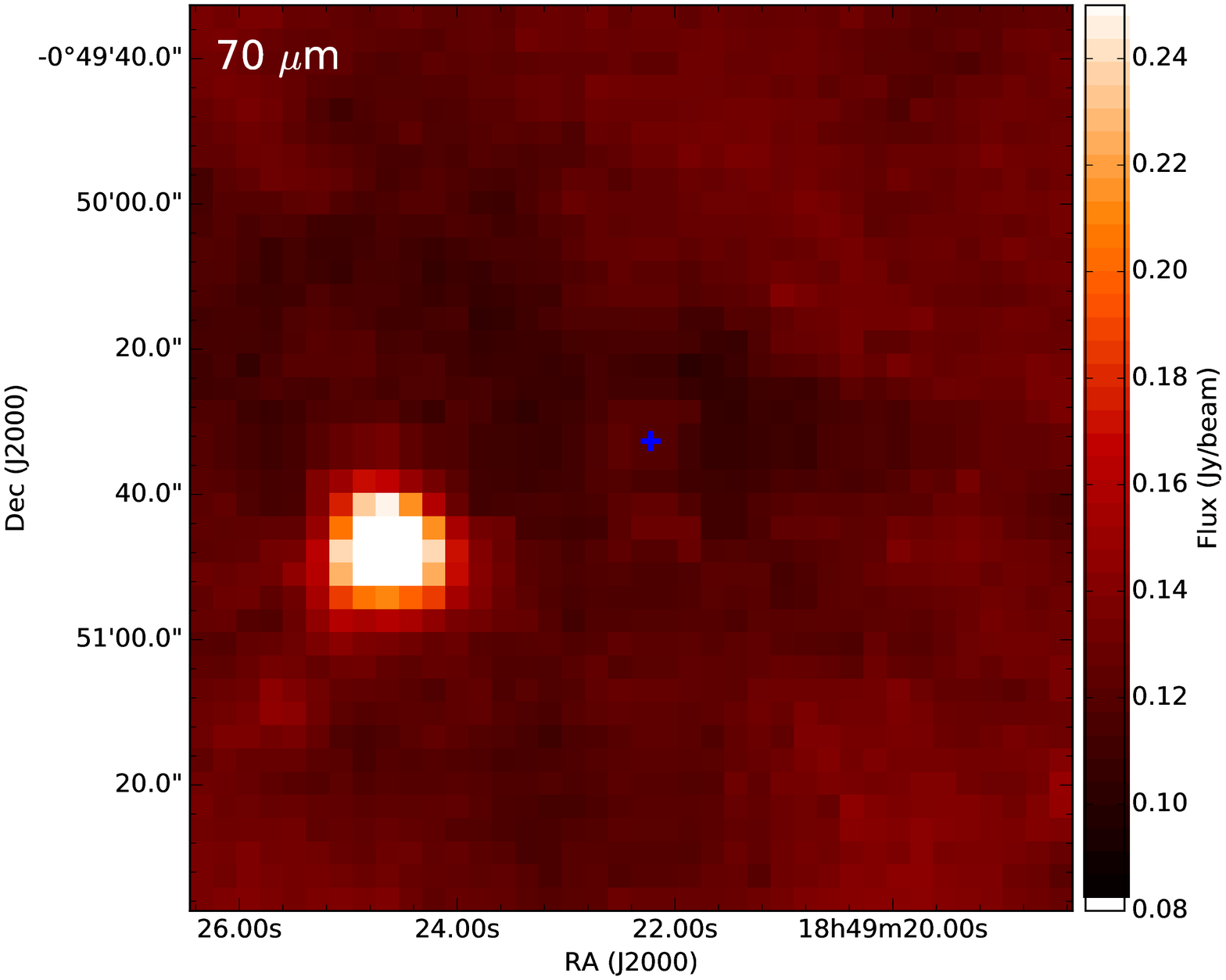} 
 \includegraphics[width=8cm]{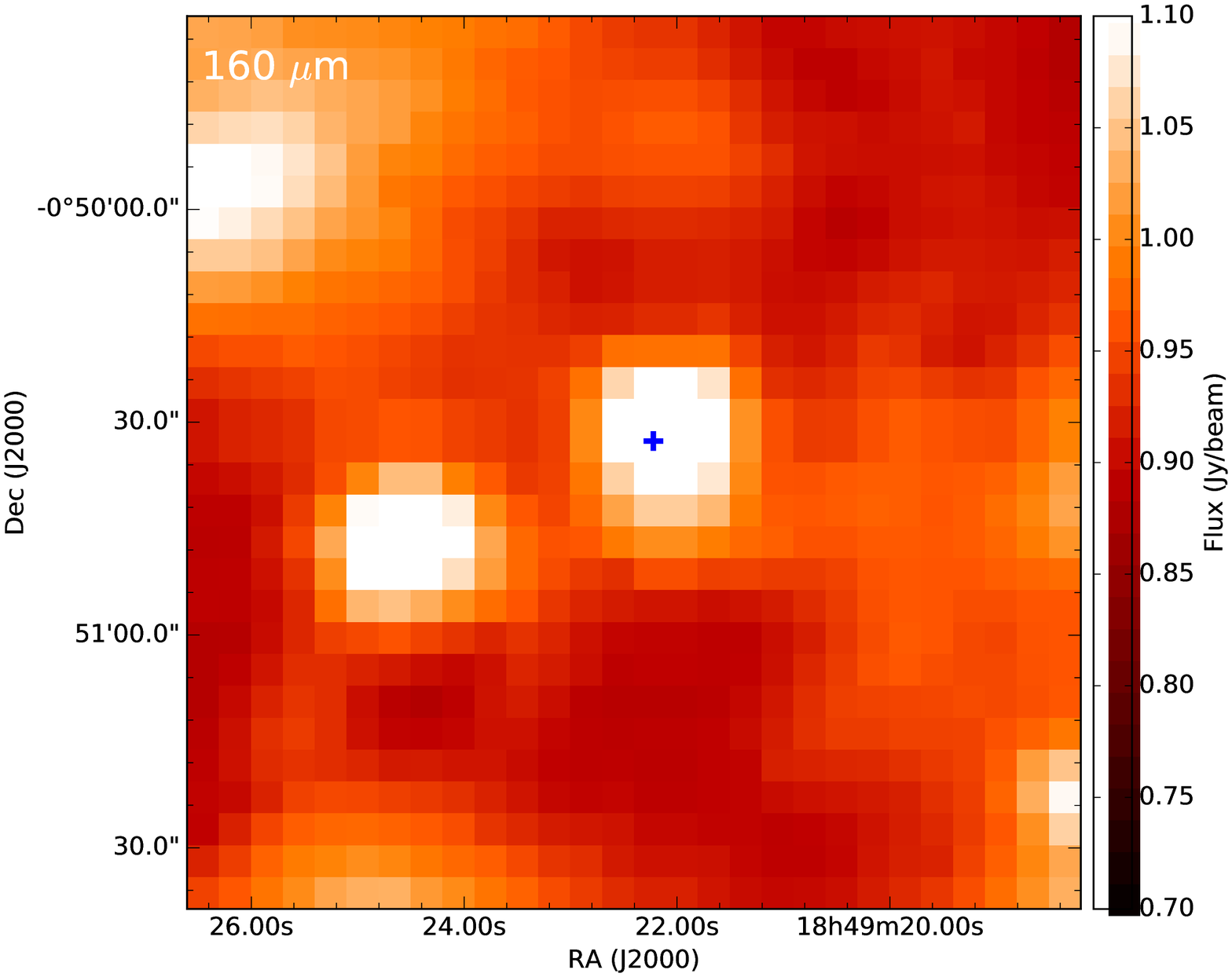}  \includegraphics[width=8cm]{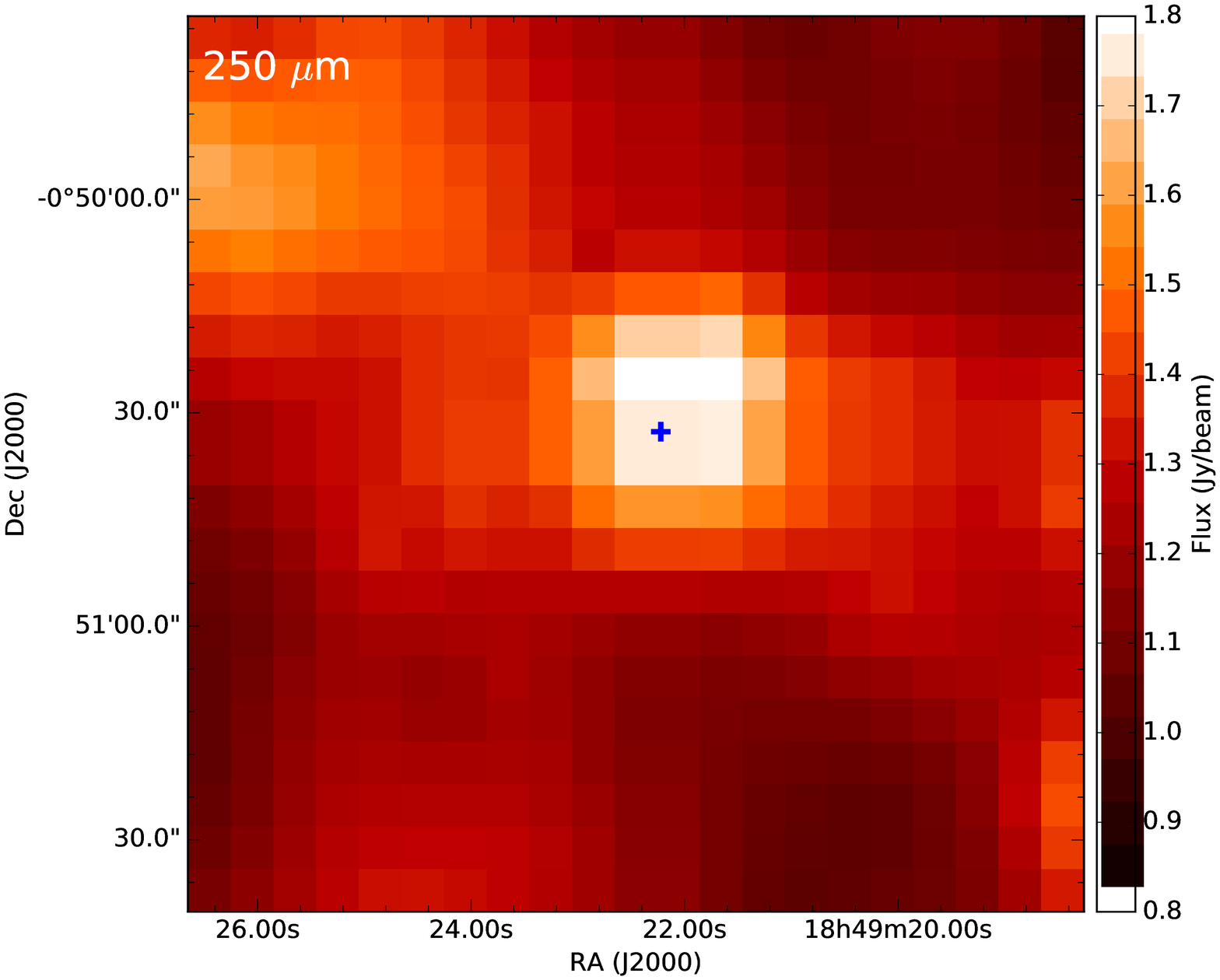} 
 \includegraphics[width=8cm]{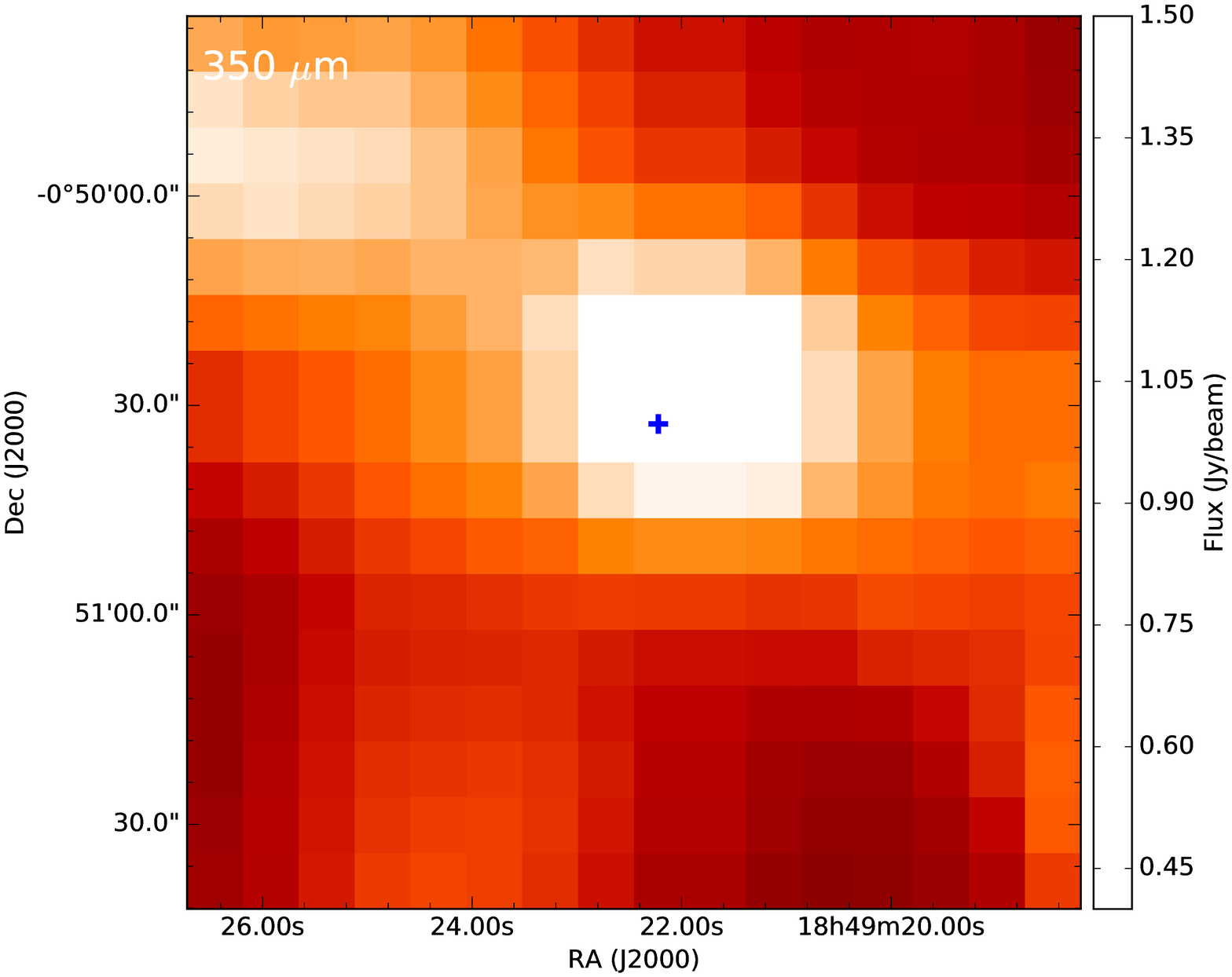}  \includegraphics[width=8cm]{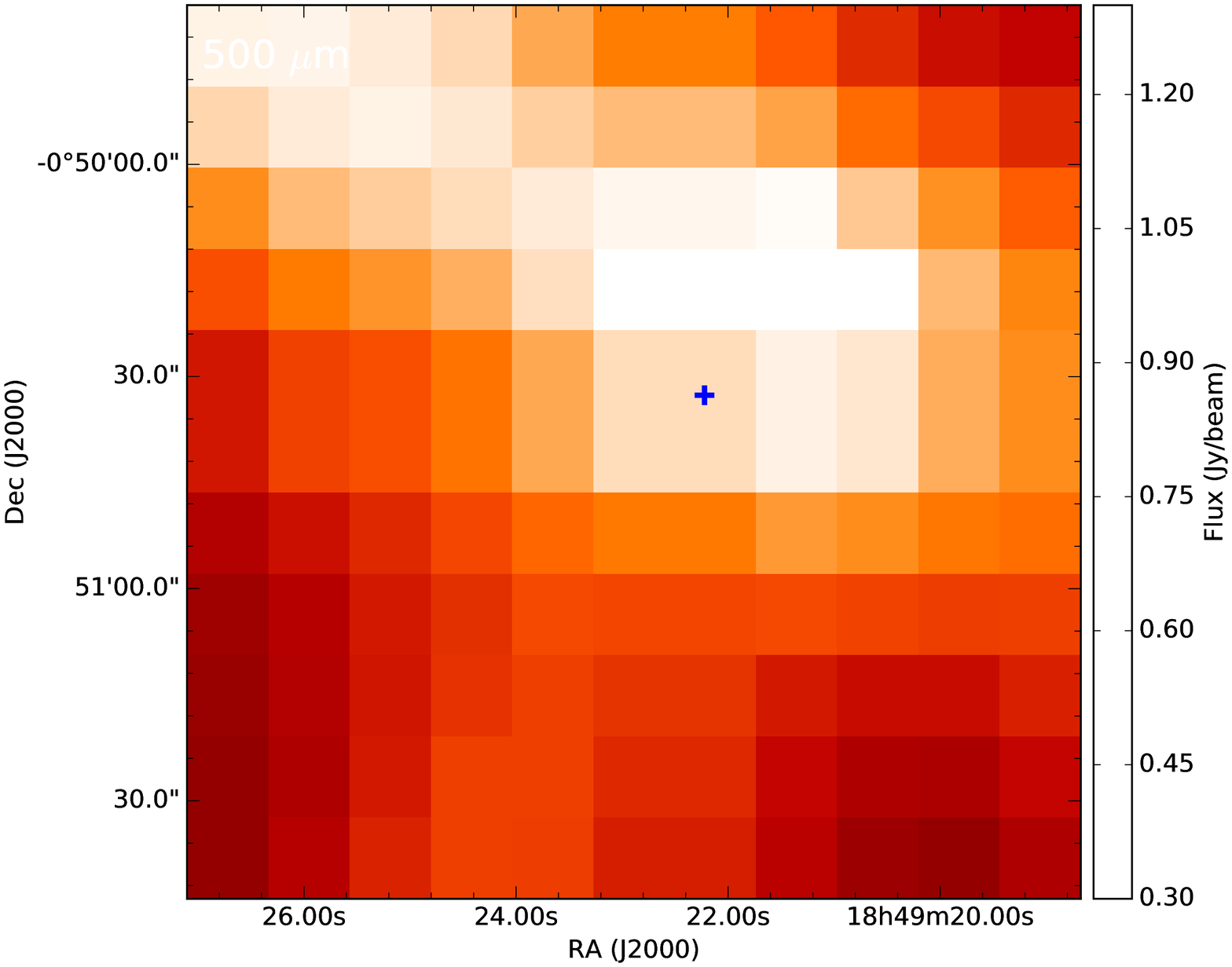} 
 \caption{31.946+0.076}
 \end{figure*}

\begin{figure*}
 \centering
 \includegraphics[width=8cm]{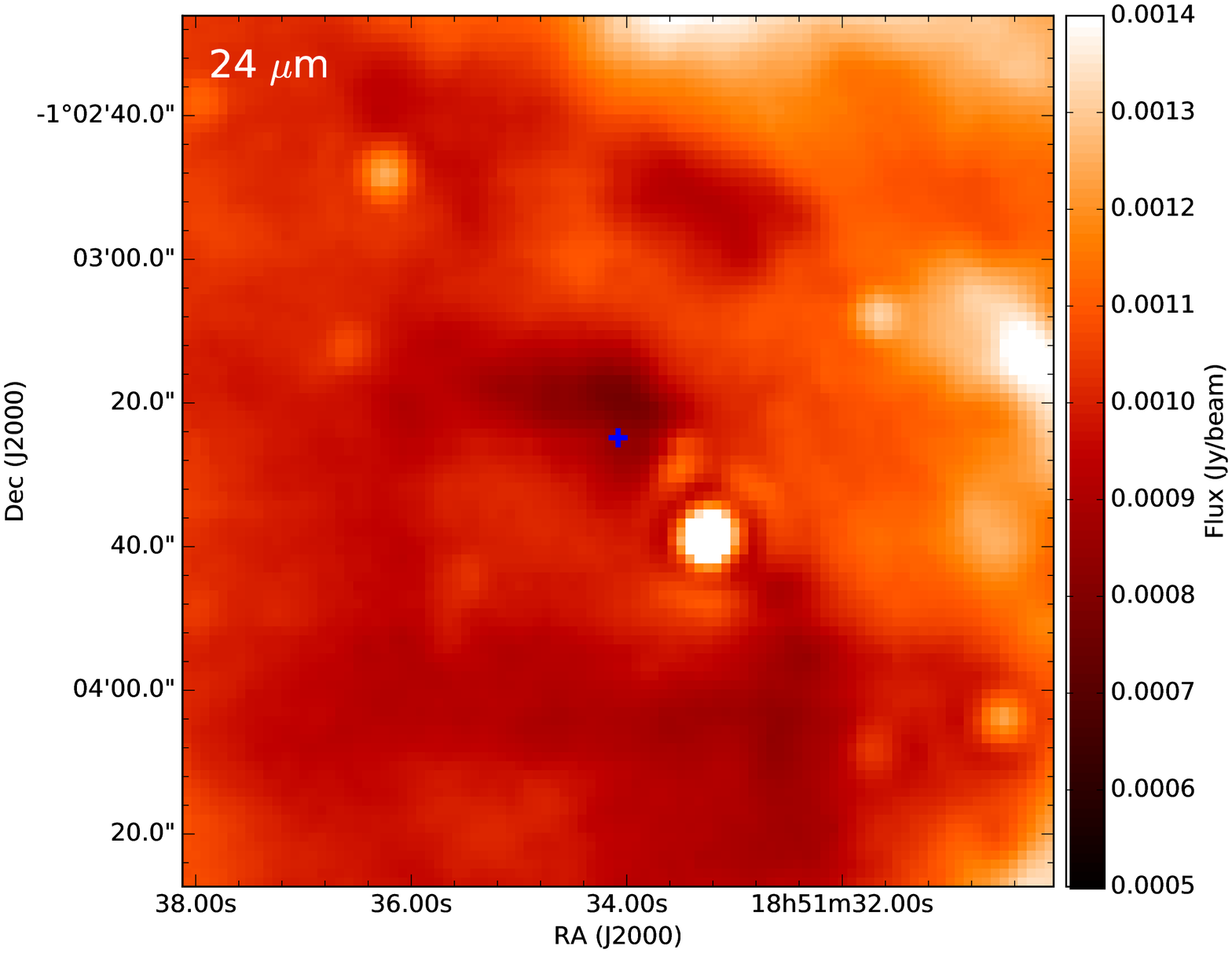}  \includegraphics[width=8cm]{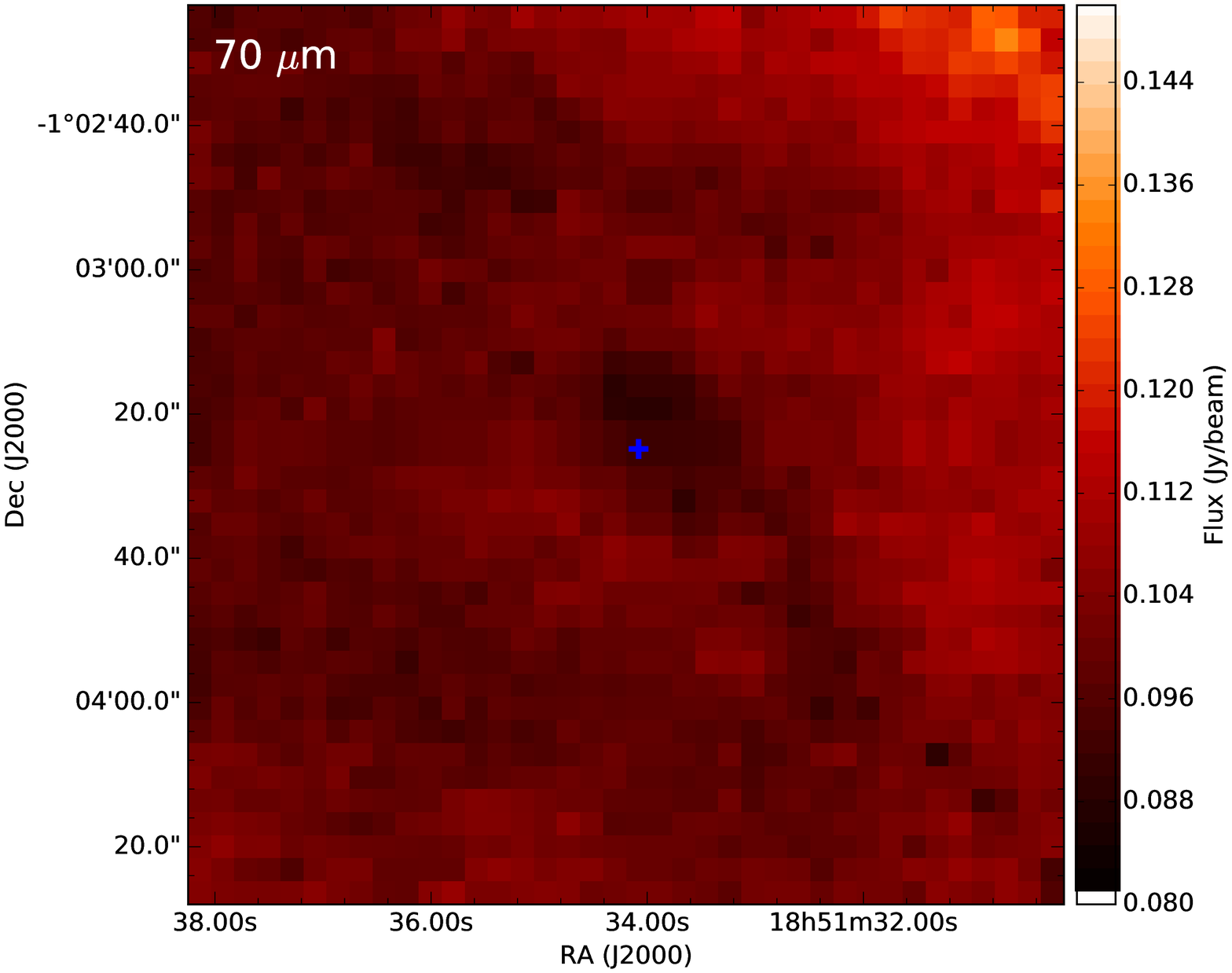} 
 \includegraphics[width=8cm]{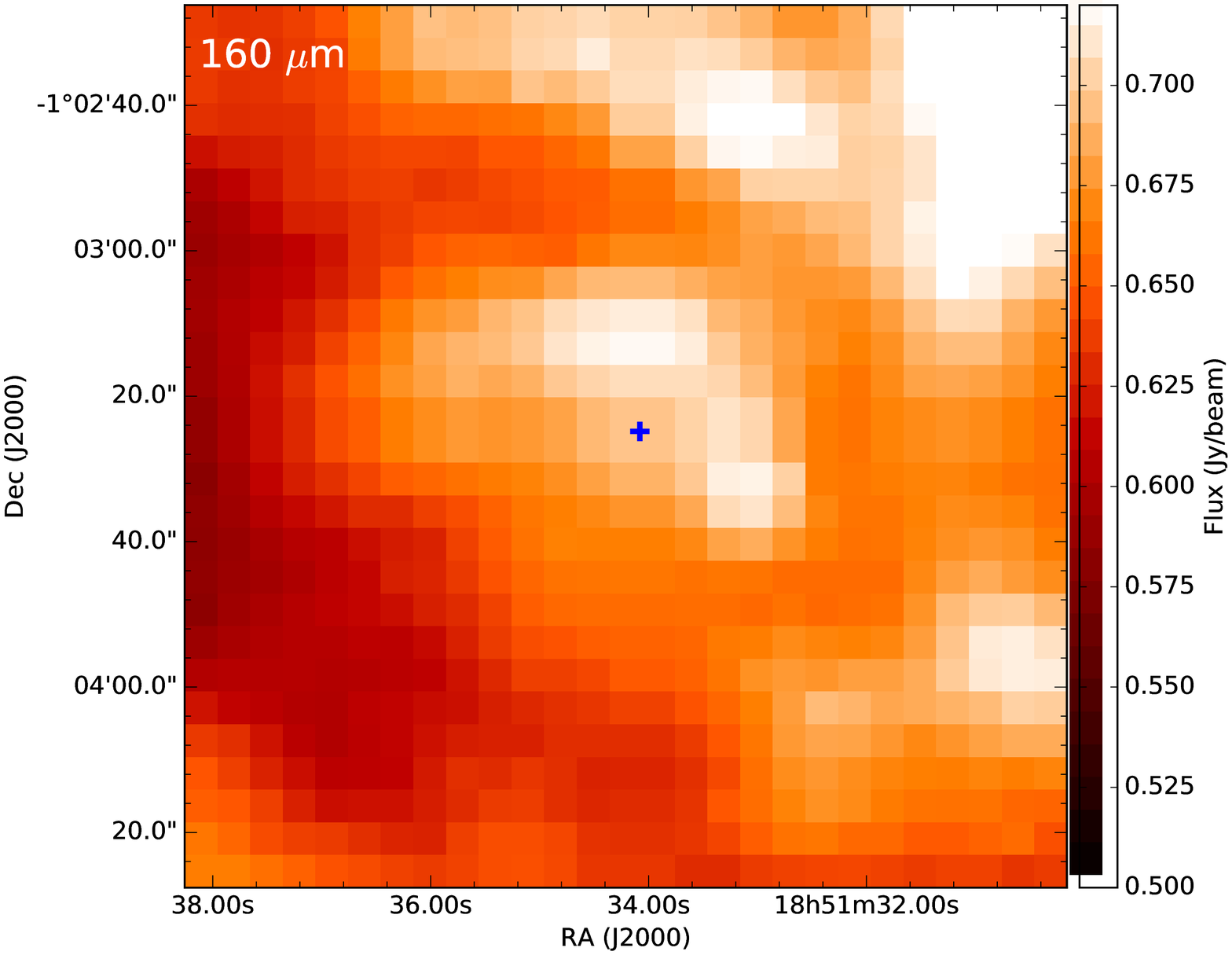}  \includegraphics[width=8cm]{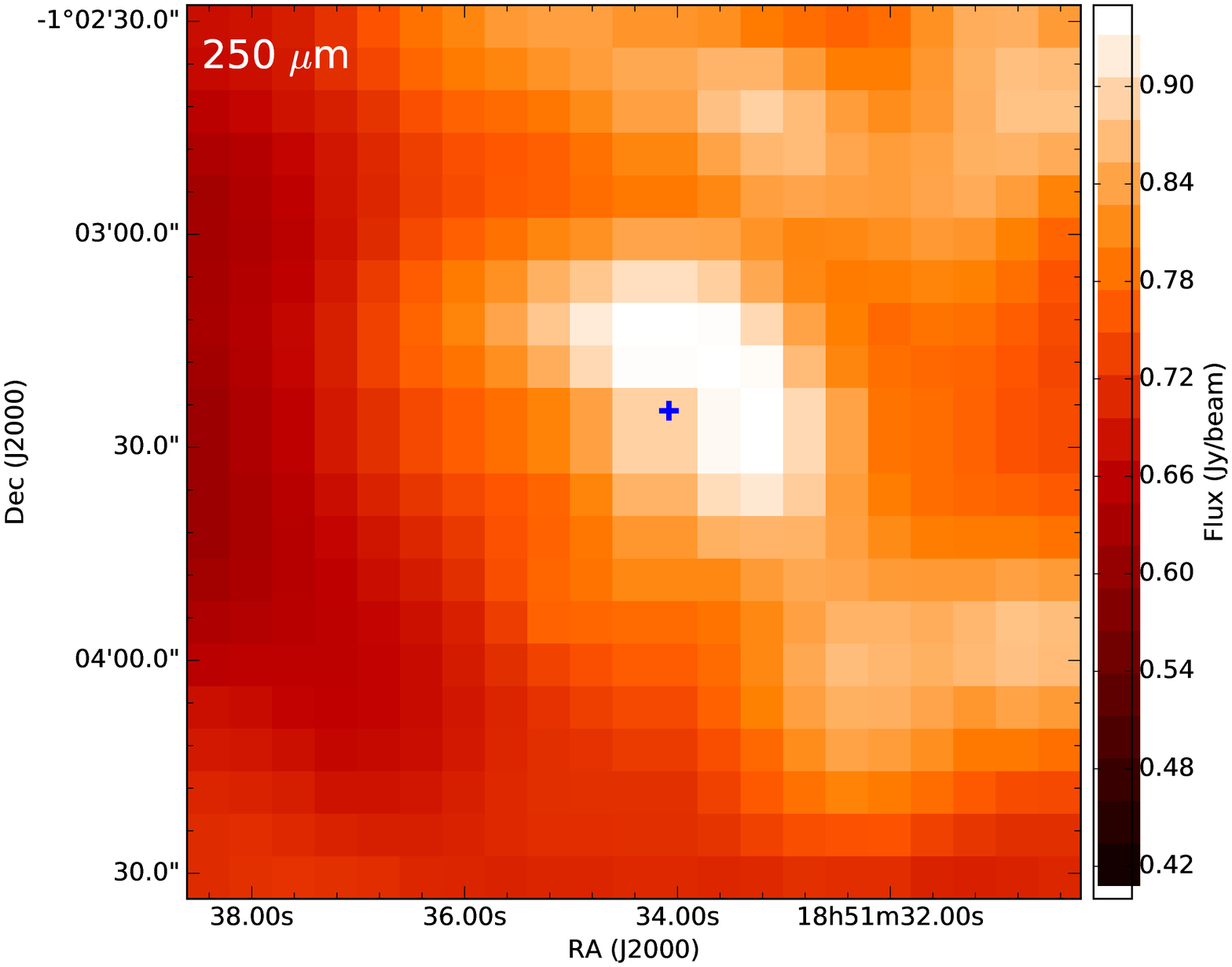} 
 \includegraphics[width=8cm]{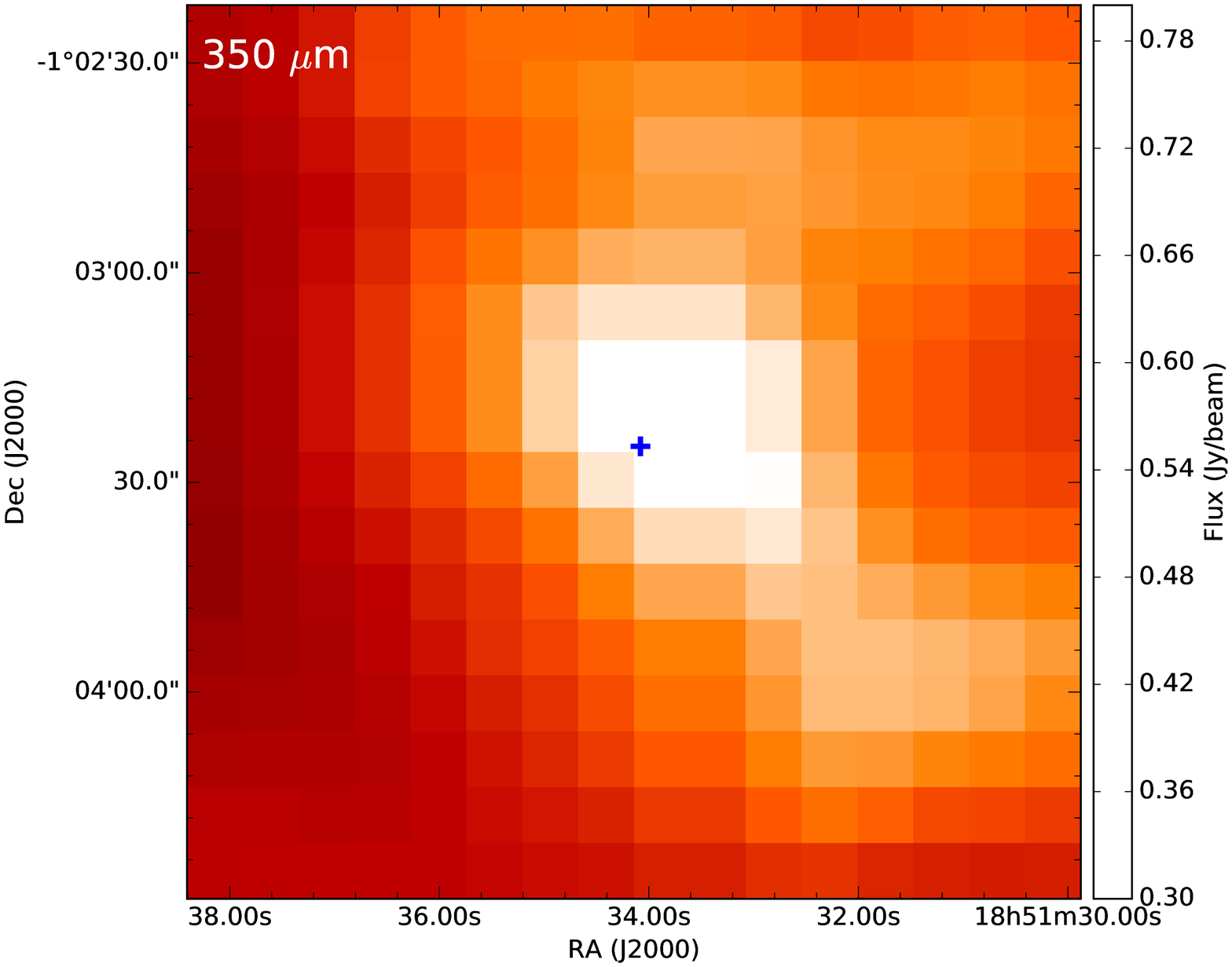}  \includegraphics[width=8cm]{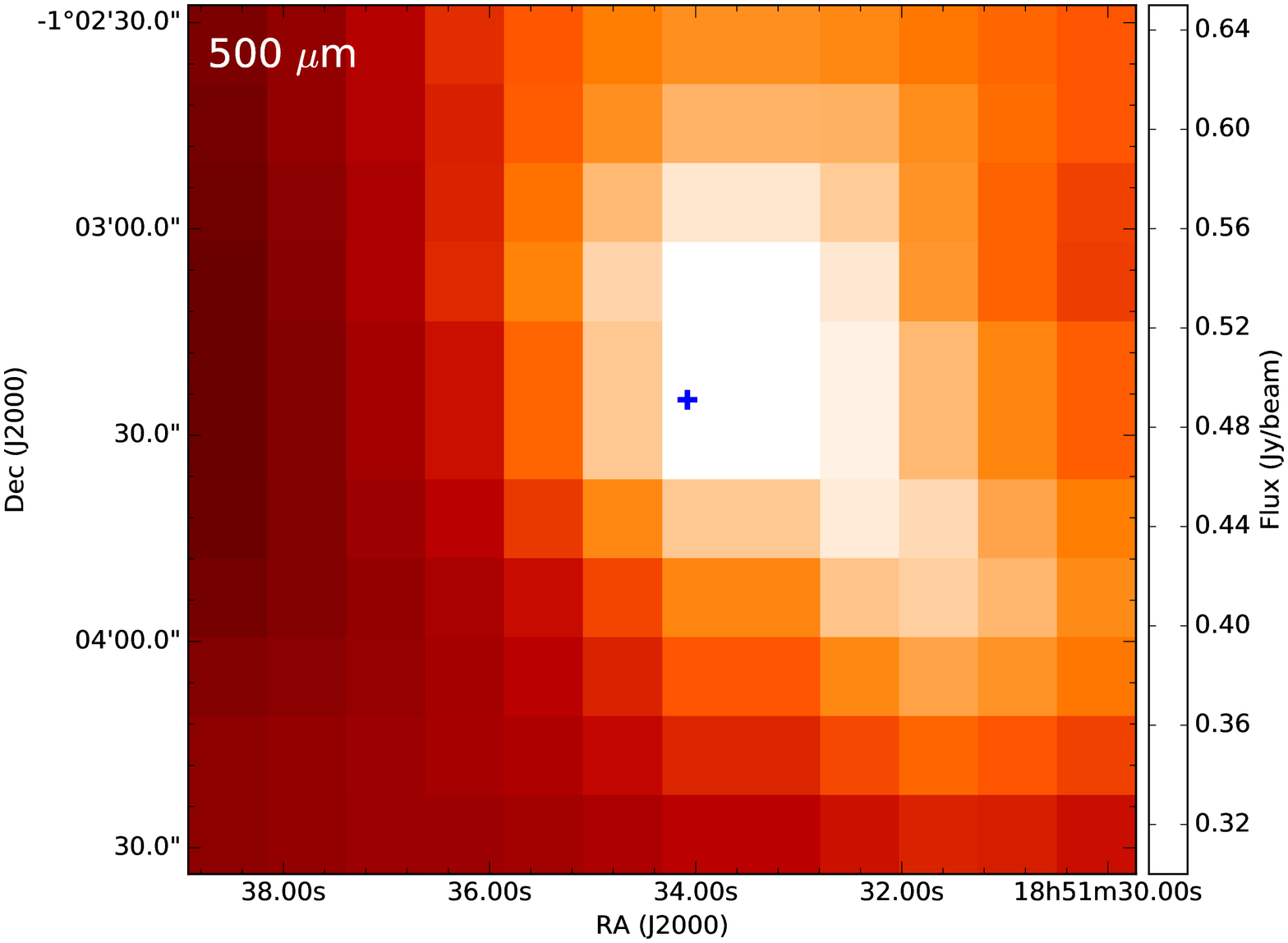} 
 \caption{32.006-0.51}
 \end{figure*}

\begin{figure*}
 \centering
 \includegraphics[width=8cm]{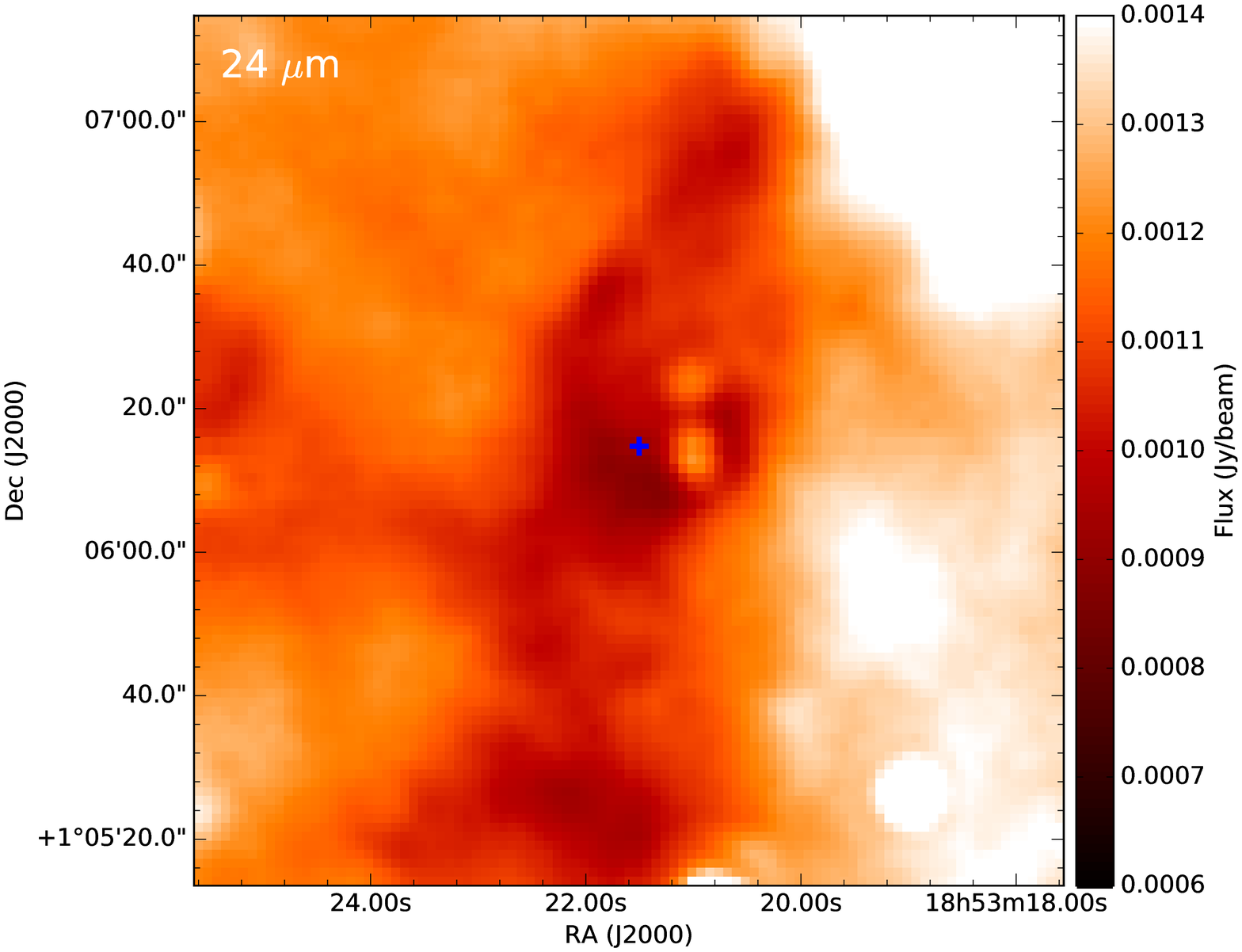}  \includegraphics[width=8cm]{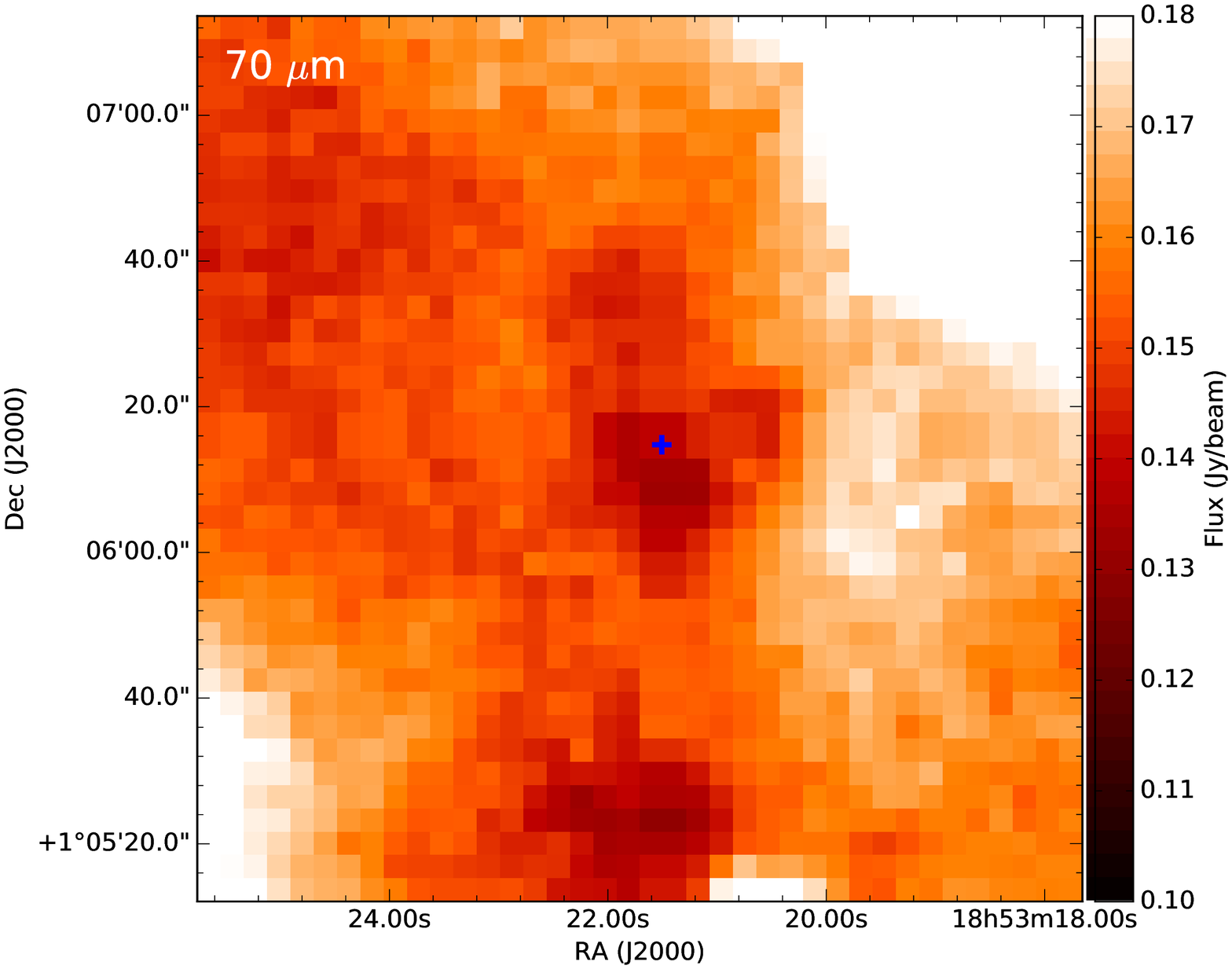} 
 \includegraphics[width=8cm]{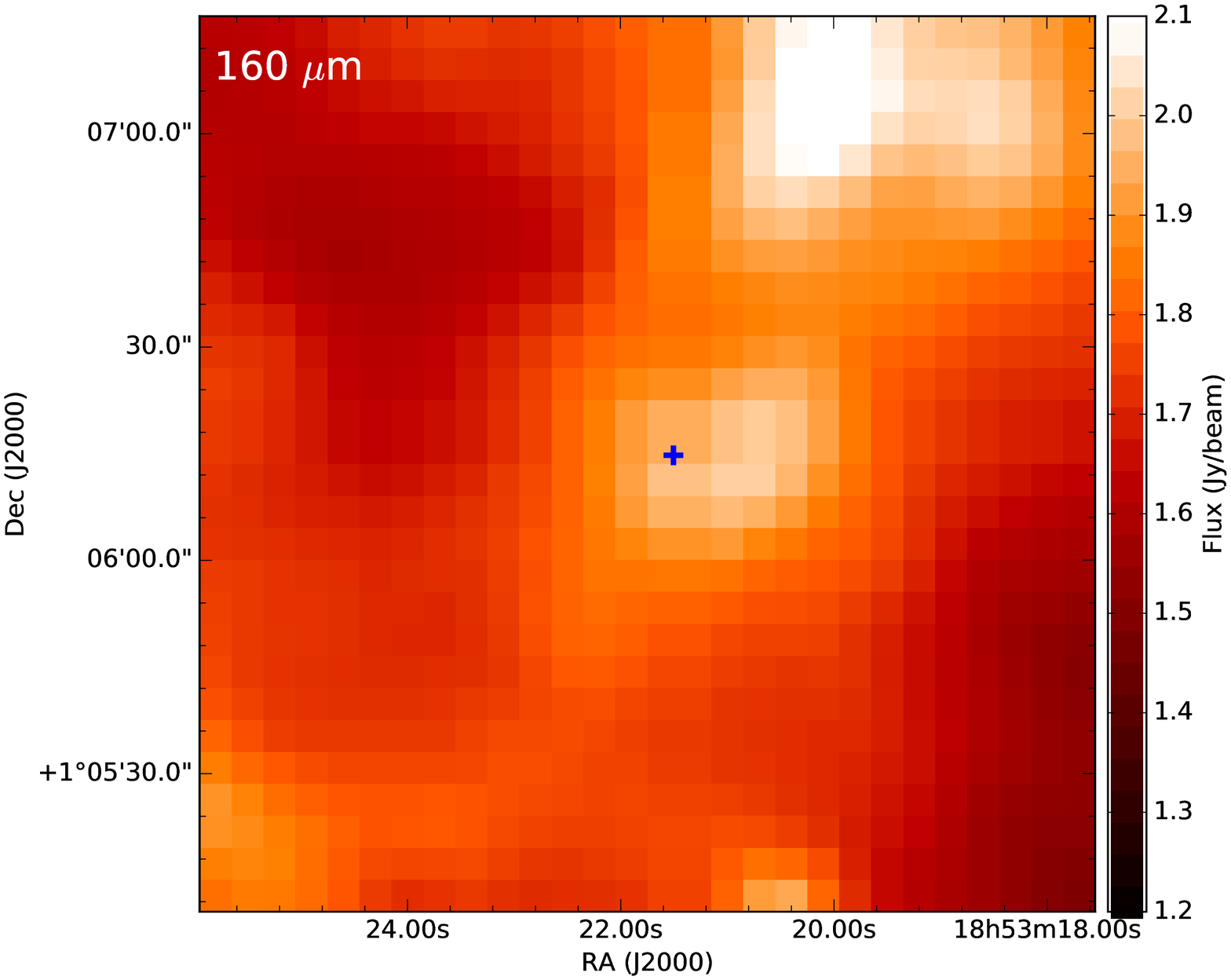}  \includegraphics[width=8cm]{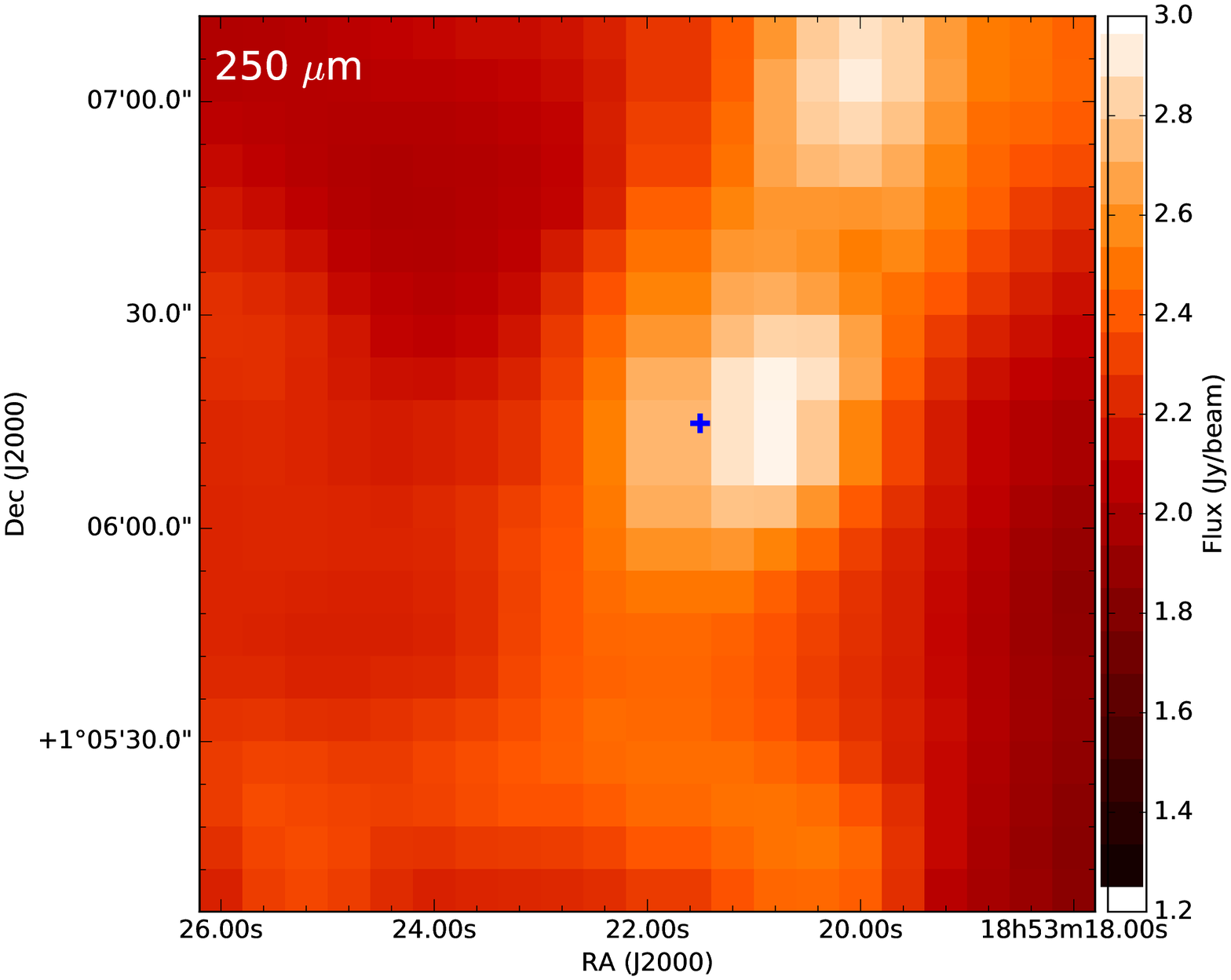} 
 \includegraphics[width=8cm]{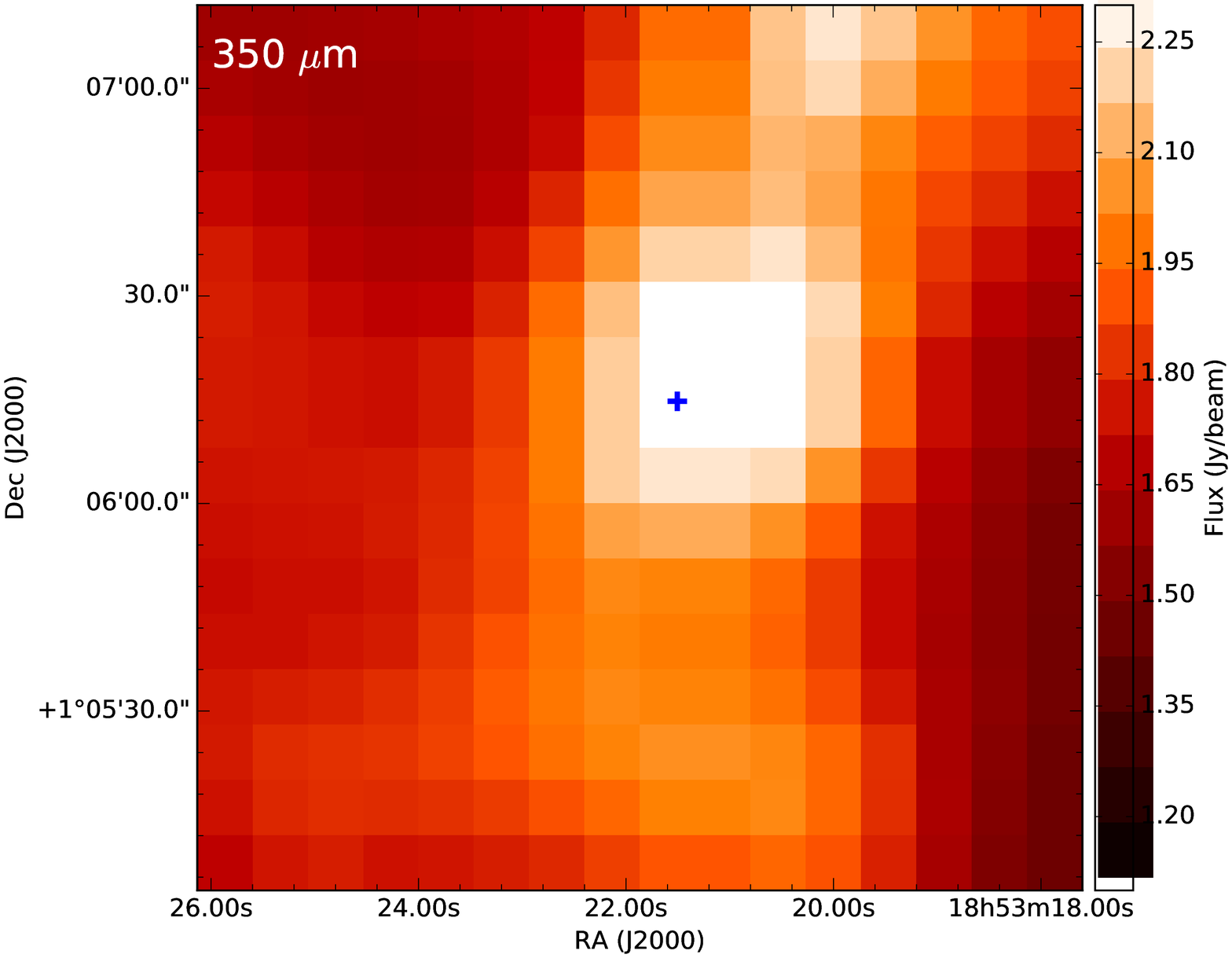}  \includegraphics[width=8cm]{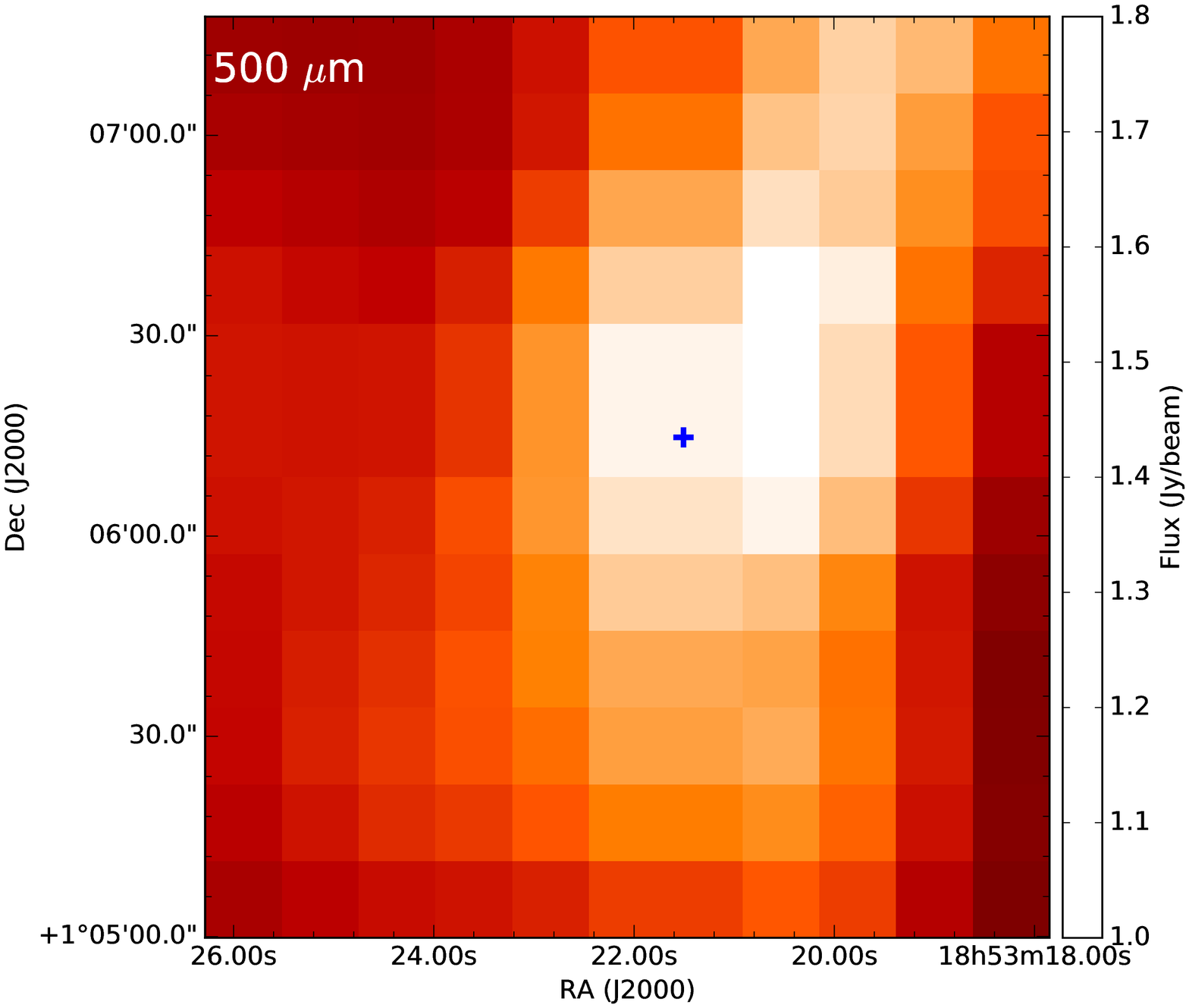} 
 \caption{34.131+0.075}
 \end{figure*}

\clearpage
 
\section{N$_{2}$H$^{+}$ (1$-$0), HNC (1$-$0) and HCO$^{+}$ (1$-$0) spectra}\label{sec:app_spectra}

\begin{figure*}
 \centering
\includegraphics[width=8cm]{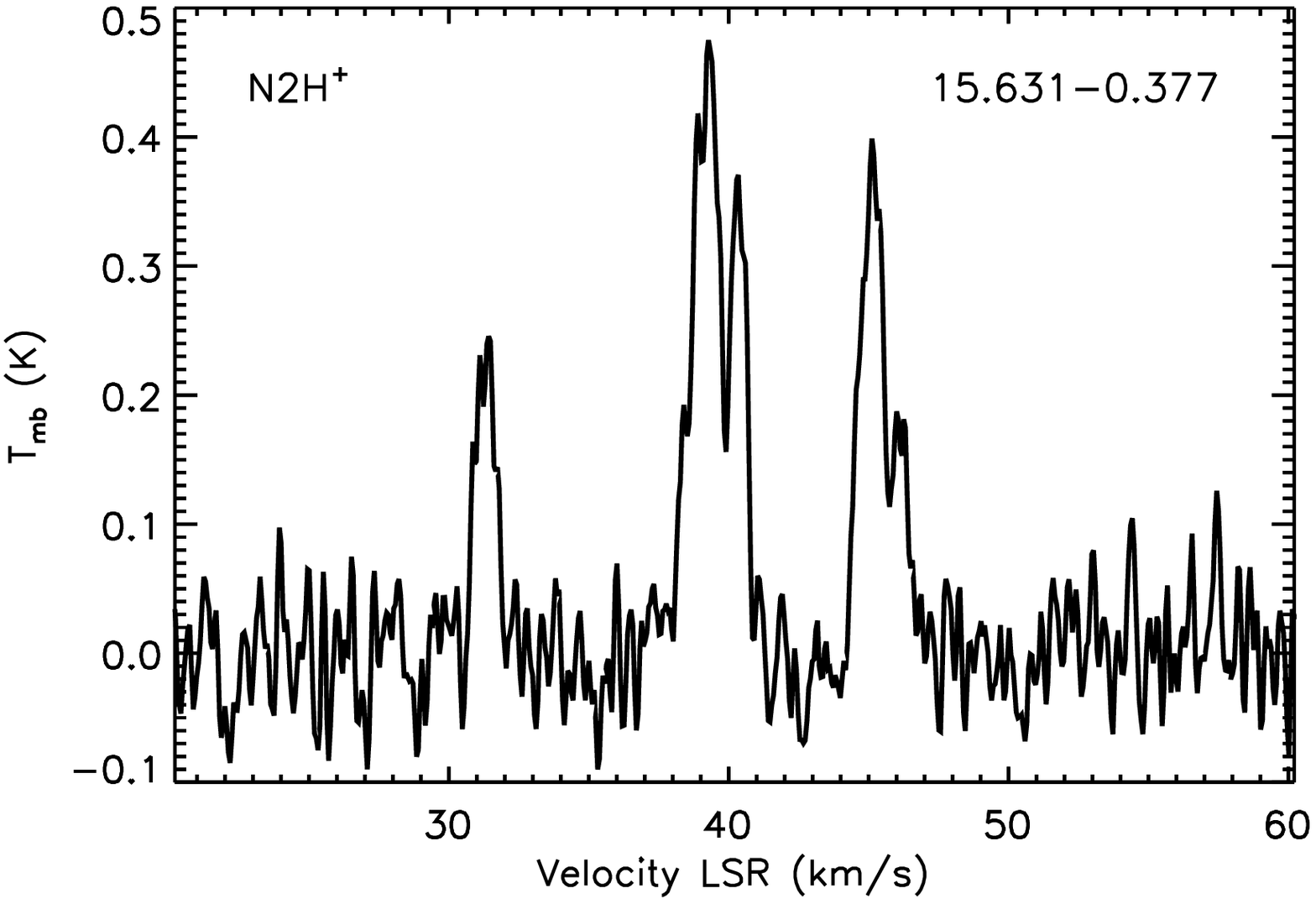}   \includegraphics[width=8cm]{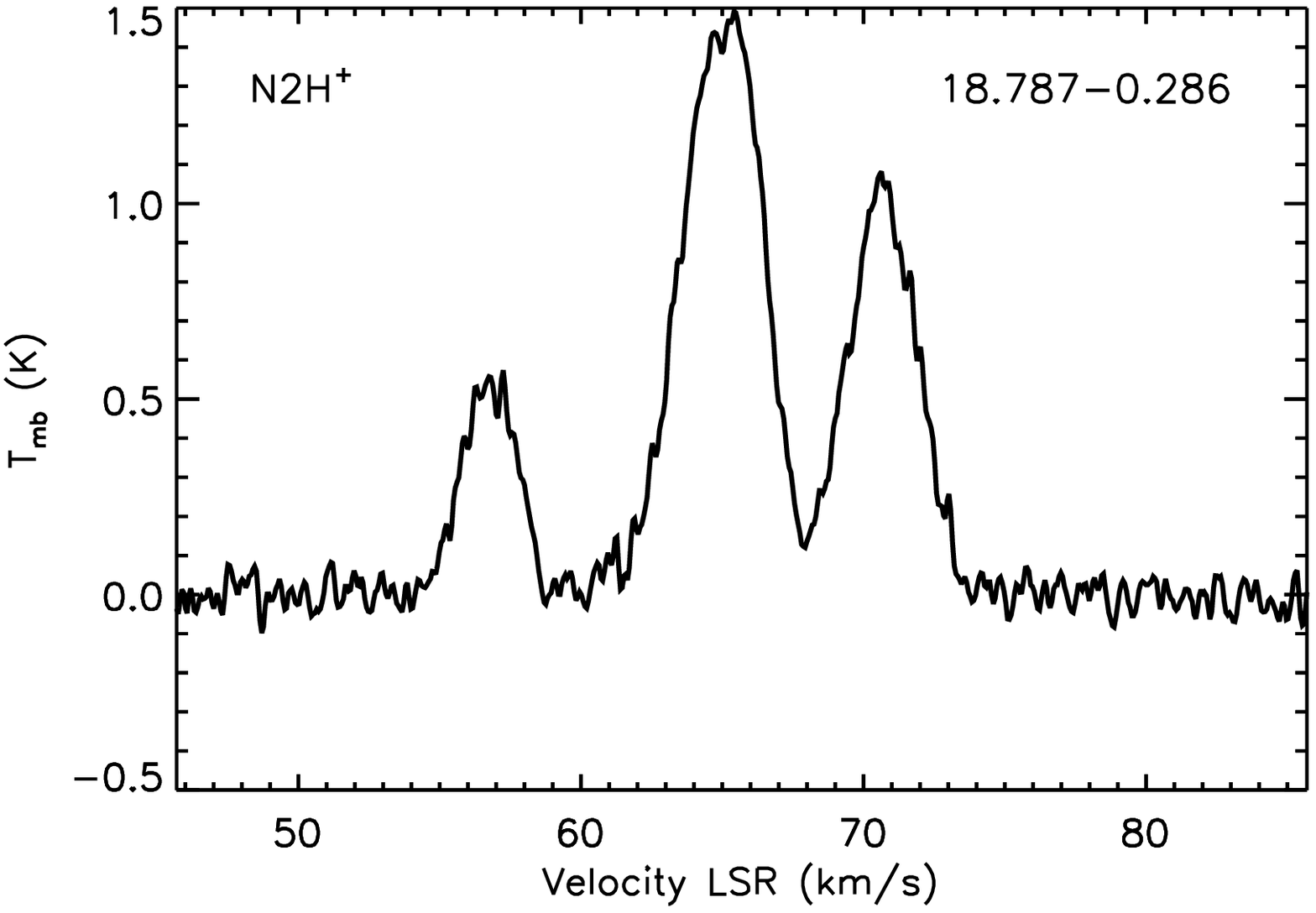}   \includegraphics[width=8cm]{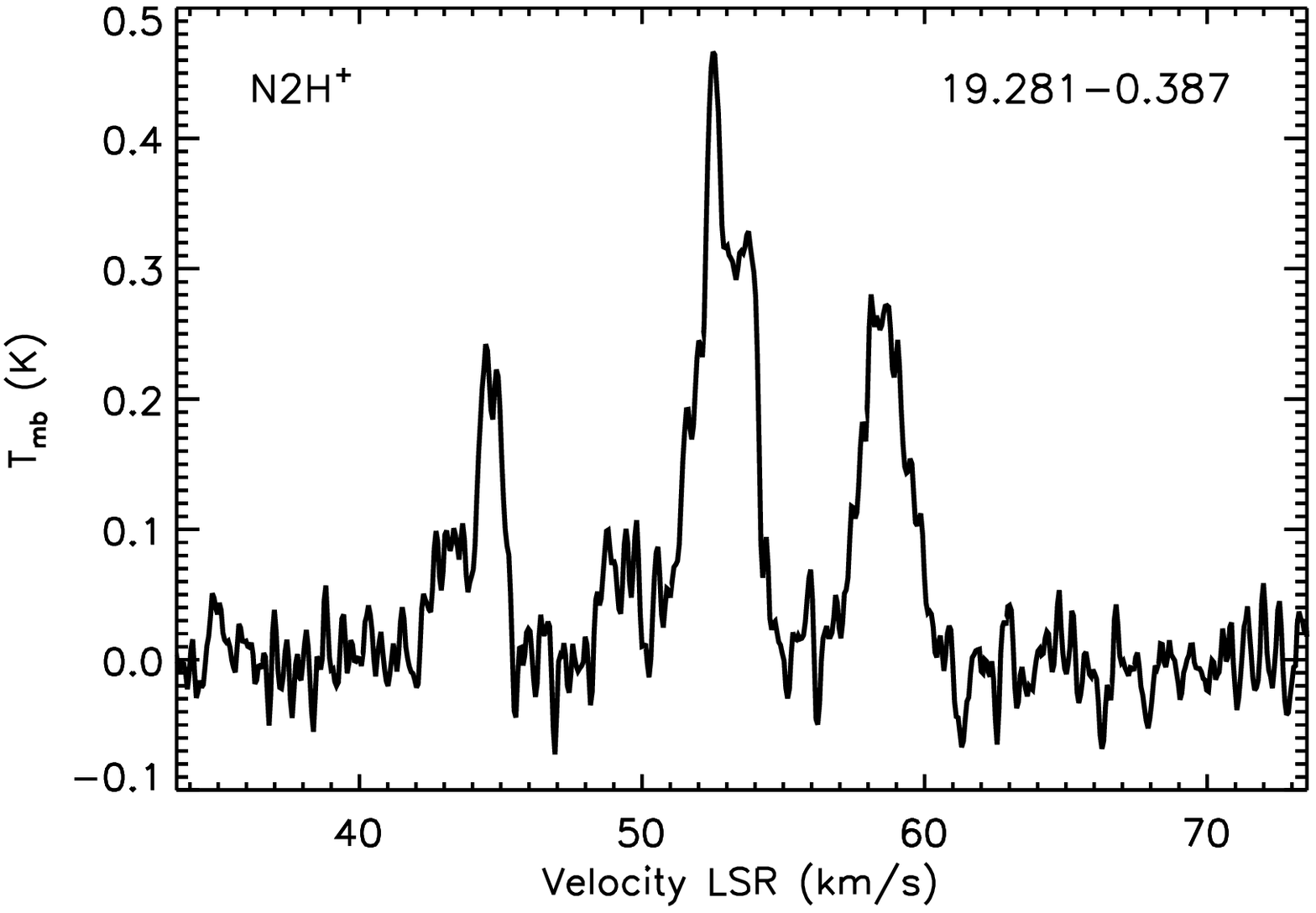}    \includegraphics[width=8cm]{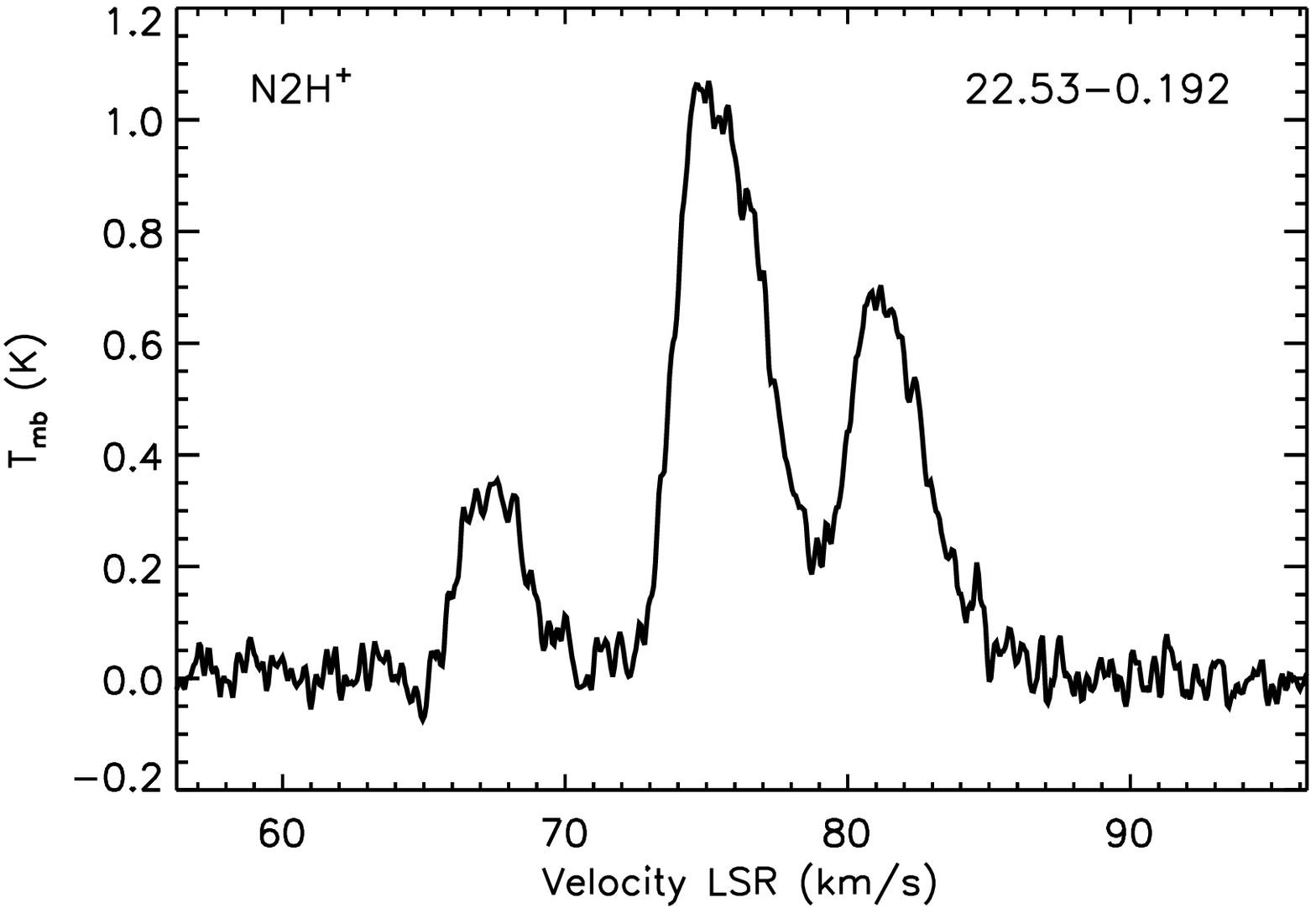} 
 \includegraphics[width=8cm]{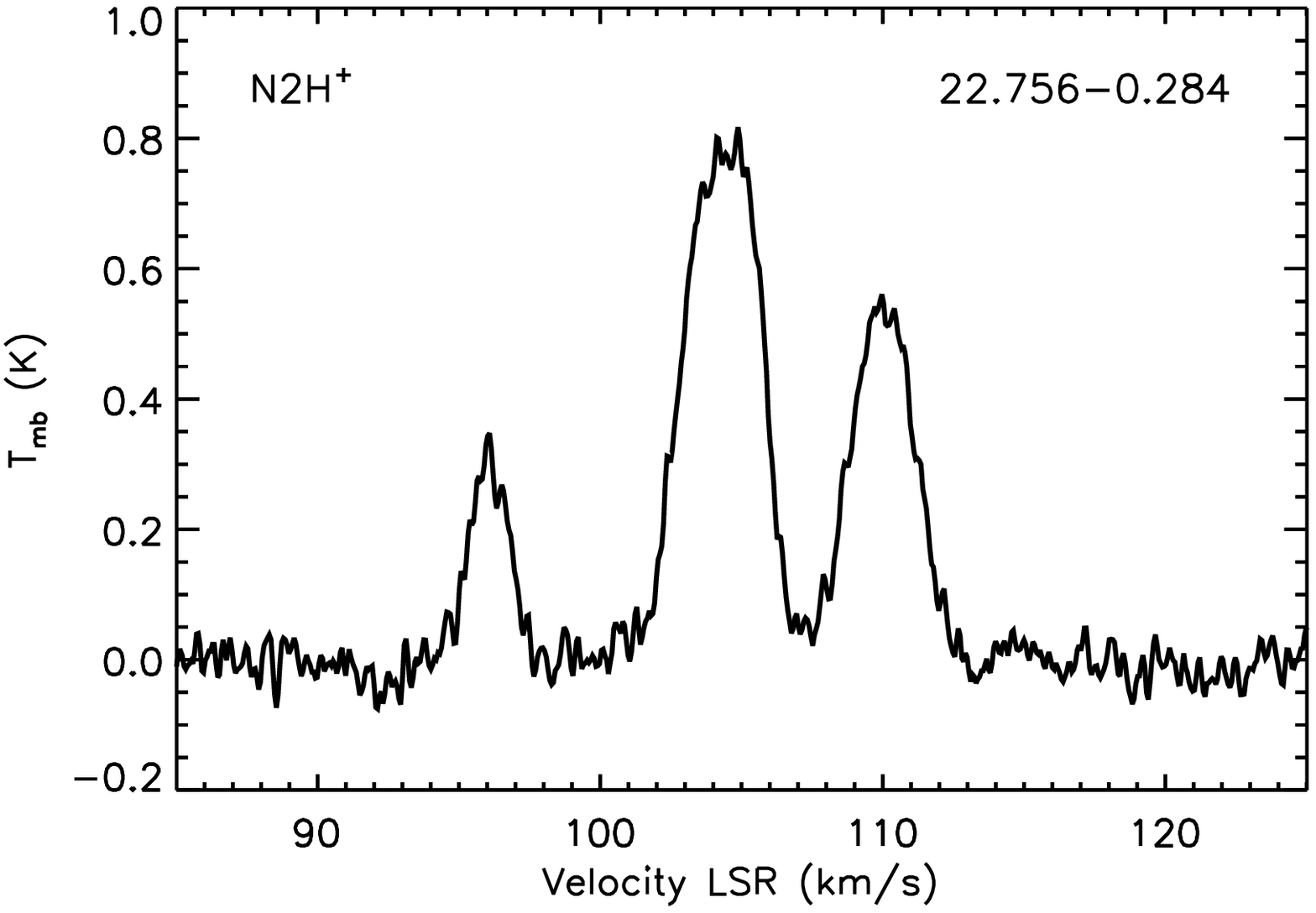}   \includegraphics[width=8cm]{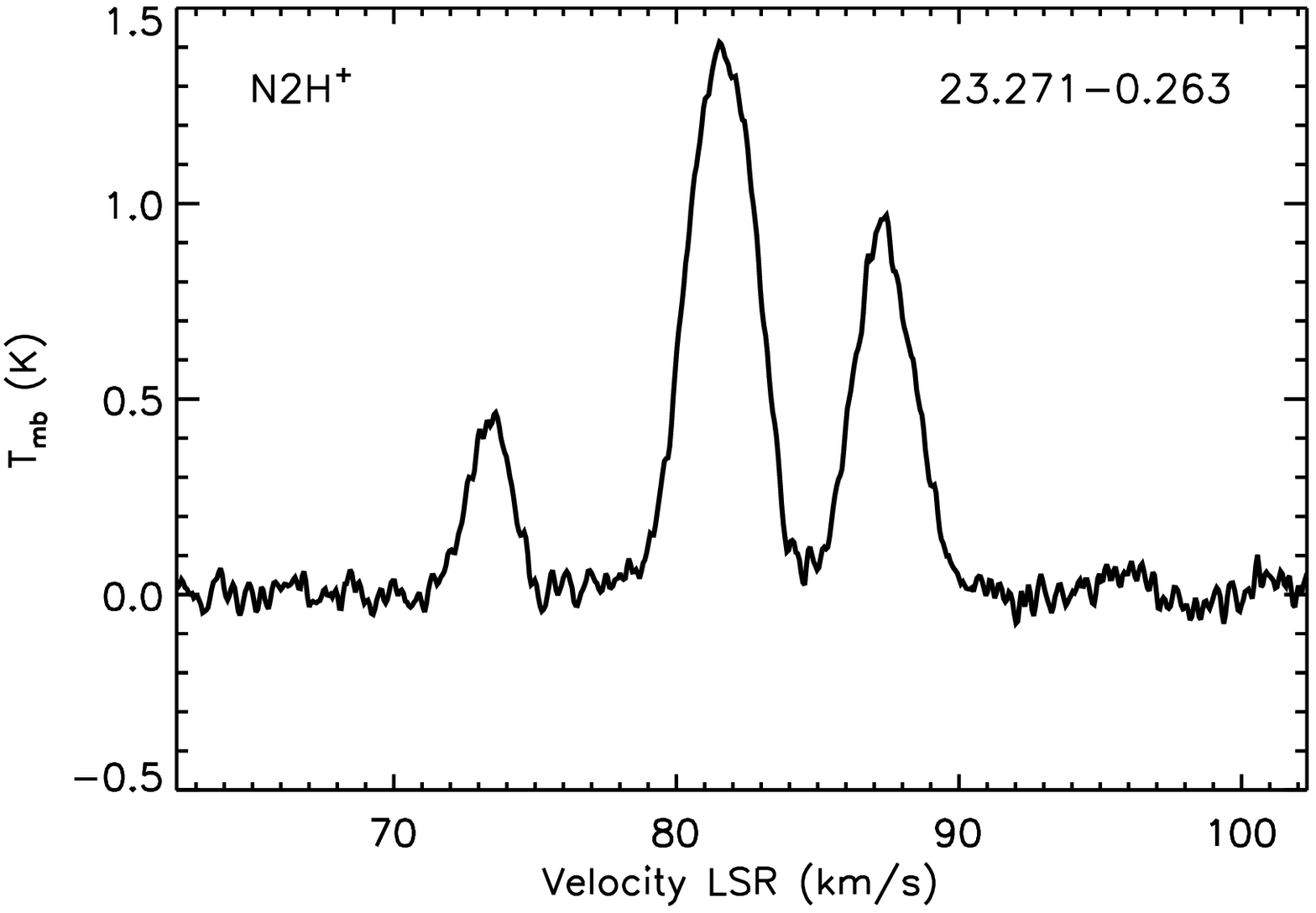}   \includegraphics[width=8cm]{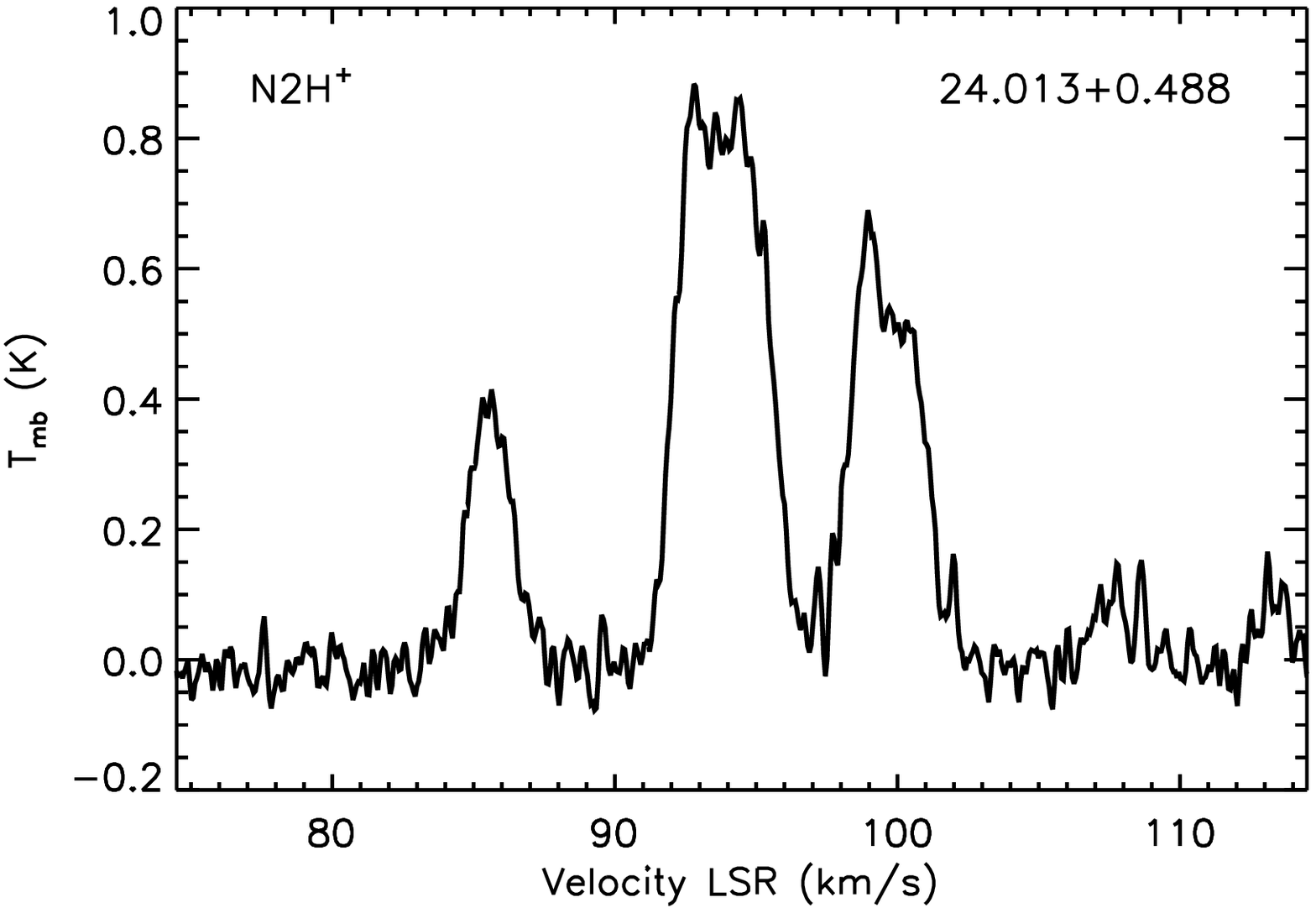}    \includegraphics[width=8cm]{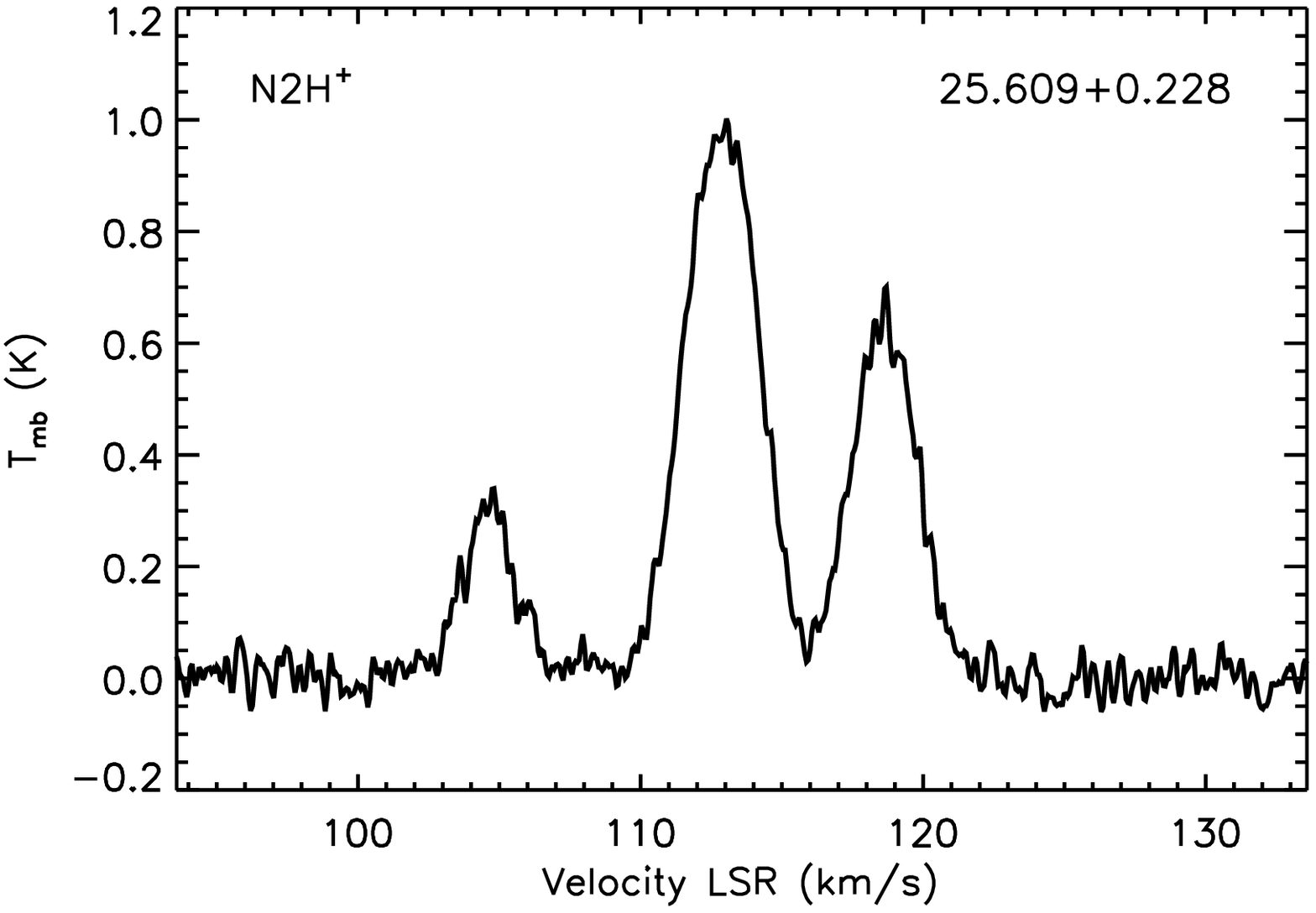}  
\caption{\n2h\ ($1-0$) spectra}
 \end{figure*}

\begin{figure*}
 \centering
 \includegraphics[width=8cm]{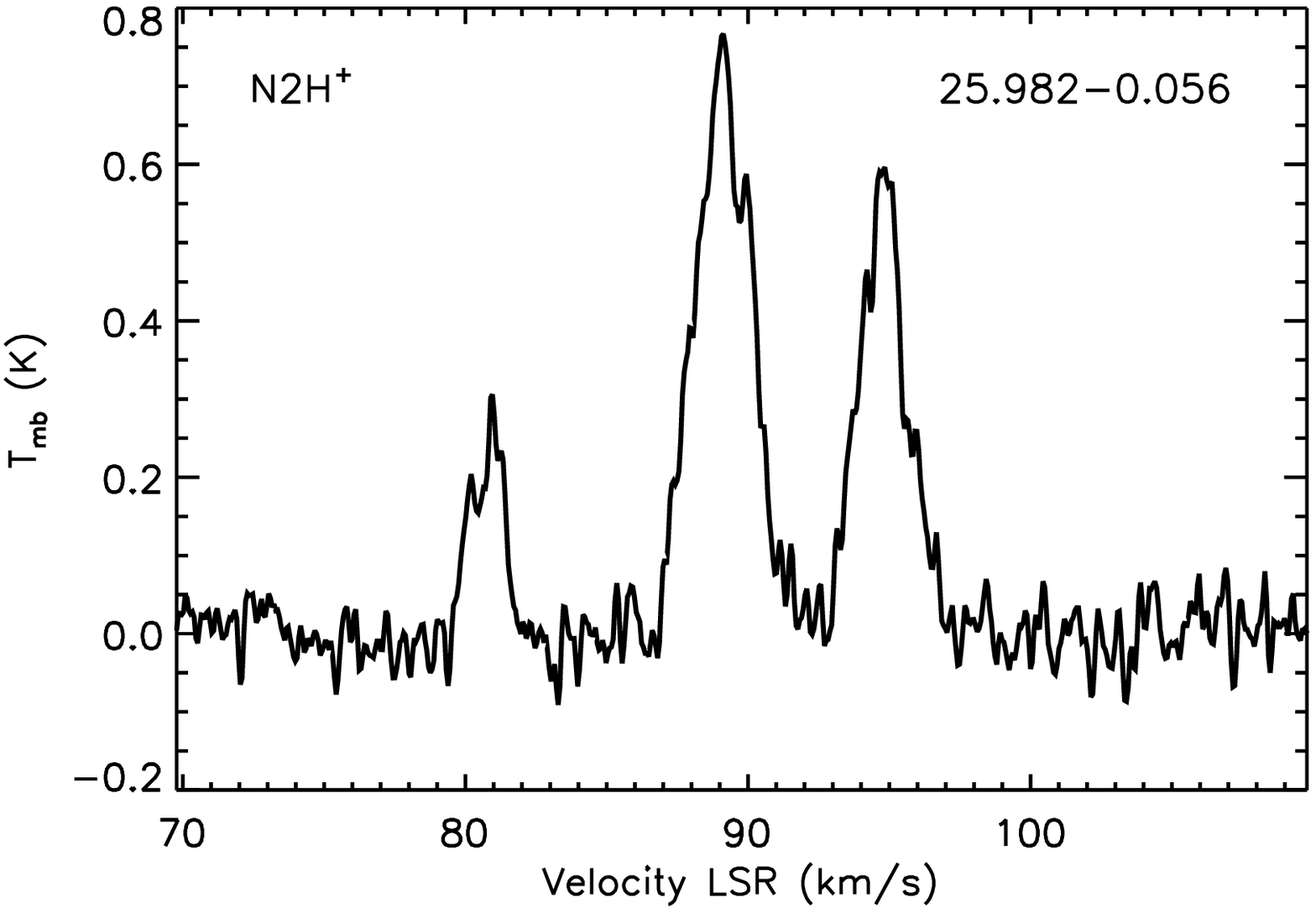}   \includegraphics[width=8cm]{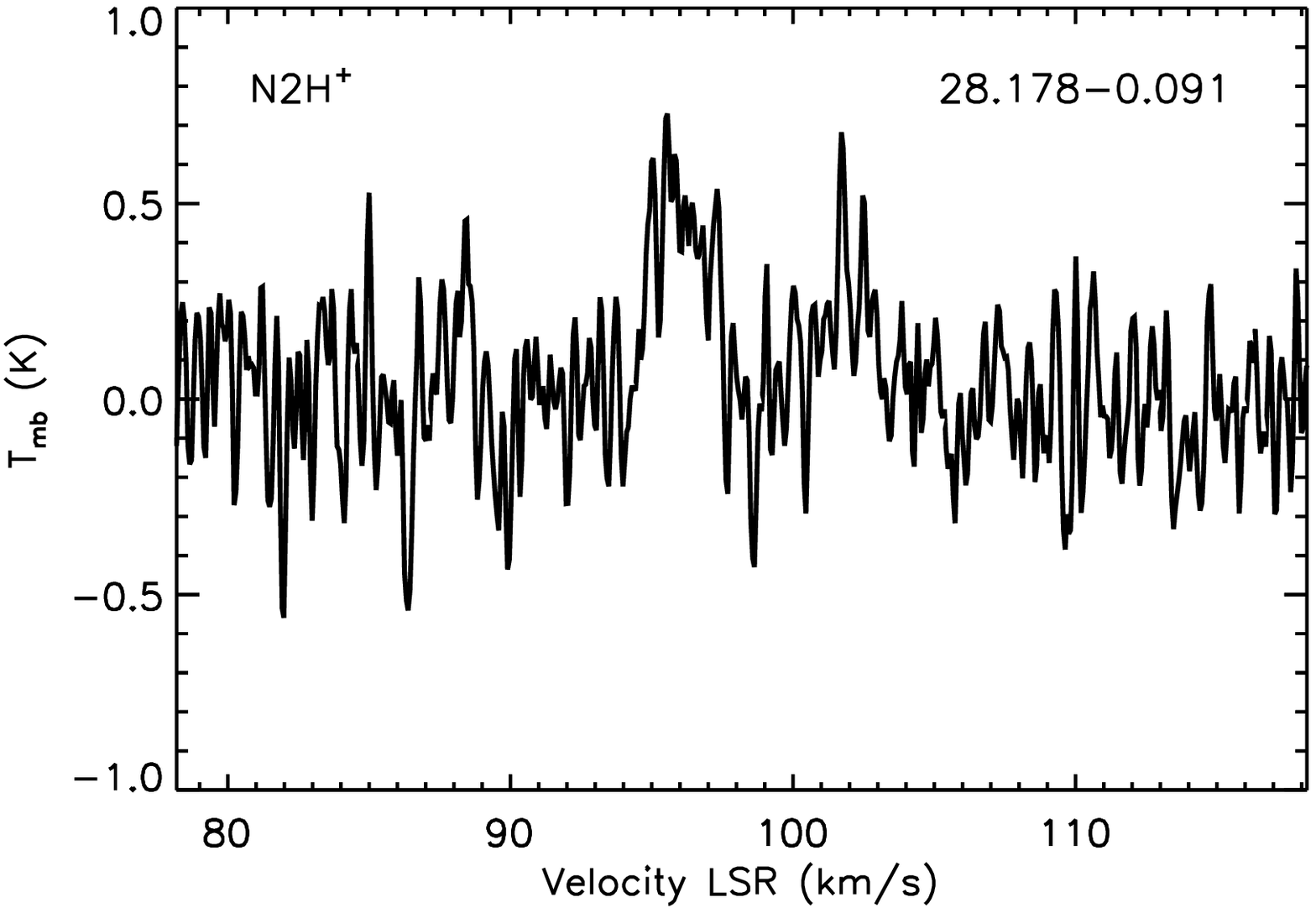}   \includegraphics[width=8cm]{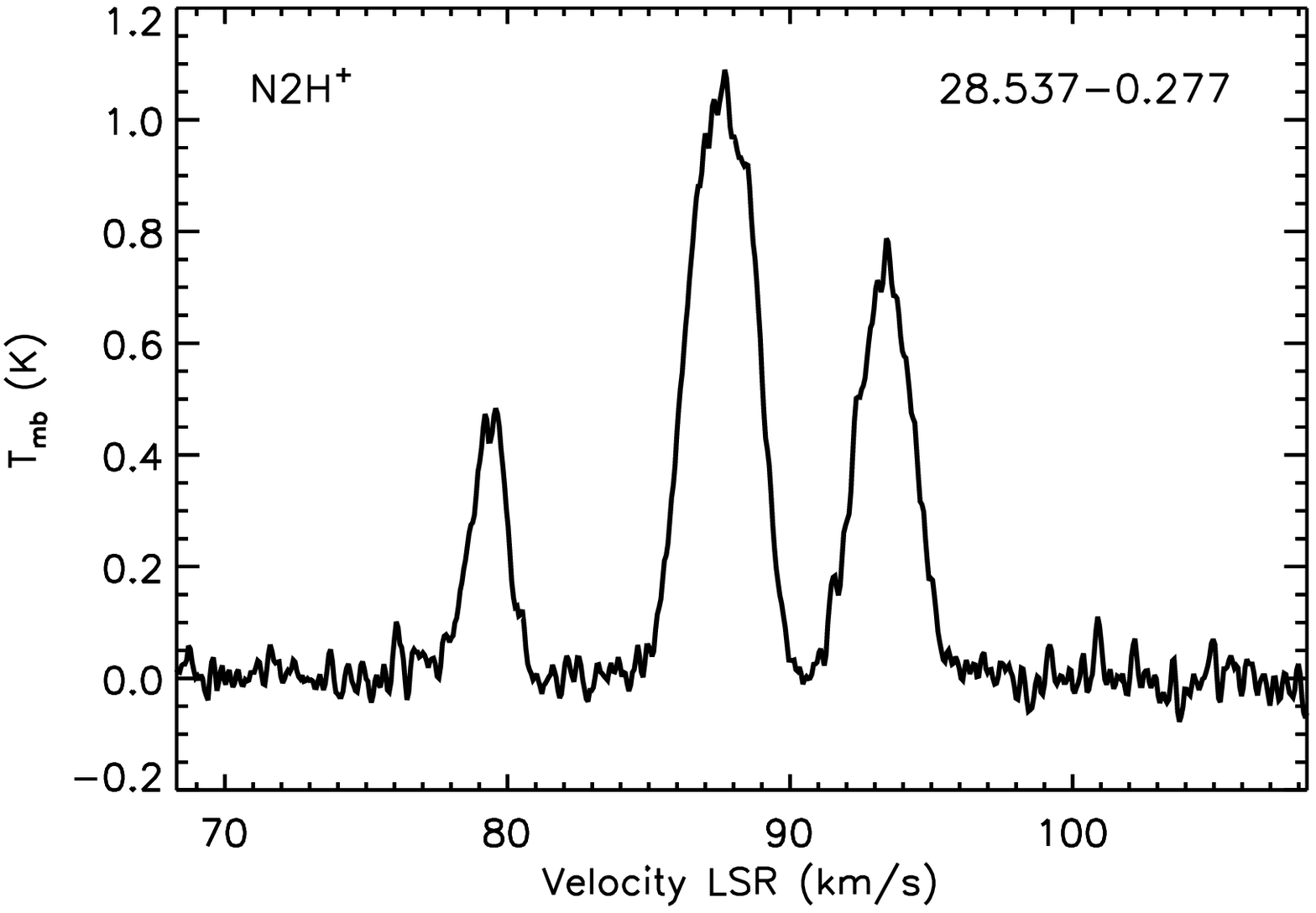}    \includegraphics[width=8cm]{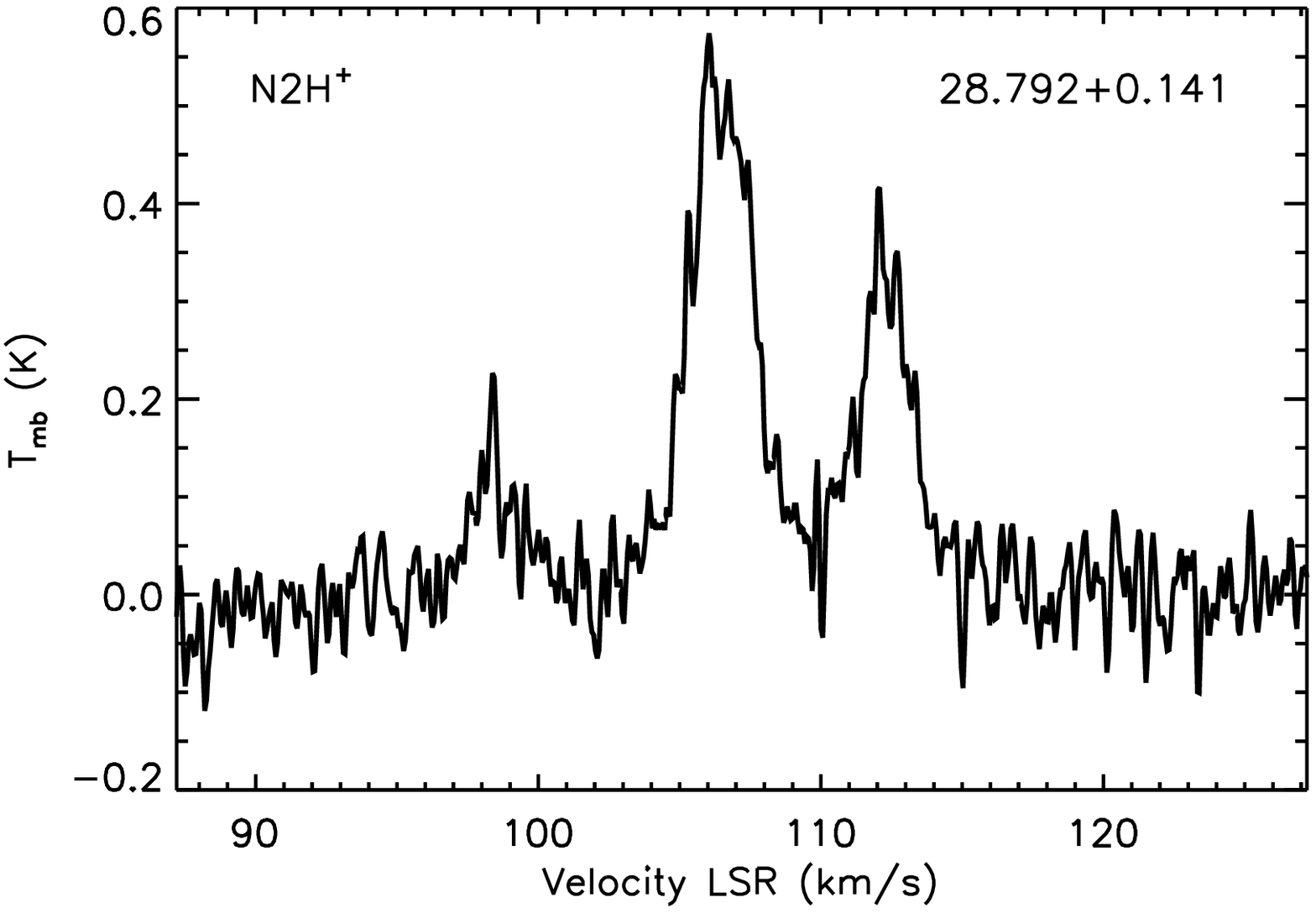} 
  \includegraphics[width=8cm]{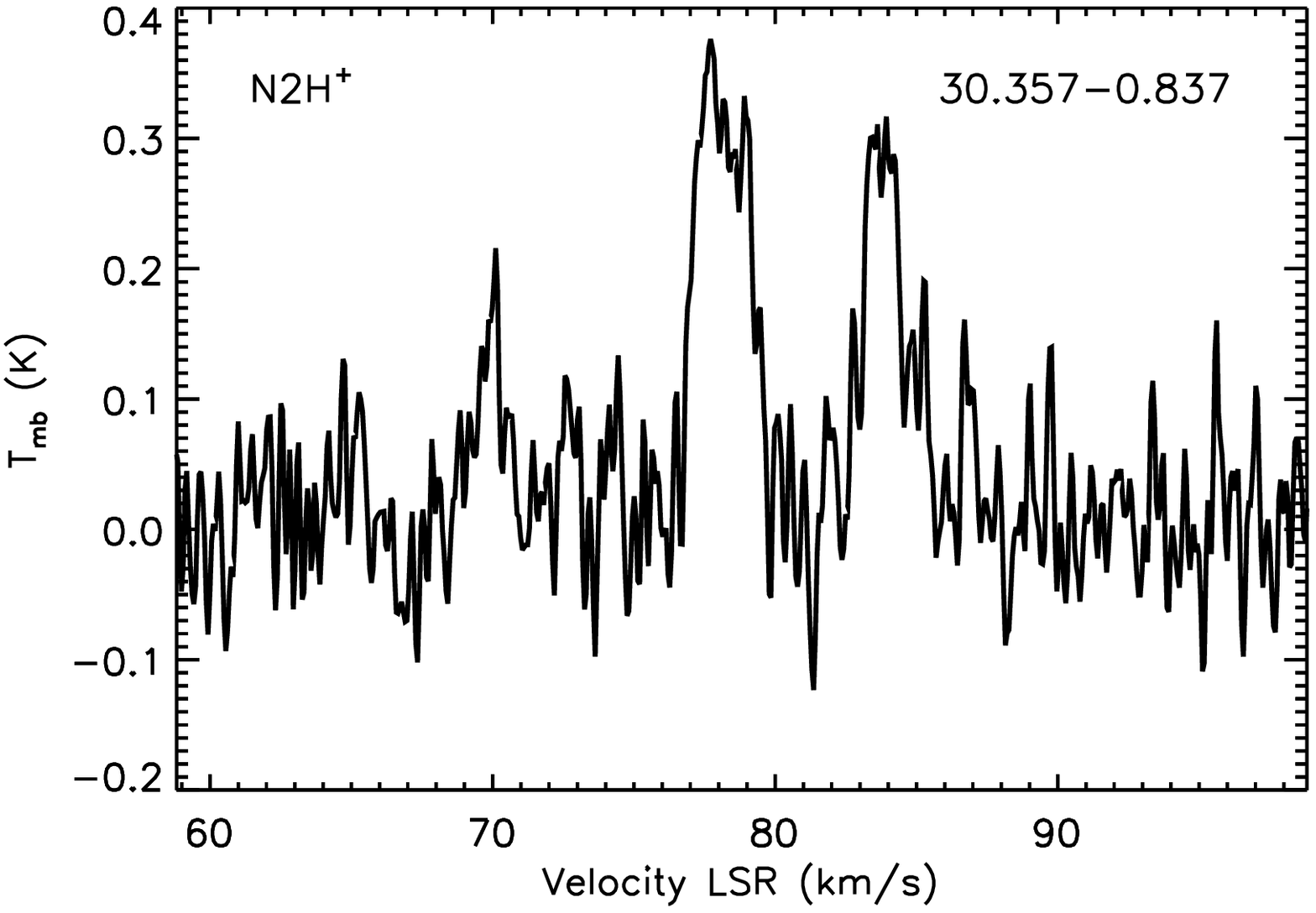}   \includegraphics[width=8cm]{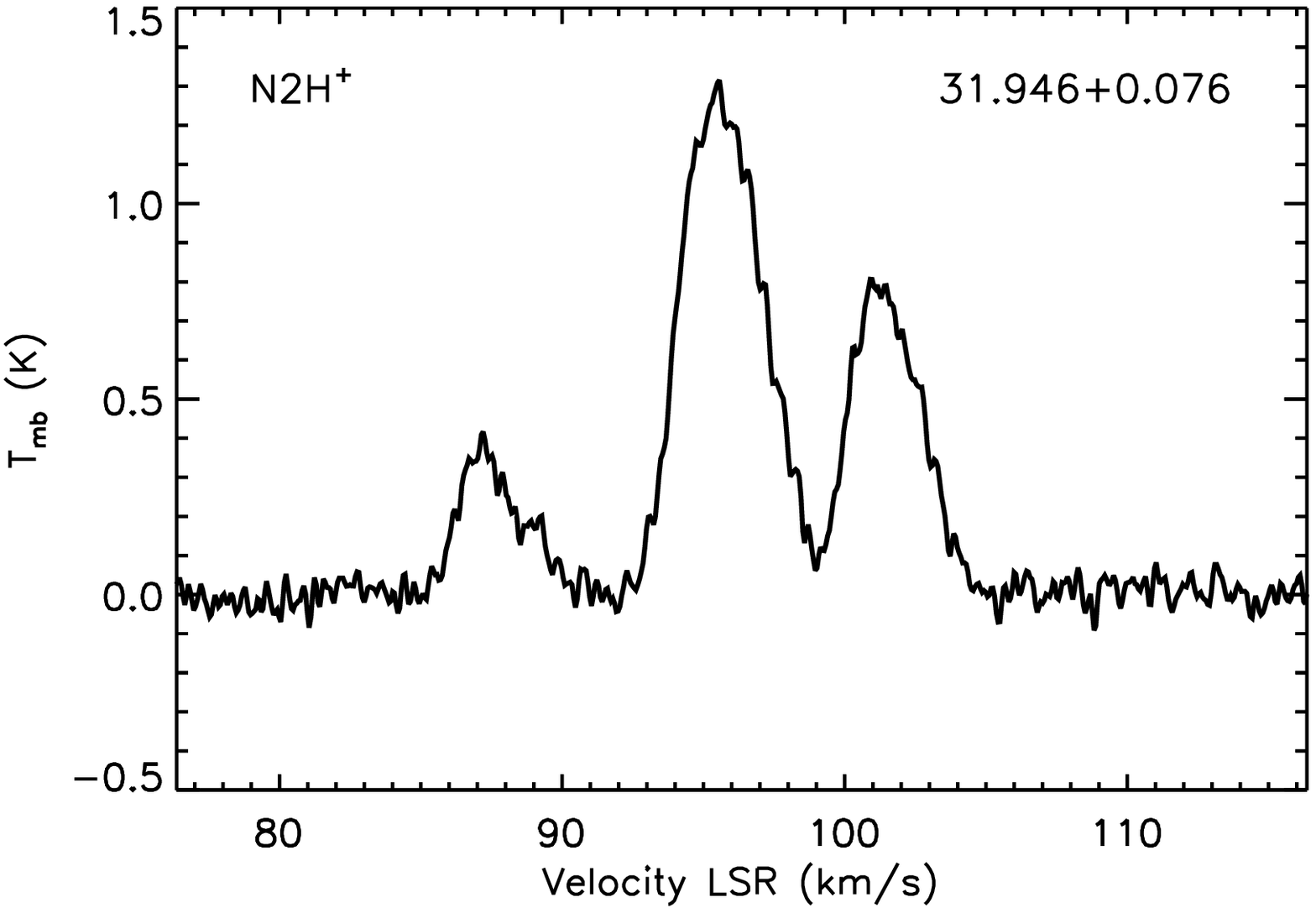}   \includegraphics[width=8cm]{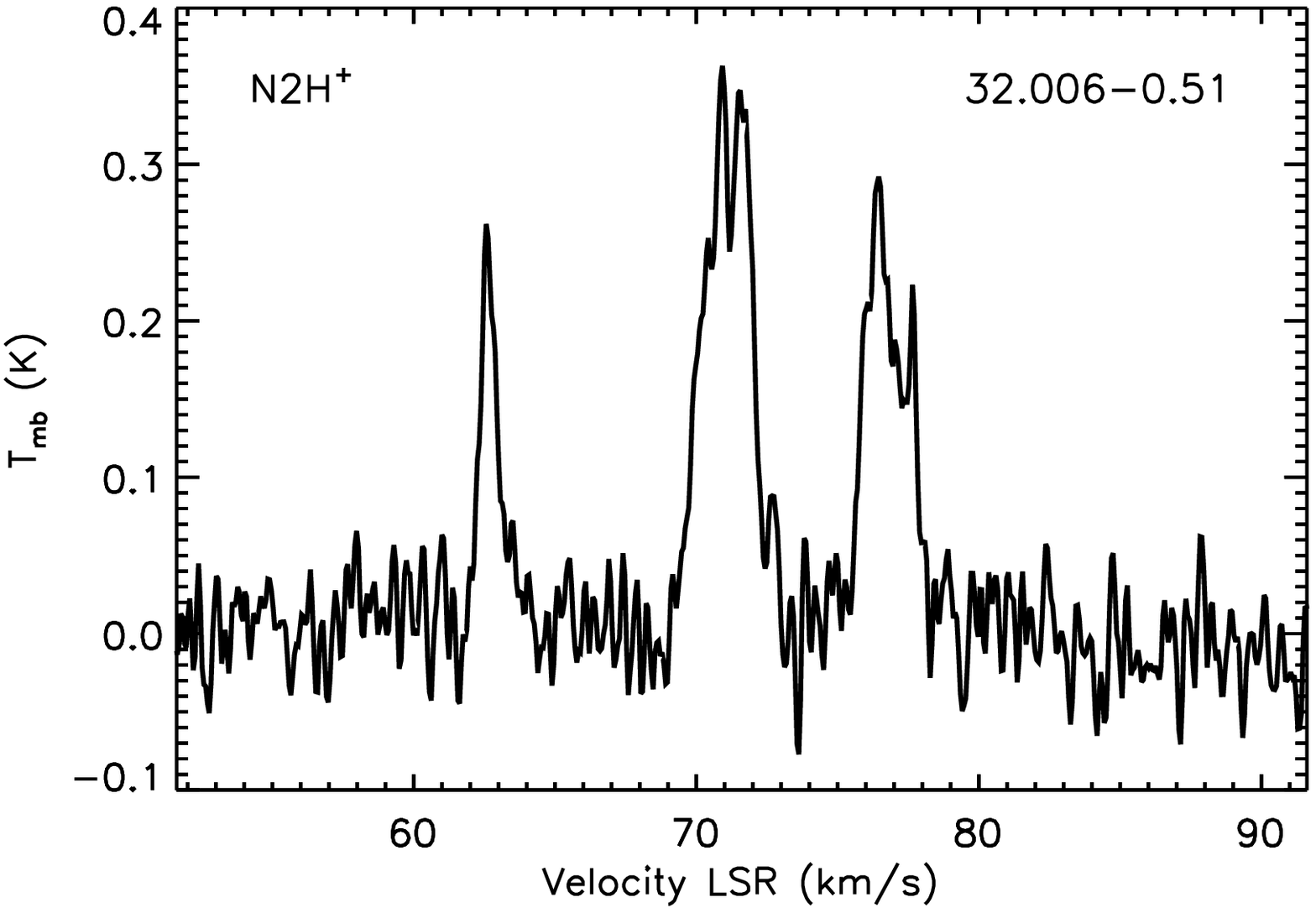}    \includegraphics[width=8cm]{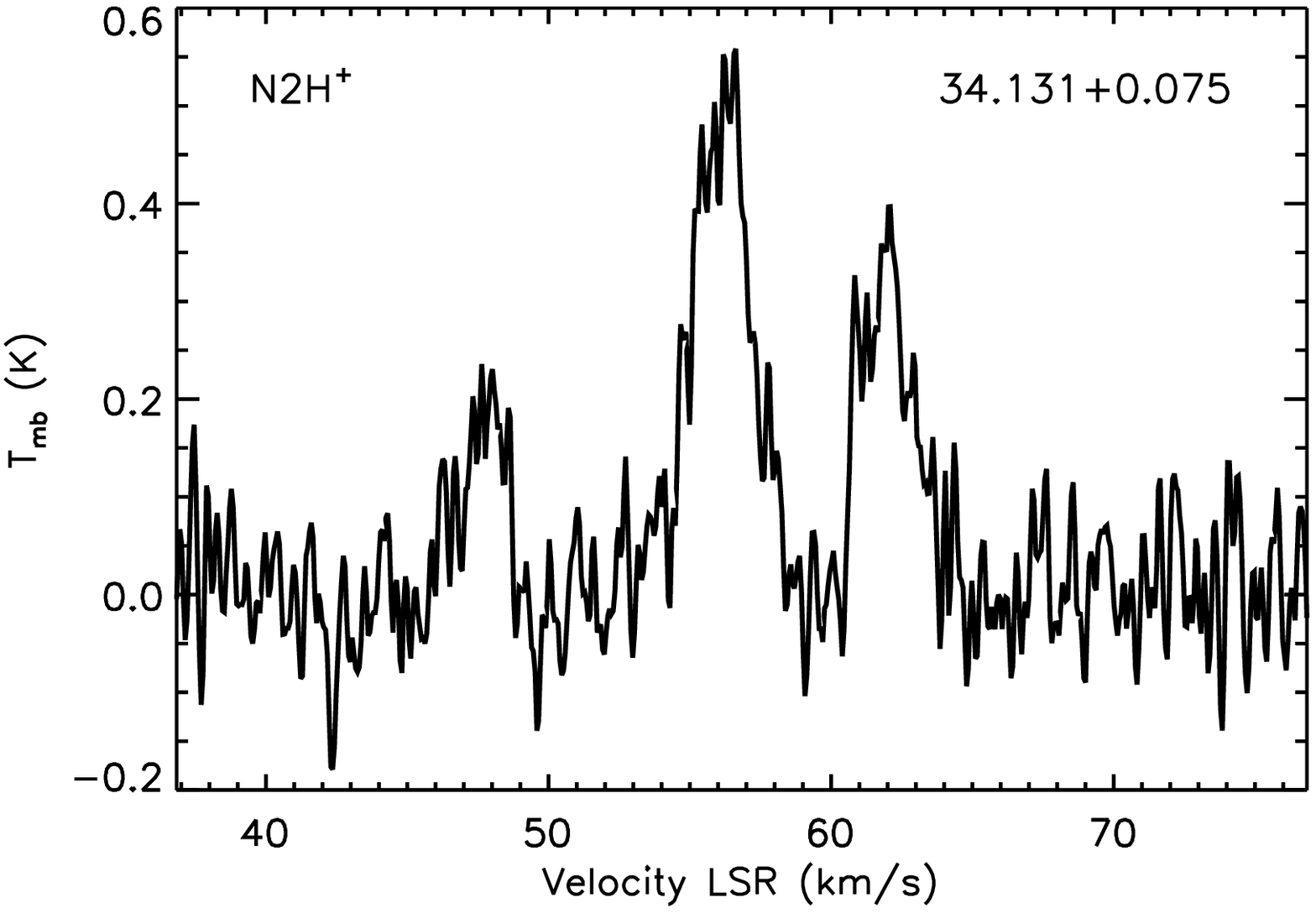}  \caption{\n2h\ ($1-0$) spectra continues}
 \end{figure*}

\begin{figure*}
 \centering
\includegraphics[width=8cm]{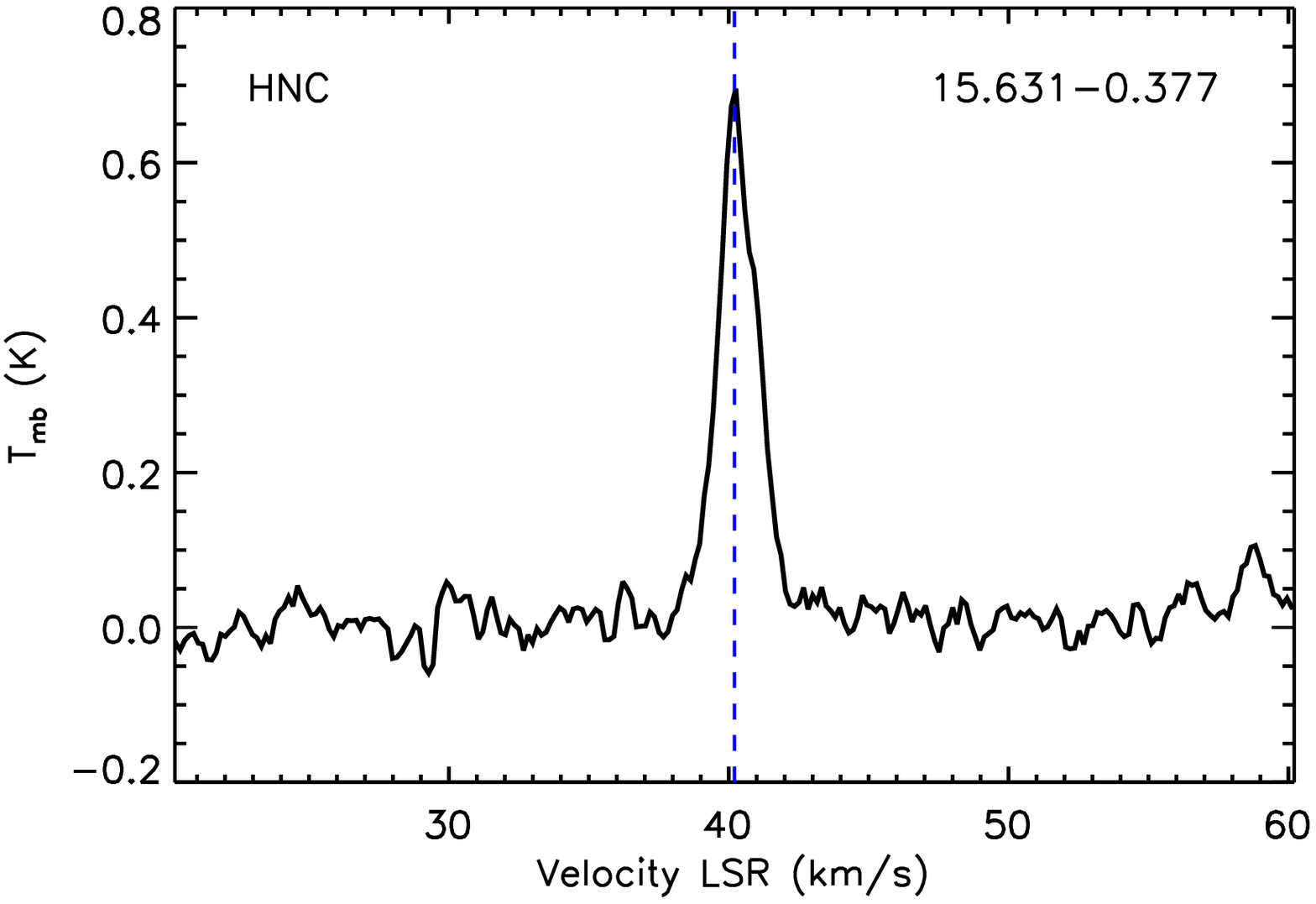} 
\includegraphics[width=8cm]{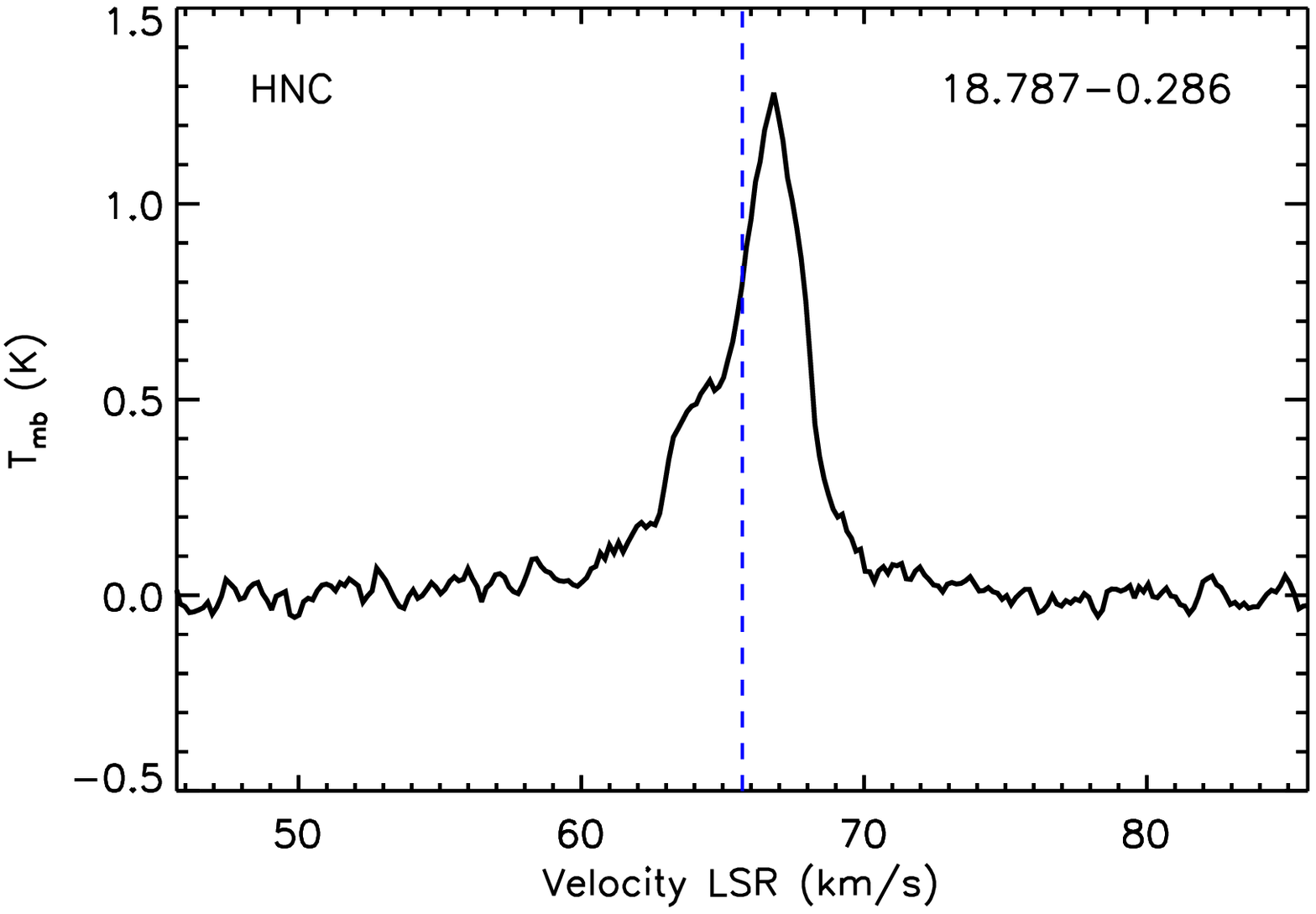} 
\includegraphics[width=8cm]{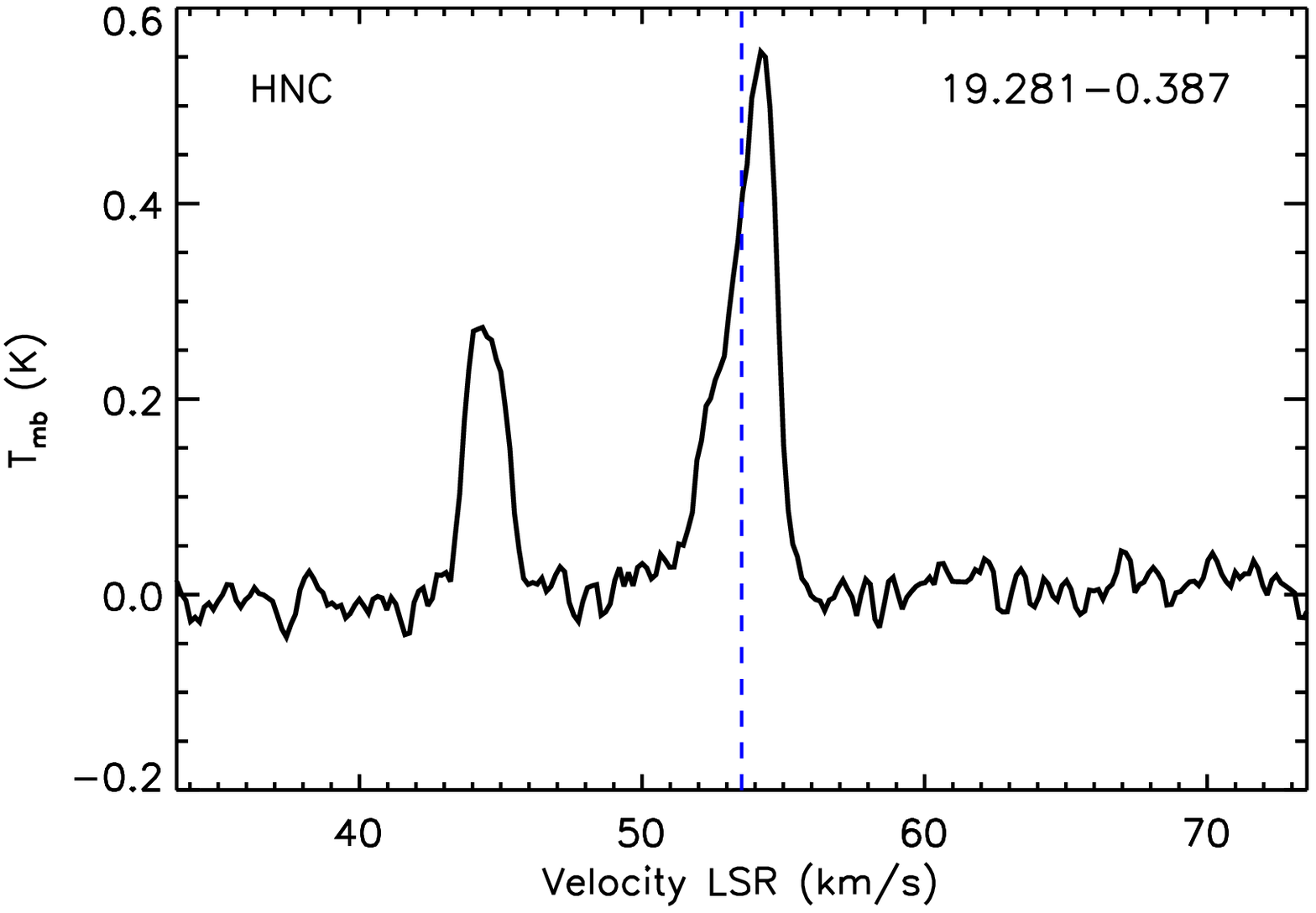} 
\includegraphics[width=8cm]{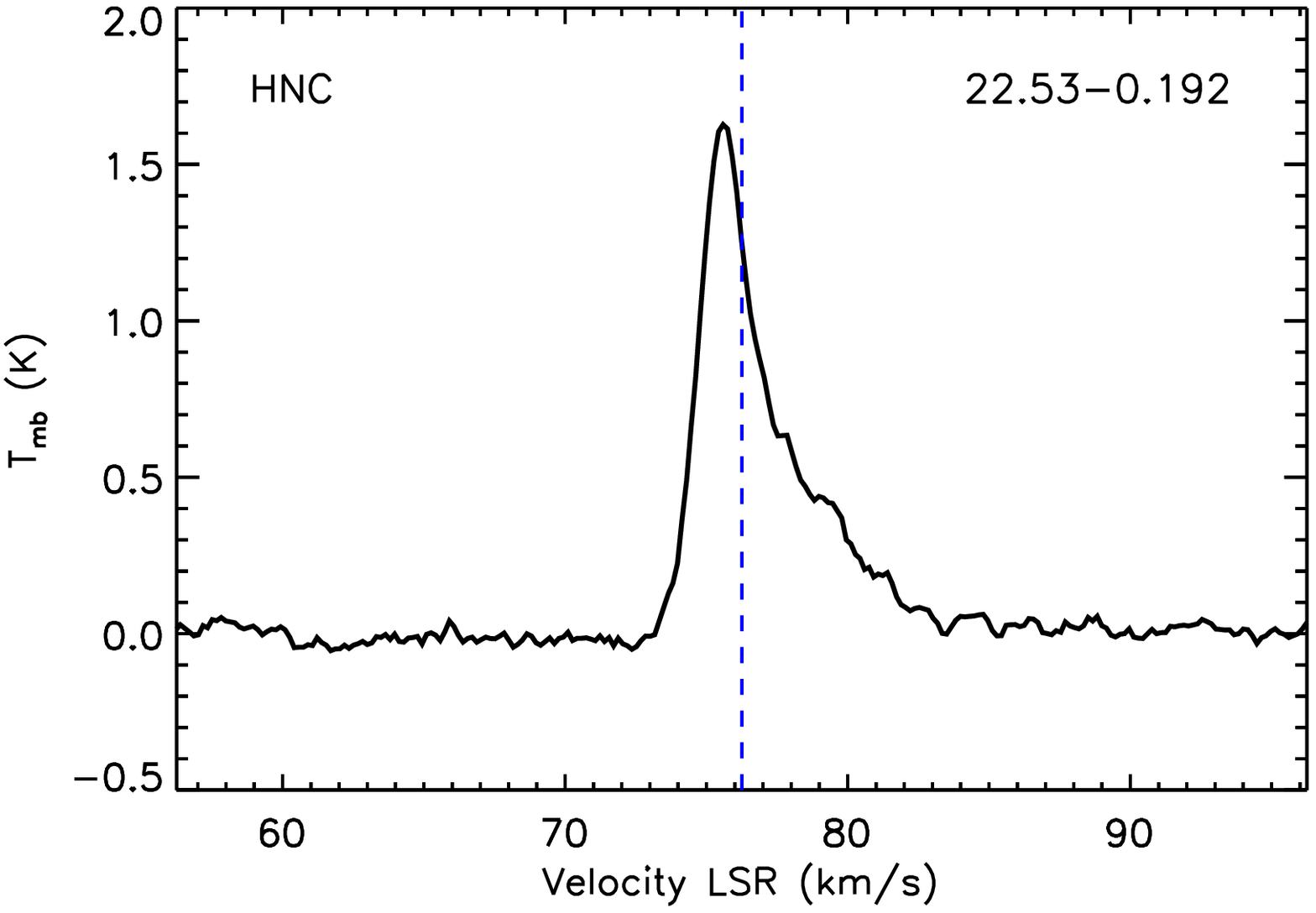} 
\includegraphics[width=8cm]{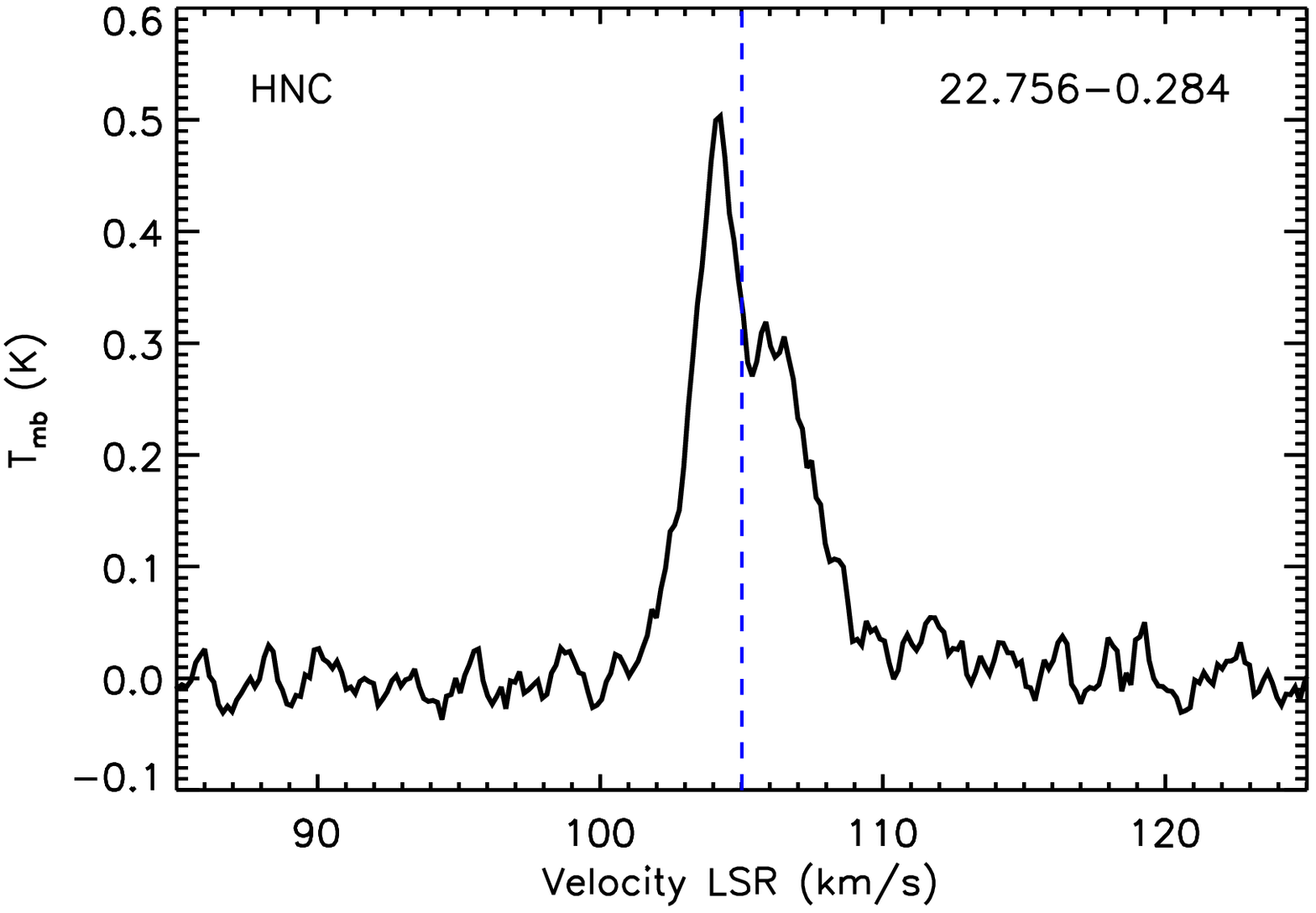} 
\includegraphics[width=8cm]{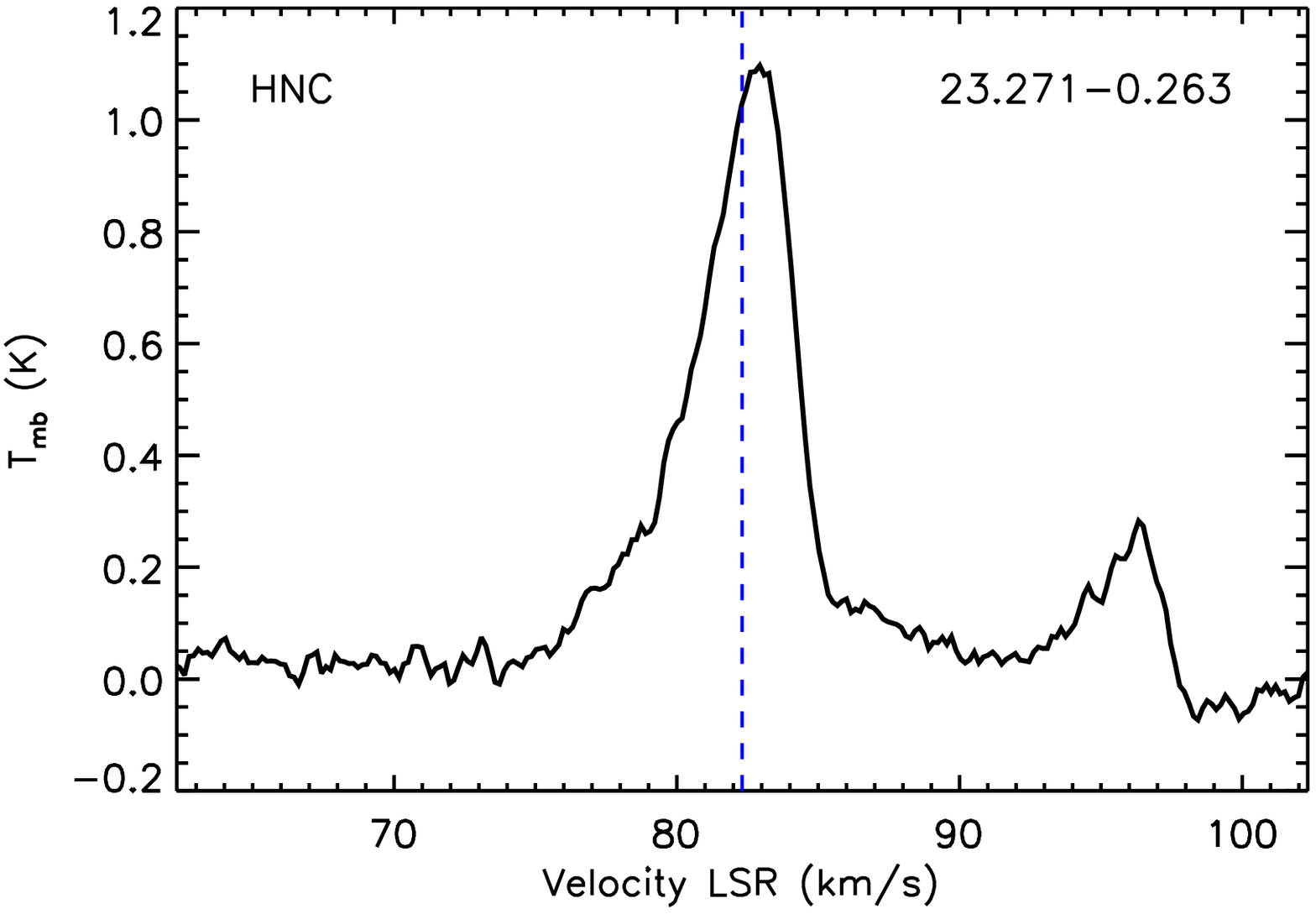} 
\includegraphics[width=8cm]{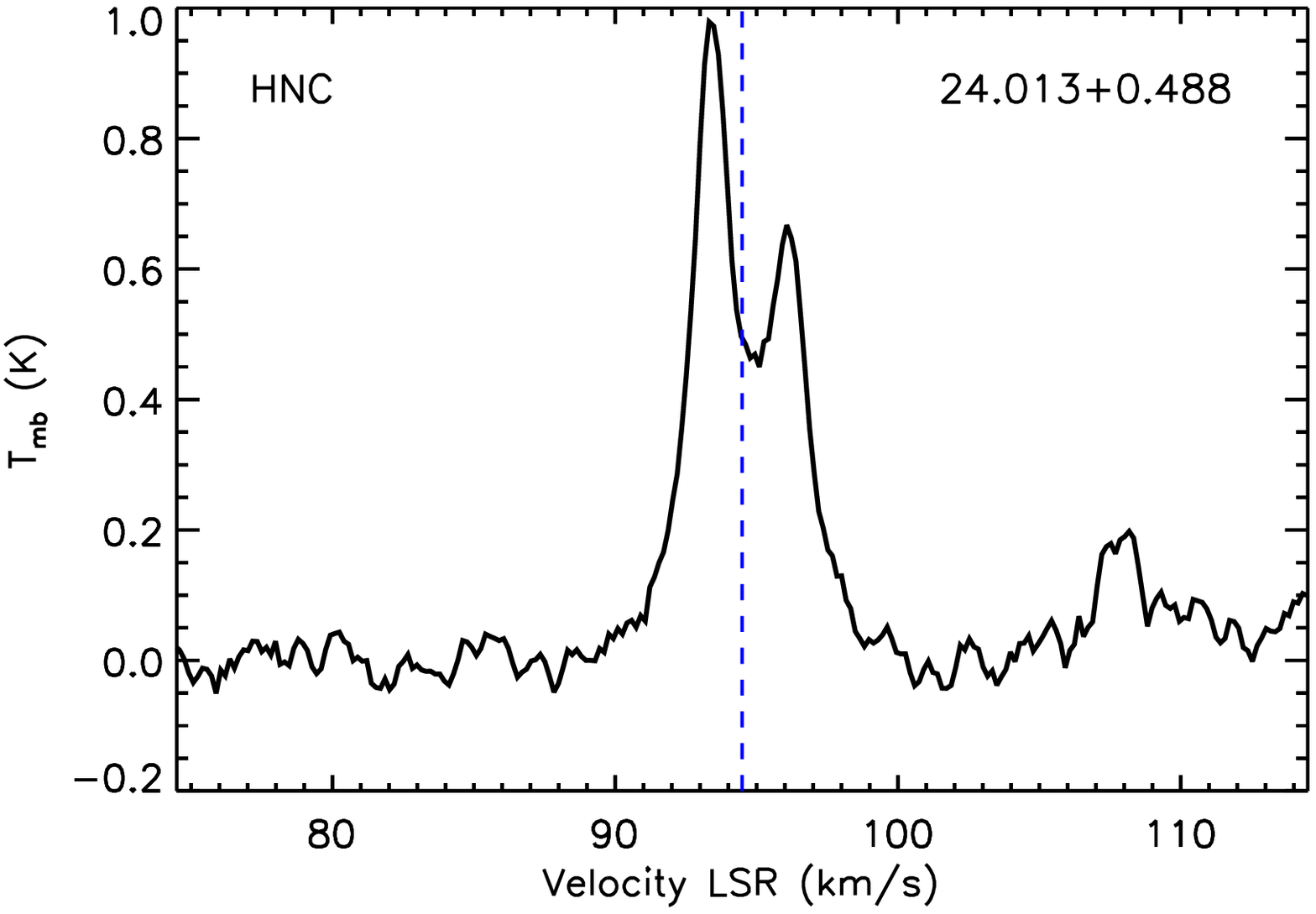} 
\includegraphics[width=8cm]{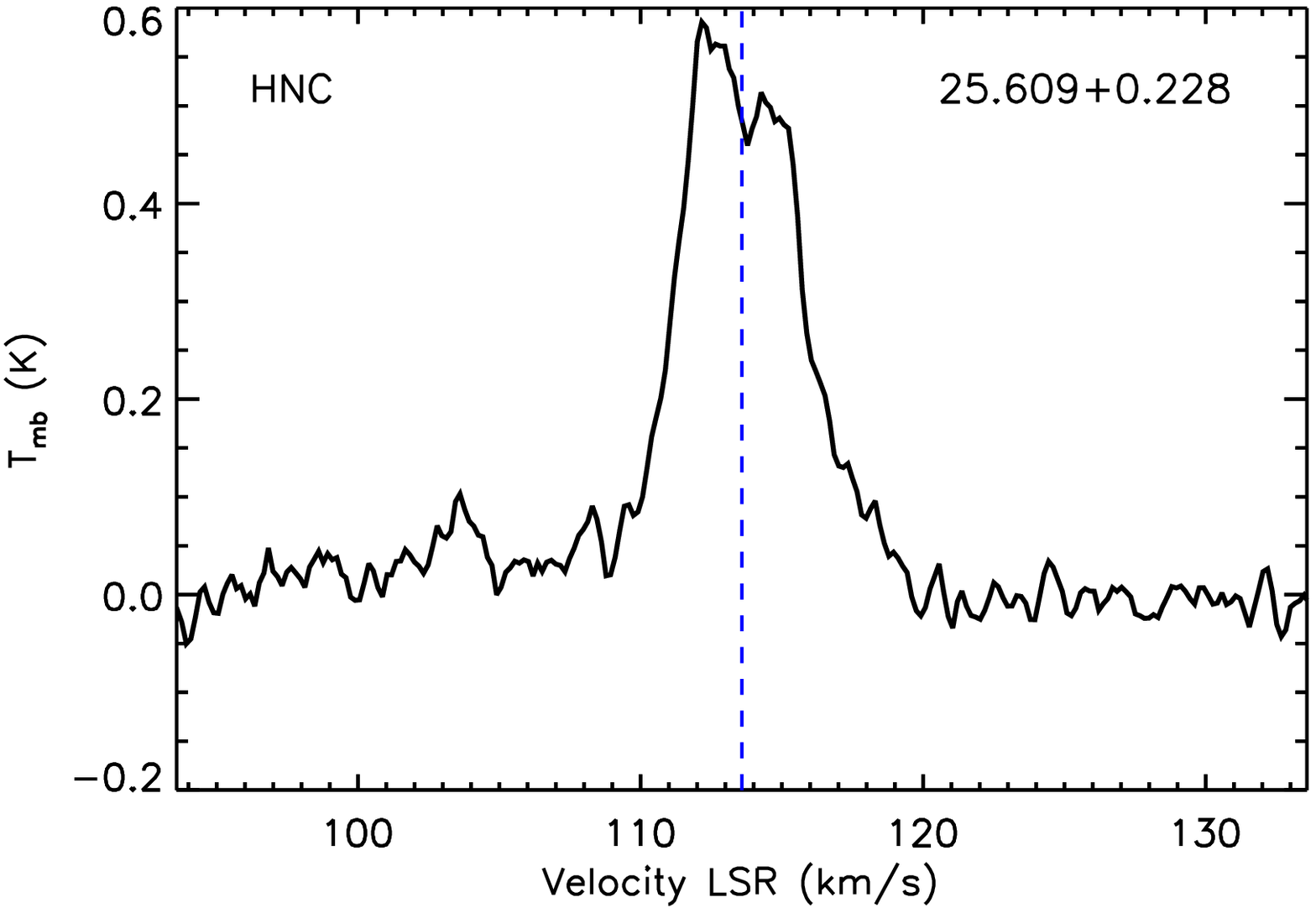} 
\caption{HNC ($1-0$) spectra}
 \end{figure*}

\begin{figure*}
 \centering
\includegraphics[width=8cm]{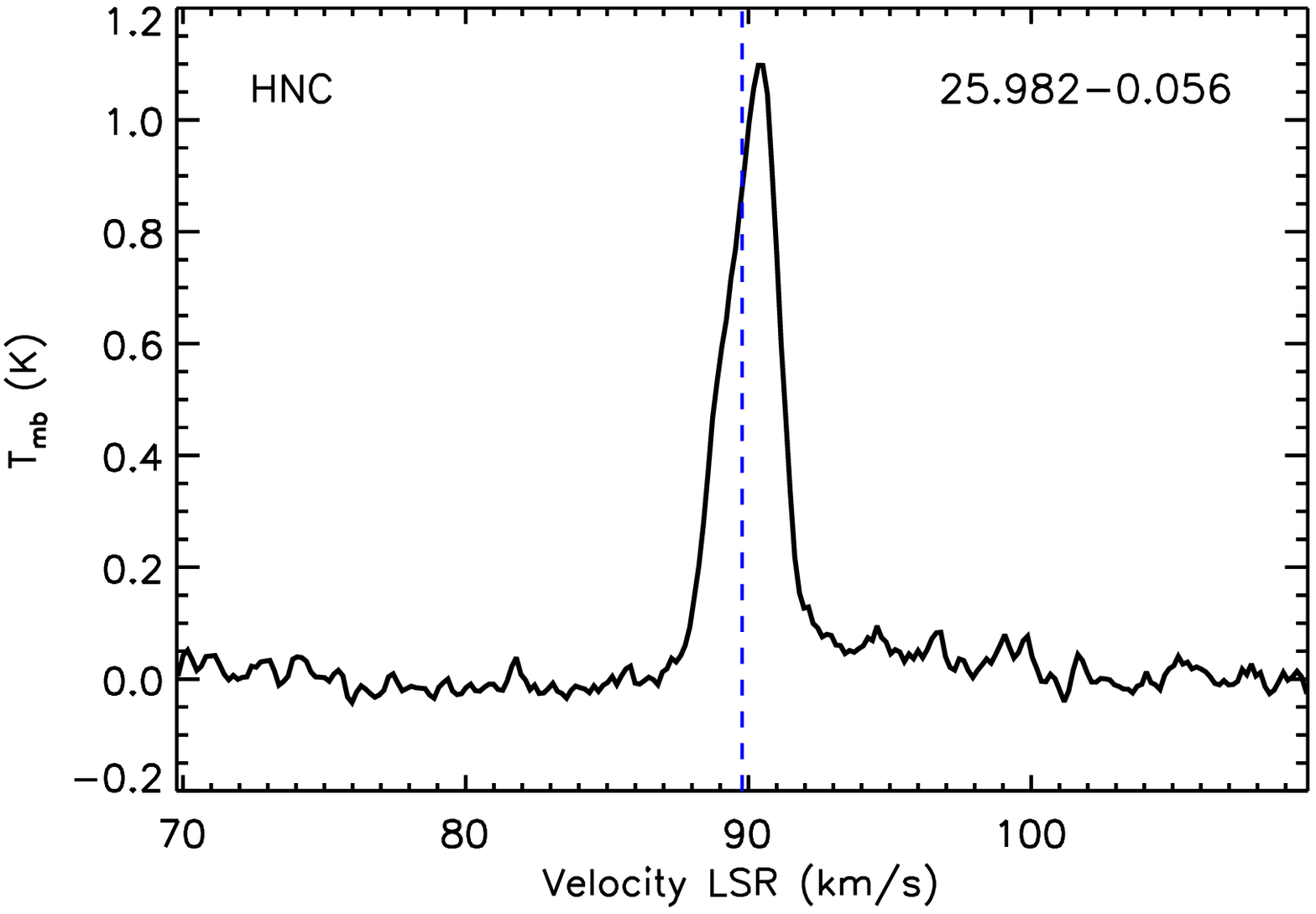} 
\includegraphics[width=8cm]{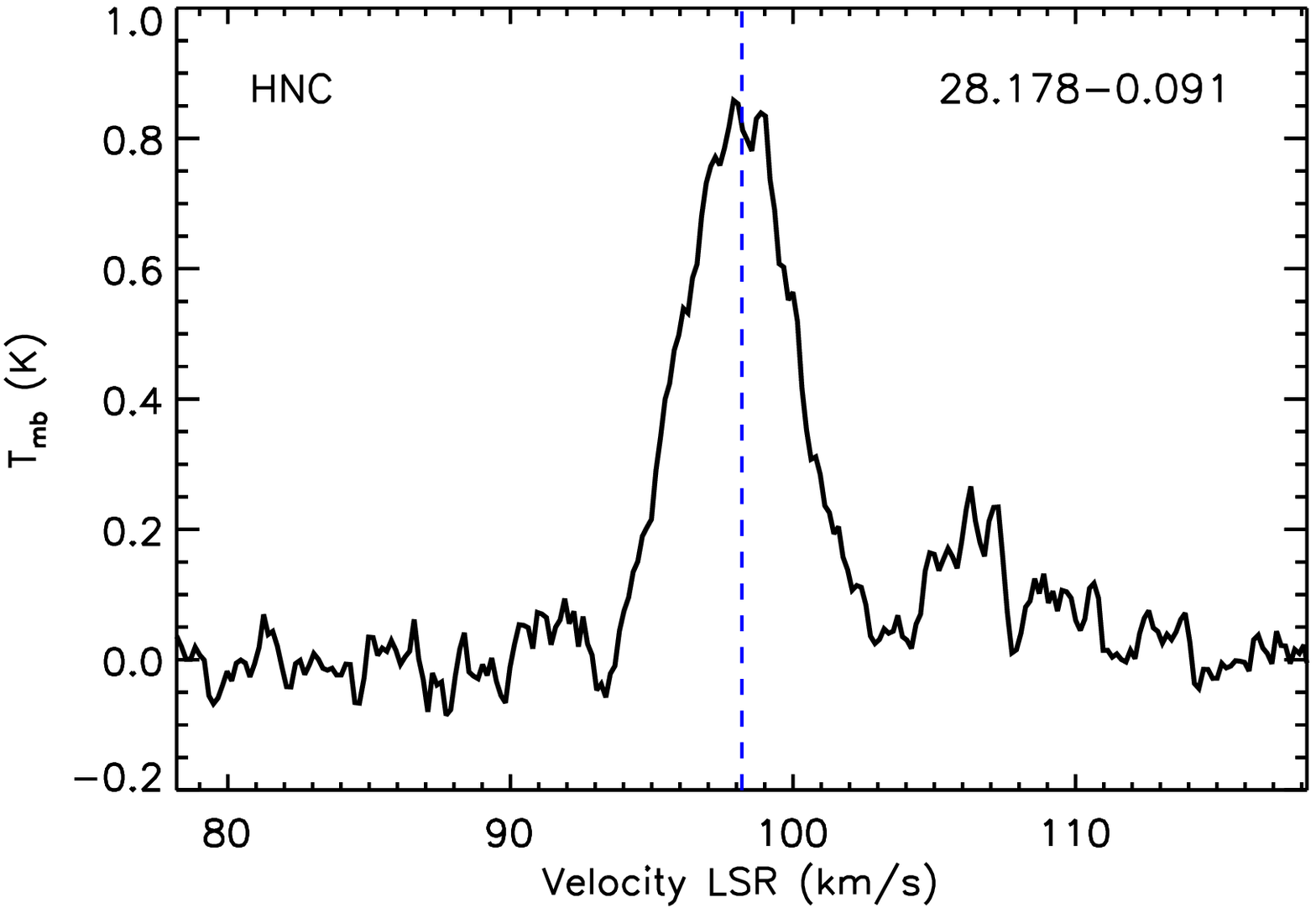} 
\includegraphics[width=8cm]{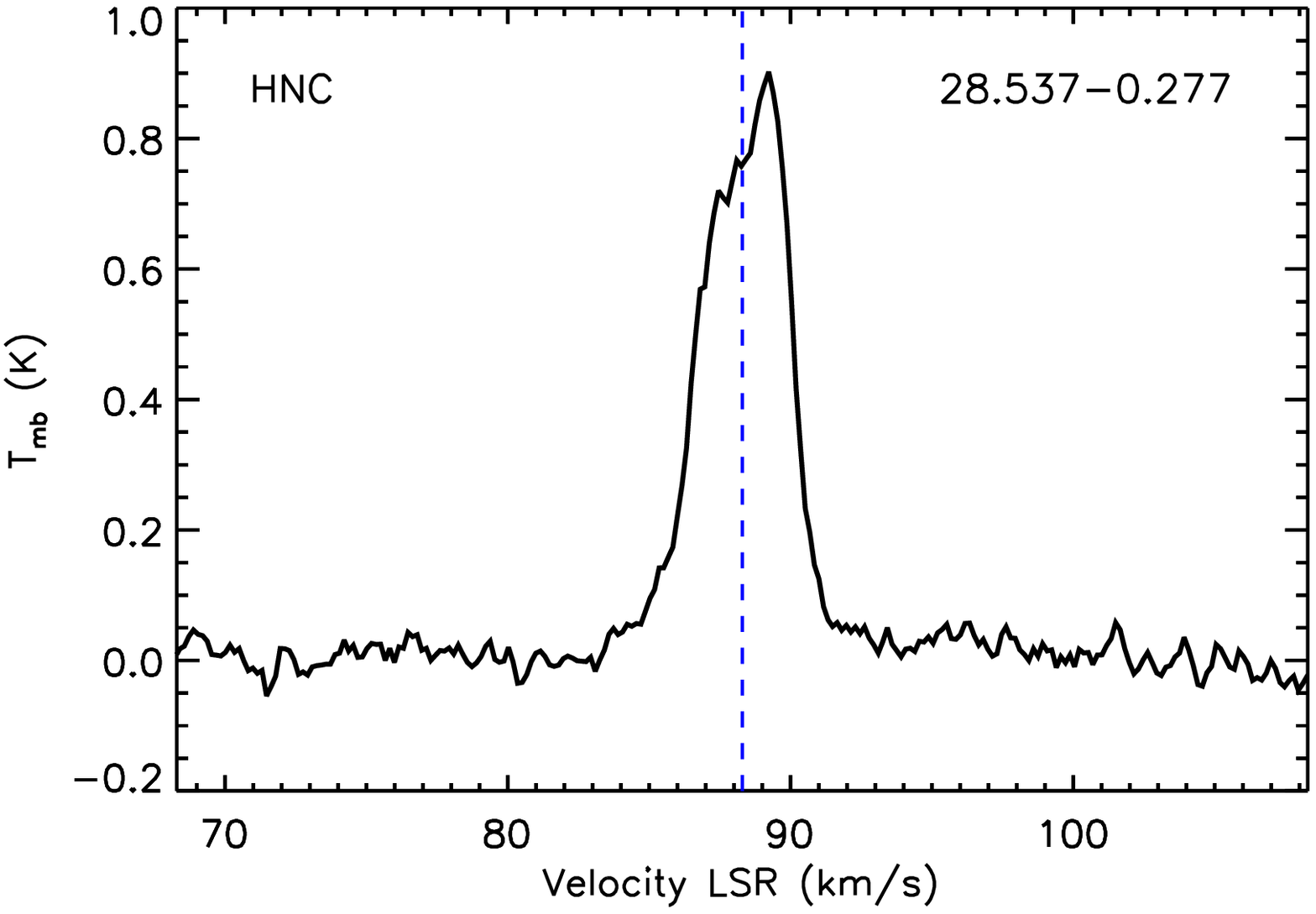} 
\includegraphics[width=8cm]{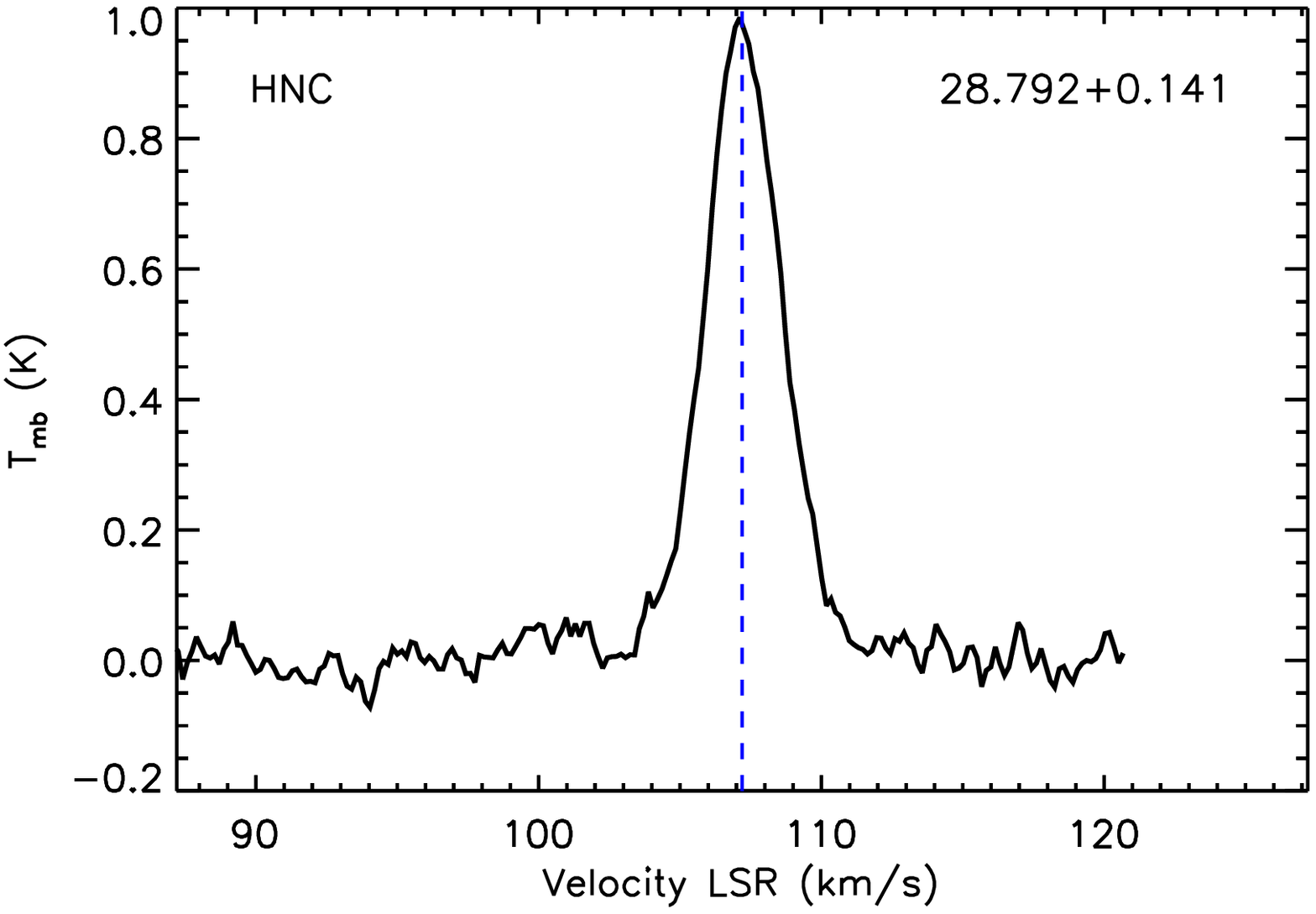} 
 \includegraphics[width=8cm]{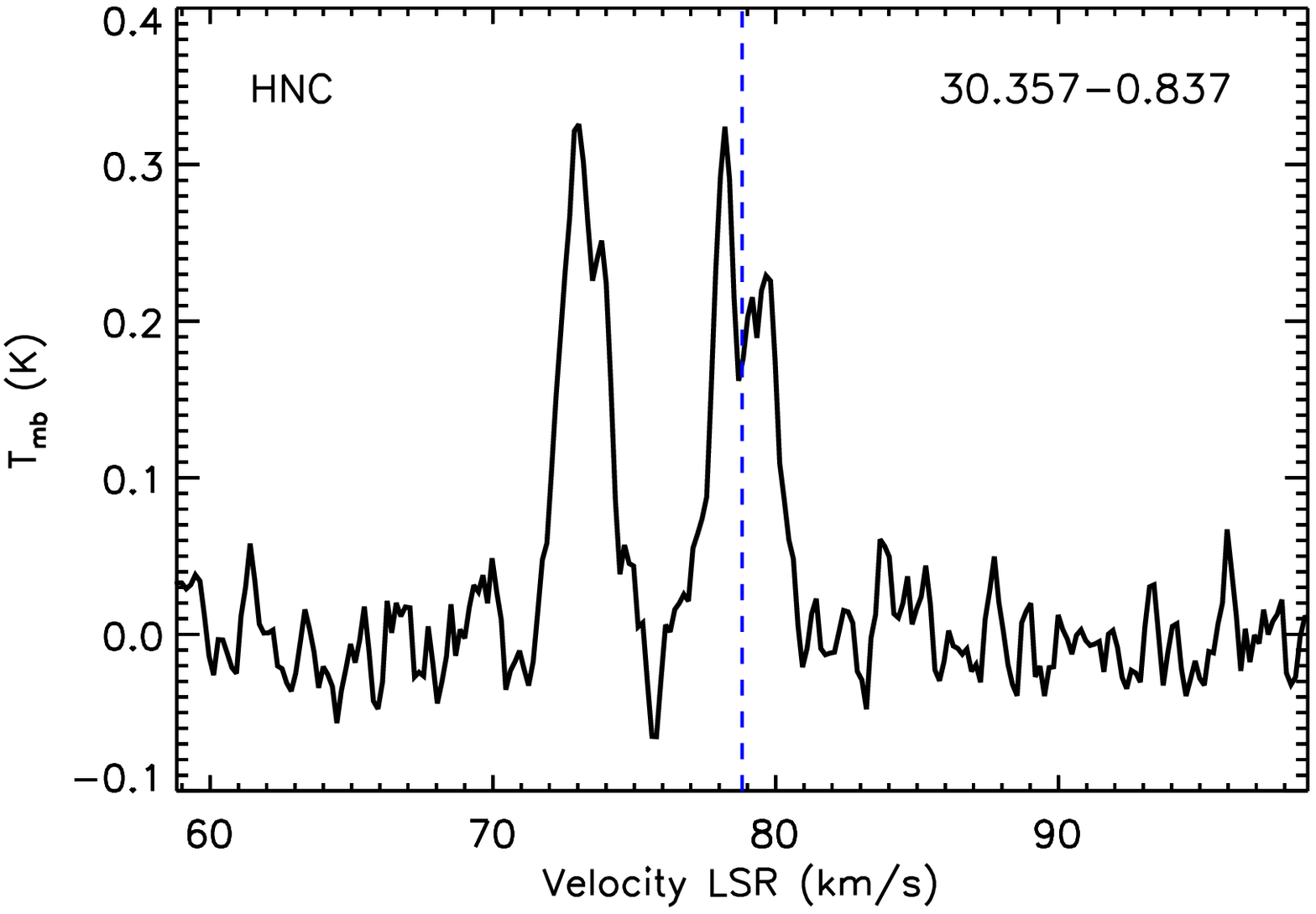} 
\includegraphics[width=8cm]{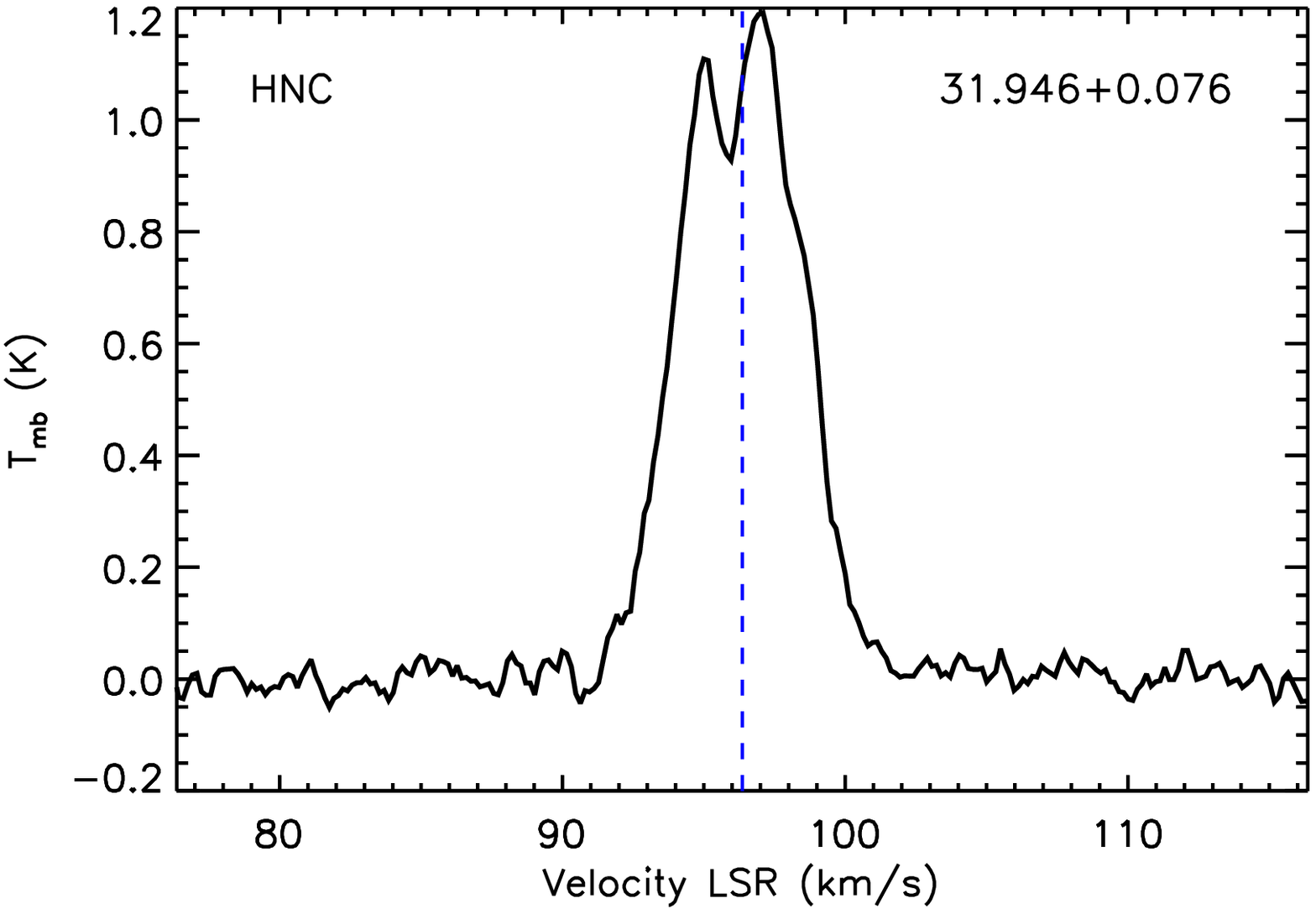} 
\includegraphics[width=8cm]{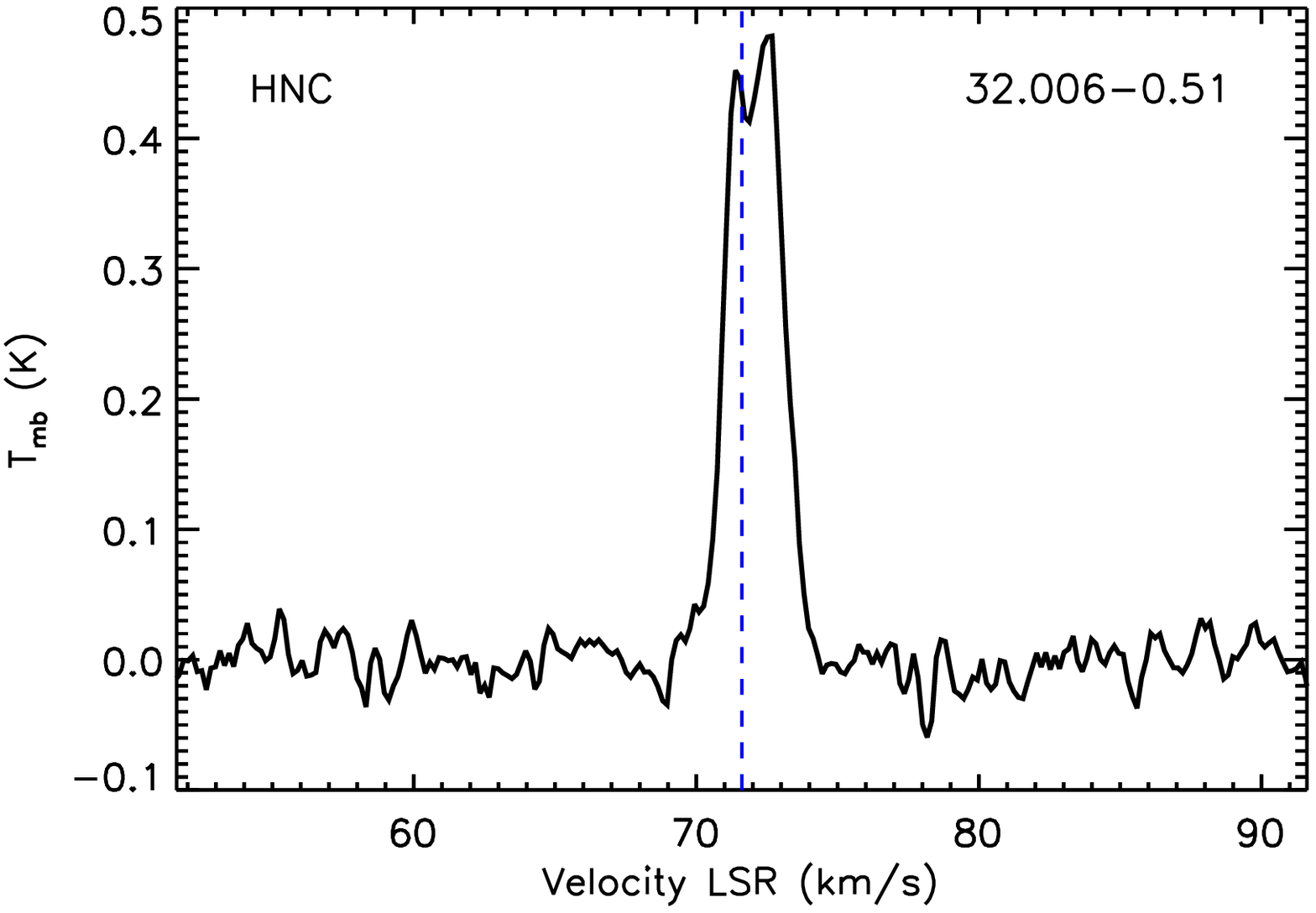} 
\includegraphics[width=8cm]{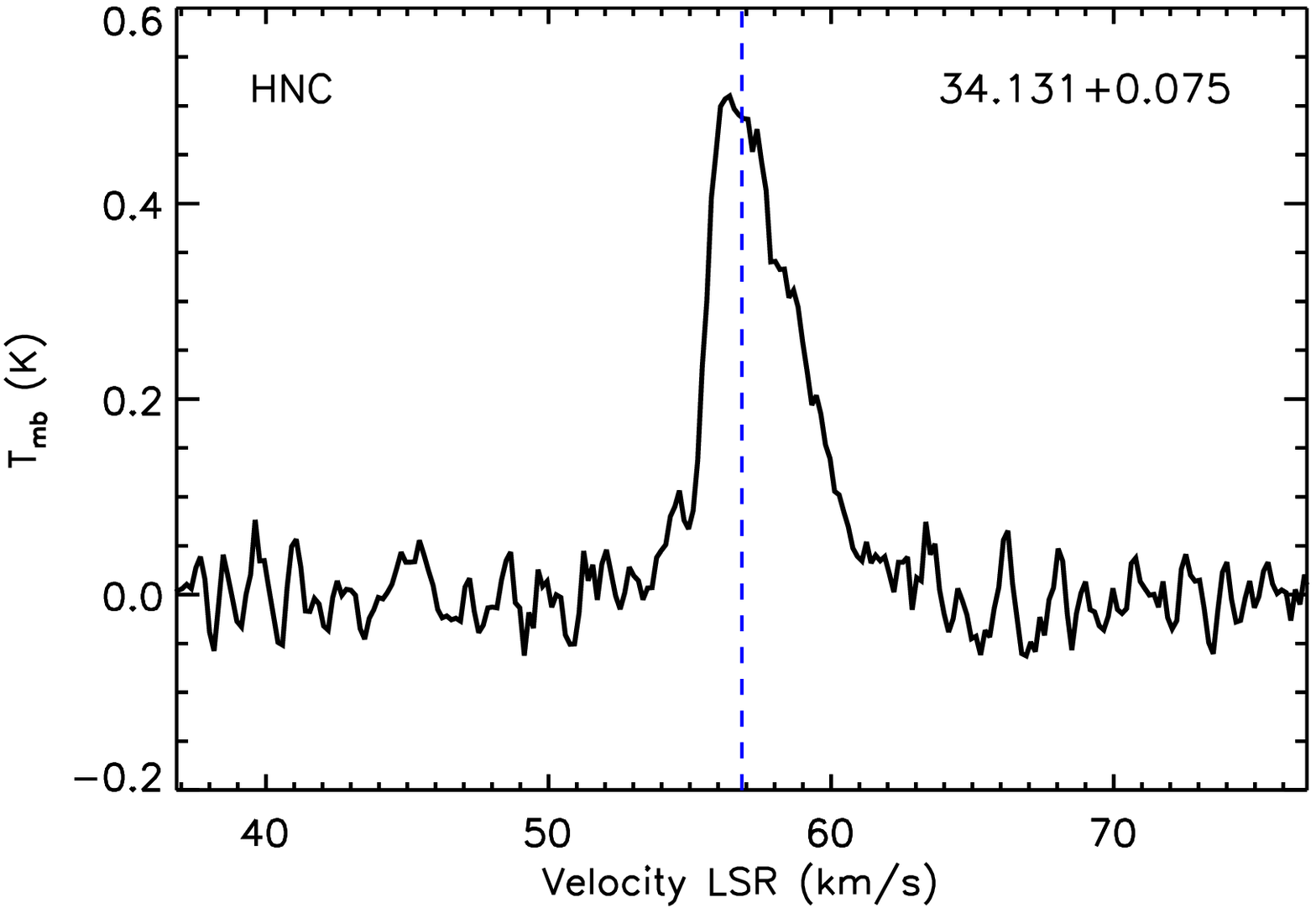} 
\caption{HNC ($1-0$) spectra continues}
 \end{figure*}

\begin{figure*}
 \centering
\includegraphics[width=8cm]{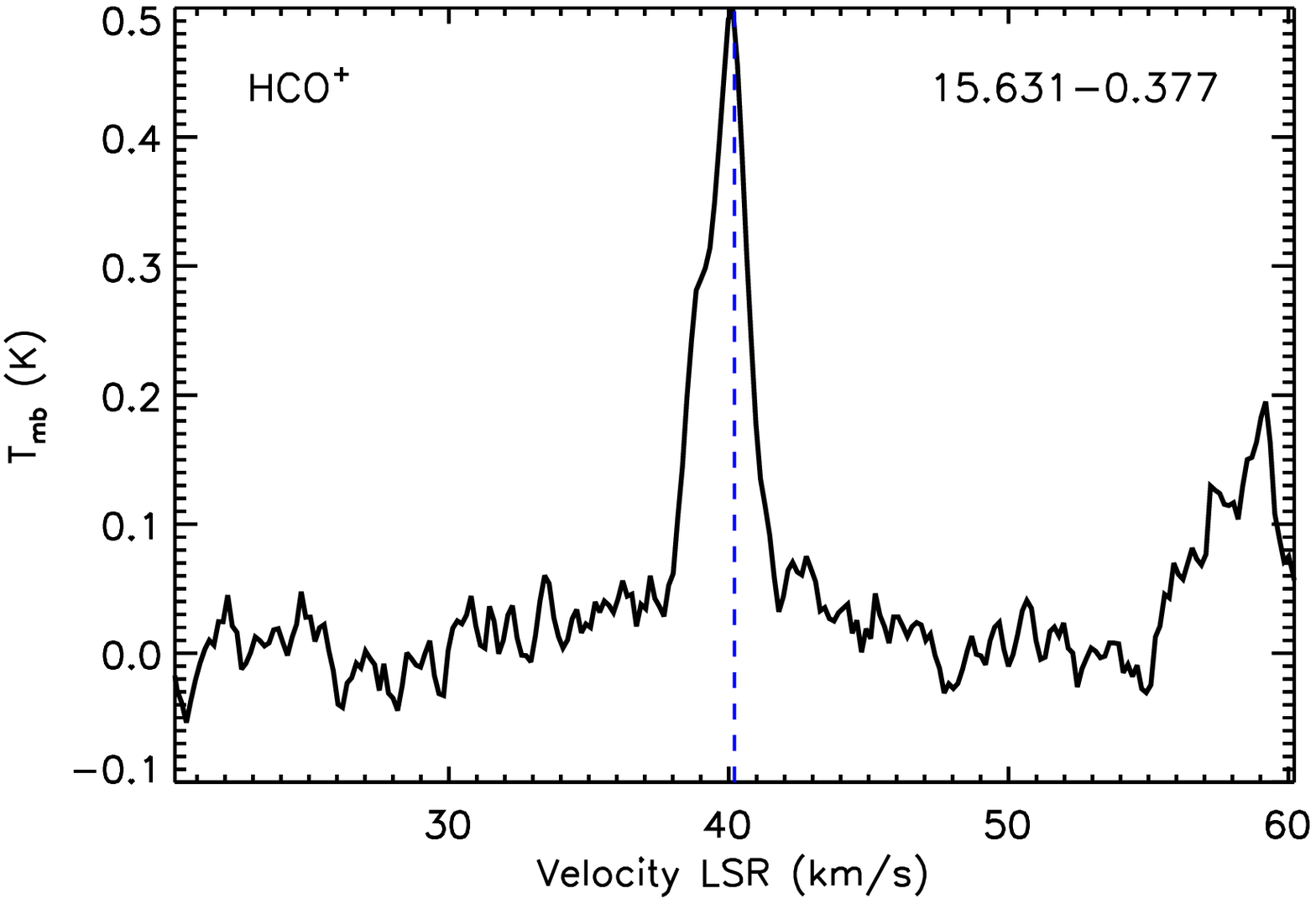} 
\includegraphics[width=8cm]{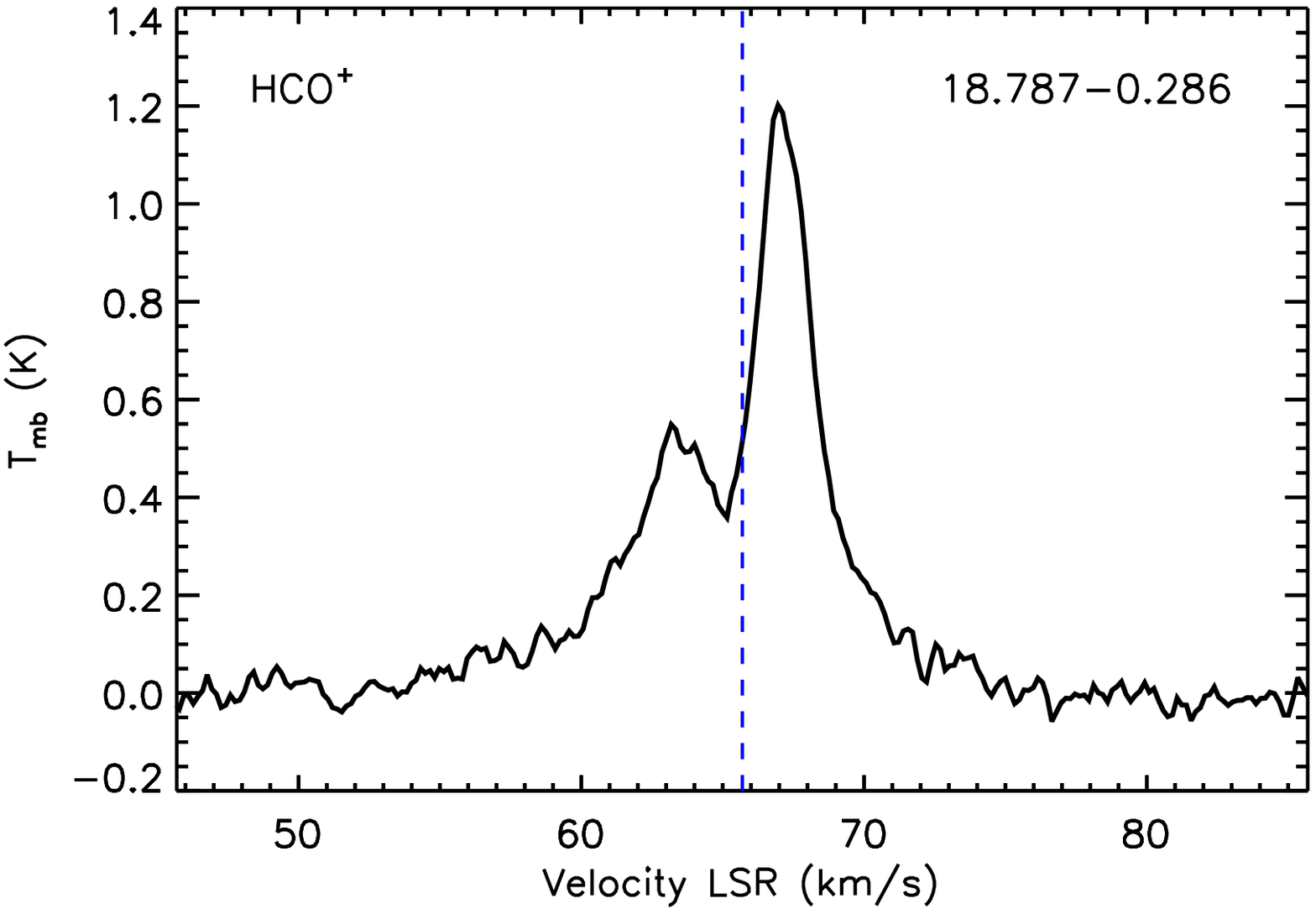} 
\includegraphics[width=8cm]{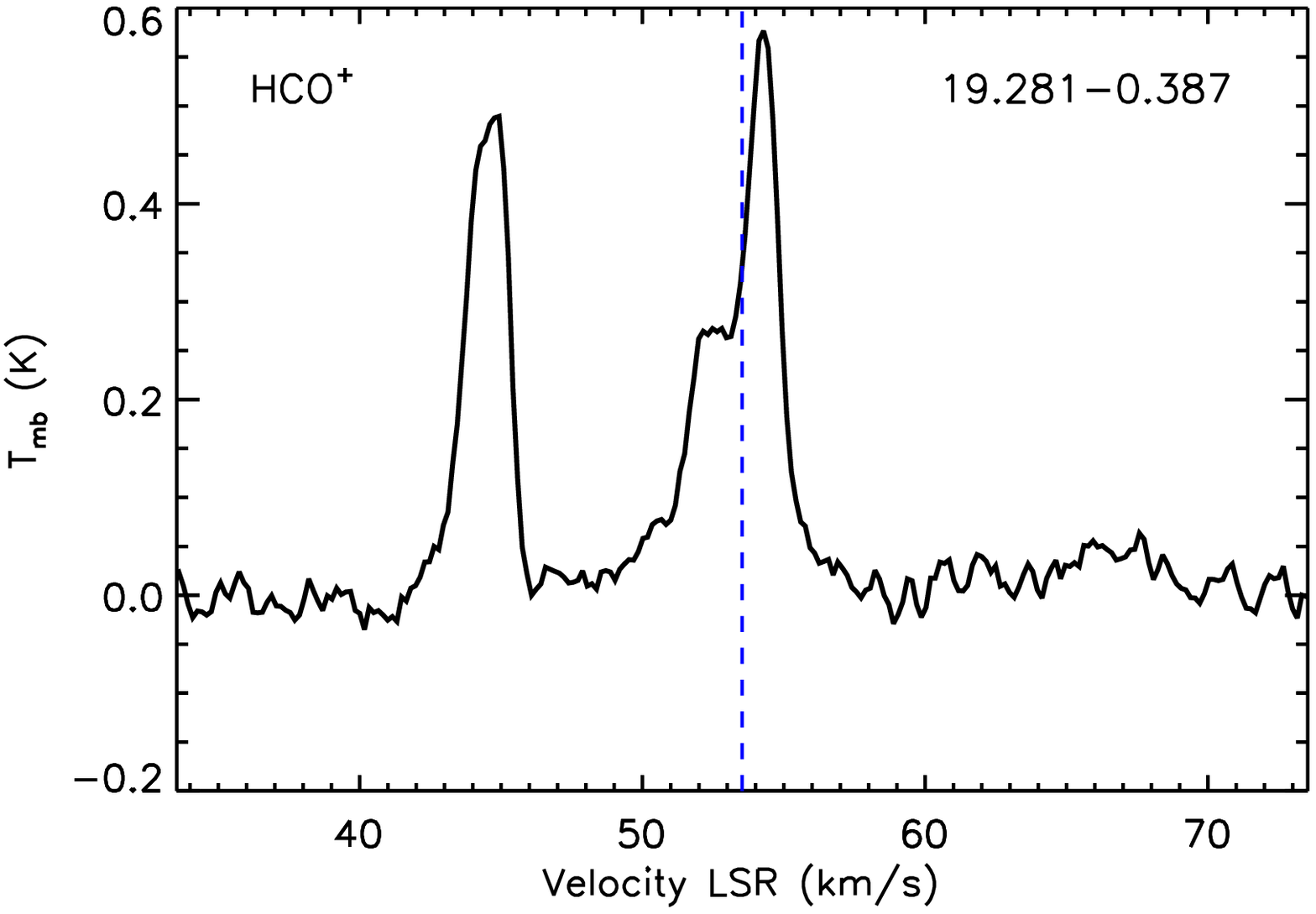} 
\includegraphics[width=8cm]{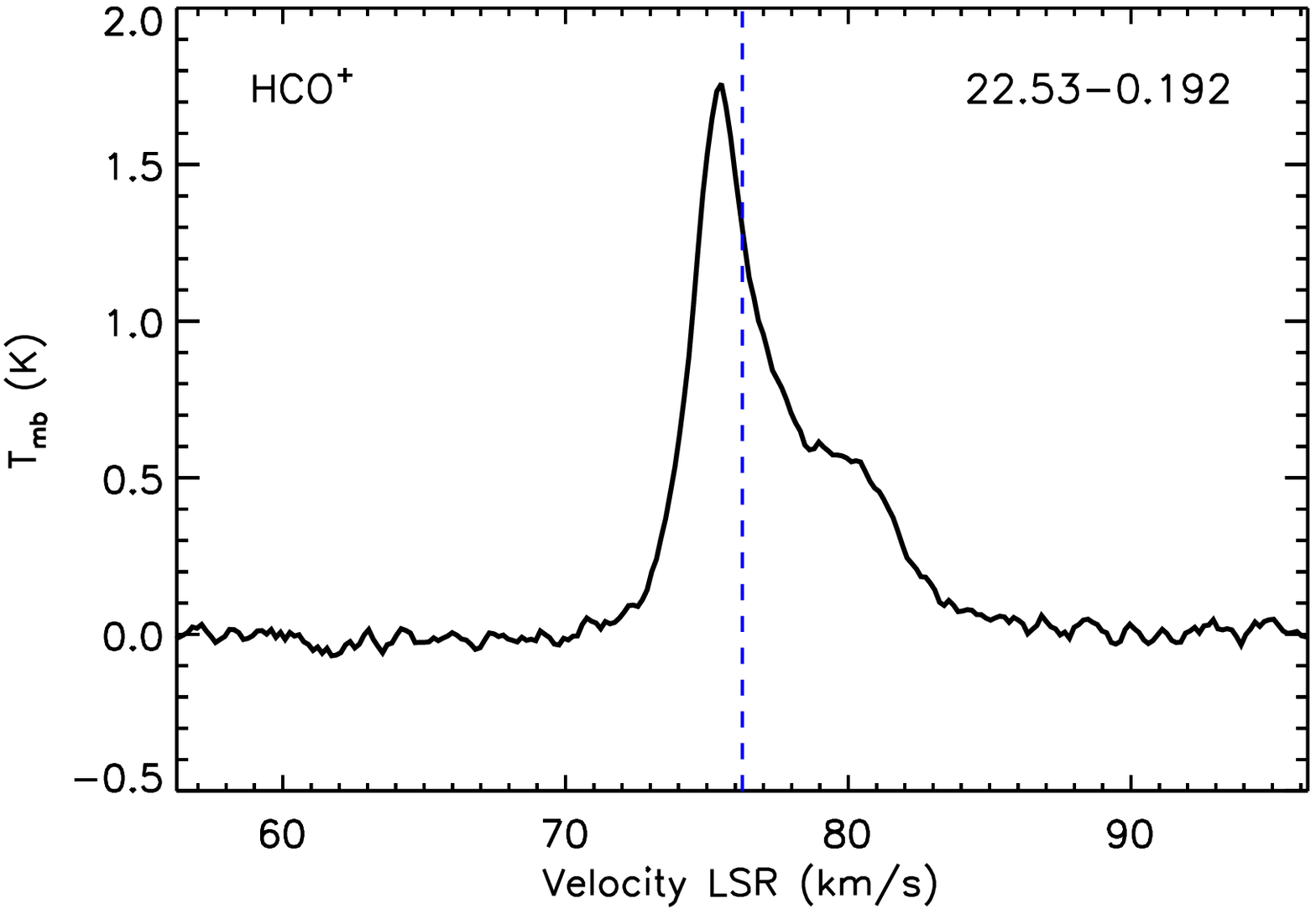} 
\includegraphics[width=8cm]{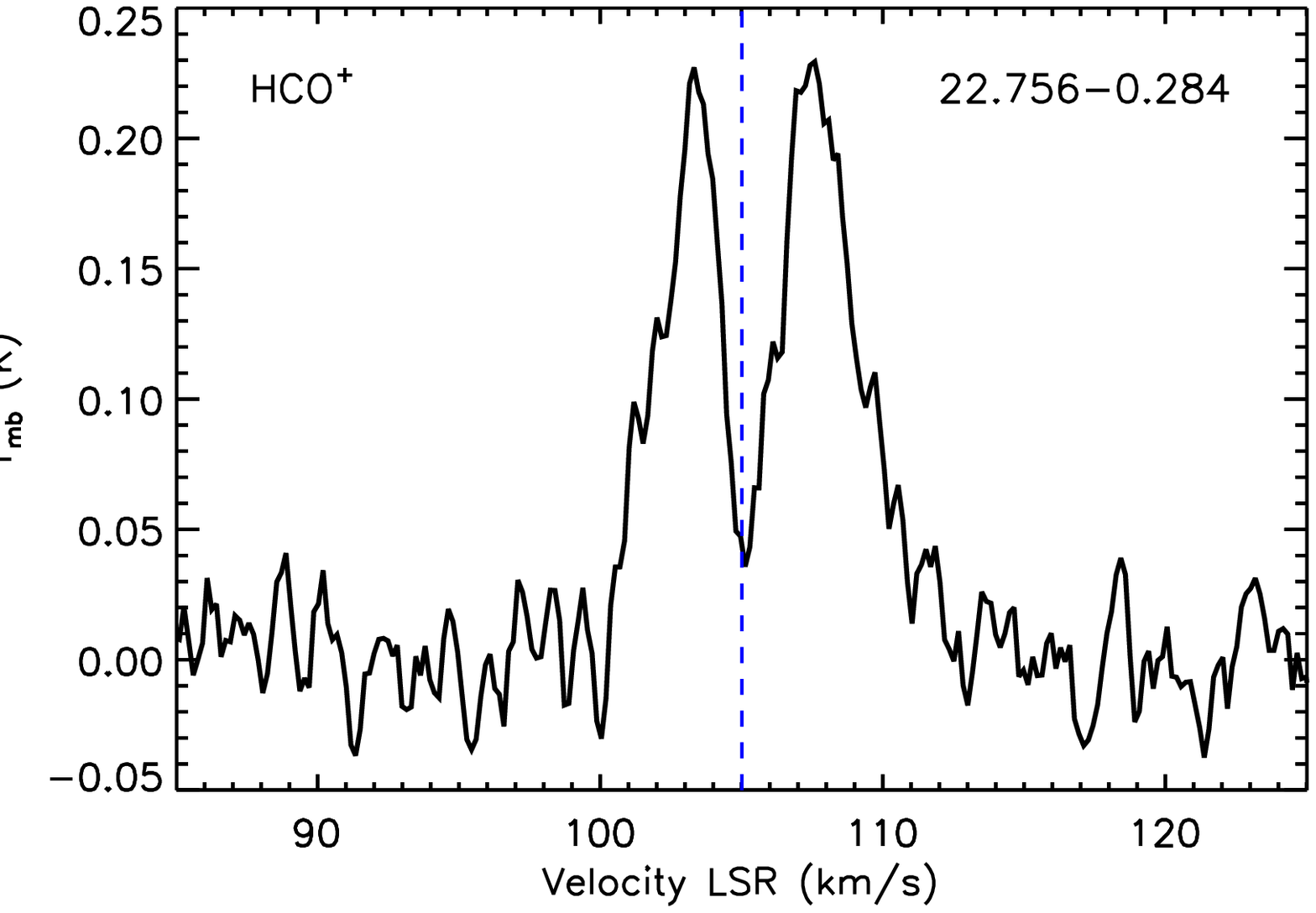} 
\includegraphics[width=8cm]{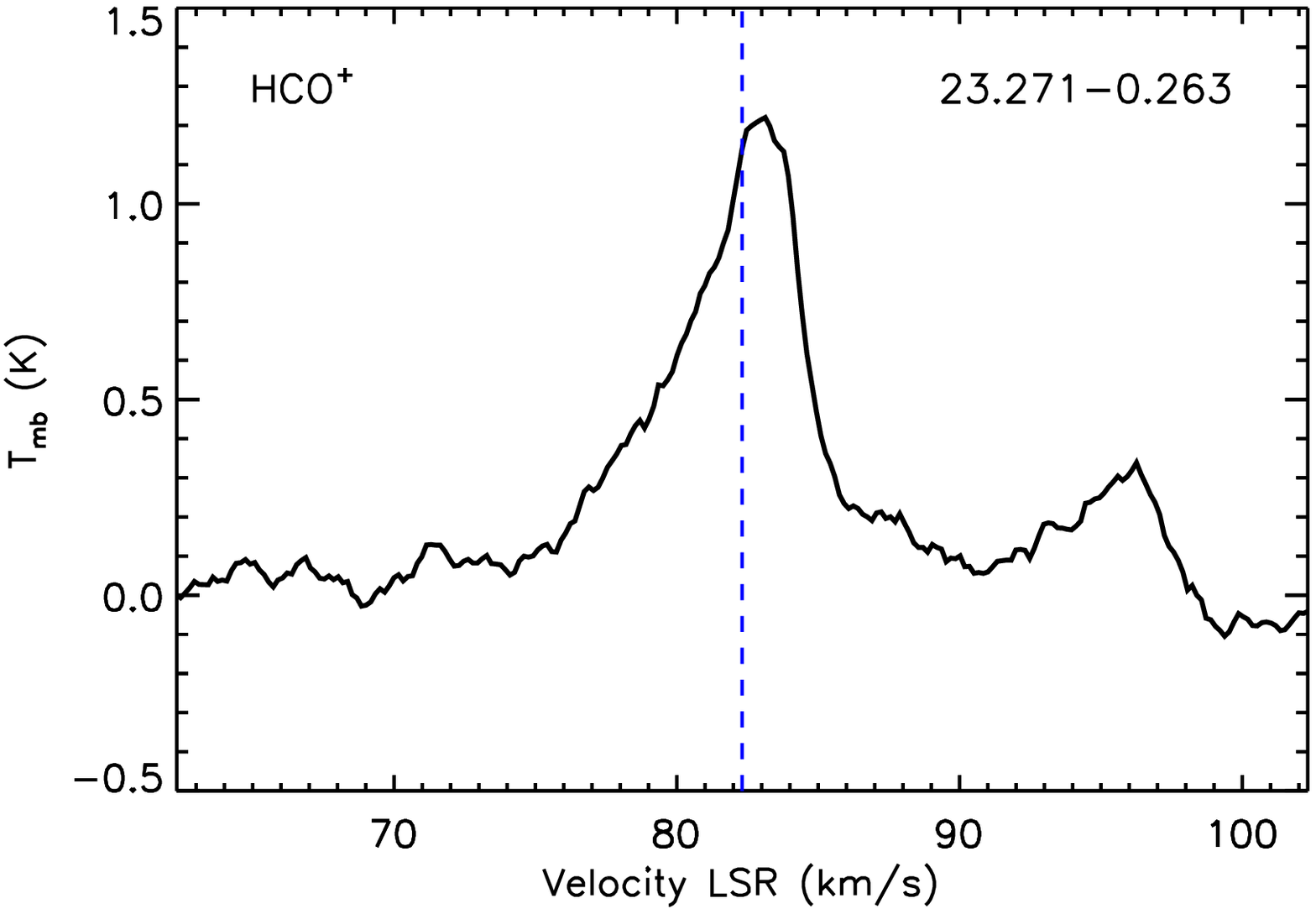} 
\includegraphics[width=8cm]{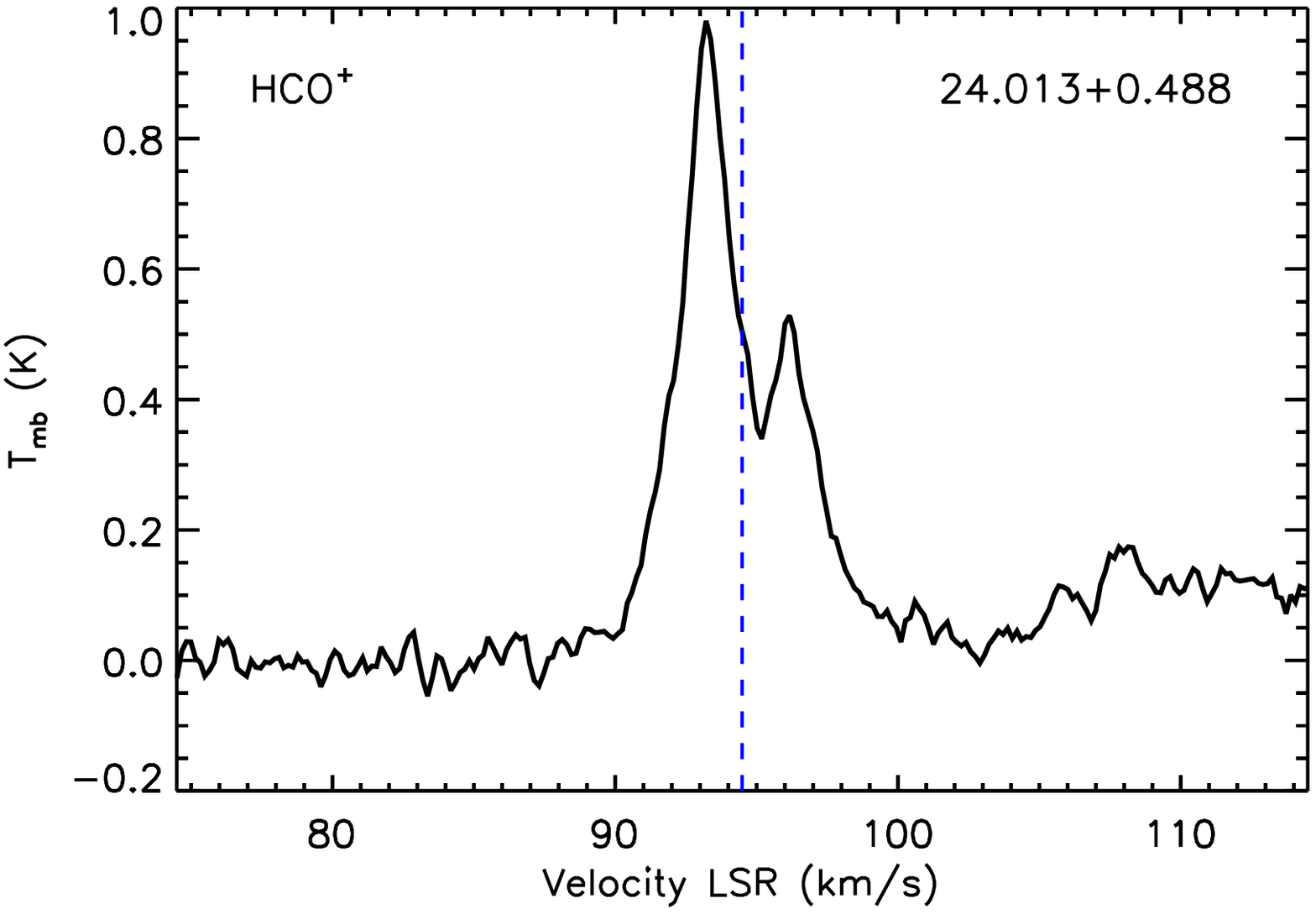} 
\includegraphics[width=8cm]{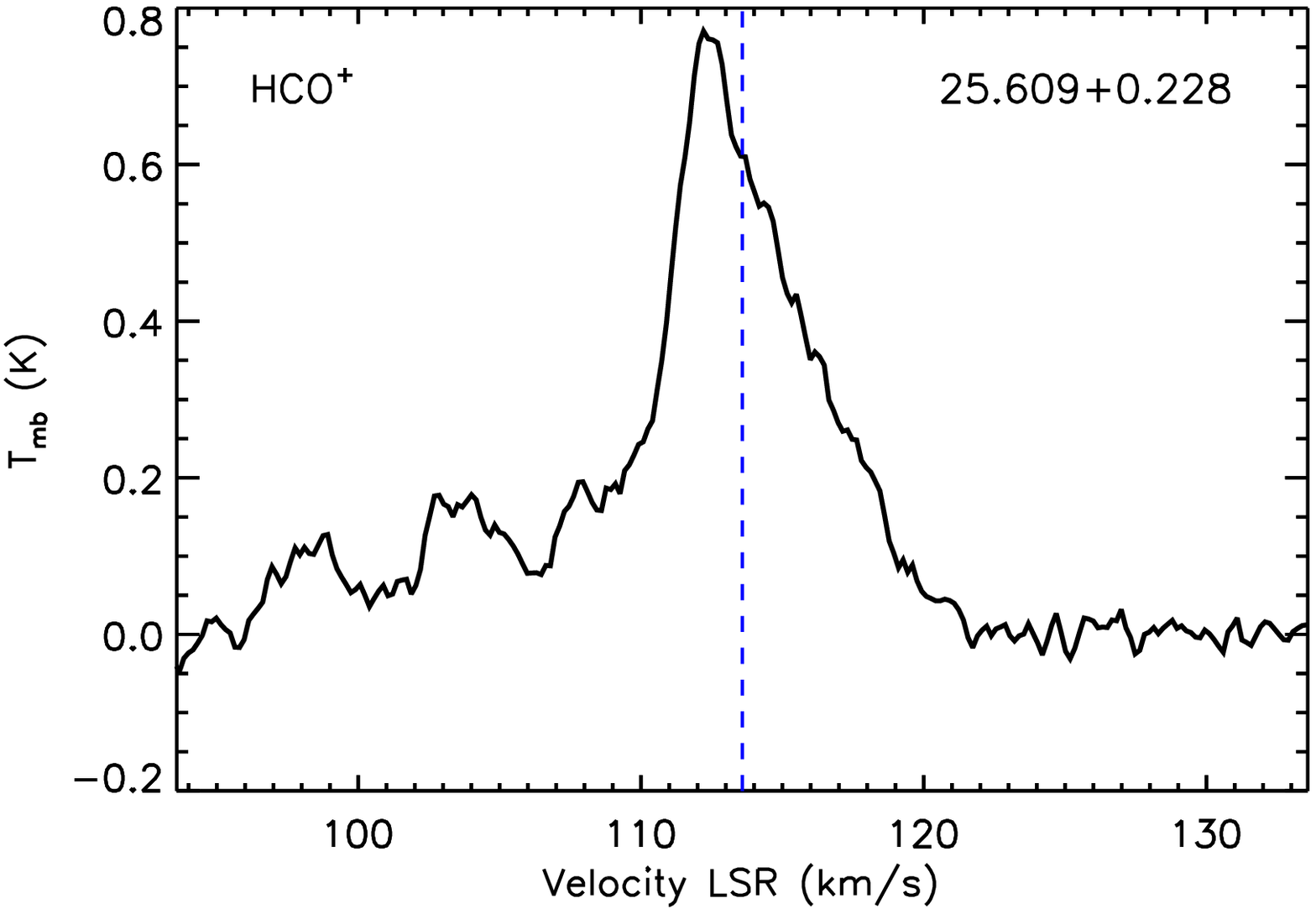} 
\caption{\hco\ ($1-0$) spectra}
 \end{figure*}

\begin{figure*}
 \centering
\includegraphics[width=8cm]{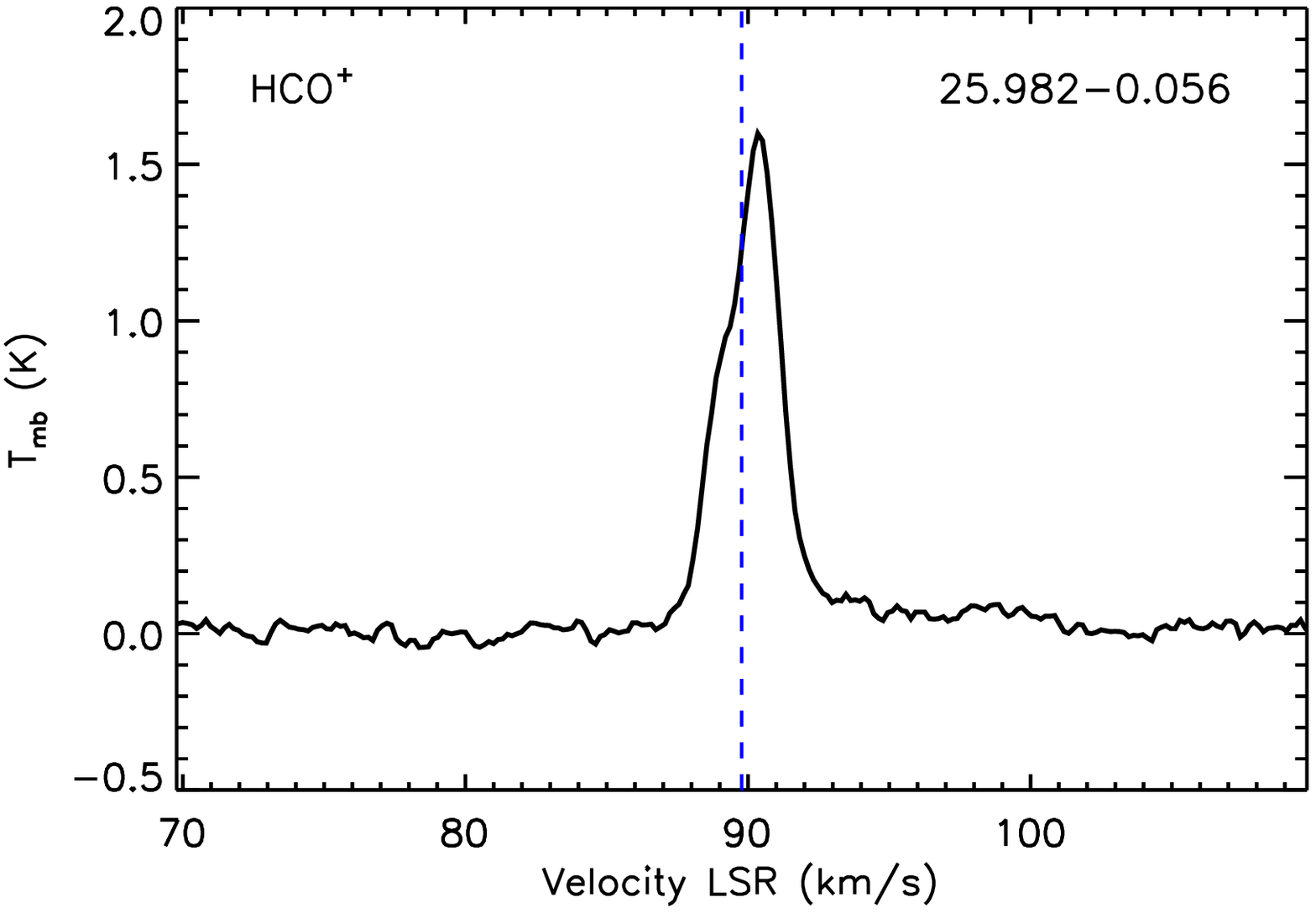} 
\includegraphics[width=8cm]{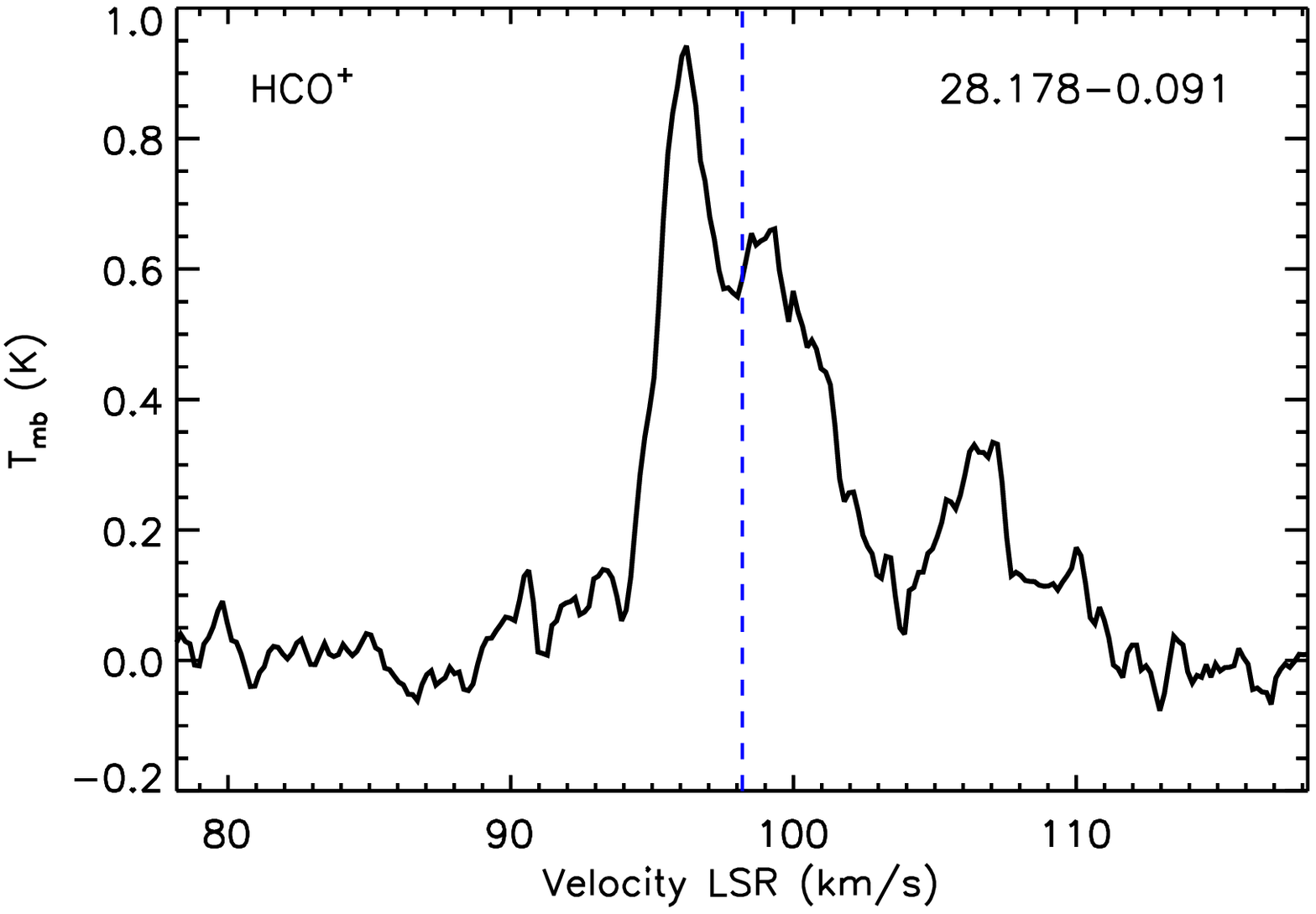} 
\includegraphics[width=8cm]{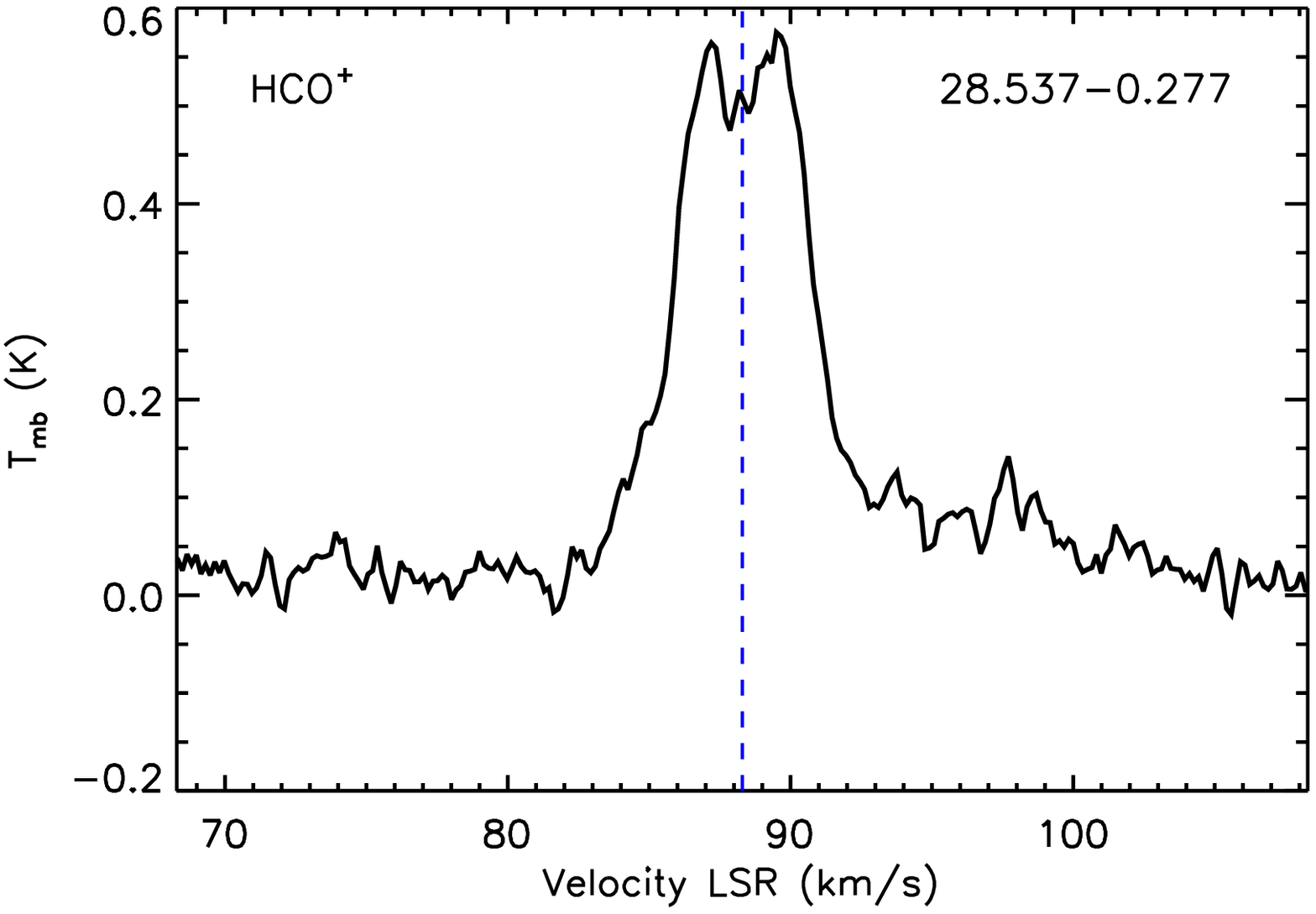} 
\includegraphics[width=8cm]{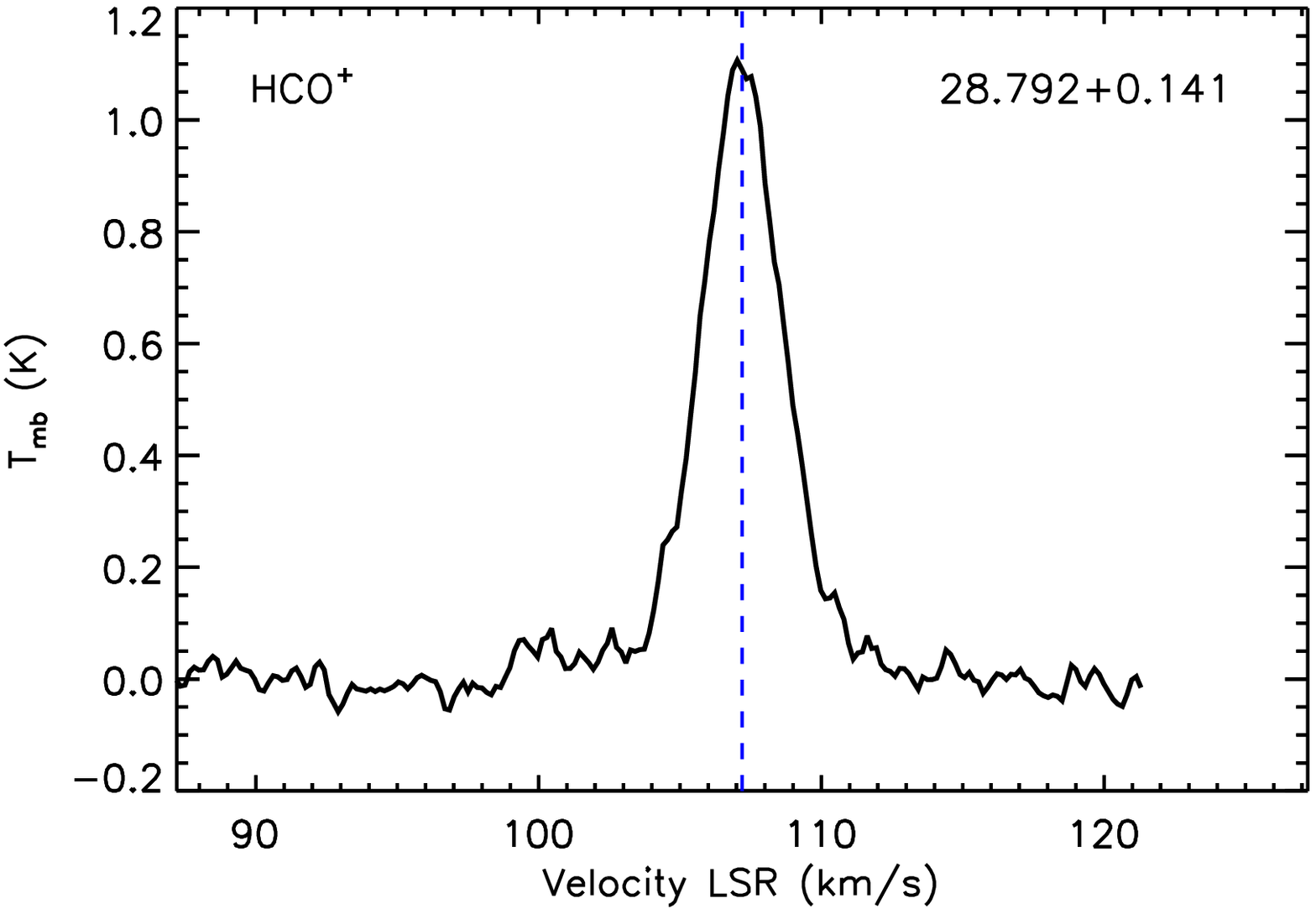} 
 \includegraphics[width=8cm]{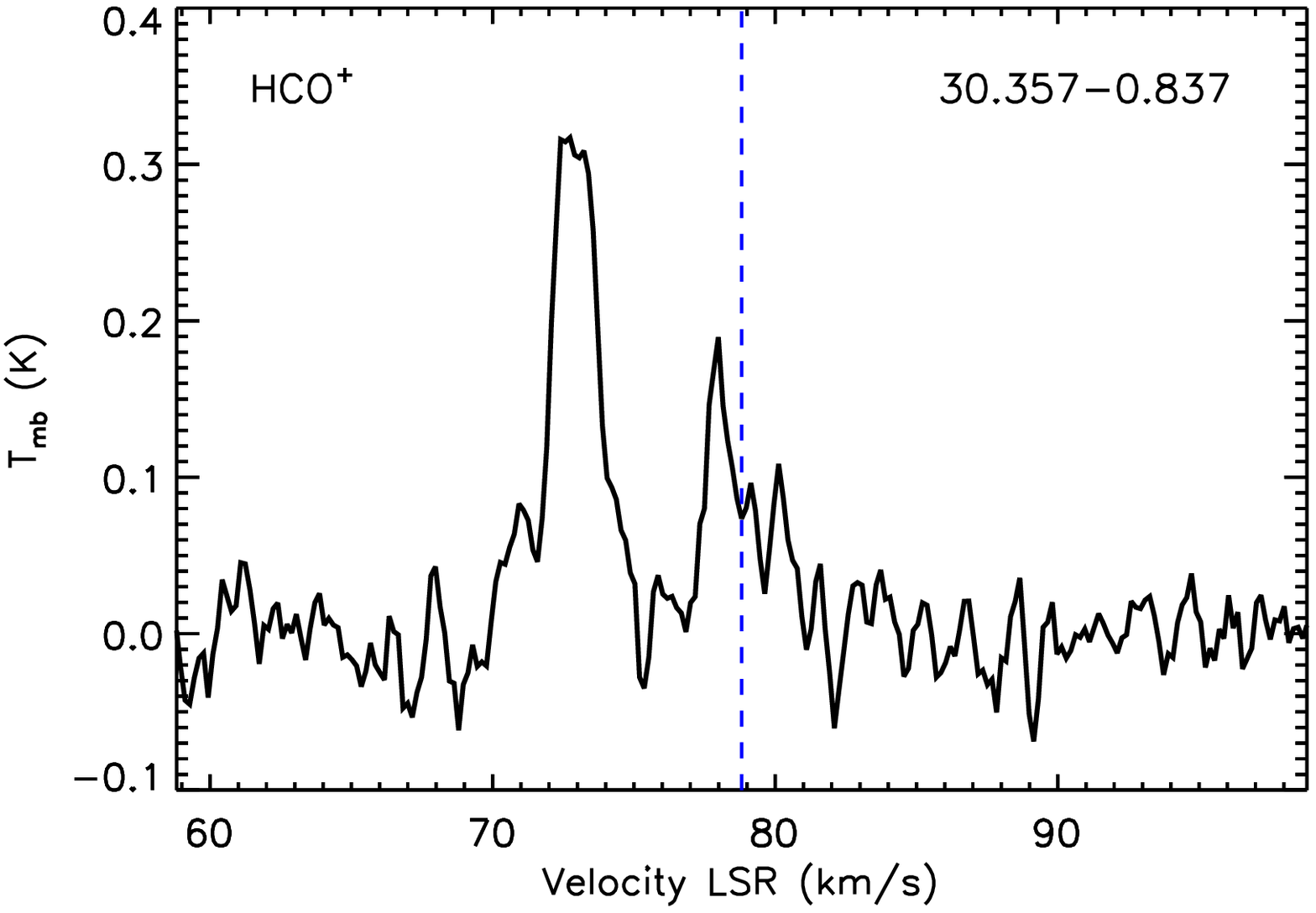} 
\includegraphics[width=8cm]{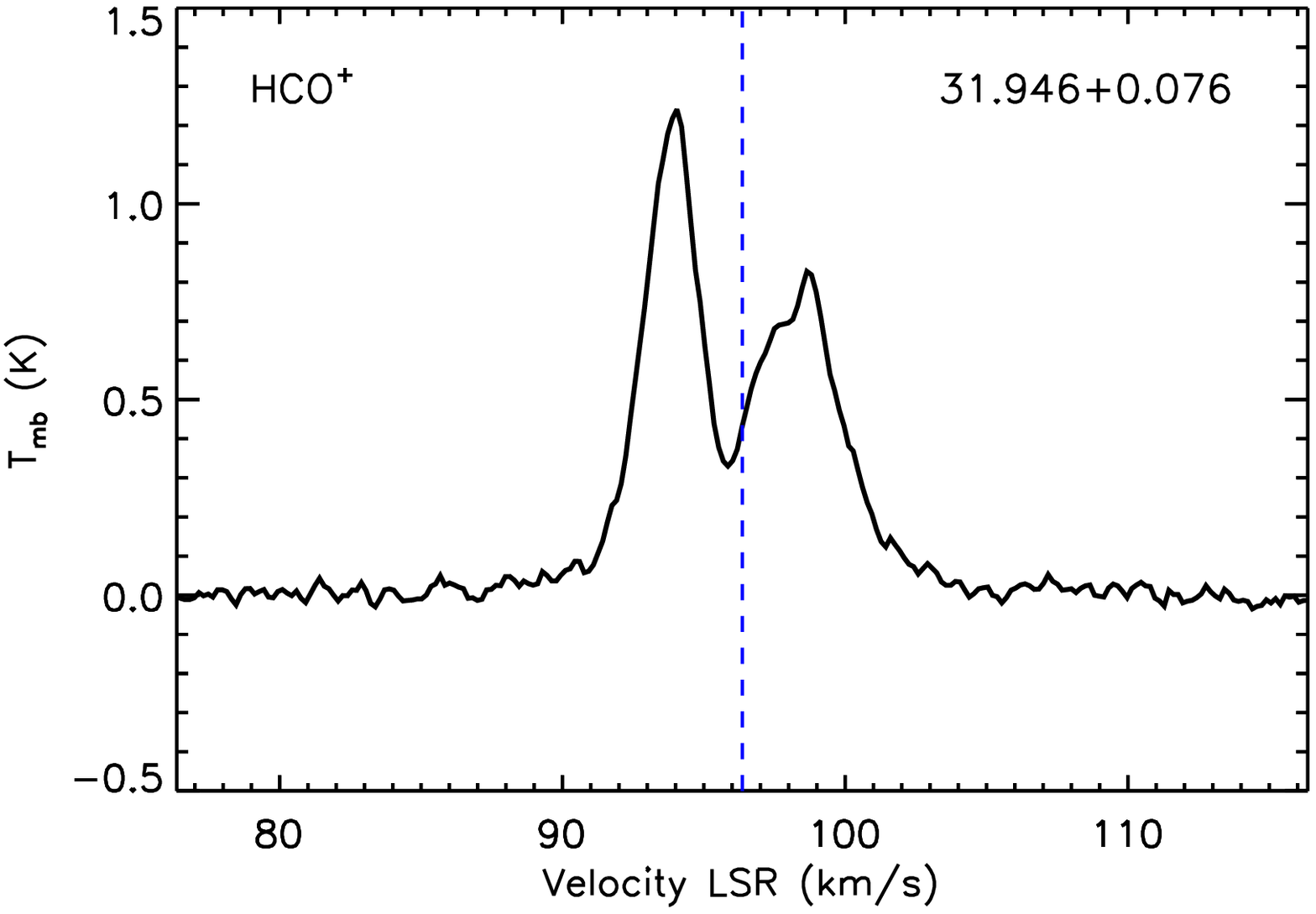} 
\includegraphics[width=8cm]{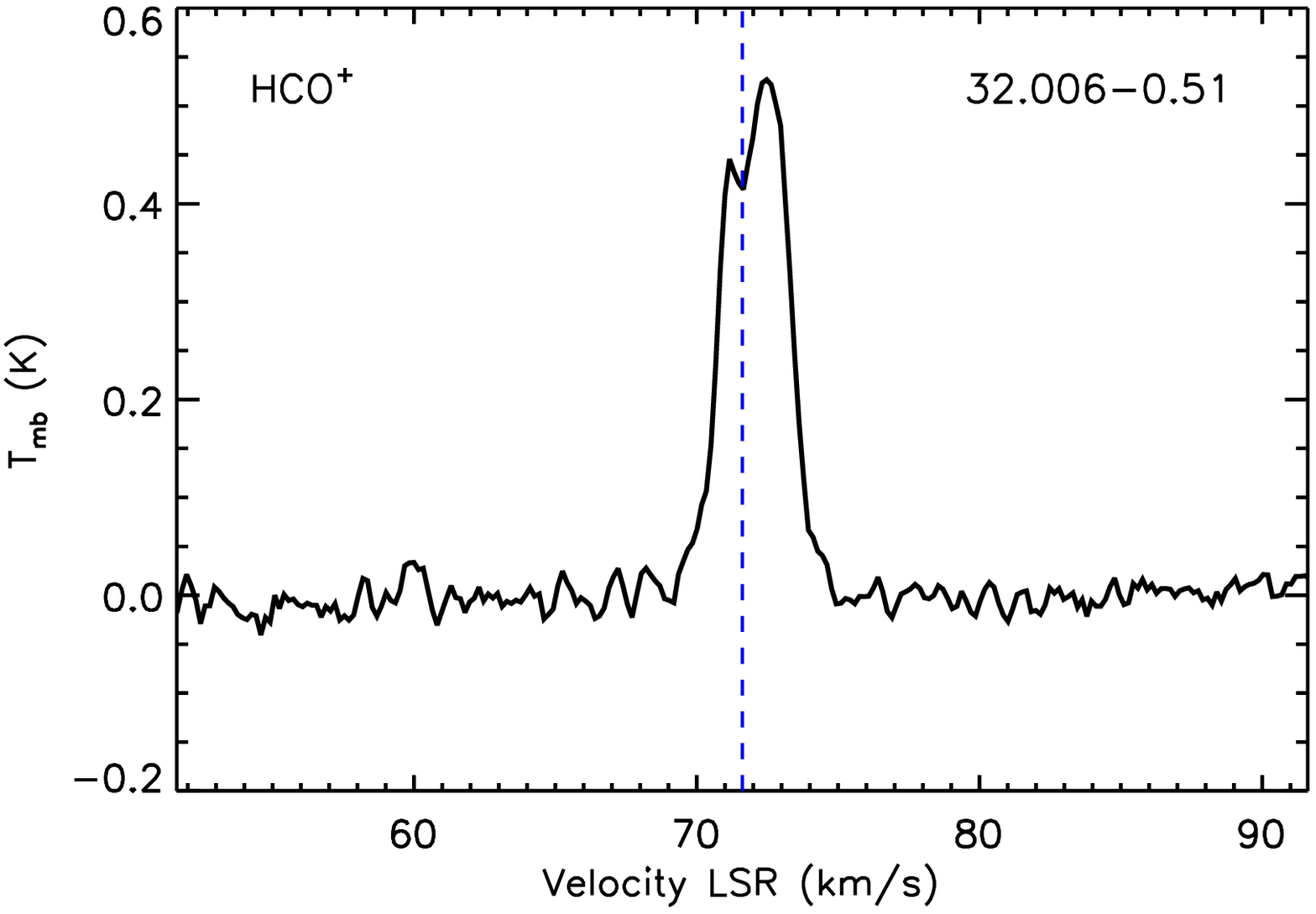} 
\includegraphics[width=8cm]{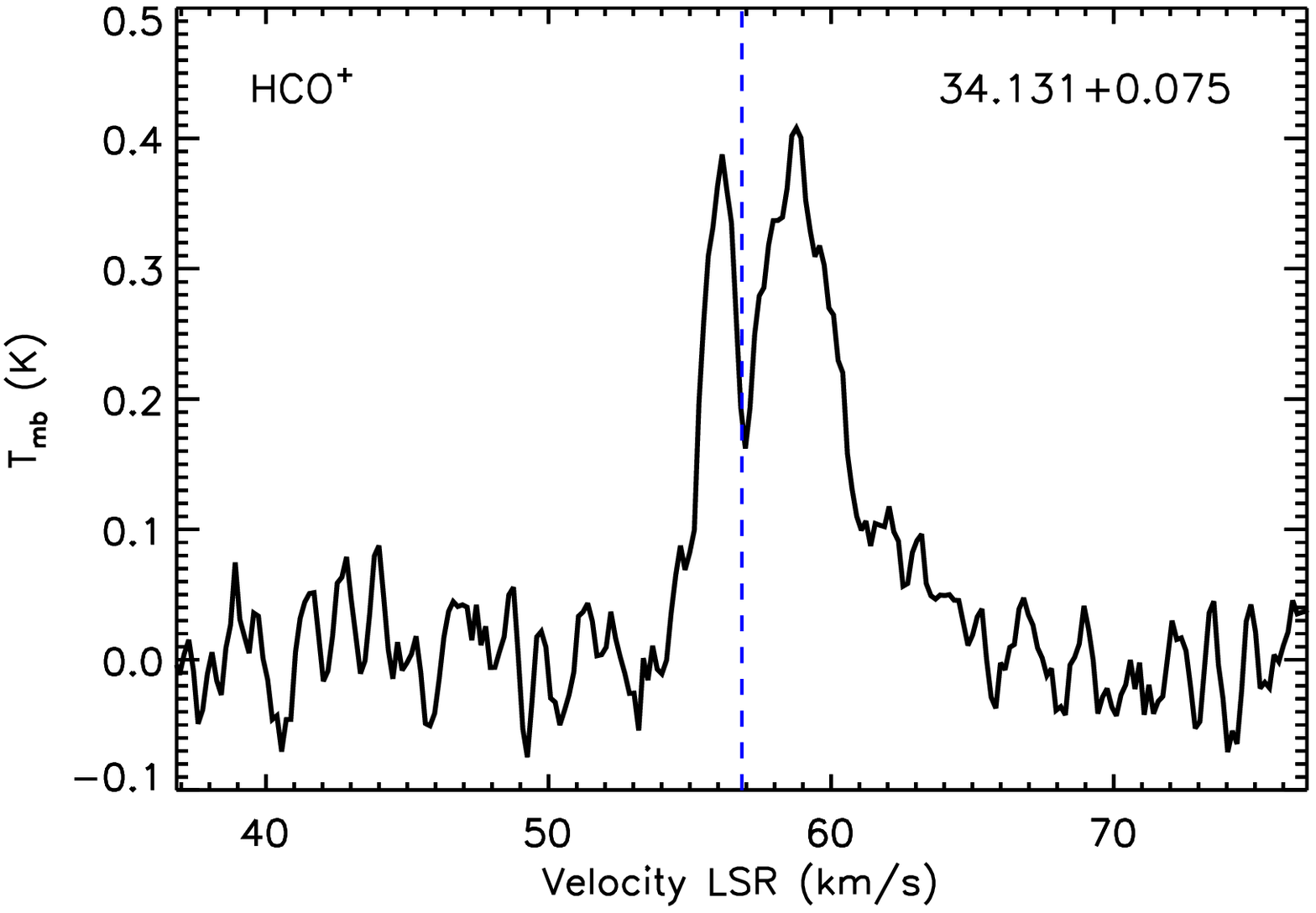} 
\caption{\hco\ ($1-0$) spectra continues}
 \end{figure*}

\end{document}